# MATHEMATICAL MODELING IN PHYSICS: DETERMINISTIC PROCESSES

<<in Info>Subject, needs change>>

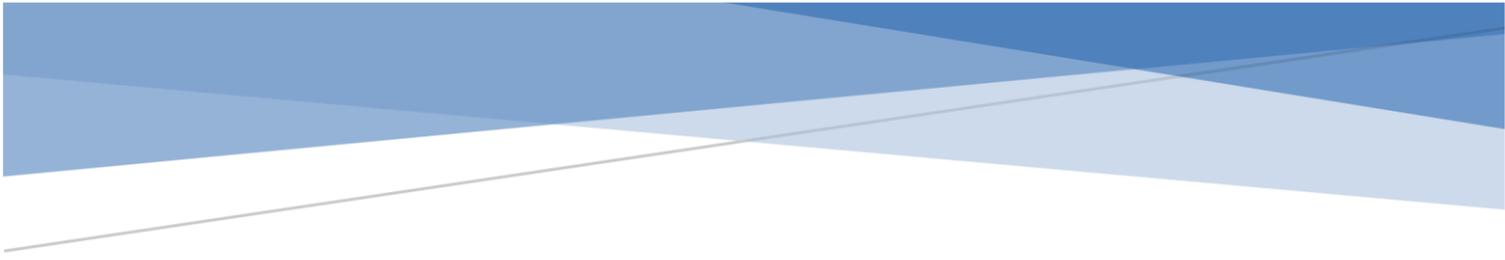

Sergej Pankratow

# Contents

## Summary

Basic principles of mathematical modeling are reviewed in this book, with the focus on physics and its practical applications, and examples of selected mathematical methods are presented. Most of the models have been imported from physics and applied for scientific and engineering problems. More specifically, the accent is placed on the theory of dynamical systems being used to produce evolutionary models. Classes of different models are presented and the relationship of mathematical modeling to other disciplines is outlined. The aim of the paper is to bridge together mathematical methods and basic ideas proved to be useful in science and engineering.

*The sciences do not want to explain, they hardly even try to interpret, they mainly make models.*

John von Neumann

## 1. Introduction

These notes can be perceived as piecing together well-known things. However, the main purpose of the present book is to find unexpected and, when possible, nontrivial aspects in apparently familiar descriptions of natural phenomena. Such unexpected features can be conveniently found if natural (or even socio-economic) phenomena are represented in the form of mathematical models – simplified entities that are more or less closed as compared to scientific theories and, thus, can be explored almost exhaustively. The model nature of many concepts becomes evident as the view changes. Thus, the familiar concepts of pressure and temperature make sense only when there is a great – macroscopic – number of degrees of freedom, the notion of a surface disappears in the fully microscopic treatment, and the notion of fluid is meaningless when one scrutinizes separate particles.

Another purpose of the book is to make a framework for getting acquainted with appropriate conceptual and mathematical structures without being lost in the details. The description of such structures is in no way exhaustive and may look hopelessly superficial to a mathematically oriented reader. Sophisticated mathematical techniques are very important, but they will not help much unless one learns how to apply them to real-life problems. We also wanted to emphasize that certain mathematical models can be constructed only for computational convenience and thus have very little to do with real-life phenomena, in particular, with physical processes.

The focus on mathematical modeling may seem a bit too methodological and sometimes even irrelevant, especially when one speaks about specific problems, yet using models brings one significant advantage: investigating models developed in one branch of science can describe the whole class of similar behavior patterns encountered in other branches that are usually considered entirely different. Well-known examples of such knowledge transfer are nonlinear dynamics and economics, evolution equations and population dynamics, fluid motion and collective behavior, critical sets of differentiable maps and social changes, etc. One of the remarkable observations in the contemporary world is the spread of mathematical models developed in physics over other fields such as ecology, economics, study of social phenomena and so on. The leap from one field of knowledge to another can be difficult unless common models are explored.

Accordingly, the book begins with the review of general properties of mathematical models and how to build such models. Despite numerous impressive results achieved by mathematical modeling, especially when using powerful computers, there is no universal viewpoint on how to understand the very notion of modeling. It would be difficult to find a generally adopted definition of mathematical modeling that would clearly delimit the field, and rather long descriptive definitions that one can find in various sources may range from semi-philosophical reflections on how to comprehend mathematical models to enumerations of solution methods to equations describing a modeled system. One can loosely define mathematical modeling as an idealized description of reality constrained by mathematical concepts. To conceptualize actually means building models – more precisely, models of external objects in the human mind, and mathematics is used to make these models as exact as possible. The human brain is just an instrument for building interior models of the exterior world that has been provided by evolution.

Probably accurate definitions (or quasi-definitions) of what a mathematical model really is would be useless anyway, but it would be desirable to have a unified consideration, even on the intuitive level of understanding. The main thing in mathematical modeling is neglecting unimportant details: any model aims at building a picture of the modeled phenomenon that would reproduce its most essential features both qualitatively and quantitatively (such a picture may be of course distorted). This purpose of modeling can only be thought of as reached when the model's outcomes, e.g., in the form of analytical formulas, are sufficiently simple. If this analytical form is too complex, it can be regarded as just a primary stage.

According to John von Neumann, an outstanding mathematician and physicist, "by a model is meant a mathematical construct which, with the addition of certain verbal interpretations describes observed phenomena." In this book, mathematical modeling will be understood in a relatively narrow sense namely as a set of techniques consistently applied to obtain the equations that can describe the system to be modeled. Besides, a corresponding number of methods designed to transform such equations into the form convenient for qualitative analysis and computer implementation is also regarded as a part of mathematical and computer modeling. In this way, compartmental fragments of reality are discussed in mathematical terms, predominantly using the language of equations and with more or less clear-cut assumptions.

In the author's opinion, the principal thing in building mathematical models is to understand what the main point is, so to say an irreducible core of the model versus the "shell" that can be varied. In this sense, it is difficult to naively brand mathematical models as "true" or "false". In general, any mental activity such as thinking is the ability to construct models, though not necessarily formalized in mathematical terms.

An important function of mathematical modeling is that it smooths over the distinction between what can be found in nature and what has been made by humans i.e., provided by the culture. Although the author takes into account the fact that no dynamical model is a perfect imitation of reality, the equations to be discussed in the book are predominantly ordinary differential equations (ODEs) in the form of dynamical systems, which only reflects the intention to treat evolutionary models. Notice that in such models we used to treat dynamical laws as timeless and acting on time-dependent (evolving) states, which can only be an approximation. Modeling with the help of partial differential equations will be touched upon only briefly since it is a vast subject bordering on field theories, which would require much more place for its exposition than the present book's framework would allow.

One of the dominant concepts exploited throughout the book is the notion of local equilibrium. Mathematical modeling of complex systems far from equilibrium is mainly reduced to irreversible nonlinear equations in differential and discretized form which are then treated by well-developed analytical methods, numerical procedures, and computer algebra manipulations. This approach is frequently encountered in various branches of science and engineering, where it is useful for modeling a lot of practically important situations. Note that in mathematical modeling (and to some extent in computer modeling, too) the explored system and its attributes are represented by mathematical variables, activities are represented by functions and relationships by equations. Two basic examples are discussed in this book in some detail in order to understand the modeling concepts: the logistic model in connection with nonlinear problems and the harmonic oscillator in the realm of linear models.

One can illustrate the concept of equilibrium with everyday observations. For example, a person may be healthy or ill, with diseases being interpreted as deviations from the state of health in a person. Such deviations can be transient when the body quickly (e.g., in the time scale of a few hours) returns to the stable mode or long-term (over several days) and runaway states i.e., when a progressive disease drives the organism away from equilibrium developing possible life-threatening physiological anomalies. The state of health corresponds to the domain of equilibrium in a state space which is a manifold formed by the admissible values of physiological parameters – recall that a manifold is, roughly speaking, a set of points that can be labeled by real numbers treated as coordinates. Using the currently fashionable language of synergetics and dynamical systems theory, one can say that the state of health is an attractor for the human body. Notice that local equilibrium is not always achieved: there exist systems that cannot reach equilibrium, for instance, Earth's atmosphere, climate, human population, probably biosphere in general. Most natural processes are non-equilibrium, and their direction is determined by entropy growth (in the simple case, by the second law of thermodynamics). In this sense, one can consider local equilibrium a happy state.

Mathematical modeling can be performed in a variety of ways, with different aspects of the situation being emphasized and different levels of difficulty standing out. This is a typical case in mathematical modeling which implies that the situation to be modeled is not necessarily uniquely described, in particular in mathematical terms. The standard metaphor portraying this situation is the tale about several blind people trying to describe an elephant. In other words, a mathematical model describes (quantitatively or qualitatively) a limited class of phenomena and does not claim to be exhaustive. By tweaking free phenomenological parameters, mathematical and computer models are adjusted to the observed reality in science and engineering. In some cases, the number of phenomenological parameters may reach several dozen (e.g., in climate models or in the famous Standard Model of high energy physics). A historical example of parameter adjustment was given by Ptolemy's geocentric astronomy, in which the epicycles of his planetary model could be fine-tuned to produce in some cases strikingly accurate predictions. In any case, Ptolemy's geocentrism dominated over rival astronomical systems up to the 16th-17th century.

Here, one touches upon the crucial issue: what is the relationship of a model to reality? Some positivistically oriented people assert that this question is meaningless, and one can only ask whether the model predictions are in accordance with specific observations and what are the discrepancies (errors). Accurate prediction is in general a very difficult business, especially if it is about future developments. Anyway, the process of modeling consists in the replacement of the object being studied by its non-unique image: a model – mathematical, computer-generated, physical, verbal or other. Such a model is always a drastic simplification of reality, with irrelevant details being ignored. Mathematically based models do not necessarily require close correlation with observations. For example, string theory models may have nothing to do with reality, at least their correspondence to reality is an open question.

The ultimate difficulty in the modeling process is related to its non-unique character: there cannot be a single recipe how to build a model, one always has to make a choice so that modeling is closer to art than to the widespread stereotype of hard science based on regular procedures. This difficulty reflects the complexity of nature: for every phenomenon, there are many possible levels of description. There exists, however, the accepted opinion that to really understand an empirical observation one needs to invent a mathematical description.

To illustrate this point, one can consider the example of gravitation which is the ubiquitous force determining nearly all structures in the human environment. It is generally accepted that gravitation is described by Einstein's general relativity, a beautiful geometric theory that has largely defined the style of thinking in contemporary physics. General relativity with its four-dimensional spacetime was one of the main drivers of modern differential geometry and the theory of smooth manifolds extremely important in today's physics. General relativity was built by Einstein as a mathematical model stemming from a centuries-old observation that all bodies fall to the ground in the same way i.e., with the same acceleration irrespective of their constitution or, to put it another way, their mass and weight are always proportional. However, general relativity is not the only theory of gravitation, there can be others, e.g., constructed like other field theories such as electrodynamics, the proponents of these field-like theories of gravitation [107] asserting that space and time are just a non-dynamic background so that Einstein's general relativity is faulty and should not be awed as orthodoxy. Anyway, one needs unequivocal analysis of cosmological observations and dedicated physical experiments to judge which model of gravity is correct, regardless of its mathematical beauty. As usual, precise measurement is the highest arbiter.

In any case, one must remember that models are not reality and should not be perceived as such. In some extreme cases, models being free from the messy details of real life may have nothing to do with reality. Sometimes, one may have an impression that the modeling approach goes under the liberal slogan "Let all the flowers blossom" which allows even the wildest fantasies to be included into the best-known collection of models called physics. As far as mathematics goes, the term "model" is often understood as the subject of a theory studying special mathematical objects such as formal languages considered as sets with a structure: composition rules, hierarchy of elements, etc.

The adjective "mathematical" before the term "modeling" manifests the striving to specifically mathematical means of cognition. Indeed, one could have used, for example, the word "physical" instead, pointing at the fact that most models are actually imported from physics and are investigated – both analytically and numerically – by the methods common in physics. Besides, physicists traditionally pursued the methodological pattern common for mathematical modeling: making models, then testing them, then scrutinizing their consequences. One may notice in passing that physicists generally dislike the term "mathematical modeling" because of its overwhelmingly general applicability. Indeed, almost everything, all approaches to study reality, both in the form of natural or behavioral sciences, which are using mathematical tools, can be declared mathematical modeling. Many people even assert that there is no mathematical modeling, but just instances of applied physics and mathematics under a fancy – and sometimes fashionable – name. At least it would be difficult to delineate what is *not* mathematical modeling, when mathematical formulas are used. There are, however, very enthusiastic proponents of mathematical modeling *per se* who tend to claim that physicists in general are lousy modelers. In any case, developing and testing mathematical and computer models of such highly complex objects as the geosphere, the Earth's climate, the universe, the human organism and social system is the best possible way to understand such objects and to appreciate the models.

One can justifiably ask: is it possible, in principle, to model mathematically – analytically or numerically – any system encountered in nature or society? It would be difficult to answer this question, in fact of metaphysical character, logically correctly. The answer can be both "yes" or "no". For instance, so-called strong emergent systems, in which completely new properties appear at each level of hierarchy so that breaking down a complex multilevel system into its irreducible components does not necessarily entail more efficient simulation or computation, even of rough macroscopic properties. Examples abound: periodic crystal lattices, climate, human organism, Universe and so on.

Over the past thirty years, there have been many advances in the field of mathematical modeling, most of the progress being connected with the rapid development of the theory of dynamical systems which is basically a mathematical theory. The general approach to analysis of dynamical systems consists in putting forward a set of hypotheses about the mechanisms of the process development. After that, the hypotheses are converted into mathematical models that can be processed analytically and/or numerically, with the modeling results being validated by empirical data. Some of the hypotheses may compete, which is typically reflected in the mathematical model as equilibrium (singular) points, bifurcations, stable and unstable branches (respectively sinks and sources) as well as attracting and repelling sets in the corresponding vector fields. The modeling scheme based on dynamical systems has led to a great success in natural sciences, primarily in physics, chemistry and biology.

However, the major advances have appeared from outside of mathematics, namely they have been produced by the explosive development of computer technology. One of the new features of mathematical modeling has become the ability to play and experiment with models in real time, and the applications allowing for such real-time simulation do not require enormously expensive supercomputers, they can run on home PCs. The advent of computers on a mass scale and especially the explosive growth of the number of PC-enabled scientists and engineers developing and using convenient mathematical platforms – in fact an extension of the ubiquitous calculator, such as Mathematica, Maple, MATLAB and Mathcad, revolutionized the whole science and engineering. These mathematical manipulators currently allow for analytical, numerical and graphical implementations that were unimaginable 20-30 years ago. After the Internet became a commodity, one was able to overcome the limits of stand-alone mode and to exchange the modeling results over long distances in real time. A rapidly increasing number of modern scientific and engineering projects are based on mathematical and computer modeling performed in the remote access mode over the Internet. Thus, without using computers, the concept of mathematical modeling would look like in the distant past, even notwithstanding all wonderful achievements in the analytical theory of differential equations and dynamical systems. Moreover, the use of computers is the key to modern nonlinear dynamics.

Nevertheless, the focus of the present book is on applied mathematical techniques, not on computer implementations. The point is that in order to learn the skills of modeling one has first to understand its crucial aspects, e.g., how to build a model, to simplify it, to dissect a complex problem into the hierarchy of interconnected chunks without sacrificing too much reality, how to produce a system of equations, to estimate the errors arising from the neglected terms, to appreciate the model's accuracy in general and so on. One may call such issues too methodological or even scholastic, but many of them should probably be mastered prior to rushing to a computer and writing codes head-on.

Since numerical simulations rarely bring new mathematical concepts, the priority in the book is given to classical mathematical techniques. Besides, computer modeling usually does not produce asymptotic regimes i.e., such when one or several parameters of the problem (for example time) tend to infinity.

Mathematical modeling as a discipline (if one might call it a discipline) is quite loose. Hence the present manuscript is – somewhat intentionally – rather defocused. This manner is closer to a journalist's approach – to write superficially on many subjects at once – than to scientific pedantry, when one knows almost everything in a very narrow domain and gladly demonstrates one's knowledge. There are many "short stories" in the book that reflect the author's intention to induce people to read more boring didactive stuff. The purpose of the present book is not to enumerate or classify specific models, but to provide a sound footing for mathematical and computer modeling; this footing is mostly related with the qualitative analysis of behavior of solutions to differential equations and their numerical counterparts.

This manuscript has a certain didactic flavor. The main pedagogical component of the manuscript consists in showing that even the most primitive models such as that of the harmonic oscillator rapidly lead to rather sophisticated scientific concepts – in this model the inherent organic unity of physics is manifested. Another pedagogical component in the text is that we are mainly using the old-fashioned local coordinate representation of vectors and tensors instead of more general and modern (although in fact about one hundred years old) language of differential forms – mostly for the sake of learning transparency.

In most parts, the book is merely a review of rudimentary mathematical concepts and standard physical principles that can be efficiently exploited for model-building in other disciplines. Basic geometric structures needed for physics are also briefly overviewed. The essential content of this manuscript could probably be stated in half of its present length, but then one would have to sacrifice comments and links between seemingly different concepts. Roots and tentacles of any given subject can be of great value for generic models that might be transferred between various disciplines.

This book does not go into any depth in covering scientific computing or numerical solutions of the mathematical models that are discussed. Where possible, we discuss the analytic solutions to the equations and techniques for simplifying the model, because this also provides insight into the nature of the mathematical model itself. For interested readers, we have also included "Supplement 4. Scientific computing and numerical modeling".

The book can be reproached for some eclecticism, but it is more mosaic rather than eclectic that is a feature of application of mathematics in a variety of quantitatively oriented disciplines. In modern times when mathematical, physical and computer knowledge is rapidly expanding, rather sophisticated concepts can be used for mathematical modeling such as, for example, various spaces, manifolds and bundles, maps, Lie groups and Lie algebras, gauge fields, instantons, non-trivial integral curves for differential equations, dynamical systems, evolution, semigroups, phase trajectories and phase flows, Green's functions as well as the methods developed in such mathematical subdisciplines as variational calculus, differential geometry, ergodic theory, stochastics, etc. It is not a priori obvious that computer-oriented students or engineers would feel comfortable with all these concepts, therefore some of them are described (mostly *ad hoc*) in this book in order to facilitate the understanding of modeling principles. Most topics are covered in this book on an introductory level and serve to help grasp the main ideas as well as to soften modern abstractions, increasingly popular today in scientific literature.

The present book, mostly based on lectures given to graduate students of mathematics, physics, computational mechanics and computer science, has been produced with some educational purposes in mind: in particular, encompassing the material supplementary to computer-oriented training in the narrow sense. The only prerequisite for that lecture course consisted in linear algebra and standard calculus, perhaps sometimes involving many variables. The same applies to the present book. Of course, not all theoretical concepts contained in the mathematical part of the book will be immediately needed for the modeling purposes, therefore, to facilitate sorting out the mathematical material serving the utilitarian purposes, the text contains examples and case studies of successful mathematical modeling. As a result, the present book is not a threaded overview or a textbook, but rather a collection of tasks from various fields of science and engineering.

## Section 2. Basic types of mathematical models

Mathematical modeling is understood here as the activity that basically consists in linking mathematical equations with a materialized physical world. In the mathematical modeling technique, we describe the state of a physical, technological, biological, economical, social or any other system by time-depending functions of relevant variables. In certain situations, time evolution is less interesting than the relationships between the system attributes in a state close to equilibrium; then one can consider *quasistatic* models. Examples of such mathematical models are abundant in engineering: all engineering thermodynamics (which would more correctly be called thermostatics) and a large part of electrical engineering including the analysis of systems and circuits (recall Ohm's law) are based on a quasistatic approach. More complex systems such as the ones encountered in life sciences (e.g., the state of health, biofeedback, stimulation of neurons, etc.) or in national economies are also modeled using the quasistatic approximation. Equilibrium states usually present a special interest: they manifest the solutions that do not change with time. From the mathematical standpoint, equilibrium states can be characterized as stationary points of a dynamical system or as some relevant functionals defined on the space of system states (see below more on equilibrium states).

Of course, the term "quasistatic" is very relative: for example, if you want to estimate the action of a microwave on a chip, you may safely use the quasistatic approach, but if you are interested in the propagation of the same wave around the Earth or between satellites (e.g., when modeling telecommunication systems) this approximation would be totally inadequate. Here we encounter the typical concept of a dimensionless small parameter, this concept will be discussed later in some detail.

Depending on the value of a relevant parameter (in the above example it is the ratio of the wavelength to the system dimension), one can judge whether it would be possible to stay within the simple quasistatic approximation or if one has to use more complicated *dynamic* models describing the variation of the model attributes as a function of time. For instance, it would be incorrect to model the spread of a disease as a quasistatic process. In general, there exist very many phenomena that are intrinsically nonstationary, weather for example.

A dynamic model lies at the foundation of the clockwork picture of the world as held by Renaissance scientists. Before the Renaissance, in the ancient period, the world had been regarded as a divinely created static or at least quasistatic construction (the "Aristotelian tradition" and the Ptolemaic model), and dynamic views of the world together with experimental findings had been suppressed and often accused of witchcraft and demonology (Copernicus, Bruno, Galileo, Kepler, Newton).

The dichotomy quasistatic–dynamic illustrates the possibility of complementary types of mathematical models. Other well-known dichotomies are:

- quantitative vs. qualitative
- discrete vs. continuous
- analytic vs. numerical
- deterministic vs. random
- microscopic vs. macroscopic
- first principle (*ab initio*) models vs. phenomenological
- hard (rigid) vs. mild models.

This list of complementarities, each defining a single dimensional subspace, is not, of course, exhaustive. Besides, one might notice that all such pure model types are in practice interpolated.

One can briefly comment on these complementary model types. Pure model types represent the opposite orientations of the corresponding "axes" in some abstract modeling space. The modeler has a certain freedom in choosing between two incompatible alternatives or making a compromise between them. Although one usually strives to produce *quantitative* models, the *qualitative* ones should not be underestimated or abandoned. Qualitative models are indispensable for understanding such phenomena as various transitions occurring in complex systems, in particular the phase transitions. The latter take place when the system (not necessarily physical in the narrow sense) surpasses the border between two phases or between two basins with different attractors. The term "phase" was coined by J. W. Gibbs, the great American physicist and mathematician, and is usually understood as the state of matter.  Moreover, qualitative models enable us to fully realize whether the system is stable or unstable and to find its attractors i.e., limit states to which the system asymptotically strives (known in mathematics as limit sets). A good quantitative model may prove very useful in engineering: one can treat any engineering device as an artificial fragment of the universe in which, however, only a limited number of physical laws, expressed by mathematical equations, are acting and, besides, they operate in an almost closed system. Moreover, a quantitative exploration, e.g., of dynamical systems, can be readily accompanied by visualization.

Discrete models are not necessarily bundled with computers: there exist physical models that are inherently discrete such as, e.g., those of crystal lattice, the Toda lattice or the Ising model. *Discrete* (lattice) models as opposed to *continuum* ones are increasingly popular today since they can sometimes be solved exactly and correspond to the idea of a natural discreteness present in microscopic physics.

In contrast with the inherent discreteness of physics, *numerical* models developed for computer simulations introduce an artificial periodic structure, with a repeated elementary pattern (step of discretization $h$) replacing the continuous representation of a variable or a function. Such a replacement results in a specific kind of error called discretization errors which are usually the principal source of errors in scientific computing, computer graphics (aliasing) and computational physics. One might say that numerical models are just like crutches, they have been invented to make difficult but natural things simpler. However, such "crutches" are not useful in all cases. For a small number of degrees of freedom, numerical models may provide a satisfactory understanding, but for macroscopic many-body systems, when the number of degrees of freedom is enormously large, the naive numerical approach can be totally inadequate, and one should not fully rely on numerical models of complex processes in order to take vital decisions. The best one can attain by numerical modeling is a highly approximate description of the statistical behavior i.e., averaged over either thermal (as, e.g., in fluid dynamics) or quantum fluctuations.

Generally, numerical methods in modeling have one essential flaw: they enable us to compute trajectories on finite time segments, $t \in I \subset \mathbb{R}$, but, as already mentioned, are almost useless for asymptotic analysis of evolutionary problems. In other words, numerical models are not designed to answer the question: what happens when time goes to infinity? For instance, it is not at all trivial to numerically explore the stability of a dynamical system since this problem has an asymptotic character and can be formulated in the geometric language as the asymptotic study of trajectories in the state space. The best one can do in long-term numerical exploration as a palliative for asymptotic analysis is to find the maximal time step still ensuring the closeness of approximate solutions to the exact one. The solar system gives a physical example: are there such perturbations of the planetary orbits when one of the planets would escape to infinity? To answer this question, one should explore the vector field trajectories in the phase space and to qualitatively understand the topological character of the phase flow (the motion along a phase trajectory). One may note in passing that modern astrophysical data testify that there exist unbound planet trajectories drifting in space.

Nevertheless, in contemporary mathematical modeling the next step after having produced a formula nearly always consists in adapting the model output to numerical treatment. This second step may require rather sophisticated techniques, so one should never think that computational modeling is inferior to analytical.

*Deterministic* models are described by a set of variables and functions that are known exactly at each spacetime point. Such models are typically based on deterministic evolution equations when the actual state is *uniquely determined* by the previous ones. It does not, however, mean that future states are fully *predictable* from the initial ones; deterministic chaos provides counterexamples: some nonlinear deterministic systems that look quite simple can behave in an unpredictable – chaotic – way.

One can distinguish between the deterministic models whose main purpose is to find out "lessons from history" and teleological models aimed at taking the right (in a certain mathematical sense) path to some goal state, sometimes elucidating the goal state itself. For instance, feedback-driven and goal-refining evolution of human organism, if one can abstract from fluctuations, include time-delay or even hysteresis terms so that they can look teleological. A well-known astrophysical model of "teleological" character is the anthropic principle whose pure and extreme version states that everything in the universe is created as to ensure the presence of an observer, e.g., a human one.

In contrast, *random* (or stochastic) models imply that their parameters are in general defined as probability distributions and, besides, future states are not exactly determined by the previous ones. In other words, the evolution of the modeled system is also described in terms of probability distributions which serve as the sole instrument of predictive modeling in random dynamics. Note that the interplay between dynamical and stochastic aspects in the description of objects placed in some environment is one of the most important problems – not only in physics. In this book, mostly the deterministic models are considered; in particular, random dynamical systems are not thoroughly discussed here. However, although the book is devoted to deterministic processes, certain mathematical models bordering on random nature are also considered, for example, models based on the Fokker-Planck equation are discussed very briefly (see 7.11. The Fokker-Planck equation). The latter can be regarded as an evolution equation for irreversible systems, and one can interpret this equation as an intermediary device between deterministic and genuinely stochastic models. Moreover, a very brief mentioning of random processes is given at the end. In effect, a system whose behavior is not uniquely determined by its initial condition is not even considered here (by definition only!) to be a truly dynamical system.

The idea of determinism came from physics (Newton) and natural philosophy (Descartes) and consists in the assumption that the evolution of an isolated (closed) system in spacetime is determined by differential equations supplemented by initial and boundary conditions. Such differential equations are declared the "laws of nature", the most prominent examples are Newton's equations in classical mechanics and the Schrödinger equation in quantum mechanics. In fact, the postulate of the existence of closed systems is a rather strong assumption mostly serving modeling purposes, since in reality a closed system is quickly destroyed by fluctuations. Furthermore, closed physical systems behave in a rather primitive manner: they tend to equilibrium. Paradoxically enough, one of the reasons for this simple trend is the assumed reversibility of dynamics in time. However, striving to equilibrium is by no means universal behavior; for example, in biology we observe absolutely different, more complex phenomena.

One can notice that the traditional interplay between philosophy and mechanics has been weakened in the course of the 20[th] and the 21[st] centuries due to the development of quantum and relativistic theories as well as of nonlinear dynamics, and especially owing to modern experimental techniques. Many philosophical or even nearly philosophical models have become irrelevant to modern physics since they bring no new outcomes. In particular, endless categorizations of philosophers concerning the ontological and epistemological status of some phenomenon as well as similar speculations bring no measurable results. In the same way, the typical either/or attitude such as building deterministic vs. indeterministic models is basically wrong since in deterministic analysis the inevitable uncertainties and fluctuations are disregarded, often without justification. For example, engineering models are mostly deterministic, largely due to inexact knowledge of material properties, phenomenological coefficients, load parameters and the like. The cost for it is the arbitrary "safety factor" whose value, although disguised as a generalization of long-term engineering experience, is actually totally arbitrary and based on intuition ("better safe than sorry"). Estimating safety factors, in particular, imposing scientifically reproducible limits (e.g., conservative worst-case or "unknown but bounded") is a branch of statistical modeling close to probabilistic risk assessment (PRA), which is especially important in nuclear safety issues.

Thus, one may observe that mathematical models can have a significant impact on human behavior and way of thinking. For instance, the Newtonian model of the world greatly simplified it, albeit temporarily. Indeed, the Newtonian world was reduced to a small number of basic elements allowing one to predict the motion of planets and such almost mystical natural phenomena as eclipses a thousand years ahead, while the model itself could be written in several lines. The Newtonian paradigm created an illusion of fundamental simplicity of the world whose natural complexity has been only recently reassessed due to the paradigm shift brought by nonlinear science and the study of many-body systems. The simplicity-complexity dichotomy is clearly manifested when one considers microscopic vs. macroscopic models.

*Microscopic* models are the ones where equations describing the behavior of a single object (particle) are used. *Macroscopic* models involve a large number of objects, typically of the order of $10^{24}$ particles. For example, interaction of laser radiation with an atom is usually described microscopically (e.g., by the Keldysh model [78]) whereas the problem of laser-matter interaction with the matter heating and destruction implies a macroscopic approach. A schoolbook concept of Ohm's law is a typical macroscopic model, but it can also be derived by using microscopic motion equations for electrons moving in a constant (or quasi stationary) electric field and colliding with atomic centers in the medium. One should not think that only "small" objects may be described by microscopic mathematical models. For instance, traffic flow models can be both microscopic (e.g., car following) and macroscopic (such as speed-density models, hydrodynamic models, phase transition models of jams, wave patterns, etc.), yet cars or trucks are in no way small in the human scale. Earth and planetary motions in the Solar System are usually also treated with the help of microscopic models since the planets may be regarded as material points whose dimensions are negligibly small as compared to astronomical distances. Thus the mean Earth radius, $R_E = 6371$ km is many orders of magnitude smaller than the mean Sun-Earth distance, 1 AU (astronomic unit) equal approximately to $1.496 \cdot 10^8$ km. Macroscopic models nearly always arise as a result of the averaging in a complex "fine grained" picture, for example in many-body systems, which are described by a reduced number of mean-value characteristics such as temperature, pressure, average density and current, electric and thermal conductivity, permittivity, permeability and so on – in short, ignoring details of the individual particle motion. For instance, the famous and increasingly popular Ginzburg-Landau model in condensed matter physics does not intend to explain the underlying microscopic mechanisms of the state (phase) transitions; it employs general equilibrium arguments.

Phenomena may radically differ on microscopic and macroscopic scales. For instance, on a microscopic level there are electrons and nuclei interacting by Coulomb forces, still deeper on a subatomic level we have the Standard Model of fundamental particles and forces, leptons, quarks and gluons, whereas on a macroscopic level we observe solid state, fluid motion, gas laws and obviously irreversible behavior. Each level of natural hierarchy implies its own patterns of behavior, conceptual approaches and research traditions.

One should not, however, think that microscopic models are necessarily *first-principle* ones whereas macroscopic models are always *phenomenological*. These are different dichotomies. For instance, the Schrödinger equation illustrates a typical phenomenological approach, although it is most often applied to microscopic objects (such as the hydrogen atom) and is thus declared microscopic. In general, phenomenological models are often produced on a microscopic level: kinetic equations provide plenty of examples. The power of phenomenological modeling can be seen on the following standard example: we need very sophisticated mathematical skills and rather complex computer simulation in order to describe the dynamics of, e.g., 20 moving and colliding gas particles in a closed volume. Furthermore, microscopic models of such a system would be asymptotically unreliable because of uncontrollable instabilities and chaotic phenomena. However, the behavior of $\boldsymbol{b}$ gas particles can be viewed as highly regular and coherent, so that it can be described by a minimal set of phenomenological equations with astoundingly good prediction capacity.

Contrariwise, the density matrix method in quantum mechanics is usually regarded as the first-principle one, although this approach is in fact macroscopic. The closely associated measurement problem in quantum mechanics may be treated both microscopically and macroscopically, depending on the postulated properties of the measuring device, and the same problem can be considered from either first-principle or phenomenological positions. So, one should bear in mind that the notions "microscopic" and "first-principle" are not synonymous alongside with "macroscopic" and "phenomenological".

Moreover, Newtonian mechanics and Newtonian gravitation, although they are often viewed as first-principle theories, are actually phenomenological models since they do not pay much attention to the profound meaning of mass and force, which stand out as typical phenomenological quantities. A macroscopic body incorporates a large number of microscopic components (molecules), all of them subordinated to quantum-mechanical laws, yet the body motion follows the classical trajectory determined from the equations of classical mechanics. It means that all the molecules change their states (positions and momenta) coherently i.e., simultaneously in the classical sense. This fact can be translated into the language of mathematical models as, for example, the statement that macroscopic motion in everyday life can be described by the stationary phase in a many-particle path integral (action functional).

The alternative *hard* vs. *mild* mathematical models proposed by the great mathematician V. I. Arnold is consistent with the concept of the stability loss of an equilibrium state. Recall that the meaning of the term "stability" implies an indifference to small external perturbations. The loss of stability is usually called "hard" if the system's trajectories in its space of states (e.g., in phase space) begin running away from an equilibrium position as soon as at least one of parameters characterizing the system (called a control parameter) passes a certain critical value, typically known as a bifurcation point. In the theory of dynamical systems, a bifurcation is commonly understood as the radical alteration of a phase portrait as the control parameter is varied, a bifurcation corresponds to the emergence of at least two equilibrium solutions. The loss of stability is called "mild" if the parameter characterizing the system has passed the critical value, yet the system paths remain in the vicinity of an equilibrium state. The corresponding mild model of the system is not necessarily fuzzy or unreliable; it may include feedback that stabilizes the system. For example, hard models of centralized planning (e.g., in micro- or macroeconomics) can easily become unstable unless the feedback accounting for the achieved state of the system are incorporated into the model. In general, hard models are usually too rigid to easily admit changes: the topological type of a hard system in its space of states would radically change if the system parameters were perturbed. We shall encounter examples of bifurcations when reviewing certain growth models (such as logistic maps).

"Hard" versus "mild" models hint at the possibility to arrange the models according to their mathematical rigidity or formalization. At one end of the formalization spectrum are the models that remind us of arithmetic, i.e., with very rigid rules whereas at the opposite end of this scale artistic or literary images are located which appeal to individual perception. When we perform arithmetic manipulations, we do not care about the content and act purely automatically, according to formal rules. These actions can be easily passed to a machine (i.e., to a computer) i.e., hard models can be viewed as computational off-the-shelf prescriptions. The next place on the formalization scale may be occupied by the interpretation of fully formalized models, which consists in finding the relationship between the model outcome and observed phenomena. In the simplest case of arithmetic, this relationship amounts to the procedure of counting objects. Ideally, the interpretation should be uniquely understood by everybody and, in principle, available for computer processing (algorithmization).

Hardly formalized models can readily become autonomized i.e., separated from interpretation or human conscience and perceived as "objective" models of the world. It is these models that serve as ideal objects of scientific study. Hamiltonian dynamics, quantum mechanics or general relativity although considered full-scale theories are in fact prominent examples of objective models of the world. It is, however, useful to recall from time to time that such formalized models as well as the concepts on which they are based stem from critical analysis of rather fragmentary experience, and ascribing to these models a perpetual and overwhelming validity may lead to dogmatic pitfalls.

# Section 3. Expected properties of mathematical and computer models

There exist "rigid" models having the property that it would be difficult to deform them without running into inconsistencies. Thus, it would be hard to modify, e.g., Newtonian mechanics or Maxwell's equations in vacuum without noticeable discrepancies within the theory (model) itself or drastic disagreements with observations. Despite the admissibility of rather wild hypotheses (such as, e.g., that the Big Bang, which is a rather speculative cosmological concept itself, stems from a collision of two brane worlds), it is highly desirable that models should not contradict fundamental laws of nature. Thus, models should be thoroughly tested (validated) against basic laws of nature. For example, the number of particles or their total mass must be conserved in simple models such as those of road traffic flow or fluid streams. It happens, however, that in some computer models the mathematical expression of mass conservation – the continuity equation – does not hold exactly in all points in order to satisfy boundary conditions or certain numerical requirements.

People claiming that they pursue some scientific or engineering purposes are just replacing a natural (or social) object by its model and studying the latter. There is an element of deceit in this process. No model fully describes the studied phenomenon, and this fact is simultaneously a shortcoming and a merit. The merit of a model is that it can be explored exhaustively. What is not necessarily required of a mathematical or a computer model is the capability to predict things; predictive power is the crucial feature that distinguishes a model from a theory. For instance, Newtonian gravity is a theory since it exactly fixes the power in the Newtonian potential ($\varphi = Gm/r$, $m$ - mass, $G \approx 6.674 \times 10^{-8} \mathrm{cm^3 g^{-1} s^{-2}} \approx 6.674 \times 10^{-11} \mathrm{m^3 kg^{-1} s^{-2}}$) which leads to a lot of mathematical consequences and predictions. One can of course build various models of classical gravitation with different dependence on distance $r$ such as $\varphi = Gm/r^\alpha$, where parameter $\alpha \neq 1$ must be adjusted to empirical data or where this dependence is not at all expressed as the power law.

Therefore, by the way, the question of the model applicability is usually highly nontrivial. Indeed, a mathematical model is not uniquely determined by the investigated object or process. Selection of a pertinent model is dictated, firstly, by intuition (which is an ill-defined concept) and secondly, by accuracy requirements (also rather vaguely understood). For example, in one of the oldest applications of mathematical modeling – ballistics – models are ranging from those totally disregarding air resistance to those with very strong influence of the atmosphere, when the resistance (drag force) is modeled as proportional to high powers $n$ of the projectile velocity (here power $n$ is the model parameter), and the choice between such models has always been based on the gunners' experience. In atomic physics, accounting for the finite dimensions of the nucleus taken as a new parameter of the model may seriously affect the numerical values obtained in the calculations of atomic spectra with the pointlike nucleus. In military planning, one should also use one's intuition to choose the right model ranging from simple linear interaction of armies to highly nonlinear models involving partisans or terrorist groups. In more mundane situations such as architecture planning, one can in some cases disregard the distinction between rectangular, square and round tables whereas in other cases, e.g., for computer planning of a specific office, this distinction can be essential.

Thus, modeling of reality does not go without penalty: some reality must be sacrificed in order to gain more understanding of the problem being modeled. Even worse: the scope of reality left outside of a mathematical model is in many cases unknown, so it is rather nontrivial to quantitatively establish the model applicability unless the model is formulated in terms of analytical functions or asymptotic expansions (when one can evaluate quantifying the corrections). The model generally constructs an abstract concept of a system or a process, focusing only on relevant aspects and ignoring details. The examples of modeling abstractions are abundant in science and engineering. Thus, in Mendeleev's periodic system, which is a table of chemical elements, details like availability of elements (that are distributed in nature very unevenly) or their combination frequencies are omitted. In astrophysics, objects are classified and identified by their spectral properties, mostly in the infrared (IR) domain. Chemical composition, volume, even location are often disregarded. In biology, classification of species ignores their behavior, while in behavioral and social disciplines modeling and classification ignores habitus. Gravitation is irrelevant for nuclear collisions or for the passage of nuclear particles through matter (reactor safety problems); quantum theory is irrelevant for the calculation of the motion of celestial bodies, in particular, for asteroid danger. Photon properties such as spin (helicity) and polarization states are irrelevant for car headlight design, at least so far.

Although it may be considered an extreme positivistic view, the book supports the concept that any physical theory is nothing more than a mathematical model, in some cases an extended one. Therefore, it does not make much sense asking if such a theory (a.k.a. model) reflects reality. The verb "reflect" is rather vague in this context, and the maximum that one might ask would be that the model outcomes agree with the experiment or observations, with the quantitative degree of such correspondence showing the model quality. It is useful to think of mathematical models of natural processes as an extension of Gedanken experiments, never forgetting that the model has little to do with the real world. Furthermore, most renowned models tend to sacrifice reality for "mathematical beauty" or "elegance", in order to impress the peers by mathematics or, lately, by computer science used. It is also useful to avoid mathematically inspired axioms since the real world does not obey axiomatization.

When a real-world system turns out to be too difficult to understand, one can substitute for it a much simpler system with a resembling behavior. These models are intended not to scrutinize the specific details of the real-life situation, but to design artificial systems that would facilitate the study of complex real-life ones. A good deal of such model substitutes may prove useless, but a few can provide an insight and lead to new and sometimes surprising ideas.

When one develops a model, one has to mercilessly throw out the features that are irrelevant for the modeling situation. For instance, the color of a projectile is irrelevant for modeling its flight, but can affect the selection of an appropriate shell. The air resistance in ballistic models can be important or not depending on the modeling setup. Probably one of the most fruitful models in physics is the one of an isolated point particle, although this concept is a complete idealization having nothing to do with reality. The model of perfect gas ("gas laws") is highly unrealistic but extremely useful in spite of the fact that it cannot produce phase states and phase transitions. Likewise, the Schwarzschild vacuum solution in general relativity is also unrealistic but useful since it describes the gravity field created by a spherically symmetric object such as a point mass, lone star, a planet or a black hole. In fact, the currently popular concept of black hole originated from the Schwarzschild solution of Einstein's field equations of general relativity – recall that black holes are just spherically symmetric solutions of the evolution (motion) equations of general relativity. For example, gravitational collapse of massive astrophysical objects such as stars results in the formation of black holes: according to the contemporary theory of gravitation, stars with the masses exceeding the limit of several solar masses end up their evolution as supernovas collapsing further into black holes. Dozens of supernova bursts are recorded each year. Note also that the reality of black holes is no longer only a theoretical possibility: experimental results justify assumptions, and today there is a strong observational support for the black hole existence (see, e.g., an overview [39]).

When speaking about the model applicability, one should recall quantum theory. It is astounding that quantum mechanics, which can be regarded as a rather artificial mathematical model and just a guesswork, has become the basis for understanding physical reality and the foundation for most modern technologies. It is considered very difficult to unite these two very important physical theories, quantum mechanics and general relativity, despite the fact that they appeared almost simultaneously. The failure to reconcile these theories contributes to a somewhat dissonant worldview characteristic of contemporary physical, philosophical and behavioral disciplines. One of the crucial differences between quantum mechanics and general relativity is the diverse treatment of time: general relativity formally does not have absolute time whereas nonrelativistic quantum mechanics does.

In this book, we practically do not touch upon quantum-mechanical models because quantum mechanics not only uses different mathematical means, e.g., self-adjoint linear operators acting on vectors (rays) in complex vector spaces, non-commuting observables and unitary representations of the Lie groups, but in general involves a different philosophy of science, which is not always reduced to differential equations and their real-valued solutions.

In general (as was noticed first by Leibniz), mathematical statements are exact but not necessarily real whereas physical assertions are real, but not necessarily exact. The crucial question is mostly of a pragmatic nature: "Does it work?" When a mathematical model is set up and processed, it is expected to produce the outcome of mathematical statements. Here, one must not forget that the validity of such statements is always limited and largely conditioned by the model. Nowadays, one can often observe the tendency to absolutize the model's outcome, which may lead to confusion and even grave mistakes. As an example, one might recall the deeply rooted stereotype that all physics should, in the final analysis, be time-reversal invariant, which is an unverifiable hypothesis. This hypothesis is just an arbitrary extrapolation of inversible models onto all observed phenomena. Nevertheless, this simple belief is sometimes used as an instrument, for instance, to discard solutions containing terms with odd powers of frequency. One can also notice a recent trend in a number of disciplines both of basic and applied character to rely more on models and computer codes rather than on observations. For example, climate studies and the associated political and economic decisions mainly rely on a vast array of mathematical texts and computer codes. Whether the climate is getting warmer or, on the contrary, colder mainly due to human activity is, strictly speaking, not the physical fact, it is a hypothesis or, at best, a computer model. Even assuming that the Earth's surface is found in the warming phase (there are indications that the average surface temperature of the Earth is higher today than it was two hundred years ago) does not prove it is moving to an imminent catastrophe due to anthropogenic carbon dioxide, although there may be modeling arguments in favor of this assertion. Climate dynamics is such a complex multiparameter system that one should not have much faith in the predictive power of climate computer models – as little as of social or economic ones. Moreover, climate dynamics can well be chaotic i.e., making all predictions beyond some temporal horizon totally unreliable.

Mathematical models and especially computer codes are sterile knowledge; they usually have little to do with real life. Models can be founded on plausible assumptions but lacking relevance to reality, although they may even be internally valid and mathematically non-contradictory. In general, models do not have to be "true", especially if truth is understood as a collection of statements that most people do not dare to argue.

One of the crucial properties that ought to be expected when building mathematical models, specifically of a dynamical character, is *causality* which is the postulate that the effect cannot precede the cause and the response cannot appear before the input signal has been applied. In mathematical modeling the term "causal" typically means that the output function (or signal) at time $t$ depends only on the history of the input prior to $t$ for all $t \in I \subset \mathbb{R}$ where time interval $I$ is a non-empty closed subset of $\mathbb{R}$. In other words, causality signifies that the output depends on previous inputs. One usually distinguishes between strictly causal models (output depends *only* on previous inputs), anti-causal models (output depends *only* on future inputs) and non-causal models (output may depend both on previous and future inputs). All physical systems are considered causal by default – we presumably live in a causal world, and relying only on signals from the future would produce weird models, probably lacking common sense. One might, however, remark that nonlocality and entanglement in quantum mechanics have lately posed new questions concerning the absolute causality requirement. Below we shall try to give a more exact meaning to the concept of causality describing it in mathematical terms.

In more mathematical language, causality means that there exists a timelike (i.e., partial) ordering in a set $\mathbb{R}^4$ corresponding to a distinction between past and future, yet spacelike coordinates are not necessarily ordered. Recall that elements of set $\mathbb{R}$ are viewed as real numbers or scalars. Hence elements of $\mathbb{R}^n, n \geq 1$, consist of all $n$-tuples $\mathrm{x} = (x^1, \dots, x^n)$, where $x^i, i = 1, \dots, n$ are scalars. Thus, elements of $\mathbb{R}^n$ can be interpreted as both vectors and points (although these two notions are in fact different), with $x^i$ being the coordinates of x. Denoting vectors by boldface letters seems to be an archaic fashion, but we still use it throughout this manuscript for pedagogic reasons. One might notice that in physical relativistic theories the vector components are counted from 0 through $n - 1$, in contrast with the usual mathematical manner $x^i, i = 1, \dots, n$. There are both historical and physical reasons for it ($x^0$ corresponds to time).

Set $\mathbb{R}^4$ incorporating time and space points, where coordinates of these points are introduced as real numbers, is usually known as Minkowski space $M^4$, but we would rather call it now the proto-Minkowski space as there is no structure on $M^4$ yet. Neither topology nor continuity (norm), nor metric (or pseudometric), nor linearity (vector space properties), nor any special kind of geometry (manifold) are required for causality – just the time ordering, where time is regarded as a canonical projection from the event space $\mathbb{R}^n$ or, more generally, $\mathbb{A}^n$, with symbol $\mathbb{A}$ denoting an affine space equipped with a scalar product, to the $\mathbb{R}$ axis. Time $t$ is considered homogeneous, and the concept "motion of a point" is actually a map $I \rightarrow \mathbb{A}^3$ where $t \in I \subseteq \mathbb{R}, I$ is an interval of the time axis. The concept of evolution that will be discussed below is defined analogously, but the target space may be different from affine space $\mathbb{A}^3$ (e.g., Hilbert space of quantum mechanics). Each motion of a point is in one-to-one correspondence with a smooth vector function $\mathrm{r}(t): I \rightarrow \mathbb{A}^3$. It is clear that the velocity of motion $\dot{\mathrm{r}}(t) \equiv d\mathrm{r}/dt$ does not depend on the choice of the coordinate origin which is one of the reasons for choosing an affine space for mathematical description of the point motion instead of vector space $\mathbb{R}^3$ or Euclidean space $\mathbb{E}^3$.

Notice that the proto-Minkowski space cannot even be considered a spacetime. Yet this is of course the question of language and interpretation: some people may say spacetime should be real, others would require stringent mathematical properties imposed on it and so on. Historically, Minkowski space was the first area where physics and mathematics were blended. From a simple mathematical viewpoint, Minkowski space served as a link between algebraic and probably more complicated, geometric structures.

There exist many mathematical models using continuous linear operators. Linear models are based on two simple principles linking output $\boldsymbol{y(t) = f(x(t))}$ with input $\boldsymbol{x(t)}$: scaling i.e. $\boldsymbol{cy(t) = f(cx(t))}$ and superposition $\boldsymbol{y_1(t) + y_2(t) = f(x_1(t) + x_2(t))}$ so that $\boldsymbol{c_1 y_1(t) + c_2 y_2(t) = f(c_1 x_1(t) + c_2 x_2(t))}$ (here we mention, for simplicity, only single-point models i.e. synchronous ones and without delay or time-dispersion). Most linear models are a priori assumed to be causal i.e. implying that a system cannot start responding to a stimulus or an excitation until they have been imposed. Alternatively, a system cannot predict how it will be driven or excited. However, the nonlinear world is different and more intricate: for instance, feedbacks that are vital for complex systems such as biological or social ones can mix up causal and anti-causal contexts.

The requirement of causality is a result of human experience: non-causal systems would allow us to receive signals from the future and to influence the past, in particular, removing the reasons for one's very existence ("killing one's grandmother before one is born"). Yet the empirical concept of causality is a normative requisition that can eventually be made obsolete.

Causality is often identified with determinism: if the present state is taken as initial data, then the future course of events is assumed to be determined (provided certain conditions for uniqueness are fulfilled). **"We may consider the present state of the universe as the effect of its previous state and the cause of the one which is to follow".** This statement of determinism implies uniqueness of evolution of the states of the universe (the flow in the respective dynamical system) i.e. a verbal translation of the corresponding Cauchy problem. Thus, determinism requires a unique solution of this problem. In other words, the principle of determinism states that the state of a system at any instant of time uniquely defines all its evolution, both in the future and in the past. However, uniqueness of the solution is not necessarily attained even in simple physical situations (for instance, the Lipschitz condition may fail), so that it would be a little naive to expect uniqueness for the whole universe.

Yet the principle of causality is wider than determinism since causality does not require the response to be unique. The effect comes after the cause but is not necessarily uniquely determined by the cause: for instance, it can be stochastic or chaotic. Conversely, it cannot be proved that each effect has a cause: one can find counterexamples i.e., events occurring with no measurable cause, e.g., due to fluctuations. Fluctuations are ubiquitous, and they can destroy a strictly deterministic course of events. The statement that all processes in nature are causal i.e., are governed by cause-and-effect relationship is a philosophical principle rather than a scientific statement; it is difficult to prove. Notice that causality cannot be measured i.e., it cannot be represented as an ordered set: more or less of causality. Accordingly, there are no units of causality.

In the primitive formulation, causality implies that when the present state of a system (i.e., of some piece of the world) is exactly known, then one can compute the future. However, this naive, mechanistic determinism fails when statistical or quantum-mechanical reasoning must be adopted. The philosophical meaning of instabilities and chaos in dynamical systems (see Section 7.6.) consists in the statement that determinism à la Laplace (given initial conditions, one can predict the future of the universe) becomes virtually impossible since any error in the initial data grows exponentially thus leading to a rather short horizon of predictability, so that the classical reality is not fully deterministic. The prerequisite of an exact knowledge of the present is wrong: one cannot in principle know the initial state in all minute details. Any attempts to specify completely the present state of a real system in order to fix its future development fail, and strictly deterministic schemes can only be treated as mathematical models. Therefore, what we observe is just some selected set of data taken out of the whole ocean of relevant information. So, the choice of future evolution trajectories is inevitably limited. Besides, sensible *mathematical* models of determinism must be based on the concept of semigroups.

The situation with determinism in the quantum world, where specific quantum fluctuations are inexorable, is still worse. The standard example illustrating the failure of determinism is the radioactive decay: can the fullest specification of the state of the universe ensure the knowledge of the exact moment of spontaneous emission of a particle (or a gamma quant) by an unstable nuclide?

Despite its philosophical (metaphysical) background, causality can be instrumental. Thus, a systematic procedure of obtaining dispersion relations in linear systems is based on the causality principle. Dispersion relations have traditionally been very important in high-energy physics as well as to study the electromagnetic response of material media. In the linear case, causality requires that the response has the form $y(t) = \int_{-\infty}^{\infty} g(\tau)x(t - \tau)d\tau$ with Green's function $g(\tau) = 0$ for $\tau < 0$. This condition leads to the following one in the dual (Fourier) domain: $\tilde{H}(\omega) = iH(\omega)$, where tilde denotes the Hilbert transform in the frequency domain. This is just another form of the dispersion relations. From a more general standpoint, causality in physics is understood as the absence of closed timelike ($ds^2 > 0$) or null ($ds^2 = 0$) curves, where $ds^2 = g_{ik}dx^i dx^k$ is the line element (interval) on the respective Lorentz manifold, $g_{ik}$ is the metric tensor (here we use the +--- signature). Note that a more correct definition should be formulated through tangent vectors, not in terms of curves, see also below.

Causality is closely connected with the non-invariance under time reversal (the "arrow of time") yet these concepts are mathematically different, although in popular expositions they are often confused. The equivalence between past and future does not necessarily coincide with the equivalence of cause and effect: here lies the difference between time reversal invariant behavior and reversible one (i.e. when all vector fields are inverted). Time reversal invariance (TRI) also known as inversibility, and time reversal symmetry (TRS) requires that direct and time-reversed processes should be identical and have equal probabilities which is not the same as to forcibly remove all non-causal (acausal) solutions. A physical theory or model can be time-reversal symmetric but acausal (or even anti-causal), as, e.g., considering the advanced solutions in electrodynamics.

Most of mathematical models corresponding to real-life processes are non-invariant under time reversal, in distinction to "microscopic" mechanical models (a film that would show the motion of the planets around the Sun might be demonstrated both backward and forward without any breach of reality). It is not at all obvious that all basic laws of nature should necessarily be symmetric under time reversal

$$t \rightarrow -t, T = T_\alpha^\beta = \begin{pmatrix} -1 & 0 & 0 & 0 \\ 0 & 1 & 0 & 0 \\ 0 & 0 & 1 & 0 \\ 0 & 0 & 0 & 1 \end{pmatrix},$$

although some physicists are still insisting on this basic symmetry, taking Newton's laws or the Schrödinger equation as the ultimate description. In some parts of physics (e.g., in weak interactions) time reversal invariance may be considered rather an exception than the norm. Even in classical electrodynamics if a charge is accelerated it radiates electromagnetic energy so that the dynamics becomes dissipative and not time reversal invariant (non-invertible). Field theories in general do not have to be invariant under time reversal (as well as under parity

$$P = P_\alpha^\beta = \begin{pmatrix} 1 & 0 & 0 & 0 \\ 0 & -1 & 0 & 0 \\ 0 & 0 & -1 & 0 \\ 0 & 0 & 0 & -1 \end{pmatrix}$$

i.e., the transformation replacing $x$ by $-x$).

Invariance under time reversal is, in fact, just an expectation, a wishful thinking; it is not necessarily a must for reality. For instance, one might imagine a classical world in which it would be necessary to know not only initial positions and velocities, but also initial accelerations to determine the law of motion. Accordingly, the motion equations would include higher order time derivatives, and not necessarily of the even order (example: reaction of radiation). Although in some restricted domains natural phenomena obey the TRI restrictions and the respective mathematical models enjoying the $t \leftrightarrow -t$ property may well reflect reality, extrapolation of this symmetry requirement on the whole nature would mean to mistake our invention for the fundamental law. Time reversal invariance is an *a priori* philosophical principle prescribing in advance what the world should be like[1].

The best models are those that accommodate three main principles: generality, simplicity, and possibility of rapid experimental verification. From this point of view, for example, many mathematical models fashionable in modern physics such as the ones based on string theories are not the best, at least so far: they are quite general, but hardly can be called simple and care very little about experiment. On the contrary, the model of the Bohr (more exactly, Bohr-Rutherford) atom was in conformity with the above three criteria. Yet the Bohr-Rutherford atom was actually a rather wild hypothesis at the time when it was devised.

There exist other lines of reasoning for model building, for example, *symmetries* should be taken into account, *scaling* can be exploited to reduce the complexity stemming from a large number of variables, and *analogies* ought to be used at all lengths to achieve maximum universality: different objects may be described by basically the same mathematical model. For instance, chemical reactions and competition processes in biology or economics are represented by very similar equations so that mathematical (and computer-based) methods of studying all such phenomena may be transferred between apparently diverse fields. A well-known example in physics is the remarkable analogy between time and inverse temperature allowing one to use the same mathematical tools in seemingly different fields: quantum field theory and condensed matter physics. Another well-known example links mechanical and electrical engineering: vibrations of the body of a car and the passage of signals through electrical filters can be investigated using very similar mathematical means. One can use the fact that analogies, associations and metaphors are all-pervasive in our thinking and language.

The use of analogies was very typical of the works of science classics. Thus, J.C. Maxwell widely used purely mechanical models to derive his famous equations of electromagnetism, Lord Kelvin (Sir William Thomson) was guided by parallels with rotational motion in perfect fluids (leading, in particular, to Kelvin's circulation theorem) when he suggested in 1867 the vortex model of atoms and then an explanation for magnetic forces. In general, hydrodynamic models were very useful for the development of electromagnetic theory. In the late 19th century, one more analogy became extremely fruitful: J.W. Gibbs built his thermodynamics and statistical mechanics using classical dynamics as a guideline.

## 3.1. Theory, experiment and models

The relationship between these three components of knowledge-attaining endeavor may be complex and even confused. The interplay between theory and experiment induces the creation of models that are used to simulate the observations and predict novel features that already can be observed in new experiments. Experimental evidence, provided that it can be independently reproduced, is hard, and such evidence can serve as an anchor. However, increasing complexity and expensive (sometimes prohibitively) experiments call for mathematical modeling and computer simulations in both theory and experiment. Nowadays we see the trend to replace or even omit experiments, even in such experimentally driven disciplines as physics. Such fancy notions as parallel universes, extra dimensions, wormholes, even strings and supersymmetry are not corroborated in experiments, but enjoy a massive social support.

---

[1] It is known, for example, that L. D. Landau, one of the best-known problem-solving physicists who created one of the most successful schools of theoretical physics in the history of science, vigorously opposed the concepts of both the parity ($P$) and time-reversal invariance ($T$) violation. At first, he simply did not want to listen to any arguments. Later, probably due to the persistence of his colleagues B. L. Ioffe, L. B. Okun and I. Ya. Pomeranchuk he softened his position a little, still considering the $CP$ conservation an exact law ($C$ is the charge), with $CPT$ being also exact and almost trivial. This implacable line of thinking left no chance for $T$-noninvariant models [22].

Nevertheless, experiment always trumps theory, mathematical models and, of course, various speculations. Falsifiable assumptions about the real world can (and should) be validated by experiments or observations whereas mathematical modeling, proofs and tools are just mental constructions. One can discuss their relevance or consistency with other assumptions, not necessarily empirically falsifiable, but *sensu stricto* should not require their possible empirical refutation. For example, can one prove that the concept of man-made climate change, which is mainly based on computer models, is falsifiable? There exists a well-known paraphrase of Einstein's terse statement: "No amount of experimentation can ever prove me right; a single experiment can prove me wrong". One can figuratively say that a single empirical falsification can outweigh the entire library of theoretical papers, computer codes and philosophical discourse, like it or not.

The conventional paradigm of science is that it starts from observations and measurements which dictate models and theories, the latter allowing one to make predictions that would be tested by further observations and measurements confirming or disproving models and theories or, in the intermediate case, helping to improve and not to reject them. That's how the mechanism of science works.

Theories are deep descriptions of the laws of the world. Theories stand on their own feet and each of them tries to provide an intellectual framework i.e., a structure comprising simple models. The latter are necessary to extract a meaning from direct observations or numerical calculations. Good theories do not rely on analogies whereas models usually do. Models tell us what the fragment of a theory is approximately like. For instance, semiclassical models occupy a very important, but also very limited fragment of nonrelativistic quantum mechanics; such models can be simpler than hard-core wave theories and therefore suitable for approximate calculations, but semi classics is by no means a full theory.

So, if a model fails, this is not a tragic occurrence: you have nothing to lose but your illusions. Yet sometimes theories and models coincide: thus, most of the theories in physics are reduced to models. For instance, one of the most successful theories across all known disciplines, quantum electrodynamics (QED) is nothing but a mathematical model acting in an imaginary world filled only with electrons (and antielectrons – positrons) and photons. All other matter which we observe is left outside of the QED world.

In the final analysis, only the models validated by the data collected from the real-world processes can be trusted. So, the ultimate confirmation of truth comes from making experiments in spite of the fact that empirical data may be incomplete, for example, because something urgently needed to validate a theory or a model has not been measured or cannot be measured in principle. For example, one has no means to observe what occurs inside a catastrophically overheated nuclear reactor, although it may be vitally important: it is hardly possible to measure the state and distribution of radioactive substances in the reactor zone – simply because no measuring device can sustain respective temperatures and radiation levels. Furthermore, being the foundation for research, experiment by itself does not necessarily guarantee correct understanding of the observed phenomena or measured effect. Indeed, there existed in natural sciences plenty of misconceptions which apparently did not contradict the collected experimental evidence. Obvious examples of the false concepts were widespread theories of phlogiston and ether, almost unanimously supported by scientists at the time. Both phlogiston and ether were regarded as material substances needed to explain the observed thermal and electromagnetic phenomena.

Therefore, it can be quite inadequate to position oneself as just a pure theoretician or a mathematical modeler, or an experimentalist, without giving due regard to complementary activities. The matter is that all main tools of acquiring knowledge – theory, laboratory experiments, observations (field experiments), and lately computer modeling – are interrelated: none of them alone can tell the whole story about a real problem. Nonetheless, mathematicians and theoretical physicists may have a greatly simplified opinion of experimental work, probably because they do not perform experiments themselves. "Theorists" typically think that it is enough to design a right experimental setup and focus on several crucial variables. Yet practical experience shows that such a simplistic view is very naive. A trivial observation is that experiments are highly specific for the field to be explored: physics, astrophysics, engineering, biology, chemistry – you name it. Even in physical a experiment, there are always complex interactions between various parts of a setup as well as between many people who are typically involved (one might recall how many industrial companies participated in the high-energy experiments carried out in CERN). Very complicated experiments are performed in biology, specifically *in vivo* experiments. Due to considerable experimental difficulties, chance and luck also play a vital role. Besides, to separate the collected data into "typical" and "special" is a very nontrivial task. Furthermore, breaking complex entities into smaller parts (hierarchical paradigm) which is regarded as a universal scientific methodology may be inadequate in sophisticated experiments because of possible feedbacks between the levels of hierarchy. Thus, thorough mathematical modeling or planning of experiments is in most cases necessary.

A theory in general may be defined as a cohesive system of concepts that was experimentally validated to the extent that there is no unclearness or contradictions within the limits of applicability for this system of concepts. A theory must be consistent anyway; an inconsistent theory gives an example of "not even wrong" (a caustic remark ascribed to Pauli). Even before an empirical falsification test (pointed out by Karl Popper) is accomplished, a consistency test ought to be performed. According to Popper, experimental proofs of correctness of a theory, strictly speaking, do not exist. However, if a single experiment contradicts a theory, it must be discarded or at least included as a limiting case into a more general theory. Besides, a good theory contains a heavy load of knowledge. For example, the mechanical theory, together with the whole set of initial values, allows one to calculate the positions of the planets for many thousands of years into the future and into the past with sufficient accuracy (limited by Newtonian approximation). Nevertheless, a physical theory may be incomplete for the time being, and the current state of the art should not be confused with inconsistency (quantum mechanics is an example).

In contrast to the broad theory, mathematical models are much more specific, illustrative and closed. The ultimate -- and the most difficult -- skill of modeling is to convert a seemingly complex problem into a much simpler one. Or to dissect a big problem into a multitude of small and sharply cut ones. More than that: a model may be understood as a set of constraints which is bound to be replaced by a new set in further study, and one can often foresee how the old model would be discarded. An essential drawback of any modeling approach is that the processes and phenomena which are not included in the model at the stage of its mathematical formulation cannot miraculously manifest themselves at the stage of the model's analytical or numerical treatment. A theory can be cleverer than its creator whereas a model cannot.

Models can sometimes look quite artificial (we shall see this below on the model of "monads" comprising a physical particle), so that building a good model becomes a real art. Model building may be considered an essential part in constructing a theory. Models are usually constructed within the framework of a certain theory, for instance, the Friedman cosmological model is built up within the general relativity theory or the BCS model of superconductivity is a limited fragment of the non-relativistic quantum theory. The keyword for a model is the outcome, the keyword for a theory is a framework.

One can also observe that theories are often closer to facts than models. On the other hand, models tend to be more intimately connected with the studied phenomena than theories. The matter is that theories are typically too general to be directly applied to the phenomenon under study. There may be a variety of ways to apply the theory to a given problem, and these ways outline specific models. The merit of a model is that it can be explored exhaustively. A theory cannot cover everything in the universe to the smallest detail and does not necessarily bring specific results; it rather provides tools to obtain them. For example, the whole solid-state theory may be considered as a particular case of quantum mechanics (also a theory but higher in the hierarchy), however, concrete results are produced when specific models within solid state theory are considered.

There may be a number of mathematical models that would claim to interpret the phenomenological data, and reliably established experimental facts can enable one to choose among the models. Yet without a solid theoretical framework one can easily get lost in countless details and in the combinatorial complexity of specific results. Thus, the traditional case-by-case approach to mathematical modeling should be supplemented by a more general scientific basis. In this connection one might recall that the famous Einstein's 1905 (annus mirabilis) paper [55] is based not on experiments, but on two general postulates (the relativity principle and the constancy of the speed of light)[2]. In this sense, special relativity is a typical mathematical model.

Mathematical and computer modeling usually gives the following results:

-   The theory is insufficient and must be modified, revised, improved.
-   A new theory is needed.
-   The accuracy of experiments is insufficient.
-   New and better experiments are needed.

There is a hierarchy both of theories and models, with inheritance of basic features down the ladder. Of course, the classification of models and theories is not absolute and may be highly subjective. Thus, the single-electron model in solid-state physics is partly a model, partly a theory, whereas the Kronig-Penney model is just a model. One of the best examples of a model related to a theory is the Ising model of ferromagnetism which is probably the simplest possible model of a ferromagnetic (e.g., in two directions). In this case, the theory provides a simple framework, and one can obtain ultimate results assuming a simple model of spin interactions. It is also interesting that the Ising model is rather universal: it serves as a pattern for a number of derived models in areas having nothing to do with a ferromagnetic or in general with physics.

So, the models can be thoroughly worked out, but they may have very little in common with real phenomena. For instance, models of Earth's climate do not describe the climate fully as a physical system, they just try to predict its certain features perceived as important. Traditional nuclear physics was almost totally comprised of such piecewise models, in some cases intuitively physical and sometimes more of mathematical nature (three types of nuclear models were specifically popular: the shell model, the liquid drop model, and the mean field i.e., collective motion model). A physical theory may often be reduced to a model, for instance, the so-called Standard Model which is in effect a profound theory, and we know, in particular from nonlinear dynamics, that models can be unstable, irreversible, time-noninvariant, or possess other apparently unacceptable qualities. In quantum mechanics, models are often counterintuitive which makes some people think that quantum mechanics is "weird".

---

[2] Einstein himself later stated that he had not seriously consider experimental facts, even the outstanding Michelson-Morley experiment.

In principle, it is not necessary that mathematical models should in all cases and right away make predictions amenable to direct observations. For instance, numerous models of Earth's climate are projected into the time domain about 100 years from now. Such models cannot be experimentally tested today. Likewise, most of cosmological and a great number of astrophysical models are at present in a highly speculative mode which fact, however, does not make them less valuable or interesting. After all, most contemporary cosmology and high-energy physics, in particular, superstring theories look more like Ptolemaic model building than traditional-style experimentally-based physics. In this new physics, there seems to be little chance of making definite predictions, with error margins being exactly indicated. Reality may only serve as a motivating factor (as, e.g., in inflationary cosmological models), but the model itself can be based on a conjecture and not fully supported by evidence. In such cases, one must study how to separate interpretation from the facts.

Notice that the main cosmological problem is to construct the model of the universe. It is generally accepted that some 13.6 billion years ago the universe was kind of "a hot soup" consisting of particles. Afterwards, the universe went through rapid expansion and accompanied cooling, with matter clumps being formed under the action of gravitation.

Today, there exist a number of cosmological models, mainly based on astronomical observations, for example, the so-called Standard Cosmological Model that embraces the major facts and values which appear to be firmly established and raise little doubts within the scientific community. The Standard Cosmological Model includes the set of fundamental constants, the notion of Big Bang, large-scale homogeneity and isotropy of the universe as well as the fact that the density of all kinds of matter to its critical value is very close to the critical value, $n \approx n_c$. According to modern astronomical observations, our universe consists of 70 percent dark energy, 25 percent of dark matter and only 5 percent of ordinary matter. A more specific model enabling one to reconstruct to some extent the evolution of the universe from the hot soup state related to a fraction of a second after the Big Bang up to the currently observed web of galaxies. Dark energy is a mysterious substance with unusual properties (such as negative pressure and repulsive gravity) that are outside of the present set of physical concepts.

In conclusion to this subsection, one can recollect that some outstanding scientists do not make much difference between a theory and a mathematical model. "I take the positivist viewpoint that a physical theory is just a mathematical model and that it is meaningless to ask whether it corresponds to reality. All that one can ask is that its predictions should be in agreement with observation." Stephen Hawking.

## 3.2. The economy principle

While building a model, one should not introduce new elements into it unless they are absolutely indispensable. Most mathematical and computer models are far from being optimal, and the possibility of further improvements is often quite obvious. Nevertheless, one should suppress the temptation to incessantly amend the model as tiny improvements may lead to the loss of precious resources such as time. One should always bear in mind the major task.

Particularly good models make as few assumptions and as use as few parameters as possible. The trend to the most economical formulation of mathematical models is often called the principle of Occam's razor. According to this principle, simple and more economical models tend to be better suited for predicting the outcome of modelled phenomena. In practice this often means that the algorithms built on the base of more economical mathematical models can be implemented in simpler codes with fewer errors. It is useful to keep in mind Einstein's prescription: "Everything should be made as simple as possible, but not simpler".

Indeed, there is a hidden danger in drastic simplification of mathematical and computer models: distortion of reality and the creation of myths. The myth is a conviction or a tenet whose trustfulness is never questioned by its carrier, at least for a long time.

## 3.3. Relation to mathematics

"As far as the laws of mathematics refer to reality, they are not certain; and as far as they are certain, they do not refer to reality" [118]. The effect of mathematical tools used to model physical reality on our conceptions of the world is still waiting for a thorough examination. For instance, calculus with analytical geometry were adequate for traditional (18th -19th centuries) classical mechanics, being totally insufficient for special and general relativity. As already mentioned, mathematical concepts and physical reality are not necessarily equivalent since mathematics can create its own structures consisting of abstract objects and study the relationships between such objects irrespective of physical existence. Observability criteria traditionally important for physics can be irrelevant in mathematics, where abstract entities would rather be cleared of any experimental baggage.

Many mathematicians hold the view that mathematics, especially in its purely deductive form, consists of models that, in contrast with those of traditional physics, may have no relation to reality. In particular, the system of axioms lying at the foundation of mathematical disciplines can be completely arbitrary and can have no connection with real-life phenomena. Equally unimportant may be the correspondence to reality of mathematical results produced by abstract disciplines based on arbitrary axiomatic systems. Historically, this view was promulgated by Descartes and gave rise to the whole stream of abstract thought expressed lately in the works under the collective pseudonym N. Bourbaki.

However, starting from the scientific revolution of the Renaissance whose central idea was the mechanical motion of matter (Galileo, Kepler, Leibniz and especially Newton), mathematics was inextricably connected with physics so that the classical mathematical constructions have a direct relevance primarily to the physical world, the strategy that was saliently pursued by the great mathematician Henri Poincaré. An essential driving force behind creating models of the surrounding world has always been to invent or discover mathematical structures that would accurately reflect the real-life phenomena. By the word "real" one typically understands "physical", but this is not necessarily true: for instance, biological or social processes are not directly linked to physics, and nobody has proved that such processes can be fully reduced to physics.

The adjective "mathematical" implies a certain rigor of exposition, at least the presence of formal definitions and proofs. Indeed, there are prevalent criteria of rigor in professional mathematics, and rigorous proofs constitute its main body. Yet in mathematical modeling, viewed as an applied discipline the accent is put not on the formal truth, but on understanding of the modeled phenomenon. The very word "understanding" suggests an intuitive connotation so that mathematical modeling is not oriented at rigid rules of ideal mathematical constructions; it is a totally different genre, free and informal despite using the honorific epithet "mathematical". Rigor is typically appreciated in mathematical modeling only insofar as it serves to deter negligent reasoning, with its typical manifestations such as claiming assumptions "obvious" or using "hand-waving" arguments to achieve a drastic simplification. This is a pragmatic approach: mathematics in modeling is perceived as a service subject, a tool, without imparting to it some principal significance.

One of the most popular *mathematical* models, although rarely acknowledged as such, is that of boundary conditions. The latter play a crucial role in many applications of mathematics, physics and mechanics. Thus, the no-slip condition in fluid dynamics vividly illustrates the usefulness of mathematically oriented scientific modeling. The practical importance of mathematics in modeling and, in particular, in physics consists in exploring the qualitative behavior of solutions as well as, to a somewhat lesser extent, in establishing their existence and uniqueness. Mathematical modeling is nearly always oriented at practical applications, and although results quite naturally must be proved and consequences deduced, an excessive rigor such as too much reliance on formalism (e.g., lengthy algebraic definitions and theorems) and abundance of technical details may obscure the substance and interfere with direct understanding. In particular, it would hardly be worthwhile (although possible) to axiomatize this discipline in the purely mathematical sense, e.g., á la Bourbaki: after such an axiomatization, modeling of even the simplest objects may become cumbersome and difficult to apply. Therefore, mathematical modeling treated as a separate discipline contains a lot of intuitive or completely fuzzy concepts which makes this discipline reproachable from purely mathematical positions. Anyway, the standard requirements of mathematical rigor are typically significantly relaxed in mathematical and computer modeling. The traditional motto of physicists in the 20th century was "First results then rigor" (possibly no longer), this principle being adopted also in mathematical modeling.

Nevertheless, one can often notice that in the mathematical literature powerful and constructive methods of modeling have drowned in elaborate technicalities employed to attain maximum generality (it is not always clear what such a generality is needed for). Unfortunately, most mathematical texts are presented in such a form that the concepts of rigor and accessibility appear mutually exclusive. Probably there exist examples when mathematical aspects of a problem are provided in a simultaneously rigorous and acceptable manner, but if such examples do exist they are exceedingly rare. One may also notice that such powerful disciplines as theoretical and mathematical physics are far from being mathematically impeccable. Even the mathematical truth (perhaps like any truth) is relative: finally, the truth is what the majority does not dare to object to. A statement of a mathematical model can be true or false, depending on the situation to be modeled and on how it is interpreted. This relativism (commonly viewed as a deficiency) is a natural consequence of the human language employed to express mathematical ideas and structures. Formulas abundantly used in mathematical modeling are indispensable to describe the relationships between the model elements.

Pure mathematicians sometimes proudly claim that they do not care which of these structures can be materialized in nature, this mundane task being left to physicists and engineers. In extreme cases (such as, for instance, followed by the Bourbaki adherents), it is even stipulated that making mathematics for physics is nearly a disgrace for the human mind. Pure mathematics resembles abstract art: one can admire or even idolize it but cannot directly apply it to any practical situation. In the Bourbaki tradition, it is customary to label everything applied as synonymous with uninteresting or trivial, which attitude leads to more restrictions than innovations. However, the term "applied" is subjective with no universal definition. For example, differential equations, both ordinary and partial, are labeled "applied mathematics" in some university curricula. The principal role of mathematics goes beyond providing technical tools for solving applied problems. More important, mathematics serves as a guide evoking our understanding of underlying structures and basic regularities.

Mathematics also enables us to link together seemingly diverse subjects – in this sense, mathematics has the same playful origin as arts.

Modeling as a specific mathematical genre has its own principles of rigor, when intuitive understanding is at least as valuable as formal proofs. Note that the same attitude is common in physics. At any rate, rigorous proofs and exhaustive definitions are not perceived as a matter of principle, and having one's own definition for conventional terms is a perfect way to create and spread confusion. In this text, in accordance with the general ideology of modeling as an essentially applied discipline, involved mathematical proofs that are not of primary importance will be omitted. The more logically minded readers will hopefully forgive the informal, discursive manner of presentation adopted in the present paper. The reader who strives for complete mathematical details should look elsewhere.

## 3.4. Equations used for modeling

To formulate a mathematical model of a phenomenon or a process means in this book to provide a full (closed) system of equations that enable one to describe the evolution or the equilibrium state of the studied object (the latter can be both an isolated item and a medium). One typically uses the following equation types to model the real-life processes.

1. Differential equations
2. Integral equations
3. Integro-differential equations
4. Difference equations
5. Algebraic equations.

Apart from equations, inequalities and constraints are used, especially in optimization problems. Each of these five equation classes is associated with a specific modeling area. Thus, differential equations are mostly related to evolutionary processes (ODE) and spatial spread (PDE). Modeling with integral equations mainly deals with inverse source problems and other kinds of incorrect problems. Integro-differential equations are typically employed for radiation transfer modeling, e.g., in reactor safety and climate variability models. Difference equations arise in numerical modeling and computational physics. Algebraic equations are extensively used to explore stability of systems and processes.

We have already seen that real systems and processes can be described by nonlinear equations, for example, by nonlinear partial differential equations with respect to the unknown function being differentiated over time and spatial coordinates. Such equations model the systems distributed in space and have an infinite number of degrees of freedom. If the equations modeling the system do not contain spatial derivatives (ODE-based models), such a system is usually called point like or having a null dimension. In general, there are two classes of objects studied in physics, fluid dynamics, engineering, biology and, lately, in a number of behavioral disciplines: particles and fields. These two large classes are usually described, respectively, by ODEs and PDEs. Fields, in distinction to particles, possess an infinite number of degrees of freedom which requires a functional-analytical approach to their study, both on the classical and quantum level of description. It is remarkable that the notion of a field as a force-carrying entity appeared rather early in human history: Newtonian gravity was actually based on the field concept.

One can notice that, in today's language, within the classical study we can deal with the fields using the familiar concept of mapped manifolds whereas in quantum theories additional difficulties arise (e.g., related to renormalizability) that require certain dedicated techniques to be dealt with. Such difficulties, however, are known today to be only relevant to the mathematical models of physics and hardly appear in the models related to other subjects.

In some cases, when modeling the processes evolving with time with the aid of ODEs, we can neglect the terms with higher derivatives (this is equivalent to the forceful contraction of the phase space). The curtailing of equations does not go without a penalty: jumps would appear so that the model ought to be supplemented with ad hoc conditions replacing time evolution at those stages, when the higher derivatives can significantly affect it. The point is that higher derivatives can essentially influence evolution, irrespective of the smallness of their coefficients.

From the physical standpoint, a dynamical system is any physical object that changes with time. A dynamical system runs through the states in its state space still preserving its identity. The state space, which is equivalently known in classical mechanics as a phase space is actually the space of solutions of the motion equations. A point in the state space gives the state of a classical dynamical system which distinguishes classical systems from quantum ones, where states are given by non-zero vectors in a complex vector space endowed with Hermitean inner product.

In the present book we shall only discuss classical deterministic systems. For instance, an oscillating pendulum, a living creature, a business entity, a country's economy, a social group or Earth's climate can be treated as classical dynamical systems whose evolution follows some deterministic (non-random) prescriptions. The state space contains all the trajectories along which the system evolves. In this book, we shall mainly consider continuous-time ($t \in \mathbb{R}$) dynamical systems, although discrete-time ($t \in \mathbb{Z}$) systems will also be treated in some models (e.g., the logistic map). Time semi-axis ($t \in \mathbb{T}_+$) can be defined as $\mathbb{T}_+ := \{t \in \mathbb{R}, t \geq 0\}$. The treatment of natural or technological processes in the form of dynamical systems is quite general and mathematically convenient since it allows us to incorporate new phenomena by appropriately enhancing the phase space dimensionality. For example, if the equations of motion had involved higher-order derivatives in variable $t$ (time), we could still have produced a first-order vector equation by increasing the number $n$ of dimensions. Likewise if the motion equations had incorporated other dependent variables $u^1(t), u^2(t), \ldots$ in addition to $x(t) \in Q$, where $Q$ is the configuration space, we also could have dealt with this situation by enlarging the dimensionality of the phase space that is usually identified with the cotangent bundle $T^*Q$. This concept generalizes the standard approach of classical mechanics (and, more commonly, of classical physics), where the state of a system is described by a point in the phase space whose mathematical structure is determined by the system's details such as the numbers of degrees of freedom, the range of the variables specifying the state of the system, etc. A well-known example of a phase space is that of a mechanical system whose configuration space is a manifold $Q$ and the phase space is given by the cotangent bundle $T^*Q$ (endowed with a canonical symplectic form $\omega = dp_i \wedge dq^i$).

One of the most important classes of equations used to model natural processes are the wave equations. They allow one to describe phenomena evolving in both time and space, often on equal footing. One can speculate that there should exist a proper wave equation for each kind of currently known interaction – gravitational, electromagnetic, weak and strong. The well-known (since the late 19th century) electromagnetic waves are conveniently long since the electromagnetic (EM) interaction is long ranged in the human scale, mostly having an infinite radius (as photons, the carriers of EM forces are massless). Likewise, the hypothetical gravitational waves are also assumed to be rather long since the corresponding gravitational interaction is also long-ranged (it is due to these great wavelengths of gravitational radiation that modern ultrasensitive devices to be used in gravitational-wave experiment such as LIGO – laser interferometric gravitational-wave laboratory have the dimensions of several kilometers). On the other hand, both the weak and strong forces are short ranged so that the corresponding waves should have miniature wavelengths on the human scale and are capable of propagating over very short, e.g., subatomic distances.

We may note that one can be thrilled by equations, specifically by symmetrical and handsome ones. Such good-looking equations can even be useful for constructing idealized models, but not entirely true. Thus, one should not be misled by the "beauty" of equations. In general, when a mathematical model of a natural process is built one should not be fooled by equations.

## Section 4. Mathematical models of physics

Physicists have developed a modeling way of thinking due to the necessity to study complex phenomena. Recall in this connection that there exists today a special discipline called physics of complex systems i.e., essentially those having a great number of degrees of freedom. To study such systems, one has to take into account a large number of factors.

We may notice that physics contains both theories and models, in contrast to weakly formalized disciplines such as, e.g., economics and finance that have only models (see Section 10.3. "Modeling of weakly formalized concepts"). In physics, there are roughly a dozen laws that describe 99 percent of reality, while in weakly formalized disciplines there are 99 laws that can be applied to a negligible part of reality. Although mathematicians tend to alienate themselves from physicists, in mathematical modeling one can hardly group the ideas separately: by mathematical as opposed to physical subjects. A serious physicist always makes *mathematical* models – otherwise these cannot be regarded as models at all. By its very idea, mathematical modeling is trying to loosely depict a patch of reality (a correct definition of the term "reality" is hardly known and would probably be unnecessary, at least in this text). The real-life phenomena typically depend on many parameters, and it may be hard to tell which of them should be taken as variables and which are external. Thus, even roughly constructing a good model may present a difficult task. When talking about a physical theory, one may ask not only whether the theory is correct but also whether it may be considered complete. Both are pertinent questions in the case of a theory. One can say that a theory is correct if it reaches good agreement with experiment. It seems to be useful to always bear in mind that an attempt to comprehend the world as a whole by contemplation alone, without proper attention to empirical verification, is usually known as metaphysics.

As the requirement of correspondence to physical reality may be weakened for mathematical models in general as well as more and often in modern physics, mathematical models can bring applications that are wider than those imposed by the traditions universally accepted in old-time physics.

Mathematical models of physical systems usually turn around two concepts: states and observables. More exactly, a mathematical model of a physical system can be based on the notion of the latter as consisting of two sets of objects, states $S$ and observables $\Omega$, with elements of these sets describing respectively the state of a system at some instant $t$ of time and the probability distribution (or exact value) for the observed quantity $A \in \Omega$ in state $\psi \in S$. This is the standard picture which is, e.g., customary in orthodox non-relativistic quantum mechanics. In particular, in quantum theory an "observable" is represented by a respective self-adjoint operator that maps the Hilbert space (i.e., the state space embracing all possible quantum states i.e., vectors in a complex vector space) into itself. Notice that in general one does not require that the sum of any two observables should also be an observable i.e., the set of observables does not reduce to an algebra.

There must be, however, one more mathematical structure reflecting how a system (physical or of other nature) evolves in time i.e., the system's dynamics. The latter is typically defined through a collection of mappings (usually diffeomorphisms) from instant $t_0$ to $t$ that are known under the names Green's function, propagator, evolution operator, flow, etc. taking state $\psi$ from $t_0$ to $t$. In the classical picture, the evolution operator is often denoted as $g_{t_0,t}$ that produces point $x(t)$ belonging to state space $S$ when acting on $x_0 = x(t_0) \in S$, $g_{t_0,t}\,x_0 = x \in S$, $S$ is usually understood as a differential manifold. In short, evolution in dynamics is a prescription for producing the next state or configuration from the actual one. The state space $S$ in classical theories is often called the phase space $P$, $S := P$. In case dynamics does not depend on time, $g_{t_0,t} = g_{t-t_0} \equiv g_\tau$. Notice that the evolution operator completely describes the dynamics, $g_\tau : P \to P$.

One can also say that a theory is complete in some area if it sustains the description of any item inside this area. Yet both questions (correctness and completeness) are often irrelevant when one deals with mathematical models. The latter can be good and beautiful and yet have very distant relationship to experiment (Thomson atom, a number of nuclear models, time-reversible dynamics, "theory of everything", wave function of the universe, certain evolutionary models, mathematical models in finance, string theories, etc.).

In physics, one always tries to reduce the complexity, in particular, by replacing a real-life system depending on many parameters and requiring a lot of data to describe it by some minimalistic model. Building such minimalistic mathematical models restricted by the "much less" and "much greater" inequalities has traditionally been the subject of theoretical physics which is progressively more and more detached from experiment: the necessary requirement (in the past) of empirical verification for both hypotheses and theories based on them has recently been drastically weakened[3]. Nevertheless, physicists tend to introduce a variety of modeling examples, in particular, to show the benefits of a certain concept. Moreover, new computational procedures are best developed, tested and refined on simple examples.

Newtonian point dynamics or Boltzmann gas in statistical mechanics are typical examples of such minimalistic models based on complexity reduction. Physics most efficiently models elementary processes such as the motion of an object influenced by gravity. In such situations, an initial cause defines a prescribed outcome. However, the real-life processes evolve in such a way as if they were capable of choosing between several alternatives. Already a rather simple physical system such as a particle moving in a potential field behaves is if it could explore – "sniff" – the route taking the most optimal integral trajectory. This observation leads to the variational formulations pervading the whole of physics. The variational formulation of a physical theory is actually a hypothesis suggesting that the equations of motion in this theory are extremals of a certain functional known as action. In sufficiently complex systems, the variational formulation of physics can be generalized to help find the best possible influence in order to take a dynamical system between two prescribed states (optimal control).

There are different approaches to the description of physical phenomena adopted by mathematicians and physicists. In particular, one can observe that mathematicians and physicists typically use different tools to describe analogous objects, although they both identify such tools with the same word "mathematics". For example, in modern mathematics mostly a coordinate-free language is increasingly used whereas calculations in physics are nearly always performed in the coordinate description. Actually, it does not seem to be easy to obtain concrete (in particular, numerical) estimates in physical applications without using a coordinate representation. More generally, mathematicians implicitly assume that if initial postulates correspond to the properties of the surrounding world, then the derived results, theorems and implications should automatically describe reality. Physicists, however, require verification and experimental testing of each theoretical result.

---

[3] This almost hypothetical status of modern physical theories leads to regarding theoretical physics as a chimeric discipline between physics and mathematics, a cross of physics without experiment and mathematics without rigor: "La physique théorique est l'alliance de la physique sans l'expérience, et des mathématiques sans la rigueur". Jean-Marie Souriau (1922-2012), a prominent French mathematician, one of the founders of symplectic mechanics who is known, in particular, by the book "Structure des systèmes dynamiques".

Coordinate-free representation is in many cases convenient since it allows one to evade tedious accounting of indices but, on the other hand, coordinate-free representation can be rather far-fetched and cumbersome. Therefore, we shall use in this text mostly the coordinate representation.

Mathematical models of physics are always constructed in some spaces which are the sets of elements, in general of any kind, endowed with an appropriate mathematical structure (recall that standard mathematical structures typically have an algebraic or a topological nature). In principle, the following topics comprise the main part of today's physics: basic classical mechanics with its variational principles; quantum mechanics, with its applications in atomic, nuclear, particle, and condensed-matter physics; relativistic quantum theory including the concept of a photon, the Dirac equation, electron-photon interaction (QED), and Feynman diagrams; quantum fields; and general relativity. Elementary working knowledge of these topics would be sufficient for a solid foothold in the physical community. However, this standard repertoire is rather limited, and one can try to list the major concepts that constitute the real bulk of mathematical models in physics. One might call these major concepts "worlds of physics" to be organized as a tree-like structure: with branches, leaves and buds as individual models. There are links between the "worlds" invoking substructures with repeatable, reusable patterns. In this book, however, we shall give only a telegraph-style account of the main items.

## 4.1. Physical models and reality

Physics can rarely be reduced to simple and handsome problems such as the two-body problem describing, in particular, the motion of a planet around the Sun. Therefore, situations that are difficult to predict are often encountered in physics, and to cope with such situations one has to resort to drastically simplified mathematical – analytical or numerical – models, primarily to get insight and not to predict experimental data. For example, when studying turbulence, the purpose of modeling by treating (mainly numerically) the Navier-Stokes equations is not to explore specific realizations of a fluid flow, but rather to comprehend the behavior of fluid in the transition to turbulence and the stochastic nature of the latter. Note that even the slow and regular motion of planets around the Sun can become chaotic and hard to predict over long time scales ($10^7$ years).

Physical facts are usually so entangled that when one starts studying just one of them on a deep level, it soon becomes obvious that one has to consider a bunch of others. In this sense, physics is a fair example of an interdisciplinary effort providing numerous examples of connections with other branches, and it is only natural that the restrictive axiomatic approach to physical phenomena almost invariably fails before model representations are attempted. Yet for many physicists, interdisciplinary subjects appear too loosely determined and perhaps a bit foreign to devote the full-scope of attention to them.

Physics contains many vivid examples of a contradiction between mathematical models and physical facts. For instance, people used to believe in symmetries, although the latter are rarely observed in reality. Here, the esthetic preferences inherent in mathematical modeling criteria trump real-life experience. One might recall that before the mid-1950s all physical processes were considered mirror (parity) symmetric, and all experimental evidence for parity non-conservation was simply ignored. This effectively led to the prohibition of parity-breaking phenomena, although a simple observation shows that nearly all biological objects are not reflection-symmetric even at the molecular level. It is curious that the faith in parity conservation was so strong at that time even among the most outstanding physicists that all Hamiltonians used in physical models were allegedly "censored" with respect to mirror symmetry, $H = PHP$, where $P = P^{-1}$ is the parity operator having the $\pm 1$ eigenvalues (this operator inverts the sign of radius-vector **r**). Parity conservation i.e., the postulate that all physical processes would be the same if seen through the looking glass remained a matter of universal belief until T. D. Lee and C. N. Yang demonstrated in 1956 that parity is not necessarily preserved, at least in weak interactions [102] (the 1957 Nobel Prize in physics).

Looking in the mirror that interchanges left and right one will notice no difference. A similar belief in the mirror left-right symmetry has been universally maintained until recently as regards already mentioned time reversal symmetry (TRS), although the evidence is abound that most processes in physics and life sciences are not time reversal ($T$) symmetric. The typical argumentation of true believers in ubiquitous time reversal symmetry was that the microscopic equations of physics, primarily Newtonian laws, are invertible in time which necessarily implies that the whole of physics, to the final analysis, should be $T$-symmetric. Many ingenious hypotheses have been invented to advocate invertibility of physical processes in time such as, e.g., that TRS would be ensured if there were a complete equality of particle and antiparticle concentrations in the universe, an assumption that would be difficult to verify. In fact, both parity and time reversal symmetries can be regarded as very special cases of Lorentz transformations (see below). Indeed, one can write matrices representing such transformations in the form

$$L = \begin{pmatrix} 1 & \mathbf{0}^T \\ \mathbf{0} & R \end{pmatrix},$$

where $\mathbf{0} \equiv (0,0,0)^T$ and $R$ are matrices of 3d rotations (preserving $\mathbf{r}^2 = x_i x^i, i = 1,2,3$). We observe that $x^i$ and $t$ can be separately changed or unchanged. In other words, parity transformations $P = $ diag$(1,-1,-1,-1)$, det $P = -1$ and time-reversal transformations $T = $ diag$(-1,1,1,1)$ are parts of the Lorentz group.

The lesson learned from considering the intuitively assumed symmetries in physical modeling is that hopes and desirable features used to build simple models may have nothing to do with crude reality that may appear "ugly". Today, one often uses the notion of "spontaneous breaking of symmetry" to designate the deviations from handsomely symmetric mathematical models, in particular, formulated through the underlying equations invariant under a wide class of symmetry transformations. At best, the "ugly" reality can be described as specific solutions to such equations.

Furthermore, if one takes quantum-mechanical calculations, the most salient contradiction (although never emphasized by informed physicists) is between the concepts of stationary states and discontinuous jumps, both being the main models studied in non-relativistic quantum mechanics. The matter is that stationary states of standard quantum mechanics cannot produce the electromagnetic radiation that we observe in the universe unless some unknown perturbation (e.g., due to vacuum fluctuations) is introduced. The probabilistic jump transitions in non-relativistic quantum mechanics are poorly compatible not only with the quasi-classical notion of continuous intermediate states, but also with spontaneous emission of radiation. In the quasi-classical picture, an electron suddenly leaves one surface of constant energy to reappear on another, without any continuous path connecting the two surfaces. Only in quantum electrodynamics (QED), where not only the particle motion but also the electromagnetic field is quantized and coupled with charged particles, atomic orbitals are no longer the eigenstates of the blended system of electrons and photons. Therefore, transitions between the "stationary" atomic states become physically understandable due to the excitation of the photonic degrees of freedom in the entire "sea" of electromagnetic oscillations. It was a great achievement of physics when fields were understood as fundamental objects, on an equal footing with particles. However, since any field, in particular the electromagnetic field $F_{\mu\nu}$, has an infinite number of degrees of freedom, the problem of correctly calculating transitions between the states of a combined electron-photon system is very difficult to treat even in the quantum field theory (QFT) [136].

It is not a rarity when scientists emphasize a particular set of data or a preferred theoretical model. It happens even in "hard sciences" such as physics or mathematics. For example, the theory of superconductivity is full of biased theories and concepts, especially in association of the high $T_c$ superconductors such as those discovered by G. Bednorz and A. Müller in early 1987[4]. In mathematics, Dirac's delta-function – an extremely useful tool – had been opposed for many years, before the theory of generalized functions was established. One can observe many conflicting "schools" and "teams" insisting rather aggressively that everything deviating from their views should be unconditionally discarded. The disdain for the concepts or models that do not support those adopted within a given community of scientists is not at all extraordinary.

## 4.2. Ten worlds of physics

Conceptually, the present book is mainly concentrated on the physical sciences, with most problems not being solved (nor even attacked) but rather identified. One might try to extract a few principal abstractions characterizing physics as a collection of mathematical models. As already commented, despite seeming abstractions, physics can be viewed as an example of a really good and practical collection of models. The abstractions mentioned initially appear as the forms of human language designating new concepts, but afterwards they tend to lose their terminological footing in science and become cultural symbols (recall such notions as relativity, quantum, evolution, spacetime, symmetry, etc.).

The history of developing mathematical models in physics began with the study of systems containing a finite (usually small) number of particles either interacting through the forces between them or moving due to external forces. This was the world of classical mechanics (see the subsection below). Later it evolved into the world of continuous mechanics and the world of statistical mechanics, when the number of particles (the considered degrees of freedom) became too large for direct mechanical (Newtonian or Lagrangian) treatment. At the beginning of the 20th century, three new generalizations (limiting cases) were considered: 1) when the particles are so small that the very act of observing them affects their motion, the world of quantum mechanics appeared; 2) when the particle velocities become comparable with the speed of light, the world of relativistic mechanics and, in general, of relativistic concepts emerges; 3) when the scale of the studied system is extremely large, the world of general relativity becomes indispensable.

---

[4] It is remarkable that the first high-temperature superconductors were not discovered by physicists - neither by experimenters nor by theorists – but by chemists.

Here we delineate the principal worlds of physics where closed mathematical models (i.e., nonintersecting with other worlds) can be constructed.

1. *The classical world*

The classical world of physics is based on the following key notions:

The Galilean group (inertial systems)

Newton's law of motion (classical limit of special relativity and quantum mechanics)

Newtonian gravity (classical limit of general relativity)

Kepler's problem (rotation of planets about the Sun)

Potential fields, classical scattering

The Euler-Lagrange equations

Variational schemes

Noether's theorems and conservation laws, conservative systems

Hamiltonian equations, Hamiltonian flows on symplectic manifolds

The Hamilton-Jacobi equation

Motion on manifolds, constraints

The Liouville theorem

Key figures in this world: Galileo, Kepler, Newton, Euler, Lagrange, Hamilton.

2. *The thermal world*

Classical thermodynamics (equilibrium)

The nature of heat, temperature, heat transfer

Interconversion of mechanical work and heat, engines and cycles

Heat capacity ($C = dQ/dT$)

Laws of thermodynamics, thermodynamic potentials, Gibbs' concept of equilibrium, entropy

Heat as the particles motion, Maxwell and Boltzmann distributions, statistical mechanics

Thermochemistry, chemical reactions

Equations of state

Phase transitions, Ginzburg-Landau model

Low temperatures, superfluidity and superconductivity.

Key figures: Boltzmann, Carnot, Clausius, Fourier, Gibbs, Ginzburg, Joule, Landau, Lavoisier, Maxwell.

3. *The nonequilibrium world*

The Liouville equation, Gibbs distribution

Open systems, reversible and irreversible processes, entropy production

Kinetic equations, Boltzmann equation, Bogoliubov's hierarchy

Diffusion, Brownian motion, Langevin equation, Fokker-Planck equation, multiple scattering theory

Fluctuation-dissipation theorem (FDT), linear response theory, Kubo formula, Onsager's reciprocal relations

Nonequilibrium phase transitions, time-dependent Ginzburg-Landau model

Classical stochastic models, nonlinear regime, branching and bifurcations, stability of nonequilibrium stationary states, attractors.

The Poincaré map, logistic model, dynamical chaos, indeterminism (impossibility of predictions)

Dissipative structures, order through fluctuations, Turing structures

Chiral symmetry breaking and life.

Key figures: Bogoliubov, Boltzmann, Gibbs, Klimontovich, Krylov, Kubo, Onsager, Poincaré, Prigozhin, Ruelle, Sinai, Stratonovich.

4. *The continuum world*

The Euler and Navier-Stokes equations

Hyperbolic flow equations, shock and rarefaction waves

Compressible gas dynamics and supersonic flows

Self-similar models and explosions

Turbulent flows and the models of turbulence

Elastic solid models

Viscoelasticity, plasticity, composites

Seismicity, seismic ray propagation and seismic ray theory

Acoustics, sound wave/pulse excitation, propagation and scattering

Detonation and flames, propagation of fires

Key figures: Archimedes, Bernoulli, Euler, Kolmogorov, Pascal, Rayleigh (J.W. Strutt), Stokes, Zel'dovich.

5. *The electromagnetic world*

Maxwell's equations

Electrostatics, Laplace and Poisson equations

Interaction of electromagnetic (EM) fields with matter, atoms and molecules in the electromagnetic field, electromagnetic response of material media, linear and nonlinear susceptibilities

Linear and nonlinear optics

Propagation of electromagnetic waves and pulses

Diffraction and scattering of electromagnetic waves

Electromagnetic radiation

Rays of light, asymptotic theories, coherence of light

Photometry, radiometry and colorimetry

Key figures: Ampère, Faraday, C. F. Gauss, O. Heaviside, J. C. Maxwell, Rayleigh (J. W. Strutt).

6. *The plasma world*

The plasma dielectric function, linear waves in plasma

Screening in plasma, correlations of charged particles

Hydrodynamic models of plasma

Distribution functions, kinetic models of plasma, collision integrals of Boltzmann, Landau, Klimontovich, Lenard-Balescu, etc.

Collisionless plasma, a self-consistent field model (the Vlasov equation)

Plasma in external fields, the magnetized plasma

Landau damping

Theory of plasma instabilities

Quasilinear and nonlinear models of plasma.

Key figures: Debye, Klimontovich, Landau, Langmuir, Vlasov.

7.  *The quantum world*

The particle nature of electromagnetic radiation, the Planck hypothesis

The duality of light, photoeffect (A. Einstein, 1905)

The Bohr atom

De Broglie hypothesis, hidden parameters discussion, a variety of interpretations

The Schrödinger equation, wave functions

Observables, measurements, probability amplitudes, entanglement

Wave packets, Heisenberg uncertainty relations

The Heisenberg-Weyl algebra, representations of compact Lie groups (H. Weyl)

The theory of unbounded self-adjoint operators (J. von Neumann), Hilbert space

Rotation group representations, spin

The Sturm-Liouville problem, discrete spectrum, eigenfunction expansions, Green's functions

Density matrix, the Wigner function (E. Wigner, 1932), Husimi and tomographic representations

Unitary evolution, semigroups

Eigenvalue perturbation theory, iterative procedures

Quantum-classical correspondence, canonical quantization, decoherence

Asymptotic expansions, semiclassical limit

Scattering theory, S-matrix, continuous spectrum

Integral equations, inverse problems

Decaying states, resonances

Periodic potentials (Bloch, Brillouin, Kramers), Floquet and Hill equations

Solid state physics, semiconductors, transistors, engineering applications of quantum mechanics

Many-body problems, second quantization, elementary excitations, condensed matter physics, Coulomb energy, thermofield dynamics

Ergodic potentials, Anderson localization

New developments, EPR and hidden parameters debates, Bell's inequalities, Bohm version, new interpretations

Quantum computing and quantum cryptography

Key figures: Bethe, Bohr, de Broglie, Dirac, Einstein, Fock, Gamov, Heisenberg, Landau, von Neumann, Pauli, Planck, Schrödinger, Wigner.

8. *The high energy world*

The Dirac equation, quantum vacuum, positron, antiparticles

Relativistic quantized fields, infinitely-many degrees of freedom, particles as field excitations, bosons and fermions, spin-statistic theorem

Quantum electrodynamics, particle-field interactions, Feynman diagrams, renormalization

Path integrals, Feynman-Kac formula

Strong (nuclear) interaction, $\pi$-mesons, exchange forces, Yukawa's model

New quantum field theories, scalar, vector, tensor fields

Resonances, supermultiplets

Barions, mesons, hyperons

CPT-theorem[5], group concepts in particle physics

K-mesons, particle mixing, C and P nonconservation, CP and T violation

Isospin, strange particles, "elementary particle zoo", SU(2), SU(3) - first attempts of Lie group classification

Early quark models (1960s), color, flavor, charm, etc.

Hypercharge, Gell-Mann - Nishijima relation

---

[5] L. D. Landau considered CPT not quite a theorem but an exact and almost trivial statement that should be satisfied by any physical Lagrangians. In fact, the CPT-theorem can be interpreted as a first attempt to unify spacetime (P, T, PT) and internal (C) symmetries. Notice that charge conservation C is an exact transformation that is not a spacetime symmetry (i.e., not generated by Poincaré's group). As to parity (P) nonconservation, Landau initially considered it nonsense asserting that "space cannot be asymmetric". Later, he began regarding the CP conservation to be an exact law of nature [22]. See also the footnote 1 (page 18).

Cross-sections, form-factors, S-matrix, current algebra

$J/\psi$-meson, confirmation of charm, quark-antiquark system

Quark-quark interaction through gluons, quark-gluon plasma

Quantum chromodynamics (QCD), confinement, deconfinement, asymptotic freedom

Electroweak interactions, the Standard Model, W- and Z- bosons, spontaneous symmetry breaking, Higgs particle

Non-abelian gauge theories and models, Yang-Mills theory

Grand Unification, new proposed theories and models: SO(10), left-right model, technicolor, SUSY, etc.

Strings, superstrings and M theories

Key figures: Faddeev, Feynman, Gell-Mann, Glashow, Goldstone, Higgs, t'Hooft, Mills, Polyakov, Salam, Weinberg, Witten, Yang.

## 9. *The relativistic world*

The Michelson-Morley experiment

Lorentz transformations, relativistic kinematics

Special relativity, Einstein's 1905 paper

Minkowski space, Poincaré group

General relativity, Einstein's 1915 paper

Redshift of spectral lines, deflection of light, time delay by gravitation

Relativistic mechanics, energy and momentum of relativistic particles, $E = mc^2$

Accelerators, relativistic nuclear physics

Gravitational radiation, gravitational wave detectors

Controversies over relativity, tests of special and general relativity

Key figures: Dirac, Dyson, Einstein, Feynman, Hilbert, Lorentz, Minkowski, Schwinger, Weinberg.

## 10. *The cosmological world*

Spacetime curvature, the spacetime of general relativity, Einstein's equations

Solutions to Einstein's equations

Early cosmological models, non-stationary metric, redshift, Hubble constant, FLRW (Friedman-Lemaître-Robertson-Walker) isotropic model

The Big Bang, relic radiation - cosmic microwave background (CMB), expanding universe, time asymmetry, evolution of the universe

Black holes, escape velocity, Schwarzschild solution, Chandrasekhar limit, event horizon, spacetime singularities

Astrophysics, radio, optical, infrared images, gravitational lenses

Early universe symmetry breaking, cosmological phase transitions, topological defects, cosmic strings and structures

Anthropic principle and other speculations (large number hypothesis, multiverse, traveling back in time, a great number of dimensions, etc.)

Cosmological constant, vacuum energy, inflationary models

Hartle-Hawking wave function and its criticism

Quantum gravity

Strings and extra dimensions

Universe: finite (Aristotle) or infinite (Giordano Bruno)

Dark energy, dark matter (hidden mass), WMAP.

Key figures: Chandrasekhar, Connes, Friedman, Hawking, Hubble, Laplace, Lemaître, Penrose, de Sitter.

Unfortunately, due to the considerations of brevity few of these "worlds of physics" will be discussed in the present book. Yet we shall allow ourselves to make a few comments in connection with the "worlds of physics" that may seem trivial to professional physicists but can be useful to possibly interested lay readers or students.

Although one might get the impression that the "worlds of physics" are presented in the form of a list of independent subjects, the latter are actually interconnected. In general, one can seldom find a pure and precisely focused topic both in physics and elsewhere (hence there are many lateral associations in this book). It is difficult to combine current physical theories into a coherent picture, despite the fact that the "worlds of physics" strongly overlap and have multiple links to each other. In particular, classical mechanics, quantum mechanics, classical field theory, quantum field theory, high energy physics (the Standard Model), general relativity, string/M theory seem essentially different, yet these theories are often based on common mathematical techniques. The word "different" means that these are a basic set of theories that can be constructed independently of one another.

One may regard "worlds of physics" as clusters of suitable intrinsic models, and one can, if desired, even find notorious contradictions between models belonging to different clusters (i.e., built up on the base of different theories). For instance, such a basic notion as particles in physics should be treated as pointlike in classical relativistic models and as extended in quantum models. Another contradiction is associated with fixed background metric of special relativity and "old" quantum field theory, which is badly compatible with the dynamic spacetime of general relativity.

## 4.3. Some remarks on special and general relativities

It is commonly thought that general relativity is a natural prolongation of special relativity. However, there are big differences between the two, and one risks falling into confusion unless such differences are properly accounted for. For instance, there are only relative particle velocities in special relativity whereas there are in fact no meaningful relative velocities in general relativities unless two particles are located at the same spacetime point. Otherwise, to compare the particle velocities one has to make a parallel transport of the corresponding vectors, and the parallel transport in a curved spacetime (in a curved manifold) depends on the chosen path.

Moreover, in special relativity one holds the concept of the global inertial system (inherited from Newtonian mechanics). This concept is, however, poorly compatible with the omnipresent gravitation field since one never can screen it. One can only locally compensate it, as, e.g., in a free fall – the "loss of gravity" in space vehicles is based on such local compensation. But one can never neutralize the inhomogeneity of the gravitation field: for example, if we declare a frame system on the Earth associated with some meridian (e.g. Greenwich 0° longitude) to be inertial, which would be an approximation of course, then a laboratory located at any other meridian (longitude ≠ 0°) would move with acceleration with respect to the lab at Greenwich due to the inhomogeneity of the Earth's gravitation field.

One can notice that general relativity is based on rather wild speculation. Recall that according to Einstein's hypothesis the main object of general relativity is the gravitation field which is regarded just as metric $g_{ik}$ in 4d spacetime $M^4$, and the evolution of the gravitation field (encoded in Einstein's motion equations) is described through the dynamics of the metric tensor field $g_{ik}$. Metric $g_{ik}$ is characterized by nonzero curvature which is proportional to the intensity of the gravitation field. In general, the curvature measures how the given connection (see below) differs from locally trivial or flat, as, e.g., in $M^n = \mathbb{R}^n$. In physical language, curvature is the field strength. Curvature usually looks like a tensor field or a differential form.

The field equations of general relativity (Einstein's equations) are, up to some constants, of the form $G_{\mu\nu}(g_{ik}) = T_{\mu\nu}$, where the left-hand side denotes a second rank tensor (the Einstein tensor) appropriately built with the aid of spacetime metric tensor $g_{ik}(x), x \in M^4$ and $T_{\mu\nu}$ is the energy-momentum tensor of the matter flow. Assuming some infinitesimal conservation laws to be valid (local preservation of stress-energy is actually a physical consideration), the energy-momentum tensor is declared to be divergence-free, $\nabla T_{\mu\nu} = 0$ or, in local coordinates, $\nabla_\mu T^{\mu\nu} = 0$. The same conservation law is required to hold in other – non-Einsteinian – relativistic theories of gravitation.

Certain physicists, mostly trained in elementary particles or high energy physics, are to these days reluctant to accept the geometric nature of general relativity, preferring to treat it as a usual quantum field theory (i.e., a Lagrangian theory on a fixed Minkowski background $M^4$ as metric tensor $g_{ik}(x)$ tends to the constant Minkowski tensor $\gamma_{ik}$ of special relativity corresponding to geometry on $\mathbb{R}^4$)[6]. This point of view is of course a hypothesis which obviously defies Einstein's guiding idea that spacetime has dynamical properties of its own and cannot be regarded as a passive background i.e., just an ambient real vector space $S^{p,q}$ on which a nondegenerate indefinite $(p,q)$ quadratic form is defined, $(p,q)$ being its index.

---

[6] Although in some versions of fixed-background theories of gravitation the underlying space can be non-Euclidean.

The Minkowski space $M^4$ representing the spacetime of special relativity is just a a pseudo-Euclidean $\mathbb{R}^4 = \mathbb{R}^{1,3}$ space characterized by quadratic form

$$s^2 = (x^0)^2 - (x^1)^2 - (x^2)^2 - (x^3)^2$$

(s is known in physics as an "interval") which is preserved by the group of linear orthogonal (pseudorotation) transformations $\Lambda = O(1,3)$ known as the Lorentz group. An invariance of the interval under the transformations of the Lorentz group is one of the most important mathematical models of physics, since Lorentz transformations connect inertial frames moving with relative velocities with respect to one another. If one adds spacetime translations $x^\mu \to x^\mu + a^\mu$, one gets an inhomogeneous Lorentz group known as the Poincaré group $P := P(\Lambda, a) = \Lambda \times \mathbb{R}^4$ which can be viewed as the group of transformations allowing one to roam over inertial systems.

Note that Minkowski space or, rather, Minkowski spacetime still seems to be a good model of the universe since its currently observable domain is nearly flat. When the "forces of gravitation"[7] can be safely disregarded, all physical processes develop in the Minkowski (i.e., flat) spacetime. For practical purposes one can treat the Minkowski spacetime $M^4$(Minkowski four-manifold) as a real four-dimensional vector space. We may also notice that such fundamental concepts of classical theories as configuration space and phase space are not relativistically invariant, alongside with the Hamiltonian and, to a large extent, the Lagrangian formulations of mechanics. Even phase spaces of the same material point with mass $m$ are different for two observers moving relatively to each other. One can observe that the geometric style of general relativity, in particular, the possibility to use arbitrary coordinate systems (general covariance) has been later adopted in other physical models and thus deeply influenced both physics and mathematics. The main point is that general relativity has evoked a keen interest in using geometric methods when dealing with problems of physics and even engineering. For example, differential geometry, formerly regarded as a separate mathematical discipline having little to do with practical needs of physics (which is wrong) is now an indispensable tool of classical mechanics and dynamical systems theory.

One can also mention the increasingly wide use of non-Euclidean geometry in many branches of contemporary physics. The concepts of differentiable manifolds, maps, diffeomorphsms, vector and tensor fields, vector bundles, connections, covariant differentiation and related concepts have become the necessary elements of the basic education of physicists. In short, special and especially general relativity has significantly changed the style of thinking in physics: already the concept of spacetime implying space + time bears the tendency to unification that has become a long-standing trend in science. In particular, the yearned for "theory of everything" (TOE) has the unification of general relativity and quantum mechanics as a prerequisite for the TOE sheer appearance. Even some specific mathematical models based on general relativity produced a considerable impact: thus, the study of the black hole (BH) model that originated from the Schwarzschild solution to Einstein's equations brought about new views on statistical physics, thermodynamics and quantum mechanics. Notice, however, that it would be hard to imagine a TOE in the spirit of the Copenhagen interpretation i.e., including the role of an observer.

---

[7] Gravity forces in the orthodox (Einsteinian) general relativity is given through the connection $\Gamma^i_{jk} = \frac{1}{2} g^{il} (\partial_k g_{jl} + \partial_j g_{kl} - \partial_i g_{jk})$. This "force" is quite different from Newton's forces defined through the second law.

One can also notice a curious "anti-causal" effect: retroactive influence of the 20th century ideas on the 18th century science, namely the impact of relativistic theories on classical mechanics. This impact is manifested, e.g., in the advancement of geometric approaches and transformation techniques. Such notions as covariance, transition between coordinate systems, connection and holonomy have become commonly accepted and almost routine in modern expositions of the old science.

There appear from time to time physical models that claim to be "theories of everything", and some of them really are. Thus, Newtonian mechanics (together with Newtonian gravity) had acquired the TOE status. The M-theory embracing such fashionable and important theoretical concepts as extra dimensions, supersymmetry and extended (non-pointlike as in classical physics) objects, membranes and strings is currently one of the leading candidates for a TOE. A ten-dimensional superstring theory has recently been nominated as a TOE. The trouble with superstrings, however, is that one can construct several (at least five) mathematically non-contradictory superstring theories. A consistent superstring theory needs a ten-dimensional spacetime in order to evade non-physical states (e.g., so-called ghosts) and negative probabilities.

In science and lately in engineering, there exist problems, partly of quasi-philosophical nature, that permeate all worlds of physics. For instance, there is a notorious irreversibility paradox, manifesting itself as a contradiction between time-reversal invariant microscopic models of physics (the so-called laws of motion) and phenomenological time non-invariance observed in our everyday experience. This fascinating paradox stirs a lot of controversy until now and does not seem to be ultimately resolved, although some people think that this is actually a pseudoproblem since irreversible behavior can be derived from physical laws by taking certain limits. It may well be that the claim that all processes in nature should be in the final analysis invertible in time (time-reversal invariant), one merely needs to reduce them to the ultimate microscopic level, is more a wishful thinking rather than the evidence-based physical fact. Strangely enough, even physicists *believe* in the microscopic time-reversal symmetry, although it can be just a modeling assumption than can bring up inconsistencies and conflicts with experiment. If we temporarily use the mathematical language, hardly anyone can prove, for example, that the Hilbert space, which can be thought of as an arena for physical evolution, should always be filled with complex conjugated states corresponding to thermally isolated systems.

Physics is distinguished from other disciplines also employing mathematical modeling by the fact that models in physics are linked. This is an important property ensuring the success of the collection of mathematical models called physics, and this is a feature that makes physics, in the words of an eminent physicist V. Weisskopf, "the splendid architecture of reason". The linked architecture appeared mostly due to the firmly established (by reproducible experiments) laws of physics that can be expressed in the form of differential equations. As soon as a model is disconnected from the main architecture, its heuristic value becomes substantially reduced. One could see this diminishing value of stand-alone models in the examples of such disciplines as economics or sociology where mainly ad hoc mathematical models are in use, characterized by a lot of arbitrary parameters and assumptions. In fact, while economics can be treated as a branch of applied mathematics, sociology may be viewed as social physics.

Even in physics, despite its generally flourishing development, there existed examples of isolated models and quickly baked concepts which practically did not use the wealth of physical results. They, however, fade away rather quickly. Thus, nowadays, after the Standard Model epoch, nobody seems to remember numerous group (multiplet) models of elementary particles, being of fashion in 1960s. Even today, one tends to *define* a free elementary particle as just an irreducible unitary representation of the Poincaré group[8] (if one abstracts oneself from the fact that such particles as neutrinos and quarks are not described in terms of mass as a constant parameter, but in terms of mass matrices).

The Standard Model (SM) is by no means the last word in fundamental physics; it is just a theory describing elementary particles (and their interactions) within certain energy limits. For example, there is no warranty that the SM can be applied to particles with very high energy (the respective experiments are lacking or insufficient). Moreover, the SM does not involve gravitation, although the particle mass is partly included in SM "by hand" due to supplementing the model by the Higgs mechanism (see more on that below).

The lack of a coherent picture can be tolerated in cookbook disciplines such as medicine, computer graphics, networking protocols compendium, even in scientific computing and numerical mathematics, but is hard to be uncritically accepted in physics which strives to provide a unified image of the world. Luckily, mathematical models in physics are not invariable: they undergo changes as new experimental facts are accumulated and novel theoretical considerations come into fashion. Besides, new mathematical structures may be developed and those long known to mathematicians may be disseminated in the physical community and thus used for modeling of physical cases. There is, however, a reverse trend, namely the absolutization of successful mathematical models. Uncritical adoption of popular beliefs is accelerated with their spread, and the barriers of self-conviction become more and more difficult to overcome. For instance, the Newtonian model for mechanical motion, which is a legacy of the mechanical philosophy from the 17th century, has been absolutized to such an extent that even today many people (usually with an engineering background) tend to denounce special relativity since it deviates from Newtonian mechanics deeply engraved in our intuition.

Originally, it was intuition that led to geometric images of the world that immediately appealed to our senses, in contrast with less vivid algebraic calculations. Geometric images can be useful in creating an illusion of understanding and, thus, in stirring the interest in geometrically pictured physical objects. Yet one should not expect from a simplified geometric description more than a heuristic tool in all cases. Thus, describing a vector as an arrow between two points in the $\mathbb{R}^3$ space or four-vector $x^\mu$ of the relativity theory as an arrow in spacetime $\mathbb{R}^{3+1}$ leaves the most important features of such objects, their transformation properties, completely neglected.

Since special relativity is extremely important for both technical calculations and the worldview, it would be pertinent to make a few remarks about this theory. Originally, special relativity had a negative meaning: this theory invalidated the notion of ether, a hypothetical omnipresent medium that allegedly remains in a state of absolute rest. Special relativity, at least as regards its mechanical part, can only be applied to describe the dynamics of a single charged particle in a given electromagnetic field. In contrast with the non-relativistic domain there seems to be no convincing relativistic many-particle theory taking into account interparticle interactions.

The attribute "special" does not imply that the theory can be only applied to inertial systems, as it is occasionally asserted in textbooks. One can introduce a momentum four-vector in a natural way and can thus explore an accelerated motion of a particle under the action of an external force. For example, one widely uses in classical electrodynamics the equation $\frac{dp_\mu}{d\tau} = \frac{e}{c} F_{\mu\nu} u^\nu$ [96], where $F_{\mu\nu}$ is the electromagnetic field tensor, $p_\mu$ is the covariant momentum four-vector, $u^\nu$ is the contravariant velocity four-vector, $e$ is the particle electric charge, $c$ is the speed of light, $\tau$ is the proper time, $d\tau = \gamma^{-1} dt = \left( dx_\mu dx^\mu \right)^{1/2} = \left( g_{\mu\nu} dx^\mu dx^\nu \right)^{1/2}$, $\gamma = (1 - v^2(t)/c^2)^{-1/2} = E/mc^2$ is the relativistic factor, $E$ is the particle energy; in special relativity the metric $g_{\mu\nu}$ is given by a diagonal tensor diag $(1, -c^2, -c^2, -c^2)$ (more customary is representation diag $(1, -1, -1, -1)$ that corresponds to $c = 1$). The term "special" means that the respective theory is limited – in the sense that it does not include the effects related to gravitation.

---

[8] This is the view of E. Wigner [159], the famous mathematician and physicist, a specialist, in particular, in group theory.

The universal constancy of the speed of light is expressed by equation $ds^2 = c^2 dt^2 - dx^2 = 0$, $x = (x, y, z)$. The main mathematical statement of special relativity is the invariance of physical laws under the Poincaré group, the transformations from this group being applied to the change of coordinate systems. It is due to this stipulated invariance that the equations of both special and general relativity have a tensor form. Here, we may note that since in relativistic theories position $x = \{x^j\}, j = 1,2,3$ and time $x^0$ (or $x^0/c$) are not independent and transform through each other, measuring separately time intervals and lengths, ascribing to these quantities totally different units, e.g., seconds s and centimeters cm (or meters m), is physically meaningless, although subsists due to historical reasons.

Here, we can recall in passing what the Poincaré group that dominates the entire relativistic physics is. On an elementary level this relativistic group is formed by the coordinate transformations $x^\mu \mapsto \tilde{x}^\mu = \Lambda_\nu^\mu x^\nu + a^\mu$, where the matrix/vector components $\Lambda_\nu^\mu, a^\mu, \mu = 0,1,2,3$ are real parameters. One usually imposes on matrix $\Lambda$ the Lorentz condition of preserving the Minkowski product $u_\mu v^\mu = \tilde{u}_\mu \tilde{v}^\mu = g_{\mu\alpha} \Lambda_\beta^\alpha u^\beta \Lambda_\nu^\mu v^\nu = u_\nu v^\nu$ or, changing indices $\alpha$ and $\nu$, $g_{\mu\nu} \Lambda_\alpha^\mu \Lambda_\beta^\nu = g_{\alpha\beta}$ or $\Lambda^{\mathrm{T}} g \Lambda = g$. The latter form is analogous to the usual orthogonality condition in Euclidean space $R^T R = I$, where $I$ is the identity matrix. It is not difficult to show that transformations $x^\mu \mapsto y^\mu = \Lambda_\nu^\mu x^\nu$ representing the Lorentz group leave the relativistic interval $s := ((x^0)^2 - \mathbf{x}^2)$, $\mathbf{x}^2 = x_i x^i, i = 1,2,3$ intact if the 4d rotation matrix $\Lambda_\nu^\mu$ satisfies the orthogonality conditions $\Lambda_\nu^\mu \Lambda_\mu^\sigma = g_\nu^\sigma$, where $g_\mu^\nu = $ diag$(1, -1, -1, -1)$. One also can represent this interval via Pauli matrices $\sigma_i, i = 1,2,3$; $\sigma_0 = I$ (identity matrix), namely $s^2 = \det A$, $A := \begin{pmatrix} x^0 + x^3 & x^1 - ix^2 \\ x^1 + ix^2 & x^0 - x^3 \end{pmatrix} = \sigma_\mu x^\mu$. Matrices $\sigma_\mu, \mu = 0,1,2,3$ form a complete set in the space of $2 \times 2$ matrices over $\mathbb{C}$, so that any such matrix $A$ can be expanded over $\sigma_\mu$. Note, however, that the set of four Pauli spin matrices

$$\sigma_\mu := \left\{ \begin{pmatrix} 1 & 0 \\ 0 & 1 \end{pmatrix}, \begin{pmatrix} 0 & 1 \\ 1 & 0 \end{pmatrix}, \begin{pmatrix} 0 & -i \\ i & 0 \end{pmatrix}, \begin{pmatrix} 1 & 0 \\ 0 & -1 \end{pmatrix} \right\}$$

form a basis for the $2 \times 2$ matrices over $\mathbb{C}$ but not over $\mathbb{R}$.

Some authors assume $\Lambda_\sigma^\mu \Lambda_\nu^\sigma = \delta_\nu^\mu$, where $\delta_\nu^\mu = $ diag$(1,1,1,1)$ is the Kronecker symbol or, in some general relativity problems, $\Lambda_\sigma^\mu \Lambda_\nu^\sigma = g_\nu^\mu$, with $g_\nu^\mu = $ diag$(1, -1, -1, -1)$. Notice that determinants of the group matrix $\Lambda = (\Lambda_\nu^\mu)$ have different signs for these two definitions. It is sometimes stated that only transformations with $\Lambda_0^0 > 0$ and positive determinant (proper and orthochronous) are physically important since only such transformations describe continuous accelerated motion or rotations. In other words, discrete transformations such as time inversion or parity change are quite arbitrarily left out. One can notice a curious coincidence: the parity transformation $t \to t$, $\mathbf{x} \to -\mathbf{x}$ is described by matrix $\Lambda_\mu^0 = $ diag$(1, -1, -1, -1)$ that formally coincides with the Minkowski metric tensor.

Since the Poincaré group is an affine one including translations, it can be conveniently represented through five-dimensional matrices

$$A = A(\Lambda, a) = \begin{pmatrix} \Lambda_0^0 & \cdots & \Lambda_3^0 & a^0 \\ \vdots & \ddots & \vdots & \vdots \\ \Lambda_0^3 & \cdots & \Lambda_3^3 & a^3 \\ 0 & \cdots & 0 & 1 \end{pmatrix}$$

in the so-called homogeneous coordinates. Introducing an extra dimension in homogeneous coordinates enables us to unify rotations and translations (also scaling and perspective projection) expressing all of them throughout matrix multiplication. More exactly, a projective space is introduced, with each spacetime point $x^\mu$ being represented as $wx^\mu, w$. For the Poincaré rotations $w = $ constant that can be put to 1, but for other transformations $w$ can be a varying parameter. It is interesting that the use of homogeneous coordinates allowing one to easily combine most geometric transformations in the form of matrix multiplication nearly revolutionized computer graphics.

An example of Lorentz transformations is given by the so-called boosts corresponding to the relative motion with velocity $v$ along the $x^1$ axis:

$$(x'^0, x'^1, x'^2, x'^3)^T = \Lambda(x^0, x^1, x^2, x^3)^T \equiv \begin{pmatrix} \cosh\varphi & \sinh\varphi & 0 & 0 \\ \sinh\varphi & \cosh\varphi & 0 & 0 \\ 0 & 0 & 1 & 0 \\ 0 & 0 & 0 & 1 \end{pmatrix} \begin{pmatrix} x^0 \\ x^1 \\ x^2 \\ x^3 \end{pmatrix},$$

where $\tanh\varphi = v/c$, see the details in [96].

One sometimes uses the Lorentz and the Poincaré groups interchangeably, which is incorrect since the Lorentz group is just a subgroup of the Poincaré group. The most crucial difference between the Lorentz and the Poincaré group is that the Poincaré transformations, in contrast to the Lorentz ones, preserve neither the inner product nor the associated $L^2$ norm. Indeed, for an inner Poincaré product we have

$$(x', y') = (\Lambda x + a, \Lambda y + b) = (\Lambda x, \Lambda y) + (\Lambda x, b) + (a, \Lambda y) + (a, b)$$
$$= (x, y) + (\Lambda x, b) + (a, \Lambda y) + (a, b) \neq (x, y),$$

where $x' = \Lambda x + a$, $y' = \Lambda y + b$. Putting $y = x$ and $b = a$ we see that the transformed norm $(x', x') = \|x'\|^2 \neq \|x\|^2 = (x, x)$ i.e. that the norm is not preserved by transformations from the Poincaré group (see more about the group properties of the Lorentz and Poincaré groups in Section 4.4. "Mass and matter").

It is important to note that the Poincaré group is the symmetry group of the well-known Klein-Gordon equation $\left(p_\mu p^\mu - m^2\right)\psi = 0$ describing the evolution of a scalar field. The Klein-Gordon equation can be regarded as the relativistic analog of the Schrödinger equation of non-relativistic quantum mechanics (here for simplicity generally adopted in relativistic quantum theories, $\hbar = c = 1$). For quantum field theorists, it is important that any particle or field theory should be invariant under the transformations from the Poincaré group.

In particular, operators of spacetime translations $T_\mu$ and rotations $R_{\mu\nu}$ leave the vacuum state $\varphi_0$ intact: $T_\mu\varphi_0 = \varphi_0$, $R_{\mu\nu}\varphi_0 = \varphi_0$. Note that the uniqueness of the vacuum is not an essential requirement: although one often considers the vacuum to be a nondegenerate local minimum of the energy function $H(\varphi)$ i.e. $H''(\varphi_0) > 0$ (or a positive-definite bilinear form $H''(\varphi_0)(X,X) > 0$ for any $X \neq 0$), there exist many physically important situations that are modeled by several vacuum states i.e. discretely degenerate vacuum. Transitions (for example, induced by the "quantum noise") between such vacuum states are typically called "vacuum instability".

Although mathematical models in physics are supposed to combine two components: the physical insight and the available (or familiar) mathematical tools, excessive intuition may prove paradoxically harmful. Mathematical modeling in physics in some sense contradicts "physical intuition" which may be just another term for replacing the mathematical procedures with less precise and often arbitrary considerations. For example, in quantum mechanics, which is basically a mathematical theory, no "physical intuition" is actually required. Nevertheless, it is in quantum mechanics that "physical intuition" seems to be especially popular (one can recall the flood of metaphysics related to the issue of interpretation of quantum mechanics).

The famous two-fluid model of superfluidity built by L. D. Landau (the 1962 Nobel Prize) is a typical mathematical model of physics appealing to intuition. Liquid Helium in this model is represented as a mixture of two components, normal and superfluid. In his main paper devoted to the explanation of superfluidity Landau himself emphasized that separating liquid Helium into two fluids, one being dissipative and transferring heat, while the other not, is just a method of intuitive representation of the astounding phenomenon of superfluidity. In reality, there are two simultaneous flows in the same fluid, one behaving as normal viscous fluid carrying entropy, the other running without friction past any surface or obstacle.

Figuratively speaking, the "physical intuition", like any strong medicine, should be added in moderate doses, otherwise it brings more confusion than awareness and understanding. One can name a number of examples showing that supplementing "physical intuition" to a mathematical model may be superfluous and unnecessary.

## 4.4. Mass and matter

Mass is a basic concept in physics. Yet it has many faces: in gravitation, mass determines the magnitude of force; in dynamics, inertial mass identifies the body's resistance to accelerate; furthermore, according to Einstein's relativity concepts, mass quantifies the energy content in the body, $E = mc^2$ (more generally, $E^2 = (pc)^2 + (mc^2)^2$). Mathematically, mass is a real number, $m \in \mathbb{R}$, in most cases $m \in \mathbb{R}_+$ i.e., mass is assumed to be a real positive parameter in motion equations. But from the physical viewpoint mass cannot be so simply defined, often requiring some intuitive interpretation. For instance, in thermophysics and in situations related to chemical transformations, mass can be conveniently interpreted as the total amount of matter in the body. The more matter, the more the mass. If we take a proton as the unit of mass (this being approximate, but the accuracy is sufficient for thermophysics), then, e.g., a uranium atom ($^{238}$U) has a mass of 238 units whereas the carbon isotope $^{12}$C has mass 12 units i.e., the amount of substance in, say, a mole of uranium is almost 20 times as massive as a mole of carbon, although both contain the same number (the Avogadro number, $N_A \approx 6 \cdot 10^{23}$ mol$^{-1}$) of particles.

Another physical interpretation of mass follows from Newton's motion equations and consists in treating mass as a measure of resistance to velocity changes: the more mass is in a body, the more difficult it would be to divert it from a given trajectory or to put it in motion. The third interpretation of mass is related to the Newtonian gravity law, in which two kinds of mass are present: active and passive. An active mass creates the field of gravitation, while the passive mass determines the force acting on a body that is moving in an external gravitation field.

The example of mass, an apparently simple coefficient, demonstrates that there is, in general, no unique (bijectional) translation from physical into mathematical language. The situation with mathematization of areas outside physics is even worse since the disciplines lying in such areas tend to be rather confused and can rarely admit an exact meaning for the objects under consideration. Thus, it would be difficult trying to compile a dictionary between physical and mathematical languages – a somewhat naïve dream of many mathematicians.

The overall concept of mass is believed to include three manifestations: inertial mass $m = m_{in}$, passive gravitational mass $m = m_{gp}$ (being acted upon by the gravity field) and active gravitational mass $m = m_{ga}$ (the source of gravity) [27]. In classical mechanics, all these are presumed equal, but there may be theories where they are not. For example, in general relativity, the principle of equivalence requires $m_{in} = m_{gp}$ whereas $m_{ga}$ is allowed to be different – actually $m_{ga}$ could remain a free parameter if one does not apply momentum conservation in its rigid mechanical form.

Is mass a variable or a numerical parameter? In the latter case one does not need to consider mass "an observable" and hence to construct for it an additional operator in the transition from classical to quantum mechanics in the non-relativistic situation. In elementary classical mechanics mass $m$ is assumed to be a positive scalar, $m \in \mathbb{R}_+ \backslash \{0\}$, sometimes defined as manifesting the matter content in a body. In quantum theories, in particular in QED, mass is allowed to be zero, as massless particles such as the photon are admitted. One can justifiably ask whether mass can belong to more general sets, e.g., $\mathbb{R}, \mathbb{C}$, etc[9]. Can mass also be negative or, more generally, can the mass matrix $m_{ik}$ in the free Lagrangian $L = m_{ik} \dot{q}^i \dot{q}^k$ or closely related to it metric tensor $g_{ik}$ be not positive definite, so that kinetic energy would not be always positive?

In more advanced theories than elementary single-particle classical mechanics, specifically in many-body theories, masses of constituent bodies $m_a$ are not necessarily scalar quantities: in certain models of physics, it may be appropriate to consider mass as a tensor $m_{ik}$ which can even play the role of the metric tensor. Indeed, in mechanics, the metric tensor is so chosen as to be defined by the quadratic form corresponding to kinetic energy $T = \frac{1}{2} m g_{ik}(q) \dot{q}^i \dot{q}^k$, where the coefficient $m$ can be viewed as a scaling factor for the mass tensor $m_{ik} := m g_{ik}$. In classical non-relativistic mechanics, the mass tensor $m_{ik}$ is always considered non-degenerate. The inverse mass tensor $m_{ik}^{-1}(\mathbf{r}, t) = \partial^2 H(\mathbf{r}, \mathbf{p}, t) / \partial p_i \partial p_k$, where $H(\mathbf{r}, \mathbf{p}, t) = T + V$ is the Hamiltonian function (see below), is usually assumed positive-definite. In other words, if the mass tensor characterizing a Newtonian system is non-degenerate, Newton's equations of motion define a dynamical system. Thus, the equations of motion in non-relativistic mechanics almost always represent a dynamical system, with solutions to such equations being known as "motion".

The concept of kinetic energy is associated with motion: it is a direct consequence of the Galilean invariance whereas elucidation of the meaning of mass is outside of classical mechanics. Within the framework of the so-called electromechanical analogy one can take the capacitors' charges $q^i$ as coordinates and magnetic inductivity coefficients $L_j$ as masses. Coefficients of the mass tensor $m_{ik}$ would correspond in this analogy to mutual inductivities $L_{ik}$; in this case the quadratic form $T$ corresponding to kinetic energy is non-diagonal. According to the rules of linear algebra, the quadratic form $T$ in a pair of forms $(T, V)$ can be reduced to a diagonal representation that would of course change the $V$-form, but in the close-to-equilibrium context such a modification is inessential since the potential energy form $V = a_{ij} q^i q^j$ is written via generic coefficients $a_{ij}$.

---

[9] The same question can be attributed to other "untouchable" physical parameters such as, e.g., the Planck constant in conventional quantum mechanics, $\hbar \approx 10^{-27} \text{erg} \cdot \text{s}$.

Recall that in mechanics the kinetic energy for a motion $\mathbf{x}(t) = \{x^1(t), \ldots, x^n(t)\} \in Q$ at time $t$ is represented by a scalar product $T^{ik}(\mathbf{v}, \mathbf{v}) = \sum_{a=1}^{N} m_a \dot{x}_a^i \dot{x}_a^k$, the family of such scalar products forming a Riemannian (even reduced to Euclidean) structure on the tangent manifold $TQ$. It is also important to notice that kinetic energy of any object depends on the reference frame in which it is defined (measured). The same, as we know, applies to velocity and momentum. In distinction to such relative quantities, the mass $m$ (and associated with it the rest energy $E_0 = mc^2$) is a characteristic of the object like the electric charge or spin i.e., a relativistic scalar that does not depend on the frame of reference. In Cartesian coordinates and in inertial systems, the kinetic energy $T$ is a quadratic form with respect to velocities in diagonal representation. In a general case $q^i = q^i(x^k)$, kinetic energy $T = \frac{1}{2} a_{ik} \dot{q}^i \dot{q}^k$ is nearly always considered positive-definite.

Newtonian mechanics quite accurately describes the motion of macroscopic objects on the Earth and of celestial bodies on comparatively small cosmic scales since this motion is smooth and its velocities are small compared with the speed of light $c$. For instance, the typical projectile velocity in ballistics $v \sim 10^3$ m/s i.e., $v/c \sim 0.3 \cdot 10^{-5}$ and parameter $\beta^2 \equiv (v/c)^2 \sim 0.9 \cdot 10^{-11} \sim 10^{-11}$ (this parameter determines the "degree of relativism" in mechanics). Notice that elucidation of the physical nature of mass is not included in the tasks of dynamics. Mass of a body in Newtonian mechanics is a purely phenomenological (i.e., taken from experience) coefficient that determines how the body responds to an applied force (the more the mass, the slower the body accelerates, the force being given) so that one can actually define the mass ratio $m_a/m_b$ that is in inverse proportion to the ratio of accelerations $\ddot{x}_a/\ddot{x}_b$ of respective bodies $a$ and $b$ reacting on a fixed applied force, rather than a mass of an isolated body; thus masses are always compared with some etalon. Quite naturally, when the body moves in an anisotropic medium, its acceleration under the action of the applied force may be different along diverse directions which means that the mass of a body in anisotropic media has tensor properties. For example, in condensed matter such as crystals the notion of a free mass of a particle often becomes irrelevant and should be replaced by the effective mass tensor. This transformation of masses (it is a particular case of a so-called renormalization) reflects a very important fact that an object moving in a medium loses the features of a free particle and becomes a quasiparticle whose properties, in particular the response to an applied force, are adapted to the characteristics of the medium. One can roughly illustrate this behavior modification caused by a surrounding matter on an example of a person in the crowd who moves (and in general behaves) in a substantially different way as on a free terrain; likewise, a car driver adjusts her/his driving regime to the road and traffic environment, etc.

Experience shows that our universe (at least its observable part) is highly isotropic. Had it been essentially anisotropic, particle masses would have probably been of tensor character since the resistance of a body moving in the universal anisotropic environment to any alteration of its state of motion, in particular defined by acceleration, would depend on direction. By the way, anisotropy of the universe would produce noticeable effects: for example, stars in different directions would have different colors. Luckily, all measured quantities (such as those characterizing the electromagnetic response of free space, permittivity and permeability) are the same in all directions. Furthermore, mass in general relativity (GR) is an even more intricate concept than, for instance, in special relativity (SR) – there are actually several types of relativistic masses. In general, to correctly define mass in general relativity would require dealing with rather sophisticated mathematics such as in the case of the so-called ADM (Arnowitt-Deser-Misner) mass and delicate physical issues such as the Mach

principle – one can, for example, ask whether mass[10] would vary if some part of the universe i.e., galaxies, nebulae, etc. were removed. According to Mach's principle taken literally, there would be neither inertial forces nor inertial masses in an empty universe. Conversely, inertia of an object rises if more and more other masses are incorporated into the background. Generalizing Mach's principle, one can state that local motion and even local physical laws are governed by the universe global structures i.e., by the large-scale distribution of matter in the universe. This theoretical statement together with Mach's principle, however, has not been experimentally corroborated. Moreover, honestly speaking general relativity has very little in common with Mach's principle.

Even without the assumption of universe anisotropy, masses of fundamental particles may be of a tensor nature. Thus, there may be a matrix corresponding to the mass in the Lagrangian (see the next subsection), and this matrix in certain models is non-diagonal. For instance, in the model of neutrino oscillations neutrino masses, considering the matrix form in which masses enter the kinetic term of the Lagrangian, can be understood as tensors. This neutrino mass matrix is non-diagonal which accounts for the fact that different neutrino types can be transformed into each other. A similar situation is encountered in a number of quark models. It is curious that even in the 1960s, when physics (at that time mostly based on linear models) flourished, only a few physicists were taking quark models seriously: the main part of scientific establishment thought such models were sheer nonsense.

The term "matter", in contrast with mass, seems to be poorly defined in spite of its ubiquitous use in physics and cosmology. One should not confuse mass and matter. For example, light is also matter, but its particles are massless. To allow the photon to have even a tiny mass $m$ would produce large discrepancies in many fields of physics. For instance, the Coulomb law would not be valid: the Coulomb forces would decay exponentially in vacuum over large distances (as, e.g., in plasmas or electrolytes). Accordingly, the standard equations of mathematical physics such as the Laplace equation would not describe electrostatics, which would entail significant experimental and technological consequences. An applied external electric field would be present within a perfect conductor, light would propagate in vacuum with a velocity smaller than the fundamental constant $c$, and frequency dispersion (i.e., dependence of the propagation velocity on the wavelength) would be observed not only in matter but also in vacuum. The so-called black hole bombs would possibly emerge i.e., strong black hole instabilities triggered by massive bosonic fields. Maybe the gravest consequence of the non-zero photon mass would be the breach of gauge invariance and the ensuing violation of charge conservation. Nevertheless, the existence or non-existence of the photon mass is not a speculative or purely theoretical question, but a subject of high-precision measurements and careful observational, in particular, astrophysical tests.

According to modern concepts, matter consists of quarks $(u, d), (c, s), (t, b)$ and leptons $(e, \mu, \tau), (\nu_e, \nu_\mu, \nu_\tau)$ that can be combined to form more complex particle systems. Particles can interact through strong $(g)$, electromagnetic $\gamma$, weak $W, \pm Z$ and gravity (graviton G) forces. Of course, this is a very crude picture of the fundamental properties of matter, but going any deeper would make this manuscript barely readable.

---

[10] Alongside the electron mass $m_e$ possibly the values of other fundamental physical constants can be related to the distribution of matter in distant parts of the universe.

Here the term "mass" is related to the constant "rest mass" which is a fundamental characteristic of the particle through one of the main equations of relativistic mechanics, $E^2 = p^2c^2 + m^2c^4$, and not to the "relativistic mass" $m(v) = m(1 - v^2/c^2)^{-1/2}$ that is allowed (mainly for didactic purposes in old textbooks) to grow with particle velocity $v$. To avoid some confusion, we can briefly comment on this point. In relativistic mechanics, which is an extension of the Newtonian model of motion, to velocities comparable with the speed of light $c$, mass is often defined as a function of the relative velocity of the observer with respect to the observed object, e.g., a particle. The "relativistic mass" grows with the relative velocity of a particle (is proportional to the relativistic factor $\gamma = (1 - \beta^2)^{-1/2}, \beta \equiv v/c$). Yet it is more convenient to treat the "rest mass" $m$ in relativistic mechanics as the particle's invariant mass that does not change during the motion. In more mathematical terms, the mass of a system in special relativity may be defined as the norm of the system's energy-momentum 4-vector $(E, \mathbf{p})$. Recall that the norm of a 4-vector $x = (ct, \mathbf{x}) \equiv (x^0, x^1, x^2, x^3)$ in special relativity is $x \cdot x = x_\mu x^\mu = x^T g x$, where $g = \text{diag}(1, -1, -1, -1)$. Lorentz transformations $y = \Lambda x$ (sometimes also called boosts) leave relativistic norms intact so that $\Lambda^T g \Lambda = g$. The total energy of a body is $E = E_0 + T$, where $E_0 = mc^2$ is the rest energy and $T$ is the kinetic energy. In Newtonian mechanics, $T \ll E_0$ and the total energy $E \approx E_0$ or $E/c^2 \approx m$ with rather high accuracy (for a projectile, as we have seen, $10^{-11}$). Therefore, from the main relationship of relativistic mechanics, $E^2 - p^2c^2 = m^2c^4$ (the Pythagorean theorem in hyperbolic geometry) we have $(E - mc^2)(E + mc^2) = p^2c^2 \approx T \cdot 2mc^2$ so that $T \approx p^2/2m$ and $\mathbf{p} = E\mathbf{v}/c^2 \approx m\mathbf{v}$ i.e., the standard relationships of Newtonian mechanics.

One can intuitively understand the existence of the universal velocity limit if, with the rising speed, inertia of the body increases so that it becomes more and more difficult to accelerate the body. Then it is only natural to assume that there will eventually be a speed limit. One might also note that although the "rest mass" of a single particle is Lorentz-covariant, the mass of a system of particles is not since the total energy-momentum 4-vector contained in the composite (multiparticle) system is in general not covariant at a fixed time instant (due to the relativistic lack of objective simultaneity), and non-covariance of a vector naturally results in non-covariance of its norm. Simultaneity in relativistic theories is not objective, it is relativized by an arbitrary choice of the coordinate frame. For instance, if we apply a boost, two events that have initially been simultaneous will be no longer such.

Note also that the rest and uniform motions are not quite trivial concepts within the classical (i.e., nonrelativistic) picture either, and they can be properly understood only by introducing the true relativistic notion of spacetime, based on the Lorentz and the Poincaré groups of symmetry transformations: transitions between inertial reference frames are produced by operations of this group. Thus, the ten-parameter Poincaré group is composed of three kinds of motion: (1) the SO(3) spatial rotations (3 parameters), (2) the Lorentz boosts (3 parameters), (3) the translations of the origin in spacetime (4 parameters). Covariant representations of the Poincaré group form a set of position-dependent operators: already mentioned spacetime translations, spatial (3d) rotations and boosts commonly known as quantum fields. The Lorentz group and its generalization, the Poincaré group that was observed to be the symmetry group of nature, can be viewed as constraints imposed on physical theories: anything not obeying Lorentz or Poincaré invariance cannot be assigned the status of a proper physical theory.

Newtonian mechanics may be interpreted as a limiting case of special relativity based on the Lorentz transformations when the speed of light tends to infinity, and then the notions of rest and uniform motion are considered in the respective inertial frames, which are related by the Galilean symmetry group. Thus, one can view the Galilean, Lorentz and Poincaré groups as special cases of relativistic groups. Accordingly, there have been several theories of relativity throughout the history of physics and philosophy based on diverse models of spacetime and differing by the manner in which physically equivalent frames of reference are connected to one another. One can illustrate that, e.g., the rotation group is a part of the Lorentz group by noting that matrices implementing the Lorentz transformations can be written in the block-diagonal form (here for simplicity $c = 1$)

$$\Lambda = \begin{pmatrix} 1 & \mathbf{0}^T \\ \mathbf{0} & R \end{pmatrix},$$

where $\mathbf{0} \equiv (0,0,0)^T$ and $R$ is the matrix of 3d rotations leaving $\mathbf{r}^2 = x_i x^i, i = 1,2,3$ unchanged. As $\det R = 1$, so is $\det \Lambda$. As already mentioned, the Poincaré transformations $P$ are just inhomogeneous Lorentz transformations, $P = P(\Lambda, a)$, so that as any inhomogeneous transformations they can be represented by matrices acting in a space with one extra dimension compared to homogeneous transformations, in the relativistic case $\dim P(\Lambda, a) = \dim \Lambda + 1$ i.e., by $5 \times 5$ matrices $P(\Lambda, a) = \begin{pmatrix} \Lambda & a \\ 0 & 1 \end{pmatrix}$, where $0 = (0,0,0,0)^T, a := a^\mu, \mu = 0,1,2,3$. One can see that such five-dimensional matrices comprise a ten-parameter set, with the Lorentz block $\Lambda$ delivering six independent real parameters (corresponding to three rotations and three Lorentz boosts) and $a = a^\mu$ the remaining four. One can also easily see that the set of matrices $P(\Lambda, a) = \begin{pmatrix} \Lambda & a \\ 0 & 1 \end{pmatrix}$ forms a group since $P_1 P_2 \equiv P(\Lambda_1, a_1) P(\Lambda_2, a_2) = \begin{pmatrix} \Lambda_1 & a_1 \\ 0 & 1 \end{pmatrix} \begin{pmatrix} \Lambda_2 & a_2 \\ 0 & 1 \end{pmatrix} = \begin{pmatrix} \Lambda_3 & a_3 \\ 0 & 1 \end{pmatrix} \equiv P_3$, where $\Lambda_3 \equiv \Lambda_1 \Lambda_2$, $a_3 \equiv \Lambda_1 a_2 + a_1$. The neutral (unit) element is $P(I_4, 0) = I_5$, where $I_n$ is an $n$-dimensional unit matrix, and the inverse element $P^{-1}(\Lambda, a) = P(\Lambda^{-1}, -\Lambda^{-1} a)$ so that all the group requirements are fulfilled. We can pay attention here to a simple but important fact: since translations $a^\mu$ are also Lorentz-rotated, translations from the Poincaré group do not commute with Lorentz transformations (see also Section 5.4. "Extrapolation").

The origin of masses of all fundamental particles is also thoroughly discussed in modern quantum field theory, where the concept of spontaneous symmetry breaking implying the so-called Higgs mechanism is one of the main ideas. What is the Higgs mechanism? Briefly, the Higgs mechanism consists in the emergence of mass (massive fields) from a gauge field. In the Standard Model of high-energy physics, the notion of Higgs mechanism can be reduced to four components: (1) existence of the Higgs field $\phi$; (2) spontaneous breaking of symmetry in gauge theories, e.g., interaction of the Higgs field with $Z$ and $W^\pm$ gauge bosons; (3) giving mass to all elementary particles; (4) forecast of the existence of the Higgs particle (which appears to have been found at the Large Hadron Collider). The Higgs particle serves as a quant of the scalar field that breaks the symmetry thus resulting in the generation of fermion (i.e., matter) masses, in particular, through acquiring masses of $W^\pm$ and $Z$ bosons.

One of the obvious shortcomings of the Standard Model is that it does not involve gravity, while gravity is inextricably connected with the concept of mass of an elementary particle. Here a fundamental question naturally arises: what is actually an elementary particle? Is it a material (mass) point having no spatial dimensions so that the density distribution within the particle can be represented by a delta-function, $\rho(\mathbf{r}, t) = m\delta(\mathbf{r} - \mathbf{r}(t))$? This model of a particle is essentially non-quantum i.e., a priori an extraneous element in the microworld so that it suits well classical and relativistic mechanics and, to some extent, classical electrodynamics, but cannot be adopted in quantum theory. In the latter one often uses the model of a wave packet, but even a free quantum-mechanical wave packet inevitably spreads, which would mean that a particle would eventually disappear if it were actually described by a wave packet. However, most free elementary particles are stable (unless not accounted for decay with the production of other elementary particles) so that the wave packet model is definitely of limited applicability: nobody has ever observed spread, gradual delocalization or eventual disappearance of elementary particles, even of photons that travel billions of years from remote regions of the universe.

All the particles treated within the Standard Model except electron, proton and neutrino (together with their antiparticles) are unstable and decay producing other particles. It is important to understand that the stability of particles is related to some conservation law: thus electron (and positron) are prohibited from decaying due to electric charge preservation, the neutrino does not disintegrate due to the conservation of angular momentum (neutrino spin $s_\nu = \hbar/2 = \text{const}$), and the proton is stable, at least in a good approximation, owing to the baryon number ($B = 1$) preservation.

One question that was previously asked quite often: why does one need the Higgs particle? Or, in related terms, what is one more scalar field for? A short answer would be that the Standard Model (or even a more complicated variant of high energy theory) would forbid, due to its symmetry, elementary particles from having mass, whereas the new field $\phi$ "spontaneously" breaks this symmetry and thus leads to the emergence of particle masses. The idea of spontaneous symmetry breaking enabled us to build the famous Standard Model of the unified weak and electromagnetic interactions. It is interesting that the same idea allowed one to demonstrate that the important phenomenon of superconductivity[11] can be regarded as spontaneously broken electromagnetism. Ultimately, the Standard Model strives to describe all the processes occurring in nature within the framework of the four known interactions: electromagnetic, weak, strong and gravitational. To understand most astronomic concepts and even cosmological models, to learn chemistry or electrical engineering, one only needs gravitation and electromagnetism. Quark-gluon models, Higgs particle or spontaneous breach of symmetry are often superfluous at this level of knowledge. Yet without strong and weak interactions underlying the respective mathematical models, one would not be able to understand what makes the stars shine nor how the Sun burns chemical elements producing the radiation power that supports life on the Earth. In general, without this knowledge one cannot grasp the principles of nuclear physics.

Current understanding of fundamental high-energy physics (formerly elementary particle physics) is based on the just mentioned Standard Model which stands on two main pillars: gauge invariance and spontaneous symmetry breaking. The original "material" foundation of the Standard Model was made up of 6 quarks × 6 leptons whereas the main goal of the Standard Model was to describe the four known fundamental forces of nature – strong, weak, electromagnetic and gravitational – on the same footing i.e., in terms of gauge concepts. This endeavor was accomplished for three out of four interactions: for the strong interaction the corresponding gauge group is SU(3), for the weak and electromagnetic interactions, respectively SU(2) and U(1) groups; within the framework of the Standard Model, weak and electromagnetic forces are unified and known as electroweak interaction. The first attempt of the unification of gravity and electromagnetism (i.e., the unification of gauge fields with gravitation) goes back to the Kaluza (1921) and Klein (1926) five-dimensional models. The unification of the three gauge forces (strong, weak, EM) with gravity is an extremely difficult endeavor, requiring special skills, and we shall not discuss this issue here. One can only note that the Standard Model does not treat gravity on an equal footing with the three microscopic gauge forces.

What is today's status of the Standard Model? It is mostly regarded as a minimal one i.e., incomplete and designed as a temporary step towards a more refined unified theory. One calls the Standard Model minimal since there are no more elementary particles in it besides six quarks, six leptons and the four gauge bosons needed to transfer interactions (the Higgs boson with high probability discovered in the LHC experiments is one more elementary – not compound – particle). The incompleteness of the Standard Model is, in particular, manifested by cosmological data. Thus, the Standard Model requires modifications to accommodate new phenomena. Such modifications are also expected to be based on gauge principles, with their geometric, Lie groups and bundles approach.

---

[11] It would hardly be possible to build the Large Hadron Collider (LHC) without using the superconducting magnets.

There is nothing particularly intricate about Lie groups and their subset – gauge groups, although the group analysis is seldom present in the university curricula of mathematical physics. The gauge language has been, for some period, monopolized by high-energy physicists and acquired a flavor of an almost esoteric tenet. In fact, the gauge group contains the transformations between different gauges such as, e.g., the U(1) symmetry group in electrodynamics. In the geometric language, gauge fields of electrodynamics are single-dimensional ($\mathbb{C}^1$) fibrations with group $G = U(1)$. The four-dimensional vector potential $A_i(x)$ represents the gauge field whereas the photon is called in this language the gauge boson. Recall that group U(1) embraces all complex numbers with unit module i.e., each element of U(1) is of the form $e^{i\varphi}$, Im $\varphi = 0$, where $\varphi$ is called a phase (a clearly overloaded term).

In general, what is known as a phase is a *complex* quantity whose absolute value is unity i.e., $\sqrt{\varphi^*\varphi} = 1$. Thus a phase can be treated as an element of the U(1) group. One can naturally generalize the latter on the concept of U($N$) groups, which is closely connected with the theory of fiber bundles.

One might also note that U(1) (i.e., transformations of the form $e^{i\varphi}$, Im $\varphi = 0$) is isomorphous to group SO(2) of plane rotations

$$\begin{pmatrix} \cos\varphi & \sin\varphi \\ -\sin\varphi & \cos\varphi \end{pmatrix},$$

a special case of transformations

$$\begin{pmatrix} a & -b \\ b & a \end{pmatrix} \begin{pmatrix} x \\ y \end{pmatrix} = \begin{pmatrix} ax - by \\ bx + ay \end{pmatrix}.$$

Classical electrodynamics has served as a model for further generalizations of the gauge concepts.

Thus, gauge groups are widely exploited in modern physics. They can be loosely interpreted as intermediate between the infinite-dimensional groups of diffeomorphisms (e.g., those preserving the volume element in a flowing fluid) and the rotation group SO(3) of a rigid body with a single fixed point. In geometric language, more fashionable today, gauge invariance merely means that some quantity is preserved under the change of connection (like vector-potential $A_i(x) \to A_i(x) + \partial f/\partial x^i$ in electromagnetism). In physics, a connection in a principal fiber bundle $P = P(M, G, \pi)$ over a manifold $M$ with structure group $G$ and projection $\pi$ is usually called a gauge field or Yang-Mills field.

In more habitual terms, the Standard Model can be translated into the mathematical language as a system of nonlinear PDEs with respect to operator-valued (in contrast with point-valued) generalized functions – quantum fields. It is also interesting to note that the ideas of acquiring mass by excitations of the medium first originated in condensed matter (solid state) physics, primarily in the works by P. W. Anderson in the early 1960s. Afterwards these ideas, mostly related to spontaneous symmetry breaking and degenerate vacua, led P. Higgs (as well as several other physicists) in 1964, then S. Glashow, A. Salam and S. Weinberg to consider analogous mechanisms in high-energy physics which resulted in the explosive development of modern fundamental physics (the Standard Model, Grand Unification, gauge theories, now LHC, etc.). This is a good example of fruitful cross-fertilization between diverse disciplines.

Today the Higgs field is used to explain the origin of mass in elementary particles (mass terms for quarks and leptons are generated by the interaction with the Higgs field), interestingly though the Higgs mechanism was historically (in 1964) introduced for other purpose, namely in connection with the newly appeared and widely exploited concept of local non-Abelian gauge symmetry. The primary intention was to preserve this symmetry in presence of massive intermediate vector bosons discovered later (1983) in CERN and known as $W^{\pm}$ and $Z$ particles https://cerncourier.com/a/finding-the-w-and-z.

Recall that the physics of elementary particles was born in the 20[th] century after the appearance of relativity theory and quantum mechanics. Respectively, the key parameters of elementary particle physics are the speed of light $c$ and the Planck constant $\hbar$. These two fundamental (relativistic and quantum) constants combined with the gravitational constant $G \approx 6.674 \cdot 10^{-8}$ cm$^3$g$^{-1}$s$^{-2}$ define a basic system of natural physical units, the so-called Planck units. Contrary to rather arbitrarily declared (mostly for historical reasons) metrological unit systems such as Gaussian, cgs or SI, Planck units, being directly derived from physical constants, are of a fundamental nature. The Planck units have an invariant physical meaning, unrelated to any actual engineering or scientific agenda. For instance, the Planck length $l_p = \frac{1}{c}\left(\frac{Gh}{c}\right)^{1/2} \approx 1.6 \cdot 10^{-33}$cm determines the spatial scale beyond which the physical notion of distance is no longer valid. Atoms, for example, are inconceivably huge on Planck's length scale: spatial grains on this scale compared with the atomic size $a_B \sim 10^{-8}$ cm gives the disparity of 25 orders of magnitude i.e., about the same as the size of an atom compared with the Sun's dimensions. Similarly, the Planck time $t_p = \frac{1}{c^2}\left(\frac{Gh}{c}\right)^{1/2} \approx 5.4 \cdot 10^{-44}$s is the smallest time interval that can be possibly measured. Contrariwise, Planck's energy $E_p = m_p c^2 = c^2 \left(\frac{hc}{G}\right)^{1/2} \approx 1.2 \cdot 10^{19}$ GeV $\approx 1$ GJ is huge in the microworld scale[12]. The physical meaning of Planck's energy is that it is the standard of energy needed to test the minuscule granularity of spacetime. In short, to observe the effects at the Planck scale, e.g., those of quantum gravity, one has to proceed well beyond the level of current physics and technology or to use some efficient (so far hypothetical) amplification techniques.

In particular, the value of the gravitation constant $G$ – and, in general, what is the gravity at Planck's scale – is unknown. There are some quite sophisticated models trying to describe physics at Planck's scale (for example, loop quantum gravity, spin foams, models based on higher-dimensional algebras, topological quantum field theories, etc.) but these models can rather be called mathematics than physics since they do not even intend to produce experimentally verifiable and falsifiable results.

The physics of elementary particles deals with the most fundamental concepts such as spacetime, matter, mass, energy, momentum and spin and studies the smallest, in most cases irreducible, fragments of matter. In the last analysis, the basic laws of motion and interaction of these tiniest fragments of matter determine the behavior of all objects on the Earth and, in general, in the universe.

---

[12] One can estimate that a car with a mass about 2000 kg accelerated to speed about 100 km/h has kinetic energy of the order of 1MJ = $10^6$ J i.e., less than Planck's energy by three orders of magnitude.

In this sense, the physics of elementary particles (now mostly referred to as high-energy physics) provides the foundation for all natural sciences. The most astounding feature of elementary particles distinguishing them from the "normal" matter fragments is that all elementary particles of the same sort, e.g., electrons, are absolutely identical everywhere in our universe. However, for reasons of space and focus, we shall not even attempt to do justice to many fascinating aspects of models of fundamental particles in this manuscript.

Let us return to the origin of mass. So, all elementary particles i.e., both fermions constituting matter (with half-integer spin $s \propto \hbar/2$) and bosons carrying forces (intermediate particles with integer spin $s \propto \hbar$) acquire mass from the Higgs field. In particular, quark and lepton masses are generated by the Higgs mechanism of coupling to the scalar field. The photon $\gamma$ which is also the intermediate boson (the photon in the Standard Model may be interpreted as the same weak interaction carrier) remains massless: within the Standard Model, the electromagnetic field $A_\mu$ does not couple to the Higgs field. Do the macroscopic bodies for which Newton's law has been successfully applied get mass owing to the Higgs mechanism? In particular, do we humans have mass due to it? Not quite. First of all, we are not elementary particles. We consist of molecules which consist of atoms, and, according to the atomic model, each atom consists of a nucleus and electrons, almost the whole mass of an atom being concentrated in its nucleus. The atomic nucleus consists of baryons such as nucleons i.e., protons and neutrons, and each of them is made up of two quarks, $u$ (up) and $d$ (down), e.g., proton $p = uud$ and neutron $n = udd$ (according to quark models, baryons are composed of three quarks). There exist also other baryons ($\Delta, \Lambda, \Sigma, \Xi, \Omega, ...$) made up of other quarks besides $u$ and $d$: $c$ (charm), $s$ (strange), $b$ (bottom) and $t$ (top)[13]. It is interesting that the total mass of $u$ and $d$ quarks constituting a nucleon ($\sim 10$ MeV/$c^2$) comprises about 1 percent of the nucleon mass ($\sim$ GeV/$c^2$) which means that the Higgs mechanism can hardly give us mass. But what can? If we recall the famous Einstein's formula $E = mc^2$, we may hypothesize that mass appears due to the energy of quarks and gluons moving inside nucleons. In other words, about 99 percent of our body mass is due to the mutual attraction of $u$ and $d$ quarks inside protons and neutrons. However, the exact nature of the strong forces mediated by gluons is still poorly understood.

Moreover, one should bear in mind that the Higgs mechanism is just a mathematical model that works perfectly as a plausible theoretical scheme, but it is still possible that the Higgs mechanism is incorrect. At least nobody, in spite of a lot of invested money and effort, has observed the Higgs particle yet. Actually, almost the entire LHC project, the modern analog of Egyptian pyramids, is devoted to the search for the Higgs particle. Why is it so difficult to find it? Primarily because the Higgs meson (provided it exists) is too heavy: its mass is estimated to lie between 120 and 180 proton masses that roughly corresponds to tin (atomic mass 119) and tungsten (184). The central value of the Higgs particle mass corresponds to atomic mass of about 140 (cerium or lanthanum).

Note that in quantum field theory it is not difficult to ensure, even without the Higgs mechanism, just by considering the scalar field models, that the mass of a particle becomes state and position dependent. Thus, already in the linear theory with $\phi$ interaction (this is the common model of elementary quantum field theory) one can have $m^2 \mapsto m^2 + \frac{\lambda}{2}\langle\phi^2(x)\rangle$. So, mass $m$ is regarded here as a function on the state space rather than a constant specifying the physical system.

---

[13] The top quark has a large mass value estimated as $m_t \sim 175$ GeV/$c^2$ (this mass is close to the atomic mass of the metal tungsten). Because of its great mass, the $t$-quark has a very short lifetime and decays earlier than it could form a hadron (in particular, a baryon).

It is clear that further discussion of relativistic, high-energy physics and quantum field concepts is outside the scope of this book, but what seems to be relevant is the very notion of a "particle". This is an illustration of the fact that things in physics that are considered the simplest are actually the most subtle. Historically, there have been many models of particles: for example, in the early days of quantum mechanics a particle was thought to be energy moving in the wave form. One might note that quantum mechanics more resembles a wave theory, in particular the classical electromagnetic theory, than classical mechanics with its point particles. To understand what a particle is, one needs to represent the basic stuff of the universe. According to the contemporary ideas of quantum field theory (QFT), the basic stuff of our universe is what might be called the vacuum grid ("grid" is the term coined by F. Wilczek, an eminent theoretical physicist). The concept of omnipresent grid[14] revitalizes the old idea of ether, but on a much more sophisticated level than the luminiferous ether of the 19th century that was vibrating to support the electromagnetic fields. In compliance with the current understanding, the basic material that fills all the space is in permanent spontaneous activity, and its adequate description is only achieved through novel mathematical techniques deployed for advanced QFT-analytics, whereas a more vulgar description is on the level of particles with mass.

Incidentally, the deficiency of the phenomenological description in terms of particles manifests itself in the discrepancies between quantum-mechanical and relativistic representations: particle in quantum mechanics is necessarily smeared, while it is impossible to introduce a non-pointlike particle in special relativity. Nonetheless, the important thing is that what looks like a totally empty space is actually a complex medium with a lot of spontaneous activity. This on-going activity of the grid is addressed in the parlance of physics as the creation and annihilation of virtual particles: they actually are fluctuations of spacetime, the transient underlying reality. What is interpreted as a real particle is the disturbance above this spontaneous activity; in other words, a real particle emerges when spacetime fluctuations are drastically amplified (in the physical jargon, "a particle appears on a mass surface"). Emergence of a particle can be crudely likened to such phenomena as a tornado or a tropical cyclone which are the moving local excitations in the constantly fluctuating atmosphere. This picture gives rise to an astounding idea: if to see something in nonrelativistic quantum mechanics we have to distort and in the limiting case even destroy it, in the more profound quantum world of QFT to see something one needs to create it.

It is curious that although QFT claims to be fully relativistic i.e., built over the Poincaré-invariant background, many QFT techniques semi-implicitly assume the presence of absolute time. In certain cases, QFT-computations are produced in absolute time and only then the results are a posteriori demonstrated to be independent of the choice of absolute time coordinates.

If the particle has an inner structure, it can be described as a point $\mathbf{x}$ in spacetime $M$ (e.g., $M = \mathbb{R}^4$) together with a set of internal states $S(\mathbf{x})$ i.e., fibration $S = \bigcup_{\mathbf{x}}\{\mathbf{x} \times S(\mathbf{x}), \mathbf{x} \in M\}$. One can treat each state in $S(\mathbf{x})$ as an element of some group $G$ (a structure group): this idea is an extension of the concept of spin or isospin. In general, the set $S$ varies from point to point as the particle moves from $\mathbf{x}$ to $\mathbf{y}$, similarly to tangent vectors. The structure group $G$ is the group of inner symmetry and acts only on the "phase" $S$ leaving points $\mathbf{x}$ of the base manifold $M$ intact. When a structured particle moves along some path $\boldsymbol{\gamma}(t) \subset M$ (the worldline), it carries with it the phase $S$ which, in geometrical terms, corresponds to parallel transport. The phase $S$ is shifted differently (by different $g \in G$) along different paths, and using again the geometrical language we may say that the phase factor $S$ can be represented through the curvature and the corresponding connection for parallel transport. This is the main idea underlying gauge theories and models, where curvature is identified with the gauge field of forces and connection with the gauge potential. Curvature is interpreted in physics as the field strength. However, going deeper into these fascinating issues would lead us far away from Newtonian models of particle motion.

---

[14] It is a very interesting fact that density of this basic stuff ("grid") seems to be the same everywhere in the universe, which is really startling since the universe is expanding.

The Newtonian model of mechanical motion $\mathbf{F} = m\mathbf{a}, \mathbf{a} := \ddot{\mathbf{r}}(t)$ has been for centuries the core of science and engineering. However, as simple as it may seem, Newton's equation ought to be further elaborated, especially when modeling the systems comprised of many interrelated parts. Indeed, in its naïve form, Newton's equation only describes the dynamical behavior of a "material point" whose only characteristics are mass $m$, position $\mathbf{r}$ and velocity $\dot{\mathbf{r}}$. The point mass occupies no volume in space and has infinite density, $\rho(\mathbf{r}) = m\delta(\mathbf{r} - \mathbf{r}(t))$. This is a pure mathematical (geometric) abstraction rather than a physical object. From the physical viewpoint, a material particle must have some minimal dimensions: for example, the minimal size of a material point in the classical continuous media may be determined by the electron radius $r_e = e^2/m_e c^2 \sim 10^{-13}$ cm. This quantity is also known as the Thomson radius since it defines the scattering cross-section in classical elastic scattering of electromagnetic radiation (photons) on free electrons (Thomson scattering). It is well-known that smoothing over the volume of an "infinitesimal" point leads to information loss and emerging irreversibility when locally reversible systems exhibit irreversible behavior. It is a common observation that a gas whose molecules obey the reversible laws of mechanics, regardless of its initial distribution, fills up the whole available volume.

The macroscopic irreversibility and its incompatibility with microscopic physical laws was tersely expressed by the famous Loschmidt paradox: why is there an inevitable growth of entropy, although microscopic laws are time-invertible? Reconciling the irreversibility of real life with the formal invertibility of classical mechanics (the one-way arrow of time) is one of the oldest problems of physics, mathematics and philosophy which still has no unique satisfactory solution. The disparity between "past" and "future" appears so pervasive and obvious to the common consciousness that the persistent attempts of scientists to reduce the time-asymmetric evolution to time-symmetric fundamental laws look troubling and unnatural to lay people. Basic laws of physics that govern elementary microscopic systems indeed have symmetries that admit the reversal of time direction, but this fact alone cannot justify the self-evident irreversibility of daily life.

The natural place to start the discussion of the asymmetry of time directions in real-life processes is Boltzmann's H-theorem (of the year 1872). The purpose of the H-theorem is to substantiate the statement that entropy in a many-particle system increases with time. Notice that, contrary to the dominating stereotype, not all microscopic laws of physics are $t \to -t$ invariant (reversible). For instance, time reversal symmetry (TRS) is broken in weak interactions. Furthermore, in orthodox quantum mechanics, the principle of observation is based on the concept of the wave function "collapse" or "reduction" which is an irreversible, time-noninvariant act, although one might note that the collapse of the wave function is the consequence of separation of microscopic observables and macroscopic observers.

For many physicists, the arrow of time does not exist or is largely irrelevant, at least at the most "basic" – microscopic – level of description. In many cases, this opinion is not corroborated by strict results, verging on a belief: it does not matter that the assumed basic level cannot be observed or correctly defined. If one defines the "basic level of description" as that of fundamental particles and their interactions, then the time reversal invariance has experimentally been proven to be broken, at least in weak – and hence electroweak – interactions. In real experiments, past and future are always different; nonetheless, the true believers in the "most basic level of description" maintain that the arrow of time in fact does not exist. This credo, if argued ad absurdum, becomes similar to medieval dogmas that treated any doubt of divine manifestations as heresy and a threat to ideal structures. Just as religious dogmas contradicted everyday experience, the statement that there exists basically no difference between past and future at some "very deep" level of description seem to contain more belief than physics.

Yet there is the obviously positive physical content in the problem of the arrow of time: one can assume that we can take time-reversal invariant mathematical models as basic microscopic laws of physics. However, this issue, which is beset with daunting mathematical and physical difficulties, is beyond the scope of the present manuscript. One might only note that the ultimate quantum mechanical reason of time irreversibility seems to lie in the impossibility to create the states that would evolve with decreasing entropy, although such states are not formally forbidden by any physical laws.

Obviously, real-life physical and technological systems incorporating many linked bodies have little in common with the point mass model described by the second law of Newton. In general, one can notice that the naive point-to-point diffeomorphism $x_0^i = x^i(t_0) \mapsto x^i(t)$ implied by the mechanical equations of motion becomes inadequate for many natural systems, which induced J. W. Gibbs to construct his statistical mechanics as a theory incorporating two mutually consistent structures: phase space and probability space or, more specifically, phase flow and probability measure. Moreover, the above form of Newton's equation only holds when the motion is viewed from the standpoint of inertial (Galilean) reference frames. The latter actually constitute a tiny subclass of realistic frames of reference due to omnipresent accelerations and gravity[15].

## 4.5. Beyond physics

The successful solution by Newton of the mysterious Kepler problem inspired thinkers and philosophers, primarily Laplace, to make bold generalizations and to develop the mechanistic model of the universe. Newton's theories of motion and gravity demonstrated the power of deterministic laws: if one had a reasonably exact knowledge of the state of the system – in particular, the position and the velocity of the body such as a planet in the solar system – at some moment usually taken as initial, Newtonian theories prescribe where the body would be (with fair accuracy) at any moment in the future. There was no room for randomness in this clockwork vision of the world. For centuries, it was believed that all events are, in principle, predictable and can be described by deterministic differential equations similar to the equations of motion in mechanics. As to apparently unpredictable and irreversible phenomena such as weather or human behavior, they were believed to be unpredictable and irreversible only due to a very large number of variables. This belief persisted until the late 20th century, when it was hoped that with the advent of powerful computers all long-range pre- and retrodictions would be possible (which is wrong).

Most real-world phenomena are beyond physics, the latter being traditionally understood as the study of non-living objects and inorganic matter. Thus, mathematical modeling, though rooted in physics, is increasingly spread beyond physical sciences. Notice that many physicists of the traditional school still regard the phenomena beyond physics, especially highly nonlinear ones in the organic world, as seemingly intractable and already therefore not worthy of serious consideration. Now, there are more and more scientists who are trying to spread the intellectual style of physical research to other phenomena.

---

[15] One might recall in this connection how difficult it was to get rid of parasite accelerations and perturbing gravity fields in certain high-precision experiments (e.g., in tests of the equivalence principle carried out by V. B. Braginsky and V. I. Panov in the Moscow University). A record sensitivity of the experimental device was needed to perform the measurement.

The biggest challenge of biology, medicine, society and economics is that great diversity of data and their randomness can nonetheless lead to fine-tuned processes (in time) and structures (in space). It means that the ideal physical notion of the world as a machine seems to be inadequate so that a paradigm shift is required. In fact, fully mechanistic nature would be incompatible with life, where evolution gains order through fluctuations. Biological evolution is typically understood as a descent accompanied by slight modifications. Diversity of biological components increases viability and resilience of a biological system. From the point of view of biological esthetics, diversity is the foundation of beauty since it produces outstanding samples against a uniform background. Variability and the ensuing diversity arise as replication errors: in the now fashionable language of code, such errors are insertion, deletion and replacement of code fragments. Anyway, biological evolution basically occurs due to errors, and this picture is poorly consistent with deterministic classical mechanics.

Classical science mostly studied systems and their states close to equilibrium, and that allowed one to construct a beautiful collection of comparatively simple physical models for the world. Such models depicted the systems that reacted to perturbations more or less predictably: these systems tend to return to equilibrium (in the parlance of statistical physics, they evolve to a state that minimizes the free energy). Remarkably, however, systems close to equilibrium can describe only a small fraction of phenomena in the surrounding world – it is in fact a linear model. Any realistic system subject to a flow of energy and matter will be driven to the nonlinear mode i.e., far from equilibrium. For example, open systems such as Earth, climate, living cell, public economy or a social group exhibit highly complex behavior that is, firstly, hard to be replicated in the laboratory and, secondly, almost impossible to model mathematically using the methods adapted mainly to mechanical patterns. In contrast with the closed mechanical models which are a drastic simplification of reality, open and nonequilibrium systems are ubiquitous in nature. Most of the processes in the open systems far from equilibrium are interrelated, nonlinear, and irreversible. Often a tiny influence can produce a sizable effect, which is a universal property of nonlinear regimes, and in the real world almost any system is nonlinear.

In non-physics, there is a group of disciplines that are increasingly trying to rely on quantification such as finance and – to some extent – medicine. However, numbers appearing in such disciplines usually cannot be attributed to an application of profound physical theories or powerful mathematical techniques; these numbers result from the use of *ad hoc* methods that are often regarded as shallow by mathematically oriented scientists. This is in most cases snobbish and unfair: employed mathematical methods should be adequate, and if the use of market or other "vulgar" variables brings the result that possesses a satisfactory accuracy, there is no need to build a deep theory or model. One only needs to honestly tell people about the shallowness of the methods in use.

Unlike in physics, there are no invariance principles, neither based on global nor on local symmetries, in "soft" sciences, in particular in socio-economic disciplines. Symmetry concepts and invariance principles pervade the whole physics distinguishing it from "softer" disciplines, where such concepts are not considered important. Accordingly, there seem to be no fundamental mathematical theories based on universal principles in socio-economic disciplines.

On the contrary, in physics-based mathematical models, it usually pays to use fundamental theories, universal principles and generic variables. There were many attempts to produce non-physical theories using the physical methodology. One might recall, for instance, Spinoza's theory of passions, the latter being built as a combination of some basic passions, just as any color can be represented as a combination of primitives in the color space such as RGB (red, green, blue) or CMYK (cyan, magenta, yellow, black). The primitive passions, according to Spinoza, are desire, pleasure, pain, and every human drive can be, using modern language, expanded on this triple. Other fundamental notions such as "good" and "evil" can, according to Spinoza's theory, explained in terms of basic passions, for example, "good" is anything that brings pleasure whereas "evil" is anything that leads to pain.

The dominant paradigm of the Enlightenment was deterministic reductionism, at first in the rather primitive form claiming that everything can be obtained from the microscopic laws of motion such as Newtonian ones. Despite being an unsubstantiated hypothesis and actually a belief, this paradigm had been almost unquestionably accepted by the scientific establishment until the 1970s when it started to be gradually and tacitly replaced by another dominant paradigm, the one of emergent behavior. It is no wonder that there was long resistance to abandoning the mechanistic paradigm: aversion to novelty is one of the main features in the majority of people, and scientists are no exception. If deterministic reductionism was focused on looking for the ultimate cause of all observable macroscopic phenomena (sometimes labeling their direct study as "phenomenology" or even "pseudo problems"), the emergent behavior is basically related to collectives and not to individual paths as in mechanical evolution. In particular, when considering the problems of stability of complex structures and, more generally, how matter is organized, unpredictable and irreversible factors such as fluctuations can play the most significant part, making a mechanistic description in terms of Hamiltonian systems and individual trajectories completely inadequate.

It is worth noting here that one sometimes confuses nonequilibrium states and irreversible processes. We shall understand a nonequilibrium state of a system as any state deviating from the complete statistical equilibrium and a nonequilibrium process as an evolution throughout nonequilibrium states (at least part of the states traversed by the system can be out of equilibrium). From the viewpoint of dynamical systems theory, irreversibility can be attributed to the action of some semigroup. Physically, irreversible processes are the ones that are accompanied by the entropy generation in the system. The difference with the nonequilibrium processes is that, in the latter, entropy can only be redistributed, in particular, when there is no dissipation. For instance, such phenomena as diffusion, heat transfer, viscosity, flow of electric current through a resistance, expansion of gases into the empty space, etc. are irreversible. From the quantum-mechanical viewpoint, the entropy growth is usually interpreted in terms of entanglement[16] with the degrees of freedom of the reservoir (thermostat).

One more typical feature of real-world systems, in particular those far from equilibrium, is that they can lose their stability and evolve to one of many states. This behavior appeared so "unphysical" from the habitual viewpoint that some time ago many orthodox physicists were inclined to despise those colleagues who were trying to consider systems far from equilibrium, especially those beyond physics. When studying such systems, one should usually look for ways to redress various imbalances. In physics, one of the most important problems is the description of objects put in some environment, the latter being usually regarded as given and modeled as a "thermostat" i.e., some generic homogeneous medium characterized by a single parameter – temperature $T$ that intuitively specifies the equilibrium of the environment. Yet to model the processes in the real world one must learn how to describe both the environment and the systems far from equilibrium. This subject is important and will be specially discussed in section 8.

---

[16] The fashionable term "entanglement" in today's quantum theory denotes a set of non-separable states. A separable state for a composite quantum system is the one that can be represented as a sum of probability (or, for pure states, amplitude distributions) over uncorrelated states i.e. $\rho(1,2) = \sum_n w_n \rho_n(1) \otimes \rho_n(2)$ or, in Dirac's notations, $\rho(1,2) = \sum_n w_n |\psi_n(1)\rangle \langle \psi_n(1)| \otimes |\psi_n(2)\rangle \langle \psi_n(2)|$. Here coefficients $w_n$ of the bilinear combination are regarded as probabilities i.e., $\sum_n w_n = 1$ and $w_n \in [0,1]$. For two pure states $|\psi(1)\rangle \in H_1$ and $|\psi(2)\rangle \in H_2$, where $H_{1,2}$ are respective Hilbert spaces, a separable state of the composite system can be represented as $|\psi(1,2)\rangle = |\psi(1)\rangle \otimes |\psi(2)\rangle$. It is in general useful to translate from the peculiar language of quantum mechanics and quantum information processing into the more habitual language of traditional functional analysis.

## Section 5. Basic modeling approaches

One can single out the following stages of mathematical and computer modeling: 1) theoretical; 2) algorithmic; 3) software development; 4) computer implementation; 5) interpretation of results. In this paper, we shall mostly focus on theoretical approaches, other stages will be only briefly discussed.

Practical steps taken to build and analyze mathematical models are typically as follows:

1. Clarify the requirements. This stage (requirement engineering) is extremely important and should not be overlooked. Actually, it is here that the modeling experience manifests itself. The requirement conflicts are quite common, they can originate in different interpretations of the same concept, dissimilar mathematical thesaurus, diverse views of the needed accuracy, even interpersonal disagreements, etc. Requirement conflicts are typically harmless as long as they are detected early and resolved, say, at the model setup phase. However, these conflicts may result in grave discrepancies and even serious modeling errors when remaining unperceived until the software development and implementation stage. In such cases, conflict resolution may become quite expensive.
2. Select the list of observable quantities. Defining variables specifies crucial assumptions of the model. Really serious errors emerge at the stage of formulating the model assumptions.
3. Determine the measurement methods needed to verify the model. Measurement is understood as establishing a correspondence between observables and numerical values that are expected to belong to some ordered set.
4. Postulate (often just through guessing) the laws governing the distribution or the evolution of observables in the space of all possible states of the modeled system or process, in particular, in the phase space. Such laws can be exact (as in classical deterministic dynamics) or probabilistic (as in quantum theory or when modeling random systems such as weather)[17].
5. Set up the mathematical problem in the form of equations, inequalities, mappings, etc.
6. Try solving the mathematical problem, estimating errors and finding asymptotics. One typically has to use rather powerful computers and advanced programming tools (such as computer graphics or computer algebra systems) to implement realistic models.
7. Test the model. One can attempt to answer the following questions: does the model describe the situation one was about to mimic mathematically? Does the model work well in relatively simple limiting (asymptotic) cases? What are the discrepancies?

---

[17]The main attention here is devoted to deterministic models, with random dynamical systems being only briefly discussed in this book because of its limited scope.

8. Try to refine the model.

One can see that building a model is an iterative process that can be represented by a graph. The process terminates when the information flow along all the paths connecting the graph nodes vanishes.

## 5.1. Hierarchical multilevel principle

Mathematical and computer models of complex phenomena can be broken down to smaller and simpler *submodels*. Thus, the *hierarchical principle* of model building with a *step-like refinement* is usually employed to manage the system's complexity. The concept of hierarchical multilevel organization is very valuable in both science and life, recall, for instance, positional systems of arithmetic, Fourier and wavelet transforms, or enterprise structure. Probably the most important concept in life – life itself – is constructed on a hierarchical principle, when more complex forms are successfully built up on the base of more rudimentary ones.

Many general things are better understood with simple examples. Let us try to illustrate the modeling hierarchy by one of the favorite models of physics – that of an oscillator. Consider the following equation of Newtonian mechanics:

$$\ddot{x} + a\dot{x} + bx = f(t), \qquad \dot{x} \coloneqq \frac{dx}{dt}. \tag{5.1.1.}$$

Here three types of processes are contained: inertial motion represented by the term $\ddot{x}$; damped motion represented by the term $a\dot{x}$ which gives the resistance force; kinematic (Galilean) motion represented by the term $bx$ (in the oscillator model this term is associated with the restoring trend), and the forced motion of the whole physical system represented by the driving force $f(t)$. Variable $x$ is the state variable whereas the pair ($x = x^1$, $\dot{x} = x^2$ also called modes) corresponds to a point in the phase space of the oscillator model. Here mass $m$ of the oscillator is absorbed in coefficients $a, b$ and force $f(t)$.

Few classical dynamical systems look as simple as (5.1.1.). But this simplicity is treacherous. Although one can build a Lagrangian and a Hamiltonian yielding the motion equation (5.1.1.), they depend on time $t$ (owing to the dissipation or, for time inversion $t \rightarrow -t$, amplification term $a\dot{x}$). This fact alone makes it hard to directly use the standard Lie algebra relations such as the Poisson brackets and commutators, e.g., for canonical quantization. The fundamental difficulty here is that dissipative systems are not closed and hence should be described through phenomenological models. Because of the difficulties while treating equation (5.1.1.) straightforwardly in standard physical situations, a rich scope of physical approaches and mathematical models have been developed to physically account for dissipation such as representing a damped oscillator as a free one linked with a continuum of harmonic oscillators with different frequencies making up a thermostat (see an overview in [36]). These thermostat oscillators can interact both with a given oscillator and with one another. The advantage of this class of models consists in the fact that one can solve the problem of coupled harmonic oscillators exactly (both in the classical and quantum case) so that one is able to trace on such models the appearance of phenomenological dissipation and, more generally, to follow the transition from reversible dynamics to irreversible statistical description, which is an extremely important subject in physics and nearby disciplines.

If $f(t) = 0$, the oscillations described by equation (5.1.1.) are said to be free, but still damped and possibly nonlinear when $a = a(x)$ and $b = b(x)$ (in general, of course, $a = a(x, t)$, but it does not matter in the present context). However, in many cases when one encounters nonlinearity, it may be considered small, for example, $a(x) = a_0 + \alpha(x)$, $b(x) = b_0 + \beta(x)$, where $|\alpha| \ll a_0$ and $|\beta| \ll b_0$ ($a_0, b_0$ are considered nonnegative). We thus may further descend the hierarchy level if we completely ignore nonlinear corrections or account for them as very small higher-order disturbances. Furthermore, in a great many situations not only nonlinearity but also damping can be totally disregarded so that the oscillator eventually is:

a)  free ($f(t) = 0$),
b)  autonomous (coefficients in equation (1) do not depend on $t$),
c)  linear (coefficients $a, b$ do not depend on state variable $x$),
d)  undamped ($a = 0$),

and as a result of consecutive removal of complex generalizations we arrive at a simple but very important model of the harmonic oscillator, $\ddot{x} + \omega^2 x = 0$, where $\omega$ is the frequency of oscillations. We shall see shortly that the general equation of free oscillations which includes both the restoring trend and nonlinear damping can be written in the autonomous case as $\ddot{x} + \omega^2 x = g(x, \dot{x})$. By making successive approximations in the right-hand side, which is the local function of the phase point, one can trace the hierarchy levels of this mathematical model imported from classical mechanics but used in many other disciplines.

Model (5.1.1.) of the damped oscillator is inspiring to study a wide class of nonlinear oscillations described by equations of the type

$$\ddot{x} + a(x)\dot{x} + b(x)x = f(t, x) \qquad (5.1.2.)$$

(the Liénard equations). To evade some inconveniences one can assume the functions $a(x)$ and $b(x)$ to be continuously differentiable over spatial coordinate $x \in (-\infty, +\infty)$. For example, one can can specify $a(x)$ and $b(x)$ to be of the simplest form, i.e., respectively $a(x) := kx$ and $b(x) := \lambda x + \mu x^3$, $\lambda > 0$, thus defining a cubic nonlinear oscillator

$$\ddot{x} + kx\dot{x} + \lambda x + \mu x^3 = f(t, x, \dot{x}) \qquad (5.1.3.)$$

subject to an external force $f(t, x, \dot{x})$ and a nonlinear damped force $kx\dot{x}$. Despite their simple look, such nonlinear equations are quite challenging even if the external (driving) force $f = 0$.

This example represents countless modeling hierarchies in physics having different degrees of generality. Most of the models created to describe physical phenomena are of a local character i.e., they are produced within the framework of a more general model (or a theory) being applied to a restricted situation. One can illustrate this nesting ("matryoshka") principle by many examples: thus, acoustics incorporate models that are specific cases of fluid dynamics and thermodynamics (under the assumption of small variations of pressure – the reverse case is shock pulses). Models of heat and mass transfer were initially constructed as an independent area with its own *ad hoc* laws. According to the postulates of the 17th-18th century scholars, heat was described as a fluid called phlogiston. Indeed, although the phlogiston model is obviously false, in some simple cases heat can be represented to behave as a conserved fluid. It was only after Maxwell, Boltzmann and Gibbs had connected the models of heat and mass transfer to mechanics, that the hierarchical structure of thermophysics became clear.

A significant manifestation of the hierarchical principle of model building is perturbation theory. There exist several main versions of perturbation techniques, but all of them essentially utilize the hierarchy concept. For instance, in the adiabatic perturbation theory, separation of scales is used to simplify and better understand the behavior of complex systems. A prominent physical example is the Born-Oppenheimer approximation in molecular dynamics and condensed matter theory; this approximation is a hierarchy of motions based on small ratios of masses, in particular those of electrons and ions.

In general, perturbation theory expresses the desired solution $u$ of a complex problem in terms of a power series in some small parameter $\varepsilon$, $u = u_0 + \varepsilon u_1 + \varepsilon^2 u_2 + \cdots$. The principal term $u_0$ in this series is the solution to a related problem – a drastically simplified version of the original one that can be solved exactly. In many situations, one can be satisfied with the first two terms of the series, $u \approx u_0 + \varepsilon u_1$, i.e., only up to the "first order" of the perturbative hierarchy. Notice that from a mathematical standpoint most perturbation methods are a manifestation of differentiability of equations over some parameter. In some complicated multiparametric cases, there may be more than one small parameter, and then one would have several expansions over a number of parameters $\varepsilon_k$, $k = 1,2, \ldots m$, obtained by a number of intertwined procedures. Perturbative hierarchies may be represented graphically, in particular with the help of Feynman diagrams, which both illustrate and exploit the hierarchical character of perturbation techniques. In quantum mechanics, and especially in quantum electrodynamics, perturbation theory has been one of the main methods of obtaining quantitative results. We shall discuss examples of perturbation techniques below in some detail.

Methods of perturbation theory are quite close to the ones used in numerical analysis. Thus, iterative methods for solving large systems of linear equations are sometimes only terminologically different from perturbation methods of physics, depending on whether they are discussed in the physical, mathematical, computer science or engineering audience. Anyway, iterative methods are also mainly based on the principle of hierarchy, and mostly due to this principle they are gaining popularity because they are comparatively easy to implement on modern computers. By the way, modern computer modeling essentially uses the hierarchical approach, for example, the inheritance hierarchy (derived classes) is one of the main principles of object-oriented modeling and design.

The most famous example of hierarchical approach in physics is the BBGKY (Bogoliubov–Born–Green–Kirkwood–Yvon) hierarchy in statistical mechanics often called the "Bogoliubov chain" (after N. N. Bogoliubov, the great Russian mathematician and physicist). This is the chain of equations for the many-particle distribution functions obtained directly from the most general model of mechanical motion in the system of $N$ identical particles (more exactly, the set of all representative points in the phase space of an $N$-particle system - statistical ensemble), the Liouville equation. The BBGKY approach allows one, in particular, to derive the Boltzmann equation (which is the equation for one-particle distribution function), to explore it in various regimes, and in general to understand the temporal evolution of many-particle systems (e.g., to their equilibrium states). We shall briefly discuss these fundamental physical issues in the small section devoted to statistical mechanics and thermodynamics.

One more important example of representing a complex problem as a hierarchy of subproblems is the multiscale analysis which is based on the introduction of "slow" and "fast" times, in general, having many different degrees of "slowness", $\tau_1 = \varepsilon t, \tau_2 = \varepsilon^2 t, \tau_3 = \varepsilon^3 t, \ldots$. Here $t$ is the "normal" physical time and $\varepsilon$ is a small parameter[18]. The physical reason for considering this hierarchy is that the system to be modeled usually exhibits a characteristic behavior over a wide range of time scales. This idea has been successfully implemented in computer modeling and scientific computing, rather prominent examples being the multigrid technology and closely related multilevel techniques. More recently, the wavelet-based methods (which are intrinsically hierarchical) have produced high expectations.

---

[18] It is curious that the separation of a complex motion into "slow" and "fast" subsystems is also associated with the name of N. N. Bogoliubov (the Bogoliubov-Krylov-Mitropolski averaging method). The multiscale analysis may be viewed as an extension of this method.

A prominent case of the hierarchical scale separation is provided by turbulence. Modeling of this phenomenon – a very complex subject – is mostly based on multiscale and multiresolution approaches, these two methodologies being related to mathematical and computer modeling, respectively. Most fluid flows observed in nature can transit to the turbulent regime so that understanding turbulence is essential for engineering applications of fluid dynamics. Turbulent velocity fluctuations spreading across the flow are characterized by a wide range of spatial and temporal scales (in the physical spacetime, $\mathbf{r}$,$t$) that is translated into the broadband spectra in the dual wave vector-frequency space ($\mathbf{k}$, $\omega$). Thus, to build a model of the turbulent flow, one has to introduce a hierarchy of scales both in time and space. Most mathematical models of turbulence are based on the Navier-Stokes system of equations (see Supplement 3) which we briefly discuss in connection with PDE-based modeling, fluid motion and scientific computing.

What we interpret as "understanding" is usually a reduction throughout the hierarchy down to the concepts that appear more elementary than the initial problem. Thus, we tend to "explain" the fluid flow by laws of continuum mechanics which can be reduced to kinetic equations on the atomic-molecular level. This level, in its turn, can be reduced to "elementary particles" such as electrons, protons, etc. Then one can descend one more level, down to the Standard Model and, later, to string theories and so on.

Finally, in association with the hierarchy principles, we can mention the self-similarity paradigms that have come into fashion in the last quarter of the 20th century. There exist, for example, models of the universe based on self-similarity ideas, assuming that all the fundamental scales of nature such as subatomic, atomic, stellar, galactic, etc. scales constitute the infinite hierarchy of nature and are interrelated by some self-similar laws (as in fractal geometry). Such models have of course a natural-philosophical character i.e., they are not suited for practical calculations and say very little about the dynamics of the modeled objects.

The term fractal introduced in science and life by Benoît Mandelbrot, a recently deceased French American mathematician, has just been mentioned; this concept is close to self-similarity (but not identical to it) and is used as a tool to model diverse complex objects in both natural sciences and humanities. Being developed from scaling maps, the fractal hierarchy has become a ubiquitous paradigm for understanding irregular natural phenomena (such as coastlines, turbulence, the highly inhomogeneous distribution of galaxies, asteroids, blood vessels, the structure of lungs, stock market behavior, growth phenomena, etc.). A highly useful aspect of the whole fractal concept was doing extensive computer experiments, and thus computer modeling has greatly benefited due to fractals.

## 5.2. Dissection and compartmentalization

An additional requirement which does not necessarily follow from modeling hierarchies is modularity and reusability of models. This property is specifically valued in computer-implemented models and is always expected from good software products. On the algorithmic level, dissection of a problem into smaller subproblems more or less of the same kind is usually known as the "divide-and-conquer" strategy. One can typically solve such subproblems easier than the parent problem; next, one ought to merge the partial results into a full solution to the problem. This strategy was traditionally employed in mathematical physics, where the techniques of variable separation with subsequent partitioning of the solution into eigenvectors in some functional space proved to be extremely successful. The same ideology permeated conventional quantum mechanics which is based on the fundamental divide-and-conquer postulate called the superposition principle, when the wave functions or "rays" i.e., single-dimensional complex subspaces of the Hilbert space form a vector space. In other words, a Hilbert space is the development of the vector space concept. Quantum mechanics, due to its modeling and computational efficiency, has today become a powerful engineering discipline (recall electronic band engineering in material science, photovoltaic device engineering, optoelectronics, nanoengineering and nanotechnology, high-temperature superconductivity, spintronics, quantum optics, quantum information science and quantum computation, etc.).

Decomposition of a modeled system into partial subsystems is a fruitful mathematical idea fully developed in such disciplines as linear algebra, matrix (operator) theory and functional analysis. The principle of superposition of states was the $20^{th}$ century extension of the old ideas about eigenvalues and eigenvectors originated in vector algebra[19]. As a result, nearly all the constructions used in complex linear algebra which is essentially based on the decomposition into partial states have transmuted into powerful tools and models of physics used to formulate fundamental laws of nature (such as the theory of atoms, nuclei, and solid state, the periodic table of chemical elements, and even the models of elementary particles). We shall see below that eigenvalues can correspond to such practically important quantities as vibration frequencies, critical values of stability parameters or energy levels in atoms.

A mathematical model often includes the idea of a "system". The term "system" in mathematical and computer modeling is usually understood as designating an assemblage of interdependent or interacting objects. For example, a physical particle interacting with its environment, in particular, with other particles or with external fields; coexistence of the populations of several interacting species; coupled attack and defense activities of two armies in the battlespace environment – all those are "systems" from the modeling standpoint regardless of their specific content: physical, ecological, military or other. The term "qualitative analysis" used above mainly implies standard analytical techniques which, though gradually becoming out of fashion, can nonetheless be very powerful and deliver both qualitative and quantitative outcomes. An important thing is that reasonable application of qualitative analysis helps to test local models and to stay away both from mathematical fairy tales disguised as complicated physical theories and from generic low-information tenets that nobody would disagree with.

## 5.3. Assumptions and estimates

Assumptions in mathematical modeling can be totally unrealistic so that the outcomes of many models should not necessarily require an empirical verification i.e., a possibility of being eventually refuted by experiment or observation. Therefore, one can appreciate mathematical models by their possible predictive abilities rather than by the close depiction of real-life processes at each step.

---

[19] Recall that an eigenvalue and eigenvector of a square matrix $A$ are a scalar $\lambda$ from some field $K$ and a nonzero vector $x$ such that $Ax = \lambda x$.

The models of real-life phenomena typically include too many parameters, and which of them are more important than the other ones is a priori unclear. Thus, selecting the appropriate parameters and variables while neglecting the rest may already be a big problem. Some parameters may be time-dependent and have to be determined by solving the equations relevant to the model. Mathematically speaking, one has to define the spaces of internal and ambient parameters as well as to discriminate between parameters and variables. In the real world it may be difficult: almost any quantity is a variable except, perhaps, several fundamental constants characterizing our universe (and even these constants can change with time in certain physical models). Moreover, the properties of the model equations, for example, the existence, uniqueness, smoothness and topological type of their solutions can be totally incomprehensible. In particular, forecasting the features of solutions to nonlinear problems may present a great challenge. It is here that one has to make assumptions, frequently even the intuition-based decisions.

For example, physicists assume that our space is isotropic i.e., all directions in it are equivalent. This is an intuitive but important assumption since it implies that one can analyze physical objects in any coordinate systems. Indeed, as there are no preferred directions, the properties of physical objects should not depend on the orientation of coordinate axes. The refinement of this idea has led to tensor analysis and general relativity.

Analogously to theorems in mathematics and applicability limits in physics, any correctly stated mathematical and computer model realizes the "if-then" paradigm, where "if" encodes the assumptions contained in the model and "then" implies the equations (or other logical procedures) as well as their computer implementation, followed by the model's outcomes. It is clear from the "if-then" scenario that neither mathematical or logical ingenuity nor software skills can create new reality so that the correspondence of a model to the situation it is designed to describe is determined by the assumptions encoded under "if".

Thus, various assumptions are used in order to simplify mathematical models. The most widespread assumption is that of linearity, since linear equations are easy to solve and explore, although most real-life systems and processes should be described by complicated nonlinear equations. In many mathematical models, linearization (i.e., replacement of a nonlinear problem by the related linear one, see section "Linear models" below) leads to a satisfactory approximate solution, although it is not always clear what accuracy can be attained by the linear approximation. Linear models occupy a special place in mathematical modeling, specifically in electrical engineering and quantum mechanics, because they possess two key features: their output is proportional to input, and the superposition principle holds. In essence, these are the vector space properties. In mathematics, linearity is the base concept of classical calculus: any smooth (or even differentiable) function $f(\mathbf{x})$ in the vicinity of each point $\mathbf{x_0}$ can be well approximated by a linear function, $f(\mathbf{x}) \approx f(\mathbf{x_0}) + \partial_i f|_{\mathbf{x_0}}(x^i - x_0^i) = f(\mathbf{x_0}) + \nabla f|_{\mathbf{x_0}}(x^i - x_0^i)$, where $\mathbf{x} = \{x^1, \ldots, x^n\}$. Differentiation can be viewed just as an exploitation of the linearity concept.

The field of nonlinear phenomena embraces such a wide variety of subjects that it is hardly possible to observe a representative part of them, even by using a crude classification. Because of the complexity of nonlinear modeling, we often treat linear models as universal laws, without realizing the underlying assumptions. Thus, Ohm's law is an example of a linear model, although many people, even some electrical engineering professionals tend to consider it the law of nature. This is wrong, and the error is probably due to the fact that since the school years we got used to associating electricity with metal wires, where it is nearly impossible to produce strong electric fields in which the linear response model does not hold. Contrariwise, in semiconductors or plasma, where internal electric fields can easily become strong[20], linear models (such as Ohm's law) usually fail and should be replaced by more sophisticated ones. The same erroneous extrapolation of linear response may be encountered in elasticity theory (Hooke's law), thermophysics (Newton's law of cooling) and heat transfer (Fourier's law), all these disciplines being essential for technology and engineering. As in the case of Ohm's law, such linear physical models are only valid for small (more correctly, infinitesimal) gradients. One may note that linear response to external perturbations is more important than just strictly linear laws of change. For example, stability of a system is determined by the sign of the system's linear response coefficients.

Making intelligent assumptions is the antithesis to relying on off-the-shelf prescriptions. To corroborate the assumptions, qualitative methods can be very useful. Such methods have been widely spread in the theory of ordinary differential equations (ODE), especially as concerns the stability theory with numerous applications in engineering, physics, chemistry and biology. In the ODE theory, qualitative methods provide the justification for many approximations that at first seem arbitrary and are merely assumed to be reasonable and valid. Below we shall use such qualitative methods to estimate the asymptotic behavior of models. It is interesting that in some cases making an intelligent assumption can already produce a workable mathematical model. For example, assuming that the local population growth (e.g., expressed throughout fertility i.e., the number of childbirths per female in a given country) is inversely proportional to the country GDP (growth domestic product) per capita, one can arrive at a simple mathematical model (curve) that roughly reflects reality.

Asymptotic estimates are, in particular, needed to answer the question: what happens when time tends to infinity? Stability of a dynamical system is a typical example. We have already mentioned that this problem can be translated into the geometric language as the asymptotic estimates of trajectories in the phase space (more generally, in the state space).

Before the advent of powerful computers, which radically changed the approach to mathematical modeling, assumptions and estimates had been the principal techniques of obtaining results with the known accuracy (quantifiable error), for example, in theoretical physics. Many physicists have mastered the art of making estimates to a virtuosic degree. Such great scientists as E. Fermi and L. D. Landau were famous for their ability to make justified assumptions and estimates without doing real calculations and even without firm data (for these and many other great physicists, ideas were obviously more important than actual numbers). The principle generally adopted in theoretical physics was that assumptions and estimates should not lead to errors exceeding one order of magnitude. Such accuracy was of course desirable, but not always attainable, besides it is not at all trivial to establish and quantify the error of a model. One more principle adopted in physics was that the calculations performed under the already made assumptions should substantiate them.

As a simple non-physical example illustrating how the technique of assumptions and estimates works, one can mention the classic Fermi estimate of the number of piano tuners operating in Chicago (Fermi chose this city because he lived and worked there on the first nuclear reactor – Chicago Pile-1). This estimate is a typical modeling approach based on a sequence of intelligent guesses. The following assumptions and calculations were attributed to Fermi in physicists' folklore (see Internet resources in bibliography to this book):

1. Chicago has approximately five million inhabitants (at about 1940).
2. Each household in Chicago is assumed to consist of two persons in average.
3. The proportion of the households with pianos to be periodically tuned is estimated as five percent.
4. Pianos should be tuned with the mean periodicity of once a year.
5. It takes approximately two hours to tune a piano, including point-to-point trips.
6. The working time of all the tuners is assumed to be eight hours a day, with five days in a week and fifty weeks a year.

---

[20] For example, an electron moving in such a field can acquire energy comparable with some critical value for the medium, e.g., sufficient for ionization.

Let us now, using these assumptions, estimate first the annual number of piano tunings.

- The average number of households in Chicago: $\frac{5 \cdot 10^6 \text{persons}}{2 \text{ persons/household}} \sim 2.5 \cdot 10^6$ households
- The average number of pianos to be regularly tuned: $2.5 \cdot 10^6 \cdot 5\% \sim 125 \cdot 10^3$ piano tunings a year in Chicago.

How many piano tunings can a single piano tuner perform per year?

8 hours/day $\cdot$ 5 days/week $\cdot$ 50 weeks/year $\sim 10^3$ piano tunings/year $\cdot$ piano tuner.
How many piano tuners are needed to perform all piano tunings in Chicago during a year?
$125 \cdot 10^3$ piano tunings/year/$10^3$ piano tunings/year $\cdot$ piano tuner $\sim 10^2$ piano tuners.

One can use assumptions and estimates to assess the mathematical expressions without performing actual computations. For example, if we need to estimate the derivative of the function whose form is $f(x) := f(x + a)$ i.e., additively depending on parameter $a$ (we may take for simplicity $f(x) = (x + a)^{-1}$) [21], then in the domain $x \gg a$ we have $f'(x) \sim f/x$ whereas in the domain $x \ll a$ the derivative is estimated as $f'(x) \sim f/a$.

The same technique can be used to estimate, say, the number of dermatologists in some locus. Making this kind of estimate prior to hard-core computations allows one to quickly appreciate the order of magnitude and thus to evade bad mistakes. Besides, qualitative estimates enable computer modelers to simplify the equations to be numerically processed by throwing out inessential terms. The qualitative approach based on estimates is especially important in such fields of physics as physical kinetics and fluid dynamics where the described processes are very complicated and depend on numerous parameters so that it is not a priori evident which terms of the equation can be disregarded (climate modeling is a prominent example).

A similar primitive example can be built regarding the sequence of significant events in world history. If we denote the current (or some future) time point as $t_*$ and start roughly counting significant events in the past as $t_n = t_* - \alpha^n T$, where $T > 0$ is some characteristic time period between subsequent events and $\alpha < 1$ is the correction coefficient accounting for the acceleration of technological progress and increasing political intensity. If we, for some slight convenience, invert this coefficient putting $\alpha \equiv 1/\beta, \beta > 1$, we shall get for two successive events $\Delta t_n = t_n - t_{n-1} = T(1/\beta^{n-1} - $ $1/\beta^n) = T\frac{\beta^n - \beta^{n-1}}{\beta^{2n-1}} = T\frac{1 - 1/\beta}{\beta^{n-1}} = T(\beta - 1)\beta^{-n} \to 0, \beta > 1$. This expression shows that new events are coming more and more densely in time, so that the model softly hints at the end of the world. However, this type of model, for which temporal ordering of events is necessary, reminds us of some arbitrary numerology which might be amusing but groundless.

---

[21] This example is essential for condensed matter physics, in particular, to estimate the electronic specific heat in metals $c = \partial \mathcal{E}/\partial T$, where $\mathcal{E}$ is electronic energy, $T$ is the temperature of the electronic subsystem (the conventional model for low-temperature specific heat is $c(T) = aT + bT^3$, where $T$ is lower than Debye and Fermi temperatures).

Note that it would be difficult to prove that all significant events and their respective consequences could be taken into account in such a primitive model. Actually, each event appearing in history can potentially produce a considerable impact on the set of human events. For instance, creation of the mechanical clock based on time-periodic process in the 13th century led at first to more or less precise time measurement, then to navigation by using visible stars (astronavigation) that made it possible for navigators to determine their position at open sea and brought with this ability the geographic revolution, later provided a scientific foundation for ballistics and so on. Newton, Kepler, Pascal, Descartes and other great scientists of late renaissance were busy with studying scientific facts, but their fundamental endeavors resulted in accelerated evolution of military technologies, the warfare evolution being based in fact on two growing and well-quantified parameters: the ability to cover larger and larger distances (range) and the destructive power.

Similar kinds of models (that may be called eschatological) can be applied to the accumulation of a certain class of life events for an elderly person, e.g. unpleasant ones or, say, health troubles. The exact knowledge of reference time point $t_*$ is usually irrelevant for this kind of model.

In many mathematical models and theories, the underlying assumptions may be hidden quite deeply. Take, for instance, conventional statistical mechanics and, hence, thermodynamics (which is fully based on statistical mechanics). In the usual textbook presentation of these disciplines, the assumption that all microscopic states that ought to be combined to form the observed macroscopic state are equally probable (under the same circumstances, e.g., having the same energy) is rarely explicitly stated. Yet this boldly assumed symmetry, being a rather far extrapolation of the simplest probabilistic test of throwing a coin, is not necessarily true: the world as we observe it does not comply with the above "democratic" assumption. Indeed, we see a lot of order in nature, especially the high degree of organization in biological objects and society, so that one does not normally expect that applying the models based on the concept of equal probability of microstates, that is of a complete chaos characterizing thermodynamic equilibrium, would always give adequate results. Even in inorganic nature, equal likelihood of states is a rare occasion: if we consider the Sun or the Earth, we shall see that these macroscopic bodies are very far from thermodynamic equilibrium (although we use the notion of temperature that is a characteristic of equilibrium).

From the perspective of traditional statistical mechanics and thermodynamics, one tends to explain the high degree of organization that we observe almost everywhere by the emergence of fluctuations, but the probability of large fluctuations accounting for a macroscopic order is exponentially small: the probability of the event corresponding to achieving the simplest order in a system of $N$ gas molecules when all of them would be found in partial volume $\Delta V$ of the entire gas volume $V$ is estimated as $p_N = (\Delta V/V)^N$. For the ratio $c := \Delta V/V = 0.2$ and $N = 1000$ (a rather small number of molecules), we have for the probability $p_N = e^S = e^{N \ln c} \approx 10^{-700}$ which number is actually zero within the commonly accepted computational accuracy (i.e. closer to zero than any possible error used in calculations). Here $S$ is entropy (dimensionless) i.e., the quantity used to characterize disorder in random systems.

More generally, one can define entropy through the number $\nu$ of microscopic states representing the macroscopic status of a system. Since entropy is conventionally defined as $\sim \ln \nu$ i.e., through a logarithmic measure for an ensemble of microscopic states, an exact definition of entropy up to each extra microstate or when we consider interchanging some of them is meaningless. Recall that the logarithm is only slightly different from a constant.

No one has ever observed that gas would not occupy the whole available volume. We shall deal with the statistical theory of macroscopic equilibrium in more detail in the section devoted to statistical mechanics and thermodynamics, in particular specifying the concepts that have just been mentioned. Here, it is only important to note that the concept of nearly universal thermodynamic equilibrium (recall the ubiquitous use of temperature), based on the implicit assumption that all microscopic states are equally likely, is just a convenient mathematical model. In many situations, this modeling hypothesis contradicts reality and should be replaced by the assumption that only special states call forth the order we observe everywhere in the world. The latter subject is studied in the discipline known as synergetics, which will be briefly discussed below in this book.

The lesson learned so far is that there may be some tacit assumptions in mathematical models that prove to be misleading even in the universal and highly successful models.

## 5.4. Extrapolation

Extrapolation is understood as a modeling procedure allowing one to expand a function or a process into an unknown domain. Thus, extrapolation is based on a rather bold assumption that all observed features remain intact and all previously employed methods are still applicable. Extrapolation, mostly understood as a kind of generalization, has always been a great source of inspiration and has resulted in significant achievements. For example, in the 18[th] century, Newtonian mechanics was extrapolated on the description of motion of celestial bodies. In the 20[th] century, similarly to Laplace who extrapolated the successful solution by Newton of the Kepler problem on the entire universe, thus postulating a mechanistic model of the world, Bohr created an anti-Laplace model by a philosophical generalization of the Heisenberg "indeterminacy (or uncertainty) relations", which are, in the simple case, just trivial consequences of the Fourier integral.

Moreover, in the 20[th] century, such fundamental physical concepts as energy, mass and temperature that were supposed to admit only positive values have been extrapolated into the negative domains: energy – to accommodate the bound states in quantum mechanics and antiparticles in quantum field theory, negative temperatures – to formally describe the active (amplifying) materials used in lasers. That mass can be negative has become immediately clear after the applications of quantum mechanics, primarily solid-state physics and semiconductor technology, were addressed. Indeed, carriers in condensed matter are not necessarily – in fact, seldom – accelerated in the direction of applied force such as exercised by an electric field. This seemingly exotic behavior is due to the fact that particles in condensed media (e.g., in crystals) lose the features of the bodies in a free space and become quasiparticles whose motion is dictated by the dispersion law (dependence of energy on quasi-momentum) defined by the symmetry and inter-particle interactions in the medium. For example, in semiconductors, electrons in pulsed high-intensity electric fields can well move in the direction opposite to the electric field force. A trivial example of a negative mass is a bubble in a fluid: it can move against the gravity force.

The concept of negative mass has lately become very popular in astrophysics, mostly in relation to the models of "dark entities" and "phantom matter". Calculations of effects accompanying the interaction of "normal" (positive) masses with "phantom" (negative) ones exhibit some peculiar features and are hoped to contribute to the explanation of the current astrophysical paradoxes. Some parts of such modeling calculations (e.g., the negative mass modification of the classical Kepler problem) are quite interesting, but unfortunately, we cannot reproduce them in this book because of the natural scope limitations.

Energy, which can of course be negative to describe finite motion and the bound states, has been also extrapolated into the complex plane in order to accommodate decay and unstable particles. There also exist attempts to extrapolate the temperature into the complex domain (e.g., in modern statistical mechanics, some condensed-matter models and, lately, in string theories), but the concept of complex temperatures does not seem to gain much popularity, at least so far.

Even such a sacral quantity as time may be complex in a number of models: for instance, in the tunneling ionization of atoms and molecules in strong electromagnetic fields of laser radiation. The complex time method of calculating the transition amplitudes in quantum mechanics (first used by L. D. Landau) has become one of the standard tools, e.g., in chemical computations and is now present in many textbooks. One usually introduces the imaginary time with the help of substitution $\tau = it, \tau \in \mathbb{R}_+$, known as the "Wick rotation" ($\mathbb{R}_+$ is the set of positive reals). Some important physical quantities, in particular, velocity, momentum, evolution operator, action and the corresponding Feynman path integral change their form following the transition to the imaginary time. For instance, velocity $\dot{x} = idx/dt$, unitary evolution operator $U(t) \rightarrow U(\tau) = \exp(-H\tau)$, where $H$ is the Hamiltonian that generates translations of time $t$, see the details in any textbook on quantum mechanics. Wick's rotation is a quantum concept, when time becomes imaginary.

Use of the Wick rotation is also one of the best-known examples of interdisciplinary cross-fertilization enabling an easy migration between two fundamental fields of physics: statistical mechanics and high energy physics. At the time when interdisciplinary barriers between these fields and their subdisciplines, solid state physics (now mostly referred to as condensed matter physics) and elementary particle physics, were drastically reduced (late 1960s – early 1970s) one did not pay much attention to the remarkable analogy between time parameter $t$ of quantum evolution and inverse temperature $\beta = 1/T$ of equilibrium statistical mechanics. Nowadays this analogy seems nearly obvious. Indeed, the partition function in statistical mechanics (also known as Gibbs' canonical integral) is $Z = \text{Tr}(e^{-\beta H}) = \sum_k e^{-E_k/T}$ whereas the evolution integral in the quantum field theory is $Q = \text{Tr } U(t) = \text{Tr}(e^{-itH}) = \sum_k e^{-itE_k}$. This integral defining the trace of the evolution operator in a basis of energy eigenstates can, of course, be represented in other bases such as, e.g., in the basis of position eigenstates $|x\rangle$ so that we get a continuum version of the trace

$$Q = Q(t) = \text{Tr } Q(x', x; t) = \int_{-\infty}^{\infty} dx \langle x|U(t)|x \rangle \equiv \int_{-\infty}^{\infty} dx K(x, x; t) \qquad (5.4.1.)$$

which is a special form of the Feynman path integral.

The analogy between statistical mechanics and quantum field theory is very deep. The extrapolated "thermal" imaginary time is basically the same variable as the Minkowski time coordinate in special relativity, $x^4 = ict$, where $c$ is the speed of light. The great mathematical physicist and cosmologist, Stephen Hawking, used the imaginary time to smear out the singularity that can be encountered in Einstein's equations for the gravitation field. In particular, Hawking built a model of a black hole whose inverse temperature coincides with imaginary time elapsed inside this fascinating object. Since the transition to imaginary time (more general, the complexification of time) makes the path integral representation of transition amplitudes in quantum mechanics formally identical with the partition function of statistical mechanics, the generalization of quantum-mechanical methods developed for zero temperatures (as in all mechanical problems) on finite temperatures (as is typical of statistical problems) becomes very simple and elegant, when one uses the so-called Matsubara time and Matsubara representation. These extrapolation methods are mathematically equivalent to analytical continuation and have a very general character (since such techniques are based on the analytic properties of Green's functions and do not use their specific forms). We shall touch upon this important issue in the short section devoted to statistical models, in particular, those of statistical mechanics. Imaginary time $\tau$ is sometimes also called the Euclidean time since the pseudo-Euclidean spacetime interval of special relativity $ds^2 = c^2 dt^2 - d\mathbf{r}^2$, $d\mathbf{r}^2 = dx_i dx^i$, $i = 1,2,3$ takes the Euclidean form $ds^2 = -(c^2 dt^2 + d\mathbf{r}^2)$ i.e., the Lorentzian signature is changed to Euclidean. Note that in the context of special relativity the split of spacetime into space vs. time is only possible for velocities $v \ll c$. For simple problems, the relativistic spacetime is manifold $M = M^4$ (Minkowski space) which looks like a four-dimensional real linear endowed with a symmetric non-degenerate quadratic form $g_{ik} = \gamma_{ik} \colon M^4 \to \mathbb{R}$ (not to be confused with the Dirac spinor gamma-matrices). We use in this book the (1,3) signature i.e., $\gamma_{ik} = \mathrm{diag}(1,-1,-1,-1)$, although some authors prefer using the $\gamma_{ik} = \mathrm{diag}(-1,1,1,1)$ signature which has the slight advantage of changing only one sign instead of three when passing to the Euclidean time. However, most physical sources discussing relativistic theories (e.g., relativistic quantum mechanics) use the $(1,3) := \mathrm{diag}(1,-1,-1,-1)$ signature, in contrast with mathematical sources putting an accent on geometry of Minkowski and Riemann spaces. In such latter sources, the metric signature $\mathrm{diag}(-1,1,1,1)$ is customarily used with, e.g., the Dirac equation being written as $\gamma^\mu \partial_\mu \psi = m\psi$ (and not as $i\gamma^\mu \partial_\mu \psi = m\psi$ as it is common in physical literature).

An interesting feature of the transition to Euclidean time is the dual role of the Hamiltonian (which the present author cannot properly explain). Note that the expression $e^{-itH}$ is the time-translation operator with Hamiltonian $H$ acting as the generator of the respective Lie group whereas the corresponding expression $e^{-\beta H}$ for for the density matrix contains the *same* Hamiltonian defining the relative presence of the system states (or the Boltzmann weight of a state) in an equilibrium statistical ensemble $\beta = 1/T$. In classical mechanics, the Hamiltonian generates infinitesimal displacements in time. In statistical physics, field theories, theoretical chemistry, probability theory and even outside of physics – in information theory, network theories, economics, game theory, etc., Hamiltonian $H$ defines the so-called partition function $Z(\beta) = \mathrm{tr}(\exp(-\beta H))$, where $\beta$ is a free parameter, not necessarily real.

A trivial but interesting physical example of extrapolation is the phenomenon of resonance. Resonances can be defined in various ways, but they nearly always correspond to the poles of analytically continued transfer operator $U(t, t')$ or its analogs (evolution operator, S-matrix, monodromy operator, etc.). Another salient example of extrapolation into the complex domain is given by instantons, which are just classical solutions for imaginary time. Although such solutions seem to be physically irrelevant, they provide convenient and powerful techniques for physical computations, e.g., to calculate tunneling probabilities. There are numerous other examples of successful extrapolations in physics; one of such examples (closely related to the just mentioned transition between pseudo-Euclidean and Euclidean structures) is the concept of tachyons, which are hypothetic particles having imaginary mass, moving with the velocities exceeding the speed of light $c \approx 2.998 \cdot 10^{10}$ cm/s and allegedly violating causality. At first, tachyons had been treated as "normal" (though conjectured) particles such as electrons, protons, mesons, etc., but later it was realized that tachyons understood in such a naïve sense can hardly exist in nature. At least, they are generally believed to be unable to interact with normal matter.

Nevertheless, the mathematical model of tachyons is widely used, often without even recognizing them: for example, tachyons manifest themselves as elementary excitations (quasiparticles) in many-particle systems when the latter are losing stability and undergo a phase transition to a more stable state. In the parlance of modern physics, one used to depict this process as the "vacuum instability". This latter notion is closely connected with such concepts of quantum field theory as spontaneous breaking of symmetry, the Standard Model of high energy physics, the Higgs boson and emergence of masses of elementary particles. Recall that the defining property of the vacuum state $\varphi_0$ is that such a state is invariant under all operations of the relativistic Poincaré group, $P(\Lambda, a)\varphi_0 = \varphi_0$ (here $\Lambda$ denotes the Lorentz transformations, $a = \{a^\mu, \mu = 0,1,2,3\}$ is the four-vector of spacetime translations).

The imaginary mass of a tachyon (it would be more correct to speak of its negative squared mass which is understood as the inverse second derivative of an effective potential at stationary point) is a particular case of the complex mass that is a symptom of the already-mentioned instability: the real part of an unstable particle represents its mass (energy) in the conventional meaning whereas the imaginary part describes the decay rate. Incidentally, it means that tachyons are difficult to describe as "normal" particles. Indeed, a particle is typically understood as a state that is invariable in time (a stable single-particle state) or at least exists long enough to be observed (an unstable single-particle state). The first case corresponds to zero imaginary part of the mass, the second to a very small imaginary part of the mass as compared to the real part. If the imaginary part grows and becomes comparable with the real one, such an object is called a "resonance" in high energy physics (it does not live long). Since tachyons do not possess the real part of mass at all, they can hardly be considered particles at all, at least from the standpoint of experimentally-oriented high energy physics. Tachyons interpreted as the states having imaginary mass also naturally emerge in modern string theory. Moreover, tachyonic instability of the spacetime metric characterizing our universe is one of the candidate mathematical models (that belong to a class of the so-called phantom models) aspiring to explain "dark components" in cosmology, in particular, dark energy which is driving the accelerated expansion of the universe[22]. As we do not know what it is, we call this energy "dark". Unfortunately, detailed exposition of all these fascinating subjects is clearly outside the scope of the present book.

The tachyon scalar field[23] is described by the equation that looks like the Klein-Gordon equation,

$$[(\partial_t^2 - c^2\Delta) \pm m^2]u = 0, \qquad (5.4.2.)$$

where the plus sign corresponds to the Klein-Gordon equation and minus to tachyons. The recent OPERA experiment in CERN indicated that neutrinos (more exactly, muon neutrinos) can be suspected to travel faster than light, which, if confirmed, would be a violation of at least two sacred physical principles, Lorentz invariance[24] and causality (time can go backward, at least in chronologically ordered theories), as people understand them today. One might note in this connection that Einstein in his famous 1905 paper "Zur Elektrodynamik bewegter Körper" did not assume that the speed of light (in vacuum) is the upper limit to be reached in nature; he only postulated that it is constant for any observer and does not depend on the state of motion of the light source. In other words, there is a logical difference between the existence of ultimate speed $c$ and invariant velocity $c$ allowing one to construct a geometric theory based on O(1,3) Lorentz rotations. Superluminal motion (i.e., the existence of faster than $c$ velocities in empty space) implies that there may be, e.g., several causal cones[25] instead of a single light cone corresponding to the conventional special relativity. Matrices

$$\Lambda(\phi) = \pm \begin{pmatrix} \cosh\phi & \pm\sinh\phi \\ \sinh\phi & \pm\cosh\phi \end{pmatrix} \qquad (5.4.3.)$$

giving the representations of the Lorentz (and Poincaré) group acquire complex entries for superluminal motion.

---

[22] Dark energy should not be confused with dark matter, the latter being hypothetical non-luminous matter needed to explain the motion of large-scale astrophysical objects.

[23] A scalar field $u$ on manifold $M$ is merely a smooth function $u(x), x \in M$. Point $x$ can be represented through its local coordinates defined on some patch $M_a, a = 1,2, \dots$ of manifold $M$ i.e., $x = \{x_a^1, \dots, x_a^n\}$. As usual, when patches $M_a$ covering the same manifold $M$ mutually intersect, local coordinates $x_a^i, i = 1, \dots, n$ are smooth functions of coordinates $x_b^j, b = 1,2, \dots, j = 1, \dots, m$ defined for another patch $M_b, b = 1,2, \dots$. Therefore, a scalar field is a smooth function on the entire manifold $M$.

[24] This is by no means a trivial issue since apart from purely geometric – kinematic – considerations such as that the hyperbolic rotation $\tanh \phi = v/c \equiv \beta$ cannot be defined for real $\phi$ at $\beta > 1$, there may also be dynamic effects reflected in the dispersion law $E(p) = (p^2c^2 + m^2c^4)^{1/2} + \Sigma(p), \mathbf{v} = \nabla_p E(p)$, where term $\Sigma(p)$ is a self-energy part containing superluminal components.

In general, complex numbers as an algebraic structure (field) was a radical extrapolation, extending real numbers together with the elementary real functions such as logarithms, exponentials, trigonometric functions, etc. into the complex domain. Later, complexification became an important tool to design engineering and scientific quantities (e.g., impedance, conductivity, susceptibility, dielectric permittivity, wave function). Extrapolation into the complex domain may be a powerful device in many branches of science and engineering: a popular saying among the 20[th] century physicists was that complex numbers were the most crucial discovery in physics. In mathematics, the fundamental concept of linear (vector) spaces can be viewed as a result of far-reaching extrapolation: such sets extend the idea of small or even infinitesimal linear deviations, which is the key concept of classical calculus, on the entire space.

We can notice that extrapolation is a ubiquitous mathematical device: thus, expanding a function into a power series near some point and truncating this series[26], which is reduced in the simplest case to linearization, amounts to extrapolating the function onto a larger domain. The series expansion technique is closely connected with analytic continuation of a given function, when its extended values can be specified in a new region.

One encounters the problem of extrapolation in real life each time one resorts to banking services, e.g., when one puts money in the bank or needs credit. In such situations, we have to assume that some conditions will remain intact in the future or at least do not change significantly. In more general terms, financial modelers are mostly required to quantitatively evaluate a variety of possible scenarios for decision makers which is often reduced to answering the "what if" questions in numerical terms or, e.g., to providing the cash flow forecasts (in this sense financial modelers attempt to predict the future). As a simple example illustrating the financial extrapolation techniques, one can take the standard deterministic model of bank accounts formulated as a first-order linear scalar ODE, $\dot{x} = a(t)x + b(t)$, where $x$ is the value of a bank account, $a(t)$ represents the time-varying interest rate and $b(t)$ the rate of withdrawals ($b(t) < 0$) or deposits ($b(t) > 0$). This model is linear provided that coefficients $a$ and $b$ do not depend on the value $x$ of the bank account (for instance, they are assumed to depend only on time). The solution to the model equation with the initial condition $x(t_0) = x_0$ can be written as (obtained, e.g., by the method of variation of constants)

$$x(t) = \exp\left\{\int_{t_0}^{t} a(s)ds\right\}\left\{\int_{t_0}^{t} d\tau b(\tau)\exp\left\{-\int_{t_0}^{t} a(s)ds\right\} + x_0\right\} = g(t_0, t)\left\{x_0 + \int_{t_0}^{t} d\tau g(\tau, t_0)b(\tau)\right\}, (5.4.4.)$$

where

$$g(t_0, t) := \exp\left\{\int_{t_0}^{t} a(s)ds\right\} \tag{5.4.5.}$$

is the fundamental solution i.e., the one for the associated homogeneous equation $\dot{x} = a(t)x$ with the same initial condition. We shall devote more attention to the concept of fundamental solution when discussing the Green functions below.

---

[25] Causal cones of special relativity are the null cones of the metric tensor $g_{ik}$.

[26] More generally, using the so-called Padé approximant.

The verbal interpretation of the expression obtained for the bank account $b(t)$ for $t > t_0$ would be quite evident: one's bank account is the sum of two terms, cash generated by one's initial capital (with the given interest rate) and cash that varies due to subsequent deposits and withdrawals. It is clear that for large negative $b(t)$ the state of one's bank account can become not only smaller than the initial deposit $x_0$, but can cross the $x = 0$ level and also become negative (which signifies running into debt).

In computer modeling, extrapolation is usually related to numerically integrating ordinary differential equations (ODE), when the data at previous points is used to appreciate the solution at the next points. Since for many models of physics, chemistry, biology, engineering, etc. an analytical solution to the evolutionary vector equation $\dot{\mathbf{x}}(t) = \mathbf{f}(t, \mathbf{x}(t))$ with initial data $\mathbf{x}(t_0) = \mathbf{x}_0$ on $t, t_0 \in I := [a, b] \subset \mathbb{R}$ can only be found in relatively rare cases, a variety of numerical procedures approximating the exact solution $\mathbf{x}(t)$ are used. Such procedures consist in computing the unknown function $\mathbf{x}(t)$ at discrete points $t_k \in [a, b]$ with a certain error and estimating this error. Thus, numerical procedures typically replace the dynamical systems describing a continuous evolution by difference equations or by maps of the form $\mathbf{x}_{k+1} = \mathbf{f}(t_k, \mathbf{x}_k)$ (or, more generally, by multistep maps $\mathbf{x}_{k+1} = \mathbf{f}(t_k, t_{k-1}, \dots, t_{k-m}, \mathbf{x}_k, \mathbf{x}_{k-1}, \dots, \mathbf{x}_{k-m})$). Notice in passing that vector sequences appear quite often in science and engineering not only in the numerical solution of ODEs by finite differences, but also in matrix eigenvalue computations or in the method of finite elements. Starting from the initial value $\mathbf{x}_0$, one can iterate the map obtaining the sequence of extrapolated values $\mathbf{x}_0, \mathbf{f}(\mathbf{x}_0), \mathbf{f}(\mathbf{f}(\mathbf{x}_0)), \dots = \mathbf{x}_0, \mathbf{x}_1, \dots$ [27]. This is the forward path generated by $\mathbf{x}_0$. In the simplest procedure of this kind known as the explicit Euler method one puts $t_k = t_0 + kh, k = 1,2,\dots$ and obtains the linearly extrapolated values $\mathbf{x}(t_k)$ in the following fashion: $\mathbf{x}_{k+1} = \mathbf{x}_k + h\mathbf{f}_k(t_k, \mathbf{x}_k)$. Although the Euler method and other closely related techniques of numerical treatment of the Cauchy problem for ordinary differential equations (the leap-frog, Verlet, Adams-Bashforth, Runge-Kutta, etc. methods) are seldom referred to in computer modeling as extrapolation, in essence they extrapolate the ODE solution $\mathbf{x}(t)$ transferring it forward along the approximate path with the increasing time-steps $t_k, k = 0, 1, 2, \ldots, n$. The basic idea is often formulated as to "follow your nose", but there are also more cunning extrapolations doing some "sniffing ahead".

---

[27] We shall also use notation $f \circ g$ to denote the product of mappings $f(g(x))$, with $g$ acting first.

One usually wants to understand the long-time behavior of the forward path corresponding to the asymptotic behavior of the exact (analytic) solution for $t \to \infty$. Here the natural question arises: how can one be sure that the extrapolated limit will be close – and in what sense – to the true limit? Moreover, the sequence of extrapolated values can be slowly convergent or at all divergent. This and similar questions reflect the already mentioned inherent drawback of numerical modeling and simulation: difficulty to find asymptotic values in the $t \to \infty$ limit.

One of the most prominent examples of a class of computer models based on extrapolation is related to the AGW (anthropogenic global warming) hypothesis. The latter subclass of general climatologic models typically predicts the runaway increase of the global mean temperature (often called also GMAST – global mean air surface temperature) during the next 50-100 years. Such a catastrophic temperature trend is assumed to be fueled by the uncontrolled emission of carbon dioxide accompanying human industrial and transportation activities. In reality, climate dynamics is determined by a complex hierarchy of various physical processes resulting in an intricate balance between the incoming wide spectrum and the outgoing long-wavelength solar radiation. Many of these physical processes are not well understood as yet, so that even scientists can be held in the grip of totally false beliefs. Mathematical and computer models of climate variability are discussed below in the "Climate variability models" section.

In general, modeling of global processes such as renewable energy deployment on a mass scale, prognoses of economic development, or population models, tend to be based on extrapolations. Such quantities as per capita growth rate which are treated as nonlinear coefficients in the corresponding evolutionary equations are usually assumed to retain their form in the unknown future domain. Many phenomenological theories and models claiming to describe the dynamics of human population are of extrapolatory character when the whole population of the Earth is treated as a monolithic dynamical system. For example, the logistic population model

$$\frac{1}{N}\frac{dN}{dt} = a - bN, \tag{5.4.6.}$$

where the case $b > 0$ describes competition due to crowding whereas $b = 0$ corresponds to the simple permanent growth. In distinction to the linear growth model $dN/dt = a$, $N = at$ and to the simple exponential growth model $dN/dt = bN$, $b = 1/\tau$, where $\tau$ is a characteristic time constant i.e., the growth rate is proportional to the actual population and the time of population doubling equals $\tau \ln 2 \approx 0.7\tau$ (which assumes that reproduction occurs independently of the presence or absence of other individuals), nonlinear models such as the just mentioned logistic model and the hyperbolic model

$$\frac{dN}{dt} = \frac{N^2}{c}, \qquad N = \frac{c}{t_0 - t}, \tag{5.4.7.}$$

which describes the so-called sharpening regime $dN/dt = c/(t_0 - t)^2$), both imply that there is an interaction between individuals. In other words, population growth is supposed to be a collective dynamic of human groups with cooperative behavior. Nonlinear models of the population growth such as the two mentioned above contain no local i.e., spatially dependent parameters and can (and probably should) thus be applied to the entire planetary population, which is an extrapolation of the idea of a universal interaction law (cooperation) between the members of human groups. One may notice that the model of the $dN/dt = \alpha N^2$ type describes an explosion in chemical physics i.e., the situation when the reaction rate in a binary mixture is proportional to the product of concentrations of both reagents (or to the product of densities of two sorts of colliding particles). The solution $N(t)$ to this model has a vertical asymptote at $t \to t_0$ which means that the population tends to infinity in finite time[28]. The presence of vertical asymptotes in general means that the solution is not defined for large $t$ values since it blows up in finite times. Translating this modeling situation into human words, it would be inappropriate to ask about the behavior of the relevant quantities (e.g., the population) for $t \to \infty$.

An explosive growth of the population at $t \to t_0$ testifies to a limited validity of the hyperbolic model. Of course, this model, being spatially uniform, cannot be applied to the regions where an intensive population exchange takes place, so it has by necessity a global character. Local population models should also contain spatially inhomogeneous convective and diffusive terms accounting for the population transfer effects. Spatially dependent extensions of the logistic model are discussed below.

Some demography experts think that one should replace the Malthus equation $dN/dt = \alpha N$ not by the currently fashionable logistic model, but by the model $dN/dt = \alpha N^{1+\delta}, \delta > 0$. Allegedly this model better describes the world population dynamics during the last 100 years. Let us compare the solutions to both equations. The first equation results in the exponential function $N(t) = N(0)\exp(\alpha t)$ whereas the second one has a solution $N(t) = A(t_0 - t)^{-1/\delta}$, where $A \equiv (\alpha\delta)^{-1/\delta}, t_0 \equiv (\alpha\delta N(0))^{-\delta}$. One can see that these two solutions are essentially different even for infinitesimal $\delta$ since the solution to the exponential model exists for all times $t \in (0, \infty)$, while the solution to the second model tends to infinity when $t \to t_0$. The time of existence of the solution $t_0$ (position of the vertical asymptote) depends on the initial data, which thus define the model applicability domain.

Notice that infinities in mathematical models usually mean that the employed model transgressed its applicability limits. For example, geometric optics submits that all rays gather in a single point called focus, where the light intensity is infinite. This is only a greatly simplified mathematical model: in reality, light forms not a mathematical point, but a finite spot (with patterns and intensity distribution), where the light intensity is high but nowhere infinite. The wrong picture arises because the ray optics ignores an essential parameter, the wavelength. In the suggested demographic model, catastrophic growth $N(t)$ at $t \to t_0$ may indicate the necessity to take into account some limiting factors (as in the logistic model) or spatially dependent processes.

One can, of course, further extrapolate the population models, e.g., to the domain $dN/dt = \alpha N^p$ with parameter $p \neq 0, 1, 2$. In general, nonzero solutions to the growth models $\dot{x} = \alpha|x|^p$ have vertical asymptotes for $p > 1$. Indeed, the solutions (that can be easily obtained by direct integration) are of the form

$$x(t) = [A - (p-1)\alpha t]^{-\frac{1}{p-1}}, t < \frac{A}{\alpha(p-1)}, x(t) = [(p-1)\alpha t - A]^{-\frac{1}{p-1}}, t > \frac{A}{\alpha(p-1)} \quad (5.4.8.)$$

---

so that there are vertical asymptotes at $t = \frac{A}{\alpha(p-1)}$. It means that one can always suspect the presence of vertical asymptotes in the growth models, when the vector field $|f(x,t)|$ (see below) may grow as $\alpha|x|^p$ with $p > 1$ at least for some $t$. For example, functions $f(x,t)$ containing exponents or high-degree polynomials are suspicious and should be tested for vertical asymptotes or sharpening regimes.

Since the nonlinear models with arbitrary $p$ which correspond to higher-order correlations between interacting agents have little relevance to the observed natural (e.g., physical, chemical, biological or demographic) effects, we shall not consider them here. Yet we shall deal more with the models of population growth while discussing the logistic equation.

When discussing such a vital engineering issue as radioactive waste management and disposal, one has to rely on extrapolated models, as no one is able to predict what will happen in $10^2 - 10^6$ years corresponding to decay periods of long-life radionuclides contained, for example, in spent nuclear fuel (Np-237, Pu-239, I-129, Tc-99, etc.). Mathematical models pertaining to radioactive management and disposal, assessing the associated risks, are mostly focused on prediction of the radionuclide migration rate in various geological repositories. Naturally, such prognoses, mainly based on numerical simulations, use rather bold extrapolations that cannot be corroborated by accurate in situ measurements.

In everyday life, we are so accustomed to extrapolations that we tend to use them rather offhandedly. For example, metaphoric use of words is nothing but an extrapolation, when a word is used in a context different from the habitual one (e.g., the term "hedge" used in the financial world). Metaphoric use of terms can be picturesque and make concepts more colorful (such as Big Bang, Big Crunch, dark energy, no hair on black holes, branes, and countless other metaphoric entities), but it is usually difficult to give them an exact mathematical meaning. Besides, ill-posed problems, i.e., the ones which are extremely sensitive to small perturbations in initial data are very hard to extrapolate, even despite significant advances in solving this kind of problems in recent years.

One must be careful with extrapolations, especially when nonlinear processes are to be considered (thus predictions of the future fail in most models). In particular, extrapolation in mathematical and computer modeling works quite well when one deals with smooth and continuous processes (as in the case of analytic continuation). If such phenomena as jumps, bifurcations, singularities and instabilities can be encountered, extrapolation of the respective quantities becomes difficult or completely impossible. This is a class of situation described in "catastrophe theory", when smooth variations of the control parameter can lead to sudden and abrupt changes in the state of the considered system or process. In everyday life we observe that most changes in biological organisms or in social institutions manifest themselves as sudden jumps: there is at first a prolonged latent period followed by a sharp discontinuous reconstruction. The mathematical image of a catastrophe reflects its main real-life feature: catastrophe always aggravates things, if it occurs, and the situation usually becomes only worse and hardly predictable.

## 5.5. Symmetry

Having a symmetry simply amounts to the statement that something looks identically from different points of view. There are abundant examples of symmetry in nature: human face, exterior body, flowers, leaves, butterflies and so on. Symmetries are not reduced to esthetics only: they are an extremely powerful tool in physics, specifically in quantum theories. This practical importance of symmetry distinguishes physics and mathematics from "softer" disciplines such as socio-economic models. Although one sometimes distinguishes between symmetry and invariance, we shall understand symmetry as an invariance under some class of operations. At least we can see that symmetry and invariance are closely related: symmetry is associated with operations that transform the system into itself so that the transformed system would be indistinguishable from the one before transformation. If the symmetry is perfect, then it would be impossible to distinguish between the initial and the final state in any experiment. For instance, rotation of a square through $\pi/4$ leaves all the properties of the square invariant under this operation. Incidentally, such a transformation may be viewed as equivalent to the permutation of vertices $(1 \rightarrow 2, 2 \rightarrow 3, 3 \rightarrow 4, 4 \rightarrow 1)$; there are 8 permutations that leave the square invariant. This is a trivial example of a finite symmetry, which also shows that there may be different ways to describe symmetries and invariance.

There are many ingenious techniques to obtain efficient model representations, for example, geometrical and graphical methods, computer vision, etc. One can observe that the most efficient models have a common feature: they maximally exploit the symmetries of the modeled object. Roughly speaking, the symmetry of an object is a transformation whose action leaves the object unchanged. For instance, an equilateral triangle after a rotation by $2\pi/3$ looks the same as before the rotation: this transformation is a symmetry. Rotations by $4\pi/3$ and $2\pi$ are also symmetries. The latter rotation is equivalent to doing nothing – it is a trivial transformation: each point is mapped to itself.

From a very general observation, one may notice that asymmetry is much more generic than symmetry, the latter being a very special feature. In general, a real physical presence of any exact symmetry is highly unlikely, and even if such exact symmetry is indeed materialized, it can be easily broken. An exact symmetry is an exception or may be used as a convenient idealization. As to the broken symmetry, this concept has become fashionable in the "new physics", when the systems with an infinite number of degrees of freedom (as opposed to traditional mechanical systems) started to be extensively studied (approximately in 1960s). One usually says that the symmetry is spontaneously broken when the Hamiltonian (or the Lagrangian) of a system is invariant under some symmetry, yet the lowest (ground) state of the system is not: the symmetry operations transform the ground state into a different one. So, the lowest state of the Hamiltonian is degenerate, and there exist multiple equivalent ground states. In case the symmetry is continuous, the degeneracy may be infinite i.e., there exists an infinite number of equivalent ground states.

Symmetries are efficiently used to classify geometric objects: thus, the equilateral triangle has 6 symmetries (3 rotations and 3 flips i.e., reflections with respect to 3 axes), an isosceles has only 2 (flip plus trivial), whereas the general triangle possesses only the trivial symmetry. This is also an example of finite symmetry transformations; one more simple example is decorating a square (Figure 1). The square is symmetric with respect to rotation about its center, reflection in diagonals and bisectors, and under an inversion through the center. If we divide a square into, e.g., four sub squares and paint these smaller squares lying along each of the main square diagonals pairwise into two colors (here green and orange), then the figure obtained no longer has the symmetry of a square: the initial symmetry will be broken. This simple symmetry can be extended by introducing a more complicated transformation: first rotation then flip, turning orange into green and vice versa. Such playing with symmetries had practical applications, e.g., in textile production (mosaic patterns of Henry John Woods, e.g., "black and white groups", "braids", etc.). Further generalizations of simple symmetric patterns include such concepts as polychromatic groups in geometry, Shubnikov groups in crystallography, spin flip in magnetism (the physical model studied by L. D. Landau, later developed in an entire branch of scientific modeling) and many others. The example of successively divided squares is close to the popular cellular automata model called the "Game of Life", where each cell in a square grid on a plane can assume two states: alive (e.g., green) or dead (orange), and there is a finite number of cells.

Basic properties of symmetries are quite obvious such as each symmetry has a unique inverse which is itself a symmetry. For example, if $G$ is a rotation of an equilateral triangle by $2\pi/3$ around its center, then $G^{-1}$ is a rotation by $4\pi/3$. The most interesting symmetries in mathematical modeling are smooth and invertible. Smoothness is commonly understood as no need to care about analytic properties. For example, let $x$ be the position of a general point of the object. If $G : x \to z(x)$ is a smooth symmetry, then the function $z(x)$ is considered infinitely differentiable. Since $G^{-1}$ is also a symmetry, it is smooth if the inverse function $x(z)$ is infinitely differentiable over $z$. Such a symmetry is called $C^\infty$-diffeomorphism i.e., a smooth invertible transformation.

Recall that the symmetry of a real physical system can be understood as the symmetry of a corresponding mathematical model i.e., of equations describing the system. In physics, the symmetry principles are of extreme importance, in particular, in quantum field theory: one can construct rather general quantum field models by using only two guiding principles: (1) symmetries and (2) the least action principle (albeit slightly generalized as compared to Lagrangian mechanics), which is, in fact, also some kind of a symmetry to be preserved under the system's dynamics.

Mathematical models of physical situations typically imply a great redundancy i.e., the same physical scene can be equivalently described by many mathematical representations. One can pass from one such representation to another, which is similar to changing the point of view, so that the physical situation remains the same, being associated not with a specific mathematical representation, but with a whole class of them. Symmetry is just another word for this equivalence of mathematical descriptions of the same physical situation. The change of "points of view" is mathematically implemented via transformations that, under certain quite natural and general assumptions, comprise a group of symmetries pertaining to a physical theory. For example, in classical mechanics if any two mathematical representations of a physical scene can be related through a Galilean group of transformations (see Section 8. "Classical mechanics and deterministic dynamical systems"), such representations correspond to the same physical situation. There may be, however, more complicated symmetries related to more general and more abstract cases when there is no global symmetry (such as local gauge symmetries).

The concept of symmetry is a generalization of perceiving repeated patterns and recognizing them is vital for our understanding of the world and is thus an important survival factor. It is a curious fact that people start feeling and using symmetry long before they actually learn the world: night and day, left-right, a ball, flowers, bird wings, a butterfly, etc.; on a more advanced level – medieval cathedrals, Egyptian pyramids, various engineering objects. Humans and possibly animals have an intrinsic ability to perceive symmetry, but it was (and remains) a challenging task to take advantage of this powerful recognition first in physical and mathematical sciences, then in modeling followed by the enforcement of symmetries in computer science and engineering, including graphical and CAD applications. In these latter classes of applications, symmetries are usually required to preserve structures, which is natural from the engineering viewpoint, whereas in general computer modeling this requirement is abandoned (e.g., in computational fluid dynamics).

The concept that symmetries are not merely an esthetic attribute but can result in actual physical forces was demonstrated in the famous paper by C. N. Yang and R. Mills [162]. The Yang-Mills theory can be viewed as an extension of electromagnetism, and although the former has been widely accepted in the physics community since the early 1970s it remains a hot subject, especially in high energy physics.

In physics, one usually considers all processes in nature to be due to local motions and interactions, on the deepest level of elementary particles. Such motions and interactions are governed by physical laws which can be formulated quite simply: they are based on some symmetry and on a semi-philosophical principle that anything compatible with the symmetry can – and even must – occur. This maxim is gaining more and more popularity among theoretical physicists. Yet not among all. Indeed, the idea that all the properties characterizing the structure of matter can be derived from spacetime symmetries may well be wishful thinking. At least it is a bold hypothesis that symmetry alone can account for the behavior of all fundamental matter constituents and their interactions. An instinctive desire that the physical phenomena should be determined by some kind of symmetry (spacetime or internal) as well as the requirement of high symmetry are mostly dictated by the quest for "beauty" or esthetic perfection in physics. Although beauty as a scientific criterion has nothing to do with physics – these two are completely disparate entities – some mathematical expressions can be perceived as "beautiful", in the same sense as visual or musical works of art. "Beauty" in mathematics typically embraces such qualities as symmetry, simplicity, compactness (i.e., economy of expression) and non-triviality, the latter being understood as unexpected truth. One often considers Euler's identity $e^{i\pi} + 1 = 0$ containing five constants connected through three basic algebraic operations as an example of mathematical beauty. Likewise, Maxwell's equations can be considered beautiful, not only because they lie at the foundation of special relativity, classical and quantum electrodynamics and even high-energy physics (due to Poincaré invariance[29]), but also because of more intricate local gauge symmetry[30] that underlies nearly all modern physical theories. The gauge concept implies that interactions between fundamental particles are mediated by vector bosons whose dynamics should be invariant under the "gauge group", usually an infinite-dimensional group of local transformations. The gauge group primarily manifested itself as a useful symmetry group for electromagnetism, being one of the symmetries of Maxwell's equations [6, 63] and later was spread over most modern physical theories.

As already mentioned, mathematical models were mainly reduced to formulating and processing equations. A symmetry of a mathematical equation is usually understood as a set of transformations taking the solutions of this equation into solutions. Incidentally, it is not necessarily required that such transformations should form a group so that symmetry may not be a group. Then one can identify the symmetry of a system or a process to be modeled with the symmetry of equations describing this system or process.

Symmetries arise quite naturally as soon as we try to describe the system in some coordinate frame: if such a frame is shifted, rotated or otherwise moved, the real-world evolution of the modeled system must remain subjected to the same physical laws as before the displacement of the reference frame. The next logical step (made by Galileo) is to state that there is no preferred frame of reference. Before Galileo pointed out that the laws of physics (actually of mechanics) are the same on the shore and on board of a ship drifting with a constant speed, one had considered these laws to be invariant under the six-dimensional Euclidean group E(3). Consequently, we arrive at a mathematical necessity to classify all dynamical variables according to their behavior under transformations (e.g., translations or rotations that are elements of the Euclidean group) of the coordinate system in which these variables are expressed. Thus, the respective group concepts emerge, formalizing such a classification process. There are basically three groups of variables:

1. Those that remain unchanged under the frame transformations are scalars.
2. Those that have components changing like Cartesian coordinates $x^i$ are contravariant vectors whereas those changing like derivatives over them, $\partial_i \equiv \partial/\partial x^i$, are covariant vectors (dual to contravariant).
3. Those that have components changing like $n$-tuples of vectors (i.e., products of $p$ Cartesian coordinates and $q$ derivatives over them, $p + q = n$) are tensors of rank $n$: $p$-times contravariant and $q$-times covariant.

---

[29] Maxwell's equations actually have higher symmetry than the Poincaré group: they are also invariant under dilations (scaling $\mathbf{r}' \to a\mathbf{r}$) i.e., under the 11-parameteric Weil group and even possess still more general conformal invariance [63]. One may notice that Maxwell's equations contain neither mass nor an intrinsic length scale which fact gives rise to their high symmetry.

[30] The term "gauge" was introduced by Hermann Weyl who suggested that most physical theories should be invariant under the change of scale i.e., of "gauge" [155].

Recall that according to the traditional definition a tensor is a collection of appropriately indexed functions, referred to as components, associated with a local chart of a manifold $M$, and these components change according to a certain rule when moving to another chart. A slightly more modern definition describes a tensor as a multilinear function of a number of variables which are either vector fields on $M$ (contravariant components) or 1-forms (covariant components). The latter are linear objects that take vectors to scalars.

The components of a vector field (more generally, of a tensor, spinor or a spin-tensor field) change when one changes the coordinate frame. The correspondence rule between the field components related to the change, e.g., while rotating the coordinate system or transforming the coordinates between moving frames is usually called the transformation law.

Here, it would be pertinent to emphasize the difference between the coordinate system and the frame of reference. In classical (nonrelativistic) mechanics, the motion occurs in three-dimensional Euclidean ($\mathbb{R}^3$) fibers and can be parameterized by the absolute (base) time $t$. The motion can be described in some local coordinates $x^i$ determining the positions of moving points in $\mathbb{R}^3$ from the standpoint of an observer associated with this local coordinate system, $\mathbf{r} = x^i \mathbf{e}_i, i = 1,2,3$, where $\mathbf{e}_i$ are local base vectors. Note that the observer's system of coordinates can be, in general, deformable or curvilinear. The coordinate system can also move with respect to some other system regarded as fixed; then the coordinates of physical objects, e.g., particles, depend on time $t$, and one tends to call the coordinate systems with time-depending coordinates the frames of reference. In practical terms, it means that the change of variables of the form $z^i = f^i(x^j)$ defines the transition to a new coordinate system whereas the transformation $\xi^i = f^i(x^j, t)$ signifies the transition into a new frame of reference. Both transformations are usually considered smooth and invertible (diffeomorphisms), but otherwise arbitrary spatial (coordinate systems) or spacetime (frame of reference) transformations. Of course, all said applies to the $n$-dimensional case. One more practical consequence of the difference between the coordinate system and the frame of reference is that the velocity vector field does not alter with the transition to a new coordinate system, but changes when a new frame of reference is introduced (the same applies to the acceleration field and to the field of higher-order time derivatives). Notice that the motion is relative even in classical non-relativistic mechanics, and that in the theories based on the spacetime concepts (such as special or general relativity), the difference between the notions of a coordinate system and a frame of reference practically disappears.

There are two trivial but very important practical implications of the abovementioned basic classification of variables namely (a) always ensure that both sides of an equation transform equally and (b) never add or subtract quantities that do not transform in the same way. These simple rules already demonstrate that symmetry is not a mere embellishment, but an operational tool. One can produce many other examples of how symmetry works: thus, a system to be modeled is known to have mirror ($P$) symmetry i.e., in the one-dimensional case is invariant under transform $x \rightarrow -x$, where $x$ is some variable, e.g., a coordinate. For instance, a butterfly is almost invariant under mirror reflection (which is an exception in the biological world, where left-right symmetry is mostly broken). Then one can build a model, disregarding the negative values of $x$. Moreover, solutions obtained for such models must possess a certain parity: they should be either odd or even with respect to $x \rightarrow -x$ transformation i.e., expressed as even or odd functions of $x$. If the solution does not satisfy this symmetry condition, then there may be some computational error or even the entire model is wrong. However, the requirement of reflection symmetry should not be overstated: much more systems in real life have no definite parity (i.e., $P = \pm 1$) than possess it[31].

Notice that if one inverts the direction of an even number of coordinate axes (as when working with $\mathbf{r} = \boldsymbol{\rho} = (x^1, x^2)$ in the $\mathbb{R}^2$ plane), this operation can be implemented through a rotation whereas the inversion of $\mathbf{r}$ in $\mathbb{R}^3$ cannot. In general, inversion in $\mathbb{R}^n$ with $n$ odd cannot be achieved by a rotation. In particular, one cannot reduce reflection in a mirror ($z \rightarrow -z$) to a rotation.

Likewise, if we assume the model to be time-reversal ($T$) invariant, we must discard all solutions that do not have a definite time-parity. This may have important consequences, for example, in time-reversible models we ought to disregard any result containing odd powers of frequency (e.g., in light scattering). Here symmetry works as a testing device: even simple symmetries provide powerful validation tools for the model. The implication of symmetry tests for mathematical and computer modeling is that symmetry of a modeled system should always be manifested in the solutions.

An important class of symmetries is represented by the coordinate transformations. The symmetry of a system may be regarded as a set of transformations that leave the model of the system (e.g., expressed in terms of differential equations) unchanged. In the simplest case, these transformations involve the local coordinates, then the symmetry is reduced to coordinate transformations. Thus, the unit circle $x^2 + y^2 = 1$ has a symmetry

$$R(\varepsilon) : (x, y) \rightarrow (x', y') = (x \cos \varepsilon - y \sin \varepsilon , x \sin \varepsilon - y \cos \varepsilon), \varepsilon \in (-\pi, \pi] \qquad (5.5.1.)$$

or, in polar coordinates, $(\cos \theta, \sin \theta) \rightarrow (\cos(\theta + \varepsilon), \sin(\theta + \varepsilon))$. Such local transformations are very important for building physical models. For example, an electron moving in the periodic field of a perfect crystal is assumed to feel the potential $V(\mathbf{r}) = \sum_a V(\mathbf{r} - \mathbf{r}_a)$, where vectors $\mathbf{r}_a, a = 1,2, \ldots$ denote the positions of atoms in the crystal (Bravais) lattice. All the equations within this model, describing the behavior of electron in crystal, should be invariant under the transformation

$$\mathbf{r} \rightarrow \mathbf{r} + n_1 \mathbf{a}_1 + n_2 \mathbf{a}_2 + n_3 \mathbf{a}_3, n_i = 1,2, \ldots, \mathbf{a}_i, i = 1,2,3, \qquad (5.5.2.)$$

being the lattice primitive periods. This statement is the essence of the Bloch model of a single electron motion in a crystal, which is of fundamental importance for semiconductor theory and technology and for electronics in general. One can notice here that symmetries of solids imply that the latter are made of rigid materials. This is a simple example of how symmetry imposes physical constraints.

In more complex cases, symmetry considerations provide crucial tests for physically based modeling. The mathematical reason for that is the inner symmetry of equations used for modeling, in particular, all main equations of physics (basic mathematical models of nature) such as Newton's, Euler-Lagrange, Maxwell's, Schrödinger, Dirac, Klein-Gordon, Einstein's, Hamilton-Jacobi, Laplace, D'Alembert, Helmholtz, Fokker-Planck, etc. can be classified according to their own symmetry. Especially important are continuous symmetries (vs. discrete ones): they correspond to arbitrarily small operations[32] and lead to conservation laws, this fact is summarized by Noether's theorems. Emmy Noether, the German mathematician, proved in 1915-1918 two important theorems about the relationship between symmetries and integrals of motion known in physics as conservation laws. The first Noether's theorem is a remarkable result, more physical than mathematical, stating that any geometrical symmetry (reflected in an invariance of the Lagrangian or Hamiltonian under some operation) leads to a conserved quantity. Based on this theorem, one can even formulate Noether's principle in mathematical physics as the statement that any continuous symmetry of the action results in a conservation law. It is, however, curious that in physics the achievements of Emmy Noether related to differential invariants in variational calculus are collectively labeled as Noether's theorem.

---

[31] Here an abuse of notations is present: $P$ is the eigenvalue of the parity operator and not the operator itself.

The fact discovered by Emmy Noether that each differentiable symmetry brings with it a conservation law is the common property of all physical theories such as mechanics (Newtonian, relativistic or quantum), high-energy physics, general relativity, etc. And vice versa, each conservation law implies a certain symmetry, this statement having an exact meaning in most cases encountered both in classical and quantum theory (see [17], §20 and [97]); the simplest and the most lucid exposition of Noether's theorems, together with accompanying notions such as Noether's current, seems to be given in [89]). Primarily, Noether's theorems have been associated with the infinitesimal transformations in Lagrangian theory; later, however, it was understood that Noether's theorems can not only be related to the correspondence between symmetries and conserved quantities arising from the associated Euler-Lagrange operator, but also to the invariance of rather general variational problems (of which Euler-Lagrange equations are just an example) under local symmetry groups involving nearly arbitrary continuous functions. This situation plays an essential role in modern gauge theories and in general relativity.

A conserved quantity or an integral of motion is typically understood as some combination of dynamical variables that does not depend on time along the evolution trajectory of the system, the time being treated as a parameter. In the language of dynamical systems, such a combination is invariant under the flow. The maximum number of conserved quantities is $n$, where $n$ is the number of degrees of freedom (in a classical mechanical system); in this case the system is known as completely integrable (in the Liouville sense). Notice that a generic mechanical system is not bound to sustain any symmetry and thus to possess conserved quantities. One should be careful, however, in not confusing conservation laws with integration constants that generally do not require any symmetry.

Notice that conservation laws can be defined not only for physical systems, but in principle for any system of differential equations, provided it possesses sufficient symmetry. One can look for such laws either by a direct computation or by applying Noether's theorem and using general symmetries of the system. However, the action functional lying at the foundation of the variational hypothesis stating that the motion equations should be extremals of this functional does not necessarily exist, which leads to the difficulty of direct application of Noether's theorems about the correspondence between symmetries and conservation laws.

---

[32] For instance, we can rotate the sphere by an infinitesimal angle and nothing changes.

The use of symmetry greatly simplifies the analysis of mathematical models, not only in physics. However, in physics this simplification is directly manifested due to just mentioned Noether's theorems which prescribe the correspondence between continuous symmetries and conservation laws. Thus, invariance under spacetime translations leads to the energy-momentum conservation, invariance with respect to spatial rotations results in the conservation of angular momentum, invariance under gauge transformations in electromagnetism leads to the conservation of electric charge and so on. Point transformations i.e., those which take points to points, $u = u(x)$, where $u = (u^1, ..., u^m)$, $x = (x^1, ..., x^n)$, are symmetry transformations of an equation if they map solutions into solutions. For instance, in the ODE case in the plane, $u = u(x, y)$, $v = v(x, y)$ is a symmetry transformation if the image $v(u)$ of a solution $y(x)$ is also a solution. In more general terms, the $n$-th order ODE $F(x, y, y', ..., y^{(n)}) = 0$ (notice that Newton's equations of mechanics are a particular case of this ODE) retains its form after the symmetry transformation, $F(u, v, v', ..., v^{(n)}) = 0$, which means that symmetry of a differential equation allows us to choose convenient variables in order to solve it. This idea implemented through one-parameter group of point transformations, e.g., $u = u(x, y, \varepsilon)$, $v = v(x, y, \varepsilon)$, brought about a powerful method of integration of differential equations using Lie point symmetries. The symmetry group for a differential equation (or a system of differential equations) is understood as a set of transformations that takes a solution to this equation (or a system of equations) into other solutions.

It is remarkable that the very existence of symmetries may result in specific difficulties in field theories. For instance, infinite-dimensional symmetry groups may look, on a classical level, as additional constraints imposed on the initial data. It appears as if the described system had a smaller number of degrees of freedom than formally required by the least action principle.

Examples related to mirror ($P$) or time-reversal ($T$) invariance demonstrate how one can simplify modeling due to symmetries. In physical modeling, this fact can be observed on the formal level, e.g., by constructing the Lagrangian and analyzing its invariance properties. One may ask: why does one need the Lagrangian at all for modeling physical processes? The answer is that the differential equations of physics are typically so complicated that they are not attacked directly but are encoded in Lagrangians from which these equations can be obtained according to certain prescriptions. The procedure of exploring the symmetry of Lagrangians leads to Noether's theorems (indispensable for physics), and the resulting integrals of motion enable us to obtain the solutions much easier than by a straightforward treatment of motion equations. Thus, it is no wonder that physics is always hunting for symmetries because they provide powerful tools that allow physicists to reduce the seemingly kaleidoscopic world to a comparatively low number of fundamental models. Physicists in this respect are somewhat different from other modelers[33] since they are mainly focused on finding new and more fundamental types of symmetry. The number of basic models of the physical world becomes smaller and smaller as new symmetries are discovered and utilized; now one can enumerate about a couple of dozens of truly fundamental models.

One can subdivide the symmetries into "obvious" and "internal". The latter are also known as gauge symmetries: the simplest of them ensures electric charge conservation and explains the zero rest mass of a photon as well as the existence of just two possible photon spiralities (two directions of polarization by a light wave). Obvious symmetries such as spacetime ones result in habitual and immediate conservation laws such as those of energy-momentum. As already mentioned, gauge symmetry first expressed itself as a useful symmetry in electromagnetism [6] and [133] and then, gradually gaining recognition, was generalized to other physical phenomena and theories, especially of non-Abelian nature – the Yang-Mills theory (1954), see e.g. [162].

---

[33] Some professional mathematical modelers, in particular those engaged in AGW (anthropogenic global warming) computer models, sometimes label physicists as "lousy modelers".

The meaning of just mentioned Noether's theorem is that all conservation laws are the consequence of some symmetry: a single-parameter symmetry group determines the first integral of a dynamical system. If the system can sustain more symmetric transformations, then several integrals of motion arise. Notice that not all of these integrals are equally important: some are due to the fundamental spacetime properties whereas others may be the consequence of the symmetry of a specific model. For instance, one can ask what quantities should be conserved when a physical or an engineering system has a symmetry of an infinite homogeneous cylinder? The obvious answer is: if the cylinder axis coincides with $z$, then $z$-components of momentum $p_z$ and of angular momentum $m_z = x\partial_y - y\partial_x$ should be integrals of motion. Of course, this is a model: there are no infinite homogeneous cylindrical objects in nature. What quantities should be conserved if the system can be modeled by a homogeneous prism whose axis is parallel to $z$? The answer: $p_z$ and, possibly, a discrete analog of $m_z$. Another example: what quantities are conserved if the system has a symmetry of an infinite $(x, y)$-plane made of homogeneous material (or is found in the field of such a plane)? The answer is equally obvious: $p_x, p_y, m_z$ should be integrals of motion (if we close our eyes on the fact that infinite planes do not exist in nature even on cosmic scales). What quantities should be preserved in the field of a cone with its axis along $z$? The obvious answer $m_z$ suits also to the field of two points located on the $z$-axis. Here at least one does not require infinite dimensions of an object that is placed into an empty space or creates the field. The two last examples hint at more robust properties of the rotation group O(3) (more exactly its proper subgroup $O^+(3)$) whose generators are $A(\varphi) \approx 1 + m_z\varphi$ than, e.g., of the translation group, which fact manifests itself in the ubiquitous concept of spin.

So nearly all such answers about conserved quantities are approximate, but the corresponding small parameter is rarely discussed explicitly. Considerations that are more intricate arise in connection with the energy conservation, which is related to time-translation invariance in closed physical systems. The matter is that ideally closed physical systems do not survive in nature, closedness is eventually destroyed by fluctuations or external influences. Nevertheless, we use energy conservation laws assuming the time translation invariance in a presumably isolated physical system: if, e.g., $x(t)$ is the law of motion, where $T \in \mathbb{R}$ is an arbitrary time translation, then we consider $x(t + T)$ to be the same law of motion. We suppose that producing measurements over a physical system yesterday, today, tomorrow or in $T$ years should give the same results (perhaps differing only by measurement errors whose statistical characteristics are time-independent). Invariance of the laws of motion under time shifts is viewed as a special case of Galilean invariance. One sometimes makes from here a rather strong inference that all the laws of nature are constant i.e., if $x(t)$ is a solution of the motion equation, then $x(t + T), T \in \mathbb{R}$, is also a solution [16], §2.

We can remark in passing that the issue of time-reversal invariance, although being probably as old as physics as a whole, still stirs considerable controversy. Time-reversal invariance is mostly taken for granted: there is a widespread belief that no real processes in nature should, in the final analysis, violate time-reversal symmetry. However, this presumption seems to be arbitrary and too restrictive. Time-reversal invariance requires that direct and time-reversed processes should be identical and have equal probabilities, but there is no compelling reason why this must always be the case. Most mathematical models corresponding to real-life processes are time-reversal noninvariant – in contrast with most mechanical models. Purely mechanical models based on Hamiltonian equations are time-reversible, whereas statistical or stochastic models, though based ultimately on classical (reversible) mechanics, are time-reversal noninvariant. Accordingly, mathematical models designed to describe real-life processes must be so constructed as to envisage irreversibility in time. Indeed, more and more evidence has been accumulated lately that time-reversal symmetry is broken on a more sophisticated physical level than simple and sterile classical or quantum models. Please note that here we are not talking about statistical problems or about open systems, where it would be almost ridiculous to require time-reversal invariance.

## 5.6. Changes and catastrophes

Things around us change all the time, and mostly we consider it normal behavior. However, changes are not necessarily smooth (as, e.g., described by the exponential function or, more generally, by a strongly continuous semigroup); a system's behavior can also alter suddenly and seemingly unpredictably. So mathematical description of the world should be based on the interplay of continuous (smooth) transformations and discrete jumps. Accordingly, one has to analyze the phenomena characterized by sharp transitions from one equilibrium state to another, such transitions being governed by small changes of some parameters specifying the environment. This analysis is the subject of "catastrophe theory" which is actually a branch of the theory of dynamical systems. V. I. Arnold [14] described catastrophes as spasmodic changes emerging in the form of a sudden response of a system to a smooth variation of external conditions.

Parameters that govern the transitions may form an $m$-dimensional control space, $m = 1,2, ...$ For example, in a mechanical model described by Newtonian equations $m\ddot{\mathbf{x}} = -\nabla_x V(\mathbf{x}, \mathbf{a})$ parameter $a = (a_1, ..., a_m)$ can form an $m$-dimensional vector space $W^m$, with equilibrium solutions $\nabla_x V(\mathbf{x}, \mathbf{a}) = 0$ corresponding to certain domains $\mathbf{a} \in A_k, k = 1,2, ...$ in $W^m$. Thus, in one-dimensional ($n = 1$) mechanical model $m\ddot{x} = \partial_x V(x, \mathbf{a})$ with potential

$$V(x, \mathbf{a}) = V(x, a) = x^3 + ax \qquad (5.6.1.)$$

we have two equilibrium points: one minimum (a stable state) and one maximum (an unstable state) for $a < 0$. For $a > 0$, there are no equilibrium states in this potential whereas the value $a = 0$ corresponds to the bifurcation point: stable and unstable points merge and disappear at this value of the control parameter. For more complicated potentials, e.g.,

$$V(x, \mathbf{a}) = x^3 + a_2 x^2 + a_1 x \text{ or } V(x, \mathbf{a}) = x^4 + a_2 x^2 + a_1 x, \qquad (5.6.2.)$$

the control space is two-dimensional ($m = 2$) and the bifurcation set is now a curve instead of a point. One can say that the motion effectively takes place in a 3d space $(x, a_1, a_2)$ instead of 1d, with different projections on the $(a_1, a_2), (x, a_1), (x, a_2)$ planes. So, the catastrophe theory (theory of singularities) discusses critical point sets when several functions of several arguments are considered. This situation is, of course, more opulent than the motion in a 1d space (i.e., for a fixed value of possible control parameters).

The projections along each of the variables or parameters may have singularities reflecting a rather intricate geometric behavior in the parental $n + m$ space (in our last example, 3d). That is why catastrophe theory is often identified with the singularity theory of smooth mappings; in fact, catastrophe theory is just a specific case of singularity theory. The latter theory treats critical points of differentiable functions (in our examples, we used polynomial models) as those for which a level set brings about singular points. Physically speaking, such geometric singularities correspond to discontinuous changes in nature following tiny variations of ambient parameters. Thus, natural catastrophes such as earthquakes or landslides suddenly occur following the gradual accumulation of stress in the Earth's crust. Economic crises and financial crashes are believed to illustrate the applied catastrophe theory, when the system is destroyed by smooth, almost unnoticeable changes. Social cataclysms typically follow the same pattern: at small to moderate social stress, smooth transitions to more irritability of the folk ensue whereas higher stress levels or provoking actions bring the control parameters into the bifurcation zone, where the society (or some of its active fractions such as the young people in large cities) at first do not exhibit any conspicuous response to changing social circumstances. Yet as the society perceives the situation to be more provoking or irritating (the actual conditions may be not necessarily deteriorating), it can discontinuously pass into the unrest mode and remain in this state of unrest even if the provoking or angering factors have been removed or drastically reduced (social hysteresis). In general, one might note that if the catastrophe starts, it is hard to stop.

To get a feeling of catastrophic behavior, we can consider a traditional example of Whitney cusp. Compare two maps of the $(x_1, x_2)$ plane onto the $(y_1, y_2)$ plane (this is a very specific case of mapping a surface onto a plane which constitutes the geometric core of the catastrophe theory):

1. $y_1 = x_1^3 + x_1 x_2, \ y_2 = x_2$    and
2. $y_1 = x_1^2 - x_2^2, \ y_2 = 2x_1 x_2$ .

The first map (the Whitney cusp) is structurally stable in the sense that if we slightly perturb the mapping, then the perturbation does not destroy the cusp, perhaps after some appropriate coordinate change (Figure 2). The second map is structurally unstable since it replaces the equilibrium point (0,0) by some orbit, e.g., a cycle. Recall that structural stability means that the phase portrait may be only slightly deformed, in particular, the perturbed map basically has the same singular points, at least locally (here near (0,0)). If we perturb slightly the second map, e.g., $y_1 = x_1^2 - x_2^2 + \varepsilon x_1, \ y_2 = 2x_1 x_2 - \varepsilon x_2$, we have the following equation for the critical points (see below, section 8.4.): $J = \det\left(\frac{\partial y_i}{\partial x_j}\right) = 0$ which signifies that critical points form a circle $x_1^2 + x_2^2 = \frac{\varepsilon^2}{4}$ (more exactly, a circular i.e., quadratic cone).

We may recall in this connection that the term "surface", in particular, a surface $S^2$ in 3d point space $\mathbb{R}^3$ denotes, from the geometric viewpoint, a specific relation between two coordinates of a point, e.g., Cartesian coordinates $(x^1, x^2, x^3) \in \mathbb{R}^3$. In fact, a surface may be regarded as a prototype of a manifold. There are three main forms representing an $S^2$ surface: explicit, $x^3 = f(x^1, x^2)$, implicit, $F(x^1, x^2, x^3) = 0$, and parametric, $x^1 = \varphi^1(u, v), x^2 = \varphi^2(u, v), x^3 = \varphi^3(u, v)$.

Recall that the phase portrait is an aggregate representation of all the phase curves passing, in general, throughout all points in the phase space $P$ for a given dynamical system. The phase portrait gives a qualitative (topological) picture of the system's behavior and depends on the parameters that are present in the evolution equations and in supplementary (e.g., boundary) conditions. If these parameters are varied, the phase portrait can, in some cases, be only slightly deformed without altering the essential features of system's behavior. Sometimes, however, the behavior of a dynamical system can change significantly depending on variations of "external" (control) parameters. Such abrupt changes accompanied by the qualitative modification of the phase portrait are known as catastrophes.

Catastrophe theory proved useful in a number of applied sciences such as chemistry, medical biology, laser physics, optics, study of phase transitions, sociology – in short, in those fields where bifurcations are observed. In many cases, the catastrophe (singularity) theory is an adequate language to describe nonlinear behavior. In modern technology, engineers are trying to figure out typical patterns of smooth evolutionary behavior both for individual components of a technological system and for interaction (communication) links. For instance, if some system had been operating and suddenly stopped working, it is because something has changed: a "catastrophe" has occurred. To find out a problem one must be capable of tracking recent changes i.e., an engineer or a user must be provided with the data allowing her/him to look for the issue. Accumulated signs of malaise typically come unnoticed, hidden behind a seemingly infallible functioning of a system, yet given due notice point at grave problems.

## 5.7. Evolutionary models and equilibrium

We have seen above many examples of evolution and of evolutionary equations. In this section we shall restrict ourselves to temporal evolution only, leaving the treatment of spatial evolution (e.g., the diffusive spread) outside. The most important class of evolutionary models deals with the case of unitary evolution. If the time evolution operator, e.g., the Hamiltonian $H(t)$ standing in Schrödinger's equation $i\hbar\partial_t\Psi(t) = H(t)\Psi(t)$ is self-adjoint for every value of parameter $t$ i.e. $H(t) = H^+(t)$, then the evolution preserves the norm[34], $\frac{d}{dt}\|\Psi(t)\|^2 = 0$. An example of unitary evolution in the classical world is the Liouville theorem about the preservation of the phase space volume (see below in Section 8.4.1. "Phase space and phase volume").

Decay and exothermic reactions (chemical or nuclear) are examples of non-unitary evolution. The latter is, in general, non-invertible. In fact, most real-world processes evolve non-unitarily and are thus non-invertible.

### 5.7.1. What is evolution?

We have seen that evolutionary processes are mathematically described in the classical domain via the theory of dynamical systems i.e., by vector fields in a state (phase) space. In such a space any state corresponds to a point, with the rate of evolution i.e., the speed of change of a given state being measured by the velocity vector attached to any point manifesting a state. We can remind the reader of the curves in the state space formed by consecutive state points belonging to the same evolutionary process; such curves are known as phase trajectories. As already mentioned, a quantitative (or at least semi quantitative) description of processes that are evolving in time is given in terms of the theory of dynamical systems.

The latter theory can be formulated in a number of equivalent versions (through vector fields, differential equations, difference equations, semigroups, evolution operators and so on). In particular, mathematical representation of evolutionary processes consists in using the concept of evolution semigroups. The latter can be of autonomous type i.e., roughly speaking, corresponding to time-independent differential operators $A$: $\dot{x} = Ax, x \in X$ ($X$ is a Banach space) acting, e.g., on the half-line $\mathbb{R}_+$ ($t \geq 0$) that is semigroups $e^{At}|_{t\geq 0}$, or of nonautonomous type corresponding to time-dependent differential equations of the form $\dot{x} = Ax, x \in X, t \geq 0$. One of the main difficulties of nonautonomous problems is that evolution can be described at least by a two-parameter family of evolution operators, one of these parameters corresponding to an initial time point, whereas autonomous problems, while being time-shift invariant, naturally depend on a single temporal parameter. Respectively, the evolution family in the autonomous case is mathematically represented by the evolution semigroup operator $g_t, t \geq 0$.

---

[34] More generally, if $\Psi(t)$ and $\Phi(t)$ are any two solutions of Schrödinger's equation $i\hbar\partial_t\Psi(t) = H(t)\Psi(t)$ and if $H(t) = H^+(t)$ for any $t$, then $\frac{d}{dt}(\Phi, \Psi) = 0$. Thus, not only the norm, but also a projection of one solution onto another is preserved in the process of evolution under the action of a self-adjoint Hamiltonian.

Speaking in general terms, the formalism for the time evolution is represented by the time-dependent Schrödinger equation. However, the resulting wave function obtained with the help of this fundamental equation (provided one could somehow solve it) would contain a tremendous number of variables, which is utterly confusing and unphysical. Moreover, the Schrödinger equation would be sufficient to quantitatively explore the time evolution if we attempt to describe that of the whole system. But in real-world situations one has to use the hierarchy of physical variables ordering them as more and less relevant. Then the Hamiltonian which determines the entire wave function will be partitioned into different subparts acting in separate subspaces of variables (degrees of freedom). The most often encountered case is the decomposition of the whole system into a relatively small subsystem $A$ with Hamiltonian $H_A \equiv H_0$ and a large subsystem $B$ (the "environment") with so many degrees of freedom that one can consider their number as infinite, so that one can view subsystem $B$ and its characteristics as "macroscopic" and unchangeable.

Evolution, in particular biological evolution, seems to be irreversible: one can hardly expect the reappearance of dinosaurs. Cosmological evolution also appears to be irreversible, although this is a highly speculative question. Nevertheless, both biology and cosmology predominantly consist of evolutionary models. On the other hand, evolution is in general poorly predictable, mostly because of bifurcations. Today's widely accepted model of biological evolution is founded on the accumulation of random mutations i.e., fluctuations in ordered genetic systems, with the most beneficial of them being preserved in the process known as natural selection. Yet, as far as the primitive mathematical modeling of this process goes, some discrepancies are encountered in describing biological evolution in terms of random mutations, which requires either attracting rather sophisticated concepts such as chaos-order transition in dissipative structures or resorting to qualitative considerations.

### 5.7.2. A remark on the concept of equilibrium

The most "comfortable" case is that of robust equilibrium when the observed system (e.g., a physical system, the human organism or an economy) is increasingly resilient when facing external disruptions. The standard approach to equilibrium in many-body theories consists in the hypothetical procedure of adiabatically (i.e., very slowly) turning on interactions at very distant past ($t \rightarrow -\infty$) and turning them off at very distant future ($t \rightarrow +\infty$). When one starts from the equilibrium state (or, e.g., from the ground state) at $t \rightarrow -\infty$, one would arrive at the same state at $t \rightarrow +\infty$, possibly gaining some phase factor $e^{i\varphi}$ in the process of evolution between the initial and the final state – this is the leading assumption.

In the situation out of equilibrium, evolution occurs in a different way as compared to the equilibrium state. For example, when one starts from some arbitrary many-body distribution, one can find the system in some unpredictable state at $t \rightarrow +\infty$, in particular, depending on the character of turning on the interactions.

## 5.8. Many ways of model simplification

Very few models admit exact solutions of the corresponding equations; therefore, one has to use approximate techniques or resort to certain tricks. If the model depends on many parameters or someone is interested in great generality, calculations may become very involved.

How can one simplify mathematical models? The first step is to reduce the model's equations to the most primitive form, retaining the essence of the model. Usual tricks of the trade include (but are not confined to):

- disregarding small terms (in fact, this is equivalent to the expansion in power series)
- using small or large parameters (this is also equivalent to series expansion, but in asymptotic series which are not necessarily convergent)
- replacing complex geometric shapes by more symmetrical ones (in most cases, it is immaterial whether a 3d-object is a sphere, a cube or a parallelepiped; thus, mountains do not have conical form, but they can be represented as cones since describing the exact shape would be an illusory precision)
- substituting constants for functions (in fact, this is an application of the mean value theorem)
- linearization
- ignoring spatial distribution of quantities, in particular, transition to the homogeneous or point-like models (when spatial gradients vanish, $\partial/\partial \mathbf{r} = 0$, one obtains ODE-based instead of PDE-based models)
- scaling
- discretization and working on lattices (lattice models)
- transition to dimensionless units (a specific case of scaling)

The dimensionless form of mathematical models plays a special role: numerical values are independent of the measurement units. Dimensionless expressions are very convenient for numerical simulation, for example, one can compare the terms not only with each other, but also with the errors in other terms. Thus, one can evade groundless assumptions of higher approximation precision. One can also consider specific cases of the model not by comparing physical quantities, but by choosing numerical limits. For example, the toy model of the harmonic oscillator $\ddot{x} + \omega^2 x = f(x, t)$ can be reduced to a dimensionless form by putting $\omega t = \tau$ or $t = \tau/\omega$ so that the oscillator equation becomes $x'' + x = f(x, \tau/\omega)$, where primes denote differentiations over dimensionless variable $\tau$ (which can be called the "slow" or the "fast" time depending on application).

Historically, scaling and dimensionless combinations appeared first in the heat and mass transfer problems which often involve combined interactions of thermodynamic, fluid dynamical and electrodynamical processes, characterized by many parameters with the area-specific dimensions. In multiphysics phenomena, multiphase system dynamics and, e.g., the study of turbulence, scaling and dimensional analysis provide indispensable heuristic tools. For instance, one can mention heavy nuclear accident modeling (such as related to Three-Mile-Island, Chernobyl and Fukushima events), where one has to study the motion of fluid with internal heat sources. Mathematical modeling and creation of the corresponding computer codes designed to simulate the molten core behavior together with the accompanying chemical transformations[35] (in particular, the Russian code "Rasplav" and its analogs, see e.g., http://en.ibrae.ac.ru) are almost completely based on the dimensional analysis and use of dimensionless combinations.

---

[35] In particular, the Russian code "Rasplav" and its analogs, see e.g., http://www.ibrae.ac.ru.
©Sergej Pankratow

Perhaps one of the most popular mathematical models, where dimensionless units were used, was the strong (nuclear) blast problem analyzed in the 1940s by J. von Neumann, L. I. Sedov, and J. Taylor. Modeling in ecology and atmospheric science (atmospheric circulation) is also largely based on dimensional analysis. The latter is an indispensable part in the models of climate dynamics on Earth and Venus.

## 5.9. Linearization

Another way to simplify a model is to linearize it, which means ignoring higher-order terms in the equations. In most cases, if one follows purely linear methods, it would be impossible to describe many natural phenomena and even to predict their existence. Yet linear methods are simple and practically always lead to a solution, in contrast with truly nonlinear models. Therefore, one often tries to find some linear system that would inherit the principal features of the corresponding nonlinear one. This replacement of a nonlinear model by a simplified linear system is known as linearization.

A good example of linearization is the study of stability of an equilibrium state. To put it simply, stability means that small perturbations do not significantly affect states or trajectories as time is running. The word "small" implies, as usual, that the linear approximation would be sufficient to describe the process. Thus, stability of a dynamical system can be analyzed by using its linearized version. This is a simple, but important idea, and to better understand it let's consider the behavior of a dynamical system near fixed points. We know that any fixed point $\mathbf{x}_m, m = 1, 2, \ldots$ gives a stationary solution to the evolution equation i.e., an equilibrium position of the modeled process, when the vectors driving the system to some eventual state vanish or compensate each other. Therefore, one can primarily explore the behavior of a dynamical system near equilibrium positions to find out whether the system, if slightly perturbed, would return to equilibrium or run away from it. If deviations from equilibrium are small, we can expand vector field $\mathbf{v}(\mathbf{x}) = \mathbf{v}(\mathbf{x}_m + \boldsymbol{\xi})$ in Taylor series retaining only the first non-vanishing term, if possible, the linear one. If all the nonlinear terms can be neglected, this procedure is called linearization. Here $\boldsymbol{\xi}$ is a vector in tangent space $T_{\mathbf{x}_m}M$, where $M$ is a manifold with local coordinates $x^i, i = 1, \ldots, n$. Note that the tangent space (manifold) is basically a rather simple object, for instance, the tangent space $T_{\mathbf{x}}Q$ to vector space $Q = \mathbb{R}^n$ at point $\mathbf{x} \in Q$ is the vector space formed by all vectors in $\mathbb{R}^n$ attached to point $\mathbf{x}$ i.e. $T_{\mathbf{x}}Q \equiv T_{\mathbf{x}}\mathbb{R}^n$ is an affine copy of $\mathbb{R}^n$, with its origin shifted to point $\mathbf{x} = (x^1, \ldots, x^n)$. Here, we may remind the curious reader that $\mathbb{R}^{n=0} = \{0\}$; this convention specifying the trivial yet non-empty vector space.

Further, we shall consider a single equilibrium point $\mathbf{x}_m = \mathbf{c}$; one can of course put $\mathbf{c} = 0$ by a proper choice of local coordinates so that $\mathbf{v}(\mathbf{c}) = \mathbf{v}(0) = 0$ and $\boldsymbol{\xi} = \mathbf{x}$. Then we have $\dot{x}^i = a^i_j x^j, a^i_j :=$ $\partial_j v^i \big|_{\mathbf{x}=0}$ or $\mathbf{v}(\mathbf{x}) = A\mathbf{x} + \mathbf{R}_2(\mathbf{x}), A = \partial \mathbf{v}/\partial \mathbf{x}|_{\mathbf{x}=0} = (a^i_j), \partial \mathbf{R}_2(\mathbf{x})/\partial \mathbf{x} \to 0$ with $|\mathbf{x}| \to 0$ (it is assumed that one can differentiate series $\mathbf{R}_2(\mathbf{x})$, corresponding to the nonlinear contribution, term-by-term). The remainder $\mathbf{R}_2(\mathbf{x})$ is a curved vector field on the phase space, while the linear term $A\mathbf{x}$ generates an auxiliary linear problem whose stability is relatively easy to explore. One can treat this auxiliary linear system and its stability as an asymptotic case of the general nonlinear problem. It is clear that if the solution to a linearized problem $\boldsymbol{\xi} = 0$ or, in an appropriate coordinate system, $\mathbf{x} = 0$, is unstable, then the same solution of the original nonlinear system is in any case unstable. This "negative" principle of linearized stability is very important for many practical applications, for example, in the theory of plasma instabilities, in chemical kinetics, in modeling fires and flame propagation, in applying the theory of dynamical systems to biology and economics, in aviation, etc. For example, linear analysis of flames (performed by G. Darrieus in France in 1938 and L. D. Landau in Russia in 1944) showed that a thin planar flame front is unstable with respect to spatial disturbances of any wavelength. This hydrodynamic instability of flames is essential to produce combustible mixtures for missile engines, energy projects, modern explosives, etc. Besides, understanding the unstable character of flames is indispensable for safety concepts, in particular, to assess and prevent the risk of spontaneous flame acceleration and transition to detonation that poses a threat to humans and structures.

Since the phase velocity field is given by $\mathbf{v}(\mathbf{x}) = d/dt(g_t\mathbf{x})|_{t=0}$, the linear operator $A$ is defined in terms of the evolution operator $g_t, t \in \mathbb{R}$ as $A\mathbf{x} = d/dt(g_t\mathbf{x})|_{t=0}$ for all $\mathbf{x} \in \mathbb{R}^n$. This expression can be understood in both directions: the evolution operator (one-parameter group $g_t$) defines the linear operator $A: \mathbb{R}^n \to \mathbb{R}^n$ and, conversely, for each $A$ one can find a one-parameter group $g_t$ translating the states in time i.e., operator $A$ defines the equation $\dot{\mathbf{x}} = A\mathbf{x}, \mathbf{x} \in \mathbb{R}^n$. For instance, if $A$ is a multiplication operator in a scalar ($n = 1$) system by some number $\mu$, then evolution $g_t$ corresponds either to contraction ($\mu < 0$) or to stretching ($\mu > 0$) of state $\mathbf{x}$ with time, $g_t = \exp(\mu t)$. In general, the solution to a linear system $\dot{\mathbf{x}} = A\mathbf{x}$ with initial condition $\mathbf{x}(t_0) = \mathbf{x}(0) = \mathbf{x}_0$ is given by matrix exponent $\mathbf{x}(t) = \exp(At)\mathbf{x}_0 = (I + At + \frac{A^2t^2}{2!} + \frac{A^3t^3}{3!} + \cdots)\mathbf{x}_0$, where $I$ is the unit matrix. Note that one can illustrate this exponentiation by direct calculation

$$\mathbf{v}\big(\mathbf{x}(t)\big) = \mathbf{v}\left[\mathbf{x}(t_0) + \int_{t_0}^t \mathbf{v}(\mathbf{x}(t))dt\right] + \mathbf{v}_0.$$

Using Galilean invariance, we can set $\mathbf{v}_0 = 0$. Integrating once more, we get

$$\mathbf{x}(t) = \mathbf{x}(t_0) + \int_{t_0}^t dt\,\mathbf{v}\left[\mathbf{x}(t_0) + \int_{t_0}^t \mathbf{v}\big(\mathbf{x}(t')\big)dt'\right] = \mathbf{x}(t_0)$$
$$+ \int_{t_0}^t dt_1\mathbf{v}\left[\mathbf{x}(t_0) + \int_{t_0}^{t_1} dt_2\mathbf{v}\left[\mathbf{x}(t_0) + \cdots + \int_{t_0}^{t_{n-1}} dt_n\mathbf{v}\big(\mathbf{x}(t_n)\big)\right] + \cdots\right] = g_t\mathbf{x}(t_0)$$

Simple series near $t = t_0$ can be written in the form:

$$x(t) = x(t_0) + v\big(x(t_0)\big)(t - t_0) + \frac{\big(v(x(t_0))\big)^2}{2!\,x(t_0)}(t - t_0)^2 + \cdots + \frac{\big(v(x(t_0))\big)^n}{n!\,(x(t_0))^n}(t - t_0)^n + \cdots$$

($x(t_0) \neq 0$) or in vector writing

$$\mathbf{x}(t) = \left[1 + \frac{v(x(t_0))(t - t_0)}{x(t_0)} + \frac{\big(v(x(t_0))\big)^2(t - t_0)^2}{(x(t_0))^2} + \cdots + \frac{\big(v(x(t_0))\big)^n}{n!\,(x(t_0))^n}(t - t_0)^n + \cdots\right]\mathbf{x}(t_0)$$
$$= \exp\big[x(t_0)_\alpha^{-1}v^\alpha\big(x(t_0)\big)(t - t_0)\big]\mathbf{x}(t_0).$$

Such expressions make clear the connection between Newtonian mechanics and classical calculus. In the complex case the solution has the same form $\dot{\mathbf{z}} = A_\mathbb{C}\mathbf{z}, \mathbf{z} \in \mathbb{C}^n$, where $A: \mathbb{C}^n \to \mathbb{C}^n$ is a complex linear operator ($\mathbb{C}$-linear) acting in the complex phase space $\mathbb{C}^n$. Note also in passing that the qualitative picture of the behavior of ODE solutions in the complex domain is very helpful for understanding the evolution described by ODEs.

A linearized vector field $A\mathbf{x}$ may be viewed as a simple pattern for the general nonlinear (curved) vector field $\mathbf{v}(\mathbf{x})$. Linearization is defined correctly: operator $A$ does not depend on the local coordinate system, in other words, linearization is an invariant operation. For small deviations from the equilibrium, the difference $\|\mathbf{R}(\mathbf{x})\|$ between the original (nonlinear) and the linearized system is small compared to the norm $\|\mathbf{x}\|$ of state $\mathbf{x}$. It means that the motions of both systems starting with the same initial conditions are close to each other, and we have seen that stability of solutions to dynamical systems is determined by the development of small fluctuations or deviations from the "regular" integral trajectories. Therefore, in most practically important cases it is sufficient to explore the much simpler linearized version of a dynamical system.

Linearization is extremely important for modeling since it leads to linear equations that can **always** be solved. A linearized dynamical system is represented by $n$ linear ODEs of the first order with constant coefficients (so that the phase space of such a system is $\mathbb{R}^n$), and treatment of this system of equations is a linear algebra problem rather than a theory of differential equations. For example, linearization i.e., the transition from $\dot{\mathbf{x}} = \mathbf{v}(\mathbf{x})$ to $\dot{\mathbf{y}} = A\mathbf{y}$ lies at the foundation of vibration theory: near the stable equilibrium point, the behavior of the potential function coincides with that of its quadratic part. One can always find a solution for small (harmonic) oscillations, which is one of the reasons why the oscillator model is so popular in physics and engineering. Moreover, small variations of the potential can only slightly shift the position of the equilibrium point, and one can infer the motion of a disturbed system with the help of perturbation techniques over these small variations, using an exact solution that can always be known as a base.

## 5.10. Linear models

Some models are linear to begin with, and many are linear because we have applied linearization to simplify the equations. In linear regimes, if the input happens to be scaled, the output is scaled accordingly. However, in the real world, this is mostly true for minute changes – this approximation lies at the foundation of classical differential calculus. Science in its contemporary form is predominantly nonlinear, and linear regimes are rare in the real world. This is especially clearly manifested in modern interdisciplinary studies, e.g., in economics, ecology, sociology, demography, climatology, etc. Recall that even modern physics historically originated in nonlinear equations of motion: already the Kepler problem contained many typical features of nonlinear systems such as dependence of the period on the amplitude and periodic orbits with many harmonics. Later, it became clear that the three-body problem should lead to a systematic study of nonlinear dynamics which effectively means that all many-body problems must be nonlinear. Simplification of the ubiquitous nonlinear models results in their linearized versions. Therefore, one has to be careful about operating in linear range or domain.

In physics, linear models typically have the form $DA(\mathbf{x}, t) = j(\mathbf{x}, t)$, where $D$ is a differential operator defined on some linear manifold and acting on field $A(\mathbf{x}, t)$ that can be of vector or tensor character obeying certain supplementary conditions (initial and boundary). The right-hand side function $j(\mathbf{x}, t)$ is usually interpreted as a source which is assumed to be known (forward or direct problems). There is an important class of inverse problems, when one has to find the source function (distribution of sources) or certain parameters defining operator $D$, given the information on field $A(\mathbf{x}, t)$. Inverse problems are in most cases ill-posed i.e., the requirements of existence, uniqueness and stability of solutions are not fulfilled. Although the study of inverse problems and models is a very interesting subject with many scientific and engineering applications (tomography, medical diagnostics based on ECG, EEG or EMG as well as geophysical explorations are good examples), we shall not discuss them here.

The success of classical electrodynamics for engineering applications, even in the pre-laser period, of antenna and wave propagation theories, the ubiquitous use of the theory of linear oscillations and the development of quantum mechanics shifted the focus from the emergent nonlinear tasks such as the problem of dynamical system stability to essentially linear approaches. The main feature of linear models is the superposition principle i.e., the possibility to represent the output (response) of the modeled system to a number of inputs as the sum of responses to each individual input or, mathematically,

$$L\psi = L\sum_n a_n\psi_n = \sum_n a_n L\psi_n = \sum_n a_n\lambda_n\psi_n \qquad (5.10.1.)$$

where $L$ is the linear operator acting in some set $V$ (vector space) containing $\psi_n$. Here elements $\psi_n$ are considered the eigenfunctions of operator $L$ constituting the basis of a vector space. In general, a system is linear with respect to any elements $\psi_n$ belonging to some set $V$ if $L\sum_n a_n\psi_n = \sum_n a_n L\psi_n$, where $a_n$ are any complex numbers (in general $a_n$ may be the elements of any field). Function $f(x)$ is linear if it satisfies the relationship $f(\sum_n a_n x_n) = \sum_n a_n f(x_n)$; thus, the following expressions are linear: $f(\mathbf{r}) = \mathbf{a}\mathbf{r} + b$, $c_i = m_{ik}x^k$, $\mathbf{v}(\mathbf{x}) = [\boldsymbol{\omega}, \mathbf{x}]$. In contrast, the following expressions are nonlinear: $f(z) = \exp z, f(z) = \log z, f(z) = \tan^{-1} z$, $f(z) = (1 + az)^m$, but can be easily linearized by using the series expansion and neglecting the higher-order terms. Linearization is in general the main idea of obtaining linear models in real-life situations.

The meaning of expressions such as (5.10.1.) is that if the response to individual inputs $\psi_n$ is known, then the total response of the system (its full output signal) is also known. This assumption drastically simplifies mathematical modeling of a great number of systems encountered in science or engineering and thus perfectly corresponds to the main purpose of modeling: to find a standpoint that enables us to view the subject as simple as possible. One can illustrate the usefulness of superposition principles on numerous examples from electrical engineering (linear circuits), mechanical engineering (combined loads), control theory (model predictive control and plant models), practical electrodynamics (computation of fields created by given charge and current distributions, design of electromagnetic structures), heat and mass transfer, hydrogeology and oil extraction problems (multiple wells pumping), etc. In the historical retrospective, linear PDE-based modeling has elaborated many useful tools whose importance spreads well beyond the classical PDE-theory, for example, the fundamental solutions, Green's functions techniques, Fourier series over eigenfunctions, Huygens' principle, Perron's method, etc. Linear systems also gave a great impetus to the development of integral transforms which were predominantly used in their study: Fourier, Laplace, Hankel, Weber, Hilbert (analytical signal) and, recently, wavelets. One can also notice that classical optics is in fact a theory of linear systems and transformations: for instance, an ordinary lens performs a two-dimensional Fourier transform. Recall that the Fourier transform is the main tool in the study of linear systems, taking an input in one domain and producing an output in another (conjugate or dual) domain. For example, time-domain signals are split into their frequency components which are represented in the frequency domain. In quantum mechanics, which is an essentially linear theory, the idea of viewing input and output as conjugate quantities allows one to migrate between different "representations" and was eventually refined to become one of the crucial devices.

In the so-called systems theory, linear systems comprise the simplest part based on the linear transforms, linear mapping or linear response: input $\rightarrow$ output. In spite of its apparent simplicity, linear modeling has brought great results in science and especially in engineering. Thus, classical linear systems theory was traditionally the main tool of electrical engineering, being successfully applied for such tasks as signal transmission, stationary noise reduction or removal, time-invariant filtering, predictive coding and many others. Note that noise is always present, and in most linear models of science and engineering it is approximated by Gaussian statistics, $v(t) = \frac{1}{\sqrt{2\pi\sigma^2}} \exp\left(-\frac{(t-\mu)^2}{2\sigma^2}\right)$. Filtering the Gaussian noise by a linear time-invariant system again produces Gaussian noise. Moreover, the mean and the variance are left intact after the passage through a linear system. Of course, some specific features of a signal while passing through a linear system may change, but the overall character of a distribution remains the same. In engineering, usually time-invariant linear systems are considered i.e., those described by linear time-invariant operators: if the input $f(t)$ is delayed by $\tau$, $f_\tau(t) = f(t-\tau)$, then the output is also delayed by $\tau$, $g(t) = Lf(t) \rightarrow g(t-\tau) = Lf_\tau(t)$. In the theory of dynamical systems, this property corresponds to an autonomous model and is closely connected with energy conservation.

In time-invariant systems and models, the behavior does not depend on time point, in particular, on the observation instant. Indeed, for any $t \in I \subset \mathbb{R}$, $g(t) = Lf(t) \rightarrow g(t-\tau) = Lf_\tau(t) = L(f(t-\tau)$, and it does not matter when you are recording the signals. Notice that, mostly due to the different behavior at various time instants, modeling techniques for time-noninvariant systems are much less developed than for time invariant ones. In this respect, the newly developed wavelet methods can help.

In the language of ordinary differential equations and dynamical systems, linear models are often expressed as $\dot{x}(t) = A(t)x(t)$ (in the scalar case) or, more generally, as $\dot{x}^i(t) = A^i_j(t)x^j(t) + F^i(t)$. Vector $\mathbf{X}(t) = \{x^1(t), \ldots, x^n(t)\}^T$ characterizes the state of the system and may be called the state vector. Function $\mathbf{F}(t)$ describes an external influence on the considered linear system. If $\mathbf{F} = 0$, then the model corresponds to an isolated system which is, in fact, an idealized situation since isolated (closed) systems are never really encountered in reality: there is always some environment. Closed or isolated systems are those for which external influence can be disregarded, which is a rare occasion. Nevertheless, the concept of isolated systems is very useful as a mathematical model, especially in the linear case, where one is mostly interested in the system's response to an external influence.

In $n$ finite dimensions, a linear system $\dot{x} = Ax$ has a solution $x(t) = U(t)x_0 = e^{At}x_0$, where $x_0 := x(t_0)$ and evolution operator $U(t)$ belongs to some semigroup $G$. Recall in this connection that an autonomous dynamical system can be fully described by a one-parameter semigroup of transformations (which is just another manifestation of determinism). Note that complex number $\lambda$ is an eigenvalue of operator $A$ if and only if $e^{\lambda t}$ is an eigenvalue of $U(t)$ i.e., Spect $U(t) = \exp\{\text{Spect}(A)t\}$.

The main mathematical tool for exploring linear models is linear algebra. In particular, differential equations describing the response to an external forcing are completely reduced to linear algebra techniques. Such techniques allow us to always obtain a solution, with certain computational difficulties arising only for large dimensionalities of the respective vector space, so that linear models are very useful for the understanding of the behavior of complex systems in the first approximation. For instance, the stability of equilibrium positions (linear stability) and the classes of equilibrium points (where vector field $\mathbf{v}(\mathbf{x}, t) = 0$) of a dynamical system can be investigated within the linear models. Note however that complex systems are necessarily nonlinear, and linear models are only used to appreciate the trends in the complex systems behavior such as long-time stability.

The value of linear models is area dependent. For example, linear models in life sciences can be used for general setup schemes, but they usually do not provide much insight into the functioning of a living organism. Modeling in biology and medicine is essentially nonlinear.

## 5.11. Contemporary trends

One can observe that modern science tends more and more towards mathematical modeling. Even physics, which has always been based on experiments, is currently developing similarly to mathematics, increasingly employing assumptions, relying on "inner beauty" more than on experimental validation and extensively using computer simulations. A number of hot subjects in science such as string theories, cosmological models, a few modern gauge theories, some climate variability models, can hardly be experimentally verified, at least in the foreseeable future. For instance, it is hard to imagine any experimental device – to test theoretical models in high-energy physics and astrophysics – operating at an energy range of about or over 100 TeV and respectively inaccessible distances. The abstract yet popular models of string theory are untestable, at least so far; therefore, it is still difficult to place them somewhere on the "illusion-reality" axis. In fact, string theory is a purely mathematical theory, a collection of mathematical models, yet intended for physicists.

So, one has to rely in more and more cases on mathematical and computer modeling, with the risk that some of the models may remain mathematical artifacts forever. Furthermore, in certain fields of engineering and military technology (e.g., nuclear engineering or weapons tests) modeling tends to replace actual experiments. One can also observe that the modeling methods employed in totally different areas are getting more and more similar. For example, mathematical and computer-based modeling in economics and finance tends to mimic the methods developed in physics. This unification of modeling techniques reflects the progressive trend towards interdisciplinarity.

## Section 6. Using fundamental laws

Mathematical and computer models are predominantly formulated in terms of equations that reflect the laws of physics or at least have their origins in physics, even when the models are related to other disciplines (such as biology or economics). It is, in principle, not at all obvious that equations must play such a principal part – laws of nature might be expressed as inequalities, long codes or concise verbal statements. Yet they are, to our current understanding, expressed by differential equations, which fact provides a hint at a primary pattern for efficient mathematical modeling.

The laws of physics have been established experimentally or through numerous careful observations, and the corresponding equations may be viewed as generalizations of empirical data. Some of the laws can be perceived to be "more fundamental" than others: for instance, Darcy's law which is a mathematical model of a highly phenomenological character – a constitutive equation describing fluid flow through a permeable (porous) medium – is intuitively less fundamental than the Navier-Stokes equation (from which Darcy's law can be derived as a very special case). Likewise, Ohm's law for electric networks is far less basic than Maxwell's equations coupled with motion equations for the electronic subsystem in conducting materials; Fourier's law for heat conduction and Fick's law for diffusion represent very particular cases of kinetic theory and irreversible thermodynamics and are thus not as fundamental as, e.g., the Boltzmann equation or microscopic expressions for fluctuating currents.

Nonetheless, the hierarchy of fundamentality seems to be totally subjective and hard to correctly establish. Can one ascertain what discipline is more fundamental, quantum mechanics or general relativity? Classical mechanics or electrodynamics? Therefore, we shall define neither an ordered set nor even an equivalence class of fundamental physical laws, leaving their list to be intuitively composed. Such a list may include the motion and evolution equations, energy and momentum (also the angular momentum) conservation laws, mass and charge conservation expressions as well as other hyperbolic PDE-systems for the conserved quantities, other major PDEs of physics, etc.

Notice that in classical physics mass and energy conservation are separate laws. In general, in classical physics mass and energy are different notions so that it is quite natural that they satisfy their own conservation laws. One can see that conservation laws in general can be interpreted as a system of constraints forbidding changes of certain quantities, e.g., energy, momentum, angular momentum, spin, charge, etc. in the course of evolution of a physical entity[36]. In fact, conservation laws can be interpreted as constraints: thus, energy conservation defines a surface $H(p, q) = 0$ in the phase space $P$, with the phase space variables being restricted to this surface.

One can approach the concept of fundamentality from the positions of experimental verification. Omitting a long quasi-philosophical discourse, one can notice that any system to be mathematically modeled or simulated on a computer should eventually rely on a comparison with its real-life prototype. Such a comparison is known as a measurement (in a general sense). Abstract, philosophical reflections about the nature of things without the restriction of measurement can be perceived as symptomatic of inability to state problems and to work hard on their solutions (think to the end) or as an indication of laziness.

The crucial problem with philosophical speculations is that they can be used to "substantiate" two directly opposite views or statements. Using only generalities i.e., ignoring details leads to misunderstandings, when the considered phenomena can be interpreted in a multitude of ways, in particular, in any *a priori* fixed manner. Besides, philosophical reflections virtually never rely on scientific measurements. The question of measurement accuracy is irrelevant to philosophy.

Scientific measurements are bound to satisfy the crucial requirement: they must be reproducible. It means that the existence of equivalence classes of identical objects of study and of identical situations (including the reference frames and measuring devices) in which these objects can be placed is implicitly assumed. Regardless of the concrete physical process of measurement i.e., the form of interaction with the measuring device the reproducibility of measurement implies obtaining the invariant outcomes. In other words, measurement of fundamental physical quantities entails the presence of some group structure (more specifically, that of a Lie group acting on a manifold). Thus, the fundamental laws are those that do not depend on a concrete measurement situation (frame of reference plus a relevant set of measuring devices) and are invariant under the transformations consisting in a transition to another fashion of measuring the same set of physical quantities. For instance, it does not matter whether one tries to verify (or disprove) special relativity by the Michelson-Morley-like experiments in the photon sector or by atomic clock tests in the matter sector, or in accelerator-based elementary particle measurements, or in other kinds of experiments (this is like a rotation in the space of all possible relativity tests). If the results are invariant i.e., independent of the experimental method used and equally corroborative (within, perhaps, certain limits), then special relativity is a fundamental theory whose laws can be used to build up reliable mathematical models. Contrariwise, Moore's law stating that the number of transistors (or, more generally, of electronical microelements) per device doubles every 18 months is not a fundamental one; it can be

---

preserved or destroyed by human efforts, and mathematical models based on Moore's law[37] are likely to be unreliable. Nonetheless, such mathematical models are still applied although they have been known to be flawed.

The same applies to many economic, sociological and natural-philosophical observations frequently called laws. There exist intermediate cases i.e., statements known as laws but formulated in such a way that their fundamentality is difficult to verify experimentally. Thus, some mathematical conjectures are called "laws", yet their basic nature is questionable. For example, Weyl's law in mathematical physics states that the largest frequencies (eigenmodes), say, in the sound of a drum are determined not by the drum shape, but merely by its area. Weyl's law relates, in general, to the normal modes of the Laplacian in a $d$-dimensional compact domain, with the Dirichlet or Neumann boundary conditions being imposed, and is of a great modeling importance, in particular, to study the propagation of seismic waves, to explore the properties of microwave resonators and waveguides, to perform acoustic testing and so on. Nevertheless, it can hardly be acknowledged as a fundamental law of nature.

As already mentioned, fundamental laws are primarily used to build mathematical models in the form of differential equations. These equations, though most of them have their origins in physics, can describe not only physical, but also biological, chemical, economic, etc. processes. We shall see further than the conservation laws can be used to model traffic flows. The universality of mathematical models derived from basic laws reflects the profound unity of the material world. In general, such mathematical models can be formulated, besides differential equations, in terms of difference, algebraic, functional, integral, integro-differential or other types of equations. In this book, however, we shall consider predominantly differential equations (ordinary and partial), other types will be just briefly mentioned. Moreover, in this book only the models derived from basic principles (maybe with the exception of competition models) are observed.

The most fundamental statements about the natural environment seem to be the conservation laws (mentioned already in the 17[th] century in the works by Galileo, Newton, Huygens, Leibnitz, and then in the early 19[th] century in those by Young). In some cases, conservation laws can be observed even when the underlying parametric motion equations i.e., time-dependent vector fields are unknown (e.g., in economics or for the motion of self-propelled objects). At first, conservation laws were associated exclusively with mechanical systems, and only in the mid-nineteenth century were they extrapolated onto the systems and processes on a non-mechanical nature such as thermodynamics and fluid dynamics. In the 20[th] century, physical laws expressing the exact mathematical conservation of some quantities were established, these exact conservation laws being based on two Noether's theorems and other facts of Lie group theory. Exact conservation laws are the first integrals of the underlying motion equations for particles or fields, corresponding to some symmetry of the physical system. We postpone for a while the discussion of this highly important subject – up to a brief overview of Lagrangian mechanics.

---

[37] Gordon Moore who was the research director of Fairchild Semiconductor, one of microelectronics pioneers, and one of the founders of Intel pointed this out in 1965. Moore's law effectively states that the computer speed doubles every 18 months.
©Sergej Pankratow

## 6.1. Balance laws

Now, we shall deal with the balance laws expressing the conservation of such quantities as mass, energy, linear and angular momentum, entropy, number of particles, and different charge-like entities enclosed in a certain volume. However, in distinction with Lagrangian mechanics and other theories based on infinitesimal continuous transformations described by the theory of Lie groups, where conserving quantities appear automatically as integrals of motion equations, the conserving quantities in balance laws are often of a phenomenological character and cannot necessarily be produced as the first integrals of some basic motion equations. It is important to understand that the balance conservation laws are in most cases just phenomenological integral relationships which serve as a starting point for deriving differential equations that describe the motion of material media – the other way round as compared to the motion of material points in classical mechanics and similarly constructed "first-principle" disciplines.

Nevertheless, the concept of balance is ubiquitously used in science, engineering, business and everyday life: one can hardly imagine accounting and financial management without this concept; power production and consumption should be balanced; balance of solar radiation determines climate dynamics (see the respective section below) as well as the behavior of Earth's ecosystems and so on. The balance conservation laws typically describe temporal variations of the density of a conserved quantity in the material media and are thus an essential issue for mathematical modeling, specifically of nonstationary transport processes i.e., transfer of substances, radiation, energy, ecological impurities, finances, vehicles and other self-propelling particles such as biological objects including humans, etc. A typical scheme of producing a mathematical model for a rate of change is to equalize growth, decline and spatial dispersal. For example, models in population biology and epidemiology are formulated as the balance of births, deaths and migration rates.

Any conservation law can be written as an expression $du/dt = 0$, where the conserving quantity (local parameter) $u = u(x, t)$ and local spatial coordinate $x$ are in general multicomponent variable (in particular, a vector), and symbol $d/dt$ expresses the total differentiation over time. In local coordinates we have

$$\frac{du}{dt} = \frac{\partial u}{\partial t} + \frac{\partial u}{\partial x^i}\frac{\partial x^i}{\partial t} = \partial_t u + v^i \partial_i u, \qquad (6.1.1.)$$

where $v^i = v^i(x, t)$ is the velocity field[38].

Note that the terms "total differentiation" and "total derivative" is mainly used in physics whereas in mathematical modeling and, e.g., fluid dynamics one usually speaks about a convective, material or substantial derivative:

$$\frac{D\mathbf{u}(\mathbf{x}, t)}{Dt} = \frac{\partial \mathbf{u}(\mathbf{x}, t)}{\partial t} + (\mathbf{v}\nabla)\mathbf{u}(\mathbf{x}, t), \qquad (6.1.2.)$$

---

[38] In general, the velocity field $\mathbf{v}(\mathbf{x}, t)$ is connected to the medium density $\rho(\mathbf{x}, t)$ by an integral relationship, e.g. in linear case, $v^i(\mathbf{x}, t) = \int_{(M)} K^i(x, x')\rho(x', t)d^3x'$.

where $\nabla \mathbf{u}$ denotes the covariant derivative[39] of vector $\mathbf{u}$. For a scalar quantity $\varphi$ this expression takes a simpler form:

$$\frac{D\varphi(\mathbf{x},t)}{Dt} = \frac{\partial \varphi(\mathbf{x},t)}{\partial t} + \mathbf{v}\nabla\varphi(\mathbf{x},t). \qquad (6.1.3.)$$

Such derivatives describe respectively the transport of a vector $\mathbf{u}(\mathbf{x},t)$ or a scalar $\varphi(\mathbf{x},t)$ function in a medium characterized by velocity field $\mathbf{v}(\mathbf{x},t)$ i.e., quantities $\mathbf{u}(\mathbf{x},t)$ and $\varphi(\mathbf{x},t)$ are dragged by the medium with velocity $\mathbf{v}(\mathbf{x},t)$. The convective derivative measures the rate of change of a quantity $\mathbf{u}$ or $\varphi$ at a fixed spatial point $\mathbf{x}$ and is taken along the path line moving with velocity $\mathbf{v}$ i.e., following the medium (e.g., fluid) flow. As such it is related to the Lie derivative $L_{\mathbf{v}}$ (which we shall encounter when discussing the flows defined by dynamical systems), in the simplest case through the equation $D\mathbf{u}/Dt = \partial_t \mathbf{u} + L_{\mathbf{v}}\mathbf{u} = 0$. The Lie derivative is one of the most important varieties of derivatives (there are many of them) since it defines the rate of change of a generic tensor field $\mathbf{u}$, in particular scalar or vector function $\mathbf{u}(\mathbf{x},t)$ along another vector field $\mathbf{v}(\mathbf{x},t)$. When local coordinates are introduced in $\mathbf{x}$-space, vector field $\mathbf{v}(\mathbf{x},t)$ is defined by a dynamical system

$$\frac{d\mathbf{x}}{dt} = \mathbf{v}(\mathbf{x},t), \qquad \mathbf{x} = \{x^i, \dots, x^n\}, \qquad \mathbf{v} = \{v^i, \dots, v^n\}. \qquad (6.1.4.)$$

The notion of a vector field simply means that at each point $x$ of manifold $M$ a vector characterizing the direction of changes is defined. For example, the motion that is directed somewhere is the change of state in this direction. Vector $\mathbf{v}(\mathbf{x})$ is a tangent vector. A set of all tangent vectors at point $\mathbf{x} \in M$ is known as a tangent space $T_{\mathbf{x}}M$. The union of all tangent spaces $TM := \bigcup_{\mathbf{x} \in M} T_{\mathbf{x}}M$ is called the tangent bundle. A vector field $\mathbf{v}(\mathbf{x})$ defined in general on a manifold $M$, $\mathbf{x} \in M$ is understood as a map from $M$ into the tangent bundle $TM$ i.e., a map assigning to each point $\mathbf{x} \in M$ a tangent vector $\mathbf{v}(\mathbf{x}) \in T_{\mathbf{x}}M$. In the mathematical language, a vector field is said to be a section of a tangent bundle $TM$ with base $M$.

Any vector field $\mathbf{v} = \mathbf{v}(\mathbf{x},t)$ can be expanded over basis vectors $\mathbf{e}_i$ (or, more generally, $\partial_i$), just as a constant vector, e.g. $\mathbf{v}(\mathbf{x},t) = v^i(\mathbf{x},t)\mathbf{e}_i$, where $v^i(\mathbf{x},t)$ can be either the Cartesian coordinates in the orthonormal basis $\mathbf{e}_i$, $\mathbf{e}_i\mathbf{e}_j = \delta_{ij}$ or the curvilinear ones. In most cases, a vector field can be identified with some differential evolution equation since the flow of a smooth vector field naturally invokes the concept of evolution[40]. Recall that the components of a vector field (more generally, of a tensor, spinor or a spin-tensor field) change when one changes the coordinate frame. The correspondence rule between the field components related to the frame change such as a rotation, Galilean or Lorentz boost or other transformation between moving frames is usually called the field transformation law.

---

[39] A covariant derivative of a vector $\mathbf{u}$ (more generally, of a tensor field $\mathbf{u}$, in local coordinates $u^{\alpha_1,\dots,\alpha_r}_{\beta_1,\dots,\beta_s}$), along a vector field $\mathbf{v}$ is a generalization of the directional derivative $\nabla_{\mathbf{v}} f(\mathbf{x}) = \mathbf{v}\nabla f(\mathbf{x})$ along a vector $\mathbf{v}$ to the case of a curved space, when the components of metric tensor $g_{ik}$ (a covariant (0,2) tensor field, $g_{ik} = g(\partial_i, \partial_k)$, generalizing inner product $\mathbf{e}_i \cdot \mathbf{e}_k$) in local coordinates are not constant as in Euclidean space, but become functions of point $x$. In this case, for example, derivatives $\partial_i u^k$ and $\partial_i u_k$ are no longer tensors (and differentials $du^k, du_k$ are no longer vectors) in the sense of their transformation properties, due to the change of the coordinate system (basis) from point to point, and vectors are transformed differently in different points. The notion of a covariant derivative is of extreme importance in theories with $g_{ik} = g_{ik}(x)$ such as general relativity, where $x$ is a spacetime point. Covariant derivatives are indispensable in electromagnetism, where $\nabla_i \equiv \partial_i + i(e/c)A_i$, $\text{Im}\, A_i = 0$ or $p_i \rightarrow p_i - (e/c)A_i$ are the covariant derivative replacing the partial derivatives. Covariant differentiation of complex vector fields has given rise to modern gauge theories (such as the Standard Model of unifying electromagnetic and weak interactions).
©Sergej Pankratow

[40] There may be also difference equations and iterated maps.

One can feel the relationship between the convectional derivative in continuous models and the Lie derivative in dynamical systems already on the level of characteristics equations, which for the balance equation of the form $\partial_t u^k + v^i \partial_i u^k = g^k$ are

$$\frac{dx^i}{dt} = v^i, \qquad i = 1, \dots, n, \qquad \frac{du^k}{dt} = g^k. \qquad (6.1.5)$$

A more accurate statement would be that the integration of a quasilinear equation $L_v u = g$ or, in local coordinates, $v^i(x, u)\partial_i u(x, t) = g(x, u)$ may be reduced to finding the first integrals of the above characteristics system (here for simplicity an autonomous case is considered). Recall that a partial differential equation is called quasilinear if it is linear in partial derivatives of an unknown function. The relationship between the theory of dynamical systems and the balance conservation laws expressed through the system of quasilinear equations is given by the following important theorem: function $\mathbf{u}(\mathbf{x}, t)$ will be the first integral of (6.1.1.) if and only if[41] it satisfies the equation

$$\partial_t u(\mathbf{x}, t) + v^i(\mathbf{x}, t)\partial_i u(\mathbf{x}, t) = 0 \qquad (6.1.6.)$$

or, in the autonomous case $\dot{\mathbf{x}} = \mathbf{v}(\mathbf{x})$, $v^i(\mathbf{x})\partial_i u(\mathbf{x}) = 0$. Theorem (6.1.6.) is nearly evident from the fact that if $u(\mathbf{x}, t)$ is some function and $\mathbf{x}(t)$ is a solution to dynamical system (6.1.5.), then $u$ will be a function of a single variable $t$ along the integral curve $\boldsymbol{\gamma}$, $u(\mathbf{x}, t) := w(t)$, and if (6.1.6.) holds, then $w(t)$ does not depend on $t$. Obviously, the reverse is also true. Thus, one can take the relation $L_v u = 0$ as a definition of the first integral of an autonomous dynamical system $\dot{\mathbf{x}} = \mathbf{v}(\mathbf{x})$, although the first integral is usually defined as function $u(\mathbf{x})$ that is constant along each trajectory of the system. Statement (6.1.6.) is almost trivial, yet it is very useful since ordinary differential equations for characteristics (of type (6.1.5.)) are usually much easier to solve than partial differential equations (6.1.6.). Mathematically, this theorem hints at the connection between the continuous media models for the system with infinite number of freedoms and ordinary differential motion equations for the systems with finite-dimensional phase space. One can observe, in passing, from this rather simple reasoning that the concept of the Lie derivative is closely connected with the theory of dynamical systems and their first integrals (to be discussed below).

In general, the method of characteristics is the main technique of solving hyperbolic systems of nonlinear equations. The physical meaning of characteristics, which represent the family of phase trajectories of the associated dynamical system, is that the dynamics of a continuous medium consisting of particles can be represented both through the field PDEs and by characteristics ODEs describing the motion of individual particles having a variety of initial conditions. Formula (6.1.6.) has the form of balance conservation laws and can be viewed as establishing the connection between them and the first integrals of dynamical systems. Condition (6.1.6.) has a very simple geometric interpretation: vectors $\mathbf{v}(\mathbf{x})$ form a tangent vector field to a hypersurface (more generally a manifold) $M_c$ defined by the condition $u(\mathbf{x}) = c = \text{const}$. Therefore, the phase curve passing through point $\mathbf{x}_0 \in M_c$ lies in $M_c$: $u(\mathbf{x}) = c$ i.e., $u(\mathbf{x})$ is constant along each phase curve of a system (the first integral or integral of motion). Recall that in general a parameterized curve in $\mathbb{R}^n$ is understood as a smooth function $\boldsymbol{\gamma}(t):[a,b] \to \mathbb{R}$, where $|\boldsymbol{\gamma}'(t)| = |\dot{\boldsymbol{\gamma}}| \equiv |d\boldsymbol{\gamma}/dt| \neq 0, t \in [a,b]$, in physical applications usually $n = 1,2$. A phase curve (phase path, phase trajectory) is usually understood as an image of a map $\boldsymbol{\gamma}(t):I \to P$ from time interval $t \in [a,b] \equiv I \subseteq \mathbb{R}$ to phase space $P$ so that $\dot{\boldsymbol{\gamma}}(t) = \mathbf{v}(\boldsymbol{\gamma}(t))$ for all $t \in I$. Note that any timelike curve in configuration manifold $Q$ or in Minkowski space $M^4$ having the topology of $\mathbb{R}$ i.e., the Euclidean line can serve as a time axis. The graph of mapping $\boldsymbol{\gamma}(t)$ i.e., the direct product $I \times P$ is called the integral curve. So, terminologically, $\boldsymbol{\gamma}(x)$ is a phase curve whereas $\boldsymbol{\gamma}(\mathbf{x}(t),t) \equiv \boldsymbol{\gamma}(t)$ is an integral curve. One can emphasize that some authors distinguish between the integral curve and trajectory: the former is associated with the dynamical system where parameter $t$ has been eliminated whereas the trajectory contains information about both the velocity direction and magnitude. Thus, according to this terminology some information is lost in the integral curve: all that remains is the slope or the direction field (any differential equation of the $\dot{x}(t) = \mathbf{v}(\mathbf{x},t)$ type defines a direction field: a curve passing through point $(\mathbf{x},t)$ has slope $\mathbf{v}$).

---

[41] The statement "if and only if" is often abbreviated as "iff".

More technically, an integral curve $\boldsymbol{\gamma}(t):I \to M$ in the vector field $\mathbf{v}(\mathbf{x},t)$ is a differential set $t \mapsto \boldsymbol{\gamma}(t)$ whose derivatives at each point $\mathbf{x}$ are given for any time $t$ by vector field $\mathbf{v}$ i.e., $\dot{\boldsymbol{\gamma}}(t) = \mathbf{v}(\mathbf{x},t) = d\mathbf{x}(t)/dt|_{\boldsymbol{\gamma}}$. We shall understand integral curve $\boldsymbol{\gamma}:I \to M$ (e.g., $M$ is a phase space $P$) as a smooth curve defined on interval $t \in I$ and lying in manifold $M$. If $\dot{\boldsymbol{\gamma}} \equiv d\boldsymbol{\gamma}/dt = g_t\boldsymbol{\gamma}$ ($I \to TM$ is smooth; $g_t$ is the flow generated by dynamics, see below), then $\dot{\boldsymbol{\gamma}} = \mathbf{v}(\boldsymbol{\gamma}(t)), t \in I$ is a vector field along $\boldsymbol{\gamma}$. One also uses the term "worldline": a worldline is a graph of a particle's position in the coordinate space as a function of time, thus the worldline specifies the particle motion more completely than the trajectory in coordinate space. A point on a worldline is called an event. This terminology is especially popular in relativistic models. Note that in many applications, it might be better not to think of $t$ as corresponding to time: it is simply regarded as an abstract parameter varying smoothly and monotonously along the path.[42]

Another way to express the same idea of conservation would be to say that for an arbitrary function $f(x,t)$, differentiable over its arguments, the expression $A\partial_t f + B\partial_x f$ with any continuous in the vicinity of some point $(x_0,t_0)$ $A$ and $B$, which are not simultaneously zero, should be proportional to the derivative of $f(x,t)$ on some smooth curve $\boldsymbol{\gamma}(\tau)$ passing through $(x_0,t_0)$ and parameterized by $\tau$. Indeed, if curve $\boldsymbol{\gamma}$ is defined by parametric equations $\boldsymbol{\gamma}(\tau): x = x(\tau), t = t(\tau), x(\tau_0) = x_0, t(\tau_0) = t_0$, then $f(x,t)$ along this curve would be the function of a single variable $\tau$, $f(x(\tau),t(\tau)) := w(\tau)$. If we now require the expression for a vector field $A\partial_t f + B\partial_x f$ to be proportional to $w'(\tau)$ regardless of function $w(\tau)$ i.e. $w'(\tau) = k(A\partial_t f + B\partial_x f)$, we shall have $dt/d\tau = kA$, $dx/d\tau = kB$ or the characteristics equations $dt/A = dx/B$ defining the direction of curve $\boldsymbol{\gamma}$ in $(x_0,t_0)$. The proportionality coefficient $k$ identifies a parameterization: thus for $k = 1$ the expression $A\partial_t f + B\partial_x f$ is reduced to the directional derivative of $f(x,t)$ along $\boldsymbol{\gamma}$ whereas for $k = (A^2 + B^2)^{-1/2}$ we shall have the natural parameterization of curve $\boldsymbol{\gamma}$, when parameter $\tau$ coincides with the arc length on the curve. Conservation of function $f(x,t)$ means that its derivative along $\boldsymbol{\gamma}$ is exactly zero. In the case of a multidimensional vector field $\mathbf{v}$, this derivative should be replaced, as already indicated, by a Lie derivative in the direction of the vector field $L_{\mathbf{v}}f = 0$.

---

[42] Examples of such applications are encountered quite often in general relativity.

In simple terms a directional derivative can be interpreted as a vector form of the classical scalar derivative. Let $p \in \mathbb{R}$ be a point, $\mathbf{v} \in \mathbb{R}^n$ be a tangent vector at $p$ and $g: \mathbb{R}^n \to \mathbb{R}$ a differentiable scalar function. Then the directional derivative of $g$ at $p$ in the direction of $\mathbf{v}$ is a number $D_{\mathbf{v}}g$

Writing the dynamical system equations in the form $d\mathbf{x} - \mathbf{v}dt = 0$ naturally leads to the study of the so-called Pfaffian forms, $\theta = P_i(x^j)dx^i$. In 2d, one has $\theta = P(x,y)dx + Q(x,y)dy$ which is an important expression for physical, engineering and economical modeling. In 3d, we have $\theta = P(x,y,z)dx + Q(x,y,z)dy + R(x,y,z)$, and equation $\theta = 0$ manifests the orthogonality of vector field $\mathbf{v}(\mathbf{r}) = \{P(x,y,z), Q(x,y,z), R(x,y,z)\}$ and the trajectories $\boldsymbol{\gamma}(t)$, in particular, integral trajectories. Thus, we see that Pfaffian forms are closely connected with the integrability of differential equations.

In physics-based modeling, the balance conservation laws are widely used in fluid dynamics and nonequilibrium statistical mechanics to describe the situations when the time rate of an extensive (i.e., proportional to the number of particles) quantity, which is stored in a closed domain, is determined by the flux of this quantity across the boundary and by its internal production or extinction. The hyperbolic regime of the balance laws describing dynamic phenomena can be represented in two forms: differential and integral, respectively the equations expressing the balance laws may be differential and integral (or integro-differential as, e.g., in nuclear reactor physics and radiation transfer models). The term "hyperbolic" in this context means that all eigenvalues of the corresponding Jacobian matrix $A_j^i \coloneqq \frac{\partial f^i}{\partial u^j}$ have nonzero real parts. Hyperbolic conservation laws generally play a central part in mathematical modeling performed through the use of nonlinear or quasilinear PDEs. In the differential form, balance laws look like time-dependent (hyperbolic) vector partial differential equations of the kind

$$\partial_t u^j(\mathbf{x}, t) + \partial_i f^{ij}(\mathbf{u}(\mathbf{x}, t), \mathbf{x}, t) = 0, \qquad x = x^1, \dots, x^n, \qquad (6.1.7.)$$

where $\mathbf{u}(\mathbf{x}, t): \mathbb{R}^n \times \mathbb{R} \to \mathbb{R}^m$ is an $m$-dimensional vector whose components $u^j, j = 1, \dots, m$ describe individual densities of the conserved quantities. A more general system would of course be the one which is not resolved with respect to derivatives i.e., $\Phi_j(\mathbf{x}, t, \mathbf{u}, \partial_i \mathbf{u}, \partial_t \mathbf{u}) = 0$. One should pay attention to the fact that vectors $\mathbf{u}$ and $\mathbf{x}$ belong to different vector spaces so that in general $m \neq n$. When $m = n$, the system is usually called definite. Quantity $\mathbf{u}$ is a vector local parameter, its components $u^j$ are often interpreted as the densities of the $j$-th state variable; thus in the one-dimensional (for simplicity) domain $[a, b]$, quantity $P^j(t; a, b) \coloneqq \int_a^b u^j(x, t)\, dx$ is the total quantity of $u^j$ contained in the compact $[a, b]$ at time $t$. Equation (6.1.7.) may be understood in the usual intuitive way: variations of a quantity $\mathbf{u}$ during the unit time interval are balanced by the flux of this quantity across the boundary surface (and possibly by the operation of sources). For example, the continuity equation $\partial_t \rho(\mathbf{x}, t) + \partial_i (\rho \mathbf{v})^i = 0$ that is a scalar equation manifesting the conservation of mass in fluid dynamics or of electric charge in electrodynamics can be interpreted as the statement that mass or charge inside a finite domain varies only if a certain amount of this quantity is gained or lost through the domain boundary. Problems where the flux function $f(\mathbf{u}, \mathbf{x}, t)$ (in the more general case, it is a tensor field, typically in physical applications of (0,2) type $f^{ij}(\mathbf{u}(\mathbf{x}, t), \mathbf{x}, t)$ ), varies in spacetime $(\mathbf{x}, t)$ and also depends on state variable $\mathbf{u}$ (the vector of conserved quantities) arise quite often both in physics and mathematical modeling. For example, nonstationary fluid flows through inhomogeneous porous media, problems of filtration and chromatography, traffic flows on the roads with spreading vehicle density variations, sound, seismic and shock wave propagation are described by the balance conservation laws (6.1.7.). Balance laws of the type

$$\partial_t u^j(\mathbf{x}, t) + \partial_i f^{ij}(\mathbf{u}, \mathbf{x}, t) = g^j(\mathbf{u}, \mathbf{x}, t) \tag{6.1.8.}$$

contain the source term that modifies the flux function $\mathbf{f}(\mathbf{u}, \mathbf{x}, t)$. Thus, the balance conservation laws prescribe that in equilibrium ($\partial_t = 0$) source and flux counterbalance one another. Note that $\partial_x f(u, x, t) = \partial_u f(u, x, t) \partial_x u(x, t) + \partial_x f(u, x, t)$ (here we took the scalar derivative for illustrative purposes).

It is clear from (6.1.7.) that the balance conservation laws may be interpreted as evolution equations. These are in fact first-order systems of quasilinear PDEs that can be written in the divergence form $A^i \partial_i u^j = g^j$, where $A^i$ are matrices whose entries depend on spacetime variables $x \equiv (t, x) := (x^0, \ldots, x^n)$ and on vector-valued unknown function $u = u(x)$ – the state variable (here, to avoid clumsiness, we temporarily do not denote vectors by bold-faced characters). A more general divergence form of a system of balance laws is $\partial_i F^{ij}(x, u) = G^j(x, u)$, where $i = 0, \ldots, n$, $j = 1, \ldots, m$, which can be written in a simplified form as $u_t + F[u] = g$, $F[u]$ being a symmetric operator. For example, $F[u] = A^\mu D_\mu u$, $D_\mu := \partial_1^{\mu_1} \partial_2^{\mu_2} \ldots \partial_p^{\mu_p}$, $\mu_1 + \cdots + \mu_p = \mu$, $\mu = 1, \ldots, p$, $p \geq 1$, $\partial_i := \partial / \partial x^i$, $A^\mu$ are $m \times m$ matrices. One of the usual simplifications to be employed when treating the balance conservation laws as a modeling tool is the assumption that the flux tensor $F^{ij}(x, u)$ does not depend on the independent variable $x$. This simplification reminds us of the transition from non-autonomous to autonomous systems in dynamical systems theory.

In physics and mathematical modeling, one usually studies a linear or quasilinear version of such a system i.e., $A u_t + B u_x = G$, where $A$ and $B$ are $m \times m$ square matrices and the source vector $G$ depends both on $x$ and $u$. When matrix $A$ is non-degenerate i.e., $\det A \neq 0$ (which corresponds to $\det A^0 \neq 0$, e.g., $A^0$ in $A^i \partial_i u^j = g^j$ is positive-definite), the quasilinear system of the balance conservation laws can be reduced to the so-called normal form

$$u_t + B(x, t, u) u_x = G(x, t, u) \tag{6.1.9.}$$

which is most frequently used for modeling purposes. As we have already mentioned, one often assumes that vector equation (6.1.9.) (which is equivalent to a system of $m$ differential equations) or a more general system $u_t + F[u] = g$ is hyperbolic. This assumption means that the Jacobian matrix $A_j^i := \frac{\partial F^i}{\partial u^j}$, $i, j = 1, \ldots, m$ has $m$ real eigenvalues and can be diagonalized for any value of $\mathbf{u}(\mathbf{x}, t)$ so that it possesses $m$ linearly independent eigenvectors making up a complete system. In particular, matrix $B(x, u)$ in (6.1.9.) can be diagonalized and all its eigenvalues are real. The property of hyperbolicity is rather important from the practical standpoint: it ensures a specific form of solutions (e.g., $u(x, t) = \varphi(x - ct)$) and thus helps to find them. We shall illustrate the assumption of hyperbolicity a bit later on low-dimensional balance law examples. Now, just to get some feeling of this property, let us consider the one-dimensional wave equation $\partial_{tt} u(x, t) = \partial_{xx} u(x, t)$ which is known to be hyperbolic as a second order PDE. With the help of a substitution $\partial_t u = u^1$, $c \partial_x u = -u^2$, the wave equation can be reduced to the hyperbolic system

$$\partial_t u^1 + c \partial_x u^2 = 0, \ \partial_t u^2 + c \partial_x u^1 = 0 \tag{6.1.10.}$$

or $\partial_t \mathbf{u} + c \sigma^1 \partial_x \mathbf{u} = 0$, where $\mathbf{u} = (u^1, u^2)^T$, $\sigma^1 = \begin{pmatrix} 0 & 1 \\ 1 & 0 \end{pmatrix}$ is the first Pauli matrix. In general, hyperbolic conservation laws are associated with the propagation of disturbances which fact leads to the possibility of reducing the expressions for conservation laws to the field and wave equations. Accordingly, the solutions of the corresponding evolution equations may be represented as waves (linear and nonlinear). Nonlinearity in the balance equations can produce jump discontinuities (such as shock waves) or non-unique profiles as in wave-breaking phenomena in continuum mechanics. Moreover, solutions of nonlinear equations, in contrast with the linear case, are characterized by an unlimited growth of the solution gradients after some time $t_c > t_0$ ($t_0$ is the initial moment in the Cauchy problem), even for totally smooth starting data. In other words, for $t_c > t_0$ the first derivatives of the desired solution become infinite i.e., the solution to a Cauchy problem no longer exists. One sometimes calls this phenomenon the "gradient catastrophe". From the physical or modeling viewpoint, the feature of unbounded growth of the solution gradients corresponds to the shock wave formation out of the pressure wave or, e.g., to traffic jams on the roads.

One can of course treat the balance conservation laws on a rather high level of mathematical abstraction, but an excessively rigorous discourse would hardly be pertinent for our purposes to present scientific facts in their most primitive form, in particular through mathematical and computer modeling. It is expedient to consider some simple examples allowing one to understand how balance laws work in practically important situations. As one such example, let us consider the balance equations (6.1.7.) in a 2d plane $(x^1, x^2)$:

$$\partial_t u^j(x^1, x^2, t) + \partial_1 f^{j1}(\mathbf{u}(x^1, x^2, t), x^1, x^2, t) + \partial_2 f^{j2}(\mathbf{u}(x^1, x^2, t), x^1, x^2, t) = 0 \qquad (6.1.11.)$$

where $\partial_i := \partial/\partial x^i$, $i = 1,2$; $j = 1, \dots, m$. For simplicity, we shall consider the case when the two-component flux function $\mathbf{f} = \{f^1, f^2\}$, $f^i: \mathbb{R}^m \to \mathbb{R}^m$, $i = 1,2$, only depends explicitly on the state variable vector $\mathbf{u}(\mathbf{x}, t): \mathbb{R}^2 \times \mathbb{R} \to \mathbb{R}^m$ i.e. does not depend on spacetime point $(\mathbf{x}, t)$. If the flux function (which is in general a tensor field) is nonlinear in $\mathbf{u}$, then it is mostly impossible to find an analytic solution to a system of equations of the (6.1.10.) type, and one needs to resort to numerical techniques. One may note here in passing that ordinary numerical methods such as, e.g., finite differences should be applied to (6.1.10.) with care because of some specific difficulties, for example the existence of discontinuous solutions (shock waves).

## 6.2. Fluids as many-body systems

The study of fluids is vital for science and life: it would be enough to recall that there are two fluids that are indispensable for life: air and water. Most mammals, in particular humans, cannot survive for more than a few minutes without air intake. Water determines equilibrium and stability of life processes on biological, ecological, economic, military and cultural levels. Furthermore, oceans seem to determine Earth's climate more than other forcings. Fluid dynamics is also essential for relativistic physics, in particular, to build astrophysical and cosmological models: of stellar cores, shells and atmospheres, of star instabilities, hydrodynamic models of large-scale structures in the universe, etc.

It is instructive to note that from the standpoint of mechanical modeling, there is no crucial difference between gases and liquids: both phases are described by the models of motion implying that the state of rest cannot be achieved unless tangent components of the stress tensor $T_{ik}$ are strictly zero. In mathematical writing, it means that $\mathbf{T_v} = -p\mathbf{v}$ or, more generally, $(\mathbf{T_v})_i = T_{ik} v^k$, where $p$ is pressure, $\mathbf{T_v}$ is stress vector and $\mathbf{v}$ is the unit normal vector to the surface occupied by a fluid. Thus, in mechanics the term "fluid" embraces both liquid and gaseous states. In physical models, however, the terms gas and liquid designate completely different objects. In liquids, the radius $a$ of intermolecular forces is of the order or even larger than the average distance $d \sim n^{-1/3}$ between molecules, $d \delta a$, where $n = N/V$ is the average particle density, $N$ is the total number of molecules in the system having volume $V$. Thus, the sphere of radius $a$ around any molecule in liquid can contain many other molecules, $4\pi n a^3/3 \gg 1$. It would be interesting, by the way, to estimate the fraction of the gas volume occupied by molecules which is $V_0 = 4\pi N a^3/3$. If we relate this to total volume $V$, we get $4\pi n a^3/3$, where $n = N/V$ is the mean gas density. Taking $a\sim10^{-8}$ cm, $N\sim6\cdot10^{23}$ and $V\sim20$ liters i.e., $V\sim2\cdot10^{-2}\mathrm{m}^3$, we have this fraction in gas to be about $4\pi n a^3/3\sim10^{-4}$ which radically differs gas from liquid. For example, in the air with the mean density $\rho\sim10^{-3}-10^{-4}\mathrm{g/cm}^3$ the average intermolecular distance would be $d\sim n^{-1/3} = (V/N)^{1/3}\sim3\cdot10^{-7}\mathrm{cm}$.

In rarefied gases, molecules are far apart so that $d \gg a$. What does it physically mean that the gas is rarefied? Each gas particle is moving freely almost all the time, interacting with other particles only through rare collisions. Thus, we can say that the average distance between gas particles (for definiteness, we shall speak about molecules of the gas, although there may be other constituent particles, e.g., neutrons or macroscopic astrophysical objects – typical distances between astrophysical objects are much greater than their diameters) $d\sim n^{-1/3}$ i.e. the average number of particles in the unit volume (e.g., in $\mathrm{cm}^3$) is much greater than the radius of interparticle interaction, $na^3 \ll 1$. Typically for gases, $a\sim10^{-8}$ cm, $n\sim10^{19}\mathrm{cm}^{-3}$ i.e., $d\sim0.5\cdot10^{-6}$ cm and $d/a\sim10^2$. Notice that one can obtain the value $n\sim10^{19}\mathrm{cm}^{-3}$ from the equation of state $PV = nk_BT$, where $k_B = 1.38\cdot10^{-16}$ erg $\cdot$ K$^{-1}$ is the Boltzmann constant.

As far as astrophysical objects – in the limit the entire universe – are concerned, one can hardly worry about the collisions between stars since the space volume occupied by the stars is much smaller than the inverse stellar density, $N_sR^3 \ll 1$, where $N_s$ is the average density of stars and $R$ is the characteristic star radius.

The radius $a$ of interparticle interaction determines the distance at which the molecular forces affect the motion of other particles so that in most cases $a$ can be identified with the correlation length between the constituent particles. The respective time scale $\tau_0\sim a/\bar{v}$, where $\bar{v}$ is some characteristic velocity which, in the systems not far from equilibrium, is of the order of thermal velocity, $\bar{v}\sim v_T$, may be called the chaotization time: after this time has elapsed correlations between the particles become drastically weakened. In the human time scale the chaotization period is very short – of the order of several picoseconds. Notice that chaotization period $\tau_0$ in gaseous fluids is of the order of a single collision time, as we do not consider here Coulomb systems such as plasmas. Therefore, $\tau_0$ practically depends neither on the interparticle force strength nor on particle density.

In gaseous fluids, one more parameter is important: the free path length $\bar{l} = (n\sigma)^{-1}$, where $n$ is the average density of gas molecules and $\sigma$ is their scattering cross-section. In the language of multiple scattering theory in molecular systems, one usually talks of scattering length $a_0$ comparing it with the mean free path $\bar{l} = (n\sigma)^{-1}$. The meaning of scattering length, if we disregard some quantum scattering technicalities, is that it ignores the details of the molecular structure in the process of intermolecular collisions. In most cases, scattering length is of the same order of magnitude as the radius of interparticle interaction. One can represent the free path length as $\bar{l} = (n\sigma)^{-1}\sim d(d/a)^2$ which makes up in rarefied gases $\bar{l}\sim0.5\cdot10^{-6}\cdot10^4\sim0.5\cdot10^{-2}$ cm. The respective mean free time is $\tau\sim\bar{l}/v_T\sim\bar{l}\sqrt{M/k_BT}\sim0.5\cdot10^{-2}\mathrm{cm}/3\cdot10^4\mathrm{cm/s}\sim10^{-6}-10^{-7}s$. Another way to express the mean free time is $\tau\sim(d/v_T)(d/a)^2$. The corresponding collision frequency is $\nu\sim n\sigma v_T\sim v_T/\bar{l}\sim(v_T/d)(a/d)^2\sim10^7 s^{-1}$. In important mathematical models of reactor physics and radiation protection, free path length as $\bar{l} = (n\sigma)^{-1}$ and the corresponding time scale $\tau\sim\nu^{-1}$mark the isotropization of the flux of ionizing particles such as neutrons.

The mean free path $\bar{l}$ and mean free time $\tau$ are the crucial parameters controlling the transition from the microscopic (kinetic) level of description to the macroscopic one. In particular, these scales determine the possibility of working with average values used in thermodynamics, fluid dynamics and other phenomenological disciplines. By the way, the "infinitesimal" quantities $d\mathbf{r}, dt$ that we use to describe the local fluid motion should satisfy the conditions $|d\mathbf{r}| \gg \bar{l}, dt \gg \tau$. Otherwise, fast microscopic fluctuations in the medium would not allow us to consider it as being in the state of regular motion (although not necessarily reversible). Quantities $\bar{l}$ and $\tau$ give the scale of microscopic relaxation ($\tau \sim \tau_r$) i.e., dying out of fast fluctuations.

Thus, we see that there exists a natural hierarchy of time scales (already mentioned in association with the hierarchical multilevel principle of mathematical modeling), this fact playing a crucial part in statistical mechanics as it determines the behavior of distribution and correlation functions. In particular, after time $\tau_0$, correlations between the particles are drastically weakened and many-particle distribution functions turn into the product of single-particle distributions $f(\mathbf{r}, \mathbf{p}, t)$ (the principle of correlation weakening was introduced by N. N. Bogoliubov) whereas for $t > \tau_r \gg \tau_0$, a single-particle distribution function tends to the equilibrium Maxwell-Boltzmann distribution[43]. In other words, the whole many-body system for $t \gg \tau_r$ reaches the state of statistical equilibrium so that the respective many-particle distribution functions tend to the canonical Gibbs distribution, $f_N(x_1, \ldots, x_N, t) \to \exp[\beta(F - H(x_1, \ldots, x_N))]$, where $\beta = 1/T$ is the inverse temperature, $F$ is the free energy and $H(x_1, \ldots, x_N)$ is the system's Hamiltonian. In other words, the single-particle distribution function $f(\mathbf{r}, \mathbf{p}, t)$ substantially changes over time scales $t\,\tau\,\tau_r \gg \tau_0$ whereas at the initial stage of evolution this function remains practically intact. Yet many-particle distribution functions can change very rapidly at short times comparable with the chaotization period $\tau_0$. Physically, one can understand this fact by considering spatially uniform systems with pair interaction between the particles, when many-body distribution functions depend on the coordinate differences of rapidly moving constituents. It is intuitively plausible that many-particle distribution functions would adjust to instantaneous values of a single-particle distribution. To translate this intuitive consideration into the mathematical language, one can say that for $\tau_r > t \gg \tau_0$ (intermediate asymptotics) many-particle distribution functions become the functionals of a single-particle distribution function

$$f_N(\,x_1, \ldots, \, x_N, t) \overset{t \gg \tau_0}{\longrightarrow} f_N[\,x_1, \ldots, \, x_N; f(\mathbf{r}, \mathbf{p}, t)] \tag{6.2.1.}$$

so that the temporal dependence of many-particle distributions is now determined by the single-particle function.

This idea (expressed by N. N. Bogoliubov) is rather important since it leads to a drastic simplification of the models describing many-body systems. In particular, although the respective distribution functions formally depend on initial data for all the particles, after a rather short time ($\tau_0 \sim 10^{-12} - 10^{-13}$ s) this dependence becomes much simpler since its relics are only retained in the relatively smooth single-particle function $f(\mathbf{r}, \mathbf{p}, t)$. One usually applies to this situation the notion of "erased memory" designating asymptotic independence of many-particle distribution functions on precise values of initial data – a huge simplification since initial values of all coordinates and momenta are never known exactly and, even if known, would be completely useless. In the modern world, the idea of fast forgotten details of microscopic instances, in fact of insensitivity to microscopics, has become especially useful for mathematical modeling of complex systems.

---

[43] One can imagine even tinier time scales in a many-body system, namely $\tau_0 \sim \tau_r/N$, where $N$ is the number of particles in the considered part of the system.

## 6.3. First principles and phenomenological models

Particular attention is devoted today to the relationship between microscopic dynamics and macroscopic behavior of complex systems in which rather unexpected phenomena may occur. We can see, for example, that the seemingly chaotic dynamics of many particles at the microlevel is capable of producing a fully deterministic behavior at the macrolevel. The "whole" typically exhibits features that cannot even be suspected while studying microscopic evolutions of its constituents. For example, one cannot (at least so far) describe the transition from laminar to turbulent fluid motion starting from the microscopic picture of interactions between fluid molecules. One can notice that economic or social systems exhibit behavior that has nothing in common with reactions of individual agents. Experience shows that complex many-body systems display the so-called emergent phenomena that are produced by cooperative, coherent behavior involving a large number of constituent elements. This cooperative behavior and the associated emergent phenomena are described by a drastically reduced set of phenomenological parameters that can be directly measured (e.g., in the simplest case temperature and pressure). Contrariwise, the underlying microscopic variables usually imply no measurable procedures, at least for "normal conditions" (such as at low energies in physics), which has a direct impact on engineering: people are designing complex devices, e.g., automobile or aircraft engines, using only the high-level view i.e., without proper understanding how such devices work on the microscopic level. Reduction of practically important engineering devices to atomic or molecular motion does not make much sense, and, for example, elementary particles are totally irrelevant for modeling of natural catastrophes.

The cardinal difference between microscopic and macroscopic levels of description is the breach of equivalence between past and future. In classical dynamics, there is in fact no genuine evolution since all dynamical paradigms (Newtonian, Lagrangian, Hamiltonian, etc.) are invariant under time inversion, $t \to -t$, and thus past and future are formally equivalent. In quantum theory, the situation is a little more complicated, although in the simplest dynamical formulation based on the Schrödinger or Dirac equations evolving states can be considered time-reversal symmetric (invertible). Thus, in dynamics treated as a microscopic theory, there is no intrinsic mechanism discriminating past and future; states corresponding to $t$ and to $-t$ are formally equivalent. An arbitrarily chosen initial state at $t = t_0$, e.g., a point (or even a domain) in the phase space unfolds in both time directions $t > t_0$ and $t < t_0$ with equal rate. Moreover, in stable and non-dissipative mechanical systems, the initial state (at $t = t_0$) can be reached by inverting the motion from any state (corresponding to $t \neq t_0$) that will be later demonstrated on an example of the harmonic oscillator, see formulas (9.1.1.)-(9.1.3.).

Contrariwise, real world macroscopic processes are never time-reversal invariant: there is always an "arrow of time". The arrow of time is a popular metaphor applied to emphasize a fundamental asymmetry of time directions. Such an asymmetry is intimately connected to causal relations manifested by time ordering of cause and effect (the causality principle). It is extremely interesting and instructive to trace how the arrow of time enters natural sciences and appears in complex systems with their emergent phenomena. In fact, there are several time arrows: in statistical mechanics, chaotic dynamics, quantum mechanics, cosmology, biology, etc., at least some of them pointing in the same temporal direction. Thus, the arrow of entropy is directed along the arrow of time in quantum mechanics or in cosmology, yet the reason for this fact is not fully understood.

In quantum theory, which we shall not consider in this book, there exist some additional reasons for the breach of time-reversal symmetry, one of them being the issue of quantum measurement that is not fully clarified – at least within the unitary scheme of orthodox quantum mechanics. However, this issue is a special problem, and its discussion would require much space so that we shall not deal with the arrow of time in the present manuscript. Notice only that in the dynamical systems theory, which can be considered as an intermediate between mechanics and macroscopic physics, the arrow of time i.e., the breakdown of symmetry between past and future appears quite naturally in view of instabilities and their limit case – chaotic behavior.

In the quantum world, the reason for the breach of time reversal symmetry is often ascribed to the process of quantum measurement, which we shall not consider here in detail. Quantum measurements disrupt continuous (unitary) evolution. Measurement in conventional quantum mechanics is a phenomenological procedure forcing the observed system to jump into a new state, with the information about its former state being completely lost and information about the new state compatible with the measurement, created. This procedure is phenomenological since it is not described by any quantum-mechanical (microscopic) equations.

When one is solely focused on unitary quantum dynamics, one typically treats quantum evolution as completely time invertible. This is, however, a somewhat naive simplifying assumption since the Hilbert space where quantum dynamics occurs is not necessarily filled with complex conjugated states (of thermo-isolated systems). Moreover, dynamics formulated in terms of transition probabilities is not in general invertible i.e., does not necessarily comply with time reversal symmetry.

Strictly speaking, it is only a hypothesis that any phenomenon in complex systems can be reduced to mechanical motion of individual particles interacting through some interparticle potentials. It would not be justified to extrapolate the medieval observations of the motion of celestial bodies (it is such observations that gave the impetus to classical physics) onto the behavior of complex systems such as economies, Earth's climate or living organisms. Persistent attempts to chase out phenomenology as an inferior, descriptive-level approach that should always be reduced to sacred "first principles" is a consequence of the medieval rationality which can be regarded nowadays as an adolescent phase of scientific modeling of the world. This phase tends to be gradually overcome by studying complex systems with inherent emergent phenomena. By the way, classical (Newtonian) and even quantum mechanics are in fact also phenomenological theories.

What is emergence? It is, simply speaking, the ability of a complex system to acquire completely new features. Thus, cognitive (in neurophysiological sense) abilities of human individuals are new and emergent with respect to animal collectives and their cooperative behavior, whereas the latter can hardly be reduced to the rules that we apply to macroscopic physical systems such as the laws of Gibbs' statistical thermodynamics. One can find much more about emergence properties and their connection with such important concepts as entropy and evolution as well as the relevant literature in a comprehensive review article by W. Ebeling and R. Feistel [58].

One of the main characteristics of complex systems is diversity (see section 12.1.). For example, in biological systems there is a diversity of genotypes, cell types, organs, phenotypes, species, populations and further up the hierarchy. In ecological systems, there is a diversity of living organisms and nonliving components such as water resources, soil, fossils, chemical compounds, etc. as well as food and water sources for living creatures. The matter is that many biochemical and physiological processes become less expensive with rising temperature, e.g., enzyme-based catalysis, metabolism rate, muscular functions, locomotion in water and, within certain limits, in air, circulation of body fluids, conduction of nerve signals, etc. Unfortunately, many physical issues of practical ecology are rather poorly understood, mainly because of their complexity and associated with it, diversity.

In socio-economic systems, there is a diversity of civilizations, cultures, nations, languages, styles, fields of knowledge, human professions and divisions of labor, industrial and trade companies, etc. It is a trivial observation that permissive societies, where diversity is easily tolerated, achieve more stable and efficient economies that each in turn support the society's development. Thus, what is known as freedom i.e., absence or mitigation of sanctions and constraints imposed on socio-economic experimentation fosters heterogeneous economic activity just as increased temperatures weakening microscopic interparticle constraints accelerate biological processes.

However, any attempt to attack a complex system accounting for all manifestations of diversity from the "first principle" perspective would probably be moonshine. One can possibly offer a plausible mechanism of how diversity emerges from uniformity, but this endeavor would be a generic framework for world vision rather than modeling a specific class of diversities. Moreover, it is difficult to measure the diversity, especially in biological, social, cultural and economic domains, at least with the accuracy (error margins) acceptable in natural sciences. Therefore, it would be hard to validate the model outcomes against empirical patterns, especially those evolving with time. It would also be difficult to correctly evaluate the factors that favor or depress diversity, thus building a causal mathematical model.

The phenomenological context that the hypothesis of reductionism strives to purge usually dictates its own laws such as the laws of thermodynamics, of population growth, of biological oscillations, of economic equilibrium, etc. It may look sad – well, there are many sad things in life – that there is such a diversity of laws and that they can be reduced to the "first principles" (i.e., to mechanical motion) only in exceptional cases – one of such cases is equilibrium thermodynamics. There are important phenomenological theories that have received their microscopic foundation, although not quite complete and not for all observed phenomena. Examples of such reducible phenomenological theories are those of phase transitions and of superconductivity. In particular, some phenomenological models of superconductivity such as the London equations, Ginzburg-Landau equations can be explained within the framework of the microscopic BCS (Bardeen-Cooper-Schrieffer) theory.

We can identify many real-life observations that might be treated as laws making a foundation for phenomenological models. As an example, we can take the obvious decrease in biological diversity (e.g., counted as an average number of species per unit area) with increasing latitude, counterbalanced by rising economic diversity that reaches its peak in high latitudes. Yet human cultures and languages seem to have a geographic gradient opposite to economic activity. One can probably attribute these gradients to the insulation ($J/m^2$) change and rising mean local temperatures, which would produce a mathematical model based on the phenomenological context. Other variables such as availability of water resources and rainfall can enter the model, e.g., as corrections. A superficial qualitative explanation can be that the natural availability of food, energy, habitat protection (shelter), etc. drops with increasing latitudes and has to be compensated for by human activities (humans are the only species who use artificial tools for survival and massive environment modification). However, it would not be easy to calculate a specific distribution of species or of economic activities starting from the first physical principles. It may well be that phenomenological models cannot be obtained from the first principles at all. One might recall in this connection somewhat naive attempts of the 1950s to attribute the time-irreversible behavior to particle-antiparticle asymmetry: allegedly there would be no death (but also no biological evolution) if there were no particle dominance in the universe.

Referring to the necessity of idealization in modeling natural phenomena can hardly be considered a proper justification for throwing away emergent phenomena that can only be represented throughout time-irreversible models as something secondary that should be reduced to the mechanical motion of fundamental particles. It is a great question in its own right whether such complex systems as living and social objects, evolving economies and the like can sustain such a reduction. Firstly, correlations in complex systems may be quite strong and play the decisive role so that reduction of a complex system to a large set of particles in mechanical motion possibly cannot take these correlations into account. Note that correlations are not just pair or even multi-particle interactions; they incorporate also memory effects. Secondly, the already mentioned instabilities, closely connected with the sensitivity to slight perturbations of the system's parameters, in particular, unknown in practice initial conditions. In general, instabilities imply the system's sensitivity to its own fluctuations (as is often observed in living, social or economic systems).

Here, we are also faced with the cardinal difference between close-to-equilibrium and non-equilibrium systems. In the systems close to equilibrium, fluctuations can be treated as unimportant "noise", they constantly appear and fade away, without bringing any significant changes of state. In physical language, equilibrium states are stable with respect to fluctuations. It means that one can describe systems close to equilibrium by their average values: as already mentioned, this idea served as a conceptual framework for statistical mechanics as founded by J. W. Gibbs. Contrariwise, in systems far from equilibrium one can no longer consider fluctuations to be insignificant noise: they readily become a major factor determining the system's evolution. In other words, the same fluctuation that could be neglected near stable equilibrium can seriously affect the system's behavior far from equilibrium.

In short, first principles help to determine the main direction of studies and they can to some extent guide us through mathematical modeling, but in most real-life cases first principles fail to give the right guess at an advanced modeling level. Then one needs to include phenomenological terms into the analysis (e.g., of a system's dynamics), and the system may become nonlinear. A simple illustration is Hooke's law of elasticity: it becomes nonlinear when one overstretches the string. Moreover, the string gets warmer which points to uncontrollable energy loss. It would be difficult to describe this process without an appropriate phenomenological model.

## 6.4. Open systems

Phenomenological models as discussed above are hardly derived from first principles, but usually based on observations. In the macroworld, that could be observations over natural or biological phenomena such as weather, population or animal behavior, the ones that constantly exchange energy and entropy. Therefore, phenomenological models are often viewed within the context of open systems.

Open systems are those that interact with the environment. Such systems are still poorly understood, often inadequately described and can be dangerously unstable, as we can see on the examples of living organisms and, in general, biological objects. When discussing open systems, one ought to consider not only the evolution of the system itself, but also the feedback provided by the co-evolution of the environment. The system under consideration can be either classical or quantum, whereas the environment is in most cases considered classical.

Strictly speaking this is wrong, since there is no classicality in the quantum world so that considering the environment classical is a flawed way to make life easier. Recall that in the physics of open systems one has to treat the system of interest as a subsystem i.e., as a part of a large closed system. In quantum theory Schrödinger's equation $i\hbar\partial_t\Psi(t) = H\Psi(t)$ (and other evolution equations) can be applied, strictly speaking, only to closed (isolated) physical systems, but one often relaxes this requirement, by using the time-dependent "effective" Hamiltonian instead of the one involving actual interaction with the environment.

An archetypal example of an open system is the thermal engine whose mathematical model is the classical Carnot heat machine discussed in any course of thermodynamics. One usually treats the Carnot heat engine as performing a periodic process ("the Carnot cycle") since the system returns to the state with the same value of internal energy. This does not, however, mean that nothing changes in the surrounding world. For instance, the steam engine burns some amount of coal and exhausts water vapor and carbon dioxide into the atmosphere. It means that the heat engine is an open system, and the Carnot cycle provides an example of a stationary dissipative process. One might remark that the Carnot heat machine is not an impeccable mathematical model: the traditionally employed notions of heater (the hot reservoir) and cooler (the cold reservoir) are abstract and devoid of physical content in the sense that their physical structure remains unclear. Note that when studying classical thermodynamics one can easily get confused with a perplexing number of differential relationships. We usually get accustomed to numerous formulas as well as to rather abstract notions of thermodynamics, without proper understanding of the underlying physical processes (such as heat fluxes needed to maintain the constant temperature of the heater and the cooler).

Other physical questions pertaining to the concept of open systems are left unanswered in the Carnot heat machine model, for instance, one might ask: which of the two heat reservoirs corresponds to a more ordered state – hot or cold? Another question is: what will be the hot and cold reservoirs when the energy of a chemical mixture is directly converted into other types of energy such as electrical (as in accumulators) or optical as in chemical lasers? The Carnot heat machine model was inspired in the 1820s by the steam engine and was useful to describe its basic functions but is hardly adequate for modeling modern processes of energy conversion.

One of the most important concepts in open systems outline is self-organization. This concept is associated with irreversible phenomena, in particular encountered in biology and social sciences. In physics, self-organization is embodied by the formation of such large-scale structures as stars, galaxies, cosmological patterns, etc. At the more mundane level, such structures as Bénard cells in fluid dynamics i.e., convective currents influenced by gravitation, reaction-diffusion systems, networks, even lasers and crystals exemplify self-organization phenomena. It is interesting that self-organization can advance in opposite directions: towards more ordered or more chaotic states. Chaotic and self-organization phenomena in open systems are related phenomena: for example, in most cities self-organized structures emerge out of chaotic and disordered patterns or the coherent structure of a tornado appears from countless chaotically moving air molecules.

Open systems cannot be conservative since they exchange energy with the environment. It means that the spacetime eigenmodes excited in the open systems have complex frequencies (energy eigenvalues). In the language of the evolution operator $U(t, t_0) = \exp{-[i(t - t_0)H]}$, this fact means that the property of self-adjointness ensuring that the energy eigenvalues are real does not hold. Since one cannot protect an open system from being affected by its environment, its time-reversed evolution is different from the time-forward one. The non-demolition physical experiment on open systems becomes very difficult.

## 6.5. Applications of the balance laws: fluid motion

All matter is composed of molecules that are in constant motion and interact – collide – with one another and with constraints, e.g., boundary objects. Fluids are no exception. Nevertheless, we view fluids as continuous media, which is actually an assumption, a mathematical idealization typical of modeling. The discrete nature of fluids is largely disregarded, with such variables as velocity, pressure, density, temperature, entropy, etc. being assumed continuous functions smoothly varying over the length scales corresponding to the mean free path and to average intermolecular distances. Deviations from the continuum hypothesis in fluid motion are caught by using statistical mechanics.

The formal geometrical description of morphisms (most frequently diffeomorphisms) associated with the motion of fluid substances is known as kinematics, while study of the laws that govern the behavior of fluid elements is called fluid dynamics. The latter mostly deals with phenomenological laws since a truly microscopic approach to fluid motion implies the study of many-particle systems, is excruciatingly difficult and rarely leads to practical results while assessing the flow patterns. The main physical question in relation to fluid motion is: why is the observable fluid behavior not so predictable as the path of a material particle or the non-chaotic motion of a rigid body? We can see this unpredictability by observing the turbulence and the wave phenomena, especially large-amplitude nonlinear waves and dangerous "freak" ocean waves.

Mathematical description of the fluid motion (the latter is typically called flow) is based on the following premises: the material fluid is considered a continuous substance occupying a positive volume ($V > 0$, $V \sim L^3$) and distributed over a domain $\Omega$ of the $3d$ Euclidean space. A fluid flow is represented by a one-parameter family of mappings of domain $\Omega$ filled with fluid into itself, $\Omega_t \rightarrow \Omega_{t+\tau}$ (Figure 3). This mapping is usually considered smooth, but in most cases, it is sufficient to only require it to be twice continuously differentiable in all variables ($C^2$), sometimes such mappings are assumed thrice continuously differentiable everywhere with the exception of some singular points, curves or surfaces, where a special analysis is required. Mapping of domain $\Omega$ into itself is also considered bijective (both one-to-one and onto), so that geometrically it is a diffeomorphism $\Omega_t \rightleftarrows \Omega_{t+\tau}$ i.e., invertible and differentiable transformation. The domain $\Omega$ assumes a differential structure resulting in a $C^k$-manifold ($k = 2, 3, ...$). As it is usual in mechanics, the real parameter $t$ designating the mapping of fluid domains is identified with time, $-\infty < t < +\infty$, and without the loss of generality point $t = 0$ is considered an arbitrary initial instant. The following primary steps are usually taken both in analytical and numerical modeling of the fluid motion: a fixed rectangular coordinate system $\{x^i\}$ is chosen, with each triple $\{x^i\}$, $i = 1, 2, 3$ denoting the $3d$ position of a fluid particle which is treated as an infinitesimal element of the fluid distribution over domain $\Omega$.

The motion of an ideal (i.e., inviscid and not heat conducting) fluid at a constant temperature in two spatial dimensions can be described by the balance laws which are often written in the form:

$$u_t + F_{x^1}(x, u) + G_{x^2}(x, u) = 0. \tag{6.5.1.}$$

Here vectors $u = u(x, t) = (\rho, \rho u^1, \rho u^2, w)^T$, $F(x, u) = (\rho u^1, p + \rho u^1 u^1, \rho u^1 u^2, \rho u^1 w + p u^1)^T$, $G(x, u) = (\rho u^2, \rho u^1 u^2, p + \rho u^2 u^2, \rho u^2 w + p u^2)^T$, $x = (x^1, x^2)$, accordingly $u^1, u^2$ are the fluid velocity components[44] along $x^1, x^2$; $\rho = \rho(x, t)$ is the fluid density, $p$ is the pressure, $w = \mathcal{E} + ((u^1)^2 + (u^2)^2)/2$ is the total energy of the unit mass (which is in many cases equal to the so-called heat function or enthalpy, $\mathcal{E} + pV = \mathcal{E} + p/\rho$), $\mathcal{E}$ is the total internal energy of the unit mass of the fluid (see below on thermodynamic potentials in the section devoted to statistical mechanics and thermodynamics). Here we restricted ourselves to the 2d case only for the writing transparency, and later, for simplicity of modeling examples, we shall consider in most cases only a single spatial dimension. To make the system (6.5.1.) closed, it is usually supplemented by the equation of state, e.g., $p = \rho \mathcal{E} (\gamma - 1)$ of the perfect gas type ($\gamma = c_p / c_v$ i.e., the ratio of specific heats of the fluid which is considered constant; typically, $\gamma > 1$). The term "gas" here and below denotes the ensemble of a great number ($\sim 10^{24}$) of particles.

---

[44] One should not confuse upper vector indices denoting contravariant components with the power symbols. We write, e.g., $u^1 u^1$ instead of $(u^1)^2$ in order to pay attention to this notorious notational difficulty.

The balance law form (6.1.6.- 6.1.7.) is mainly used in mathematical modeling and computational fluid dynamics, as these expressions are more convenient for numerical treatment than the vector form of the balance equations commonly used in physics. The motion of an ideal fluid is described in physics by the following system:

$\partial_t \rho + \partial_i j^i = 0$, $\mathbf{j} = \rho \mathbf{u}$ – the continuity equation (that can also be written as $\frac{d\rho}{dt} + \rho \partial_i u^i = 0, \rho = \rho(\mathbf{r}, t), \mathbf{u} = \mathbf{u}(\mathbf{r}, t)$: Euler's description)

$\partial_t \mathbf{u} + (\mathbf{u}\nabla)\mathbf{u} = -\frac{\nabla p}{\rho}$, (more correctly, $\partial_t u^i + u^k \partial_k u^i = -g^{ik} \partial_k p / \rho$, where $g^{ik}$ is the inverse metric tensor) – the Euler equation,

$\frac{dS}{dt} = \partial_t S + (\mathbf{u}\nabla)S = 0$ – the adiabaticity condition (isoentropic flow).

Here $S = S(p, \rho)$ is the specific entropy (i.e., entropy per unit mass). The Euler equation, describing an inviscid flow, manifests the momentum conservation law and can thus be written in a divergent form $\partial_t \rho u^k = -\partial_i \Pi^{ik}$, where $\Pi^{ik} = p g^{ik} + \rho u^i u^k$ is the momentum density tensor. The energy conservation law may be written in the following way

$$\partial_t \left(\rho \mathcal{E} + \frac{\rho \mathbf{u}^2}{2}\right) = -\partial_i q^i, \qquad (6.5.2.)$$

where $q^i = \varrho u^i (\frac{\mathbf{u}^2}{2} + w)$ is the energy flux vector. The natural question here will be: assume that the density $\rho$ (understood as a local parameter, $\rho = \rho(x^i, t)$, see also below) of each individual fluid particle of an incompressible medium is constant, can this density vary with time at some point $x^i$ of the fluid domain $\Omega$? The answer is, of course, yes, as $\partial_t \rho = -\text{div}\rho \mathbf{u}$ which is in general non-zero.

One can model rather complex situations using the balance laws and the associated hyperbolic equations. For example, one can write the system of equations governing the coupled fluid dynamics and chemical kinetic processes as

$$\frac{\partial \mathbf{F}}{\partial t} + A\mathbf{F} = \mathbf{g}, \qquad \mathbf{F} = (\rho, \rho v_i, \rho \mathbf{u}, w)^T, \qquad \mathbf{g} = (0, \alpha_i, 0, 0, 0)^T, \qquad (6.5.3.)$$

where $A$ is some spatially dependent nonlinear operator, $\rho$ is the flow density, $\mathbf{u}$ is the velocity vector, $v_i$ is the mass fraction of the $i$-th species, $w$ is some macroscopic energy characteristic such as specific energy, temperature or enthalpy related to pressure $p$ through the equation of state. This system, under the assumption of chemical reactions running independently of the flow of substances, can be split into two blocks: fluid dynamics ($4d, v_i = 0$) and chemical kinetics (ODE, $v_i \neq 0$). Both initially independent blocks can be processed numerically, e.g., via finite-difference schemes using parallel computations; at the subsequent stages coupling of the two processes is explicitly accounted for, in particular, by calculating the temperature and mass fraction changes in the spacetime domain, see, e.g., [71].

The primary focus areas in the mathematical modeling of real (viscous) fluids are the Euler and the Lagrange descriptions of fluid flows, the continuity equation and the Navier-Stokes equation. The coordinate frames corresponding to these two manners of description are known as Euler's (or Eulerian) coordinates. Euler's coordinates are fixed in space whereas the Lagrangian coordinates follow the path of a fluid particle. Euler's coordinates are therefore also called spatial, while Lagrangian coordinates are called material (Figure 4).

The idea of the Navier-Stokes equations, the most general set of equations describing fluid motion, is to represent the motion of fluid substances resulting from the balance of forces acting on any volume of fluid (or stresses occurring in any fluid domain).

Due to diffeomorphism $\Omega_t \rightleftarrows \Omega_{t+\tau}$, the fluid motion is completely determined either by transformation $\Omega_0 \to \Omega_t$ i.e. $x^i = x^i(\xi^j, t)$ (Euler's picture) or by its inverse $\Omega_t \to \Omega_0$ i.e. $\xi^j = \xi^j(x^i, t)$ (Lagrangian picture). The crucial technical questions arising in both descriptions are respectively: how to correctly describe the state of motion at a given position in the course of time and, alternatively, how to characterize the flow in terms of moving mathematical objects defined at $\{x^i, t\}$? The simplest object to perform the transition from one description to another is the instantaneous fluid velocity $u^i$ since the $u$-functions can be defined in both descriptions, $u^i(x^j, t)$ and $u^i(\xi^j, t)$. From the physics viewpoint, in the Lagrangian description trajectories of medium particles are scrutinized, while in the Euler's picture, density $\varrho(\mathbf{r}, t)$ and velocity $\mathbf{u}(\mathbf{r}, t)$ play the role of dynamical variables.

One can observe that the fluid particle that initially was in position $\xi^j$ has moved to position $x^i$ that corresponds to the fluid transformation $x^i = x^i(\xi^j, t)$. If $\xi^j$ remains fixed while parameter $t$ varies, then such a transformation specifies the path of a fluid particle (a pathline or a particle line): $x^i = x^i(t) = x^i(\xi^j, t)|_{\xi^j=\text{const}}$, $x^i(0) = \xi^i$. If $t$ is fixed, then this transformation of the fluid domain $\Omega_0 \to \Omega_t$ is exactly the fluid flow, $x^i(0) = \xi^i \mapsto x^i(\xi^i, t)$. Functions $x^i(\xi^j, t)$ are single-valued (one-to-one) and are assumed to be at least twice differentiable in all their variables so that the Jacobian

$$J \equiv A_j^i = \frac{\partial(x^1, x^2, x^3)}{\partial(\xi^1, \xi^2, \xi^3)} \equiv \det\left(\frac{\partial x^i}{\partial \xi^j}\right) \neq 0 \qquad (6.5.4.)$$

and the flow is invertible i.e., there exists a set of inverse functions $\xi^j = \xi^j(x^i, t)$ defining the initial position of a fluid particle that occupies any position $\{x^i\}$ at time $t$. This fact means that a fluid flow can be equivalently described by a set of initial positions of particles as functions of their positions at later times. Notice that the inverse transform of fluid positions has the same properties as the direct one, which is a nontrivial feature of fluid flow models since in distinction to the most part of mechanics, where only time-reversal invariant problems are considered, the motion of fluids is, in general, a dissipative system. Thus, we have an important consequence of invertibility namely two equivalent descriptions of fluid flows:

1. The Eulerian description i.e., in terms of spacetime variables $(x^i, t)$, when a given set of coordinates $\{x^i\}$ remains attached to a fixed position throughout all the time. Spatial coordinates $\{x^i\}$ are called the Euler coordinates; they serve as an identification mark for a fixed point. This is actually a field-theoretical description.

2. The Lagrangian description which chronicles the history of each fluid particle, the latter being uniquely identified by variables $\{\xi^i\}$. Throughout the time, a given set of coordinates $\{\xi^i\}$ remains attached to an individual particle, they are called the Lagrangian coordinates.

These two manners of description[45] imply different physical interpretations: in Euler's picture, an observer is always located at a given position $\{x^i\}$ at time $t$ observing the fluid particles that are passing by, whereas in the Lagrange picture an observer is moving with the fluid particle, which was initially at the position $\{\xi^i\}$, and views changes in the flow co-moving with the observer's own particle. In other words, fluid particles in the Lagrangian framework are distinguished and individually followed throughout their motion, $x^i = x^i(\xi^j, t) = x^i(x_0^j, \dot{x}_0^j, t), \dot{x}_0^j = u_0^j$. Simply speaking, Euler's coordinates are taken with respect to a fixed frame, while the Lagrangian ones are in a comoving frame with the matter in motion. The concepts of the trajectory of each particle and of the pathline are associated with the Lagrangian picture. In the Eulerian framework, the velocity field at any given time $t$ in the whole fluid occupying domain $\Omega_t$ is the main concept, and the family of curves describing the flow of "anonymous" fluid particles are called streamlines. Euler's perspective is important for the standard computational fluid dynamics (numerical modeling of flows), while the Lagrange approach is particularly useful in modeling ecological processes such as motion of passive tracers, spread of contaminations and the like.

For example, if we consider function $f$ characterizing the fluid in Eulerian or Lagrangian frameworks i.e., some local parameter (it may be, e.g., temperature $\vartheta$), then quantity $f$ can be viewed as a function of spatial variables, $f = f(x^i, t)$ and also of material variables $f = f(\xi^i, t)$ i.e. $f = f(x^i(\xi^j, t), t)$ or $f = f(\xi^j(x^i, t), t)$. The geometric meaning of these representations is: either $f = f(x^i, t)$ gives the value of local parameter assumed by the fluid particles *instantaneously* located at $\{x^i\}$ or $f = f(\xi^i, t)$ is the value of local parameter at time $t$ *initially* (i.e., at time $t = 0$ located at $\xi^i$. Then the question naturally arises: why do we need fluid velocity $u^i$ (and other objects) as the tool for transition from Eulerian to Lagrangian picture, if we already have functions $x^i = x^i(\xi^j, t)$ and their inverses $\xi^j = \xi^j(x^i, t)$? The answer is that in most situations the real motion of fluid particles is not given explicitly i.e., neither $x^i(\xi^j, t)$ nor $\xi^j(x^i, t)$ are actually known. What we really observe is two different time derivatives giving the variation of local quantity $f$ when the fluid flows: (6.1.1.) $\frac{\partial f}{\partial t} = \left.\frac{\partial f(x^i, t)}{\partial t}\right|_{x^i = \text{const}}$ that is merely a partial derivative i.e., the rate of change of $f$ at a fixed position $\{x^i\}$ and (2) $\frac{df}{dt} = \left.\frac{\partial f(\xi^i, t)}{\partial t}\right|_{\xi^i = \text{const}}$ which is a material (or convective) derivative i.e., the rate of change of $f$ for a moving fluid particle. In particular, we can choose position $\{x^i\}$ as a local parameter, then, by definition, for $f = x^i$

$$\frac{dx^i(\xi^j, t)}{dt} = u^i(\xi^j, t) = \left.\frac{\partial x^i(\xi^j, t)}{\partial t}\right|_{\xi^j = \text{const}} = u^i(\xi^j(x^i, t), t) = u^i(x^i, t) \qquad (6.5.5.)$$

is the velocity of a fluid particle. Usually, it is this fluid velocity that is a measurable quantity. We see that this velocity identifies the fluid particle in terms of its initial position $\xi^i$.

One can notice that in general the notion of velocity has a clear geometric significance, and coordinate transforms always affect the values representing velocity as a set of numbers i.e., vector or matrix (see also the next section, "Geometric aspects of fluid flows"). The fluid velocity constitutes a vector field with regard to transformation properties. Throughout the time, a set of velocity components $u^i(x^i, t)$ remains attached to the fixed position: they are the field variables. In numerical modeling (and in experiment), fluid velocity components $u^i$ for $d$-dimensional problems, $d = 1,2,3,4$, being considered constant throughout time interval $\Delta t, \Delta x^i \approx u^i \Delta t$, are used to determine $x^i \approx x^i(\xi^j, t)$ (Figure 5). From the analytical standpoint, the state of a fluid in motion is fully represented by the velocity field $u^i(x^j, t)$ i.e., by velocities of fluid particles at $\{x^j, t\}$. Indeed, by (6.4.5.) we have

$$\frac{dx^i}{dt} = u^i(x^j, t), \; x^i(0) = \xi^i \qquad (6.5.6.)$$

i.e., a dynamical system which is, in the language of classical mathematics, the Cauchy problem. This form of dynamics is important because it establishes the connection between two seemingly different disciplines: fluid dynamics and the theory of dynamical systems. One can say that a dynamical system transforms its initial (Cauchy) data into a final state which, in some cases, manifests the system's asymptotic behavior. From the famous Cauchy-Lipschitz-Peano theorem[46] we know that a unique solution to the initial value problem (6.5.6.) exists on domain $D \times I$, where $D \subset \mathbb{R}^n, I \subset \mathbb{R}$, provided vector-function $\mathbf{u} = u^i(x^j, t)$ defined in $D \times I$ satisfies the Lipschitz condition with respect to $\mathbf{x} = \{x^j\}$: $|\mathbf{u}(\mathbf{x}_1) - \mathbf{u}(\mathbf{x}_2)| \leq L|\mathbf{x}_1 - \mathbf{x}_2|$ (it would be more correct to write the norm symbol instead of absolute value in the multidimensional vector case, but it does not really matter here).

---

[45] One might note that though the descriptions of fluid flows in terms of spatial and material coordinates are attributed, respectively, to Euler and Lagrange, both pictures were in fact due to Euler who was apparently very fond of migrating between fixed and comoving frames of reference (see a historical remark by J. Serrin in *Handbuch der Physik,* Band 8/1, Springer, 1959).

Moreover, the solution continuously depends on initial values $\xi^i$ which means that the fluid motion is characterized by (6.5.6.) completely i.e., the description in terms of fluid velocity (vector field or, synonymously, a dynamical system) is equivalent to the Euler or Lagrange pictures. We have seen in the preceding section that the fundamental importance of the field description is that it provides the possibility to formulate fluid dynamics in terms of PDEs (recall that PDEs i.e., partial differential equations are some expressions between spacetime derivatives). Indeed, we have also seen how to calculate the material derivative of any scalar local quantity $f(x^i, t)$:

$$\frac{df}{dt} = \frac{\partial f(\xi^i)}{\partial t}\bigg|_{\xi^i = \text{const}} = \frac{\partial f(\xi^i(x^j, t), t)}{\partial t}\bigg|_{\xi^i = \text{const}} = \frac{\partial f(x^j(\xi^i, t), t)}{\partial t}\bigg|_{\xi^i = \text{const}}$$
$$= \frac{\partial f(\xi^i)}{\partial t}\bigg|_{\xi^i = \text{const}} + \frac{\partial f}{\partial x^j}\frac{\partial x^j(\xi^i, t)}{\partial t}\bigg|_{\xi^i = \text{const}} \qquad (6.5.7.)$$

or $\frac{df}{dt} = \partial_t f + u^j \partial_j f$. This formula, expressing the rate of change of any local parameter $f(x^i, t)$ with regard to a moving fluid particle located at spacetime point $(x^i, t)$, interconnects the material and spatial derivatives through the velocity field $u^i(x^j, t)$. Such procedures lead to the systems of quasilinear PDEs, in particular, of the balance conservation type. As a simple example, we can calculate the material derivative of fluid acceleration field $a^i(x^j, t)$ in terms of velocity field: $a^i = \frac{du^i}{dt} = \frac{\partial u^i}{\partial t} + u^j \frac{\partial u^i}{\partial x^j}$. Introducing the common notation $a^i = D_t u^i$, we shall have

$$D_t u^i = \partial_t u^i + u^j \partial_j u^i \equiv u_t^i + u^j u_j^i \tag{6.5.8.}$$

---

[46] This theorem is also known as the Picard-Lindelöf theorem.

Looking at this formula one might ask: can the fluid particles move with acceleration, if, say, velocities of all particles are equal? The answer is, of course, yes, since the whole fluid can be accelerated (which actually occurs on the Earth's surface). One can give the same positive answer to the question: can the fluid particles move with acceleration, if the velocity does not change with time at each point $x^i$ of domain $\Omega$ occupied by a fluid? Indeed, although the term $u_t^i$ in (6.5.8.) is zero, the convective term $u^j \partial_j u^i$ does not necessarily vanish.

One can remark here that there is a salient contrast between Newtonian mechanics and fluid dynamics as far as acceleration is concerned. In classical mechanics constructed by Newton, acceleration plays the central part (in distinction to the Aristotelian picture) whereas in fluid dynamics, although formally it is just a branch of mechanics, acceleration is much less important. The main reason for it seems to be that in most physical and engineering settings the acceleration $a^i$ of a fluid particle is, according to (6.5.7.), uniquely determined by its coordinate $x^i$ and velocity $u^i$. Therefore, acceleration is not viewed as a state parameter, and one usually formulates the following description framework: fluid motion can be characterized in terms of "objects" i.e., local parameters of the media defined at spacetime point $(x^i, t)$, the simplest of such objects being the fluid velocity $u^i$.

## 6.6. Geometric aspects of fluid flows

Let us now pay attention to other important objects characterizing fluid motion. A natural question would be: what are the geometrical features of the flow? The primary geometric characteristics of the flow are trajectories of fluid particles and so-called streamlines. One tends to identify them, but this is, in general, not true. The trajectory (path line) of a fluid particle is defined as a locus of the particle's positions at all moments $t > 0$ i.e., it can be represented as $(x^i, t)$. A trajectory is also called a particle line or a path line. A streamline in fluid kinematics is defined as a curve for which the tangent is collinear to the velocity vector $u^i(x^j, t)$ for any given $t > 0$:

$$\frac{dx^1}{u^1(x^j, t)} = \frac{dx^2}{u^2(x^j, t)} = \frac{dx^3}{u^3(x^j, t)}, \tag{6.6.1.}$$

and we see that streamlines are defined by the characteristics equations or, which is the same, by dynamical system (6.5.6.) equivalent to the quasilinear system of the (6.5.5.) type. We shall see shortly that the theory of dynamical systems is an essentially geometric theory. In particular, geometry associated with dynamical systems enables us to describe the global properties of families of solution curves filling up the entire phase space. Thus, the dynamical system (6.5.6.) corresponding to fluid motion can be visualized as a vector field in the phase space, on which the solution is an integral curve. The modern geometric approach to fluid motion, originally suggested by V. I. Arnold in the late 1960s [16], appears to be a natural framework to describe it.

Note that, in general, streamlines are non-stationary: they depend on time $t$ (or on other vector field parameter) so that they do not necessarily coincide with fluid particle trajectories. Only for a stationary flow, $u^i(x^j, t) = u^i(x^j)$, i.e., for the autonomous equivalent a dynamical system, one can guarantee that streamlines coincide with particle lines (trajectories). Note also that the inverse is not true: streamlines can coincide with particle lines, yet the flow may be non-stationary, for example, in spherically symmetric flow $u^i = (1/4\pi) Q(t) x^i / r^3$. In other words, the property of the flow to be stationary is a sufficient, but not necessary condition for the identification of streamlines with trajectories. One might rightfully ask; can one find the velocity field (i.e., $\mathbf{u}(x^j, t)$) if the streamlines are given? At first sight it appears that the answer is affirmative, but it is wrong due to the obvious reason: at each point, the straight line along which the velocity is directed (i.e., $\mathbf{e}(x^j, t) = \mathbf{u}(x^j, t)/u(x^j, t)$, where $u \equiv |\mathbf{u}|$), may be known, but the velocity values can be different. As to particle lines, one may pose a similar question: can one find the law of motion i.e., $x^i(x_0^i, t) \equiv x^i(\xi^i, t)$ of a continuous medium, if particle lines are known? The answer is, of course, also "no" since although a curve (trajectory) along which a particle moves is known for each fluid particle, the velocity of the motion can be different – it is not defined.

Note that a dynamical system "of the hydrodynamic type" can be written for the one-dimensional case in the form

$$\frac{\partial u^i}{\partial t} = A_j^i(\mathbf{u}) \frac{\partial u^j}{\partial x}, \mathbf{u} = \{u^i(x, t)\}, i = 1, \ldots, m. \qquad (6.6.2.)$$

Here vector variable $\mathbf{u}$ denoting the collection of physical components $u^i$ is usually assumed a slowly varying function of coordinates $x$ (for $t = 0$). For example, in the one-dimensional fluid dynamics described by the Euler system of equations (i.e. in Euler's approximation for the inviscid fluid), vector field $u = \{u^i(x, t)\}, i = 1, \ldots, 4$ may be comprised of $u^1 = v$ (velocity), $u^2 = p$ (pressure), $u^3 = \rho$ (density) and $u^4 = s$ (entropy). In the spatially $n$-dimensional $m$-component case, we have

$$\frac{\partial u^i}{\partial t} = A_j^{i,\alpha}(\mathbf{u}) \partial_\alpha u^j, \qquad \partial_\alpha = \frac{\partial}{\partial x^\alpha}, \mathbf{u} = \{u^i(x, t)\}, i = 1, \ldots, m, \alpha = 1, \ldots n. \qquad (6.6.3.)$$

Operator $A_j^{i,\alpha}(\mathbf{u})$ is in general nonlinear; in Euler's type of equations $A(\mathbf{u})$ is of quasilinear character (e.g., $A_j^{i,\alpha}(\mathbf{u}) = u^\alpha$) so that the equations for a perfect fluid are quasilinear ones:

$$\partial_t u^i + u^k \partial_k u^i = -\frac{g^{ik} \partial_k p}{\rho}. \qquad (6.6.4.)$$

The convective terms $u^k \partial_k u^i$ in Euler's (and in the Navier-Stokes) equations are in fact quasilinear rather than fully nonlinear. Notice that these terms appear in the equations of fluid dynamics due to the geometric perspective i.e., Euler's description of fluid motion focusing on the features in a fixed spacetime point. The form (6.6.2) or (6.6.3.) of Euler's equations for ideal fluid (often called the Whitham form) allows one to introduce the Hamiltonian (geometric) structure, e.g., the Poisson brackets accompanied by metric $g_{ik}$ and connection (the Christoffel symbols) $\Gamma_{ik}^j$. Moreover, the powerful averaging method can be conveniently applied to such equations (see [161]).

In contrast with Euler's quasilinear system of equations, the Navier-Stokes type of equations include a diffusive term $B(\mathbf{u}) = \xi \Delta \mathbf{u}$ or, in some cases, tensor term $B_j^i(\mathbf{u})$ so that one can write the Navier-Stokes system expressed in local coordinates

$$\partial_t u^i + u^k \partial_k u^i + \xi \Delta u^i = -\frac{g^{ik} \partial_k p}{\rho} + F^i, \qquad (6.6.5.)$$

where $F = \{F^i\}$ represent external forces, in an abstract form

$$\partial_t u^i + A(\mathbf{u})u^i + B^i(\mathbf{u}) = -\frac{g^{ik}\partial_k p}{\rho} + F^i. \qquad (6.6.6.)$$

The main physical meaning of the Navier-Stokes equation is the competition between advection and diffusion mathematically reflected in the dominance of $A(\mathbf{u})$ and $B^i(\mathbf{u})$ terms in (6.6.6.). In case the diffusive term can be written as $\mathbf{B}(\mathbf{u}) = \xi\Delta\mathbf{u}$, factor $\xi$ represents the relative contribution of diffusive stray processes in the flow.

One might notice that autonomous systems are analogous to a stationary fluid flow when a fluid particle drifts from point $x(t)$ to point $x(t+\tau) = g_\tau x(t)$. This is one of the reasons why continuous-time dynamical systems are often called "flows": evolutionary transformation $x(0) \mapsto x(t)$ defines a mapping $g_t : M^n \to M^n$ that associates to each initial condition $x(0)$ solution $x(t) = g_t x(0)$. Mapping $g_t$ is, as we know, called a flow. Flow $g_t$ generated by a dynamical system $\dot{x}^i(t) = v^i(x^j, t)$, $\mathbf{x} = \{x^j\} \in M^n, t \in \mathbb{R}$ is a diffeomorphism from $M^n$ to $M^n$ i.e., a smooth invertible transformation. Note that the analogy between dynamical systems and fluid flows is incomplete since the phase (Liouville) fluid that drifts through the phase space consists of non-interacting elements, in contrast with the real moving fluid particles.

## 6.7. What is a particle?

One of the fundamental notions in physics to which we will often refer in the course of this book is a particle. As it turns out, there is a very important difference between classical and quantum mechanics when the idea of a particle is defined, in terms of points and waves, so it will be worth taking a closer look at this.

Quantum-classical monadology is an amusing conjecture giving us the occasion to discuss the important question: what is actually a physical particle? The catch is that nobody has ever observed the deeply hidden monads in any high-energy experiments. At today's level of experimental knowledge, e.g., ~200 GeV for electrons achieved by 2000 in LEP, the model of a compound quantum particle such as electron looks totally unrealistic. Recall, however, that there is no working model of electrons in quantum electrodynamics (QED), where it is sometimes controversially regarded as a point particle (since special relativity requires it). In fact, point particles in QED, in particular fermions (such as electrons), are regarded as an idealization that is not encountered in nature. Therefore, point particles are mainly considered devoid of physical interest. In strong interactions, mainly treated in quantum chromodynamics (QCD), inadequacy of the model of point particles such as nucleons and other baryons becomes nearly obvious, but one also observes the deviations from this model in the case of leptons such as electrons (provided one can ensure high-energy measurements i.e., small-distance resolution).

The model of electron intrigued many outstanding scientists, yet no final solutions have been obtained. One of the first physicists who tried to build the classical model of electron was H. A. Lorentz, the founding father of relativity (see, e.g., [108] and later papers by Lorentz). One could notice that scientific journals of the early 20th century were full of papers on models of the electron, and such renowned scientists as M. Abraham, P. Langevin, H. Poincaré, K. Schwarzschild and A. Sommerfeld were among the authors. In fact, the models of electron prepared the ground for relativity, in particular because in most of them mass depended on velocity. Of course, the language and mathematical arrangements have drastically changed in the course of more than one hundred years, but many ideas remained the same. It seems that fascination with the model of an extended (not pointlike) electron greatly contributed to obtaining classical solutions in general relativity. We have mentioned, however, that particles having a finite radius are poorly compatible with special relativity. On the other hand, a point charge would possess infinite Coulomb energy (self-energy). One often assumes the classical electron radius to be provided by fundamental constants such as elementary charge $e = 1.6 \cdot 10^{-19}$ C $\approx 4.8 \cdot 10^{-10}$ CGSE, electron mass $m = 9.1 \cdot 10^{-28}$ g, speed of light $c \approx 3 \cdot 10^{10}$ cm/s i.e. $r_0 \sim e^2/mc^2$. This estimate corresponds to an assumption that the electron mass has a purely electromagnetic origin, which is difficult to substantiate.

In quantum field theory (QFT), the electron can be regarded for many purposes as a quantum field corresponding to a point particle. To produce a point[47] particle in the quantum theory of electrons one has to sum up infinite series in the dual **k**-space. We have no idea about the internal structure of the electron (that was the primary particle to have been studied in quantum theory), but we intuitively understand that electrons or other fermions, while interacting with photons or other bosons, are always surrounded by such particles, albeit virtual ones i.e., transient excitations not appearing on the "mass shell" $m^2c^4 = E^2 - p^2c^2$. If, however, the internal structure of electrons does exist, we do not know how the mass, charge, angular momentum and, possibly, currents are distributed inside the electron. There are, however, some techniques developed in QED for how to assess the spatial distribution of charge or current (e.g., in the form of magnetic moment) inside an "elementary" particle. The information about deviations from being represented by a mathematical point is contained in the so-called form-factor characterizing the distribution of mass, charge and current within the particle. The electromagnetic form-factor of the electron is considered to be determined by the cloud of virtual particles such as electron-positron pairs over the distance $r_0 \sim 10^{-13}$ cm (classical electron radius). In the simplest case, the form-factor is reduced to a Fourier transform.

An atomic form-factor is a function characterizing the spatial distribution of electrons in atom, the mean radius of this distribution being $\sim 10^{-8}$ cm. Atomic form-factors are, in particular, important for quantum chemistry. The nuclear form-factor characterizes the distribution of nucleons in the atomic nucleus, with the mean radius $\sim 10^{-12}$ cm. The hadronic formfactor characterizes the distribution of quarks inside hadrons (radius $\sim 10^{-13}$ cm). One suspected long ago that there are smaller (presumably pointlike) particles inside hadrons, named partons by R. Feynman; today these partons are identified with quarks. There exist a number of models trying to elucidate how the nucleon is composed of quarks (such as the Skyrmion or the bag model), but to obtain the real distribution of quarks inside a nucleon one has to solve the QCD dynamical equations, which is only feasible today by numerical techniques (e.g., lattice QCD) using rather powerful computing facilities.

It is curious that the main equation of the relativistic quantum theory, the Dirac equation, being produced from the mass-shelf form

$$p_\mu p^\mu = m^2 c^2 = \frac{E^2}{c^2} - |\mathbf{p}|^2 \qquad (6.7.1.)$$

evokes an analogy with the D'Alembert-Euler (also known as Cauchy-Riemann) conditions for analytical functions. By following the common procedure, one obtains the canonical Dirac equation (see, e.g., [6], [22], and [46]):

---

[47] We do not distinguish here point and pointlike particles: both are zero-dimensional.

$$\left(i\gamma^\mu \partial_\mu - \frac{mc}{\hbar}\right)\psi = 0 \tag{6.7.2.}$$

and then, separating the four-component spinor $\psi$ into a pair of two-component spinors, $\psi = \begin{pmatrix} \chi \\ \phi \end{pmatrix}$, we get two equations

$$i\left(\partial_0 - \sigma^j \partial_j\right)\phi = \frac{mc}{\hbar}\chi \tag{6.7.3.}$$

and

$$i\left(\partial_0 + \sigma^j \partial_j\right)\chi = \frac{mc}{\hbar}\phi, i = 1,2,3, \tag{6.7.4.}$$

where $\sigma^j$ are the Pauli matrices (one typically works in relativistic quantum theories using the system of units with $\hbar = c = 1$). The latter two adjoint equations for $\chi$ and $\phi$ are similar to the equations for conjugate harmonic functions $u(x,y), v(x,y), f(z) = u(x,y) + iv(x,y), z = x + iy$, where the "left-handed" $\chi$ and "right-handed" $\phi$ components of the total four-component spinor $\psi = (\chi, \phi)^T$ being far generalizations of $u, v$.

It is interesting to notice that if one considers a charged quantum particle, e.g., an electron, there is an option that its center of mass ($\mathbf{r}_m$) does not coincide with the center of charge ($\mathbf{r}_c$). In this case dynamical evolution of a quantum particle would be essentially different when the electromagnetic field is present or absent. Moreover, both a dipole moment and a non-zero intrinsic angular momentum should appear since the spatial isotropy would be broken and there would exist a special direction uniquely defined by vector $\mathbf{r}_c - \mathbf{r}_m$. However, no dipole moment of the electron has been observed yet (although the current experimental techniques might be not sensitive enough), and the electron's intrinsic angular momentum (spin) is characterized by the features (Dirac spinors) that have nothing to do with the assumption about non-coinciding centers of mass and charge of an electron.

Quantum particles are by default elementary, and unless annihilation processes of QED are taken into account, particles cannot change their inner states. In particular, quantum particles have no excited states so that allowed variables to describe a quantum particle are of a kinematic nature[48]. Nevertheless, we can roughly think of quantum particles as being extended (not pointlike) since their wave function, that is the solution of the Schrödinger wave equation, seldom coincides with the delta-function $\delta(\mathbf{r} - \mathbf{r}(t))$, where $\mathbf{r}(t)$ is the law of the particle's classical motion. For example, the lowest-order (angular momentum $l = 0$) solution to the force-free Schrödinger equation in 3d is of the form $\psi_0 = \sin kr / kr$, where $k^2 = 2mE/\hbar^2$, $E$ is the particle energy (the higher-momentum states can be obtained from $\psi_0$ by simple differentiation which results in the phase shift at $\sin kr$). The above delta-function corresponds to mass density $n(\mathbf{r}, t) = m\delta(\mathbf{r} - \mathbf{r}(t))$ or mass current for a single particle, $\mathbf{j}(\mathbf{r}, t) = m\mathbf{v}\delta(\mathbf{r} - \mathbf{r}(t))$ of idealized classical objects having no (zero) dimensions. Analogously, charge density and current for many point particles can be defined as, e.g., $n(\mathbf{r}, t) = \sum_{a=1}^{n} e_a \, \delta(\mathbf{r} - \mathbf{r}_a(t))$ and $\mathbf{j}(\mathbf{r}, t) = \sum_{a=1}^{n} e_a \mathbf{v}_a \, \delta(\mathbf{r} - \mathbf{r}_a(t))$. The point particle is a good example of a mathematical model; real physical events occur in finite spacetime domains. In the same way, we can treat the Moon and the Earth as point particles in many astrophysical problems.

---

[48] It would be very interesting, for example, to register internal oscillations due to coherent excitation of the hidden degrees of freedom within an elementary quantum particle channeled in a crystal lattice.

We may recall that QED and some other quantum field theories suffer from serious diseases such as divergences, the gravest of them at short distances that can be traced back to the infinite self-energy of the classical pointlike electron. In the standard calculation techniques based on Feynman diagrams, such divergences are removed through so-called renormalization, which is the cunning artificial procedure to cancel infinite terms in the diagrams by equally infinite counter-terms (in the parts of diagrams corresponding to infinite terms, virtual particles of QED appear as closed loops i.e., cycles). As a final result, finite values remain which can be compared with experimentally obtained quantities. One might remark that few mathematicians would endorse subtracting one infinity from another as being a legitimate operation (see, e.g., [6]).

Thus, as already commented, the existence of extended (not pointlike) particles in quantum mechanics is poorly compatible with special relativity. In this respect, special relativity contradicts quantum mechanics, where, on the contrary, the model of particles as mathematical (0d) points is difficult to accept. One usually considers for crude estimates a quantum particle having momentum $p = |\mathbf{p}|$ to be spread over distances $\sim \hbar/p$; in the relativistic area, the characteristic dimensions of a particle are usually taken to be of the order of the Compton length, $l_C = \hbar/mc \sim 4 \cdot 10^{-11}$cm i.e., approximately two orders of magnitude larger than the classical electron radius $r_0 = e^2/mc^2 = \alpha l_C$ and about the same two orders of magnitude smaller than the Bohr atomic radius $r_B = \hbar^2/me^2 = \alpha^{-1} l_C \approx 5.3 \cdot 10^{-9}$cm, where $\alpha = e^2/\hbar c \approx 1/137$ is the fine structure constant. Notice that the Compton length is determined by the particle mass. It is remarkable that, due to the fact that a quantum particle has wave properties, and is effectively extended in space (i.e., its wave function is spread over some finite distance), when moving through matter it can "feel" the environment over considerable spatial scales – much larger than the mean interatomic distance – so that the particle can interact simultaneously (coherently) with many atoms of matter. A number of interesting phenomena have been discovered such as, e.g., the Landau-Pomeranchuk-Migdal (LPM) effect when the processes of Bremsstrahlung and pair production in matter are suppressed as well as coherent scattering in crystals (see, e.g., [105, 98, 115]).

## 6.8. Motion of a fluid particle

A fluid particle moving along its pathline (trajectory) changes its volume, shape and orientation (e.g., rotates) with respect to its initial position. These are geometric transformations of a fluid particle that can be described as the time evolution of the flow governed by the geodesics equation defined on the group of volume-preserving diffeomorphisms. Like any free mechanical system, fluid moves along geodesic curves (see more below about mechanical motion), however, in distinction with the motion of lumped (rigid) bodies fluid motion involves infinitely many degrees of freedom which makes the description of fluid behavior necessarily field-like and accounting for deformations, even when we try to stick to the Lagrangian ("material") picture. The simplest geometric object describing spatial variations of a fluid particle is tensor $\tau_{ij}(x^k, t) := \partial u_i/\partial x^j$ constructed from the fluid velocity $u_i = g_{ij} u^j = \delta_{ij} u^j$. In general, $\tau_{ij}$ has no a priori symmetry, however, it can be decomposed into a symmetric and antisymmetric parts:

$$\tau_{ij} = \theta_{ij} + \omega_{ij}, \qquad \theta_{ij} = \theta_{ji}, \qquad \omega_{ij} = -\omega_{ji}, \qquad (6.8.1.)$$

where $\theta_{ij} := \frac{1}{2}(\partial_i u_j + \partial_j u_i)$, $\omega_{ij} := \frac{1}{2}(\partial_i u_j - \partial_j u_i)$. These quantities are usually called strain rate (deformation tensor) and vorticity, respectively. Note that vorticity is defined by different authors in a variety of ways, which can cause some petty inconveniences, e.g., as a covariant tensor (2-form) $\omega_{ij} = \frac{1}{2}(\partial_j u_i - \partial_i u_j)$, or as a vector field $\omega^i = \frac{1}{2}\varepsilon^{ijk}\omega_{jk}$, or simply as the curl of velocity field $\boldsymbol{\omega} = \nabla \times \mathbf{u}$. Here $\varepsilon^{ijk}$ is the unit antisymmetric pseudotensor of rank 3. These definitions are extensively used while treating from different viewpoints the Navier-Stokes equation, the main mathematical model of fluid dynamics (see Supplement 3).

Using strain rate and vorticity, one can represent the velocity field ($u_i = g_{ij}u^j = \delta_{ij}u^j$) translation by vector $\mathbf{h}$ as $\mathbf{u}(\mathbf{x} + \mathbf{h}) \approx \mathbf{u}(\mathbf{x}) + \theta\mathbf{h} + \boldsymbol{\omega} \times \mathbf{h}$ (the Cauchy-Helmholtz formula) or, in local components,

$$u_i(x^j + h^j) = u_i(x^j) + \theta_{ij}h^j + \varepsilon_{ijk}\omega^j h^k + \mathcal{O}(h). \tag{6.8.2.}$$

This formula admits a simple physical interpretation: the generic fluid motion is a sum of three effects, translation, rotation and deformation. The first two describe the motion of a fluid element as a rigid body. Note that vorticity physically plays an essential part in inviscid fluids since there are no intrinsic dissipative mechanisms to stop the rotation (apart from boundary effects and external forces). One can read many useful things about vorticity in [92] (see, e.g., §§ 8,14 of this book): how the vorticity is transported by the velocity field, the motion equation for $\boldsymbol{\omega}$, the circulation theorem and so forth. We shall not reproduce this classical material here.

The physical meaning of strain rate and vorticity becomes transparent when we consider the behavior of a liquid drop that models the fluid particle in the course of its motion. Evolution of a liquid drop is an important mathematical model not only in fluid dynamics, but it has also influenced the analysis of swarming and globular behavior of diverse objects such as two- and three-dimensional media consisting of both non-interacting and interacting particles in nuclear physics, electronics, astrophysics, ecology, biology, etc. Assume, for simplicity, the drop to be initially of spherical shape and let $\mathbf{x}_0 = (x_0^1, x_0^2, x_0^3)$ be the coordinates of the particle center, while vector $\mathbf{x} = (x^1, x^2, x^3)$ denotes points (e.g., molecules or molecular clusters) belonging to the fluid particle. In other words, points $\mathbf{x}$ lie within a sphere of "physically infinitesimal" radius $\rho$. During some small time $\Delta t$, point $\mathbf{x}_0$ will be displaced into $\mathbf{z}_0 \approx \mathbf{x}_0 + \mathbf{u}(\mathbf{x}_0)\Delta t$ whereas points $\mathbf{x}$ are displaced to $\mathbf{z} \approx \mathbf{x} + \mathbf{u}(\mathbf{x})\Delta t$, where $\mathbf{u}(\mathbf{x})$ is, as before, the velocity vector field. One may notice that such a representation repeats Euler's method in numerical mathematics, which is quite natural. If we consider the difference vector $\mathbf{a} := \mathbf{x} - \mathbf{x}_0$, then we observe that it transforms into

$$\mathbf{b} := \mathbf{z} - \mathbf{z}_0 \approx \mathbf{a} + \big(\mathbf{u}(\mathbf{x}) - \mathbf{u}(\mathbf{x}_0)\big)\Delta t \approx \mathbf{a} + a^j\Delta t\partial_j\mathbf{u} \tag{6.8.3.}$$

or

$$b^i = \big(\delta_j^i + \Delta t\partial_j u^i\big)a^j = \big(\delta_j^i + \Delta t g^{ik}\partial_j u_k\big)a^j \equiv \big(\delta_j^i + \Delta t g^{ik}\tau_{jk}\big)a^j, \tag{6.8.4.}$$

where $g^{ik}$ is the inverse metric tensor that we may, in Euclidean space, replace by the Kronecker tensor $\delta^{ik}$. Physically, this means that a fluid particle is displaced by $\mathbf{u}(\mathbf{x}_0)\Delta t$ and is deformed, if tensor $\tau_{ij} \neq 0$. Notice that curved spaces ($g_{ik}(\mathbf{x}) \neq \delta_{ik}$)[49] introduce additional deformation and torsion of a moving fluid particle. Representing tensor $\tau_{ij}$ by the sum of its symmetric and antisymmetric parts $\tau_{ij}(\mathbf{x}) = \theta_{ij}(\mathbf{x}) + \omega_{ij}(\mathbf{x})$, just as before, we get

$$b^i \approx \left(\delta_j^i + \Delta t \delta^{ik}(\theta_{jk} + \omega_{jk})\right)a^j \approx \left(\delta_j^i + \Delta t \delta^{ik}\theta_{jk}\right)(\delta_j^i + \Delta t \delta^{ik}\omega_{jk})a^j \qquad (6.8.5.)$$

or, in matrix form,

$$\mathbf{b} \approx (\mathbf{I} + \Delta t \boldsymbol{\theta})(\mathbf{I} - \Delta t \boldsymbol{\omega})\mathbf{a}, \qquad (6.8.6.)$$

where $\mathbf{I}$ is the unit matrix, $\boldsymbol{\theta}$ and $\boldsymbol{\omega}$ are matrices with entries $\theta_{ij}$ and $\omega_{ij}$. Thus, vector $\mathbf{b}$ expressing the distance from the center within the fluid particle, transformed in process of fluid motion, is obtained from its initial value $\mathbf{a}$ by two successive infinitesimal transformations: $U_{\boldsymbol{\omega}} = (\mathbf{I} - \Delta t \boldsymbol{\omega})$ and $U_{\boldsymbol{\theta}} = (\mathbf{I} + \Delta t \boldsymbol{\theta})$. We shall shortly see from here that the fluid particle participates in three evolutions: it is translated by $\Delta \mathbf{r} = \mathbf{u}(\mathbf{x})\Delta t$ (Galilean motion), being deformed along three orthogonal axes and rotated with angular velocity $\boldsymbol{\omega} = \nabla \times \mathbf{u}(\mathbf{x})$.

---

[49] One might notice that general invariance requirements are fulfilled only for the $\delta_j^i$ Kronecker tensor and not for $\delta^{ij}$ or $\delta_{ij}$ symbols (specifically when working with unitary SU(n) groups).

An experienced reader will immediately see here the generators of Lie groups defining the evolution of a fluid particle. Quantities $U_{\boldsymbol{\theta}}$ and $U_{\boldsymbol{\omega}}$ are two families of matrices depending on parameter $\Delta t \equiv \tau$ i.e., one can imagine each of them to represent a curve $U(\tau)$ passing through unity ($U(0) = \mathbf{I}$). For each $\tau$, the flow $g_\tau$ generated by $\boldsymbol{\theta}$ and $\boldsymbol{\omega}$ is a diffeomorphism from $\Omega$ to $\Omega$ (one can consider domain $\Omega$ to be a manifold). This flow is a commutative one-parameter group $g_\tau = U(\tau)$ transforming distance $\mathbf{a}$ into $\mathbf{b} = (\mathbf{I} + \tau X)\mathbf{a} \equiv U(\tau)\mathbf{a}$, where $X$ is the respective vector field (in this case $\boldsymbol{\theta}$ or $\boldsymbol{\omega}$) being the generator of group $U(\tau)$. Notice that matrix $U(\tau)$ is invertible as long as $|\tau X| < 1$. The tangent (velocity) vector at $\tau = 0$ is $\left.\frac{dU}{d\tau}\right|_{\tau=0} = X$ so that we have a pushforward of tangent vector $X$ at $\tau = 0$ by $g_\tau = U(\tau)$. If mapping $g_\tau = U(\tau)$ (i.e., flow) is a diffeomorphism, then such a pushforward is invertible and its inverse defines a pullback from $\Omega$ to $\Omega$ (transforming distance $\mathbf{b}$ into $\mathbf{a} = U(\tau)\mathbf{b}$). Note that this inversion only holds when the vector field does not depend explicitly on time: the flow generated by a time-dependent vector field is no longer a one-parameter group of diffeomorphisms.

One can interpret a diffeomorphism simply as a coordinate change $x \to z = f(x)$ implemented by a smooth function $f(x)$ that has an inverse, and it is also smooth. One can also treat a diffeomorphism as a one-to-one mapping of a differentiable manifold (see more on that below) with itself – in simple words, a diffeomorphism is a smooth deformation of a differentiable manifold. Note that the requirement of simultaneous smoothness or even of continuous differentiability ($C^1$-diffeomorphism) both of direct and inverse mapping is rather strong. As we know from classical calculus, differentiability of a function implies that in the vicinity of each interior point of the function's domain of differentiability one can approximate its behavior by a linear expression. One might thus note that differentiability requires exact notions of the distance (norm) specifying the term "vicinity of a point" and of linearization, yet such notions cannot be applied for an arbitrary set.

One should also be careful with the notion of a diffeomorphism: many elementary functions are not diffeomorphisms, for example, the map $f(x) = x^3, \mathbb{R} \to \mathbb{R}$ is not a diffeomorphism since $f'(x)|_{x=0} = 0$. Notice also that diffeomorphism does not have to preserve angles and distances – the crucial characteristics of Euclidean spaces. Therefore, diffeomorphism does not, in general, preserve the area (volume). Volume-preserving diffeomorphisms play a key role in Hamiltonian mechanics; in this context they are known as symplectomorphisms. For example, canonical transformations of classical mechanics as well as the Hamiltonian flow give examples of a symplectomorphism. Recall that canonical transformations of mechanics are, in fact, changes of variables leaving intact such important structures as Hamiltonian equations and Poisson brackets (see also below).

As already mentioned, any dynamical system can be visualized as a vector field on the system's state space; in physics, a dynamical system and a vector field are just synonyms. According to the mathematical model of Hamiltonian mechanics based on the concept of a symplectic structure (see Supplement 1), a Hamiltonian vector field $X = X_H$ is sometimes called a symplectic gradient (recall that the gradient of a differential function $f(x^i), i = 1, \ldots, n$ is a vector field defined by $\partial f / \partial x^i \equiv \partial_i f = (\nabla f)_i$). Slightly generalizing one can say that for any real function on a symplectic manifold (the Hamiltonian) one can find a vector field with the flow preserving both the given function (i.e., the Hamiltonian) and the symplectic form (i.e., a non-degenerate closed two-form).

A nearly obvious property of Hamiltonian vector fields $X_H$ is that they point tangentially to surfaces of constant energy in the phase space i.e., a Hamiltonian field is simply the vector field $\mathbf{v}(z)$ in the phase space, where $z$ denotes a symplectic canonically conjugated pair $(p, q)$. Thus, the dynamic equation $\dot{z} = v(z)$ indicates at what direction a particle or a system would move. The latter equation preserves a symplectic structure i.e., a geometric structure formed by canonically conjugated pairs of variables.

In the language of matrix theory, condition $|X| < 1, X = (x_i^j)$, is a unit sphere without borders. After exponentiation we have $\mathbf{b} = e^{\tau X} \mathbf{a}$ or, in the guise of dynamical systems, $\frac{db^i}{d\tau} = X^i, b^i(0) = a^i$, where $X$ is a vector field on $\Omega$ and $\mathbf{b}(\tau)$ can be interpreted as an integral curve of $X$ in $\Omega$. Recall that solving the autonomous ODE-system is geometrically expressed as determining the integral curve (see section 6.1.) of vector field $X = X^i \partial_i$. Specifically, tensor quantities $\theta_j^i = (U_j^i - \delta_j^i)/\tau$ are local coordinates of matrix $\boldsymbol{\theta}$ in the vicinity of unit matrix $\mathbf{I} \in \mathrm{GL}(n, \mathbb{R})$ (the general linear group of the set of $n \times n$ invertible matrices). Notice that one can introduce local coordinates near any point (i.e., matrix) $\mathbf{C} \in \mathrm{GL}(n, \mathbb{R})$, e.g., through multiplying all matrices by $\mathbf{C}^{-1}$ which is equivalent to the change of variables.

One can also rewrite the expressions for the flow as

$$\frac{d\mathbf{b}(\tau, \mathbf{a})}{d\tau} = \frac{d}{d\tau}(e^{\tau X}\mathbf{a}) = Xe^{\tau X}\mathbf{a}, \qquad \mathbf{b}(0, \mathbf{a}) = e^{0 \cdot X}\mathbf{a} = \mathbf{a} \qquad (6.8.7.)$$

and notice the following Lie group properties:

$$\mathbf{b}(\tau, \mathbf{b}(\sigma, \mathbf{a}) = \mathbf{b}(\tau, e^{\sigma X}\mathbf{a}) = e^{\tau X}e^{\sigma X}\mathbf{a} = e^{(\tau+\sigma)X}\mathbf{a} = \mathbf{b}(\tau + \sigma, \mathbf{a}). \qquad (6.8.8.)$$

These are examples of standard reasoning when working with flows and one-parameter groups of continuous transformations often encountered in modeling evolution processes. One must, however, bear in mind that parameter $\tau \equiv \Delta t$ should, in general, be kept within a certain interval so that the uniqueness and existence theorem for ODEs could ensure a unique local solution for $\mathbf{b}(0) = \mathbf{a}$.

Now, let us return to our moving fluid particle. We see that matrix $U_{\boldsymbol{\theta}}$ is symmetric, therefore it can be brought to a diagonal form i.e., there exists a triad $(\mathbf{e}_1, \mathbf{e}_2, \mathbf{e}_3)$ of orthonormalized basis vectors consisting of eigenvectors of $U_{\boldsymbol{\theta}}$. Vector $\mathbf{b}$ is obtained from $\mathbf{a}$ by extension or contraction along the directions defined by $(\mathbf{e}_1, \mathbf{e}_2, \mathbf{e}_3)$ with the respective factors $(\lambda_1 = 1 + \mu_1, \lambda_2 = 1 + \mu_2, \lambda_3 = 1 + \mu_3)$, where $\mu_i, i = 1,2,3$ are the eigenvalues of $\Delta t \boldsymbol{\theta}(x)$. It means that a spherical fluid particle with radius $\rho$ is transformed, in the course of its motion, into an ellipsoid with three elliptic radii (semiaxes) $\rho\lambda_1, \rho\lambda_2, \rho\lambda_3$. Note that in general ellipsoids are produced by an invertible linear transformation of a sphere, this transformation being represented by symmetric $3 \times 3$ matrices (in 3d coordinate space). Now, we can easily calculate the volume variation of a fluid particle: $\frac{V}{V_0} = \lambda_1\lambda_2\lambda_3 \approx 1 + \mu_1 + \mu_2 + \mu_3 = 1 + \Delta t \operatorname{Tr} \boldsymbol{\theta}(\mathbf{x}) = 1 + \Delta t \partial_i u^i(\mathbf{x}) = 1 + \Delta t \nabla \mathbf{u}(\mathbf{x})$ i.e. volume change per unit time is

$$\frac{\Delta V}{V_0 \Delta t} = \frac{V - V_0}{V_0 \Delta t} = \operatorname{Tr} \boldsymbol{\theta}(\mathbf{x}) \qquad (6.8.9.)$$

One can treat the vorticity in the same way: $\mathbf{b} = U_{\boldsymbol{\omega}}\mathbf{a} = (\mathbf{I} - \Delta t\boldsymbol{\omega})\mathbf{a} = \mathbf{a} - \frac{1}{2}\Delta t\mathbf{a} \times (\nabla \times \mathbf{u}(\mathbf{x}))$. This transformation is a rotation by angle $\Delta\varphi = \frac{1}{2}\Delta t|\nabla \times \mathbf{u}(\mathbf{x})|$ which leaves the volume intact for infinitesimal $\Delta t$ and $\rho$ up to higher-order corrections. Here $\dot{\boldsymbol{\varphi}} = \boldsymbol{\omega} = \nabla \times \mathbf{u}(\mathbf{x})$ is the instantaneous angular velocity of the rotational vector field $\mathbf{u}(\mathbf{x})$.

We can make the following remark in connection with the deformation of a fluid particle: a deformable body $D$ in $\mathbb{R}^3, D \subset \mathbb{R}^3$, is in fact identified with the respective domain $D$. Any displacement of this domain regardless of its deformation is a diffeomorphism in $\mathbb{R}^3$ generated by vector field $\mathbf{u}(\mathbf{x})$ that characterizes the displacement. In elasticity theory mostly dealing with rigid bodies and providing a foundation for mechanical engineering, the strain tensor $\varepsilon_{ij}(\mathbf{x})$ is usually defined as the difference of infinitesimal distances before and after deformation i.e.

$$ds^2 - ds_0^2 \approx \theta_{ij}dx^i dx^j + \frac{1}{2}\partial_k u^i \partial_l u^j dx^k dx^l, \quad \varepsilon_{ij}dx^i dx^j$$
$$= \left\{ \left[ \partial_i(g_{jk}u^k) + \partial_j(g_{ik}u^k) \right] + \frac{1}{2}\delta_i^k \delta_j^l \partial_k u^i \partial_l u^j \right\} dx^i dx^j. \qquad (6.8.10.)$$

Here, $u^i$ is understood as a displacement $u^i = x^i - x_0^i$ and not as velocity vector field i.e., formally $\Delta t = 1$. In Euclidean spaces $g_{ik}(\mathbf{x}) = \delta_{ik}$, and one can identify $u^i$ and $u_i$. By the way, notice that the authors of many valuable sources usually do not make a difference between upper and lower indices i.e., between contra- and covariant components of vector and tensor fields (specifically while treating fluid dynamics and deformable bodies). Although in many cases such a difference may be immaterial, an implicit identification of vectors and covectors is in general wrong: they have a significantly different geometric nature (belong to dual spaces). The displacement $u^i$ due to strain results in change of Euclidean metric $ds^2 = dx_i dx^i$ in continuous media i.e. (in finite differences)

$$\Delta s^2 = \sum_{i=1}^{i=3}\left[\left(x^i + u^i\right) - \left(x_0^i + u_0^i\right)\right]^2 = \Delta s_0^2 + 2\sum_{i=1}^{i=3}\Delta x^i \Delta u^i + \sum_{i=1}^{i=3}\left(\Delta u^i\right)^2 \qquad (6.8.11.)$$

with $\Delta u^i = u^i(\mathbf{x}) - u_0^i(\mathbf{x}_0)$, $\Delta x^i = x^i - x_0^i \approx u^i$.

One might note in passing that we can use another metric in this procedure, but the Euclidean metric is the simplest one in physics. For infinitesimal $\Delta x^i$, we have the displacement vector field $\Delta u^i \to du^i \approx \partial_j u^i dx^j$ and

$$ds^2 = ds_0^2 + 2\partial_j u^i dx^i dx^j + \partial_k u^i \partial_l u^j dx^k dx^l$$

so that after symmetrization $2\partial_j u^i dx^i dx^j = \left(\partial_j u^i + \partial_i u^j\right) dx^i dx^j$ we arrive at (6.6.10.) (with $g_{ik} = \delta_{ik}$). One can also obtain formula (6.8.10.) from the preceding considerations, namely by heuristically setting $ds_0 = |\mathbf{a}|$, $ds = |\mathbf{b}|$. Then in the linear approximation we have $ds^2 - ds_0^2 = |\mathbf{b}|^2 - |\mathbf{a}|^2 = |(\mathbf{I} + \Delta t \boldsymbol{\theta}(\mathbf{x}))\mathbf{a}|^2 - |\mathbf{a}|^2 \approx 2\Delta t(\boldsymbol{\theta}(\mathbf{x})\mathbf{a}, \mathbf{a}) = 2\Delta t \theta_{ij}(\mathbf{x}) dx^i dx^j$     or     $ds^2 = ds_0^2 + \Delta t \left(\partial_i u_j + \partial_j u_i\right) dx^i dx^j$. Putting formally $\Delta t = 1$ and interpreting $u^i$ as strain, we get $ds^2 = ds_0^2 + \varepsilon_{ij} dx^i dx^j$ with strain tensor $\varepsilon_{ij} = \partial_i u_j + \partial_j u_i$. We see that the strain tensor characterizes the change of the distance $ds = (g_{ij} dx^i dx^j)^{1/2}$ between two points when a displacement $u^i$ is produced. Mathematically, a strain tensor is the Lie derivative of the metric $ds = (g_{ij} dx^i dx^j)^{1/2}$ over the deformation.

We may note in conclusion to this subsection that such considerations represent a standard geometric (kinematic) setting of linear elasticity theory. Transition to dynamics in this theory is based, in accordance with the general idea of linearity, on the assumption that small strains produce stresses in the medium, linearly dependent on strains. This statement is known as Hooke's law, $\sigma_{ij}(\mathbf{x}) = E_{ij}^{kl} \theta_{ij}(\mathbf{x})$, where tensor $E_{ij}^{kl}$ (modulus of elasticity, stiffness or elasticity tensor having in general $3^4 = 81$ components) can be reduced to rather simple forms, often merely scalars, under some natural assumptions about the media (isotropy, homogeneity, etc.). Hooke's law is one of the main mathematical models in elasticity, material science and mechanical engineering; one can recall that the ubiquitous oscillator model exemplified by a string is a very particular case of Hooke's law (the restoring force, $F = -kx$, relating the force appearing in a stretched string to small displacement in it, factor $k$ is the elastic spring constant).

## 6.9. Advection

A physical manifestation of the difference between the volume-preserving i.e., incompressible fluid with divergent-free (div $\mathbf{v} = 0$) flow and volume-changing (e.g., compressible) fluid with divergent (div $\mathbf{v} > 0$) or convergent (div $\mathbf{v} < 0$) flow is observed, in particular, in ecological problems of water basins, ocean and atmosphere. More specifically, one can consider the behavior of impurities in a fluid flow characterized by velocity field $\mathbf{v}(\mathbf{x}, t)$. In this class of problems, field $\mathbf{v}(\mathbf{x}, t)$ often has a random component $\boldsymbol{\xi}(\mathbf{x}, t)$, e.g., with zero mean value, $\langle \boldsymbol{\xi}(\mathbf{x}, t) \rangle = 0$, but in our dynamical treatment we shall pay little attention to random fields. The dynamical system corresponding to low-inertial motion of impurity (such as silver) in the fluid flow, usually called advection, can be written in the Lagrangian form as

$$\frac{d\mathbf{x}(t; \mathbf{x}_0, t_0)}{dt} = \mathbf{u}(t; \mathbf{x}_0, t_0), \quad \frac{d\mathbf{u}(t; x_0, t_0)}{dt} = -\beta\left(\mathbf{u}(t; \mathbf{x}_0, t_0) - \mathbf{v}(\mathbf{x}, t)\right), \tag{6.9.1.}$$

with initial conditions $\mathbf{x}(t = t_0) = \mathbf{x}_0$, $\mathbf{u}(t = t_0) = \mathbf{u}_0 = \mathbf{u}(\mathbf{x}_0)$. Dynamical system (6.8.1.) is known as the advection equation. In many ecological problems, it is interesting to consider the advection of inertialess tracers i.e., of passive scalar impurities, when the trajectory of an impurity coincides with the path of fluid particles. The term "passive" designates in this context that the influence of an impurity transferred by the flow on the latter characteristics is negligible. A good example of such passive impurity is the spreading smoke particles (or other scalar particles) transferred by the atmospheric flows. This physical picture corresponds to the asymptotic ($\beta \to \infty$) case of model (6.7.1.), when the impurity exactly and without delay follows fluid particles, $\mathbf{u}(t; \mathbf{x}_0, t_0) = \mathbf{v}(\mathbf{x}, t)$ so that we obtain a purely kinematic model

$$\frac{d\mathbf{x}(t;\mathbf{x}_0,t_0)}{dt} = \mathbf{v}(\mathbf{x}(t;\mathbf{x}_0,t_0),t), \qquad \mathbf{x}(t=t_0;\mathbf{x}_0) = \mathbf{x}_0. \tag{6.9.2.}$$

The Jacobian determinant for the transformation $x^i(x_0^j,t)$, $J(t,t_0) = \det\|\partial x^i(t;\mathbf{x}_0,t_0)/\partial x_0^j\|$ characterizes the fluid volume change and satisfies the equation $dJ(t,t_0)/dt = \operatorname{div}\mathbf{v}(\mathbf{x},t)J(t,t_0)$ with initial condition $J(t=t_0)=1$. For an incompressible fluid, $\operatorname{div}\mathbf{v}(\mathbf{x},t)=0$ and the volume is preserved by the fluid motion, $J(t,t_0)=1$ for all $t>t_0$. In local coordinates, the Jacobian $\partial x^i(t;\mathbf{x}_0,t_0)/\partial x_0^j$ of a smooth map $x^i(x_0^j,t),\Omega\to\Omega$ (more generally, mapping $\varphi\colon\Omega_0\to\Omega,\mathbf{x}_0\in\Omega_0,\mathbf{x}\in\Omega$ for each point $\mathbf{x}_0\in\Omega_0$ which is not necessarily a bijection) is a matrix representation of the map between two tangent space manifolds $T_{\mathbf{x}_0}\Omega_0\to T_{\varphi(\mathbf{x}_0)}\Omega$. One can take the time evolution of the Jacobian $J(t,t_0)$ as a definition of incompressible fluid: when the fluid flow does not entail the change of the fluid volume i.e., $dJ/dt=0$ for all $(x^i,t)$. Otherwise, the fluid is compressible, $dJ/dt\neq0$. Thus, the model of an incompressible fluid is an example of a Lagrangian system (see below) with (a very large) number of degrees of freedom – one can even think of their infinite number.

Notice that for steady flow, $\mathbf{v}(\mathbf{x}(t),t)=\mathbf{v}(\mathbf{x})$, dynamical system (6.8.2.) becomes autonomous, $d\mathbf{x}(t)/dt=\mathbf{v}(\mathbf{x}),\mathbf{x}(t_0)=\mathbf{x}_0$ . For a stationary, incompressible, and inviscid flow, the motion of tracers coincides with the streamlines of the flow. Moreover, the autonomous system may have stationary equilibrium points $\tilde{\mathbf{x}}_k$ , $k=1,\dots,m$ that can be stable and attract the trajectories or unstable and repel them. One can get a feeling of the tracer's behavior by observing the gathering of tealeaf fragments at the center of the cup when the tea is stirred (this observation is known as tealeaf paradox or Einstein's problem [53]).

Advection usually stirs the concentration field in such a way that sharp concentration gradients appear, hence the diffusion naturally becomes more and more noticeable. Thus, the advection-diffusion connection is natural. The advection equation (6.7.1)-(6.7.2.), even in its simplified inertialess form $\dot{\mathbf{x}}=\mathbf{v}(\mathbf{x},t)$, where $\mathbf{v}$ is the velocity field that defines the flow $g_t\colon x(t)=g_t x(0)$, is a nonlinear differential vector equation corresponding to a dynamical system that in general depends on some parameters $\mu_1,\dots,\mu_s$ and for some values of them (and for certain initial conditions) may exhibit stochastic (chaotic) behavior. In fluids, manifestations of stochasticity are generally called turbulence, in this case chaotic behavior corresponds to the so-called Lagrangian turbulence (see, e.g., [92] and [116]). It is, however, important that both the advection and, more generally, the fluid flow dynamics within the Lagrangian framework can, due to the possibility of representing them in the dynamical systems form, be treated as Hamiltonian systems. The role of the Hamiltonian in fluid dynamics is in the two-dimensional case played by the deterministic stream function $\Psi(\mathbf{x}_\|,t)$ ([16], Supplement 2), where $\mathbf{x}_\|=(x^1,x^2)\in\Omega_2$. In a particular case when the 2d domain $\Omega_2$ occupied by the flow lies in the Euclidean plane with Cartesian coordinates $x^1=x, x^2=y$ on it, we have the Hamiltonian (symplectic) form of the motion equations $\dot{x}=v_x(x,y,t)=\partial\Psi/\partial y, \dot{y}=v_y(x,y,t)=-\partial\Psi/\partial x$. Here the domain $\Omega_2\equiv\Omega_{2+1}$ of variables $(x,y,t)$ corresponds to the extended phase space. In other words, the nonstationary planar flow is described by the Hamiltonian dynamical system with 1.5 degrees of freedom. Notice, however, that the stream function $\Psi(x,y,t)$ is a somewhat strange Hamiltonian: a pair of "canonical" variables $(x,y)$ are just local coordinates, and they are not canonically conjugate as, e.g., coordinate and momentum in mechanics.

## 6.10. Summing up fluid kinematics

Although all liquids are fluids, not all fluids are liquids. Fluid motion is modeled analytically and numerically in a macroscopic discipline known as fluid dynamics whose main subject of study is continuous media. The meaning of the word "continuous" is in fact already a model: any "small" fluid element – a fluid particle – is still big enough to contain $N \gg 1$ molecules. Dimensions (e.g. diameter) of a fluid particle $\delta x$ are supposed to satisfy the condition $d \ll \delta x \ll L$, where $d$ is a typical intermolecular distance, $d \sim n^{-1/3}$ and $L \sim V^{1/3}$ is a characteristic system dimension (an external macroscopic parameter, e.g., the pipe diameter, the wing length, the car body size, etc.). These are primary physical aspects of fluid motion modeling.

Fluid kinematics describes the motion of fluids ignoring the forces and moments which cause this motion. There exist various techniques to visualize the flows: streamlines, streaklines, particle lines, timelines as well as optical methods and computer graphics. All such techniques are beautifully presented in numerous textbooks on fluid mechanics and computational fluid dynamics (CFD) which is the collection of numerical methods allowing one to obtain solutions for the flow, therefore we shall not dwell on them.

Geometry of fluid flows mainly deals with their kinematics, when the action of external forces driving the flow between geodesics is ignored, nevertheless fluid kinematics, despite its apparent simplicity, has many intricate aspects that are important to understand when modeling fluid motion in countless engineering problems. For the reader's convenience, we sum up the elementary facts about fluid kinematics:

- All particles are individualized (classical mechanics!).
- Lagrangian coordinates $\{\xi^i\}$ are particle identifiers.
- Motion of continuous media and accompanying processes are described by physical fields (velocity, pressure, temperature, etc.). In case these fields are considered as functions of $\xi^i$, such description is called *Lagrangian* or *material.*
- Events in the Lagrangian picture occur with individual particles.
- The law of motion of a continuous medium: $x^i = x^i(\xi^j, t)$.
- Velocity of fluid particles: $u^i(\xi^i, t) = \partial_t x^i(\xi^j, t)$ constitutes the velocity field i.e., the velocity of particles located at $\{x^i\}$ at time $t$.
- Acceleration of particles in Lagrangian description: $a^i(\xi^i, t) = \partial_t u^i(\xi^i, t)$.
- *Euler's description*: fields of local variables considered as functions of $\{x^i, t\}$.
- Time derivative of a local scalar quantity $f = f(x^i, t)$ (for example, temperature or density): $D_t f = (\partial_t + u^k \partial_k) f$.
- Acceleration in Euler's description: $a^i(x^j, t) \equiv D_t u^i(x^j, t) = \partial_t u^i + u^k \partial_k u^i$.
- Transition from Lagrangian to Euler's description: $x^i = x^i(\xi^j, t) \mapsto \xi^i = \xi^i(x^j, t)$ so that $f(\xi^i(x^j, t)) \equiv \varphi(x^j, t)$.
- Transition from Euler's to Lagrangian description: to solve the Cauchy problem $\dot{x}^i = u^i(x^j, t), x^i|_{t=0} = \xi^i$. The solution, if it could be found $x^i(\xi^j, t) = \psi^i(\xi^j, t)$. Then for any $f(x^i, t)$ whose Euler's description is known, $f(\psi^i(\xi^j, t), t) = \mathrm{g}(\xi^j, t)$.

## 6.11. Transition to fluid dynamics

While describing fluid motion, there was only kinematics so far and no real physics i.e., no information about the nature or physical structure of the fluid was required. However, one needs this information in order to find the equations governing the fluid behavior. Apart from purely mathematical axioms and theorems, which are mostly related to the geometrical aspects of fluid motion, mathematical and computer modeling of fluid flows is based on certain physical hypotheses about structure, interaction law, etc. Besides, to produce really valuable mathematical and computational models of such complex systems as fluids one should also append experimental evidence to theoretical constructs. The main specificity of fluids is that, in distinction to rigid bodies, a liquid mass $M$ is contained within the domain $\Omega_t \equiv \Omega(t)$ whose shape may change in the course of motion. Mass $M = \int_{\Omega_t} \rho(x^i, t) d\Omega > 0$ for any fluid domain $\Omega_t$, hence $\rho(x^i, t) > 0, \{x^i, t\} \in \Omega_t$. Transition to dynamics is enacted by taking inertia into account (in classical mechanics, this property is phenomenologically accounted for by virtue of the 2nd law of motion).

In fluid dynamics, two principal states of motion are known: laminar and turbulent. Laminar flows are smooth and predictable, while the turbulent ones are of a complex nature characterized by interacting vortices. Turbulent flows are wider spread but much less understood. There are some heuristic models of turbulence, yet no theory exists enabling one to describe this phenomenon from first principles. Turbulence is often described not in terms of a random ensemble of waves (in general, of nonlinear ones), as the physical approach to fluid dynamics would suggest, but rather as a "soup" of metastable structures such as eddies, clumps, vortices, blobs, holes and similar phenomenological objects. It is interesting to notice that the phenomenological nature of such localized formations is manifested in the fact that they are not obliged to correspond to zeros of the collective response function such as a dielectric function $\varepsilon(\omega, \mathbf{k})$ in electromagnetic media [92] (e.g., for the case of plasma turbulence), whereas waves should correspond to $\varepsilon(\omega, \mathbf{k}) = 0$.

Turbulence, due to its stochastic nature, makes flows unpredictable, which is the usually cited reason for weather forecast failures even despite rapidly growing computational capabilities. Nonetheless, one can to some extent forecast weather in conditions of turbulent atmosphere, control oil and gas streams, design airplanes and missiles, etc. Turbulence as a physical phenomenon is often assumed to emerge when the flow passes through an instability regime (e.g., large eddies become unstable in the famous Kolmogorov model of turbulence), although general criteria of flow instability and the underlying physical mechanisms [154], [116], [127] are not quite clear despite tremendous research efforts such as extensive experimentation, theoretical work and numerical simulations. The flow is merely observed to become unstable at high enough Reynolds numbers (Re $\sim \rho u l / \eta$, where $\eta$ is viscosity), and such observations give rise to many phenomenological theories of turbulence. Yet a mathematical theorem accurately stating the conditions for the flow mixing, instability, chaoticity or increased complexity resulting in turbulence in (3+1)d spacetime does not seem to exist. More specifically, there is hardly any theorem stating the relationship between the Reynolds number and characteristics of turbulent flows such as correlation functions of the velocity field. The domain of Re between $10^2$ to $10^4$ is empirically viewed as transitional between laminar and turbulent flow regimes. At low Reynolds numbers (Re $< 10^2$) there is no turbulence, as viscous forces strongly dominate over inertial effects and any induced vortices are rapidly damped. At high Reynolds numbers, it is practically impossible to return a fluid to its initial state whereas it can be done for the laminar mode at low Re (provided all the forces, stresses and pressures acting on the fluid are reversed) (Figure 6). The flows in which turbulence plays a significant part are usually considered the most challenging ones to be analytically [92] or numerically [71] explored.

Thus, turbulence is a still only partially understood problem of classical physics – not many of such conceptually unsolved problems are left. Turbulent fluid is an example of an easily observable, ubiquitous physical system to which formal mathematics cannot be successfully applied – at least so far. Main difficulties in the theory of turbulence are related to the hierarchy of macroscopic fluctuations and to the abundance of strongly coupled degrees of freedom. One can consider a specific feature of turbulence that the traditional approaches of fluid dynamics proved to be nearly useless. For this reason, turbulence is currently treated more as a stochastic problem rather than that of fluid dynamics. Due to its stochastic properties, turbulence usually shuffles the concentration field of the passive scalar in an advection flow, but does not necessarily decrease its fluctuations.

Besides, maybe because of the notorious failure of the traditional fluid-dynamical techniques, turbulence has been largely neglected by the physics community for about half a century. Quantum electrodynamics, particle physics and gravitation attracted much more attention, although the engineering importance of turbulence was (and still is) incomparably higher. It is only due to the explosion of interest towards dynamical systems and chaos in the 1970s-1980s that the study of turbulence became popular again. Furthermore, turbulence is an essentially nonlinear process, and linearization so favored by physicists (see Section 5.9.) is in most cases totally inadequate to tackle the problems with a developed turbulence. Besides, another simplifying trick, building a lower-dimensional model, is poorly adapted to the study of turbulence: for example, the 2d turbulence is not *sensu stricto* a turbulence. Note that 2d fluid dynamics radically differs from the 3d case, where particle trajectories are not constrained to the plane and, therefore, can never be closed or self-intersecting. One consequence of this topological fact is that the developed 3d turbulence is a genuine spatially chaotic motion, whereas in the fluid restricted to a plane, there may be at maximum the "sea" of stochastic motion mixed with "islands" of regular (stable) quasiperiodic flow.

It is interesting that the concept of turbulence seems to be much more general than the solution of the Navier-Stokes equation for large Reynolds numbers. For example, chaotic turbulence-like behavior of human crowds following panic or alarm during mass events may easily lead to wide-spread traumatism and even massive loss of life. It is one of the urgent tasks of modern security planners to achieve the laminar motion of people in the process of evacuation, e.g., in emergency situations at the football stadiums.

We shall briefly discuss some approaches to modeling turbulence below, after we get acquainted with the equations used to describe fluid dynamics. The main among such equations is the famous Navier-Stokes equation (see Supplement 3 for its derivation).

The dimensionless form of the Navier-Stokes equation is

$$\frac{D\mathbf{u}}{Dt} = -\nabla p + \frac{1}{\mathrm{Re}}\Delta\mathbf{u} + \mathbf{f}, \qquad (6.11.1.)$$

where $D/Dt \equiv \partial/\partial t + (\mathbf{u}\nabla)$, $\mathrm{Re} = \rho\bar{u}L/\mu$ is the Reynolds number, $\mu$ is viscosity (internal friction) in the fluid, $L$ is a characteristic linear dimension of the fluid motion, $\bar{u}$ is the typical (average) flow velocity, $\mathbf{f}$ denotes external forces. Notice that, for incompressible fluid, density $\rho = \mathrm{const}$ and one usually replaces $\mu$ with $\nu = \mu/\rho$. If $\mathrm{Re}^{-1} = 0$ i.e., $\mathrm{Re} \to \infty$, the fluid is inviscid, and we have Euler's equation for incompressible inviscid flow. Viscosity in the Navier-Stokes equation is physically attributed to stress gradients in fluids leading to the dissipative motion between the fluid domains with different velocities. Since stresses are actually the forces, the momentum transfer due to viscosity is additional to that resulting from the pressure forces acting in fluids.

Viscosity actually plays a rather complex part in fluid flows: it can both damp their perturbations and bring about instabilities [92], [127]. The boundary conditions on boundary $\partial\Omega$ of domain $\Omega$ occupied by the fluid are either $\mathbf{u}|_{\partial\Omega} = 0$ which means that the flow stops at the boundary (typical for the Navier-Stokes equation with $\mathrm{Re}^{-1} \neq 0$) or $\mathbf{u}_n = \mathbf{nu}|_{\partial\Omega} = 0$ which means that the fluid cannot penetrate through the boundary so that its velocity normal to the boundary must vanish (typical for Euler's equation). The kinetic energy of the flow can be defined as

$$E = \frac{1}{2}\int u^i u_i d^3x = \frac{1}{2}\int g_{ik}u^i u^k d^3x, \ d^3x = dx^1 \wedge dx^2 \wedge dx^3 \qquad (6.11.2.)$$

Now, using the dynamical equations (Navier-Stokes or Euler's) we can compute the rate of change of this energy:

$$\frac{dE}{dt} = \int_\Omega \mathbf{u} \frac{D\mathbf{u}}{Dt} \, d^3x = \int_\Omega (-\mathbf{u}\nabla p + \mathrm{Re}^{-1}\mathbf{u}\Delta\boldsymbol{u}) \, d^3x = \mathrm{Re}^{-1}\int_\Omega \mathbf{u}\Delta\mathbf{u} \, d^3x = -\mathrm{Re}^{-1}\int_\Omega (\nabla\mathbf{u})^2 \, d^3x,$$

where the first term vanishes due to boundary conditions (e.g., $\mathbf{u}|_{\partial\Omega} = 0$) after integration by parts and external forces are assumed to be absent. Thus, if $\mathrm{Re}^{-1} = 0$ and integral $\int_\Omega (\mathrm{div}\,\mathbf{u})^2 \, d^3x$ exists, energy is preserved. In other words, Euler's equation does not imply dissipation: in this model molecular motion and macroscopic motion of the fluid characterized by velocity $\mathbf{u}$ are totally uncoupled.

Fluid kinematics and dynamics are, despite their somewhat intimidating mathematical gestalt, very practical disciplines, and the art of modeling fluid flows (especially by using numerical techniques developed in computational fluid dynamics) gains broadening demand. It is because of this persistent need for expertise in mathematical and computer modeling of fluids that we have devoted so much space to fluid motion.

## Section 7. Foundations of statistical mechanics

In this section, we discuss some aspects of statistical mechanics and thermodynamics in terms of their meaning for mathematical modeling. To be precise, the theory of thermodynamics is older, based on phenomenological principles and the work of Carnot, Kelvin and Clausius in the nineteenth century, covered concepts such as temperature and pressure, and the basic laws of thermodynamics. Later, the work of Boltzmann, Maxwell, and Gibbs led to the theory of statistical mechanics, based on first principles (Newtonian dynamics of molecules), and provided a basis for all of thermodynamics. Today, we are tempted to put all of these two fields into one basket, and often think of thermodynamics concepts, such as temperature and pressure, in terms of the motion of molecules.

Since many issues in the present manuscript *nolens volens* touch upon the concepts traditionally covered in statistical physics, we decided to introduce some of them in a very brief way. Besides, despite multiple years of research and considerable efforts of many outstanding scientists, the foundations of statistical physics remain a fairly obscure and controversial issue (although one can, as in quantum mechanics, ignore the foundations adopting the usual "shut up and calculate" prescription-based approach).

Most statistical theories in physics, in particular both classical and quantum equilibrium statistical mechanics, are constructed according to the same pattern [95] invented by J. W. Gibbs [67]. One considers a closed system $S$ that can be separated into two subsystems, $S = A \otimes B$, where it is usually assumed that $A \ll B$. The following terminology seems to be standard: the greater subsystem $B$ is called a heat bath (heat reservoir) or a thermostat, and the smaller one $A$ simply a subsystem or a system. Both $A$ and $B$ have too many degrees of freedom to be exactly known, controlled or even counted, but the influence of the smaller subsystem $A$ on the greater $B$ is assumed negligible. This asymmetry between the subsystems is manifested by the fact that the temperature $T$ of the heat bath $B$ is considered constant, with the smaller subsystem $A$ adopting the same temperature.

The closed system $S$ is described by Hamiltonian $H$ and, respectively, its evolution is unitary and described by a state vector $\Psi(t) = U(t)\Psi(0)$, where $U(t) = \exp(-itH)$ is the evolution operator (see below, in particular, Section 7.9. "The Liouville phase fluid"). In the general case, subsystem $A$ is described not by a pure state $\psi(t)$ (a vector in Hilbert space), but by the density operator (density matrix) $\rho_A(t) = \text{Tr}_B |\Psi(t)\rangle\langle\Psi(t)|$ which describes the unitary evolution of an open system. The density matrix of subsystem $A$ obeys the evolution equation

$$\frac{d\rho_A(t)}{dt} = i\text{Tr}_B[\rho(t), H],$$

where $\rho(t)$ is, in particular, $|\Psi(t)\rangle\langle\Psi(t)|$.

Density matrices $\rho(t)$ can be averaged in time

$$\bar{\rho} = \lim_{\tau \to +\infty} \frac{1}{\tau} \int\limits_0^\tau \rho(t)dt$$

to produce the equilibrium (or quasi-equilibrium) statistical states. It is usually assumed in the equilibrium statistical mechanics that the interaction between subsystem $A$ and heat bath $B$ is relatively weak so that $H \approx H_A + H_B$. In the limit, subsystems $A$ and $B$ can be totally independent at the outset so that initial state $\Psi(0)$ is factorized, $\Psi(0) = \psi(0)\Phi(0), \psi \in A, \Phi \in B$. If, in addition, the initial state is characterized by a small energy dispersion, then the equilibrium (quasi-equilibrium) state of subsystem $A$ is described by the Gibbs distribution, $\bar{\rho}_A = Z^{-1}exp(-\beta H_A)$, where $Z$ is the so-called partition function, $\beta \equiv 1/T$ is the inverse equilibrium (quasi-equilibrium) temperature (more exactly see, e.g., in [95]).

There exist a few facts that lie at the foundation of equilibrium statistical mechanics. These facts are mostly related to virtual insensitivity of the equilibrium mixed state[50] $\bar{\rho}_A$ to the choice of a specific initial state $\Psi(0) \approx \psi(0)\Phi(0)$. Namely for any (almost any) initial state $\Psi(0)$ the instantaneous state density matrix $\rho_A(t)$ tends to $\bar{\rho}_A$ and subsequently remains close to the latter[51]. It means that equilibrium state $\bar{\rho}_A$ of subsystem $A$ only weakly depends on the initial state $\psi(0)$ (more generally, $\rho_A(0)$) of subsystem $A$. Moreover, $\bar{\rho}_A$ weakly depends on the value of microscopic variables determining the initial state $\Phi(0)$ (more generally, $\rho_B(0)$) of heat bath $B$ (but only on its macroscopic functionals such as $\beta(\Phi)$).

We have seen above on the example of Bogoliubov's (BBGKY) approach to many-particle systems that exploiting the naturally occurring hierarchy of characteristic times may be of crucial importance (see Section 5.1. "Hierarchical multilevel principle"). Indeed, the BBGKY approach bridging the gap between first-principle (mechanical) approach and phenomenological models is based on taking into account the hierarchy of time intervals. The concept of time hierarchy seems to stem from nonlinear dynamics and characterizes the rate of processes occurring in the system. The existence of time hierarchy can be qualitatively understood from very general considerations that are close to Poincaré's ideas of return to any a priori chosen microscopic state. In particular, in the theory of dynamical systems that can be put in the foundation of modern statistical mechanics, the so-called Poincaré recurrences characterize the time-repetitive behavior of a system.

---

[50] A mixed state is understood as a weighted probabilistic mixture of pure states. A more "physical" depiction of a mixed state would be that the latter describes a subsystem (A) of a larger system (B) which is found in a pure state. Note, however, that it is inessential whether such a larger system (B) does really exist.

[51] This statement may be viewed as too vague since it inevitably leads to the question: how long it takes for subsystem $A$ to reach the state close to $\bar{\rho}_A$ and what kind of stability is thus attained.

The most general microscopic description of any physical system is achieved through using the density operator $\rho(t)$ (density matrix). If we write the latter for an evolving system in the energy representation i.e., with matrix elements

$$\big(\rho(t)\big)_{jk} = \langle j|\rho(t)|k\rangle = \exp\left\{-\frac{i}{\hbar}\big(E_j - E_k\big)t\right\}\rho(0)_{jk} = \exp\left\{-\frac{i}{\hbar}\big(E_j - E_k\big)t\right\}\langle j|\rho(0)|k\rangle, \quad (7.1.)$$

we shall see that there is an ordered set (hierarchy) of time scales $\tau_{jk} \sim \hbar/\Delta_{jk}, \Delta_{jk} \equiv E_j - E_k$. Notice by the way that the density operator considered as a time-evolving entity may be considered an almost-periodic function i.e., the initial state will be periodically reproduced with any arbitrary accuracy (recall that a complex continuous function $f(t), -\infty < t < \infty$ is known as almost-periodic if for any $\varepsilon > 0$ there exists interval $s(\varepsilon) > 0$ containing at least one translation period $\tau \subset s$ such that $|f(t + \tau) - f(\tau)| < \varepsilon$ for any $t$). Of course, one can write the density operator in any other representation corresponding to other self-adjoint operators, not necessarily the Hamiltonian, but then the establishment of the temporal hierarchy would not be so straightforward. We have chosen the energy representation since it is organically connected with time evolution through the spectral expansion of any process evolving in time. In Hamiltonian-based quantum mechanics, the time evolution operator, traditionally denoted as $U(t)$ ($U$ standing for "unitary")[52] is $U(t) = \exp(-itH)$, where $H$ is the Hamiltonian of a quantum-mechanical system, and can be represented as

$$U(t) = \exp(-itH) = \big(\psi, e^{-itH}\psi\big) = \int_{\mathbb{R}} e^{-it\lambda} d(\psi, E(\lambda)\psi). \quad (7.2.)$$

Here $(\psi, H\psi) \equiv \langle \psi, H\psi \rangle = \int_{\mathbb{R}} \lambda d(\psi, E(\lambda)\psi)$ is the inner product. The latter relation is known in the theory of linear operators as the spectral theorem, although it is almost obvious to physicists and widely spread in orthodox quantum mechanics. If $\|\psi\| \coloneqq (\psi, \psi)^{1/2} \equiv \langle \psi, \psi \rangle^{1/2} = 1$, then $d\mu \coloneqq d\langle \psi, E(\cdot)\psi \rangle$ is the probability measure. Since all quantum representations are mathematically equivalent, the transition between them has just a geometric meaning corresponding to the change of basis.

The time hierarchy typical of many-body systems also entails other hierarchies, in particular, of length, momentum or energy scales. In accordance with the time hierarchy, computer modeling of kinetic and transport processes, e.g., intended for handling directional flows, typically dissects the problem into two regimes: collisional and ballistic. At the collisional stage, particles are locally redistributed over directions and energies, with momentum and energy being preserved at each phase space mesh point. At the ballistic stage, the propagating particles are tracked in a phase space mesh (e.g., $m \times m$ in 2d or $m \times m \times m$ in 3d) until they suffer a transition due to a collision. Notice that in most cases one implicitly considers only the processes on background manifolds whose curvature length scales (in the simplest situation inverse curvatures) are much greater than the mean free path of the fluid at temperature $T$. Under this assumption, a manifold can be represented as a union of almost flat patches, the whole geometric construction resembling a soccer ball. Recall that a set of points $M$ in some Euclidean space $\mathbb{E}^m$ is usually understood as an $n$-dimensional manifold $M = M^n$ if every point $x$ of $M$ has a neighborhood that is homeomorphic to Euclidean space $\mathbb{E}^n$ i.e., a manifold is a locally Euclidean space. As an illustration one can recall that when a construction project is designed one usually assumes the earth to be flat over the building area.

---

[52] The reader could notice that in some cases we are using notation $g_t$ to express the evolution process; this was to emphasize that temporal evolution is not necessarily unitary.

As far as the abovementioned collisions between astrophysical objects are concerned, e.g., between stars in the universe, one can almost never worry about such events today, although there are about $5 \cdot 10^{11}$ stars in a typical galaxy. Indeed, the space volume occupied by the stars is exceedingly small compared to the inverse stellar density ($N_s R^3 \ll 1$, where $N_s$ is the average density of stars and $R$ is the characteristic star radius). The density of stars scattered in the intergalactic void is estimated as $N_s \sim 10^{-9}$ stars/ly$^3$ or 1 star per cube with a side of one thousand light years (ly). Therefore, one has to wait long enough to observe colliding stars, the respective time being of the order of $t\tau(N_s R^2 V)^{-1} \sim (N_s R^2)^{-1}(GM/N_s^{-1/3})^{-1/2} \sim N_s^{-1/6} R^{-2}(GM)^{-1/2}$, where $M$ is a typical stellar mass (say, $M \sim 10 M_\odot$, $M_\odot$ is the solar mass), $G \approx 6.67 \cdot 10^{-8} \text{cm}^3 \text{g}^{-1} \text{s}^{-2}$ is the gravitational constant. This crude estimate gives for the time between star collisions about $10^{17}$ years.

## 7.1. Statistical reasoning and stochastic considerations

Although the present book is devoted predominantly to deterministic dynamical systems, we shall have to cover some elementary facts about statistical reasoning and stochastic considerations used in the modeling of natural processes. In many cases, one simply cannot evade bringing statistical considerations. There are many other salient examples naturally involving statistical concepts such as dynamical systems, wave propagation, fluid motion (especially turbulence), performing and analyzing measurements, quantum theory – you name it. For instance, we used the well-known Fokker-Planck differential equation which can be regarded as an example of an effective modeling approach to natural probabilistic processes. Thus, the operation of sensitive measuring devices can be modeled by the one-dimensional version of the Fokker-Planck equation.

One mostly employs statistical methods to process raw data gathered through observations. The amount of observational data can be enormous (usually designated as Big Data). Big Data can be useful, especially when dedicated software techniques and powerful computers are employed, yet it is not a substitute for classic ways of understanding processes in the world. One still needs a theory, a model or to apply intuition in order to detect a cause. Metaphorically speaking, classical and, to some extent, modern science is permanently waging a war against the enormity of data about the world lest human intelligence would be drowned in data or smothered by it.

Statistical methods are also frequently used for the study of random processes. We have seen that random processes naturally emerge even within a completely deterministic framework. For example, chaotic phenomena arise as a direct consequence of instabilities in dynamical systems. Deterministic behavior is actually a lucky coincidence, and one can rarely avoid considering random effects that seriously affect determinism. We have already mentioned the crucial role of noise and fluctuations in practical problems, when the exact structure of mappings, operators, initial data and other quantities is either hypothesized or known very approximately. Hence there is often no reason to increase the accuracy of computations or numerical precision, as the initial modeling assumptions and measurement errors can be much rougher or even unknown.

For example, many mathematical models can be cast, as we have seen, to the form $Af = g$, where $A$ is an operator (linear or nonlinear) describing the behavior of a simulated system, $f$ being the latter state vector and $g$ being interpreted either as input data or as a forcing that drives the system, or as some other external source term. The increasing computing power can ensure rather good numerical evaluations of $f = A^{-1}g$, provided $A$ is non-degenerate and can be exactly defined together with source term $g$. In practice, however, coefficients of operator $A$, if expression $Af = g$ is represented by differential or integro-differential equations, or entries of the respective matrix as well as source function $g$ are rarely known exactly. Therefore, it may become reasonable to treat the source term and the parameters of operator $A$ (in general including its domain $D$) as random variables or fields. Then solution $f$ automatically becomes a random field, too, and we arrive at the problem of stochastic modeling which a priori implies the probabilistic description. The latter can be regarded as a more general technique than deterministic modeling. In particular, expression $Af = g$ becomes a stochastic differential, integral, integro-differential or matrix equation generalizing deterministic dynamical systems. Treatment of stochastic equations usually requires special approaches and approximation techniques that are not employed in deterministic models (e.g., stochastic calculus, homogenization, Karhunen-Loève approximation, etc.).

## 7.2. Some notes on Statistical Thermodynamics

The study of statistical systems, i.e., consisting of a large number of particles, starts with the question: what are the right variables to describe the behavior of statistical systems? Thermodynamics provides the simplest answer to this question.

One can notice a certain frustrating paradox in the second law of thermodynamics: on the one hand, all the systems should tend to the morgue state of absolute equilibrium while on the other hand there are plenty of examples in nature demonstrating in no way the continuous disorganization or progressive decay, but just the opposite, evolution and self-organization. All physicists seem to be aware of this paradox, sometimes formulated as the "heat death of the universe", but probably each physicist has her/his own version of treating it. The matter is that if the universe (or multiverse) has existed for a very long time, in some models infinitely long, then it must have reached the equilibrium state. Nevertheless, the world as a whole as well as its separate subsystems are apparently in a state rather far from thermal equilibrium, and one cannot observe the processes that would bring the universe into the equilibrium state. One possible solution to this paradox is that statistical physics and thermodynamics cannot be applied to the world as a whole. Probably, this inapplicability is related to the fundamental role played by gravity forces, for which even the Gibbs distribution can be introduced only with certain difficulties due to emerging divergences[53]. However, one cannot be sure that the declaration of non-validity of the usual statistical physics and thermodynamics for gravitating systems is an ample explanation of the absence of equilibrium state in the universe. Another popular explanation of the heat death paradox consists in the consideration that the metric of general relativity depends on time and therefore the universe is submerged into a non-stationary field [95].

---

[53] Free energy F/N counted per one particle diverges in the presence of gravitational interaction due to the long-range character of the gravitational forces (see, e.g., [19], see also [164]). Moreover, this quantity diverges also on small distances, but one can get rid of these divergences by restricting the possibility for the particles to condense, e.g., by introducing the exclusion principle at zero separation. Long-range forces imply a strong coupling of distant domains of the system, even widely separated. It is interesting to notice that similar divergences could be observed in the presence of Coulomb interaction, but they are removed due to the neutrality of Coulomb systems containing charges of opposite sign. This is impossible for gravity because mass is always positive.

The real implication of the "heat death" paradox is that one can easily transgress the limits of applicability of classical thermodynamics, without even noticing it. The problem with thermodynamics is that it tends to be extrapolated over the areas where it cannot be applied. Strictly speaking, classical thermodynamics is valid only for equilibrium states of the explored systems. However, the equilibrium conditions are only an idealization, a mathematical model and can be seldom encountered in nature. Thermodynamics is in essence a branch of physics studying the collective properties of complex steady-state systems. The main part of thermodynamics is purely phenomenological and may be treated completely independently from mainstream – microscopic – physics. Phenomenological thermodynamics historically served the industrial revolution, mostly driven by steam engines. These alluring machines produced a euphoria in the then existing society, similar to that created by today's digital gadgets. Steam machines fascinated not only the engineers, even such outstanding scientists, mainly physicists, as N. S. Carnot, E. Clapeyron, R. Clausius, J. B. J. Fourier, H. Helmholtz, J. Joule, Lord Kelvin (W. Thomson), J. C. Maxwell who were trying to formulate the laws governing the work of heat engines in a mathematical form. The most challenging task was to correctly describe the conversion of heat into mechanical work. Later L. Boltzmann, J. W. Gibbs, and M. Planck provided links from thermodynamics to the major body of physics.

Although time is not included in the set of thermodynamics, classical thermodynamics states that we live in a universe that becomes more and more disordered with time. The second law of thermodynamics, despite its apparently soft – statistical – character, is one of the hardest constraints ever imposed by the science on life and technology Historically starting from heat engines, classical thermodynamics still investigates the processes when energy, heat, and mass can be exchanged. There are two main laws governing such processes. While the first law of thermodynamics postulates the strict energy balance, the second law states that for all thermodynamic processes some portion of energy is necessarily converted into heat and dissipates in the environment. Such dissipation is irreversible: to undo it, one must spend some energy. Thus, losses are unavoidable. This is the informal gist of the second law of thermodynamics (without introducing a rather counterintuitive concept of entropy), which leads to some grave consequences of the second law such as the limited efficiency of heat engines or irreversibility of erasing the data in computer memory: this latter process generates heat [99].

The thermodynamical – in fact thermostatistical – approach to many-particle systems is simple but unsatisfactory because the crucial question: "how can the values of thermodynamical variables be derived from the equations of mechanics, classical or quantum?" is irrelevant. Thermodynamics was constructed as a totally independent discipline, an engineering isle whose study does not imply any knowledge of physics. Quite naturally, physicists were frustrated by such isolated standing of thermodynamics, and the first expansion of physicists into the field of thermodynamics was in the form of attempts to interpret such quantities as heat, work, free energy, entropy on a level of individual moving molecules.

Derivations of thermodynamic quantities from molecular models of physics are not at all easy, even for such a comparatively simple system made up of many particles as a rarefied gas. Three great physicists, all of them very proficient in mathematics, were the first to apply probabilistic concepts and, accordingly, to use the term "statistical" in the context of thermodynamics. Those were (in chronological order) J. C. Maxwell, L. Boltzmann, J. W. Gibbs.

Due to a greatly reduced number of variables, the behavior of statistical systems may look simple, but this is deceptive. Statistical irreversibility is a clear symptom of serious difficulties.

The apparent time reversibility of Newtonian dynamics formally allows us to observe many dramatic events such as the gas leaving the whole volume empty (say $1\,m^3$) it had just occupied. Time reversibility simply requires the gas molecules to follow trajectories back in time in the phase space However, nobody has ever observed such events, at least we believe so. The matter is that these time reversed paths as compared to "direct in time", those that result in filling up a prepared vacuum, though in principle possible, are characterized by such a tiny probability as to render any observable macroscopic effect unfeasible. Similarly, we would not expect spontaneous heating of some macroscopic body being initially at room temperature (even though there are not so few people who continue designing engines working on the principle of heat transfer from cold to warm bodies). We could perhaps expect this to occur in a nanoscopic system consisting of a small number ($N \leq 10$) of molecules – for a short period of time, but the experience shows that a spontaneous heating or cooling of a considerable piece of matter ($N \gg 1$), which would persist for a long period, e.g., sufficient to produce mechanical work, is considered an abnormal state and has never been observed. Such phenomena are of a fluctuative nature and can be modeled mathematically using an assortment of fluctuation theorems [10, 82, 57] (Jarzynski, Wojcik, Evans, Searles).

One might ask, why should we be concerned with rare events that can happen only on molecular time and distance scales and become almost totally improbable when we try to observe macroscopic phenomena? The answer is that these rare microscopic events are important for understanding the whole statistical thermodynamics. Besides, new experimental techniques have made microscopic and nanoscopic scales accessible to observation so that fluctuations and their probabilities which had been irrelevant before are becoming important for understanding experiments.

## 7.3. Statistical Equilibrium

In the foundation of equilibrium statistical mechanics lies some formal expression called Hamiltonian which must help us to find the probability distribution for any random physical process – and most physical processes are in fact random. For example, gas in some closed volume gives an elementary example of a random physical process; there may be less trivial systems, for instance, with interaction between the constituents. Since no physical system, probably except the entire universe, is isolated from the environment, one must account for its influence, e.g., by introducing some "physically natural" supplementary conditions. For example, initial or boundary values must be fixed, subjugation to which turns the physical system into a model.

We see that there exists a natural hierarchy of time scales (already mentioned in association with the hierarchical multilevel principle of mathematical modeling), this fact playing a crucial part in statistical mechanics as it determines the behavior of distribution and correlation functions. In particular, after time $\tau_0$, correlations between the particles are drastically weakened and many-particle distribution functions turn into the product of single-particle distributions $f(\mathbf{r}, \mathbf{p}, t)$ (the principle of correlation weakening was introduced by N. N. Bogoliubov) whereas for $t > \tau_r \gg \tau_0$, a single-particle distribution function tends to the equilibrium Maxwell-Boltzmann distribution[54]. In other words, the whole many-body system for $t \gg \tau_r$ reaches the state of statistical equilibrium so that the respective many-particle distribution functions tend to the canonical Gibbs distribution, $f_N(x_1, \ldots, x_N, t) \to \exp[\beta(F - H(x_1, \ldots, x_N))]$, where $\beta = 1/T$ is the inverse temperature, $F$ is the free energy and $H(x_1, \ldots, x_N)$ is the system's Hamiltonian. In other words, the single-particle distribution function $f(\mathbf{r}, \mathbf{p}, t)$ substantially changes over time scales $t \tau \tau_r \gg \tau_0$ whereas at the initial stage of evolution this function remains practically intact. Yet many-particle distribution functions can change very rapidly at short times comparable with the chaotization period $\tau_0$. Physically, one can understand this fact by considering spatially uniform systems with pair interaction between the particles, when many-body distribution functions depend on the coordinate differences of rapidly moving constituents. It is intuitively plausible that many-particle distribution functions would adjust to instant values of a single-particle distribution. To translate this intuitive consideration into the mathematical language one can say that for $\tau_r > t \gg \tau_0$ (intermediate asymptotics) many-particle distribution functions become the functionals of a single-particle distribution function

$$f_N(x_1, \ldots, x_N, t) \xrightarrow{t \gg \tau_0} f_N[x_1, \ldots, x_N; f(\mathbf{r}, \mathbf{p}, t)]$$

so that the temporal dependence of many-particle distributions is now determined by the single-particle function.

This idea (expressed by N. N. Bogoliubov) is rather important since it leads to a drastic simplification of the models describing many-body systems. In particular, although the respective distribution functions formally depend on initial data for all the particles, after a rather short time ($\tau_0 \sim 10^{-12} - 10^{-13}$ s) this dependence becomes much simpler since its relics are only retained in the relatively smooth single-particle function $f(\mathbf{r}, \mathbf{p}, t)$. One usually applies to this situation the notion of "erased memory" designating asymptotic independence of many-particle distribution functions on precise values of initial data – a huge simplification since initial values of all coordinates and momenta are never known exactly and, even if known, would be completely useless. In the modern world, the idea of fast forgotten details of microscopic instances, in fact of insensitivity to microscopics, has become especially useful for mathematical modeling of complex systems.

## 7.4 Statistical Ensembles

Let us start by recapitulating the basic notions of the idea of statistical ensembles. In conventional statistical mechanics, there has been a strong expectation that an ensemble average can correctly describe the behavior of a single particular system. Despite numerous attempts, there seems to be no rigorous mathematical proof for applying statistical ensembles to an individual observed system. The ensemble methodology lying at the very foundation of statistical physics still has the status of a logical presumption rather than of a compelling mathematical fact. A variety of good samples for concrete ensembles can be satisfactory from the physical point of view but require a more ample treatment for a cultivated mathematical taste. In physical terms, the default of ensemble methodology would mean that some quantities for an observed system may substantially deviate from the ensemble averages. Nonetheless, it would not be a tragedy; on the contrary, such deviations can provide important physical information about the observed system.

---

[54] One can imagine even tinier time scales in a many-body system, namely $\tau_0 \sim \tau_r/N$, where $N$ is the number of particles in the considered part of the system.

In the model of a microcanonical ensemble, one basically considers an isolated system, which is in fact not very interesting from a practical point of view. Explaining this requires a reminder of basic thermodynamics, which probabilists call large deviation theory. In the microcanonical ensemble one considers the number of microstates of an isolated system at some fixed value of internal energy $U$, volume $V$ and other extensive conserved quantities. The logarithm of this number of microstates is the entropy $S$ (it is convenient to set the Boltzmann constant $k_B = 1$) and by inverting this function one obtains the internal energy $U(S, V, \dots)$. From this so-called thermodynamic potential all other thermodynamic quantities (temperature, pressure, heat capacity and so on) can be computed as a function of the extensive quantities.

## 7.5. A brief distraction: differential forms and thermodynamics

While studying classical thermodynamics, many people experience some discomfort because one has to get accustomed to an unconventional mathematical style, when both differentials of functions and infinitesimal increments of the quantities that are not functions enter the same equations and are denoted by the same symbol $d$. This discrepancy is encountered already in the first law of thermodynamics, $dU = \delta Q - \delta A$, where $U$ is the internal energy (a function), $dU$ being its exact differential, whereas $\delta Q$ and $\delta A$ are elementary heat conveyed to a thermodynamic system and elementary mechanical work produced on it, respectively. Quantities $\delta Q$ and $\delta A$ are path-dependent and cannot be considered differentials of a function since integrating them does not produce a unique function so that they are sometimes called inexact differentials. To distinguish between these two different entities (exact and inexact differentials), exotic notations are introduced such as $\delta, \Delta$ or $đ$ demonstrating the "inexactness".

In fact, thermodynamics (at least its equilibrium part) is naturally treated using differential forms. The state of a thermodynamic system in equilibrium is determined by its temperature $T$ and a set of external parameters $a_1, \dots, a_m$. The inflow of heat can be described by 1-form $\theta = dU + \sum_{i=1}^{m} A_i \, da_i$, where $U$ is the internal energy of the system and $A_i, i = 1, \dots, m$ are "thermodynamic forces" corresponding to parameters $a_i$. Quantities $U$ and $A_i$ are functions of parameters $a_i$, and relationships $A_i = A_i(T, a_k)$ are usually called the equations of state. Parameters $a_i$ can often be treated as generalized coordinates, just as in Lagrangian mechanics. A typical example is $a = V$, where $V$ is the volume of a gas. Then the generalized force corresponding to volume $V$ is the gas pressure, $A = p$, and, e.g., for the model of a perfect gas the equation of state is $p = nk_B T$ or, in the customary Clapeyron-Mendeleev form, $pV = Nk_B T$, where $k_B = 1.38 \cdot 10^{-23}$ J.$\text{K}^{-1} = 1.38 \cdot 10^{-16}$ erg $\cdot \text{K}^{-1}$ is the Boltzmann constant, $N$ is the number of randomly moving and non-interacting particles in volume $V$, $n$ is their average density.

The model of perfect gas implies that the gas particles do not interact with one another, but are allowed to interact with boundaries or, e.g., with a piston inside the gas container. In the latter case there exists some effective interaction between the gas particles through the piston since it is a mechanical body that moves due to the difference of the pressure $p = nT$, $n = N/V$, $k_B = 1$ at two opposite sides of the piston or, one can say, due to the difference of the average collision frequency $\nu$ since $p = nT \sim (\nu / \sigma v_T) m v_T^2 \sim \nu m v_T / \sigma$, where $v_T$ is the thermal velocity and $\sigma$ is the scattering cross-section. One might note that the perfect (or ideal) gas model which corresponds to a greatly simplified many-particle system can be viewed as physically unrealistic. Indeed, only the interaction between particles which by definition is absent in a perfect gas can ensure that the thermodynamic system would be found in a certain phase state and can undergo phase transitions. Accordingly, it may be very difficult to implement the perfect gas in a physical experiment. A more realistic model for a rarefied and nearly perfect gas is the model of Knudsen's gas in which the particle's free path length $\bar{l} = (n\sigma)^{-1}$ is of the order of the system dimensions $L$ so that particles collide at times $\tau_c \sim L/v_T$.

Nevertheless, it is remarkable that the perfect gas model, apart from being a successful mathematical illustration of thermodynamic relationships, is in many cases quite close to reality. Most real gases under "standard conditions" can be regarded as ideal gases, this model only becomes invalid at low temperatures and high pressures when the particle size and some details (cross-section) of interparticle interactions become important. Besides volume $V$, such physical quantities as density $n_j$ of the $j$-th component (or concentration $c_j$ in a multicomponent system), electrical polarization $\mathbf{P}$, magnetization $\mathbf{M}$, charge density $\rho_e$, strain tensor $\epsilon_{ik}$, etc. can be used as parameters or "generalized coordinates" $a_i$. Then the corresponding "generalized forces" will be, respectively, chemical potentials $\mu_j$, electric field $\mathbf{E}$, magnetic field $\mathbf{H}$, scalar electromagnetic potential $\varphi$, stress tensor $\sigma_{ik}$, etc. Notice that thermodynamics can be constructed in almost complete analogy with classical mechanics (which was done by J. W. Gibbs and C. Carathéodory), with 1-forms $\theta = A_1(x^1, x^2, x^3)dx^1 + A_2(x^1, x^2, x^3)dx^2 + A_3(x^1, x^2, x^3)dx^3 \equiv P(x, y, z)dx + Q(x, y, z)dy + R(x, y, z)dz$ in $\mathbb{R}^3$ or $\theta = P(x, y)dx + Q(x, y)dy$ in $\mathbb{R}^2$ playing the most important role.

## 7.6. The heat and diffusion equations

The irreversible behavior of propagating heat as it spontaneously flows from hotter to colder places looks quite natural if we recall that the temperature is defined in kinetic theory as the average energy density, $T(r, t) = \int (\mathbf{p}^2/2m) f(r, p, t) d^3p$, so that it will pass from the regions of higher energy density to those of lower energy density, as in any diffusion process with density gradient. On a deeper level this fact is the consequence of the drastic discrepancy between the volumes of the different domains in the "coarse grained" phase space $\Gamma$ of the system which is just another description of the second law of thermodynamics: the phase point of a dynamical system passes with overwhelming probability into the larger and larger volumes. The observation of the heat flow from hot to cold led to the phlogiston concept: the temperature was assumed to be a measure of the density of some hypothetical fluid named phlogiston (from the Greek word "phlox" – flame) that is contained in any combustible substance. Accordingly, this caloric fluid would naturally tend to flow from high to low temperatures. Of course, the phlogiston model does not correspond to any physical reality, although one can find cases when this model can be applied for simple thermophysical calculations.

Nearly everything that is associated with the heat equation constitutes a simple linear classic theory, yet there are some links of this theory protruding into rather sophisticated areas of modern mathematical physics. The same (and even to a greater degree of sophistication) could be said about the diffusion equation and diffusion processes in general. We may remark in passing that, until the 1960s-1970s, mathematical physics was almost completely associated with the theory of linear PDEs such as heat transfer, diffusion, Laplace, Helmholtz, wave and other classical equations. Now mathematical physics has been extended to include new – mostly nonlinear – structures and models, yet even in the 1960s mathematicians mainly looked at these attempts to modernize mathematical physics with an almost demonstrative indifference, while physicists viewed them with scornful irony.

## 7.7. Potentials

Classical, in particular Newtonian, mechanics served for Gibbs as an efficient mathematical model for thermodynamics [67]. Specifically important was the concept of potentials transferred from mechanics. We have seen that forces in mechanics acting in a conservative system can be obtained from a scalar function known as the potential by simple differentiation (taking the gradient). For example, in the system with a single degree of freedom, force $F$ is obtained as $F = -d\phi/dx$, where $\phi = \phi(x)$ is the potential energy or simply potential (we do not denote the potential by the usual $V(x)$ in this context in order to avoid confusion). The name "potential" is quite natural since it indicates that there exists the possibility to convert $\phi$ into the energy of motion or into the work produced by force $\mathcal{F}$. In thermodynamical systems such as, e.g., fluids, the mechanical force $\mathcal{F}$ is translated into pressure $p$ (more generally, stress tensor $\sigma_{ij}$) that defines the force per unit area. One can then assume that there exists some effective potential $U$ giving pressure i.e., the force applied on a surface (at least its scalar, isotropic part) through differentiation, $p = -dU/dV$, where $V$ is the volume of a thermodynamic system. Here pressure (or force) is understood as an average momentum transfer or, using the ergodic assumption, over time. However, thermodynamic systems are more complex than mechanical ones, being necessarily multi-parametric. In particular, the form $\delta A = -pdV$ is not in general equal to the change of internal energy $dU$. Indeed, thermodynamic work can be produced due to one more source of energy, namely owing to the heat $\delta Q$ received by the system (e.g., by fluid) from the environment. This is the essence of the first law of thermodynamics, $dU = \delta Q + \delta A$, where $\delta A$ is the work produced on the system, or in thermodynamic variables, $dU = TdS - pdV$, where $S$ is the entropy. Today, we perceive the first law as an obvious – almost tautologic – statement: one more energy balance formulation, but originally the first law of thermodynamics was a bold step (made by Clausius around 1850 and presumably even earlier – around 1824 – by Carnot). The matter is that it was not at all obvious at that time that heat can be equivalent to mechanical energy and to mechanical work. Recall that the dominating model of heat was the one of phlogiston, with the latter being difficult to imagine as a source of mechanical work.

We have already mentioned that classical thermodynamics can be easily cast in the language of differential forms. In 3d, for example, differential form $\theta = A_1(x^1, x^2, x^3)dx^1 + A_2(x^1, x^2, x^3)dx^2 + A_3(x^1, x^2, x^3)dx^3 \equiv P(x, y, z)dx + Q(x, y, z)dy + R(x, y, z)dz$ is known as *exact* on domain $\Omega \subseteq \mathbb{R}^3$ if there exists on $\Omega$ a scalar function $U = U(x^1, x^2, x^3)$ such that $dU = \partial_i U dx^i = A_i dx^i, i = 1,2,3$ in $\Omega$.

The internal energy $U$ can be viewed as a potential energy only in thermal isolation i.e., when heat $\delta Q$ supplied to a thermodynamic system vanishes, $\delta Q = TdS = 0$. Accordingly, one has to write in thermodynamics not merely $p = -dU/dV$, analogously to $F = -d\phi/dx$ in mechanics, but $p = -(\partial U/\partial V)_S$, where subscript $S$ designates that the partial derivative is taken under the condition of the entropy being kept fixed. One also uses the Jacobian form for such partial derivatives, e.g.,

$$p = -\frac{\partial(U, S)}{\partial(V, S)} = -\begin{vmatrix} \partial U/\partial V & \partial S/\partial V \\ \partial U/\partial S & \partial S/\partial S \end{vmatrix} = -\frac{\partial U}{\partial V} + \frac{\partial U}{\partial S}\frac{\partial S}{\partial V} = -\left(\frac{\partial U}{\partial V}\right)_S.$$

One might note that more often in thermodynamics (and in general in macroscopic physics) one deals with the systems that are not in thermal isolation from the environment, but are in thermal equilibrium with it. For instance, such macroscopic systems are sustained at fixed temperature $T$ (in "heat bath"). In such thermally reversible case, there also exists an effective potential $F$ giving force $F$ (expressed as pressure $p$) by the formula $p = -(\partial F/\partial V)_T$ i.e., by fixed $T$. Potential $F$ is called free energy and can obviously be written as $F = U - TS$. One sometimes calls the term $TS$ in the additive combination $U = F + TS$ the "bound energy", in distinction to the free energy $F$. It is easy to see that $dF = -pdV - SdT$ so that free energy is the potential with respect to variables $V$ and $T$ and entropy $S = (\partial F/\partial T)_V$, a very useful formula in thermodynamics and statistical physics. The usefulness of the free energy in statistical physics is mostly due to the expression for the partition function $Z = \sum_i \exp\left(\frac{E_i}{T}\right), F = -T \log Z$ (here we put the Boltzmann constant $k_B = 1$, measuring temperature in energy units).

Thus, for $T = $ const (thermal equilibrium) i.e., when $dT = 0$, the pressure is proportional to the rate of decrease of potential $F$, just as the force in mechanics is proportional to the rate of decrease of mechanical potential $\phi = \phi(x)$.

One might observe that in many practically important situations, e.g., in thermal engineering, keeping the pressure fixed, in particular when applied from outside, is more realistic than confining the medium to a fixed volume. In such situations, one more thermodynamic potential is used, namely enthalpy $H = U + pV$, so that $dH = TdS + VdP$ and $T = (\partial H/\partial S)_p$. Enthalpy is obviously the potential with respect to entropy $S$ and pressure $p$, in complete analogy with the internal energy that is the potential with respect to entropy $S$ and volume $V$. If the pressure remains constant, elementary heat $\delta Q$ received (or exchanged) by the system is equal to $(\delta Q)_P = d(U + pV) = dH$, just as $(\delta Q)_V = dU$ in case the volume is kept constant. In this latter case temperature is similar to some effective force obtained from the potential $U$. There are some more useful potentials introduced by J. W. Gibbs, in particular for the case when the macroscopic system can exchange with the environment not only by heat and energy, but also by particles. One can read about thermodynamic potentials from the physical positions in classical textbooks [95].

Classical thermodynamics is based on the concept of equilibrium in macroscopic systems. Using this concept can be a rather strong assumption since it is in most cases a priori not obvious whether a macroscopic system in question has reached equilibrium. This especially concerns open systems which can exchange energy, entropy and matter with the environment. Besides, in certain parts of any thermodynamic system considerable deviations of local values of, e.g., energy and particle densities from the average level permanently occur. For macroscopic systems, equilibrium is understood as the state which is defined by the fixed external parameters $\mathbf{a} = \{a_1, ..., a_m\}$ and temperature $T$ of the homogeneous system $(\partial T/\partial x^i = 0)$. Physically, it signifies that there exist time intervals $\tau$ during which all internal parameters $\mathbf{b} = \{b_1, ..., b_n\}$ (in particular, "thermodynamic forces" $A_i, i = 1, ..., m$ corresponding to parameters $a_i$) have reached their relaxation values $\tau_r \ll \tau$, but the variations of external parameters $\mathbf{a}$ and temperature $T$ are negligibly small (maybe up to unavoidable small-scale fluctuations). Eventually, a single parameter survives, characterizing the equilibrium state that is temperature[55] $T$.

As will be discussed in Section 7.9. ("The Liouville phase fluid"), the thermodynamic limit corresponds to an infinite number of the Hamiltonian system of equations and, thus, of the respective phase space. The motion of an infinite number of particles described by an infinite number of coupled differential equations is hardly possible to find. Therefore, only average values and asymptotic properties (particularly for $t \to \pm\infty$) are of practical interest. When the system strives to equilibrium, the number of parameters defining the many-particle distribution function (phase density) diminishes i.e., the dynamics of a non-equilibrium system becomes simpler. Correlations between individual particles and different parts of the system become weaker, fluctuations die out, specific initial conditions are forgotten, and finally, as the equilibrium state has been reached, the system will be settled to a constant and uniform temperature $T$.

---

[55] It is useful to remember that temperature as a measure of an average energy can be expressed both in Kelvins and in grams, e.g., 1 GeV approximately equals $10^{13}$ K and $2 \cdot 10^{-24}$ g.

## 7.8. The Bogoliubov Chain

In this section, we are going to study mainly the stationary solutions of the Bogoliubov hierarchy equations. In literature this chain of equations – an infinite system of integro-differential equations for the many-particle distribution functions – is generally known as Bogoliubov-Born-Green-Kirkwood-Yvon (BBGKY) hierarchy. For systems in a finite volume these equations are equivalent to the dynamical Liouville equation and characterize the time evolution of the probability measure on the phase space of a finite number of particles. Performing the thermodynamic limit, one obtains the infinite chain of the Bogoliubov hierarchy equations which are related to a system of particles in the whole space. As near as I know, the problem of existence and uniqueness for this chain of equations has not been solved so far. It is natural to connect the stationary solutions of the Bogoliubov hierarchy equations with the states of an infinite system of particles (i.e., probability measures defined on the phase space) which are invariant with respect to time evolution. In the cases where the dynamics on the phase space has been constructed it is possible to demonstrate that any invariant measure satisfying further conditions of a general type generates a stationary solution of the Bogoliubov hierarchy equations. On the other hand, an immediate analysis of stationary solutions of the Bogoliubov hierarchy equations (unlike the invariant measures) does not require, in general, the use of such delicate dynamical properties as clustering. Apparently, the point is that only functions of a finite (although not bounded) number of variables enter the Bogoliubov hierarchy equations. One can consider these functions (the correlation functions) as integral characteristics of a measure and their behavior must not necessarily show the influence of singularities arising from the motion of individual configurations of an infinitely large number of particles. Thus, the approach based on the Bogoliubov hierarchy equations seems not only to be more general but also more natural from the physical point of view.

We shall also discuss the derivation of the kinetic equation for a classical system of hard spheres based on an infinite sequence of equations for distribution functions in the Bogoliubov (BBGKY) hierarchy case. It is known that the assumption of full synchronization of all distributions leads to certain problems in describing the tails of the autocorrelation functions and some other correlation effects with medium or high density. We shall discuss how to avoid these difficulties by maintaining the explicit form of time-dependent dynamic correlations in the BBGKY closure scheme.

The question usually is how to obtain hydrodynamic equations (Euler, Navier-Stokes) from the Liouville-type equations of Hamiltonian mechanics, classical or quantum. The original idea was due to Ch. Morrey (1956) who introduced a concept of a hydrodynamic limit and was able to formally derive a Euler equation from the classical Liouville equations (more precisely, from the corresponding BBGKY hierarchy). However, Morrey had to make some assumptions about the long-term behavior of the motion, and this included a statement on ergodicity, in the sense that all 'reasonable' first integrals are functions of the energy, linear momentum and the number of particles. Since then, the idea of a hydrodynamic limit became very popular in the literature and has been successfully applied to a variety of models of (mostly stochastic) dynamics relating them to non-linear equations. However, in the original problem there was no substantial progress until the work by S. Olla, S. R. S. Varadhan and T. Yau (1992) where Morrey's assumptions were replaced by introducing a small noise into the Hamiltonian (which effectively kills other integrals of motion), and a classical Euler equation was correctly derived. In some quantum models (e.g., of the Bohm-Madelung type) the hydrodynamic limit can be rigorously demonstrated. The resulting Euler-type equation is similar to the one that arises for the classical counterpart of these models. This suggests that perhaps classical and quantum hydrodynamic equations must look similar if they are written for local densities of 'canonical' conserved quantities (the density of mass, linear momentum and energy).

## 7.9. The Liouville phase fluid

The "physical" balance equations discussed above are of phenomenological character and should be obtained from more profound principles. Such principles have been developed in kinetic theory and, interestingly enough, also have the meaning of balance conservation laws. Thus, one of the main equations of kinetic theory, the Boltzmann equation for the one-particle distribution function, has the form of the balance equation in the phase space of a single particle (sometimes called $\mu$-space). As it has already been mentioned, the Boltzmann equation itself can be obtained by a hierarchical chain procedure, the Bogoliubov-Born-Green-Kirkwood-Yvon (BBGKY) sequence from the most fundamental Liouville equation. The latter equation, which is a starting position for deriving kinetic and subsequently fluid dynamics equations, is also of the balance type, expressing the conservation of the number of phase points characterizing the system of $N$ particles in process of its temporal evolution. The mathematical model corresponding to the Liouville equation is based on the concept of the flow of "gas" of phase points $x = (x^1, \dots, x^n)^T, x^i = (\mathbf{r}_i, \mathbf{p}_i)$ [56] moving in a $n = 6N$ phase space ($\Gamma$-space). This is an important idea so let us expand on it a little further. Each point in the $\Gamma$-space corresponds to the whole set of locations and momenta of all $N$ particles together. For example, motion of the planets in the Solar system can be represented as a flow in 54-dimensional (or 48, if Pluto is excluded) phase space. Thus, evolution of a dynamical system is represented by the motion of a single point in the phase space and so a phase path emerges. The solution of dynamic equations $\dot{\mathbf{x}} = \mathbf{f}(\mathbf{x}, t)$, e.g., in mechanics, $\mathbf{r}_i = \mathbf{r}_i(\mathbf{r}_{i0}, \mathbf{p}_{i0}, t)$, $\mathbf{p}_i = \mathbf{p}_i(\mathbf{r}_{i0}, \mathbf{p}_{i0}, t)$ determines the phase path. If we consider some domain $\Gamma_0$ in the phase space and regard all the points inside this domain as initial conditions, then we can think of these points as comprising some "phase fluid". The phase flow (that makes the phase fluid flow $g_t$) is given by the motion equations i.e., in the operator form $x(t) = U(t, t_0)x(t_0)$, where the operator of unitary evolution $g_t = U(t, t_0)$ is a volume-preserving diffeomorphism which ensures that $\Gamma_0 \rightleftarrows \Gamma_t$. Recall that flow $g_t$ generated by a dynamical system $\dot{x}^i(t) = v^i(x^j, t), \mathbf{x} = \{x^j\} \in M^n, t \in \mathbb{R}$ is a diffeomorphism from $M^n$ to $M^n$ i.e., a smooth invertible transformation (Figure 3). In other words, phase flow is a map $g_t: M \times \mathbb{R} \to M$, $x(t) = g_t x(t_0) \equiv g_t x_0, x_0 \equiv x(t_0) \in M, x(t) \in M$ is a point in $M$ provided it started at $x_0$ at $t = t_0$ moving for time $t$ in vector field $\mathbf{f}(\mathbf{x}, t)$. Contrariwise, if map $x(t) = g_t x(t_0), M \times \mathbb{R} \to M$ is known, one can restore vector field $\mathbf{f}(\mathbf{x}, t)$ that generates this flow by differentiation.

Since there are no sources or sinks in the $N$-particle ensemble (although relative positions of the phase points do change in process of evolution), one can write the continuity equation for the $N$-particle distribution function in the phase space, $f_N(x, t)$:

$$\partial_t f_N(x, t) + \nabla_x(\dot{x} f_N(x, t)) = \partial_t f_N(x, t) + \sum_{i=1}^{N}\left(\frac{\partial}{\partial \mathbf{r}_i}(\dot{\mathbf{r}}_i f_N)\right) + \sum_{i=1}^{N}\left(\frac{\partial}{\partial \mathbf{p}_i}(\dot{\mathbf{p}}_i f_N)\right) = 0. \quad (7.9.1.)$$

Or, using the identity

$$\frac{\partial}{\partial \mathbf{r}_i}(\dot{\mathbf{r}}_i f_N) + \frac{\partial}{\partial \mathbf{p}_i}(\dot{\mathbf{p}}_i f_N) = f_N\left(\frac{\partial}{\partial \mathbf{r}_i}\dot{\mathbf{r}}_i + \frac{\partial}{\partial \mathbf{p}_i}\dot{\mathbf{p}}_i\right) + \left(\dot{\mathbf{r}}_i\frac{\partial}{\partial \mathbf{r}_i}\right)f_N + \left(\dot{\mathbf{p}}_i\frac{\partial}{\partial \mathbf{p}_i}\right)f_N, \quad (7.9.2.)$$

---

[56] Here, the lower index denotes simply the particle number having coordinate $\mathbf{r}_i$ and momentum $\mathbf{p}_i$, in distinction to the upper vector index.

where the first term is zero when Hamiltonian mechanics holds (see below), we get the standard – Hamiltonian – form of the Liouville equation

$$\partial_t f_N(x,t) + \{H, f_N(x,t)\} = 0, \tag{7.9.3.}$$

where symbol $\{.\,,.\}$ denotes the Poisson brackets:

$$\{H, f_N(x,t)\} := \sum_{i=1}^{N} \left( \frac{\partial H}{\partial \mathbf{p}_i} \frac{\partial f_N}{\partial \mathbf{r}_i} - \frac{\partial H}{\partial \mathbf{r}_i} \frac{\partial f_N}{\partial \mathbf{p}_i} \right) = 0. \tag{7.9.4.}$$

Here $H = H(x) = H(\mathbf{r}_i, \mathbf{p}_i)$ is the Hamilton function of the physical system or the Hamiltonian that plays a crucial role both in classical and quantum mechanics and will be discussed in the respective sections. Now, it is only important to understand that the form of the Liouville equation is considered as a balance conservation law and is, accordingly, the first-order partial differential equation is determined by the system Hamiltonian. In a mathematical sense, this equation differs from the above one- or two-dimensional examples only by the number of variables. For instance, to describe the processes in fluids and sometimes even in solid state the following model Hamiltonian is usually chosen:

$$H(x) = H(\mathbf{r}_i, \mathbf{p}_i) = \sum_{i=1}^{N} \left( \frac{\mathbf{p}_i^2}{2m} + U(\mathbf{r}_i) \right) + \frac{1}{2} \sum_{i,j=1}^{N} V(|\mathbf{r}_i - \mathbf{r}_j|) = 0. \tag{7.9.5.}$$

One often uses the notion of the Liouville operator defined by the relationship

$$\partial_t f_N(x,t) + i\mathcal{L} f_N(x,t) = 0. \tag{7.9.6.}$$

The Liouville operator is defined on the phase space $\Gamma$ and performs time translations along the phase trajectories (Hamiltonian flow). The respective dynamical (Hamiltonian) system generates the transformation group on the phase space. The Liouville operator $\mathcal{L}$ (or, rather, superoperator since it acts on other operators) being defined through (7.9.6.) and determining the evolution is Hermitian, which implies that eigenvalues of $\mathcal{L}$ are real. Evolution of the $N$-particle distribution function (a volume-preserving diffeomorphism $\Gamma_0 \rightleftarrows \Gamma_t$) is given by $f_N(x,t) = U_{\mathcal{L}}(t, t_0) f_N(x, t_0)$ or, more generally, the evolution of phase density $\rho(t) = U_{\mathcal{L}}(t, t_0)\rho(t_0)$. The respective evolution operator $U_{\mathcal{L}}(t, t_0) = \exp - [i(t - t_0)\mathcal{L}]$ is unitary: physically, it means that there is no irreversibility in the Hilbert state description of time evolution of many-particle systems – in complete accordance with Hamiltonian mechanics. The operator exponent is defined as usual by

$$\exp - [i(t - t_0)\mathcal{L}] = \sum_{k=0}^{\infty} \frac{(t - t_0)^k}{k!} (-i\mathcal{L})^k. \tag{7.9.7.}$$

Notice that if we consider the system of $N$ particles to be a subsystem of a larger system $M$ (e.g., the case of an open system), operator $\mathcal{L}$ can depend both on spatial coordinates and time $t$: in the latter case evolution is non-autonomous, and the spatial and time variables can only be approximately separated.

There is, however, a fine point here. Unitary evolution represented through the Liouville operator is $\rho(t) = \exp - [i(t - t_0)\mathcal{L}]\rho(t_0)$ whereas representing the same evolution via the Hamiltonian evolution operator gives $\rho(t) = U(t, t_0)\rho(t_0)U^+(t, t_0)$, where $U(t, t_0) := \exp -[i(t - t_0)H]/\hbar$. The latter expression is in fact a formal solution of the von Neumann equation

$$\frac{\partial \rho(t)}{\partial t} = \frac{i}{\hbar}[\rho, H], \qquad (7.9.8.)$$

which may be considered the quantum analog of the classical Liouville theorem of statistical mechanics. Here, square brackets $[.,.]$ denote, as usual, a commutator $[F, G] := FG - GF$. The von Neumann equation is often written as $\frac{\partial \rho(t)}{\partial t} + i\mathcal{L}\rho = 0$, in the case superoperator $\mathcal{L}$ is given by $\mathcal{L}G = \frac{i}{\hbar}[H, G]$. Recall that the general quantization condition (see, e.g., [149]) states $[F, G] = i\hbar\{F, G\}$.

One must note the sign difference in the von Neumann equation with the Heisenberg equations of motion: the matter is that phase density $\rho$ is the Schrödinger and not the Heisenberg operator. In general, we tend to treat dynamics as performed by timeless operators acting on time-dependent (evolving) states, but such a representation in many cases can only be an approximation.

It is important to bear in mind that even in the classical case the $N$-particle distribution function (probability density) contains far more information that is really needed for any practical purpose. The observables such as currents, kinetic coefficients, expectation values of some quantities or correlators between them typically require one-particle, maximum two-particle distribution functions. The Hamiltonian function (7.9.5.) containing three terms – kinetic energy of particles, their potential energy in an external field and energy of interparticle interaction – corresponds to a very specific case of the pair interaction between the bodies constituting the physical (or any other) system, this interaction depending solely on the distance between the bodies (particles). In condensed matter theory, when particle density is high, this model is often inadequate and should be replaced by more realistic ones.

The transition to fluid dynamics i.e., to the flow of matter, given the assumption that the constituent particles of the matter interact according to the laws of classical mechanics[57], is nonetheless a mathematically complicated problem. Indeed, one can formulate this problem as follows [26]. Let parameter $\varepsilon > 0$ describe the transition between the microscopic and macroscopic scales. One can assume that the initial distribution (at $t = 0$) is locally homogeneous in the scale $1/\varepsilon$ i.e., this distribution is almost translationally invariant under spatial shifts by $l \ll l_0/\varepsilon$ where $l_0$ is a characteristic size of the initial distribution (correlation length). Recall that in statistical mechanics the distribution of the system of $N$ particles is understood as the probability law for a random field on phase space $\Gamma \equiv \Gamma_N$. In other words, if we look at volumes $V \sim l^3 \ll (l_0/\varepsilon)^3$, we require that the main statistical characteristics (such as, e.g., one-particle correlation functions or integrals of the number of particles, momentum, energy, etc.) should be almost translation-invariant within volume $V$; this would be a macroscopically small volume or "physically infinitesimal" one.

One should not, however, take the analogy between the phase fluid and usual liquid or gas too literally, as there is no interaction between the constituent particles of the phase fluid. Besides, the phase space, where the "phase fluid" flows, is a symplectic manifold (see below) of an enormous number of dimensions: for a classical system of $N$ particles with no internal degrees of freedom, it is a space of $6N$ dimensions. The volume of the phase space is given by the "Liouville measure" $\prod_{i=1}^{N} dp_i dq^i$. For instance, in a liter (1000 cm³) of gas we have at the standard atmospheric pressure $N \sim 10^{22}$ molecules so that the respective phase space contains $\sim 6 \cdot 10^{22}$ dimensions[58]. When the number of particles $N$ corresponds to a macroscopic body, any attempt of finding a solution of the respective Hamiltonian system is hopeless, and one has to resort to probabilistic models. The fact that a phase space of a macroscopic system has a tremendous number of dimensions plays a crucial role in statistical mechanics, where the notion of entropy characterizing the degree of randomness, variability and disorganization in a many-body system is one of the most fundamental. Thus, statistical mechanics arises from the lack of knowledge, and its models study not the individual (nor even $N$-particle) trajectories $\boldsymbol{\gamma}(t)$ as in mechanics or in the theory of dynamical systems, but the evolution of probability measures on the phase space. This evolution is, however, defined by the underlying $2Nd$ Hamiltonian system ($d$ is the dimensionality of the model; in most physical problems $d = 1, 2, 3$). Quite naturally, for $N \sim 10^{22} - 10^{24}$ one might be interested in asymptotic regimes, in particular at $N \to \infty$. The $N \to \infty$ limit is often called the thermodynamic one. Here, one of the most intriguing and principal issues emerges: that of the relationship between dynamical and statistical description of the evolution of macroscopic (in particular, complex) systems.

---

[57] This is a drastic simplification since classical mechanics usually cannot be applied at small distances when quantum correlations become noticeable.

We should not, however, forget that mathematically such a limit corresponds to an unbounded expansion of the Hamiltonian system of equations and, thus, of the respective symplectic structure (the phase space). Indeed, motion of a system of $N$ particles of dimension $d$ is described by $2Nd$ coupled first-order differential equations. A correct mathematical description of the Hamiltonian system for an infinite number of particles, albeit of the same sort i.e. just $N \to \infty$, is very difficult since this limit corresponds to a mechanical model described by an *infinite* number of coupled differential equations. Finding exact dynamics of such a system is equally hopeless and useless already because one cannot fix all initial conditions for an infinite number of trajectories. Therefore, only average values and asymptotic properties (particularly for $t \to \pm\infty$) are of practical interest.

Many-particle systems differ by the states of individual particles. When the particles are structureless, these states are those of motion. In classical (Gibbs) statistical mechanics the distribution of $N$-particle systems is given by the phase density $f_N(x, t) \equiv f_N(x_1, \ldots x_N, t)$, $x_i = (\mathbf{r}_i, \mathbf{p}_i)$ in $(6N + 1)$-dimensional space $\mathbb{R}^{3N} \times \mathbb{R}^{3N} \times \mathbb{R}$. Phase density gives the probability $f_N(x_1, \ldots x_N, t) dx_1 \ldots dx_N$ that the $N$-particle system can be found at time $t$ in the state when phase variables (coordinate and momentum) of the first particle ($i = 1$) are in infinitesimal volume $dx_1 = d^3 x_1 d^3 p_1$ near $x_1 = (\mathbf{r}_1, \mathbf{p}_1)$, of the second particle ($i = 2$) in $dx_2 = d^3 x_2 d^3 p_2$ near $x_2 = (\mathbf{r}_2, \mathbf{p}_2)$, etc. It is clear that the Liouvillian phase density can be normalized

$$\int_{(P)} f_N(x_1, \ldots x_N, t) dx_1 \ldots dx_N = 1, \qquad (7.9.9.)$$

where $P$ is the phase space of the $N$-particle system, since the latter must be found in some state. The states, according to Gibbs, are no longer points of the phase space, but measures.

Note that one usually takes for granted that the Liouville phase fluid is continuous in each point $(x, t) \in \Gamma$ which enables us to write the continuity equation, to prove the Liouville theorem (volume-preserving diffeomorphism, $\Gamma_t = \Gamma_0$ under $g_t$) and to develop the theory of kinetic equations. Strictly speaking, continuity of the phase fluid is an assumption or a postulate. For instance, one cannot guarantee the absence of singular points for dynamical systems deviating from the Hamiltonian ones. Moreover, both for Hamiltonian and non-Hamiltonian systems, one can find the portion of the ensemble points within any small phase space area by merely counting the ensemble members. Neither continuity (with respect to some topology or a norm) nor more stringent analytic properties of the phase space density are *a priori* provided, although it has always been a basic tenet of classical physics that everything is continuous in nature. The Liouvillian phase density $f_N(x_1, \dots x_N, t)$ is not necessarily continuous on every compact subset of phase space $\Gamma$; more than that, it is not necessarily defined at every point $(x, t) \in \Gamma$. Intuitively speaking, there can be some regions of the phase space from which $f_N(x_1, \dots x_N, t)$ does not map into certain intervals $\Delta f_N$ of $f_N$.

---

[58] The number of particles in the universe is roughly estimated as $10^{79} - 10^{80}$.

## 7.10. An example of toy Liouvillian models: monads

What follows in this section may be regarded as a science fantasy so that the busy or pedantic reader can rightly turn the respective pages. As already mentioned, mathematical modeling, in contrast with physics, is not necessarily restrained by observability criteria and can cherish rather wild fantasies. Models, as already mentioned, do not need to coincide with physical reality, although they might reflect some of its features. Here, we shall consider an example of such "unphysical" modeling based on the concept of phase fluid and the Liouville equation. The below model is in fact a physical fairy tale having little to do with reality. Specifically, one can naively try to compose a quantum particle which, contrary to the particles in classical and relativistic mechanics, cannot be pointlike as it is described by the wave function that does not reduce to a delta-function. One of the simplest models of a spatially spread particle is a sack containing a myriad of subquantum classical particles like a football being full of air molecules[59]. One can, for simplicity, consider the number $N$ of these classical objects underlying the quantum particle to be large but finite. Then such subquantum objects would hypothetically compose any material entity. This fancy scheme is, in fact, a version of the Leibniz doctrine of monads – the final but unobservable constituents of physical reality. Monads, according to Leibniz, are the primary and irreducible elements of any compound object. Leibniz's monads are, however, of purely speculative nature since they presumably cannot be detected in any experiments. Yet these hypothetical indivisible particles can be considered ultimate foundational elements of reality.

Within the framework of monadology, a quantum particle is no longer elementary, but possesses an internal structure and is spread in space (in volume $\Omega$) embracing $N$ identical classical particles i.e., point masses $m_0$ that can move and interact with one another according to the principles of Newtonian (Lagrangian, Hamiltonian) mechanics. In this classical picture, a quantum particle has $3N$ hidden degrees of freedom. In other words, a quantum particle of mass $m$, which is considered fundamental in today's physical paradigm, is no longer elementary but is represented by a volume $\Omega$ of colliding point masses (monads) and can thus be described by the methods of physical kinetics and fluid dynamics [92], [105], and [31]. It is interesting that if hypothetic monads comprising a quantum particle had really existed, the monadology would have represented a physical model of truly closed systems. However, we have already noticed that completely closed systems do not exist in nature since they are destroyed by fluctuations within rather a short time (on the macroscopic scale).

---

[59] Recall in this connection that according to Stephen Hawking the universe as a whole may be viewed as a quantum particle that can be found in an infinite variety of states bringing about innumerable possible worlds.

Let us get back to monadology. The physical world appears to have another hierarchical ordering, not based on indivisible monads. Therefore, the monadology scenario can probably exist only as a mathematical model having nothing to do with reality. By the way, can such classical particles be further subdivided (similar to fractals), and if they can, then endlessly or not? Thus, one is compelled to think that monads possibly must have a hidden inner structure. One might also notice that the model of monads is not quite consistent since the full description of many-particle evolution in the phase space is only reached through quantum-statistical methods. In short, monadology appears to be a physically meaningless model, but it can invoke productive associations and useful techniques.

Monadology is in reality a rather primitive concept, and it is reproduced here only for illustrative purposes. Moreover, we consider this model here for tutorial purposes as it provides an opportunity to get acquainted with models and concepts from diverse fields of physics. Generally, mathematical models with interdisciplinary links can be of a high heuristic value irrespective of their perceived physical adequacy (one can recall the Ptolemaic model-building, the phlogiston theory, Thomson's "plum pudding" atom, Bohr's atom, radiation reaction, even the Schrödinger and Dirac equations, many modern cosmological models, etc.). One of the most insightful scientists pursuing an interdisciplinary discourse was Herrmann von Helmholtz who, being a physician and physiologist, obtained outstanding results in mathematics and physics while exploring human sense organs such as the eye and ear. Hardly any comparable examples of fruitful interdisciplinary research are currently known despite all the hype about interdisciplinary endeavor.

As to monadology, we can turn our fantasy loose and develop the concept of a quantum particle consisting of a large number of classical particles to somewhat exotic predictions (such as excited states of an individual electron, the possibility of a particle to deform, etc.). So even unrealistic models can be of some value since they stimulate the development of the existing context.

Let us, nonetheless, briefly consider the monadology model as an illustration of the classical phase space methods and of using the BBGKY hierarchy. One can build a model describing the quantum particle as a statistical ensemble of $N$ identical particles (point masses $m_0$) with the one-particle distribution function $f_1(\mathbf{r}, \mathbf{p}, t) \equiv f(\mathbf{r}, \mathbf{p}, t)$ whose evolution is governed by the kinetic (e.g., Boltzmann) equation

$$\frac{Df}{Dt} = \frac{\partial f}{\partial t} + \frac{\mathbf{p}}{m} \frac{\partial f}{\partial \mathbf{r}} + \mathbf{F} \frac{\partial f}{\partial \mathbf{p}} = \mathcal{I}[f], \qquad (7.10.1.)$$

where $\mathcal{I}[f]$ is the collision integral, $\mathbf{F}$ is the total force acting on a gas particle. Operator $\mathcal{L} \equiv D/Dt$ is known as the single-particle Liouville operator defining the evolution, while $\mathcal{I}[f]$ accounts for the change of the phase density of monads per unit time due to interactions between them. If one considers a more sophisticated model when the monads possess some internal structure, then the distribution function $f$ would have a set of additional arguments $\{s\}$, e.g., corresponding to diverse spin states. One can, in principle, view the particles having different values of a discrete argument $\{s\}$ as different particles. As commented above (see section 7.3.), from the exact Liouville equation for an $N$-particle ensemble

$$\frac{\partial f_N}{\partial t} + \sum_{a=1}^{N} \mathbf{v}_a \frac{\partial f_N}{\partial \mathbf{r}_a} + \sum_{a=1}^{N} \left( \mathbf{F}_a - \frac{\partial}{\partial \mathbf{r}_a} \sum_{\substack{b \neq a}}^{N} V_{ab} \right) \frac{\partial f_N}{\partial \mathbf{p}_a} = 0, \qquad (7.10.2.)$$

where $\mathbf{F}_a$ is an external force acting on the $a$-th particle and $V_{ab} \equiv V(|\mathbf{r}_a - \mathbf{r}_b|)$ is the potential energy of pair interaction, one can produce the BBGKY hierarchical chain of kinetic equations of the evolutionary form

$$\frac{\partial f_n}{\partial t} + \sum_{a=1}^{n} \frac{\mathbf{p}_a}{m_0} \frac{\partial f_n}{\partial \mathbf{r}_a} + \sum_{a=1}^{n} \mathbf{F}_a \frac{\partial f_n}{\partial \mathbf{p}_a} = \sum_{a=1}^{n} \mathcal{I}[f_n], \, n = 1, \dots, N. \tag{7.10.3.}$$

Here functionals $\mathcal{I}[f_n]$ are the "collision integrals" stemming from the last term in (7.10.2.). Since it is assumed that individual monads move according to classical mechanics, we can use simple expressions for the forces: force $\mathbf{F}_a$ can be, for example, $\mathbf{F}_a = -\partial U(\mathbf{r})/\partial \mathbf{r}|_{\mathbf{r}=\mathbf{r}_a}$ when the system of $N$ monads (i.e., the parental particle) is placed in the potential field $U(\mathbf{r})$ or $\mathbf{F}_a$ can be the Lorentz force $\mathbf{F}_a = e_a(\mathbf{E} + c^{-1}[\mathbf{v}_a B])$, if monads carry charge $e_a$. Note that if the monads are all of the same sort, in particular, they are not characterized by different values of intrinsic parameter $\{s\}$, then their $N$-particle phase density $f_N(x_1, \dots x_N, t) dx_1 \dots dx_N$ is symmetric with respect to permutations of particles in the phase space. If we introduce a $6N$-dimensional phase space $P$ with points $x = (\mathbf{r}, \mathbf{p}) = (\mathbf{r}_1, \dots, \mathbf{r}_N, \mathbf{p}_1 \dots, \mathbf{p}_N)$, $3N$-dimensional force $\mathbf{F} = \{\mathbf{F}_a\} \equiv \{\mathbf{F}_1, \dots, \mathbf{F}_N\}$ and $3N$-dimensional operators $\partial_\mathbf{r} \equiv \partial/\partial \mathbf{r} = \{\partial/\partial \mathbf{r}_1, \dots, \partial/\partial \mathbf{r}_N\}, \partial_\mathbf{p} \equiv \partial/\partial \mathbf{p} = \{\partial/\partial \mathbf{p}_1, \dots, \partial/\partial \mathbf{p}_N\}$, we can write the kinetic (balance) equation for the $N$-particle distribution function $f_N(\mathbf{r}, \mathbf{p}, t)$ in a compact form similar to the Boltzmann equation

$$\frac{\partial f_N}{\partial t} + \mathbf{v} \frac{\partial f_N}{\partial \mathbf{r}} + \mathbf{F} \frac{\partial f_N}{\partial \mathbf{p}} = \mathcal{I}[f_N]. \tag{7.10.4.}$$

Notice the abuse of notation: here symbols $\mathbf{r}$ and $\mathbf{p}$ denote $3N$ vectors whereas in the Boltzmann equation they are just coordinates and momenta of a material point in the Euclidean space. If we integrate (7.10.2.) (or (7.10.3.)) over all variables $x_b$ except $x_a$ assuming that $\lim_{|\mathbf{p}| \to \pm \infty} f_N = 0$ and $f_N|_{\mathbf{r}_a = \mathbf{r}_0}, a = 1, \dots, N$, where $\mathbf{r}_0$ is the radius of the observable (parental) particle.

Using the standard techniques of physical kinetics we can obtain the single-particle kinetic equations and hydrodynamical quantities. Integrating (7.10.2.) (or (7.10.3.)) over all variables $x_a = (\mathbf{r}_a, \mathbf{p}_a)$ except $x_1 = (\mathbf{r}_1, \mathbf{p}_1)$ using the just mentioned assumptions about limits, we get the first equation in the hierarchy

$$\frac{\partial f_1(\mathbf{r}, \mathbf{p}, t)}{\partial t} + \mathbf{v} \frac{\partial f_1(\mathbf{r}, \mathbf{p}, t)}{\partial \mathbf{r}} + \mathbf{F} \frac{\partial f_1(\mathbf{r}, \mathbf{p}, t)}{\partial \mathbf{p}}$$
$$= \rho \frac{\partial}{\partial \mathbf{p}} \int d^3 r' d^3 p' \, f_2(\mathbf{r}', \mathbf{p}', \mathbf{r}, \mathbf{p}, t) \frac{\partial V(|\mathbf{r} - \mathbf{r}'|)}{\partial \mathbf{r}}, \tag{7.10.5.}$$

where $\rho = N/\Omega$ is the mean density of monads and function $f_1$ is defined on the phase space of just one particle (the $\mu$-space). Notice that in order to find the single-particle distribution function one has to know the two-particle function $f_2$ and so on. We can also define average quantities over the entire phase space of the $N$-particle system (the $\Gamma$-space). For instance, the $N$-particle average density can be defined as $\rho_N(r_1, \dots r_N) = \int d^{3N} p f_N(x, t)$ and the respective functionals of any multiparticle physical quantity $A(x, t)$ as

$$\rho_N \bar{A} = \int d^{3N} p A(x, t) f_N(x, t). \tag{7.10.6.}$$

Multiplying the distribution function $f_N$ sequentially by 1, $\mathbf{p}_a - \mathbf{p}_0$ and $(1/2m)(\mathbf{p}_a - \mathbf{p}_0)^2$, where $\mathbf{p}_0(\mathbf{r}, t) = (\rho(\mathbf{r}, t))^{-1} \sum_a n_a \mathbf{p}_a$, $\rho(\mathbf{r}, t) = \sum_a n_a$ is the mean momentum, and integrating over momenta $\mathbf{p}$ (which is the standard procedure) we shall get the hydrodynamic equations both for the classical and quantum (Bohmian) case. One can find the calculation details in [26] so that we shall not expand on them.

## 7.11. The famous kinetic equations: the Fokker-Planck equation

A kinetic equation is a mathematical model allowing one to find (approximately!) the distribution function for the statistical ensemble. Kinetic equations in general serve to describe macroscopic phenomena through the motion of a large number of molecules. We have seen that such equations generally have the form of balance laws, but unlike simple balance relationships expressing conservation of some material quantities in 3d space, kinetic equations manifest more sophisticated balances in other spaces (e.g., in the phase space or in the operator space of quantum statistics). In the kinetic description one often assumes that the many-particle (molecular) system only slightly deviates from the equilibrium state. In particular, the Boltzmann kinetic equation mainly describes the rarefied gas near equilibrium. The linear versions of the Boltzmann equation i.e., corresponding to the systems close to equilibrium are successfully applied to the radiation transfer problem (for example, in climate modeling) and to the analysis of neutron fields in nuclear reactors.

A very important application of balance laws is the Fokker-Planck equation which plays a crucial role in studying fluctuations in physical, biological, economic and social systems. The Fokker-Planck equation (sometimes also known as the Fokker-Planck-Kolmogorov equation) often appears in the analysis of dynamical systems perturbed by noise. The importance of the Fokker-Planck equation for mathematical modeling of a wide variety of processes, particularly in life sciences and socio-economic studies, is related to its capability to describe the evolution of a complex subsystem immersed in the environment. The Fokker-Planck equation in a single spatial dimension has the form

$$\frac{\partial F(x,t)}{\partial t} = L_{FP} F(x;t), \qquad (7.11.1.)$$

where $L_{FP} = \frac{\partial}{\partial x} D_1(x,t) + \frac{\partial^2}{\partial x^2} D_2(x,t)$ is the Fokker-Planck differential operator. The first term in $L_{FP}$ corresponds to the drift, including deterministic and noise-induced (sometimes called spurious) drift, whereas the second term accounts for diffusion. In some exceptionally simple cases one can solve the Fokker-Planck equation exactly, e.g., by separating the variables. Notice that the Fokker-Planck operator determines the temporal evolution of a system; in this sense it is similar to the generic Liouville operator $\mathcal{L}$ (see below). In (7.11.1.) the Fokker-Planck equation is represented in the relaxation form which is convenient to describe the behavior of a system far from equilibrium. Recall that in most important situations the behavior of systems far from equilibrium can be described by linear equations of the relaxation form

$$\frac{\partial f}{\partial t} = Lf, \qquad (7.11.2.)$$

where $f$ is the distribution function of a set of physical or other modeling quantities (comprising some manifold $M$) and $L$ is some linear operator which is defined on $M$ and can depend both on the modeling variables belonging to $M$ and on time $t$. In the latter case the problem becomes non-autonomous and therefore more complicated.

In general, one can represent the operator of evolution as $U(t,t_0) = \exp\left[-i(t-t_0)\mathcal{L}\right]$, where $\mathcal{L} = -iL$ is the Liouville operator and factor $i, i^2 = -1$ is introduced for convenience to emphasize the unitary character of evolution (when relevant). That is if $\mathcal{L}$ is self-adjoint, maps $U(t,t_0)$ sending initial state $f(t_0)$ to solution $f(t)$ form a unitary representation $U: \mathbb{R} \to U$.

We can observe that the Fokker-Planck operator $L_{FP}$ can depend both on spatial coordinates and time $t$ (non-autonomous evolution): in the latter case variables can only be approximately separated and the boundary value problem (BVP) for operator $L$ that can be the Liouville operator $\mathcal{L}$ or the Fokker-Planck operator $L_{FP}$ may become meaningless. The boundary value problem is usually understood as an expansion over eigenfunctions i.e., when $F$ can be represented in the form $F(x, t) = \sum_n a_n F_n(x) \exp(-\lambda_n t)$, where $\lambda_n$ and $F_n$ are eigenvalues and eigenfunctions of the Fokker-Planck operator.

The Fokker-Planck equation is closely connected with the model of Brownian motion which is the random motion of a test particle under sudden impacts of much smaller medium molecules, e.g., in a fluid. To reveal this connection, we shall first try to describe the motion of a comparatively big test particle (e.g., that of dust) in the medium consisting of smaller particles that can randomly collide with the test particle. The equation of motion for such a case within Newtonian model can be written as

$$m\ddot{\mathbf{x}} + \gamma \dot{\mathbf{x}} + \nabla V = \mathbf{F}_r(\mathbf{x}, t), \qquad (7.11.3.)$$

where $m$ is the particle mass, $\gamma$ is the damping coefficient, and $V = V(\mathbf{x})$ is the external potential such as of gravitation. One can estimate the friction coefficient $\gamma$, e.g., for the model of dissipative motion of a spherical particle having radius $a$ in a fluid with viscosity $\eta$ using Stokes' formula [92], §20 $\gamma = 6\pi\eta a$. Random force $\mathbf{F}_r(\mathbf{x}, t)$ is usually assumed to have a zero mean i.e., $\langle \mathbf{F}_r(\mathbf{x}, t) \rangle = 0$, where angular brackets symbolize averaging with some probability distribution, e.g., Gaussian.

A "big" test particle means in this context that its size greatly exceeds both the size of medium molecules and the average intermolecular distance. Recall that in the air this distance is typically $\sim 10^{-7}$ cm (see Section 6.2. "Fluids as many-body systems"), while the size of a Brownian particle is $\sim 10^{-4}$. When external deterministic forces are absent $\nabla V = 0$ and, for simplicity, $\mathbf{F}_r(\mathbf{x}, t) \equiv \mathbf{F}_r(t)$ and $m = 1$ (or $\gamma/m \to \gamma$), we can write the formal solution to "shortened" equation (7.4.3), $\dot{\mathbf{v}} + \gamma \mathbf{v} = \mathbf{F}_r(\mathbf{x}, t)$:

$$\mathbf{v} = \mathbf{v}_0 e^{-\gamma t} + \int_0^t dt' e^{-\gamma(t-t')} \mathbf{F}_r(t'). \qquad (7.11.4.)$$

In a viscous medium or for high values of the collision frequency the motion of the big test particle can be "overdamped" i.e., the inertial term $m\ddot{\mathbf{x}}$ can be disregarded in comparison with the viscous term $\gamma\dot{\mathbf{x}}$ and deterministic force $\nabla V$ generating the deterministic velocity $\mathbf{v}(\mathbf{x}) \equiv -\gamma^{-1}\nabla V$. In the case of negligible inertial force (the Aristoteles model) the equation of motion under the random force takes the Langevin form

$$\dot{\mathbf{x}} + \mathbf{v}(\mathbf{x}) = \boldsymbol{\xi}(\mathbf{x}, t), \qquad (7.11.5.)$$

where $\boldsymbol{\xi}(\mathbf{x}, t) = \gamma^{-1}\mathbf{F}_r(\mathbf{x}, t)$ is the random velocity (with a zero mean $\langle \boldsymbol{\xi}(\mathbf{x}, t) \rangle = 0$). When external forces are absent ($\nabla V = 0$) the particle position for time $t > 0$ can be calculated as a random variable $\mathbf{x}(t) = \mathbf{x}(0) + \int_0^t \boldsymbol{\xi}(\mathbf{x}, t') dt'$. Here, for simplicity, we can omit the dependence of random velocity on space point (disregarding inhomogeneity) and take the initial time instant $t_0$ to be 0.

If random velocity $\boldsymbol{\xi}(\mathbf{x}, t)$ is absent (i.e., random force $\mathbf{F}_r(\mathbf{x}, t) = 0$) equation (7.11.5.) becomes fully deterministic, e.g., $m\ddot{\mathbf{x}} + \gamma\dot{\mathbf{x}} = 0$ or $\dot{\mathbf{u}} + \gamma\mathbf{u} = 0$ where, $\mathbf{u} \equiv \dot{\mathbf{x}}$ and, for simplicity, $\gamma/m \to \gamma = \tau^{-1}$. The physical meaning of this deterministic model is that the particle velocity is damped in collisions with the medium molecules exponentially degrading to zero with relaxation time $\tau$, $\mathbf{u}(\tau) = \mathbf{u}(t_0)e^{-\gamma(t-t_0)}$. This is a deterministic process since the test particle velocity is fully determined by its initial value $\mathbf{u}(t_0)$ and fixed model parameter $\gamma$.

If we denote as $w(\mathbf{x}, t|\mathbf{x}_0, t_0)$ the probability density to find the Brown (test) particle at point $\mathbf{x}$ at time $t$, provided that the particle initially (i.e., at $t = t_0 = 0$) was at $\mathbf{x} = \mathbf{x}_0 = 0$, we can regard the probability of finding the particle in spatial volume $d^3x$ obtained through density $w(\mathbf{x}, t|\mathbf{x}_0, t_0)$ to be normalized for all $t$ i.e., $\int w(\mathbf{x}, t|\mathbf{x}_0, t_0) \, d^3x = 1$ with initial condition $w(\mathbf{x}, t_0|\mathbf{x}_0, t_0) = \delta(\mathbf{x} - \mathbf{x}_0)$. One might note that function $w(\mathbf{x}, t|\mathbf{x}_0, t_0)$ provides no information about intermediary states or positions via which the particle came into point $\mathbf{x}$ from point $\mathbf{x}_0$ during time interval $t - t_0$. Moreover, function $w(\mathbf{x}, t|\mathbf{x}_0, t_0)$ does not depend on trajectories, collisions nor on any other events that occurred to the particle prior to time point $t_0$. Processes that have such properties are known as having the Markov property (Markov processes).

The Brownian motion had been experimentally known since 1827, and, moreover, the Brown particles are easily observable, e.g., through a common microscope. This fact could have ensured the rapid emergence of a thorough theoretical description, given the high level of development of analytical mechanics in the 19$^{th}$ century. Paradoxically though, qualitative estimates and mathematical models related to evolutions of Brownian particles appeared only in the early 20$^{th}$ century.

To connect the Brownian motion, in particular its inertialess model (7.11.5.), with the Fokker-Planck equation, we can use the recursive approach (often employed in physics) that would allow us to find the function $w(\mathbf{x}, t|\mathbf{x}_0, t_0)$. Indeed, the probability $w(\mathbf{x}, t + \tau|\mathbf{x}_0, t_0)$, where $\tau \ll t$, can be represented as

$$w(\mathbf{x}, t + \tau|\mathbf{x}_0, t_0) = \int d^3y \, w(\mathbf{y}, t|\mathbf{x}_0, t_0) \, w(\mathbf{x}, t + \tau|\mathbf{y}, t). \qquad (7.11.6.)$$

The first function under the integration sign denotes the transition from the initial spacetime point $(\mathbf{x}_0, t_0)$ into an intermediary state or location $(\mathbf{y}, t)$ whereas the second function corresponds to the transition from $(\mathbf{y}, t)$ into the final state $(\mathbf{x}, t + \tau)$. Intuitively, the final probability $w(\mathbf{x}, t + \tau|\mathbf{x}_0, t_0)$ can be interpreted as the sum of probabilities related to intermediary spatial points $\mathbf{y}$, with $w(\mathbf{x}, t + \tau|\mathbf{y}, t)$ playing the role of the probability transition function, analogous to Green's function. Equation (7.11.6.) is well-known in physics and was originally written by M. Smolukhowski in 1906 [150], but it is more often called the Chapman-Kolmogorov equation.

One should not be deceived by the simplicity of equation (7.11.6.). In fact, it is a nonlinear integral equation for which the customary existence-uniqueness theorems habitual for differential equations are not known.

When the dependence of the distribution function $w(\mathbf{x}, t|\mathbf{x}_0, t_0)$ on $t$ is homogeneous, i.e., all time instants are equivalent, then this function depends on time in combination $t - t_0$: $w(\mathbf{x}, t|\mathbf{x}_0, t_0) \equiv w(\mathbf{x}, t - t_0|\mathbf{x}_0, 0)$ and the Chapman-Kolmogorov-Smolukhowski equation takes the form

$$w(\mathbf{x}, t + \tau|\mathbf{x}_0) = \int d^3y \, w(\mathbf{y}, t|\mathbf{x}_0) \, w(\mathbf{x}, \tau|\mathbf{y}). \qquad (7.11.7.)$$

Equation (7.11.7.) usually serves as raw material for obtaining the Fokker-Planck equation whose derivation is quite simple and easily available (see, e.g., [95] and [130]) so that we shall not reproduce it here.

Modeling based on the Fokker-Planck equation can be very efficient; however, in order to keep the size of this book within reasonable limits, we have to omit the discussion of the Fokker-Planck equation.

## Section 8. Classical mechanics and deterministic dynamical systems

Classical dynamics is probably the most developed part of science since it studies the evolution of the simplest objects – finite systems of material points. This model is a drastic idealization: material points correspond to bodies that are so small that their inner structure is disregarded and the only remaining characteristic is their position in space, $\mathbf{r} = \mathbf{r}_i(t)$, $i = 1, \dots, N$ ($N$ is a finite number). Parameter $t$ defining the variability of point positions is usually interpreted as an absolute (i.e., universal) time. In classical mechanics (that has served for over two centuries as a model for all other disciplines), the state of a system is determined by a finite set of variables, and their temporal evolution determines the system's dynamics. The main task of classical mechanics is to explore the dynamics of a system for the given initial data, and in this sense classical mechanics studies deterministic dynamical systems. However, it is not always possible to represent a given dynamical system as a manifestation of mechanical motion: in this sense, dynamical systems theory and classical mechanics are not completely equivalent.

Classical mechanics is represented in a number of spaces, the principal of them being the configuration space $Q$ with dimensionality $N$ i.e., coordinates $q^i \in Q$ run from $i = 1$ to $i = N$ and phase space $P := T^*Q$ with dimensionality $2N$, i.e., $q^i, p_j, i, j = 1, \dots, N$, where $p_j, j = 1, \dots, N$ are momenta of the considered classical system. Usually, momenta $p_j \in M$ ($M$ is the momentum space) are defined through coordinates $q^i$ as $p_j = \partial L / \partial q^j$, where $L = L(p_j, q^i, t)$ is the system Lagrangian. In addition to these two spaces three others are used as supplementary for the motion description, namely the energy-momentum space $M \times H$ of dimensionality $N + 1$ and with coordinates $p_j, j = 1, \dots, N$ and $H = H(p_j, q^i, t) = \dot{q}^\iota \partial L / \partial q^i - L(p_j, q^i, t), \dot{q}^\iota = \dot{q}^\iota(p_j, q^i, t)$; the event space $Q \times T$ of dimensionality $N + 1$ and with coordinates $q^i, t$; and the state space $P \times T$ of dimensionality $2N + 1$ and with coordinates $q^i, p_j, t$. This latter space is useful for treating the collision processes.

Classical mechanics is based on a certain relativity concept – the Galilean relativity. This concept amounts to the postulate that it is impossible to distinguish the phenomena occurring in the state of rest from those in the state of uniform motion: the laws of dynamical evolution are exactly the same when referred to any frame uniformly moving with respect to any other. Such a picture naturally leads to the mathematical model of Galilean spacetime as a fiber bundle with time as the base manifold, $t \in \mathbb{T} \cong \mathbb{R}$, and spatial positions belonging to the fibers $\mathbb{R}^n$; more exactly, these are affine spaces $\mathbb{A}^n$ endowed with an inner or, for ordinary vectors from $\mathbb{R}^n$, a dot product[60] and distance providing the

---

[60] One can regard an inner product as a generalization of the dot product of the $\mathbb{R}^n$ vectors which assigns a scalar (0-vector) to each pair of vectors from the linear space. Therefore, due to the scalar result of multiplication, in elementary vector algebra a dot product is known as a scalar product yielding the projection of one vector along another. In the language of differential forms, the dot product (as well as the inner product) is a 1-form. As a generalization of a scalar product, an inner product is defined as a map $(\mathbf{a}, \mathbf{b}) \equiv (\mathbf{a} \cdot \mathbf{b}): V \cdot V \to K$; $\mathbf{a}, \mathbf{b} \in V$, where $V$ is some vector space (not necessarily isomorphic to $\mathbb{R}^n$) over field $K$. In contrast to a scalar product, inner product is traditionally denoted by diamond brackets, $\langle \mathbf{a}, \mathbf{b} \rangle$. The inner product holds for spaces with infinite dimensions, complex spaces, functional spaces,

Euclidean structure of Euclidean space $\mathbb{E}^n$. For example, distance $\rho(\mathbf{a}, \mathbf{b}) = \|\mathbf{a} - \mathbf{b}\| = \sqrt{(\mathbf{a} - \mathbf{b})(\mathbf{a} - \mathbf{b})}$ (Euclidean length) defined through the inner product converts affine space $\mathbb{A}^3$ and linear space $\mathbb{R}^3$ into 3d Euclidean space $\mathbb{E}^3$ that we used to observe. Both Euclidean and affine spaces are geometric structures on $\mathbb{R}^n$, and it is important to remember that $\mathbb{R}^n$ is the linear (or vector) space where the presence of Euclidean structure, i.e., of the Euclidean norm (used to define a metric on $\mathbb{R}^n$) and of the inner product $(\mathbf{a}, \mathbf{b}) = a^1 b^1 + \cdots + a^n b^n$, is not required. As to affine spaces, one might say that an affine space $\mathbb{A}^n$ can be upgraded to vector space $\mathbb{R}^n$ by fixing the origin $0 = (x^1 = 0, \ldots, x^n = 0)$. Recall that unlike a vector space $\mathbb{V}$ which has a distinguished element – the zero vector – an affine space does not possess such element (a singled-out point in space or spacetime such as the Garden of Eden or the First Day of Creation). Formally, it is reflected in the definition of an affine space $\mathbb{A}$ as a pair $(\mathbb{P}, \mathbb{V})$ including a set $\mathbb{P}$ and vector space $\mathbb{V}$ (over an arbitrary field $\mathcal{K}$), together with an action of an abelian group $\mathbb{G}: \mathbb{V} \otimes \mathbb{P} \to \mathbb{P}$ i.e., a map $\mathbb{G}(x + y, a) = \mathbb{G}(x, \mathbb{G}(y, a))$ for any $x, y \in \mathbb{V}, a \in \mathbb{P}$. In classical mechanics, field $\mathcal{K}$ is typically taken to be the field of the reals i.e., $\mathcal{K} = \mathbb{R}$ so that e.g., three positions in space as well as their differences can be described by three real numbers (coordinates). Our world is modeled in mechanics by a 4d affine space $\mathbb{A}^4$ whose elements are events, usually called in physics world points. Parallel transport of world $\mathbb{A}^4$ by vectors $\mathbf{a} \in \mathbb{R}^4$ gives linear space $\mathbb{R}^4$ (recall that in affine spaces $\mathbb{A}^n$ contains both points and vectors only the difference of two points is defined, this difference being a vector from $\mathbb{R}^n$). One might also note that from the topological viewpoint vector space $\mathbb{R}^n$ is homeomorphic to an open $n$-sphere.

Spacetime is a basic concept of physics and is not necessarily reduced to the simple Galilean spacetime serving as an arena for classical motion. Spacetime is a generic mathematical model that somehow unifies space and time, therefore spacetime is usually understood as a bundle $S$ consisting of "copies" of manifolds $S^{n+1} = \mathbb{T} \times Q^n$, where fibers $Q^n$ correspond to a physical space and $\mathbb{T}$ is a "time manifold" usually identified with time axis $\mathbb{R}$ playing the role of base manifold. In particular, configuration space $Q$ of classical nonrelativistic mechanics is a fiber bundle $Q \to \mathbb{R}$ over the time axis $t \in \mathbb{R}$. The Lagrangian equations of classical mechanics are second-order differential equations (over time $t$) on $Q \to \mathbb{R}$. In principle, nothing precludes using other objects as time, e.g., zeros of some differential form $dt = \alpha_1 d\xi_1 + \cdots + \alpha_m d\xi_m$ or foliation into isochrones $t(\xi) = \text{const}$, here $\xi$ is assumed to lie in some parameter space (it can be angle $\varphi$ or, e.g., the inverse frequency of atomic transitions, etc.) This question is related to the "nature of time" issue and the possibility of its measurement.

In the classical nonrelativistic (Newtonian) picture, any event can be attributed to a single value of time $t$. Thus, $t$ is a projection from $\mathbb{R}^3$ (or $\mathbb{R}^{3N}$, where $N$ is the number of particles or other irreducible components of the observed scene). We also imply that there exists a 1d temporal order in the Newtonian picture, when an event corresponding to $t_2 > t_1$ is said to occur later than that corresponding to $t_1$. We may remark that time in physics is always oriented, as it flows non-invertibly from past (e.g., infinite past $t = -\infty$) to future (e.g., infinite future $t = +\infty$).

One can imagine the classical time in simple cases as a linear mapping on the real time axis of the parallel transported Euclidean world: $t: \mathbb{E}^4 \to \mathbb{R}$, where $x, y \in \mathbb{E}^4$ are events and function (projection) $t(x, y) = t(y - x)$ is the time interval between them. If $t(x, y) = 0$, events are simultaneous, and a set of simultaneous events constitutes 3d affine space $\mathbb{A}^3 = \mathbb{E}^3$. The three described elements: classical world $\mathbb{A}^4$, time $t \in \mathbb{T}$ and Euclidean length $\rho(\mathbf{a}, \mathbf{b}), \mathbf{a}, \mathbf{b} \in \mathbb{A}^3$ i.e., the distance between

---

Hilbert spaces, etc. The notion "inner" distinguishes the scalar type of multiplication giving as a result a 0-vector from the "outer" product which yields a matrix ($m \times n$ bivector) or a tensor of rank (1,1).

simultaneous events (located on the same fiber) constitute the Galilean spacetime structure for which fundamental symmetries of classical and nonrelativistic quantum mechanics can be defined.

A group of all transformations preserving the Galilean structure is known as the Galilean group whose elements are affine transformations leaving the time intervals and the distances between simultaneous events unchanged. Three typical Galilean transformations are: (a) $G_1(\mathbf{r}, t) = (A\mathbf{r}, t)$, rotation in Euclidean coordinate space $\mathbb{E}^3$, $A: \mathbb{E}^3 \to \mathbb{E}^3$ is an orthogonal transformation preserving Euclidean length $\rho(\mathbf{a}, \mathbf{b})$ i.e., $x^i = A_{ij}x^j$, $A^T = A^{-1}$; (b) $G_2(\mathbf{r}, t) = (\mathbf{r} + \mathbf{v}t, t)$, corresponding to the motion with constant velocity $\mathbf{v}$; (c) $G_3(\mathbf{r}, t) = (\mathbf{r} + \mathbf{a}, t + s)$, spacetime shift of the coordinate origin. One can see that the general Galilean transformations have the form $\mathbf{r} \mapsto \mathbf{r}' = A\mathbf{r} + \mathbf{v}t + \mathbf{a}$, $t' \mapsto t + s$ and depend on 3+3+1+3=10 parameters $A_{ij}$, $v^i$, $s$, $a^k$ so that the dimensionality of the Galilean group is 10. The group parameters of the Galilean group are the variables determining classical motion of a point particle i.e., $x = (t, \mathbf{r}, \mathbf{v}, \boldsymbol{\alpha})$, where $\boldsymbol{\alpha}$ describes the orientation of a local frame at point $\mathbf{r}$. Any Galilean transformation can be uniquely represented as the product $G = G_1 G_2 G_3$, and the identical Galilean transformation corresponding to the unit element of the group can be symbolically written as $\mathbf{e} = (1,0,0,0)$. One can write a matrix representation of the Galilean group $G$ in the form

$$G = \begin{pmatrix} A & \mathbf{v} & \mathbf{a} \\ 0 & 1 & s \\ 0 & 0 & 1 \end{pmatrix}, \mathbf{a}, \mathbf{v} \in \mathbb{R}^3, s \in \mathbb{R}, A \in O(3)$$

acting on elements $(\mathbf{r}, t, 1)^T$ of $\mathbb{R}^4$, $t \in \mathbb{R}$ as

$$\begin{pmatrix} \mathbf{r}' \\ t' \\ 1 \end{pmatrix} = \begin{pmatrix} A & \mathbf{v} & \mathbf{a} \\ 0 & 1 & s \\ 0 & 0 & 1 \end{pmatrix} \begin{pmatrix} \mathbf{r} \\ t \\ 1 \end{pmatrix} = \begin{pmatrix} A\mathbf{r} + \mathbf{v}t + \mathbf{a} \\ t + s \\ 1 \end{pmatrix}, \qquad (8.1.)$$

where vector space $\mathbb{R}^4$ is represented through homogeneous coordinates i.e., as a projective space[61].

The fact that the Galilean transformations really comprise a group (a 10-dimensional Lie group) is readily illustrated by identities

$$\begin{pmatrix} A_1 & \mathbf{v}_1 & \mathbf{a}_1 \\ 0 & 1 & s_1 \\ 0 & 0 & 1 \end{pmatrix} \begin{pmatrix} A_2 & \mathbf{v}_2 & \mathbf{a}_2 \\ 0 & 1 & s_2 \\ 0 & 0 & 1 \end{pmatrix} = \begin{pmatrix} A_1 A_2 & A_1\mathbf{v}_2 + \mathbf{v}_1 & A_1\mathbf{a}_2 + \mathbf{v}_1 s_2 + \mathbf{a}_1 \\ 0 & 1 & s_1 + s_2 \\ 0 & 0 & 1 \end{pmatrix}$$

and

$$\begin{pmatrix} A & \mathbf{v} & \mathbf{a} \\ 0 & 1 & s \\ 0 & 0 & 1 \end{pmatrix}^{-1} = \begin{pmatrix} A^{-1} & -A^{-1}\mathbf{v} & A^{-1}(\mathbf{v}s - \mathbf{a}) \\ 0 & 1 & -s \\ 0 & 0 & 1 \end{pmatrix}.$$

Slightly generalizing, we can define the Galilean group as the affine group of all coordinate transformations $(x^0, x^i)^T \to (z^0, z^i)^T = (x^0 + a^0, A_i^j x^i + v^i x^0 + a^i)^T$, where $A_i^j \in SO(n, \mathbb{R})$, $a^\mu = (a^0, a^i)$, and $v^i$ are constant vectors. Here, as usual, Latin indices run from 1 to n whereas Greek indices from 0 to n. If $\det A_i^j \geq 0$ (in particular, $\det A_i^j = +1$), this group is known as proper inhomogeneous, if $\det A_i^j < 0$ (in particular, $\det A_i^j = -1$), the group is improper which means that its transformations do not preserve space orientation (time inversions can be included if we consider $\det A_\alpha^\beta, \alpha, \beta = 0, \ldots, n$). The group becomes homogeneous when $a^\mu = 0, \mu = 0, \ldots, n$ i.e., when the origin is left intact. One may consider the matrix representation of the Galilean group; then the latter can be viewed as a subgroup of the general linear group $GL(n+1, \mathbb{R})$ of $(n+1) \times (n+1)$ matrices with entries from $\mathbb{R}$ defined on $\mathbb{R}^{n+1}$, $GL(n+1, \mathbb{R}) \times \mathbb{R}^{n+1}$: $z^\alpha = G_\beta^\alpha x^\beta$ so that group $GL(n+1, \mathbb{R})$ can be interpreted as containing all linear transformations of Euclidean space $\mathbb{R}^{n+1}$. It might be interesting to explore the geometric properties of the Galilean group treating it as a standard set of transformations acting on vectors $\mathbf{x} = x^\alpha \mathbf{e}_\alpha = z^\mu \boldsymbol{\eta}_\mu$, where $\{\mathbf{e}_\alpha, \alpha = 0, \ldots, n\}$ and $\{\boldsymbol{\eta}_\mu, \mu = 0, \ldots, n\}$ are different bases in vector space $V^{n+1}$ (in particular, in $\mathbb{R}^{n+1}$). Then, according to the standard rules of linear algebra, transformations of vector coordinates $x^\alpha \mapsto z^\mu = G_\beta^\mu x^\beta$ is associated with the basis transformation $\mathbf{e}_\alpha = G_\alpha^\beta \boldsymbol{\eta}_\beta$. Again, by using the standard linear algebra techniques, one can introduce the dual space $V_{n+1}^*$ with the adjoint basis $\boldsymbol{\epsilon}^\alpha$ dual to $\mathbf{e}_\alpha$: $\boldsymbol{\eta}^\beta = G_\alpha^\beta \boldsymbol{\epsilon}^\alpha$. Specifically, $G_0^0 = 1, G_i^0 = 0$, $G_0^i = v^i, G_i^j = A_i^j, i, j = 1,2,3$ so that $\eta^0 = \epsilon^0, \eta^i = v^i \epsilon^0 + A_j^i \epsilon^j$ and $\eta_0 = e_0 - (A^{-1})_j^i v^j e_i, \eta_i = (A^{-1})_i^j e_j$ (see, e.g., [16]).

The Galilean group is the group of classical mechanics: already electromagnetic phenomena expressed in economical form by Maxwell's equations are not invariant under the Galilean group. Recall in this connection that special relativity appeared due to the study of invariance of Maxwell's equations with respect to inertial motion of the frames of reference [55]. One may, however, notice that Galilean invariance, which actually summarizes such empirical facts as uniformity of time, homogeneity of space and isotropy of the world, directly leads to nontrivial physical consequences in the non-relativistic domain. One of such consequences is the notion of inertial systems. For example, if we differentiate over time relationship (b) for uniform motion $\mathbf{r}' = \mathbf{r} + \mathbf{v}t$, we get $\dot{\mathbf{r}}' = \dot{\mathbf{r}} + \mathbf{v}$ i.e., the rule of additive velocities in non-relativistic mechanics. This rule means that the same point has different velocities in different coordinate systems: there exist neither absolute velocity nor absolute rest. Yet acceleration of the uniform motion in Galilean-Newtonian mechanics does not depend on the coordinate system, $\mathbf{w}' \equiv \ddot{\mathbf{r}}' = \mathbf{w} \equiv \ddot{\mathbf{r}}$, the model-dependent statement that may not hold for more general evolution laws. Anyway, all inertial systems are invariant under the Galilean group. By the way, one of the manifestations of the Galilean invariance is the fact that atoms, molecules and subatomic particles are the same in a variety of reference frames in the universe.

Note that we can use the concept of inertial systems only in the case of uniform motion ($A = 0$). If we differentiate the transformed point $\mathbf{r}' = A\mathbf{r} + \mathbf{v}t + \mathbf{a}$ over time, we get $\dot{\mathbf{r}}' = \dot{A}\mathbf{r} + A\dot{\mathbf{r}} + \dot{\mathbf{v}}t + \mathbf{v} + \dot{\mathbf{a}}$ and $\ddot{\mathbf{r}}' = \ddot{A}\mathbf{r} + 2(\dot{A}\dot{\mathbf{r}} + \dot{\mathbf{v}}) + A\ddot{\mathbf{r}} + \ddot{\mathbf{v}}t + \ddot{\mathbf{a}}$. An arbitrary element $A$ of the rotation group $O(3)$, more specifically of $SO(3)$, can be produced, e.g., by multiplying three rotations $A_1, A_2, A_3$ about, respectively, $x^1, x^2, x^3$ axes:

$$A_1 = \begin{pmatrix} 1 & 0 & 0 \\ 0 & \cos\theta_1 & \sin\theta_1 \\ 0 & -\sin\theta_1 & \cos\theta_1 \end{pmatrix}, A_2 = \begin{pmatrix} \cos\theta_2 & 0 & \sin\theta_2 \\ 0 & 1 & 0 \\ -\sin\theta_2 & 0 & \cos\theta_2 \end{pmatrix}, A_3 = \begin{pmatrix} \cos\theta_3 & \sin\theta_3 & 0 \\ -\sin\theta_3 & \cos\theta_3 & 0 \\ 0 & 0 & 1 \end{pmatrix},$$

$$A = A_1 A_2 A_3$$
$$= \begin{pmatrix} \cos\theta_2 \cos\theta_3 & \cos\theta_1 \sin\theta_3 - \sin\theta_1 \sin\theta_2 \cos\theta_3 & \sin\theta_1 \sin\theta_3 + \cos\theta_1 \sin\theta_2 \cos\theta_3 \\ -\cos\theta_2 \sin\theta_3 & \cos\theta_1 \cos\theta_3 + \sin\theta_1 \sin\theta_2 \sin\theta_3 & \sin\theta_1 \cos\theta_3 - \cos\theta_1 \sin\theta_2 \sin\theta_3 \\ -\sin\theta_2 & -\sin\theta_1 \cos\theta_2 & \cos\theta_1 \cos\theta_2 \end{pmatrix}.$$

Differentiation of this expression is rather tedious and gives almost incomprehensible results in terms of $\dot{\theta}_i$ and $\ddot{\theta}_i$, $i = 1,2,3$. The general implication is that the apparently simple Galilean invariance results in a rather complicated geometry of non-uniform motion as soon as curvilinear trajectories ($A \neq 0$) enter the picture.

One might notice one more important physical consequence of Galilean invariance. For a material point, the Lagrangian $L$, which is a function characterizing the state of a physical system, can only depend on variables $\mathbf{r}, \mathbf{v}, t$, but due to Galilean symmetry requirements (homogeneity and isotropy of spacetime) it cannot be an arbitrary function of these variables. Namely $L = L(\mathbf{v}^2)$, and in the simplest case $L = \alpha \mathbf{v}^2$, where the phenomenological coefficient $\alpha$ is for further convenience denoted $m/2$, $m$ being called mass of a material point (particle). If one can neglect interaction between material points and with the environment (which is a drastic idealization, of course), the Lagrangian of a system of points becomes $L = \sum_{i=1}^{N} \frac{m_i}{2} \mathbf{v}_i^2$. This quadratic form is called kinetic energy (usually denoted by $T$).

Yet physical space $Q$, in general, does not necessarily have the structure of vector space $\mathbb{R}^n$: often physical space $Q^n$ is a Riemannian manifold i.e., only locally looks like a vector space. For example, our world affected by the ubiquitous gravitation is non-Euclidean. Nevertheless, in many cases, especially in mechanics, one can put $Q = \mathbb{R}$ i.e., the Euclidean model is very useful for many practical purposes unless very high accuracy is required (such as needed for GPS, satellite distancing, missile navigation, astrophysical observations, etc.). In general, a manifold in physics is usually understood as a set of points that can be represented by real numbers (coordinates). It is important for the interpretation of physical events that $S^{n+1}$ admits a just-mentioned fibration over the time axis (Figure 7). We shall briefly explain in the following subsection why the fiber bundle is an adequate model for spacetime in physics.

In principle, nothing precludes using other objects as time, e.g., zeros of some differential form $dt = \alpha_1 d\xi_1 + \cdots + \alpha_m d\xi_m$ or foliation into isochrones $t(\xi) = $ const, here $\xi$ is assumed to lie in some parameter space (it can be angle $\varphi$ or, e.g., the inverse frequency of atomic transitions, etc.) This question is related to the "nature of time" issue and the possibility of its measurement.

A standard example is relativistic spacetime: all possible motions of particles with mass $m \neq 0$ are represented by "worldlines" i.e., graphs corresponding to velocities $v < c$ and lying, at each point of manifold $S^{3+1}$, inside the quadratic cone $\|dx\| = |cdt|$. This cone contains all the motions compatible with the concepts of special relativity (the Lorentz cone) and is "attached" to each spacetime point, the latter being known in special relativity as an "event", which can be interpreted as a flash of light. For massless particles, e.g., photons, worldliness touches the surface of the Lorentz cone at each point; this picture is typical of geometric optics and other shortwave approximations in classical electrodynamics.

Note that the concept of the affine structure of spacetime is not only compatible with relativistic and even cosmological concepts, but actually invokes them. For example, Hubble's law that has led to the Big Bang model is naturally linked with the affine spacetime. Indeed, Hubble's law states that distant galaxies are receding with velocity proportional to their distance from the observer, $\mathbf{v}_k = H \mathbf{r}_k$, where $\mathbf{v}_k$ and $\mathbf{r}_k$ are the velocity and position of the $k$-th galaxy with respect to an observer located at some coordinate origin. However, this origin does not imply that the observer (e.g., a scientist or God) is at the center of the universe. If we take another origin displaced by vector $\mathbf{b}$ from the previous one, we shall see that Hubble's law is still valid and the old origin has been forgotten. Notice that general relativity says nothing about the Big Bang phenomenon: it does not elicit its origin, nor how can one describe this event in physical terms. Just a "singularity" has been declared, presumably completely determining further evolution of the universe. Yet the Big Bang model has been generally considered the best one for the origin of the universe, especially after the discovery of cosmic microwave background variation in 1964.

## 8.1. Models of motion

Motion in classical physics is usually understood as a continuous map $\mathbf{r}(t)$ from the interval of parameter $t \in I \subseteq \mathbb{R}$ to the $n$-dimensional real linear space $\mathbb{R}^n$; if $n = 3N$, this is a mechanical motion of $N$ particles in the coordinate space, this mapping is commonly assumed to be differentiable. The graph of such a mapping is called a trajectory, a path, a curve or, for $n = 3$, a world line. Then the motion of $N$ points is defined by $N$ world lines, $\mathbf{r}_i(t), i = 1, \dots, N$; one usually calls a direct product of $N$ copies of material points, each moving in $\mathbb{R}^3$, a Euclidean configuration space for a system of $N$ particles. The motion of such a system can be defined by a single mapping $\mathbf{x}(t): I \to \mathbb{R}^n$, where $\mathbf{x}(t) = x^1(t), \dots, x^n(t), n = 3N$. Here, we denote by $x^i(t)$ the Cartesian (sometimes also called Euclidean or Galilean) coordinates which are simply the $n$-tuples of real numbers i.e., $\mathbf{x}(t) = (x^1(t), \dots, x^n(t)) \in \mathbb{R}^n$. The class of Cartesian coordinates is characterized by the fact that the length of a vector $\mathbf{a} = (a^1, \dots, a^n)$ is expressed by the formula $|\mathbf{a}|^2 = a_i a^i = \sum_{i=1}^{n} |a^i|^2$ i.e., by the Pythagorean theorem. Such a coordinate system can be used to describe comparatively simple situations, for instance, the motion of lumped masses modeled by Newton's equations in the Euclidean 3d space (see below examples from elementary ballistics).

However, using the Cartesian frame becomes rather inconvenient in more complex problems such as the rigid body motion when there are three coordinates to describe the position of the center of mass and three more for the rotational orientation. Modeling of the rigid body motion evokes the concept of a configuration space which generally has a nontrivial topology of a manifold (as compared to a trivial topology of Euclidean space). For example, when modeling the motion of a realistic projectile or the reentry of a spaceship one has to deal with at least a six-dimensional manifold $\mathbb{R}^3 \times \mathrm{SO}(3)$, where $\mathrm{SO}(3)$ is the rotation group of a rigid body about some fixed center (in this case, center of mass). It is not so hard to see that, in general, the dimension of the rotation group $\mathrm{SO}(n)$ treated as a submanifold (it's a Lie group) in $n \times n = n^2$ dimensional space is $m = n(n + 1)/2$. In particular, the rotation group $\mathrm{SO}(3)$ over $\mathbb{R}^3$ (i.e., the Euclidean group of special orthogonal matrices) is three-dimensional that is its dimensionality, if this group is treated as a continuous topological manifold, equals to the dimensionality of the underlying real vector space $\mathbb{R}^3$. This is an exceptional property since in general for $\mathrm{SO}(n)$ with $n \neq 3$ the dimensionality of the group differs from that of the underlying field $K$ or real vector space $\mathbb{R}^n$.

Physical experience shows that different events are arranged in series so that they can be ordered and consecutively numbered. Thus, one can associate these numbers (belonging to $\mathbb{R}^1$) with events calling the respective set $\mathbb{R}^1$ the time axis. However, the event-to-number association can be rather arbitrary which might result in a variety of definitions of "time", due to a different choice of available mappings.

Parameter $t$ in classical mechanics is nearly always interpreted as time (although in geometrical terms it may be just a variable to ensure the parametric representation of the curves, e.g., in the plane $(x(t), y(t))^T$). In the Galilean-Newtonian interpretation of natural laws, each event is inherently characterized by a time. It may be thus worthwhile to note that the motion through Galilean spacetime does not necessarily imply that the *same* vector space $\mathbb{R}^n$ (or affine space $\mathbb{A}^n$) as a scene for various physical and possibly other interactions evolves in time i.e., one cannot state that for a body at rest on Earth's surface its location would occupy the same point in space at time $t$ and $t + 1$ seconds, as this body unintentionally participates in a number of different motions (Earth's spinning rotation, revolution around the Sun, motion together with the entire solar system, etc.). Hence in the Galilean-Newtonian model of motion one has different location spaces for each moment of time, with a set of simultaneous events forming a $3n$-dimensional vector or, more exactly, affine space. Therefore, an adequate mathematical model for the Galilean-Newtonian spacetime $S$ would be a fiber bundle with the base space $\mathbb{R}$ (time) and fibers $\mathbb{R}^n$ (locations), rather than a direct product space $\mathbb{R} \times \mathbb{R}^n$. In the fiber bundle picture, time is obtained as a canonical projection from $S$ to $\mathbb{R}$ and has an absolute character (Figure 7). It is important to bear in mind that when discussing evolution in the Galilean-Newtonian spacetime the time dependence is quite often not written explicitly (e.g., in the Hamiltonian $H(p, q)$).

We may notice that Galilean-Newtonian mechanics tries to reduce all the processes to motion. One assumes that the evolution of any dynamical system may be described by a set of generalized coordinates $q^i(t), i = 1, \ldots, n$, as functions of time $t$. For mechanical systems, to fully characterize its behavior it is enough to know $n$ values of coordinates $q^i(t_0)$ and $n$ values of velocities $\dot{q}^i(t_0)$ or $n$ values of momenta $p_i(t_0)$, in generalized coordinates $p_i(t) = \partial L(q, \dot{q}, t)/\partial q^i, q \equiv \{q^i\}$. The Galilean-Newtonian model has been very useful in a civilization based on engineering applications, and there were numerous attempts to reduce all processes in nature to mechanical motion (Descartes) or at least to build other sciences using the same pattern. Such attempts mostly failed because it is not at all obvious that every phenomenon follows the elegant scheme of mechanical motion. For instance, biological and social phenomena can hardly be reduced to mechanical motion, although the respective evolutionary models do exist in biological and social sciences. One of the possible manifestations of inadequacy of the models of mechanical motion for complex systems is the problem of irreversibility and the related issue of time reversal non-invariance.

One can naturally ask if the time understood as an image $t: S \simeq \mathbb{R}^n \to \mathbb{R}$ i.e., a collection of parallel translations of the world on the real "time axis" is a finite or infinite set. In other words, does time protrude ad infinitum so that for any event there is always an "after"? Recall that an infinite ordered set is the one when for each element $n$ there is always the $(n + 1)$-th element. Time, however, in distinction to mathematical sets is related to physical measurements so that there may exist a time point in future after which no physical event will take place. Indeed, in physics, despite the fact that it is difficult to uncontroversially define the time operator, it is a measurable quantity i.e., time is always observed through material objects[62] (the "clock", albeit reduced to a single atom). If the matter ceases to exist so does the time. This means that if time is endless, the universe must eternally exist which nobody can guarantee. In modern theoretical astrophysics, there are models of both eternal and finite-life universes. Yet since there are hardly any corroborative observational tests to prefer one choice of models to another, the question of the end of the universe (Big Crunch) will probably remain perennial and of purely academic nature, and the same applies to the end of time.

Notice that the term "motion" being applied to a dynamical i.e., evolving system is generally understood in a broad sense, e.g., evolution of a country's economy is also a motion of a bunch of phase points in a properly chosen phase space; competition of two interactive species or a military contact of two (or more) adverse armies is a motion, etc.

---

[62] Recall Einstein's joke: time is what the clock shows. There is a profound meaning in this joke since time (e.g., a second) is actually not measured, but *defined* by what the atomic clock measure i.e., through the measurement of other physical quantities such as frequency. Thus, the standard second is defined as the duration of 9.192.631.770 oscillation periods corresponding to the transition between two hyperfine levels of the ground state of $^{133}$Cs atom.

We have already noticed that the so-called laws of nature nearly always have the form of equations. So, assume that we must select some equations to describe processes or phenomena. As a rule, real – not too much idealized – systems and processes are described by nonlinear equations, e.g., nonlinear partial differential equations (PDE) with respect to time and spatial coordinates. Such systems are distributed in space and correspond to an infinite number of degrees of freedom. Nonlinear PDEs are usually very complicated, and only very specific types of such equations admit physically observable analytical solutions. However, models – by their very definition – can be drastically simplified, for instance, by considering first spatially homogeneous situations. In this case, the equations modeling the system do not contain spatial derivatives and become ordinary differential equations (ODE-based models). Such a system is called point-like or having null-dimension (0d). In other words, it is the transition to homogeneous (point) models or, more exactly, ignoring spatial distribution of quantities, $\partial/\partial \mathbf{r} = 0$, that leads to ODE instead of PDE. It is always a favorable situation when one can reduce the equations to the most primitive form, retaining the essence of the model (a good example of such reduction to ODE-based (0d) modeling is the treatment of atmospheric processes in terms of radiation balance and global near-surface temperature dynamics).

The theory of ordinary differential equations may be regarded as an engine of dynamical systems analysis, driving both qualitative and quantitative study of evolution and prediction. The connection between the theory of differential equations and that of dynamical systems has been known since the middle of the 18[th] century (although the notion of a dynamical system hardly existed at that time) and is usually associated with the name of d'Alembert. In particular, if an $n$-order ordinary differential equation is traditionally written in the form $\frac{d^n x}{dt^n} = F\left(x, \frac{dx}{dt}, \dots, \frac{d^{n-1}x}{dt^{n-1}}, t\right)$, then putting $x(t) = y_1(t), \frac{dx}{dt} = y_2(t), \dots, \frac{d^{n-1}x}{dt^{n-1}} = y_n(t)$ we shall have a dynamical system $\frac{dy_1}{dt} = y_2, \dots, \frac{dy_{n-1}}{dt} = y_n, \frac{dy_n}{dt} = F(y_1, y_2, \dots, y_n, t)$ or, in a more general form, $\frac{dy_1}{dt} = f_1(y_1, y_2, \dots, y_n, t), \dots, \frac{dy_{n-1}}{dt} = f_{n-1}(y_1, y_2, \dots, y_n, t), \frac{dy_n}{dt} = f_n(y_1, y_2, \dots, y_n, t)$. If $t$ is regarded as time, these equations define the motion of a point over an $n$-dimensional manifold. Parameter $t$ is often, but not always, interpreted as the time; anyway $t$ may be viewed as an "intrinsic variable" that can be expelled to obtain $\frac{dy_1}{df_1} = \frac{dy_2}{df_2} = \dots = \frac{dy_n}{df_n}$. Note that although quantities $y_1, y_2, \dots, y_n$ are vector components, we do not write them here in the usual contravariant form $\{y^j\}, j = 1, \dots, n$ – for notational simplicity and unimportance, in the current context, of the transformation properties with the change of basis or, using the algebraic language, of the difference between vectors and linear functionals.

Note also that the conversion of a differential equation of the $n$-th order into a dynamical system in $n$-dimensional phase space is, in general, not unique. The non-uniqueness of the transition from a differential equation to the corresponding dynamical system can be illustrated on a trivial example of the free harmonic oscillator $\frac{d^2 x}{dt^2} + \omega^2 x = 0, \omega^2 := k/m$, where $k$ is an elastic (Hooke's) constant, $m$ is the oscillator mass. This equation can be written either as a system $\frac{dx_1}{dt} = x_2(t), \frac{dx_2}{dt} = -\omega^2 x_1(t)$ or as $\frac{dx_2}{dt} = x_1(t), \frac{dx_1}{dt} = -\omega^2 x_2(t)$.

Many equations of motion for classical mechanical systems have the form of ODEs. Such systems are point-like or lumped. In the modeling of a lumped mechanical system with ODEs, each degree of freedom is described by a second-order ODE. Recall that the number of degrees of freedom is defined as the dimensionality of the configuration space $Q$ (manifold) of a physical system. For example, the system $\ddot{\mathbf{x}} = \mathbf{F}(\mathbf{x}, t)$, where $\mathbf{F}$ is a plane vector field and $\mathbf{x} \in \mathbb{R}^2$, has two degrees of freedom. Accordingly (see also about tangent and cotangent bundles below), a degree of freedom is understood in classical mechanics as a pair "position plus conjugate momentum". Such a pairwise description is closely related to the concept of symplectic space, see below. Differential equations of the first order correspond to 1/2 degrees of freedom. Here one must be careful, since in the theory of dynamical systems the notion of degrees of freedom may be defined differently: each degree of freedom in some texts on dynamical systems corresponds to a phase space coordinate (which is more convenient since no half-integer degrees of freedom arise) so that the classical notion of a "degree of freedom" becomes redundant. Accordingly, in classical mechanics, the phase space dimensionality is just the doubled number of state variables $N$ understood in the spirit of dynamical systems theory. For example, instead of denoting a practically important case in the chaos theory as having 1.5 degrees of freedom one can say, "a system having 3 degrees of freedom". Nevertheless, this terminological discrepancy usually does not lead to confusion.

Spacetime events can be regarded as instances in the history of a physical object, and likewise objects of another, "non-physical" nature develop their history in some other spaces, not necessarily spacetime. The whole family of events composing the history of the object can be represented as a trajectory of a system in a relevant space: for instance, one can talk of a political trajectory of a country. If we limit ourselves to physical systems, the simplest case will be the trajectory of a material point, i.e., of a dimensionless particle with mass $m \neq 0$. Such a trajectory can be treated as consisting of local points, but one should remember that the requirement $m \neq 0$ is necessary for spatial (and spacetime) localization: one can localize a massless point ($m = 0$) in the momentum space only. In geometrical terms, the solution of the differential equation for $N$ point particles defines a curve which is a smooth map $\boldsymbol{\gamma}: I \rightarrow M, \mathbf{x}(t) = (x^1(t), \ldots, x^n(t)) \in M, t \in I \subseteq \mathbb{R}$, with $\mathbf{x}(t)$ taking values in the configuration space $M$ that is a smooth manifold, $M \subseteq \mathbb{R}^n, n = 3N$.

In a slightly more general setting than the Newtonian model, curve $\boldsymbol{\gamma}$ corresponds to the motion of a mechanical system when (and only when) points $\mathbf{x}$ belonging to this curve satisfy a certain second-order differential equation $\Phi(\mathbf{x}, \dot{\mathbf{x}}, \ddot{\mathbf{x}}, t) = 0$ which may be also called the motion equation (but not necessarily resolved with respect to the second derivative). This expression is a level set of an implicit function $\Phi$ that may be considered continuous and having continuous partial derivatives, and whose domain is a nine-dimensional affine space (or, for $N$ particles, $9N$) formed by quantities $\mathbf{x}, \dot{\mathbf{x}}, \ddot{\mathbf{x}}$ for each value of parameter $t$ (time-based fibers). Recall that according to the implicit function theorem (e.g., in its shortened form) equation $\Phi(\mathbf{x}, \dot{\mathbf{x}}, \ddot{\mathbf{x}}, t) = 0$ uniquely defines function $\ddot{\mathbf{x}} = \varphi(\mathbf{x}, \dot{\mathbf{x}}, t)$ if $\partial\Phi/\partial\ddot{\mathbf{x}} \neq 0$ which is actually the requirement that the mass of a system should be non-zero. Moreover, it is a priori not at all obvious that the motion equations should be universally limited up to the second order: this restriction is an astonishing fact implying that it is sufficient to know only the positions and velocities of the particles to fully describe their motion, but it could well be that one ought to know also accelerations (torsion of a curve) and even higher time-derivatives – in addition to positions $\mathbf{x}$ and velocities $\dot{\mathbf{x}}$ – to study the mechanical motion. Notice that ODEs in the explicit form are just the expressions for higher time derivatives via lower time derivatives, which is not true for the implicit form.

In other words, equations of motion for mechanical evolution would have the form $\Phi(\mathbf{x}, \dot{\mathbf{x}}, \ddot{\mathbf{x}}, \dddot{\mathbf{x}}, \ldots, t, b) = 0$, where dots signify the differentiation over time and $b$ denotes a set of parameters (characterizing inertia, constraints, etc.). An example of such a situation in physics is given by the model of a particle motion in classical electrodynamics taking into account the radiation force (reaction of radiation). The radiation reaction term is proportional to $\dddot{\mathbf{x}}$ and has intrigued many physicists since the beginning of the 20[th] century, it still is one of the oldest unsolved mysteries. Thus,

Abraham, Lorentz, Poincaré, and later Dirac tried to describe the motion of electron for small – nonrelativistic – velocities with the help of Newtonian mechanics i.e., in the form (here, for simplicity, the 1d motion is considered)

$$m_{in}\ddot{x} = \frac{a}{r_0}\ddot{x} + b\dddot{x} + cr_0 x^{(IV)} + \cdots, \qquad (8.1.1.)$$

where $m_{in}$ denotes the inertial mass of the electron i.e., the coefficient defining the particle inertial response to an applied force $F = F(t)$ and $\frac{a}{r_0}$ is the electromagnetic contribution to the mass. One can see that expansion in the right-hand side of (8.1.1.) can be interpreted as the abridged Laurent series in variable $r_0$ i.e., $m_{in}\ddot{x} = F$, where

$$F = F(t) = \sum_{n=-\infty}^{\infty} a_n r_0^{n-1} x^{(n+2)} = \frac{1}{r_0} \sum_{n=-\infty}^{\infty} a_n r_0^n x^{(n+2)}.$$

At the beginning of this book, we pointed out the usefulness of simple models. Thus, the naive Lorentz model of the electron representing it as a sphere of radius $r_0$ filled with electricity was very useful for developing the theory of electromagnetic fields and of particle behavior in them.

It has already been mentioned that in general the state of a system, e.g., expressed by the positions of its mass points may be defined not necessarily through their Cartesian coordinates, but also by $n$ parameters $q^i = q^i(t)$ belonging to some space $Q$. This space has in general the topological structure of a manifold and is called a configuration manifold, and parameters $q^i$ are known as the generalized coordinates. The generalized velocities are naturally defined as $\dot{q}^i = dq^i(t)/dt$. For example, if one considers a particle moving on a surface of the two-sphere $\mathbb{S}^2$ (one can model the motion of a satellite or of a missile in such a way), then the coordinates typically chosen on $\mathbb{S}^2$ are the azimuthal angle $\varphi \in [-\pi, \pi]$ and the polar angle $\theta \in [-\pi/2, \pi/2]$ (or sometimes $0 \leq \varphi \leq 2\pi, 0 \leq \theta \leq \pi$). Variable $\varphi$ corresponds to the longitude and variable $\theta$ to the latitude. Notice that these coordinates on a sphere contain multivalued points: each point having $\varphi = \pi$ and any value of $\theta$ coincides with the point having $\varphi = -\pi$ and the same value of $\theta$. Analogously, for $\theta = \pm\pi/2$ any value of $\varphi$ corresponds to the same point.

In general, the configuration space of a system with $n$ deqrees of freedom is a manifold $Q = Q^n$ on which coordinates $q^i$ have been chosen. Accordingly, one can introduce the notion of a configuration spacetime with coordinates $(q^i, t)$. The cotangent bundle $T^*Q$ is identified with the phase space of the system, where natural coordinates are $(q^i, p_j)$, while natural coordinates on tangent bundle $TQ$ are $(q^i, \dot{q}^i)$ as in Lagrangian mechanics (see below). When the Lagrangian depends on acceleration $L = L(q^i, \dot{q}^i, \ddot{q}^i)$, we shall have for the momentum $p_i = \partial L/\partial q^i - d/dt(\partial L/\partial \dot{q}^i)$ or $\mathbf{p} = \partial L/\partial \dot{\mathbf{q}} - d/dt(\partial L/\partial \ddot{\mathbf{q}})$.

## 8.2. From Aristotle to Newton

Ancient thinkers were enchanted by the celestial clockwork and devised a number of interesting models, the most well-known among them was that of Aristoteles (Aristotle), which can be formulated as the statement that the motion of bodies is possible only in the presence of external forces produced by other bodies (around 350 B.C.). Thus, in Aristoteles' mechanics, the state of rest was a natural and preferable one, to which a body strives when forces or their momenta are removed. In other words, the natural state of a body is to be at rest, and it can move only when driven by a force like a cart is pulled by a donkey. This verbal construct of Aristotle can be translated into the mathematical language as the first-order differential equation describing the change of the state of a moving body (or particle) under the influence of an external force $\mathbf{f}$

$$\frac{d\mathbf{r}}{dt} = \mathbf{f}(\mathbf{r}), \qquad \mathbf{r} \coloneqq (x, y, z), \qquad \mathbf{f} \coloneqq (f_x, f_y, f_z),$$

where vector $\mathbf{r}$ describes the state of the body. This appears to be the simplest model of motion which was supported by the "obvious" observation: no object could move unless acted upon – pulled or pushed, i.e., without a constant supply of energy. In this picture, Aristoteles' mechanics relates to viscous media that are resistant to motion.

In contrast with the model of Aristoteles, in Newtonian mechanics, forces $\mathbf{f} \coloneqq (f_1, f_2, f_3, \dots)$ change not the position, but the velocity of a body so that there is no absolute state of rest: "initial coordinate" $\mathbf{r}_0$ is an arbitrary integration constant. The concept of the absence of absolute rest, as first postulated in Newtonian mechanics, has later become one of the milestones of modern physics. One can immediately see that the crucial difference of Aristotle's model based on the first-order vector equation from the second-order system corresponding to the Galileo-Newton's model is, primarily, in the irreversible character of the motion. Aristotle's considerations were rooted in everyday experience: the trolley should be towed to be in motion; if the muscle force stops acting, the trolley comes to rest. In such a model the state of a system would be given by positions alone, and velocities could not be assigned freely. One might observe in this connection that even the most fundamental models reflecting everyday reality tend to be not unique and often controversial. Incidentally, the difference of Aristotle's and Galileo's models of mechanical motion is clearly manifested in gravitation: Aristotle thought that heavy objects fall faster while Galileo asserted (1636), based on real physical experiments, that all objects fall with the same acceleration that does not depend on their mass (once we may regard air resistance as negligible). The contrast of Aristotle's model to that of Newton (1686) is readily seen when one starts thinking about acceleration.

It is, by the way, not a priori obvious that any motion in the world should be described by a second-order differential equation so that to find the law of motion $\mathbf{r}(t)$ we must know initial positions and velocities. This is actually an astonishing fact. One can imagine a universe where one must also know initial accelerations (this will be an explicitly time reversal non-invariant universe) or even higher-order time-derivatives in order to determine the law of motion. Newtonian mechanics is a phenomenological mathematical model, a very successful one, but having a limited and not always clear applicability (small velocities, smooth motion, large masses). It is not at all obvious that Newton's differential equations are valid for small accelerations, which consideration has given rise to the so-called MOND (modified Newtonian dynamics) theory designed to understand the "dark matter" effects: the MOND model seems to work well in describing the observed galactic rotation curves ascribed to the action of some invisible substance. More specifically, according to Newton's law of gravitation, attraction $GmM(r)/r^2$ must result, for the spherically symmetric motion, in the appearance of "centrifugal force" $mv^2/r$ (see formula (S2.5) below) which automatically leads to the $1/r$ law of squared velocities of rotating stars and galactic clusters. Yet astrophysical observations do not confirm this simple rule related to the flat rotation of astrophysical objects, which has become a challenging problem in cosmological dynamics. One of the ways to interpret the contradiction between Newtonian theory and observations is to assume that mass $M(r)$ is distributed in some exotic manner, e.g., linearly grows with the distance $r$ from a gravitating object. However, this assumption immediately leads to contradiction with the observations of luminous mass distributions in the universe so that there must be large amounts of unseen – non-luminous or "dark" – matter. It is in order to evade this physically uncanny hypothesis that the MOND theory has been proposed.

So, one might ask: was Aristotle always wrong? Consider a trolley that stops due to friction: $\mathbf{f}_- = -\alpha\mathbf{v}$. Newton's equations for this case may be written in the form

$$\frac{d\mathbf{r}}{dt} = \mathbf{v}, \qquad m\frac{d\mathbf{v}}{dt} = \mathbf{F} - \alpha\mathbf{v},$$

where $\mathbf{F}$ is the towing force and $m$ is the mass of the trolley. When $\mathbf{F}$ is nearly constant, i.e., the towing force, as is frequently the case, slowly varies with time, the mechanical system is close to equilibrium, so the inertial term $m\dot{\mathbf{v}}$ is small compared to other terms. In this case, we get the equilibrium (stationary) solution $\mathbf{v} = d\mathbf{r}/dt = \mathbf{F}/\alpha$ which has the Aristotle's form. This solution is valid only when the motion is almost uniform i.e., the acceleration is negligible and the friction is sufficiently large, $|d\mathbf{v}/dt| \ll |\alpha\mathbf{v}|$. One can, in principle, imagine the world constructed on Aristotle's principles: the Aristotle model would correspond to the Universe immersed in an infinite fluid with a low Reynolds number. However, it is worth noting that Aristotle's model, had it been generally accepted, would have been badly compatible with modern physics. Indeed, neither quantum mechanics in today's form nor Einstein's general relativity could be reconciled with this model. The Schrödinger equation, which is a phenomenological extension of the Hamiltonian energy integral for Newton's equations, would not have existed in its current form as the energy integral does not exist for the dissipative Aristotle's picture. The statement that all bodies fall equally in the gravity field irrespective of their masses (the equivalence principle) would have been wrong since a heavy body in Aristotle's model should fall faster than a light one: it is pulled stronger by the Earth and hence has a greater velocity.

It is, however, curious that Aristotle's model is in fact extensively used in contemporary physics, engineering, and even in everyday life. An example is the Ohm's law, $\mathbf{j} = \sigma\mathbf{E}$ or $\mathbf{v} = \mathbf{E}/en\rho$, where $e$ is the electron charge, $\mathbf{E}$ is the electric field (acting force), $n$ is the charge density, $\rho = \sigma^{-1}$ is specific resistance and $\mathbf{v}$ is the average velocity of the charge carriers. Ohm's law is a typical macroscopic stationary model, when the driving force is compensated by resistance. Stationary models are ubiquitous in classical physics: one may note that classical physics almost exclusively dealt with slow and smooth motions such as planetary movement. Models describing rapid and irreversible changes, resulting in multiple new states, appeared only in the 20th century. Stationary and quasistationary models serve as a foundation of thermodynamics whose principal notion – the temperature – may be correctly defined only for equilibrium. Another example of a quasistationary model is the steady-state traffic flow simulation to be discussed later.

Notice that according to the Newtonian model of motion, forces symbolizing an external influence on the considered subsystem (a selected by us, human observers, part of the universe) produce acceleration of a body, therefore in mechanics the law of motion $\mathbf{r}(t)$ is in most cases obtained in two stages: at first one finds how the velocity varies with time and only then one determines the law of motion. In the theory of dynamical systems, this separation into stages becomes immaterial as well as the order of differential equation employed to describe evolution.

## 8.3. Newtonian mechanics

We shall see below that Newtonian mechanics, despite its old-fashioned look, can be a source of many useful models. As already mentioned, Newtonian mechanics can be regarded as a special form of a dynamical system represented by a system of ordinary differential equations (ODE) for a moving state (phase) point $x(t) = (\mathbf{q}(t), \dot{\mathbf{q}}(t), t)$, where $\mathbf{q} = \{q^i\}, \dot{\mathbf{q}} = \{\dot{q}^i\}, i = 1, \ldots, N; \ t \in \mathbb{R}$ (flow) or $t \in \mathbb{R}_+$ (semiflow) is the time parameter. Recall that a dynamical system can be viewed as describing any time of evolution. The system of ODEs corresponding to Newtonian model of evolution i.e., the mechanical motion has the form

$$\frac{dq^i}{dt} = v^i, \qquad \frac{dv^i}{dt} = F^i(\mathbf{q}(t), \dot{\mathbf{q}}(t), t) = F^i(x),$$

where functions $F^i(x)$ are proportional to the agents (forces) driving point $x$ in the i-th direction. These equations can be interpreted as describing a single moving particle in a state space of $N$ dimensions; in classical mechanics $N = 3n$, where $n$ is the number of degrees of freedom. In the habitual 3d space one usually considers the motion of a system of points positioned at $\mathbf{r}_a(t)$, where $\mathbf{r}_a(t): I \to \mathbb{R}^{3n}$ (more exactly $I \to \mathbb{A}^{3n}$), $a = 1, \dots, N$.

For a single point particle, Newton's equation that has the form $\ddot{\mathbf{r}} = \varphi(\mathbf{r}, \dot{\mathbf{r}}, t)$, where $\varphi: \mathbb{R}^3 \times \mathbb{R}^3 \times \mathbb{R}$ is a smooth function and can be interpreted as merely the consequence of the existence and uniqueness theorem known in the theory of differential equations. Or, to put it differently, Newton's equation is equivalent to the principle of determinism (see Section 3. Expected properties of mathematical and computer models.) When studying mechanics, one commonly starts from a simple 1d picture when it is assumed that both the force $F$ acting on a mechanical system (particle) and its motion $x(t)$ are one-dimensional quantities, $x(t) \in \mathbb{R}$. Besides, it is often assumed that the force $F$ is a function of position only, $F = F(x)$. Then we have a second-order nonlinear differential equation $F(x(t)) = m\ddot{x}(t)$ which is known as Newton's second law. Appropriate supplementary (initial) conditions to this equation are usually formulated as some starting position $x(t_0)$ and velocity $\dot{x}(t_0) = v_0$ at time $t = t_0$. Assuming that $x(t)$ is a smooth function we can deduce from the theory of differential equations that for each pair of initial conditions we shall get a unique solution in some neighborhood of time $t = t_0$.

Historically, classical mechanics originated in the beginning of the 17[th] century through astronomical observations and, at that time, controversial heliocentric concepts by Galileo. Almost simultaneously with Galileo, in the early 1600s, came Kepler who ingeniously articulated three laws of planetary motion. However, Kepler's laws were utterly phenomenological[63]: they depicted the patterns of motion but did not consider its causes. It was Newton who endeavored to ascertain the universal reasons for motion, and that was a major triumph marking the beginning of modern science: understanding the roots of motion and gravitation.

In contemporary terms, Newtonian mechanics was the "theory of everything" (TOE) before 1900, and as such, as a theory, it permeated almost all fields of science – not only physics. However, as already mentioned, the Newtonian version of mechanics, although it was historically the main mathematical formulation of classical mechanics and engineering, is a rather limited mathematical model. It is not accidental that despite the fact that Newton's law of motion may be considered a complete statement of classical mechanics, there have been many attempts to provide other mathematical formulations for this discipline. It would be interesting and useful to better understand why one would desire to have other models.

The utmost concern of mechanics is to provide a full description of the motion of particles and bodies under an external influence. Newton's laws are used to find the position $x(t)$ of a material point of mass $m \neq 0$ as a function of time $t$. The so-called Newton's second law is the most important primarily because it provides a framework for building simple models of motion. The mathematical content of Newtonian mechanics is reduced to the second-order differential equations for $x(t)$, which are nonlinear and, therefore, do not in general admit an explicit solution, especially for the most interesting problems.

---

[63] And based on religious grounds.

Notice that in Newtonian mechanics or in general in the non-relativistic (or, rather, pre-relativistic) paradigm, time $t$ is a special quantity that "flows" and physical laws express the change of the considered systems with time, the latter is a real independent variable whose value can be measured with the help of periodic physical processes declared as "clocks". In general, any parametric family of diverse states can be treated as a clock, with the time value being encoded in these states. The quality of the time measurement would be characterized by the time resolution i.e., determined by the ability to distinguish between such states while estimating their distinctive parameters.

In relativistic physics, different notions of time appear. Thus, alongside with the local time coordinate $t = x^0/c$ that partly characterizes the spacetime point $x^\mu = (x^0, x^1, x^2, x^3)^T \in M$, where $M$ is four-dimensional spacetime endowed with Riemannian (semi-Riemannian) metric $g_{ik}(x^\mu)$, and enters, e.g., in the set of arguments for relativistic fields such as the electromagnetic tensor field $F_{ik}(x^\mu)$ or metric tensor $g_{ik}$, there exists the proper time $\tau(\boldsymbol{\gamma}) = \int_{\boldsymbol{\gamma}} d\tau \left( g_{ik}(\gamma(\tau))(d\gamma^i/d\tau)(d\gamma^k/d\tau) \right)^{1/2}$, where $\boldsymbol{\gamma} = (x^1(x^0), x^2(x^0), x^3(x^0)): [\tau_0, \tau_1] \to M$ is the world line. The proper time $\tau$ has a nonlocal – integral – character and yields the arclength measured along the world line. Note that time $t$ in non-relativistic mechanics is a directly observable quantity produced by physical measurements (clocks), whereas in relativistic physics it is the proper time $\tau$ that is measured along the world line of the clock so that the coordinate parameter $t = x^0/c$ can be obtained from $\tau$.

It is well known that the general problem in Newtonian mechanics amounts to solving the system of equations

$$m_a \frac{d^2 \mathbf{r}_a}{dt^2} = \sum_i \mathbf{F}_i(\mathbf{r}_a, \dot{\mathbf{r}}_a, t), \qquad (8.3.1.)$$

where index $a = 1, \dots, N$ enumerates the particles and the sum goes over all the forces acting on the $a$-th particle. This expression may serve as a definition of Newtonian mechanics.

A mathematical solution of the finite system (8.3.1.) of second-order differential equations is a $3N$ vector-function[64] $\mathbf{x}(t) = \mathbf{r}_a(t), a = 1, \dots, N$. If one considers the simplest case of the motion of a point mass (material particle) in 3d space, the position vector $\mathbf{r}$ may be expressed as $\mathbf{r} = x^1 \mathbf{e}_1 + x^2 \mathbf{e}_2 + x^3 \mathbf{e}_3$, where $x^1, x^2, x^3$ are projections that are varying when the particle moves, the three quantities $\mathbf{e}_1, \mathbf{e}_2, \mathbf{e}_3$ are unit vectors assumed to be constant in the laboratory system. In general, the second-order system of ODEs (8.3.1.) is difficult to solve and even to explore unless the associated linearized equation $\ddot{\mathbf{x}} = \mathbf{F}_1 \dot{\mathbf{x}} + \mathbf{F}_2 \dot{\mathbf{x}} + \mathbf{F}_3(t), x \in \mathbb{R}^n$, where $\mathbf{F}_1, \mathbf{F}_2$ are matrices with real entries, $\mathbf{F}_3(t) \in \mathbb{R}^n$, can be thoroughly investigated.

---

[64] Variables $\mathbf{r}$ denote a vector in $\mathbb{R}^n$ whereas variables $\mathbf{x}$ denote the point corresponding to vector $\mathbf{r}$ in an associated affine space $\mathbb{A}^n$; this clumsy difference of notations is due to the fact that curves and surfaces live in affine spaces, and we eventually have to deal with trajectories. For practical purposes of physicists, such a difference is inessential.

We may note that mechanical systems considered in the Newtonian paradigm can be nonlocal, and in such cases, they are described not by the second-order systems of ODEs but by integro-differential equations. We, however, shall not touch upon nonlocal systems in this book, although they are widely used in nature.

Notice that any solution to Newtonian equations (8.3.1.) depends on masses $m_a$ that enter the equations as parameters. In many courses of classical mechanics and probably in the entire elementary physics, these parameters are taken for granted and considered nearly a godsend. According to Newton the meaning of mass is determined by his own equation: if a body is accelerated under the influence of some force, the body's mass is simply the ratio of the force to acceleration. Of course, one assumes that masses are measured in the inertial (Galilean) system of reference, where bodies are at rest or are moving uniformly along straight lines with respect to one another (unless driven by some force). Recall that according to Galileo's relativity principle all mechanical laws are the same in all inertial systems.

In modern physics, the Newtonian understanding of mass is no longer valid. Actually, the concept of mass is one of the most sophisticated in nature. One can encounter different kinds of mass in physics: inertial, gravitational (both passive and active), effective mass, reduced mass, electromagnetic mass, rest mass (invariant, $m$), relativistic (variable, $\gamma m$), transversal and longitudinal masses, etc. The reason for the existence of diverse masses seems to be twofold: firstly, that mass is in general not the measure of the quantity of matter in an object, as we were taught at school, but depends on the distribution of the surrounding particles and fields (as well as on their states) and, secondly, that each kind of mass emerges in a given context and corresponds to a specific model. For instance, if the particle's background fluctuates as often occurs in many-body systems, then the mass involving the particle dynamics may acquire a stochastic component. In other words, each particle's mass may vary as the particle changes its location with respect to other particles comprising the background. See more in "Mass and Matter" above.

Forces $\mathbf{F}_i$ are regarded as given functions of positions and velocities of each particle and of time $t$ at every point. An important specific case is represented by potential forces; recall that a vector field $\mathbf{F}$ is potential if and only if its work $A_\Gamma = \int_\Gamma \mathbf{F} d\mathbf{r}$ does not depend on contour $\Gamma$. It is clear that in general the contour integral $A_\Gamma$ depends on trajectory $\Gamma$ described by position vector $\mathbf{r}(t)$ so that one cannot obtain elementary work i.e., infinitesimal increment $\delta A = \mathbf{F} d\mathbf{r}$ simply by differentiating $A_\Gamma$. In other words, one cannot in general find function $V(\mathbf{r})$ (a scalar potential) such that $\mathbf{F} = \nabla V$ or $\mathbf{F} = -\nabla V$, as conventional in mechanics. When, however, such a function of point $\mathbf{r}$ can be correctly defined, $V(\mathbf{r}) = -\int_{\mathbf{r}_0}^{\mathbf{r}} \mathbf{F}(\mathbf{r}) d\mathbf{r}$ and $\mathbf{F}(\mathbf{r}) = -\nabla V(\mathbf{r})$, then mechanical work $A$ does not depend on trajectory $\Gamma$, but is only a function of its initial and final points $\mathbf{r}_0, \mathbf{r}$. In the language of differential forms, which seems to be the most appropriate technique to represent systems of ODEs, in particular dynamical systems, i.e., to handle vector fields, especially on Riemannian or pseudo-Riemannian manifolds, a potential vector field in 3d corresponds to an exact 1-form and a scalar potential is identified with a 0-form i.e., a scalar function. Recall that a differential form $\theta$ is called exact when it can be represented as an exterior derivative of another form $\omega, \theta = d\omega$, giving a differential form of higher degree. In 3d, the exterior derivative of 1-form $\theta \equiv \theta_{\mathbf{v}} = v_1 dx^1 + \cdots + v_n dx^n$ for vector field $\mathbf{v}$ on $\mathbb{R}^n$ ($n = 3$ in our case) is the 2-form $\eta_{\mathbf{v}} = d\theta_{\mathbf{v}} = \text{curl } \mathbf{v}$. Locally $\theta_{\mathbf{v}}$ is the inner (dot) product with vector field $\mathbf{v}$; the integral of $\theta_{\mathbf{v}}$ along a path gives a scalar quantity analogous to mechanical work, see more details, e.g., in [16].

## 8.4. Dynamical systems

The notion of a dynamical system appeared as a generalization of Newtonian mechanics, where motion i.e., the evolution in time is described by the second-order differential equations resolved with respect to an unknown function. The extension of Newton's equations to a more general structure represented, in particular, by a dynamical system can be used to describe the entities of any nature: not only material points (particles) which are the main objects of study in Newtonian mechanics, but also much more complex physical, chemical, biological, economic, social, etc. systems. Dynamical systems can model a wide range of behavior: the spread of diseases over a population, the physiological processes that regulate the heart rhythms, oscillations in electronic circuits, interactions of optical pulses and laser generation, the motion of planets in the solar system, climate and weather, variations of stock prices, formation of traffic jams, even the shape of plants and their growth. In general, the dynamical systems theory is used to model the processes governed by a set of evolutionary laws, with the emphasis on long-term behavior.

Consider for example the dynamics of astrophysical objects such as stars, planets, galaxies and their clusters. Will they always move regularly as we used to think of the planets in the solar system or may something unpredictable happen? For example, can some planets collide with each other or be repelled from their smooth paths around the Sun and travel freely through space? To answer these questions one ought to analyze a dynamical system formed by Newton's equations for $n$ gravitationally interacting bodies and find the asymptotic (topological) patterns of trajectories. The study of planetary $n$-body motion, where the dynamical system describes the concerted evolution of point masses, one of which being much larger than any other ($n = m + 1$), with all the masses interacting only by gravity forces (a mechanical model of the solar system), has motivated an extensive mathematical research and resulted in the well-known KAM (Kolmogorov-Arnold-Moser) theory.

Starting from the 1970s, with the exploding interest in chaos and strange attractors, dynamical systems theory has gained much popularity and many unexpected connections have been discovered with various areas of science and technology. In mathematics, for example, topology and even number theory are using the results produced by studying dynamical systems (the generic features of vector fields) whereas in engineering, economics, meteorology and other applied disciplines the theory of dynamical systems serves as a foundation of mathematical and computer modeling of evolving processes. When one speaks about dynamical systems, one typically implies that some part of the world can be viewed as a quasi-closed entity characterized by an intrinsic temporal behavior, the system's own evolution. The hypothesis of closedness, though ubiquitous, is rather strong and in many cases does not hold since most of the systems considered strictly closed are eventually destroyed by external fluctuations always penetrating the system. Fluctuations are understood as an uncontrolled process that cannot be reproduced in all detail.

A mathematical model for an entity (an isolated part of the world) evolving in time consists in defining its actual state $\mathbf{x}$, initial state $\mathbf{x}_0$ and dynamics $\dot{\mathbf{x}} = \mathbf{f}(\mathbf{x}, t)$. Function $\mathbf{x}(t)$ is a solution to this differential equation if substituting $\mathbf{x}(t)$ in $\dot{\mathbf{x}} = \mathbf{f}(\mathbf{x}, t)$ produces an identity $d\mathbf{x}/dt = \mathbf{f}(\mathbf{x}(t), t)$ at any point $(t, x)$. In other words, solutions are the trajectories that everywhere have the right velocity. We see that the actual state $\mathbf{x}(t)$ of a dynamical system at any time $t > t_0$ is, under very natural conditions, uniquely produced from an initial state $\mathbf{x}_0 = \mathbf{x}(t_0), (t_0, \mathbf{x}_0) \in I \times \mathbb{R}^n$, according to a certain rule $\mathbf{f}$ established for a given entity. Here $t_0$ is some initial time instant; in most cases one can put $t_0 = 0$, this is by all means justified in the autonomous case $\dot{\mathbf{x}} = \mathbf{f}(\mathbf{x})$.

The physical meaning of the existence-uniqueness theorem (Cauchy-Lipschitz-Peano or Picard-Lindelöf theorem) is that if the vector field $\mathbf{f}(\mathbf{x}, t)$ i.e., the law of evolution is given, then the state of the system at any future moment $t > t_0$ is completely determined by its initial state (at $t = t_0$). This property is known as "determinism". For most applications, the dynamical system is equivalent to a Cauchy problem $\dot{\mathbf{x}} = \mathbf{f}(\mathbf{x}, t), t > t_0 \geq 0, \mathbf{x}(t_0) = \mathbf{x}_0$. In the global theory of dynamical systems, one is typically interested in asymptotic regimes i.e., what will be the ultimate behavior of trajectories when $t \to \pm\infty$. One can also represent the dynamical system as an explicit function of some parameter seriously affecting the system's behavior – a control parameter $\mu$, $\dot{\mathbf{x}} = \mathbf{f}(\mathbf{x}, t, \mu), t > t_0 \geq 0, \mathbf{x}(t_0) = \mathbf{x}_0$. Analysis of such systems is simplified if they admit a polynomial representation, $\mathbf{f}(\mathbf{x}, t, \mu) = \sum_{k=1}^{m} f_k(t, \mu)(x - x_k)^{\nu_k}, \ \nu_k > 0$. More generally, a control parameter is represented by a function or even by a vector field so that a dynamical system $\dot{\mathbf{x}} = \mathbf{f}(\mathbf{x}, t, \mathbf{u}), \mathbf{x} \in Q, t \in I \subset \mathbb{R}, \mathbf{u} \in U \subset \mathbb{R}^s$ can be interpreted as a mathematical model of a control process.

Although the concept of dynamical system has its origins in Newtonian mechanics, it is unimportant whether the considered system has anything to do with physics. The essential thing is evolution, qualitatively described by the asymptotic behavior of phase trajectories. The physical nature of such an evolving system is immaterial; it is just a mathematical abstraction that can be applied to the objects of any nature (physical, technological, biological, social, economical and so on). Thus, a dynamical system gives in general a model for the evolution of a phenomenon or a process of any nature. This generality of dynamical systems is due to reliance on geometric information in the form of a vector field. Interpretation of this geometric information is the primary means of grasping the process development and the corresponding evolutionary equations. One can recall the metaphor by R. Abraham and C. Shaw "Dynamics – the geometry of behavior" [3]. Notice, however, that economics, social organizations, biological assemblies and other complex systems may have their own fundamental laws of evolution that cannot be necessarily derived from the microscopic physical laws of motion.

In mathematics, a dynamical system is usually understood as a triad $(M, \mu, g_t)$, where $M$ is a measurable space or manifold with measure $\mu$ (a state or a phase space) and $g_t$ (measurable on $M \times \mathbb{R}$) is the group of measure-preserving automorphisms, $\mu(g_t x) = \mu(x)$ for all $t \in \mathbb{R}$ (continuous dynamical system) or $t \in \mathbb{Z}$ (discrete systems) and all measurable $x \subset M$. Recall that here symbol $\mathbb{R}$ as usual denotes a set (group, field) of reals and $\mathbb{Z}$ is a set (group, ring) of integers. One usually assumes that the one-parameter group $g_t$ expressing the evolution of state $x$ ensures differentiability of $g_t x$ over $x$ (i.e., locally over all $x^i, i = 1, \ldots, n$) and $t$. So, a dynamical system can be interpreted as the action of some group (or semigroup) $g_t$ on a manifold $M$ which is regarded as a state space or, in problems similar to mechanics, as a phase space[65]. Hence dynamical systems can be classified according to the symmetry group $g_t$ implementing the evolution of system states in state space $M$. Thus, the theory of dynamical systems is largely a geometric theory (as, incidentally, also classical and relativistic mechanics).

One should, however, bear in mind that dynamical systems cannot be always modeled by differential, difference, integral or integro-differential evolution equations. In some cases, a time series obtained from observations or experiments describes a dynamical system. In general, there exist situations when there is no substitute for raw data. For example, while studying financial markets or evolution of the climate what we really have is chaotic time series. The respective dynamical equations (such as the Black-Scholes equation describing the time-dependent value of financial derivatives, e.g., options, or the system of fluid dynamics equations combined with radiation transfer equations), which are derived under rather simplified assumptions and can thus be used to model a sterile artificial world, may have little to do with real-life processes. Recall that mathematical models should not be expected to adequately represent reality.

---

[65] Sometimes the state space is defined as the phase space extended by the time coordinate i.e., as the manifold $T^*Q \times \mathbb{R}$.

### 8.4.1. Phase space and phase volume

We can write a dynamical system locally in components, $\dot{x}^i = f^i(x^j, t), i, j = 1, \dots, n$. Here quantities $x^1, \dots, x^n$ determining the state of a dynamical system are viewed as coordinates of point $\mathbf{x}$ in an $n$-dimensional space called phase space $\Gamma$. In other words, the domain $D, x^j \in D$ on which $f^i$ are defined is called, in the theory of dynamical systems, a phase space (i.e., $D := \Gamma$), and $n$ is its dimensionality. Notice the difference with physics, where we are used to define the phase space as a direct product of coordinate and momentum spaces (this is a consequence of the tacit assumption that all systems either are Hamiltonian or can be described by symplectic pairs, which is far from truth). The term "phase space" originates from the classical works of physicists of the late $19^{\text{th}}$ – early $20^{\text{th}}$ century, when scientists called the state of a physical system its "phase", see e.g., the papers [67, 68] by J. W. Gibbs. Point $\mathbf{x} = (x^1, \dots, x^n) \in \mathbb{R}^n$ is usually referred to as a phase point (in physics sometimes representing point), and the change of state with time can be described as the motion of the corresponding phase point along some path in $\Gamma$, this path being called a phase curve or phase trajectory. In plane words, the phase space is a set of all possible states of a process modeled by a dynamical system.

More generally, the state $\mathbf{x} = (x^1, \dots, x^n)$ of a dynamical system is characterized not only by its configuration, but also by the latter's speed of change $\mathbf{v} = \mathbf{f}(\mathbf{x})$ which is understood as evolution or motion $\dot{\mathbf{x}} = \mathbf{v}(\mathbf{x})$. Phase point $\mathbf{x}(t)$ coincides at moment $t$ with point $\mathbf{x} \in \mathbb{R}^n$ and has velocity $\mathbf{v} = \mathbf{f}(\mathbf{x})$ at this point, which is often described with the help of the delta-function, $\rho(\mathbf{x}, \dot{\mathbf{x}}, t) = \delta(\mathbf{x} - \mathbf{x}(t))\delta(\dot{\mathbf{x}} - \mathbf{v}(\mathbf{x}))$, where $\rho$ is the phase density. Thus, the state of a dynamical system is represented geometrically as a moving point (a phase point), and a set of all such points is called, as we know, the phase space. It is important to note that in general $\Gamma$ is not a Euclidean vector space $\mathbb{R}^n$. For example, in continuous-time dynamical system theory, phase space is usually understood as a differentiable (smooth) manifold. Recall that a vector space $U$ endowed with local coordinates $x^i, i = 1, \dots, n$ (a coordinate chart) may be thought of as a set of linear combinations of the form $\mathbf{v} = v^i \partial_i \equiv v^i \partial / \partial x^i$, where $v^i$ represents a tuple of functions (a vector field) on $U$ and $\partial_i \equiv \partial / \partial x^i$ is the basis of $U$ dual to $x^i$. Note that the coordinates or coordinate functions can be understood as homeomorphic maps onto charts (or open sets) in some model manifold or, simply speaking, the coordinate functions are representing the manifold charts.

The phase flow is a map $g_t \colon M \times \mathbb{R} \to M$, $x(t) = g_t x(t_0) \equiv g_t x_0, x_0 \equiv x(t_0) \in M, x(t) \in M$ is a point in $M$ provided it started at $x_0$ at $t = t_0$ moving for time $t$ in vector field $\mathbf{v}(\mathbf{x})$. Contrariwise, if map $x(t) = g_t x(t_0), M \times \mathbb{R} \to M$ is known, one can restore vector field $\mathbf{v}(\mathbf{x})$, generating flow $g_t$, by differentiation. To construct the flow of vector field $\mathbf{v}$ means to solve the ODE $\dot{x}^i(t) = v^i(x(t))$ in a coordinate chart of manifold $M$, where vector field $\mathbf{v} = v^i \partial / \partial x^i$ can be found; coordinates $x^i(t)$ spann the phase space. A more general concept of a state space can also be discrete such as in lattice or probabilistic models, where states are only counted when the system has reached a local equilibrium (description of "heads or tails" experiment is a simple example).

In physics, most of the models are related to the motion in potential fields, the archetypal problem is a particle submerged in such a field. The respective modeling equations are $\dot{\mathbf{x}} = m^{-1}\mathbf{p}, \dot{\mathbf{p}} = -\nabla V(\mathbf{x}), \mathbf{x} = \{x^1, x^2, x^3\}, \mathbf{p} = \{p^1, p^2, p^3\}$. In this simple case, the phase space is just a six-dimensional Euclidean space yet in general it is not the Euclidean space, but a differentiable manifold, e.g., a surface of constant energy. Recall that each smooth manifold can be embedded into some Euclidean space. This embedding produces associations with a surface in the Euclidean space which is a graphic representation of a manifold that is useful for many practical purposes.

The practical value of differentiable manifolds is that one can apply the rules of classical calculus on them, at least locally – calculus i.e., mathematical analysis of infinitesimals was initially developed in Euclidean space $\mathbb{R}^n$ (at first for $n = 1$). So, when performing differential operations in the phase space, one can ignore the fact that it looks like Euclidean only locally. In still more practical terms, one can treat mappings between differentiable manifolds $M$ with local coordinates $\{x^i\}, i = 1, \ldots, m$ and $N$ with local coordinates $\{y^j\}, j = 1, \ldots, n$ as a usual smooth ($C^\infty$) vector function $y^j = f^j(x^i)$ of $m$ variables. This property is actually used when we change variables, e.g., while performing a canonical transformation in classical mechanics.

In most situations, the phase space is defined by many numbers i.e., phase space is multidimensional. For instance, the phase space of two interacting populations (usually denoted as predators and preys in a class of competition models) is two-dimensional. There exist, however, numerous processes and physical systems that cannot be described in terms of a finite-dimensional state or phase space, for example, flows in fluid dynamics, wave propagation in acoustics, optics and seismics, elementary excitations in condensed matter, fields in classical and quantum electrodynamics, etc.

One must bear in mind that the concept of a deterministic dynamical system is an idealization just like the Newtonian model of motion. In physical reality, such phenomena as fluctuations and noise are always present, and even if one considers physically unrealistic "closed systems", sources of random motion tend to destroy such closedness rather quickly. Systems can be subject to external disturbances and internal fluctuations. The latter pervade the whole universe, yet in physics and especially in engineering it is customary to consider the situations when the role of fluctuations is negligible. For instance, if one deals with a macroscopic stream of gas characterized by mass current $\mathbf{j} = \rho\mathbf{u}$, then this quantity is understood as the average value of the flow consisting of a great number of gas molecules.

One typically assumes that the role of fluctuations is insignificant when the characteristic size $L$ of a considered system is large compared to the average intermolecular distance, $L \gg d \sim n^{-1/3}$. This is, however, a very intuitive ("physical") assumption that does not necessarily hold, specifically when the kinetic description is required. Moreover, although one can, in such situations when condition $L \gg d$ is fulfilled, disregard the fluctuations and describe the system with fully deterministic equations, the microscopic (grainy) structure of a fluid still can manifest itself as, e.g., for the case of a rarefied (Knudsen) gas, when the mean free path $\bar{l} = (n\sigma)^{-1} \sim d(d/a)^2$ becomes comparable with the characteristic system dimension $L$. Thus, in most engineering problems one assumes that a more stringent condition is satisfied, $L \gg \bar{l}$; then the microscopic structure of a fluid disappears completely, and its properties can be described fully macroscopically, e.g., by the equations of fluid dynamics. Recall that such equations always contain phenomenological parameters such as viscosity, heat conductivity, diffusion coefficients, etc., which are regarded as given (in particular, from experiment). This is a payment for the simplified macroscopic treatment. Calculations and estimates of such phenomenological parameters, which are also known as kinetic coefficients, is the task of kinetic theory that shall not be considered here.

The theory of dynamical systems in its naive pointwise setting as a Cauchy problem for vector differential equation $\dot{\mathbf{x}} = \mathbf{f}(\mathbf{x}, t), \mathbf{x}_0 = \mathbf{x}(t_0)$ is a typical modeling task. If it is well-posed, this model may allow a crude prediction of states $\mathbf{x}(t)$ at all later times $t > t_0$ within some interval $I, t \in I \subseteq \mathbb{R}$, determined by the uniqueness theorem, but such predictive modeling is of purely theoretical character. The matter is that a simple pointlike initial condition $\mathbf{x}_0 = \mathbf{x}(t_0)$ giving rise to a single-phase trajectory in most cases does not make sense from the practical standpoint. Indeed, to completely specify the pointlike path $\boldsymbol{\gamma}\colon \mathbf{x}(t) = (x^1(t), \dots, x^n(t))$ emanating from $\mathbf{x}_0$ and giving the history of the flow in phase space $P$, one has to perform measurements with an infinite precision which is physically meaningless. Not only physically: in biology, economics, sociology, etc., i.e., in any discipline, where one has to resort to predictive modeling based on deterministic evolution equations, infinite precision of measurements or, nearly equivalently, an infinite number of measurements, can never be attained. Even if it could – purely hypothetically of course – such excessive accuracy would be meaningless anyway since at the next moment the initial state (or any other real-life state) would randomly change. The measurement process is *always* limited by finite accuracy (not only in quantum mechanics, where this fact is regarded as a fundamental principle).

A more adequate modeling approach than describing evolution with pointlike trajectories, from the experimental standpoint, would be to consider small domains $\Delta\Gamma$ in the phase space $\Gamma$ instead of phase points. These small domains correspond to a finite and not to an infinite precision. All mathematical points $\mathbf{x} \in \Delta\Gamma$ represent in this scheme the same observationally accessible state so that mathematically distinct initial conditions give rise to a non-zero measure (the Gibbs measure)[66] and constitute the so-called statistical ensemble. We already mentioned that the states, according to J. W. Gibbs, are measures rather than points of the phase space. The concept of statistical ensemble introduced in 1902 by J. W. Gibbs [67] and [68] radically changed the entire science, not only physics. A statistical ensemble is comprised of a very large (physically infinite) number of identical systems evolving according to the same evolution law $g_t$, subordinated to the same constraints and to be found, in general, in the same circumstances. In the simplest setting, the only difference between all such clones – the ensemble members – is in diverse initial conditions. This approach naturally emerging from practical considerations leads to introducing the probability $w(\Delta\Gamma, t)$ for a system to be found in the phase space domain $\Delta\Gamma$ at moment $t$. Without probabilistic concepts, it would be hardly feasible to describe complex or stochastic systems (such as chaotic regimes), even when the starting equations are completely deterministic. The phase space in the Gibbs picture looks like a collection of cells rather than points, these cells may have different size, dimensionality and topology (the concept of coarse graining). The probability density of finding the system at time $t$ in a phase cell $\Delta\Gamma$ is given by $w(\Delta\Gamma, t)$ proportional to phase volume $\Delta\Gamma$ when this volume becomes infinitesimal. In the context of dynamical systems, if $\mathbf{v}(\mathbf{x}, t)$ is a smooth vector field on an $n$-dimensional manifold and $g_t$ is a corresponding group (semigroup) of translations along the vector field trajectories (respectively, flow or semiflow), then a continuous measure $\mu$ is given in any local coordinate frame as $d\mu = \rho(x^1, \dots, x^n) dx^1 \dots dx^n$. The Liouville theorem mentioned above (see 7.9.) states that measure $\mu$ is invariant under the action of flow $g_t$ generated by vector field $\mathbf{v}(\mathbf{x})$ (or $\mathbf{v}(\mathbf{x}, t)$) on a manifold, provided the probability density $\rho$ satisfies the Liouville equations div $\mathbf{v} = \partial_i(\rho v^i) = 0$ or $\partial_t\rho + \partial_i(\rho v^i) = 0$ in the non-autonomous case (see (7.11.1.)-(7.11.2.)). Measure $\mu$ is an invariant of a dynamical system which in practical terms means that the number of phase trajectories stemming from a domain of initial data $\mu(\mathbf{x}_0) = \mu(\mathbf{x}(t_0))$ is conserved, with density $\rho$ playing the part of the density of such trajectories. Rewriting the Liouville equation in the form

---

$$\frac{d\rho}{dt} = \frac{\partial \rho}{\partial t} + v^i \frac{\partial \rho}{\partial x^i} = -\rho \, \mathrm{div} \, \mathbf{v},$$

we have the ODE $d \ln \rho \, / dt = -\mathrm{div} \, \mathbf{v}$, and formally integrating it from $t_0$ to $t$ we get

$$\frac{1}{T} \ln \frac{\rho(t)}{\rho_0} = -\frac{1}{T} \int\limits_{t_0}^{t} \mathrm{div} \, \mathbf{v} \, dt = -\langle \mathrm{div} \, \mathbf{v} \rangle_T,$$

where $T \equiv t - t_0$, $\rho_0 = \rho(t_0)$ and angular brackets denote the time averaging. One can interpret the obtained relationship in the following manner. The meaning of probability density $\rho$ is the number of states $\Delta N$ in phase cell $\Delta \Gamma$, related to the total number $N$ of states in the phase (state) space i.e., $\rho \approx \frac{1}{N} \frac{\Delta N}{\Delta \Gamma}$ in the limit of infinitesimal $\Delta \Gamma$ ($\Delta \Gamma \to 0$). Then, assuming that the number of states in each cell $\Delta \Gamma$ remains constant throughout the evolution, we have $\frac{1}{T} \ln \frac{\Delta \Gamma(t)}{\Delta \Gamma_0} = \langle \mathrm{div} \, \mathbf{v} \rangle_T$ or $\frac{S(t) - S_0}{T} = \langle \mathrm{div} \, \mathbf{v} \rangle_T$, where $S_0 \equiv S(t_0) = \ln \Gamma_0 \equiv \ln \Gamma(t_0)$, $S(t) = \ln \Gamma(t)$ is the entropy of phase volume (for a given phase space partitioning). Thus, the entropy change per unit time in a dynamical system is proportional to its vector field divergence, which naturally leads us to the concept of non-conservative systems. For conservative systems, $\langle \mathrm{div} \, \mathbf{v} \rangle_T = 0$, i.e., the flow $\mathbf{v}$ preserves the phase volume, which is, in particular, by virtue of the Liouville theorem the feature of Hamiltonian systems, then entropy does not change. Note that the phase space volume $\Delta \Gamma$ and hence its logarithm – the entropy[67] – are conserved quantities even when energy is not preserved, e.g., in the expanding universe treated as a whole (the energy is not a conserved quantity in a non-stationary universe).

Recall that for Hamiltonian systems vector field $\mathbf{v}(\mathbf{x})$, $\mathbf{x} = (q^i, p_j) \in M$, $i, j = 1, \dots, m$, where $M$ is $n = 2m$-dimensional symplectic manifold (with symplectic form $d\omega = dq^i \wedge dp_i$), is defined by the Hamiltonian equations $\dot{q}^i = \partial_{p_i} H, \dot{p}_i = -\partial_{q^i} H$. Here $H$ is a time-independent Hamiltonian function and measure $\mu$ has density $\rho(q, p) = 1$ which is an obvious invariant. For a linear autonomous system $\dot{\mathbf{x}} = A\mathbf{x}$, where $n \times n$ matrix $A$ has only constant entries, we have $\mathrm{div} \, \mathbf{v} = \mathrm{div} A\mathbf{x} = \mathrm{Tr} A$ so that $\Delta \Gamma(t) = \exp(t \mathrm{Tr} A) \Delta \Gamma_0$ – a simple but important evolution formula. We may note in passing that the initial conditions $\mathbf{x}_0$, in particular, $\mathbf{x}_0 = (\mathbf{q}_0, \mathbf{p}_0)$ for a mechanical system may be regarded as integrals of motion.

When the dynamical system is defined on a smooth manifold $M$ and diffeomorphism $g_t: M \leftrightarrow M$, $\mathbf{x}_0 \equiv \mathbf{x}(t_0) \leftrightarrow \mathbf{x}(t)$ is a one-parameter group of translations along the trajectories of smooth vector field $\mathbf{v}(\mathbf{x})$ on $M$, then $\Delta \Gamma(t) = g_t \Delta \Gamma_0$. If $\mathrm{div} \, \mathbf{v} < 0$, then the system is called *dissipative*. In the case of dissipative systems, phase volume $\Delta \Gamma(t)$ diminishes and may tend to zero as an $n$-dimensional volume. Yet this does not mean that the corresponding measure $\mu_t = \mu(\Delta \Gamma(t))$ becomes degenerate, rather $\mu_t$ strives to the measure of a subset whose dimensionality is lower than $n$ (the phase space dimension). This set of smaller dimensionality is known as an *attractor* which is a very important notion, implying that the evolution of a dynamical system by itself gives rise to a natural (invariant) measure. In most cases, an attractor can be understood as an asymptotically contracting solution to the evolution equation (e.g., vector differential equation $\dot{\mathbf{x}} = \mathbf{f}(\mathbf{x}, t)$, $\mathbf{x}_0 = \mathbf{x}(t_0)$) i.e., for the flow $g_t \colon \mathbf{x}(t, t_0, \mathbf{x}_0)$. In other words, when averaged over long times the solution will be confined within some limits which means that the flow should be restricted to some domain – an invariant manifold.

---

[67] One might note that there seems to be no unique definition of entropy.

The study of attractors as invariant manifolds is one of the crucial issues in dynamical systems theory. An attractor is understood as a closed invariant set $C$ if it has a neighborhood $D_0$ such that for $t > 0$ $D(t) = g_t D_0 \subset D_0$ and $\bigcap_t D(t) = C$. In other words, flow $g_t \colon \mathbf{x}(t, t_0, \mathbf{x}_0)$ is restricted to $C$ for large enough time $t - t_0 > T$, $g_t(C) = C$. A stable fixed point $\mathbf{x}_0$ (i.e., $\mathbf{v}(\mathbf{x}_0) = 0$) and a stable periodic path (limit cycle) give simple examples of attractors, $\mathbf{x}_0$ being zero-dimensional whereas a limit cycle is a one-dimensional attractor (see [92], §§30-32). Physically, a limit cycle implies a stationary oscillation.

Dissipation in dynamical systems is rooted in the drastic reduction of the phase space dimensionality (i.e., the number of degrees of freedom) in the idealized first-principle Hamiltonian system. This is basically the transition from microscopic to macroscopic description. The penalty for this forceful phenomenological transition is irreversibility so that the simplicity of macroscopic description is often spurious. A more precise formulation of a dissipative dynamical system is in terms of a so-called wandering set of positive measure, but we shall not use it here.

The opposite to the dissipative case, div $\mathbf{v} > 0$, i.e., when the phase volume expands, is comparatively seldom discussed in mathematical modeling of real-life systems since this case corresponds to an eventual blow-up of the phase volume. Such a behavior can hardly be expected of a physical (or other) system containing a finite amount of energy, mass, entropy, enthalpy, information and other macroscopic quantities. In general, asymptotic behavior of dynamical systems is determined by matrix $A_j^i := \partial_j v^i$ and not only by its trace div $\mathbf{v} = \mathrm{Tr} A_j^i$. These questions will be discussed when exploring the stability of dynamical systems (see Section 8.4.4.).

Thus, the theory of dynamical systems becomes closely related to statistical mechanics. Notice that randomness in classical statistical – Boltzmann-Gibbs – mechanics may be attributed in a macroscopic system to an enormous number of degrees of freedom that cannot be fully known and controlled. It is actually a fact of crucial importance: the invariant measure that naturally emerges in smooth dynamical systems can be identified with the Gibbs measure which is the main concept of statistical mechanics.

### 8.4.2. Flows and vector fields

There are many ways to treat dynamical systems, yet irrespective of the manner of presentation, there is a simple physical meaning underlying all possible descriptions: a dynamical system specifies a rule of evolution on the phase (more generally, state) space. Evolution is a key word for dynamical systems: one can interpret a dynamical system as a recipe to ensure the time evolution. If we restrict ourselves to considering only deterministic continuous-time dynamical systems (also called flows), then we can represent them through the evolution operator $g_t$ i.e., $\mathbf{x}(t) = g_t \mathbf{x}(t_0)$. In other words, the formal solution, if it exists, can be symbolically represented by this expression, which is actually known as a flow of the vector field. One can say that the evolution of an autonomous dynamical system can be described by a one-parameter semigroup of time transformations. In coordinate representation, finite-dimensional autonomous systems are described by dynamic equations, e.g. $\dot{x}^i = v^i(x^j)$, $i, j = 1, \ldots, n$, on a manifold $X \equiv M^n$ i.e. dim $X = n \geq 1$, where $v^i$ is a vector field on $X$. This is a system of first-order differential equations on a fiber bundle $X \times \mathbb{R} \to X$.

The flow or the evolution operator (we shall consider these two notions to be synonymous – operator $g_t$ gives the law of evolution anyway) is a map $g_t = g_t(\mathbf{x}) \colon X \times \mathbb{R} \to X$, $\mathbf{x} \in X$ (in most interesting cases set $X \equiv M^n$, where $M^n$ is a compact manifold) such that $g_t \mathbf{x}(t_0) = \mathbf{x}(t)$ i.e., it transforms some initial state of the system, $\mathbf{x}(t_0) \equiv \mathbf{x}_0$, into the current state $\mathbf{x}(t)$. The evolution operator $g_t$ is assumed to depend continuously on parameter $t$ (time) and moreover is differentiable over time as needed. So, one can write in most cases $g_t = e^{At}$, where $A$ is an operator of time translation.

One might think that a dynamical system can be defined as the one for which the evolution operator satisfies the relation $g_t g_s = g_{s+t}$ (more exactly, defines $g_t \cdot g_s = g_{s+t}$). This property, however, not always ensures that the vector field $\mathbf{v}(\mathbf{x})$ determines the flow $g_t$ as a diffeomorhism: a well-known example is $\mathbf{v}(\mathbf{x}) = v(x)\mathbf{e}_x = x^2 \mathbf{e}_x$ giving $g_t x = x/(1 - xt)$. The reason why this vector field does not generate a flow as a diffeomorphism is that the phase space (manifold) is not compact. A smooth vector field on a compact phase space always defines such a flow so that flow $g_t$ generated by a dynamical system $\dot{x}^i(t) = v^i(x^j, t), x = \{x^j\} \in M^n, t \in \mathbb{R}$ can be regarded as a diffeomorphism i.e. a smooth invertible transformation from $M^n$ to $M^n$ (Figure 8).

Each flow can be characterized by its critical elements, primarily by equilibrium (fixed) points and closed orbits (limit cycles). Equilibrium points correspond to rest states of the system whereas closed orbits manifest periodic processes. The phase portrait of a dynamical system (if we could draw it in the space with dimensionality more than two or, maximum, three) is determined by the distribution of these critical flow elements. The latter are also of paramount importance for stability properties of a dynamical system.

We may interpret property $g_t g_s = g_{s+t}$ as the statement that time is additive, and the evolution operator is multiplicative. Moreover, if one understands parameter $t$ as time, then it would be natural to require $[g_t, g_s] = 0$, where square brackets denote a commutator. The meaning of this condition is that evolution operators corresponding to different temporal intervals commute.

Note that evolution operators may appear under various names, e.g., under the guise of transfer operators (mostly in the theory of dynamical systems) or as push-forward transformations (in the theory of differential manifolds). The concept of flow is closely related to evolution operators, yet it underlines somewhat different features of evolution dynamics, closer to intuitive representation of evolution such as through phase trajectories or orbits. Two natural properties of the flow $g_0 \mathbf{x} = \mathbf{x}$ and $g_t g_s \mathbf{x} = g_{t+s} \mathbf{x}, t, s \in \mathbb{R}$ demonstrate that it is an Abelian group, a one-parameter group of diffeomorphisms ([16], §5 and [123]) generated by the flow (the inverse element to $g_t$ is $g_{-t}$). In other words, the mapping operations performed by flow $g_t$ corresponding to a dynamical system $\dot{x}^i(t) = v^i(x^j, t), \mathbf{x} = \{x^j\} \in M^n, t \in \mathbb{R}$ is a diffeomorphism from $M^n$ to $M^n$ i.e., a smooth invertible transformation. One may be tempted to think that this one-parameter group of transformations is isomorphous to $\mathbb{R}$ which is an Abelian group under addition, but this is wrong: in general, there is no global isomorphism, which can be seen already on the simple example of a harmonic (linear) oscillator i.e., flow $g_t = g_t(x^1, x^2) = (x^1 \cos \omega t - x^2 \sin \omega t, x^1 \sin \omega t + x^2 \cos \omega t)$ generated by vector field $\mathbf{v}(\mathbf{x}, t) = -(k/m)x^1 \partial/\partial x^2 + x^2 \partial/\partial x^1$ on $M = \mathbb{R}^2, \omega = (k/m)^{1/2}$.

Below we shall see concrete realizations of the flow on some well-known examples: oscillator, logistic model, competition models, etc. Instant $t = t_0$ separates the time axis into past and future. Since a flow represents deterministic dynamics on a manifold $M$ and is assumed at least continuously differentiable with respect to time, we can immediately produce differential equation $\dot{\mathbf{x}} = \mathbf{f}(\mathbf{x}, t)$ from this evolution rule. Here vector-function $\mathbf{f}$ defines a vector field or a velocity field i.e., a vector directed along the instantaneous velocity at each point $\mathbf{x} = \{x^i\}$ of the phase space. Therefore, we shall often denote $\mathbf{f}(\mathbf{x}, t)$ as $\mathbf{v}(\mathbf{x}, t)$.

The meaning of a dynamical system can be verbalized as follows: if you find yourself at point $\mathbf{x} = \{x^i\}$, a dynamical system or the corresponding vector field tells you where to go next. In other words, the differential equation of a dynamical system is synonymous with a vector field. Definition of a dynamical system in terms of a differential operator is convenient in modeling special cases since it allows one to reduce the general representation of dynamical systems to specific equations, ODE, PDE, integro-differential, difference, etc. In this manuscript, we only deal with the simplified presentation of vector fields; the discussion of more general tensor fields and differential forms is outside the scope of this book.

It is convenient to treat vector field $\mathbf{v}(\mathbf{x})$ (we now suppress the parametric dependence on time) as a system of functions $f^i = v^i(x^j)$ defined on an open domain $C \subset \mathbb{R}^n$, giving a continuous one-to-one mapping of $C$ on some target area $D \supset \mathbf{v}(\mathbf{x})$. Recall that such a mapping is known as a homeomorphism of domain $C$ on $D$. For many applications, one has to impose a more stringent requirement: mapping $\mathbf{v}(\mathbf{x})$ is assumed smooth i.e., one may have as many continuous derivatives of $\mathbf{v}(\mathbf{x})$ over $x^i$ as one needs to process the task. For simplicity, the notion "smoothness" implies that the relevant functions are continuously differentiable an infinite number of times i.e., $f^i \in C^\infty$. Recall that the corresponding smooth and invertible ($\mathbf{x} = \mathbf{v}^{-1}, D \to C$) mapping is usually referred to as a diffeomorphism. If the Jacobian of mapping $\mathbf{v}(\mathbf{x}), J(\mathbf{v}) \neq 0$ in all $\mathbf{x} \in C$, then there is an open neighborhood at any point $\mathbf{x}$ where $\mathbf{v}(\mathbf{x})$ locally defines a flow. This statement is, in fact, a consequence of the implicit function theorem. Yet the set of functions $v^i(\mathbf{x})$ may fail to define the flow globally i.e., inverse mapping $\mathbf{x} = \mathbf{v}^{-1}$ might not exist. A standard example is the plane flow defined everywhere on $\mathbb{R}^2$ except at the origin, $C = D = \mathbb{R}^2 \backslash \{0\}$, and mapping $v(z) = z^2$ ($z = x + iy = re^{i\varphi}$). One can see that for this mapping $J \neq 0$, but $v(z)$ does not have a one-to-one inverse. Two-dimensional dynamical systems are convenient for mathematical modeling because they admit the complex-domain representation, $dw/dz = f(w, z)$ with the analytic right-hand side. One might, however, feel something special about two-dimensional flows: in this case phase trajectories cut the phase space (manifold) into two separate parts, which is not possible for $n \geq 3$. The generalization on higher dimensions is based on the so-called foliations that we shall not discuss in this book.

A two-dimensional autonomous dynamical system leads to a pair of equations: $dx^2/dx^1 = f^2(x^1, x^2)/f^1(x^1, x^2), f^1(x^1, x^2) \neq 0$ and $dx^1/dx^2 = f^1(x^1, x^2)/f^2(x^1, x^2), f^2(x^1, x^2) \neq 0$. Using, for writing simplicity, the $(x, y)$ coordinates and denoting $f^1(x^1, x^2) = P(x, y), f^2(x^1, x^2) = Q(x, y)$, we get the form $d\omega := Pdy - Qdx = 0$ that defines the tangent to trajectory at each point $(x, y)$. The trajectory determined by this form can be visualized with undirected segments at each point (mark the information loss: the original dynamical system defines a vector field). Therefore, one says sometimes that the above differential form defines a field of "linear elements". Accordingly, curves $P(x, y) + c_1 Q(x, y) = 0$, $Q(x, y) + c_2 P(x, y) = 0$, where $c_1, c_2$ are constants, are called isoclines of form $d\omega$ or of the initial dynamical system $\dot{\mathbf{x}} = \mathbf{f}(\mathbf{x}), \mathbf{x} = (x, y)$ since tangents at each point of such curves are the same. In particular, for $c_1 = 0$ we get vertical isocline $P(x, y) = 0$ and for $c_2 = 0$ horizontal isocline $Q(x, y) = 0$.

The interpretation of vector fields as mappings allows one to build the products of flows, e.g., $\mathbf{w} = \mathbf{w} \circ \mathbf{v} = \mathbf{w}(\mathbf{v}(\mathbf{x}))$, as the Jacobi matrix for a composition of mappings can be represented as a product of Jacobi matrices for individual mappings. Moreover, within the mapping approach one can easily define smooth trajectories corresponding to the constant flow values, $v^i(\mathbf{x}) = b^i$, where $b^i, i = 1, \ldots, n$ are the components of some constant vector. Solution to this system of equations i.e., the inverse $\mathbf{x} = \mathbf{g}(b^i)$ gives the surfaces in $C$ on which the $i$-th component of the flow is fixed. For example, if we fix the angle in the cylindrical geometry, we shall have a vector field in plane $(r, z)$ or, conversely, we may define curves in $C$ along which only one component varies (e.g., in radial direction), all others being fixed. This is similar to coordinate lines and surfaces. If the flow is explicitly time-dependent (non-autonomous system), then the whole picture "breathes". Non-autonomous Hamiltonian systems are sometimes referred to as having $n + 1/2$ degrees of freedom.

In order to better understand a general theory of $n$-dimensional nonlinear conservative or dissipative dynamical systems on continuous vector fields, it is useful to familiarize oneself with the systems having one-dimensional phase space (i.e., with vector fields on the line). Recall that the phase space is a set in which the behavior of variables $\mathbf{x} = \{x^i\}$, $i = 1, \ldots, n$, parameterized by $t$ is described in terms of evolution equations $\dot{\mathbf{x}} = \mathbf{f}(\mathbf{x}, t, \sigma)$, $\sigma$ is some parameter of the problem usually called the control parameter. A point with coordinates $x^i(t)$ for a fixed parameter $t$ is called a phase point which, with increasing $t$, moves through the phase space along some curve $\boldsymbol{\gamma}$. In case the phase space has a single dimension, the phase point moves along the $x$-axis. The 1d equation $dx/dt = f(x, t, \sigma)$, where $x \in \mathbb{R}, f \in \mathbb{R}, t \in I \subseteq \mathbb{R}$, gives a simple example of a dynamical system i.e., the one whose behavior is uniquely determined by its initial state (deterministic behavior)[68]. The physical meaning of function $\mathbf{f}$ is a phase velocity, therefore in many cases it is denoted by $\mathbf{v}$. One-component dynamical systems i.e., when the phase space has a single dimension, reflect a rather poor situation since certain phenomena cannot be observed in just one dimension, e.g., chaotic attractors cannot be present in 1d continuous systems or turbulence cannot fully develop. More generally, quasiperiodic and chaotic attractors can only be observed in dynamical systems with phase space dimensionality $n \geq 3$ since this is the minimal number of dimensions required to embed a 2d torus $S^2 \times S^2$. Nevertheless, one-component models are very useful to grasp the main features of evolution with the focus on scalar quantities such as in population growth or climate variability models.

Two- and three-dimensional flows are very important for practical purposes: thus, two-dimensional (plane) models correspond to a second-order differential equation $\ddot{x}(t) = F(x, \dot{x}, t)$ often encountered in mechanics, physics, chemistry, technology, automatic control, airspace and other fields, i.e., $\dot{x}^1 = x^2, \dot{x}^2 = F(x, \dot{x}, t)$ $(x = x^1, \dot{x} = x^2)$. This latter system is usually easier to explore than the second-order ODE (e.g., Newtonian model of mechanics). For instance, the equilibrium states can be more efficiently explored on the $(x^1, x^2)$ plane. Notice that two-dimensional dynamical systems can be defined not only on the plane, but also on two-dimensional surfaces, e.g., on a cylinder or torus.

### 8.4.3. A note on uniqueness

A modeling problem formulated in mathematical terms is said to be "well posed" if there exists a solution and the latter is unique. Consider a dynamical system described by an autonomous equation $\dot{x}(t) = v(x, \sigma), x(t_0) = x_0$. Trying to explore at first the simplest setting, we disregard possible dependence of control parameter on time $t$, even of the slow (adiabatic) character. When one-dimensional vector field $v(x)$ satisfies the conditions of the classical Cauchy-Lipschitz-Peano (or Picard-Lindelöf) theorem that in fact only requires that $v(x, \sigma)$ should be Lipschitz-continuous in $x$, i.e., in a general vector case $|\mathbf{v}(\mathbf{x}, \sigma) - \mathbf{v}(\mathbf{x}, \sigma)| \leq L|\mathbf{x} - \mathbf{y}|$ ($L$ is the Lipschitz constant, $|\mathbf{x}|$ denotes

---

[68] Strictly speaking, one may also consider such dynamical systems whose future states are not uniquely determined by the evolution of initial ones, but we shall not discuss them in this book.

the Euclidean norm), then the above Cauchy problem has a unique solution in some domain $|t - t_0| \leq I \subseteq \mathbb{R}$ and this solution is given by the formula

$$t - t_0 = \int_{x_0}^{x(t)} \frac{dx'}{v(x')}, (v(x_0) \neq 0); \; x(t) = x_0 \; (v(x_0) = 0) \qquad (8.4.3.1.)$$

Notice that the usually required smoothness of vector field $v(x)$ is in any event sufficient for the uniqueness of solutions $x(t)$, this requirement can be too harsh. Smoothness is not actually used in the proof of uniqueness: the latter can even be ensured when the derivatives (gradients) of field $v(x)$ do exist but are not continuous. In other words, an exact solution exists, is unique and continuous on right-hand side (vector field variations) and initial conditions under very general assumptions.

The physical meaning of the Lipschitz condition is that we compare the motion in a curved velocity field $v(x)$ with the faster motion in linear field $Lx$, where field lines are straight and $\dot{x} = Lx$. Then $x = x_0 \exp(Lt)$ serves as an etalon for comparison, which is the model of exponential growth or decay whose outcome is that singular points $x = 0$ or $x = \infty$ can only be attained in infinite time.

Stipulating initial conditions $x_0^i, i = 1, \ldots, n$ at $t = t_0$ uniquely defines past and future. In physics, one typically uses the data related to some initial time $t_0$, e.g., for the present to find the solution for $t > t_0$ and thus to predict the future. The only well-established mathematical model that attempts to find the solution back in time is the Black-Scholes equation when one specifies not the present, but the future value of the options so that the Black-Scholes model may be interpreted as using the future to predict the present.

Although in physics (and to some extent in mathematical modeling), in distinction with mathematics, the issue of uniqueness is often ignored and even viewed with a certain contempt, it is really important. Non-uniqueness signifies, for example, that by reversing the time, which is a natural operation, e.g., in models based on Newtonian dynamics, one can reach the initial state from two or more positions in finite time so that one can arrange rather weird loops. The same applies to equilibrium points of a dynamical system: if there is no uniqueness, they can be attained in finite time along different trajectories from various positions. The Cauchy-Lipschitz-Peano uniqueness theorem forbids intersections of any two trajectories and self-intersections on a single path so that phase and integral trajectories can only conflate at singular points ("hotspots" of a vector field). In other words, the uniqueness theorem imposes rather severe topological restrictions on phase flows and on the complexity of the phase portrait.

Recall that the flow on an $n$-dimensional manifold $M$ (phase space) is defined by the integral trajectories, and the configuration i.e., topology of the flow lines determines the phase portrait. One can understand the phase portrait of a dynamical system as a partitioning of the phase (state) space into trajectories. In practice, however, it is rarely possible to draw all phase trajectories in a single picture, especially when the motion has a complex character. In particular, such a complex phenomenon as deterministic chaos in continuous-time dynamical systems only becomes possible starting from three-dimensional phase space (see more on that below, in relation to stability and chaos) so that depicting this motion as a phase portrait requires a lot of effort, if at all feasible.

### 8.4.4. Stability of solutions

We now turn to a class of problems that are of utmost importance for any scientific discipline: stability of states and solutions. Such historically great figures as Newton, Laplace, Lagrange, Poisson, Dirichlet, later Lyapunov, Kolmogorov, Arnold and other outstanding scientists were busy with the question of stability of the solar system. Now we know, mainly due to numerical computer modeling, that the motion of the planets comprising the solar system can be unstable and even chaotic. The term "chaotic" qualitatively means, for example, that the forecasts of the planets' paths become very inexact or even downright impossible to forecast after some time from now. Beyond this time horizon any prediction, even qualitative, breaks down. Physically speaking, planets can suffer collisions or be ejected from their orbits due to nonlinear interactions.

Imagine that a measurement, e.g., of a planet position/velocity or of some quantity in a more down-to-earth process, is made with a finite accuracy having error $\varepsilon_0$ – recall that no measurement is perfect. Then after time $t$, the error becomes $\varepsilon(t) = \varepsilon_0 e^{\lambda t}$, where $\lambda$ is the so-called Lyapunov exponent that is positive for a given system, and chaotic systems always have at least one positive Lyapunov exponent $\lambda$. Then the knowledge of a measured quantity, e.g., of position or velocity of a planet becomes hopelessly inexact after time $t_* \sim \lambda^{-1} \log \delta / \varepsilon_0$, where $\delta$ is the observation tolerance. Improvement of either accuracy (i.e., initial error) or tolerance brings very little since the prediction horizon logarithmically depends on these quantities, and the logarithm can be regarded as almost constant in rough estimates.

Stability of physical and, in particular, of mechanical systems is essentially described by the Lyapunov stability theorem. This theorem holds not only for mechanical motion, but for differential equations on Banach spaces in general relating the stability of their solutions (with respect to perturbations) to spectral properties of the operator of the corresponding dynamical system, $\dot{\mathbf{x}} = A\mathbf{x}$. More specifically, if $A$ is a bounded operator on a Banach space $X$ and the spectrum $\sigma(A)$ of $A$ lies in the open left $C_-$ of the complex plane $C$, then the autonomous vector differential equation $\dot{\mathbf{x}} = A\mathbf{x}$ is uniformly exponentially stable on $\mathbf{x} \in X$. Solutions to this equation starting at point $\mathbf{x}_0 \in X$ at time $t_0$ are given for $t - t_0 \in I \subset \mathbb{R}$ as $\mathbf{x}(t) = e^{A(t-t_0)}\mathbf{x}_0$ i.e., function $t \mapsto \|e^{At}\|$ tends to zero exponentially fast for $t \to +\infty$.

The prediction horizon in a few-body nonlinear system can, in general, be hard to find in a concrete process; one only can roughly appreciate its order of magnitude using physical – mostly intuitive – considerations. Nevertheless, one qualitative thing is very important: unstable and especially chaotic systems always involve imperfect knowledge, with their prediction horizon being much shorter than one could expect – owing to exponentially rising errors in unstable or chaotic dynamical systems, $\varepsilon(t) = \varepsilon_0 e^{\lambda t}$. In other words, we cannot make fair forecasts beyond a few multiples of $\lambda^{-1}$, no matter how hard we try improving our measurement technologies. The weather forecast is a standard example.

Chaotic systems are extremely sensitive to disturbances: even a tiny perturbation of parameters, e.g., of initial conditions, leads to drastic changes. In this sense, chaotic systems are almost indeterministic: one cannot define similar causes that would produce the same effects i.e., coarse graining mappings (morphisms) in the phase or the state space. This negative property of chaotic phenomena qualitatively distinguishes them from stable and even from unstable but regular ones. In this context, the term "regular" designates the absence of exponential instability with respect to the variation of initial conditions (or perhaps other parameters). In the case of such instability, if we require that the system should remain predictable during at least time $t = a\tau$, where $\tau$ is a predictability horizon, $a$ is some positive constant, we should increase the accuracy of measuring the initial state parameters i.e., diminishing the measurement errors by $e^a$. If, e.g., $a = 10$ the quality of measurements should be improved by factor $e^{10} = e^{2\cdot 5} \approx 10^5$. For example, if we want to increase the weather prediction horizon from the customary three days to a month, we will need to provide the input data with the 100000 better accuracy. If, for instance, temperature $T$ and pressure $p$ sensors have the accuracy (relative error) ${\Delta T}/{T}$ and ${\Delta p}/{p}$ about one percent, then one would need to have at least the $10^{-5}$ percent accuracy to extend ten-fold the predictability horizon, which is hardly possible not only for engineering reasons, but also for physical ones.

So exponential instability results in the emergence of stochastic properties in even rather simple deterministic systems (see, e.g., [41], [153]). A basic example of exponential instability is given by transformations "extending" the [0,1] segment i.e., $x \mapsto f(x)$ with $|f'(x)| > 1$. One can see that successive application of such transformations exponentially increases the distance between two initially close points $x_1, x_2 \in [0,1]$.

We can observe that the main concepts of dynamical systems theory are actually quite simple. First, we find the fixed (or critical, or equilibrium) points of a vector field, which are defined as steady-state solutions $\mathbf{v}(x, t, \sigma) = 0$, and then, once the fixed points $x = c_k, k = 1, 2, \ldots$ have been found, the next thing to do is to explore their stability. Stability of solutions and, in particular, of equilibrium points is a crucial notion in mathematical and computer modeling because no model is exact and, moreover, each solution to a mathematical model, the latter being expressed by equations, practically always suffers from some deviations in the integral paths. Such deviations can distort initial conditions or, more generally, displace any point on a trajectory. Primarily, there may be fluctuations, which are in fact physically unavoidable. Indeed, the laws of motion or of any evolution are formulated in terms of a reduced number of macroscopic quantities (defining the vector field $\mathbf{v}(\mathbf{x}, t)$ i.e., the right-hand side in dynamical system models), these quantities may have a diverse physical nature such as the population of humans or other species, impurity concentration, temperature (e.g., of the Earth surface), energy (e.g., in generators), pressure, number of micro-organisms, of neutrons in a nuclear reactor, etc., but any macroscopic variable is a quantity averaged over countless microscopic states (see section 7 devoted to statistical physics and thermodynamics). In the phenomenological equations used for modeling, all individual microscopic states (realizations) are totally ignored and only the average dynamics is considered. Molecular interactions and temperature fluctuations (in physical modeling) or practical impossibility of knowing or measuring the precise numbers (in modeling outside physics) make us consider all the parameters of mathematical models as being affected by perturbations or noise. Such perturbations can produce sudden changes of the modeled system's behavior, in particular, causing its jumps to nearby trajectories. If stability is lacking, modeling can be haunted by indeterminacies. In effect, unstable states are mostly not manifested in nature. One can illustrate the concept of stability by comparing the upper and the lower equilibrium positions of a pendulum ($\dot{x}^1 = x^2, \dot{x}^2 = -(g/l)\sin x^1$) in the gravity field which is directed downwards.

The concept of stability is associated with the asymptotic – for very large times – behavior of perturbed trajectories. For example, if we slightly disturb a particular trajectory or make a small change in the initial conditions, will the new trajectory converge to the initial one? Alternatively, will a perturbed trajectory stay forever asymptotically close to the initial one? The first case is stronger and is known as stable, while the second as asymptotically stable. Conversely, if a perturbed trajectory is flying away from the given orbit, then the dynamical system is unstable. In general, one can think of an unstable manifold as of a set of points whose evolution back in time converges to a fixed point. In the above one-dimensional illustration, the procedure of exploring the stability is very intuitive. For $x$ near fixed point $c$ we can expand $v(x) = v(c) + v'(x)|_{x=c} + \mathcal{O}(x - c)^2 \approx v'(c)(x - c)$ (see above more about linearization). Making the coordinate shift $x - c = z$, we have in the first order (linear theory!) $\dot{z} = v(c)z$ which gives $z(t) = z(0)\exp(v'(c)t)$. We see that the stability of an equilibrium point $c$ ($v(c) = 0$) is determined by the first derivative of the vector field on the line: if $v'(c) < 0$, the deviation $z = x - c$ fades out in time and the equilibrium point $c$ is stable, whereas if $v'(c) > 0$, the deviation grows exponentially with time and the equilibrium point is unstable i.e., the nearby integral trajectories are "repelled" from it. Notice that in the autonomous case one can interpret the derivative $v'(x)$ as acceleration $a(x)$ in point $x$ divided by velocity in this point, $v'(x) = \dot{v}(x)/v(x) \equiv a(x)/v(x)$ which means that if $v(c) > 0$ and $a(c) > 0$ motion is obviously unstable near $x = c$ since a phase point is running away from the equilibrium position with increasing velocity. This form of vector field gradient produces natural associations with the definition of the curvature of a plane curve (see also below).

Thus, we may conclude that the vector field gradient $\partial v(x,t)/\partial x$ characterizing the rate of spatial variation (more generally, tensor $A_j^i \equiv \partial v^i(x^j,t)/\partial x^j$) is a measure of divergence of solutions $x(t)$ or, in other words, a criterion of their stability. One can illustrate these considerations by introducing the potential function for vector field $v(x)$: $\dot{x} = -\partial V(x)/\partial x$, whenever possible. Notice that in many situations one can formally define a potential also for non-autonomous dynamical systems, $\dot{x} = -\frac{\partial V(x,t)}{\partial x}, V(x,t) = \int_{x_0}^{x} v(x',t)dx'$, in this sense potential and conservative vector fields are not synonymous. Recall in this connection that potential and potential energy is an essentially Newtonian i.e., non-relativistic notion since it implies instantaneous interaction (potential depends only on the actual relative positions of interacting bodies).

Using the language of potentials, equilibrium points i.e., constant solutions where vector field $\mathbf{v}(\mathbf{x})$ vanishes correspond to their local minima or maxima, and equilibrium stability is gauged by the potential function curvature (i.e., the measure of the extent to which a curve deviates from a straight line), $k = \partial^2 V/\partial x^2|_{x=c} = -v'(c)$. If $k > 0$, then the potential $V$ has a minimum at $c$ and this equilibrium point is stable; if $k < 0$, then $V(c)$ is a local maximum and equilibrium at this point is unstable.

Now, suppose that, while observing a stable point $c$, we slightly disturb the initial conditions $x_0$ within a certain interval $\delta = (\alpha, \beta)$. Then the maximal interval $\delta$ giving rise to the paths that will end up in the stable equilibrium point $c$ or very near to it is usually called the domain of attraction. One can easily see that in the one-dimensional case each two stable equilibrium points must be separated by an unstable point and, vice versa, any two unstable points must be separated by a stable one. Indeed, let $c_1$ and $c_2$ be two successive stable fixed points and let $x$ be a point on an integral curve near $c_1$. Then since $c_1$ is stable $x(t)$ must approach $c_1$ with time, $x(t) \to c_1$ as $t \to +\infty$ and therefore $\dot{x} = v(x) < 0$ ($x$ decreases with time). In the other case, when $x$ is near $c_2$, $x(t) \to c_2$ as $t \to +\infty$ so that $v(x) > 0$ ($x$ is rising). Therefore, since $v(x)$ is continuous there must be a point $x^*$ where $v(x) = 0$ so that $c_1 < x^* < c_2$ separates domains of attraction of $c_1$ and $c_2$. This equilibrium ($v(x^*) = 0$) point should be unstable since we have assumed that there are no stable fixed points between $c_1$ and $c_2$. There are many illustrations to this important intermittency principle. For example, such favorite mathematical models of physics as the double-well potential and one-dimensional periodic lattice (H. A. Kramers) may be regarded as its manifestations. The same intermittency is exploited in one-dimensional climate variability models like the popular Budyko-Sellers model.

It is interesting to trace the origins of the terminology generally accepted in mathematical modeling based on dynamical systems theory. Why are equilibrium points also called singular and fixed? The term "equilibrium" seems to be clear: such points correspond to steady-state solutions, $d/dt = 0$. It is also more or less clear why the same points are called singular: apart from singularity $v(x) = 0$ (more generally, $\mathbf{v}(\mathbf{x}) = 0$) in the integral (8.4.3.1.) defining the time of dynamical motion, singularity appears when we consider the angles that the tangent vector of $\mathbf{v}(\mathbf{x})$ makes with the phase trajectory. If we formally write for the phase curve orientation cosines $\alpha_i(\mathbf{x}) = ds/dx^i$, where $ds = (dx_j dx^j)^{1/2}$ is the line element along the phase path, then

$$\alpha_i = \frac{(dx_j dx^j)^{1/2}}{dx^i} = \left(1 + \sum_{j \neq i}^{n} \left(\frac{dx^j}{dx^i}\right)^2\right)^{1/2} = \left(1 + \sum_{j \neq i}^{n} \left(\frac{v^j}{v^i}\right)^2\right)^{1/2}$$

and we have singular points in the phase space with $\alpha_i$ undefined at points $\mathbf{x}_m(\sigma), m = 1,2,\ldots n$ produced as the solutions to the system of algebraic equations $v^i(\mathbf{x}, \sigma) = 0$. Quite naturally, such points are called singular. If we consider an autonomous system, the singular points do not drift with time; they remain fixed in the phase space and thus are justifiably called fixed points.

All such considerations about stability are also important for computer modeling. To feel the taste of stability concepts, consider the simple scalar ($n = 1$) equation $\dot{x}(t, \mu) = \mu x(t)$ with initial condition $x(0) = x_0$. Its solution is $x(t) = x_0 \exp(\mu t)$. Obviously, $x = 0$ is also a solution; in the language of vector fields, condition $\mathbf{v}(\mathbf{x}) = 0$ (in our one-dimensional case, $\mu x(t) = 0$) gives singular points $\mathbf{x}_m, m = 1,2,\ldots$, where the vector field vanishes. Assume that the modeled system has reached the equilibrium state $\mathbf{v}(\mathbf{x}) = 0$. Theoretically, the system will remain there forever, but in practical situations such as engineering or economics either an unexpected perturbation will arise, or fluctuations will destabilize the system. These influences can displace both the equilibrium location and the system's position so that the solutions will fly away from it. Then we have to analyze the algebraic equation $\mathbf{v}(\mathbf{x}, \mu) = 0$ which can, in general, be only performed numerically (mostly by the Newton method and related iterative procedures).

### 8.4.5. Dynamical systems on a plane

One often refers to a dynamical system with the two-dimensional phase space as simply to a plane or two-dimensional system. The general form of such a system,

$$\frac{dx}{dt} = F(x, y, t), \quad \frac{dy}{dt} = G(x, y, t), \tag{8.4.5.1.}$$

corresponds to a direct generalization of a mechanical system with a single degree of freedom affected by a time-dependent force.

Probably the simplest example of instability in a two-dimensional system is the free one-dimensional mechanical motion, $\ddot{x} = 0$. Indeed, we have the dynamical system $\dot{x}^1 = x^2, \dot{x}^2 = 0$ (in coordinates $x^1, x^2, x = x^1, v = \dot{x} = x^2$) so that the corresponding vector field is $v(x) = (x^2, 0)^T$. For initial conditions $(x_0, v_0)^T$ at $t = t_0$, we have the solution $x^1(t) = x_0 + v_0(t - t_0), x^2(t) = v_0$ i.e., the trajectories are straight lines $x^2 = \text{const}$. If we produce a small disturbance of initial conditions, e.g., $v_0 \rightarrow v_0 + \eta, \eta > 0$, then for large $t$ we get a completely different state. Thus, cars with slightly different velocities on a highway arrive at totally different cities, located at distance $d = \eta(t - t_0)$ from one another.

In general, autonomous dynamical systems with two-dimensional phase space play an important role. Apart from describing a single-dimensional mechanical motion in some time-independent potential field i.e., conservative systems with one degree of freedom, 2d dynamical systems provide a foundation for the quantitative and qualitative study of mathematical models in physics, electronics, airspace and automotive engineering, chemistry, biology, ecology, epidemiology, medical sciences, etc.

As already noted, one must distinguish $n$-dimensional dynamical systems (flows) and mechanical systems with $n$ degrees of freedom. For example, mechanical models with just two degrees of freedom corresponding to motion in the Euclidean plane $(x^1, x^2)$ i.e., this plane is a configuration space of such models which have respectively a four-dimensional phase space. Note that models with already two degrees of freedom are much more complex than those with one degree of freedom. For instance, an autonomous mechanical system with two degrees of freedom can be described by the differential equation $\ddot{\mathbf{x}} = \mathbf{f}(\mathbf{x}), x \in \mathbb{E}^2$, where $\mathbf{f}(\mathbf{x})$ is a 2d vector field. This system is potential if there exists a function $V: \mathbb{E}^2 \to \mathbb{R}$ (potential or potential energy) such that $\mathbf{f} = -\partial V / \partial \mathbf{x}$; in this case equations of motion are $\ddot{\mathbf{x}} = -\partial V / \partial \mathbf{x}$. It is important that one can always find a potential in the one degree of freedom system: $V(x) = -\int_{x_0}^{x} f(\xi) d\xi$. For two-degree systems, one cannot in general find $V(\mathbf{x}) = V(x^1, x^2)$; for example, vector field $f^1 = x^2, f^2 = -x^1$ is not a potential one. The motion equations in the plane can be written in the dynamical system form as

$$\dot{x}^1 = x^3, \qquad \dot{x}^2 = x^4, \qquad \dot{x}^3 = -\frac{\partial V}{\partial x^1}, \qquad \dot{x}^4 = -\frac{\partial V}{\partial x^2}.$$

This system defines a flow i.e., the velocity in 4d phase space of the system with two degrees of freedom. The $(x^1, x^2)$ plane is the configuration space of this mechanical system containing its trajectories (orbits). Phase curves of the corresponding dynamical system lie in the phase space; they are actually the subsets of the phase space. Notice that phase curves cannot intersect whereas the trajectories (which are projections of the phase curves onto the $(x^1, x^2)$ plane), in general, can. Variables $x^3, x^4$ span the velocity space, also a subspace of the phase space. From the mathematical standpoint, the configuration space (formed by coordinates) and the velocity (momentum) space are equivalent; their equivalence is especially important in statistical physics and quantum mechanics.

An example of a plane vector field in classical mechanics is given by a central force $\mathbf{f}$ that sets a particle in a circular motion. Let force $\mathbf{f}$ lie in the Euclidean $(x^1, x^2)$ plane and be perpendicular to radius-vector $\mathbf{r}$ at each point; we may assume that $\mathbf{f}$ is directed counterclockwise i.e., $\mathbf{f} = \text{const } \mathbf{e}_\varphi$, where $\mathbf{e}_\varphi$ is the unit vector in the polar system of coordinates. Let this force field be everywhere proportional to $r = |\mathbf{r}|$ i.e., $|\mathbf{f}| = \alpha r, 0 < r < +\infty, \alpha \in \mathbb{R}$. The elementary mechanical work $dA = \mathbf{f} d\mathbf{r} = |\mathbf{f}| r d\varphi = \alpha r^2 d\varphi$ gives the total work for the whole closed circle, $A = 2\alpha\pi r^2$, proportional to the encircled area. Therefore, work $A$ is nonzero for the closed path $(r > 0)$, and the field $\mathbf{f}$ is not potential, $\mathbf{f} = -\alpha x^2 \mathbf{e}_1 + \alpha x^1 \mathbf{e}_2$. This plane field has a rotational character: curl $\mathbf{f} = 2\alpha \mathbf{e}_3$, $\mathbf{e}_1, \mathbf{e}_2, \mathbf{e}_3$ are unit vectors for the $x^1, x^2, x^3$ axes. If the medium is rotating as a whole in the $(x^1, x^2)$ plane with angular velocity $\omega = \dot{\phi}$, then the velocity vector field is $\mathbf{v} = -\omega x^2 \mathbf{e}_1 + \omega x^1 \mathbf{e}_2$ and curl $\mathbf{v} = 2\omega \mathbf{e}_3 = 2\boldsymbol{\omega}$. For plane fields, complex analysis seems to be the most convenient tool, with any plane vector $\boldsymbol{\xi}$ being represented as $\boldsymbol{\xi} = \xi^j \mathbf{e}_j$, where $\xi^j = x^j + iy^j$ are complex coordinates, $\mathbf{e}_j$ are the basis vectors (here $j = 1, 2$, in general $j = 1, \dots, n$). The scalar product is defined as $\langle \boldsymbol{\xi}, \boldsymbol{\eta} \rangle = \sum_{j=1}^{n} \xi^j \bar{\eta}^j = \xi_j \bar{\eta}^j$ (in physics more often $\sum_{j=1}^{n} \bar{\xi}^j \eta^j = \bar{\xi}_j \eta^j$) where the overline bar here denotes complex conjugation.

On the example of 2d dynamical systems we can see how linearization (see above) enables one to determine the local behavior of trajectories. Let point $(x_0, y_0)$ be an equilibrium (critical) point of system (8.4.5.1). Linearizing this system near $(x_0, y_0)$, we get

$$\frac{dx}{dt} = \frac{\partial F}{\partial x}\bigg|_{(x_0,y_0)} x + \frac{\partial F}{\partial y}\bigg|_{(x_0,y_0)} y, \qquad \frac{dy}{dt} = \frac{\partial G}{\partial x}\bigg|_{(x_0,y_0)} x + \frac{\partial G}{\partial y}\bigg|_{(x_0,y_0)} y \qquad (8.4.5.2.)$$

This is a linear homogeneous system with constant coefficients, and its phase trajectories can always be found. We can rewrite linear system (8.4.5.2) near the critical point as

$$\dot{x} = a_{11}x + a_{12}y, \qquad \dot{y} = a_{21}x + a_{22}y, \qquad (8.4.5.3.)$$

where, to simplify the notations, we made the affine translation $x - x_0 \to x, y - y_0 \to y$. System (8.4.5.2.) has a nontrivial solution when the characteristic equation $\lambda^2 - \text{Tr}A + \det A = 0$ holds, where matrix $A$ is the linear system matrix, $A = (a_{ij})$, $i, j = 1,2$. The solution vector $(x, y)^T$ can be found by a standard procedure using linear transformation $\tilde{X} := \begin{pmatrix} \tilde{x} \\ \tilde{y} \end{pmatrix} = S\begin{pmatrix} x \\ y \end{pmatrix} \equiv SX$, $S = (s_{ij})$, $i, j = 1,2$ that allows one to bring system (8.4.5.2.) to a diagonal form, $\dot{\tilde{x}} = b_1\tilde{x}$, $\dot{\tilde{y}} = b_2\tilde{y}, d\tilde{y}/d\tilde{x} = b_2\tilde{y}/b_1\tilde{x}$. One can find matrix $S$ by a number of ways, for example, by differentiating system $\begin{pmatrix} \tilde{x} \\ \tilde{y} \end{pmatrix} = S\begin{pmatrix} x \\ y \end{pmatrix}$ and using (8.4.5.2.). Taking into account $\dot{\tilde{x}} = b_1\tilde{x}$, $\dot{\tilde{y}} = b_2\tilde{y}$, we have

$$s_{11}(a_{11}x + a_{12}y) + s_{12}(a_{21}x + a_{22}y) = b_1(s_{11}x + s_{12}y),$$
$$s_{21}(a_{11}x + a_{12}y) + s_{22}(a_{21}x + a_{22}y) = b_2(s_{21}x + s_{22}y).$$

Since this matrix equation must hold for all $(x, y)$, indeterminate coefficients in both parts of equation should coincide, and we get two systems of linear homogeneous equations for $s_{ij}$, $AS_i = b_iS_i$, $i = 1,2, S_i \equiv (s_{i1}s_{i2})^T$ or, in explicit form,

$$(a_{11} - b_1)s_{11} + a_{21}s_{12} = 0, \qquad a_{12}s_{11} + (a_{22} - b_1)s_{12} = 0;$$
$$(a_{11} - b_2)s_{21} + a_{21}s_{22} = 0, \qquad a_{12}s_{21} + (a_{22} - b_2)s_{22} = 0$$

These systems have nontrivial solutions when their determinants are zero i.e., coefficients $b_i$ should satisfy the equation

$$(a_{11} - b_i)(a_{22} - b_i) - a_{12}a_{21} = 0$$

or $b_i^2 - b_i\text{Tr}A + \det A = 0, i = 1,2$. We see that this equation for coefficients $b_i$ realizing the diagonal form of system (8.4.5.1.) is the same as the characteristic equation (which is not a coincidence) so that $b_i = \lambda_i$ and $s_{11}/s_{12} = a_{21}/(\lambda_1 - a_{11}) = (\lambda_1 - a_{22})/a_{12}$; $s_{21}/s_{22} = a_{21}/(\lambda_2 - a_{11}) = (\lambda_2 - a_{22})/a_{12}$. Then we have the solution for trajectories in the affine-transformed coordinates, $d\tilde{y}/d\tilde{x} = b_2\tilde{y}/b_1\tilde{x}$: $\tilde{y} = C|\tilde{x}|^{\lambda_2/\lambda_1}$. If the roots of the characteristic equation are real, this solution determines the families of hyperbolic or parabolic curves. If the roots $\lambda_i, i = 1,2$ are complex, the affine-transformed coordinates are, in general, also complex (since the transformation coefficients are expressed through complex ratios $s_{11}/s_{12}, s_{21}/s_{22}$). One can see that if eigenvalues $\lambda_{1,2}$ are complex-conjugated, then $\tilde{x}, \tilde{y}$ are also complex-conjugated, and we may write $\lambda_1 = \mu + i\sigma, \lambda_2 = \mu - i\sigma$; $\tilde{x} = u + iv, \tilde{y} = u - iv$. Then $\dot{\tilde{x}} = \dot{u} + i\dot{v} = (\mu + i\sigma)(u + iv)$, $\dot{\tilde{y}} = \dot{u} - i\dot{v} = (\mu - i\sigma)(u - iv)$ and $\dot{u} = \mu u - \sigma v, \dot{v} = \mu v + \sigma u$. In the $(u, v)$-coordinates, we get the following equation defining the phase trajectories:

$$\frac{dv}{du} = \frac{\mu v + \sigma u}{\mu u - \sigma v}, \qquad \frac{u}{v} \neq \frac{\sigma}{\mu}. \qquad (8.4.5.4.)$$

When $\mu u - \sigma v = 0$, we shall have $\dot{u} = 0$ and the velocity $\dot{\tilde{x}}$ (the rate of change of coordinate $\tilde{x}$) is purely imaginary. If we introduce polar coordinates $u = \rho \cos \varphi, v = \rho \sin \varphi$, then after simple but tedious algebra we have $\rho \, d\varphi (\alpha \rho - d\rho/d\varphi) = 0$ or $\rho = \rho_0 \exp(\alpha \varphi)$. Here exponent $\alpha$ of the logarithmic spiral represents the ratio of real and imaginary parts of the eigenvalues, $\alpha \equiv \mu/\sigma$. If the eigenvalues (i.e., roots of the characteristic equation of the system) are real, then $v$ and $u$ are proportional to each other, $v = Cu$, so that $\tilde{x} = u(1 + iC)$, $\tilde{y} = u(1 - iC)$ and polar angle $\varphi = (1/\alpha) \log(\rho/\rho_0) = 0$ for any $\rho$ (semi-direct line). In case the roots are imaginary ($\mu = 0$), the logarithmic spiral degenerates into circle $\rho = C$. Notice that if one of the roots, e.g., $\lambda_1$, of the characteristic equation is zero, equations (8.4.5.4) are linearly dependent and are in fact just one equation $dy/dx = k$. In this case, solutions are represented as a family of straight lines on the $(x, y)$ plane. If the characteristic equation has multiple roots (eigenvalues) $\lambda_1 = \lambda_2 \equiv \lambda$, then one can use the standard change of coordinates $(x, y) \to (\xi, \eta)$: $\xi = a_{11} x + (1/2)(a_{12} - a_{21}) y, \eta = y$ and the original system can be reduced to the form $\dot{\xi} = \lambda \xi, \dot{\eta} = \xi + \lambda \eta$ or $d\eta/d\xi = (\xi + \lambda \eta)/\lambda \xi = 1/\lambda + \eta/\xi$. The solution is $\eta = A\xi + \lambda^{-1}\xi \log \xi$, where $A$ is constant. One can see that this function on the $(x, y)$ plane is close to the linear one, as in the case of $\lambda_1 = 0$; for practical (numerical) purposes one can represent $y(x)$ through iterations, the first approximation being

$$y^{(1)} = \frac{a_{11}\left(A + \dfrac{L}{\lambda}\right) x}{1 - \dfrac{\Delta}{2}\left(A + \dfrac{L}{\lambda}\right)},$$

where $\Delta \equiv a_{12} - a_{21}, L$ is some value of $\log \xi$ (this function is nearly constant).

An important question is: can one appraise the behavior of the initial nonlinear system by the structure of the phase trajectories corresponding to (8.4.5.2.)? The answer is yes, but not in all cases. One can prove that if the equilibrium point is 1) a node, 2) a focus, and 3) a saddle, then phase trajectories of the linearized system (8.4.5.2.) have the same topology as those of the nonlinear problem. Equilibrium points of the 1)-3) type are known as hyperbolic: all eigenvalues of the corresponding Jacobian matrix have non-zero real parts (Figure 9). Hyperbolic equilibria are said to be structurally stable or robust which implies that small perturbations of (8.4.5.1.), $\delta x = \varepsilon f(x, y), \delta y = \varepsilon g(x, y)$ do not qualitatively change the behavior of phase trajectories near equilibrium points. Critical points of the center type are not robust: if the phase portrait of system (8.4.5.2.) has an equilibrium point $(x_0, y_0)$ of this type i.e., the phase portrait consists of an infinite sequence of closed concentric orbits going around $(x_0, y_0)$, then the initial nonlinear system may have a center as an equilibrium position, but the latter may also be a focus i.e., an expanding or collapsing spiral. Centers are located on the line Tr $A = 0$ and are thus the borderline case: any small perturbation displaces the line of zero real parts and closed orbits and converts them into spiraling sources or sinks.

Note that in contrast with the trajectories on the plane ($d = 2$), which cannot intersect, in the multidimensional case ($d \geq 3$) a dynamical system can have attractors that differ significantly from singular points in the plane (e.g., foci and limit cycles). For instance, there exist the so-called strange attractors that can point to a chaotic regime. By the way, the famous Poincaré section method was invented in order to cope with the difficulty of studying the phase trajectories for $d \geq 3$.

In conclusion, one may note that in contrast with the dynamical systems on a plane (2d phase space, $\mathbb{R}^2 \to \mathbb{R}^2$), e.g., those corresponding to mechanical systems with one degree of freedom, linearization techniques and analysis of symmetries for a $n$-dimensional phase space $\mathbb{R}^n$ (i.e., for systems of $n$ ODEs) remains for arbitrary $n$ a rather difficult problem. The mentioned study of symmetries in dynamical systems is important since the systems of nonlinear ODEs can have the same topological structure of phase trajectories near the critical points as the respective linear systems, if they possess the same symmetry. In other words, the phase portraits of the nonlinear system can be locally mapped to the one of the corresponding linear system.

### 8.4.6. First integrals of dynamical systems

In contrast with the balance conservation laws discussed above, the "real" conservation laws are produced not as intuitively obtained phenomenological balance equations, but by integration of differential equations of motion. These latter conservation laws are represented by the first integrals $u(\mathbf{x})$ of an autonomous dynamical system $\dot{\mathbf{x}} = \mathbf{f}(\mathbf{x})$. Function $u(\mathbf{x})$ is called the first integral of this system if it is constant along each path defined by such a system i.e., if $\mathbf{x} = \boldsymbol{\varphi}(t)$ is a solution to a dynamical system, then $w(t) := u\big(\boldsymbol{\varphi}(t)\big) = \text{const}$ for all $t \in I \subseteq \mathbb{R}$. Recall that $\mathbf{x} = \boldsymbol{\varphi}(t), t \in I \subseteq \mathbb{R}$ is a solution to $\dot{\mathbf{x}} = \mathbf{f}(\mathbf{x}, t)$ when $\dot{\boldsymbol{\varphi}}(t) = \mathbf{f}(\boldsymbol{\varphi}(t), t)$ is identically satisfied for all $t \in I$. If $\boldsymbol{\varphi}(t_0) = \mathbf{x}_0$, the solution passes through $(\mathbf{x}_0, t_0) \in \mathbb{R}^n \times \mathbb{R}$; in this case point $(\mathbf{x}_0, t_0)$ belongs to integral curve and may be regarded as an initial condition.

A closely related subject is that one should distinguish between a constant of motion and an integral of motion (or the first integral), the latter being a subset of the former. Indeed, each integral of motion remains constant along the orbit in the phase space of an autonomous dynamical system – it is a combination of state space parameters (in mechanics, a real function of phase space coordinates) that is preserved during the evolution, whereas a constant of motion can be related to non-autonomous systems (in mechanics, time-dependent Hamiltonian or Lagrangian) with the only requirement that this constant should be preserved throughout the trajectory $\boldsymbol{\gamma}(t) \in \mathbb{R}^n \times \mathbb{R}$, just as the combination $\mathbf{A}(\mathbf{r}, \mathbf{r}_0, \mathbf{v}, t, t_0) = \int_{\mathbf{r}_0}^{\mathbf{r}(t)} d\mathbf{r} - \int_{t_0}^{t} \mathbf{v}(t) dt$ can remain constant ($\mathbf{A}(\mathbf{r}, \mathbf{r}_0, \mathbf{v}, t, t_0) = \mathbf{r} - \mathbf{r}_0 - \mathbf{v}(t - t_0)$) for $\ddot{\mathbf{r}} = 0$).

A very important fact is that the first integrals of dynamical systems are related (in the necessary and sufficient sense) to the first order PDEs which typically have the form of balance conservation laws. The first integrals are always the consequence of some symmetry: if a symmetry can be observed during the motion, then the invariants of motion exist i.e., quantities or combinations of variables that do not change in the process of the system's evolution. In other words, for any symmetry operation there is a corresponding first integral. A more accurate formulation of this fact is known as Noether's theorem: *if a dynamical system admits a one-parameter group of diffeomorphisms, then there exists a first integral of this system.* One should pay attention to the fact that Noether's theorem holds not only for Lagrangian and Hamiltonian systems: in fact, it can be applied to any dynamical system to be described by the action principle, which may also include some dissipative systems. For example, the Lagrangian conservation laws can be produced through Noether's first theorem. It is interesting that the relationship between symmetry and conservation laws requires the existence of some minimum principle. In physics, it is the principle of least action. To put it simply, one can construct the Lagrangian function and notice its symmetry properties which immediately lead to the integrals of motion. One can even state that the relationship between symmetries and conservation laws is trivial because it is a direct consequence of the Lagrangian formalism.

All well-known conservation laws are just specific cases of Noether's theorem. For example, the conservation of energy is related to the invariance under shifts in time, the conservation of momentum is due to the homogeneity of space, angular momentum conservation is related to isotropy of space and so on. When physicists carry out an experiment at some location (e.g., in CERN), then try to repeat the experiment with the same equipment at another place (e.g., somewhere in the US), nobody cares much about the change of location: the physical phenomena must be invariant under spatial translations. Likewise, if one rotates an experimental setup by a fixed angle, the experiment must yield the same result: if this were not true, reproducibility of experiments would be questionable as the Earth rotates. Thus, space has the same properties in any direction, and physical phenomena should be invariant under rotations. Finally, repeating an experiment under the same conditions must give the same results i.e., basic physical laws should not change when time is translated.

## 8.5. Autonomous and non-autonomous systems

### 8.5.1. Autonomous systems

Let us now consider a particular but very important case when the variable $t$ is not explicitly present in the vector equation of a dynamical system. Such systems are called autonomous[69]. We have already seen, in connection with energy conservation issue, that autonomous systems are invariant with respect to time translations, viz. if $x(t)$ is a solution in $D \subset \mathbb{R}^n$ then $x(t - t_0)$ is also a solution, in general a different one (as, e.g., $\sin t$ and $\cos t$). Here $t_0$ is assumed to be a constant time shift.

In the theory of dynamical systems, the domain $D \subset \mathbb{R}^n$ is regarded as the phase space for the vector equation $\dot{x} = f(x), x \in D$. This concept may be considered as a certain generalization of the traditional physical terminology where phase space is understood as a direct product of coordinate and momentum spaces. In modern classical mechanics, phase space is typically defined as a cotangent bundle $T^*M$ where $M$ is a configuration manifold (see section on Hamiltonian mechanics for some comments). However, when dealing with dynamical systems there are some other features and accents which are important as compared to the traditional exposition of classical mechanics. For instance, in mechanics it is often convenient to consider the phase space as a Euclidean space whereas in the theory of dynamical systems the phase space is, in general, not a Euclidean space but a differential manifold, on which a vector field corresponding to the vector equation $\dot{x} = f(x)$ is defined. This equation means that, in the process of evolution described by the dynamical system, to each point $x$ a vector $f(x)$ is ascribed determining the velocity of the phase point (the vector $f(x)$ belongs to the tangent space of the manifold at point $x$). This is a kinematic, in fact a geometric, interpretation of the above vector equation.

Nevertheless, all these points are just nuances: one usually knows exactly in what space one finds oneself and operates. In all cases, phase space is a geometric concept embracing the total number of all states of a dynamical system and convenient to describe the evolution of state points $x = (x^1, \ldots, x^n)^T$ parameterized by variable $t$ (usually time). The state points $x(t) = \left(x^1(t), \ldots, x^n(t)\right)^T$ for a fixed $t$ are, as we know, called phase points[70], and they move through the phase space with changing $t$, each of them traversing the phase manifold along its own phase trajectory. The term "phase" was probably coined by J. W. Gibbs who referred to the state of a system as its phase. In physical literature, one often talks about ensembles of dynamical systems – the set of non-interacting systems of the same type differing from one another only by their state at any given moment, that is by initial conditions. There is an illustrative analogy that is customarily exploited in physics namely the picture of a stationary flow of some fluid, in which every fluid particle moves from point in phase space to another during time $t - t_0$ according to equation $\dot{x} = f(x), x(t_0) = x_0$. Later we shall see that this equation may be interpreted as a mapping of the phase space into itself so the "flow" of the "phase fluid" implements a transformation (in fact a family of transformations) of the phase manifold into itself, in the considered case of a smooth vector field described by differential equations – a diffeomorphism i.e., continuously differentiable invertible map. One may notice that the analogy between the "phase fluid" and some physically observable continuous media – liquid or gas – is not complete: there is no interaction between particles of the phase fluid.

---

[69] Scalar equations of the $n$-th order corresponding to the autonomous case take the form $\frac{d^n x}{dd^n} = F\left(x, \frac{dx}{dt}, \ldots, \frac{d^{n-1}x}{dt^{n-1}}\right)$

[70] Rarely, representing points.

Excluding $t$ from parametric equations $x^i(t), i = 1, \ldots, n$ (if we can do it), we get a projection onto phase space $D$. The domain $S = T \times D$ is conventionally called (mostly in old sources) an extended phase space. Although it is not always easy or at all feasible to get rid of parameter $t$[71], it may be possible to obtain differential equations directly describing the trajectories in the phase space. Indeed, from the equation $\dot{x}^i = f^i(x), i = 1, \ldots, n$ we get, for example, the following system of $n - 1$ equations:

$$\frac{dx^2}{dx^1} = \frac{f^2(x)}{f^1(x)}, \ldots, \frac{dx^n}{dx^1} = \frac{f^n(x)}{f^1(x)}$$

whose solution gives the trajectories parameterized by $x^1$. Conversely, we can turn any non-autonomous system into an autonomous one by introducing a new variable $x^{n+1}$, thus producing a system of $n + 1$ equations instead of $n$ which corresponds to increasing the dimensionality of the phase space. In this sense, autonomous systems may be considered general enough to focus mainly on their study.

Applying the uniqueness theorem to the above system of $n - 1$ equations, we may conclude that the phase trajectories do not intersect. Of course, here it is assumed that $f^1(x) \neq 0$. If $f^1(x_a) = 0$ in some points $x_a, a = 1,2 \ldots$, we can take any other $f^j(x), j = 1, \ldots, n$ instead of $f^1(x)$ provided $f^j(x) \neq 0$, which means that we are taking $x^j$ as a parameter. There may be, however, difficulties in the so-called critical points – zero points $\bar{x} = (\bar{x}^1, \ldots, \bar{x}^n)$ where $f(\bar{x}) = 0$ i.e., $f^i(\bar{x}) = 0$ for all $i = 1, \ldots, n$. We shall discuss this problem as soon as we learn a little more about dynamical systems in general.

The above-mentioned uniqueness theorem[72] states that under rather weak assumptions about the properties of vector field $f(x)$ (usually it is assumed differentiable or just Lipschitz-continuous) there exists for each point $x \in D$ exactly one solution $x(t)$ of the law of motion $\dot{x} = f(x)$ with initial value $x(t_0) = x_0$. In other words, the evolution of a dynamical system – its future states at $t > t_0$ – is completely determined by its initial state.[73] It is a fundamental model of determinism: what we can say about tomorrow (more correct – about the properties the system in question will have tomorrow) is uniquely determined by what we can say about the system today. This is not true for quantum or statistical mechanics, although some philosophers contend that quantum mechanics is a fully deterministic theory (because of time evolution features). In my opinion, this is just a cunning word usage typical of philosophers since it is hard to imagine a deterministic scheme where observations affect the system.

When speaking about a dynamical system, one can totally forget about its mechanical origin. It is completely irrelevant whether the vector equation for a dynamical system describes a mechanical or

---

any other evolution. Mechanical systems are commonly used in textbooks as convenient examples of dynamical systems. More important, however, is the fact that some mechanical systems possess specific properties narrowing the entire class of dynamical systems to a clearly defined distinguished subclass, e.g., that of Hamiltonian systems. Nevertheless, it would be a mistake to think that only the Hamiltonian systems are considered in classical mechanics. For instance, non-holonomic systems of mechanics are also dynamical systems of course. The notion of a dynamical system is a generalization of classical mechanics.

Thus, the term "dynamical system" can be applied to any vector field described by a first-order vector differential equation of the form $\dot{x} = f(x), x(t_0) = x_0$ (or any equivalent form, see below), irrespective of its natural or behavioral content. This abstraction serves as a background for mathematical modeling based on dynamical systems.

Similarly, for all $r > 1$. See also Topological dynamical system; Bendixson criterion (absence of closed trajectories); Poincaré-Bendixson theory that has the unfortunate tendency to explode. A rough system is sometimes called a structurally stable system or a robust system. A well-documented survey on (mainly differentiable) dynamical systems is in [76].

In general, an attractor of a dynamical system is a non-empty subset of the phase space such that all trajectories from a neighborhood of it tend to this subset when time increases. An attractor is also called a domain of attraction or basin of attraction. A repelling set, or repellor in a dynamical system is a subset of the phase space of the system that is an attractor for the reverse system. If an attractor, respectively repellor, consists of one point, then one speaks of an attracting, respectively repelling, point. For details (e.g., on stability of attractors) see [76]. It should be noted that in other literature the definition of an attractor is what is called a stable attractor in [76]. For discussions on the "correct" definition of an attractor see [112], and [138].

A general autonomous system is a system of ordinary differential equations (ODE) of the form $\dot{x} \equiv \frac{dx}{dt} = \mathbf{F}(\mathbf{x}), \mathbf{x} = (x^1, \ldots, x^n)$, i.e., it is an ODE system in which variable $t$ (usually interpreted as time) is not explicitly present.

A special case of general autonomous systems are linear autonomous systems, which are described by systems of linear ordinary differential equations (ODE).

The linear autonomous systems are specifically valuable for modeling of countless systems described by differential equations with constant coefficients, for example, the systems where small oscillations can be observed (equation (8.1.1.)), problems in electrical and electronic engineering, etc.

The theory of linear autonomous systems can be almost completely reduced to linear algebra manipulations and can also be formulated in the geometric language as the theory of one-parameter groups of linear transformations. Consider an autonomous system with $n$ dimensional phase space, $\dot{\mathbf{x}} = \mathbf{v}(\mathbf{x})$ and its linearization $\dot{\mathbf{x}} = A\mathbf{x}$, where matrix $A \coloneqq \left(a_j^i\right) = \partial_j v^i(\mathbf{x}_0)$, $\mathbf{x}_0$ is an equilibrium point.

### 8.5.2. Non-autonomous systems

When time enters the problem explicitly as, e.g., in a forced system, we have a more general form of a time-dependent vector field i.e. $\dot{\mathbf{x}} = \mathbf{f}(\mathbf{x}, t)$ which can formally be reduced to an autonomous system $\dot{x}^i = f^i(x^1, \ldots, x^n, x^{n+1})$, $\dot{x}^{n+1} = 1, i = 1, \ldots, n$ at the cost of increasing the phase space dimension.

However, the idea of extending the phase (state) space by including time in it has some defects (see below) since time plays a double role in the non-autonomous case: the base manifold and the variable belonging to the time axis. A prototype of a non-autonomous system is the Newtonian equation for a particle moving in a time-dependent force field, $m\ddot{\mathbf{r}} = \mathbf{F}(\mathbf{r}, t)$. Although this equation seems to have a very primitive form, it may be in general very hard to integrate and is typically solved by specially invented numerical techniques (such as Störmer, Verlet, Runge-Kutta, leapfrog, etc. methods). When describing evolution in the non-autonomous quantum-mechanical or semiclassical case i.e. when the Hamiltonian $H = H(\mathbf{r}, \mathbf{p}, t)$, one has to replace the propagator $U(t) = U(t - t_0) = e^{-iH(t-t_0)/\hbar}$ by a more general evolution operator $U(t, t_0)$.

A generic form of a dynamical system is $\dot{x}^j = f^j(x^i, t)$, $i, j = 1, \dots, n$ or, in vector notations, the just mentioned equation $\dot{\mathbf{x}} = \mathbf{f}(\mathbf{x}, t)$. Here, symbol $\mathbf{x} = \{x^i\}$ denotes elements of the vector space $V^n = \mathbb{R}^n$ (more generally, phase space $P$ or manifold $M$), and tangent vectors $\mathbf{f}(\mathbf{x}, t)$ to manifold $M$ at $\mathbf{x} \in M$ may be interpreted as operators acting on $\mathbb{R}^n$. The time-dependent vector field $\mathbf{f}(\mathbf{x}, t)$ is the derivative of the flow $g_t$, as previously for autonomous systems. It is important to notice that the usual group properties for the Hamiltonian flow $g_t^H$ playing the role of evolution operator $U(t)$ in classical mechanics, $g_t^H \circ g_s^H = g_{t+s}^H$, $(g_t^H)^{-1} = g_{-t}^H$ generally hold if the Hamiltonian $H$ does not depend on the parameter $t$ (time), although the initializing group identity $g_0^H = I$ is still valid. The reason for this complication is that for non-autonomous systems, in particular for time-dependent Hamiltonians, the vector field $\mathbf{v}(\mathbf{x}, t) = \mathbf{f}(\mathbf{x}, t)$ defining the dynamical system is in fact not just a single vector field but a family of them on phase space $P$ (phase manifold $M$). This family of vector fields depends on the parameter $t$, smoothly for continuous-time systems – as if the stationary field were breathing. So, the flow generated by a time-dependent vector field (in particular, generated by a time-dependent Hamiltonian, see below more on Hamiltonian mechanics) is no longer a one-parameter group.

Non-autonomous dynamical systems correspond to the time-dependent external influence. In principle, the non-autonomous case is always a phenomenological model since it always results from artificially closing of the system or throwing out some part of it. In such situations, part of a system is replaced by the time-dependent external forcing. Therefore, invariance under time translations and, correspondingly, energy conservation does not hold for non-autonomous systems. In particular, non-autonomous systems break Galilean invariance of nonrelativistic classical mechanics: recall that all classical $\mathbb{R}^3$-scenes are invariant under affine translations along the time axis. Although the non-autonomous dynamical systems can be formally reduced to the autonomous ones by introducing one more phase space dimension, non-autonomous systems are typically treated by dedicated techniques, being either explored by qualitative methods or solved numerically. The qualitative methods are mostly reduced to finding the principal features of the direction field in $\mathbb{R}^n \times \mathbb{R}$ vector space. For single-dimensional systems ($n = 1$), one can easily sketch the direction field on the $(x, t)$ plane and then produce the integral curves following the direction field i.e., tangential to the segments marking the slopes. Note that qualitative analysis of solutions can provide a better understanding of the process modeled by a dynamical system than numerical or even analytical calculations.

The direction field for a differential equation $\dot{\mathbf{x}} = \mathbf{f}(\mathbf{x}, t)$, is defined as a vector field when each spacetime point $\{x^0, x^1, \dots, x^n\} \equiv (\mathbf{x}, t) \in U \subseteq V^n \times I, I \coloneqq \{t \in \mathbb{R}, \ a \le t \le b\}, (a = -\infty, b = +\infty$ are also admitted) corresponds to a vector $(\mathbf{f}(\mathbf{x}, t), 1) = (f^1, \dots, f^n, 1)$. Here $\mathbf{f}(\mathbf{x}, t)$ is defined on the direct product domain $U = P \times I$ which is known as the enhanced or extended phase space (for simplicity, one can identify phase space $P = V^n$ with $\mathbb{R}^n$ and $I$ with $\mathbb{R}$)[74]. We may assume that $\mathbf{f}(\mathbf{x}, t)$ is differentiable over its arguments as many times as needed i.e., each component $f^i(x^1, \dots, x^n, t)$, $i = 1, \dots, n$ belongs to class $C^p, p \geq 1$. Recall that when $p = \infty$ one usually says that $\mathbf{f}(\mathbf{x}, t)$ is smooth. Each integral curve is the graph of mapping $\boldsymbol{\gamma} \colon I \to P$ satisfying the relationship $\dot{\boldsymbol{\gamma}} = \mathbf{f}(\boldsymbol{\gamma}(t), t)$ for all $t \in I$. In other words, each integral curve is the solution to dynamical system $\dot{\mathbf{x}} = \mathbf{f}(\mathbf{x}, t)$ and, conversely, the graph of such a solution is an integral curve of a direction field in domain $U = P \times I \subseteq \mathbb{R}^n \times \mathbb{R}$. One usually assumes that the direction field is nowhere parallel to $\mathbb{R}^n$ i.e., vectors at every point $(\mathbf{x}, t)$ of the enhanced phase space $U$ always possess a nonzero $t$-component. If the system is non-autonomous, any equilibrium point is required to be a zero of vector field $\mathbf{f}(\mathbf{x}, t)$ at any time $t \in I$, e.g., for all $t \in \mathbb{R}_+$ or $t \in \mathbb{R}$ (if the system is defined for $t < 0$ and can be invertible).

Notice that the trick of regarding time $t$ as one more phase (state) space variable i.e., $\dot{x}^i = f^i(x^1, \dots, x^n, x^{n+1})$, $\dot{x}^{n+1} = 1, i = 1, \dots, n$ is only of limited value since there are no equilibrium states for a dynamical system in the enhanced phase space. Moreover, such a dynamical system, strictly speaking, does not have bounded solutions since all of them, even equilibrium points, are necessarily time-dependent; therefore, it is difficult to consider limit sets and attractors. For example, an infinite number of diverse phase trajectories may pass, in the non-autonomous case, through every point in the phase space, which makes the phase portrait very complex and hardly useful for analysis, in contrast with the autonomous case. This tangled situation can be illustrated by closed orbits that are bound to remain closed at all times as, for example, a periodically driven damped oscillator should have a closed orbit regardless of approaching this cycle individual non-stationary trajectories.

When exploring stability of a non-autonomous dynamical system, we, as in the autonomous case, ought to consider gradients of vector field $\mathbf{f}(\mathbf{x}, t)$ (the partial derivatives for one-dimensional fields). Note that vector field gradients are taken at fixed $t$ so that if we go back in time (e.g., to the initial conditions, $t \to t_0$ from the right), we may encounter three situations: (1) integral trajectories are narrowing ($\partial f / \partial x > 0$) which means that the distance between the solutions widens as $t \to +\infty$; (2) integral trajectories are running away from each other ($\partial f / \partial x < 0$) i.e., solutions fly apart as $t \to -\infty$; and (3) integral trajectories stay more or less close to each other ($\partial f / \partial x \approx 0$). For illustrative purposes, we consider one-dimensional phase space here, extension to an $n$-dimensional vector space is trivial. One can notice that while exploring stability of autonomous systems one is mostly focused on the stability of fixed points, stability of non-autonomous systems relates to time-dependent solutions. Thus, the equilibrium positions $\mathbf{x}_0 \colon f(\mathbf{x}_0) = 0$ may be stable or unstable. In non-autonomous systems, equilibrium positions are defined by the condition $f(\mathbf{x}, t) = 0$, which gives $\mathbf{x}_0 = \mathbf{x}_0(t)$ i.e., equilibrium points drift with time and form a trajectory[75]. It means that when considering the stability of non-autonomous systems, the accent is displaced from points (which are actually time-independent solutions) to time-dependent solutions. For instance, a solution $\mathbf{x} =$

<hr/>

[74] Previously, we occasionally denoted vector field $\mathbf{f}(\mathbf{x}, t)$ as $\mathbf{v}(\mathbf{x}, t)$ in order to emphasize its affinity to the physical velocity field. Here, while discussing dynamical systems, a more general (and standard) notation for a tangent field seems to be more pertinent. Likewise, we denote the evolution operator in classical dynamical systems by symbol $g_t$, while reserving the more general notation $U(t - t_0) = e^{-iH(t-t_0)/\hbar}$ to designate the universal unitary evolution.

[75] Notice that one can define equilibrium of non-autonomous systems in a stronger sense, e.g., $\mathbf{x} = \mathbf{x}_0$ is an equilibrium point of a dynamical system if $\mathbf{f}(\mathbf{x}_0, t) = 0$ for all $t \geq t_0 > 0$. Then $\mathbf{x}_0$ is said to be stable at $t_0$ if given $\varepsilon > 0$ there exists $\delta(\varepsilon, t_0) > 0$ such that $|\mathbf{x}(0) - \mathbf{x}_0| < \delta$ implies $|\mathbf{x}(t) - \mathbf{x}_0| < \varepsilon$ for all $t \geq t_0$.

$\boldsymbol{\varphi}(t, t_0; \boldsymbol{\xi})$ to $\dot{\mathbf{x}} = \mathbf{f}(\mathbf{x}, t)$ with initial condition $\boldsymbol{\varphi}(t_0) = \boldsymbol{\xi}$ is Lyapunov stable if for any $\varepsilon > 0$ there exists a $\delta > 0$ so that $|\boldsymbol{\xi} - \mathbf{x}_0| < \delta$ implies that $|\boldsymbol{\varphi}(t, t_0; \boldsymbol{\xi}) - \boldsymbol{\varphi}(t, t_0; \mathbf{x}_0)| < \varepsilon$ for all $t > t_0$ (Figure 10.

One can analogously construct a more stringent definition of asymptotic stability ($\lim\limits_{t \to \infty} |\boldsymbol{\varphi}(t, t_0; \boldsymbol{\xi}) - \boldsymbol{\varphi}(t, t_0; \mathbf{x}_0)| = 0$). For simplicity, we are using the absolute value symbol instead of the norm $\|.\|.$ This is, of course, a very sketchy résumé of the theory of nonautonomous dynamical systems which can be rather intricate.

## Section 9. Oscillations

In many cases, the forces appearing in a mechanical system, when it is driven from the equilibrium, strive to return the system to the equilibrium position, when there are no forces producing motions in the system. Return to equilibrium is the general tendency of almost all natural systems. For small deviations from equilibrium, the returning forces are proportional to such deviations (Hooke's law), which is simply a manifestation of the fact that each smooth (or analytical) function is locally linear. This linearity leads to the equation of small oscillations in a $1d$ mechanical system, $m\ddot{x} + kx = 0, k > 0$, which is the simplest possible model of a finite analytical motion. This motion is obviously periodic.

One may notice that although the motion equations are purposed to describe evolution, systems in purely periodic motion are difficult to call evolving since they repeat the same states a countless number of times. In an evolving system, each state is, in general, unlike any other. A little later we shall specially discuss time evolution in classical mechanics in connection with dynamical systems theory.

The case of small oscillations with a single degree of freedom is such a common mathematical model that it is rather boring to discuss it once again. Nonetheless, we shall do it in order to have a possibility of returning to this ubiquitous model in future chapters and, perhaps, to find some not so common features in it. One can of course find careful exposition of a $1d$ oscillation theory in any textbook on mechanics so that we do not need to worry much about details. Let $x = x_0$ be a stable equilibrium i.e., $\partial U(x_0)/\partial x = 0, \partial^2 U(x_0)/\partial x^2 > 0.$[76] Then the solution to the mechanical system with kinetic energy $T(p, x) = T(\dot{x}, x) = (1/2)x(a)\dot{a}^2, U = U(x)$ is periodic for $(p, x)$ near $(p = 0, x_0)$. The first question that arises here is: what is the period of the motion? The answer is: the period $T_0$ near equilibrium position with the decrease of the amplitude $x(t)$ tends to the limit $T_0 = 2\pi/\omega_0$ where

$$\omega_0^2 = \frac{b}{a}, \qquad b = \frac{1}{2}\frac{\partial^2 U(x_0)}{\partial x^2}, \qquad a = a(x_0),$$

since for a linearized system $U(x) = (1/2)b(x - x_0)^2$ and the solution is $x(t, \omega_0) = A\cos\omega_0 t + B\sin\omega_0 t$ . The qualitative picture of $1d$ small oscillations (e.g., in the $(\dot{x}, x)$-space) can be plotted even without solving differential equations of motion: the relationship for the "invariant manifold" (see above the section on dynamical systems)

$$\frac{m\dot{x}^2}{2} + \frac{bx^2}{2} = const$$

represents the ellipse.

---

[76] I write these expressions for brevity not in a fully correct form; one should write, e.g., $\partial U(x)/\partial x \mid_{x=x_0}$ and analogously for second derivatives.

The most common example of a periodic, i.e., oscillating, $1d$ mechanical system is a pendulum. The simple mathematical pendulum consisting of a point mass $m$ and of a massless thread (constraint) of length $l$ is described by the Hamiltonian $H = p_\varphi^2/2ml^2 - mgl\cos\varphi$ where $p_\varphi = ml^2\dot{\varphi}$ is the angular momentum and $\varphi$ is the declination angle. In this section, we shall mostly deal with small oscillations. For small oscillations $\varphi \gg 1$ around the equilibrium position $\varphi = 0$, the pendulum motion is described by the linearized (harmonic) Hamiltonian

$$H = \frac{p_\varphi^2}{2ml^2} + \frac{mgl}{2}\varphi^2 = \frac{ml^2\dot{\varphi}^2}{2m} + \frac{mgl}{2}\varphi^2 = \frac{m\dot{x}^2}{2} + \frac{m\omega^2 x^2}{2},$$

where $\omega = (g/l)^{1/2}$.

One can of course describe oscillations in all possible versions of mechanics: Newtonian, Lagrangian, Hamiltonian, with the help of Hamilton-Jacobi equations and all other possible methods. We prefer to attach oscillations to the most intuitive version of mechanics, the Newtonian one, because geometric language, more adequate in other formulations than in Newtonian, may obscure the basic notions of the oscillation theory. Oscillations are usually understood as finite motions that occur in the vicinity of equilibrium points (Figure 11). Recall an equilibrium point.[77] If, for instance, point $a$ is a local minimum of the potential $U(x, \lambda)$, where $\lambda$ is some parameter, then $x = a(\lambda)$ brings the Lyapunov stability (see the section on dynamical systems), i.e., for initial conditions $\{p(0), x(0)\}$ sufficiently close to $\{0, a\}$ the whole phase trajectory $\{p(t), x(t)\}$ is close to $\{0, a\}$. Mathematical models of almost all multidimensional vibrating systems are often just generalizations of the $1d$ case: $x_{n+1} = Q(x_1, \ldots, x_n)$, and if $Q$ is positive definite, then any small motion can be represented as a superposition of oscillations along the main axes.

## 9.1. Harmonic oscillator

The model of the harmonic oscillator is undoubtedly one of the most favorite mathematical models in physics and engineering. One might recall the well-known statement by Sidney Coleman, a prominent theoretical physicist, that "the career of a young theoretical physicist consists in treating the harmonic oscillator in ever increasing levels of abstraction". There are, however, a number of surprising nuances in the apparently primitive model of harmonic oscillator. It is remarkable that this model immediately leads to very sophisticated concepts: the obvious example is the quantum field theory (QFT). A little less obvious is the fact that the harmonic oscillator serves as a bridge between classical and statistical mechanics. Even in biology and physiology, the oscillator model is of everlasting value: for instance, such a phenomenon as tremor is oscillatory movement of body parts resulting from periodic contraction of opposing muscle groups.

---

[77] One more time trying to infuriate the mathematicians, We shall make in the present context no distinction between equilibrium, fixed, stationary and critical points. In general, however, these notions may be different.

A perfect harmonic oscillator can also be used for time measurement. A warning: one should not confuse the model of harmonic oscillator with reality, where there are no precisely periodic phenomena, but only quasiperiodic so that the oscillator will not return to exactly the same point.

The harmonic force (Hooke's law) $F = -kx$ has almost universal applicability, as the motion in any analytic potential near its local minimum is governed by the linear restoring force. In short, despite its apparent primitiveness, the model of oscillator, even in the harmonic approximation, is very rich and can be approached from different perspectives. Thus, from the Hamiltonian viewpoint, the harmonic oscillator is just a linear representation of symplectic gradients

$$\dot{p}_i = -\frac{\partial H}{\partial q^i} = -A_{ij}q^j, \qquad \dot{q}^i = \frac{\partial H}{\partial p_i} = B^{ij}p_j.$$

This linear model can of course be extended to nonlinear operators $A$ and $B$. The above representation, by the way, allows one to easily consider some generalizations of a linear oscillator, in particular, when matrices $A_{ij}$ and $B^{ij}$ are slowly changing in time. The corresponding approximately preserved quantities are known as adiabatic invariants. In other words, adiabatic invariants can be regarded as approximate constants of motion whose preservation in the process of evolution is the more accurate the slower the system parameters (in this case assembled in the $A$ and $B$ matrices) are changing. The theory of adiabatic invariants is well developed and one of its important achievements is replacing actual dynamical systems by their means obtained by averaging over "fast" variables i.e., the variables corresponding to the shortest time scales.

The separation of the motion into "slow" and "fast" time scales or, more generally, an interference between slow and fast modes of evolution can be traced on a simplified example of a one-dimensional harmonic oscillator.

An oscillator with a single degree of freedom (one-dimensional configuration space) is a two-parameter model described by the Hamiltonian

$$H(p, q) = \frac{p^2}{2m} + \frac{m\omega^2 q^2}{2},$$

where parameters $m$ and $\omega$ are assumed positive and are interpreted, respectively, as mass and oscillation frequency. This model is called oscillator in both classical and quantum theory. It is interesting that in the quantum case this problem can be solved purely algebraically using only the Heisenberg commutation relations, $[p, q] = i\hbar$. The oscillator model, despite its apparent simplicity, connects two main classes of physical systems: particles and fields, and thus it permeates the whole of physics. On the other hand, the harmonic oscillator model may be considered a very special case of linear mathematical models described by the second-order ordinary differential equations. We have seen that many problems – not only in mechanics and physics – lead to the models represented by equation $\ddot{x} - f(x, \dot{x}, t) = 0$ (which is in general nonlinear) or, denoting $x = x^1, \dot{x} = x^2$, we have a dynamical system with a two-dimensional phase space

$$\dot{x}^1 = x^2, \qquad \dot{x}^2 = f(x^1, x^2, t).$$

The choice of $f$ as a linear function, $f(x^1, x^2, t) = -(k/m)x^1$, where $k \equiv m\omega^2 > 0$ leads to the model of harmonic oscillator. Of course, a more general case of the two-dimensional dynamical system can be written in the form $\dot{x}^1 = F(x^1, x^2, t), \dot{x}^2 = G(x^1, x^2, t)$, where $\dot{x}^1$ (or $F$) does not necessarily coincide with velocity, $\dot{x}^2$ with acceleration and $G$ with physical force. This is an archetypal system with a low-dimensional phase space. Recall that in case variable $t$ is absent (it has already been mentioned that $t$ is perceived as a parameter, mostly identified with time), the dynamical system is called autonomous. This reminder is to emphasize that one should not confuse autonomous and conservative systems: a system may be autonomous but non-conservative, for example, a damped oscillator. The right-hand side $(F, G)^T$ defines a vector field on the phase plane $(x^1, x^2)$ i.e., the field of phase velocity. One can naturally denote a plane phase velocity vector $\boldsymbol{w} = (\dot{x}^1, \dot{x}^2)^T = (F, G)^T$ and a radius-vector $\boldsymbol{\rho} = (x^1, x^2)^T$ and form, e.g., their scalar product $\boldsymbol{\rho w} = Fx^1 + Gx^2$. In the special case of the harmonic oscillator, this scalar product is equal to $x^1 x^2 (1 - k/m)$, and if we choose the system of units where $k = m\omega^2 = 1$ (this can be achieved by properly scaling mass or time), then this scalar product becomes identically zero i.e., the phase velocity vector is everywhere normal to the radius vector. Geometrically, this means that the phase trajectories for $k = 1$ are concentric circles and integral curves are circular helices (screw lines). If $k/m \neq 1$, then phase trajectories are ellipses defined by the oscillator energy $\frac{m(x^2)^2}{2} + \frac{k(x^1)^2}{2} = E = T + V$ i.e., since energy in such a system is conserved, by initial condition $x^1(t_0) = A, x^2(t_0) = B$ (it is easy to verify that quantity $E$ is constant along each phase trajectory). Since the system is autonomous, we can freely choose the starting time $t_0 = 0$, and obtain the harmonic solution

$$x^1(t) = A\cos\omega t + \frac{B}{\omega}\sin\omega t, \qquad x^2(t) = -A\omega\sin\omega t + B\cos\omega t, \omega = \left(\frac{k}{m}\right)^{\frac{1}{2}}, \qquad (9.1.1.)$$

which of course can be written in a number of equivalent forms, most of them being also expressed in a parametric form $\mathbf{x}(t) = (x^1(t), x^2(t))$. Equations (9.1.1.) give a concrete realization of a flow generated by the dynamical system corresponding to the harmonic oscillator.

One can notice that the phase point moves along the phase path with angular velocity $\dot{\chi} = -\frac{k(x^1)^2 + m(x^2)^2}{k[(x^1)^2 + (x^2)^2]} = -\frac{2E}{k\rho^2}$, where $\chi \equiv \tan^{-1}\left(\frac{x^2}{x^1}\right)$ and $\rho^2 = (x^1)^2 + (x^2)^2$. In case $\omega^2 \equiv k/m = 1$, the absolute value of angular velocity of a phase point is 1 everywhere; that is obvious already from the harmonic oscillator equations $\dot{x}^1 = x^2, \dot{x}^2 = -x^1$. Namely, by taking squares and adding up, we obtain $\rho^2 = w^2$ or $|\rho| = |w|$ which means that angular velocity $\dot{\chi} = \left|\frac{w}{\rho}\right| = 1$ i.e., the phase point is moving uniformly. The minus sign symbolizes the motion in the clockwise direction. It would be more correct from the physical standpoint to denote axes on the phase plane as, e.g., $(x^1, \frac{x^2}{\omega})$ and, respectively, $\chi \equiv \tan^{-1}\left(\frac{x^2}{\omega x^1}\right), \rho^2 = (x^1)^2 + \left(\frac{x^2}{\omega}\right)^2 = \frac{2E}{k}$ so that the angular velocity in phase space $|\dot{\chi}| = 1$ automatically, but in the theory of dynamical systems the dimensionality of space variables does not really matter. It is less trivial that angular acceleration of the phase point $\ddot{\chi}$ over elliptic curves $k \neq 1$ is also zero due to energy conservation:

$$\ddot{\chi} = \frac{d}{dt}\left(\frac{2E}{\rho^2}\right) = -\frac{4E}{\rho^4}(x^1\dot{x}^1 + x^2\dot{x}^2) = -\frac{4E}{\rho^4}(x^1 x^2 - x^2 x^1) = 0.$$

One can also notice that in order to obtain the law of motion $\mathbf{x}(t)$ one has to solve the second-order differential equation (or, equivalently, the system of two first-order equations) whereas the phase curves are defined by a single first-order equation

$$\frac{dx^2}{dx^1} = -\frac{\omega^2 x^1}{x^2} \qquad (9.1.2.)$$

that is produced if we simply divide the first equation of the dynamic system for harmonic oscillator by the second. However, equation (9.1.2.) contains less information than the initial dynamical system: it gives the phase trajectories but does not provide the direction of motion along them (figuratively speaking, (9.1.2.) defines the vector field without arrows). This slight loss of information reflects the fact that parametric definition of a function is more general than explicit or implicit.

Recall that dynamical systems generally describe the evolution i.e., they allow us to determine future from the past (in non-chaotic cases), in particular to compute the law of motion $\mathbf{x}(t)$ for any $t > t_0$, provided $\mathbf{x}(t_0)$ is assumed to be known. Less popular is determining initial data $\mathbf{x}(t_0)$ from the current state $\mathbf{x}(t)$, yet this problem may also be important (compare, e.g., Eulerian and Lagrangian pictures in fluid motion modeling). One can, by the way, pay attention to the fact (mainly overlooked) that initial coordinates and velocities in Newtonian mechanics are integrals of motion, corresponding to some Noether's transform. One can invert (9.1.1.) to obtain

$$x_0^1 = x^1(t)\cos\omega t - \frac{x^2(t)}{\omega}\sin\omega t, x_0^2 = x^1(t)\omega\sin\omega t + x^2(t)\cos\omega t, \omega = \left(\frac{k}{m}\right)^{\frac{1}{2}}. \quad (9.1.3.)$$

Formally, this expression is produced by replacing $t \to -t$; mathematically, it is a natural consequence of the flow diffeomorphism. We may note that initial conditions $\mathbf{x}_0 = (x_0^1, \ x_0^2)^{\mathrm{T}}$ or, more generally $\mathbf{x}_0, \mathbf{p}_0$, may be regarded as integrals of motion for a mechanical system.

This "naive" diffeomorphism can be interpreted as the equivalence of past and future i.e., as easy switching of cause and effect, which is the key point of Newtonian (time-reversible or invertible) dynamics. Time reversibility leads to the famous "time arrow" paradox emphasizing the contradiction between time-reversible mechanical equations at the "microscopic" level and macroscopic irreversibility which is omnipresent in real life. The macroscopic irreversibility can be concisely expressed in the form of the so-called Loschmidt paradox: why is there an inevitable growth of entropy, although the microscopic physical laws are invariant under time inversion?

Note that the equivalence between past and future is not necessarily the same as the possibility of swapping cause and effect – here rests the difference between time reversal noninvariant and irreversible systems. We have already mentioned in connection with the causality requirement imposed on realistic mathematical models (section 3) that time asymmetry is not identical with causality violation since the underlying mathematical structures are different (time invertibility of microscopic dynamics vs., e.g., the condition in linear case that Green's function is zero for a negative value of its argument).

One might notice here the immediate connection with quantum evolution of position $x(t)$ and momentum $p(t)$ operators:

$$\frac{d}{dt}x(t) = \frac{i}{\hbar}[H, x(t)] = \frac{p(t)}{m}, \frac{d}{dt}p(t) = \frac{i}{\hbar}[H, p(t)] = -m\omega^2 x(t). \qquad (9.1.4.)$$

Solution of these equations corresponding to initial conditions $\dot{x}(0) = p_0/m, \dot{p}(0) = -m\omega^2 x_0$ has the "classical" form (9.1.1.)

$$x(t) = x_0 \cos \omega t + \frac{p_0}{m\omega} \sin \omega t, \quad p(t) = p_0 \cos \omega t - m\omega x_0 \sin \omega t, \tag{9.1.5.}$$

where, however, quantities $x(t), p(t)$ are operators for which we can compute commutation relations $[x(t_1), x(t_2)] = \frac{i\hbar}{m\omega} \sin \omega(t_2 - t_1)$, $[p(t_1), p(t_2)] = i\hbar \sin \omega(t_2 - t_1)$, $[x(t_1), p(t_2)] = i\hbar \cos \omega(t_2 - t_1)$. For coinciding time points, $t_2 = t_1$, we get the standard commutation relations $[x, x] = [p, p] = 0, [x, p] = i\hbar$ (more generally, $[x^j, p_k] = i\hbar \delta_k^j$, where $j, k = 1,2,3$; $\delta_k^j$ is the Kronecker symbol). Note that in a primitive sense, classical functions (c-numbers, in the jargon of physicists) differ from the respective (Heisenberg) operators by the property that operators related to different time points in general do not commute.

An obvious generalization of the scalar harmonic oscillator i.e., moving only along the $x$-axis is matrix equation $\frac{d^2\mathbf{x}}{dt^2} + \Omega^2 \mathbf{x} = 0$, where $\Omega^2 = \left(\frac{k_{ij}}{m_i}\right), i, j = 1, \dots, n, m_i \neq 0$ is a nonsingular square matrix. Notice that $\Omega = \sqrt{\Omega^2}$ does exist if $\Omega^2 \neq 0$. This matrix equation is equivalent to a system of $n$ second-order linear differential equations $m_i \ddot{x}^i + k_{ij} x^i = 0$ and corresponds to $n$ coupled harmonic oscillators. The matrix equation can be supplemented with initial conditions $\mathbf{x}_0 = \mathbf{x}|_{t=t_0}, \dot{\mathbf{x}}_0 = \dot{\mathbf{x}}|_{t=t_0}$ (these initial data are vectors in $n$-dimensional linear space). Obviously, for $n = 1$ i.e., when $x$ and $\Omega^2$ are scalars ($\Omega \equiv \omega = (k/m)^{1/2}$), we get the above results. If we define $\cos \Omega t = \mathrm{I} - \frac{1}{2!}(\Omega t)^2 + \frac{1}{4!}(\Omega t)^4 - \cdots$ and $\Omega^{-1} \sin \Omega t = t - \frac{1}{3!}\Omega^2 t^3 + \frac{1}{5!}\Omega^4 t^5 - \cdots$, where I is the unit matrix then we get the same kind of solutions for $\mathbf{x} = (x^1, \dots, x^n)^T$. It is easy to rewrite the matrix equation for $\mathbf{x}$ in the form of a dynamical system with $2n$-dimensional phase space: $\dot{\mathbf{x}} = \mathbf{y}, \dot{\mathbf{y}} = -\Omega^2 \mathbf{x}$. A number of important engineering applications lead to considering oscillations in a damped mass-spring system with $n$ degrees of freedom, when the $i$-th mass is linked with the $(i + 1)$st one by a spring having elastic constant $k_i$ and damping constant $\gamma_i$. Oscillations in such a system are described by a second-order matrix equation for vector $\mathbf{x} = (x^1, \dots, x^n)^T$ (for writing simplicity we shall denote vectors simply by $x$)

$$M\ddot{x} + \Gamma \dot{x} + Kx = F(t), \tag{9.1.6.}$$

where mass matrix $M$ in most models is assumed diagonal, $M = \mathrm{diag}(m_1, \dots, m_n)$ whereas damping $\Gamma$ (see below) and stiffness $K$ matrices are typically assumed tridiagonal and symmetric (the nearest neighbor approximation); $F(t)$ is the driving force vector, here we shall consider free oscillations, $F = 0$.

The corresponding eigenvalue problem for $n \times n$ matrices $M, \Gamma$ and $K$ is $(\lambda^2 M + \lambda \Gamma + K)x = 0$ and is known in engineering as the quadratic eigenvalue problem (QEP). This problem was encountered in aerodynamics, in particular, to explore flutter in supersonic aircraft, in electrical engineering, in railway engineering, e.g., to analyze vibrations of the rail tracks, in bridge construction and maintenance. Even nuclear power plants are modeled by 8 degrees of freedom QEP, when the whole plant is broken into four main elastically interconnected blocks: basement, building, pressurized vessel and reactor core, with each block having two degrees of freedom corresponding to possible swaying and rocking. The quadratic eigenvalue problem, $P_2(A, \lambda)x = (\lambda^2 A_2 + \lambda A_1 + A_0)x = 0, A \equiv A_k \in \mathbb{C}^{n \times n}, k = 1,2,3, x \in \mathbb{C}^n$, looks rather innocent, yet it is actually quite intricate and rapidly becomes numerically complex with growing $n$. There exists a vast amount of literature on QEP, but the main results are dispersed among applications; see, however, a comprehensive survey by Tisseur, F., Meerbergen, K. *The quadratic eigenvalue problem*. SIAM Review, v.43, no.2, pp. 235-286 (2001). One of the common approaches to QEP is reducing it to a linear form

$$\left[ \lambda \begin{pmatrix} A_2 & 0 \\ 0 & I \end{pmatrix} + \begin{pmatrix} A_1 & A_0 \\ -I & 0 \end{pmatrix} \right] \begin{pmatrix} \lambda x \\ x \end{pmatrix} = 0$$

which is identical to

$$\begin{pmatrix} \lambda^2 A_2 + \lambda A_1 + A_0 \\ -\lambda I + \lambda I \end{pmatrix} x = 0$$

and then applying some known algorithms (such as QR or QZ transformations). One might note that $\lambda = -\frac{1}{2}(x^*, A_2 x)^{-1} \left[ (x^*, A_1 x) \pm \sqrt{(x^*, A_1 x)^2 - 4(x^*, A_0 x)(x^*, A_2 x)} \right]$. If $(x^*, A_1 x)^2 > 4(x^*, A_0 x)(x^*, A_2 x)$ for any $x \neq 0$ ($P_2(A, \lambda)$ is hyperbolic), the eigenvalues of $P_2(A, \lambda)$ are real. However, verifying hyperbolicity for large $n$ can be a nontrivial problem of its own. In general, any eigenpair of polynomial $P_2(A, \lambda)$ satisfies $(x^*, P_2(A, \lambda)x) = 0$, where $x^*$ is a left eigenvector, $y P_2(A, \lambda) = 0$. The QEP is naturally generalized to the polynomial eigenvalue problem (PEP) when $P_2(A, \lambda)$ is replaced by $P_m(A, \lambda) = \lambda^m A_m + \cdots + \lambda A_1 + A_0$ and the problem is to find an eigenpair i.e., complex scalar $\lambda$ and vector $x \in \mathbb{C}^n$, $x \neq 0$ such that $P_m(A, \lambda)x = 0$, where $A := (A_0, \ldots, A_m)$ is an $(m + 1)$-tuple of complex $n \times n$ matrices. Still more generally, one needs to find both right $x$ and left $y$ nonzero eigenvectors, $x, y \in \mathbb{C}^n$, satisfying $P_m(A, \lambda)x = 0$, $y^* P_m(A, \lambda) = 0$, and generalizing the problem further, we arrive at the homogeneous eigenvalue problem, $Q_m(A, \lambda, \mu)x = 0$, $y^* Q_m(A, \lambda, \mu) = 0$, where $Q_m(A, \lambda, \mu) \equiv \sum_{k=0}^{m} \lambda^k \mu^{m-k} A_k$ is a homogeneous polynomial in $\lambda, \mu \in \mathbb{C}$ of degree $m$. For example, the homogeneous QEP is $(\lambda^2 A_2 + \lambda \mu A_1 + \mu^2 A_0)x = 0$. So, one can see how far the elementary oscillator model can be developed.

Notice that nearly all said above about the harmonic oscillator is true only in the nonrelativistic approximation. Indeed, the solution to nonrelativistic equation $\ddot{x} + \omega_0^2 x = 0$, $\omega_0^2 = k/m$, e.g., for initial conditions $x(0) = A, \dot{x}(0) = 0$ is $x(t) = A \cos \omega_0 t$ so that the respective velocity $\dot{x}(t) = -A\omega_0 \sin \omega_0 t$ can exceed the light speed $c$ in absolute value for $\omega_0 > c/A$. In relativistic mechanics, the simplest possible variant for the motion equation would be[78] $dp/dt = d(\gamma m \dot{x})/dt \equiv d/dt \left[ m\dot{x}(1 - \dot{x}^2/c^2)^{-1/2} \right] = -kx$.

## 9.2. Damped harmonic oscillator

One of the main difficulties in treating a one-dimensional damped oscillator

$$\ddot{x} + a\dot{x} + bx = f(x, \dot{x}, t), \dot{x} := \frac{dx}{dt} \, , \qquad (9.2.1.)$$

with the conservative "returning" force $mbx := m\omega_0^2 x \equiv kx$, dissipation $f_d := -ma\dot{x}$ (the friction force), applied force $mf(x, \dot{x}, t)$, where $m$ is the oscillator mass or inertial analogs of the latter, is the impossibility to represent equations of motion in the Lagrangian form and thus to directly apply Noether's theorem. The mathematical reason for it is that equation (9.2.1.) taken as an ordinary differential equation is not self-adjoint[79] because of the friction term [16]. Notice in passing that the presence of damping also leads to significant difficulties in the transition to quantum theory, since classical damped systems do not, in general, produce self-adjoint quantum Hamiltonians.

---

[78] Another option would be, for example, $\frac{dp}{d\tau} = -kx$, where $d\tau = dt(1 - \dot{x}^2/c^2)^{1/2}$ is the proper time element.

One can, however, see that dissipative equation (9.2.1.) can be obtained from the explicitly time-dependent Lagrangian

$$L = \frac{1}{2} m e^{at} (\dot{x}^2 - bx^2). \qquad (9.2.2.)$$

Indeed, the Lagrange equation $\frac{d}{dt} \left( \frac{\partial L}{\partial \dot{x}} \right) - \left( \frac{\partial L}{\partial x} \right) = 0$ gives $m e^{at} (\ddot{x} + a\dot{x} + bx) = 0$ which amounts to (9.2.2.). The rather exotic looking Lagrangian (9.2.2.) can be regarded as that of an oscillator with the mass exponentially depending on time: $m \to m e^{at}$. Another way to express the same result i.e., to bring (9.2.1.) to a self-adjoint (Lagrangian) representation is multiplying equation (9.2.1.) by $e^{\sigma t}$, then changing variable $x(t)$ by $\xi(t) := x(t) e^{\sigma t/2}$ and putting $\sigma \to a$. Then we would have in the new variables the equation $\ddot{\xi} + \left( b - \frac{a^2}{4} \right) \xi = 0$.

More generally, one can always eliminate the explicitly present friction term in the second-order ODE with dissipation

$$\ddot{x} + a(t)\dot{x} + b(t)x = 0$$

by making the substitution $x(t) = y(t)z(t)$. Then $y\ddot{z} + (2\dot{y} + ay) + (\ddot{y} + a\dot{y} + by)z = 0$ and if we require $2\dot{y} + ay = 0$, we shall have

$$x(t) = \exp \left( -\frac{1}{2} \int_{t_0}^{t} a(t')dt' \right) z(t)$$

and

$$\ddot{z} + \left( b(t) - \frac{a^2(t)}{4} - \frac{\dot{a}(t)}{2} \right) z = 0.$$

Notice that functions $x(t)$ and $z(t)$ have the same zeros i.e., more or less the same analytic behavior. So, for dynamical modeling one can consider in most cases equations of the Schrödinger type $u'' + Q(x)u = 0$ instead of the three-term dissipative ODE.

The energy dissipation per unit time in a 1d damped oscillator (9.2.1.) is $-\frac{d\mathcal{E}}{dt} = f_d \dot{x} = ma\dot{x}^2$. An obvious extension of this model, e.g., on a system of coupled, damped and driven oscillators is a matrix equation

---

[79] This fact seems to have been first noticed by H. von Helmholtz in the paper „Ueber die physikalische Bedeutung des Princips der kleinsten Wirkung", J. für die Reine und Angewandte Mathematik, v.**100**, pp. 137-166 (1887). It is curious that Helmholtz uses the notion of energy quanta in this paper (ibid. p.138) – thirteen years before Max Planck.

$$A_{ik}(t)\ddot{q}^k + B_{ik}(t)\dot{q}^k + C_{ik}(t)q^k = F_i(q^k, \dot{q}^k, t), i, k = 1, \ldots, n, \qquad (9.2.3.)$$

where mass matrix $A$ is symmetric and positive, $A = A^T > 0$, $B$ is the $n \times n$ dissipation matrix (usually $B = -B^T$), $C = C^T$ is the $n \times n$ matrix of internal (potential) forces, $F_i(q^k, \dot{q}^k, t)$ is the $1 \times n$ matrix of external forces, $\mathbf{q} = \{q^k\}^T, k = 1, \ldots, n$ is the coordinate vector. One can note that if off-diagonal entries in mass matrix $A_{ik}$ in (9.2.3.) are non-zero, oscillators are also coupled via accelerations.

Equation (9.2.3.) appears to be a trivial generalization of a single-dimensional damped oscillator (9.2.1.) with scalar coefficients being replaced by the respective matrices, but to solve matrix equation (9.2.3.) i.e., to find vector $q(t), t > t_0$, for arbitrary initial vector $q(t_0)$ and for any (in particular, nonlinear) forces $F_i(q^k, \dot{q}^k, t)$ seems to be a task far from being trivial. This problem appears to have not been exhaustively explored, although matrix equation (9.2.3.) has many important applications.

The ideal – nondissipative – harmonic oscillator is a simple example of a Hamiltonian system with the Hamiltonian function of the type $H(x, y) = \frac{1}{2}(x^2 + y^2)$. The motion equation can be written (in the commensurate units) as the Hamiltonian dynamical system, $dx/dt = -\partial H/\partial y$, $dy/dt = \partial H/\partial x$ i.e., a symplectic vector field. One can easily see that function $H$ is preserved on the system's trajectories. However, the simple harmonic motion is structurally unstable: small perturbations can destroy the Hamiltonian character of the system, together with its phase portrait consisting of concentric circles (more generally, ellipses) and the conservation law. In particular, dissipative perturbations result in energy loss.

Thus, a more realistic model is the oscillator with dissipation (friction). The corresponding dynamical system can be written as $\dot{x}^1 = x^2, \dot{x}^2 = -ax^1 - bx^2$ (see (9.1.1.), section 9.1.). One can of course scale this system in order to reduce the number of parameters, for example, putting $a = 1$. Then we have linear evolution equation $\dot{\mathbf{x}} = A\mathbf{x}$, where matrix $A = \begin{pmatrix} 0 & 1 \\ -1 & -b \end{pmatrix}$, $\det A = 1$, $\operatorname{Tr} A = -a$. The characteristic equation $\lambda^2 + b\lambda + 1 = 0$ has real different roots when $|b| > 2$. In this case both roots are negative, and all the solutions tend to zero with $t \to \infty$. By a simple linear transformation, one can bring matrix $A$ to the diagonal form $\dot{z}^1 = \lambda_1 z^1, \dot{z}^2 = \lambda_2 z^2$ so that the solution is $z^1(t) = e^{\lambda t}z^1(0), z^2(t) = e^{\lambda t}z^2(0)$, and the damped oscillator in fact does not oscillate but strives to the equilibrium position. Recall that in the linear case each flow (evolution operator) $g_t, t \in \mathbb{R}$ i.e., one-parameter group of linear transformations of $\mathbb{R}$ is of the form $g_t = e^{At}$, where $A: \mathbb{R} \to \mathbb{R}$ is a linear operator defining the dynamical system $\dot{\mathbf{x}} = A\mathbf{x}$. The solution to this equation with initial condition $\mathbf{x}(0) = \mathbf{x}_0$ is given by $\mathbf{x}(t) = e^{At}\mathbf{x}_0$.

We can easily transform the dynamical system corresponding to the harmonic oscillator ($dx/dt = -y$, $dy/dt = x$, characteristic matrix $\begin{pmatrix} 0 & -1 \\ 1 & 0 \end{pmatrix}$, eigenvalues $\lambda = \pm i$, general solution through two basic solutions $x(t) = c_1 \begin{pmatrix} \cos t \\ \sin t \end{pmatrix} + c_2 \begin{pmatrix} -\sin t \\ \cos t \end{pmatrix}$) to polar coordinates $(r, \varphi)$, obtaining from the resulting linear system $dr/dt = \Delta_r/\Delta = 0, d\varphi/dt = \Delta_\varphi/\Delta = 1$ ($\Delta_r = 0, \Delta_\varphi = \Delta = r$). Here, $\Delta$ is the determinant of the plane rotation matrix, $R_\varphi = \begin{pmatrix} \cos \varphi & -\sin \varphi \\ \sin \varphi & \cos \varphi \end{pmatrix}$, times $r$. One can easily see that matrix $A = \begin{pmatrix} 0 & -1 \\ 1 & 0 \end{pmatrix}$ gives $e^{At} = \begin{pmatrix} \cos t & -\sin t \\ \sin t & \cos t \end{pmatrix}$. In the parlance of dynamical systems theory, the flow on a set $P$ (the phase space) is simply a rotation of the set so that it preserves the area in the phase space. If the system of units with inhomogeneous position and velocity scales ($a \neq b, \omega \neq 1$) is used, one gets equation $R_\varphi \begin{pmatrix} \dot{r} \\ \dot{\varphi} \end{pmatrix} = \begin{pmatrix} -a \sin\varphi \\ b \cos\varphi \end{pmatrix}$ which leads to the following dynamical system in parametric form $\dot{r} = (b-a)r \sin\varphi \cos\varphi$, $\dot{\varphi} = a \sin^2\varphi + b \cos^2\varphi$. Notice that parameter $t$ over which the differentiation is performed does not, in general, coincide with the angle (times a constant) of point $(x(t) = a \cos t, y(t) = b \sin t)$ with the $x$-axis, see, e.g., http://mathworld.wolfram.com/about/author.html and "Ellipse", http://mathworld.wolfram.com/Ellipse.html.

We can use the plane (2d) oscillator to demonstrate various possibilities to enhance and generalize the seemingly primitive oscillator model. Equations of motion for a two-dimensional harmonic oscillator i.e., a system with Lagrangian $L(\rho) = L(x^1, x^2) = \frac{1}{2}(m\dot{\boldsymbol{\rho}}^2 - \alpha\boldsymbol{\rho}^2)$, $\rho = (x^1, x^2)^T$, $\boldsymbol{\rho}^2 = (x^1)^2 + (x^2)^2$, $\dot{\boldsymbol{\rho}}^2 = (\dot{x}^1)^2 + (\dot{x}^2)^2$ can be written as $m\ddot{x}^1 = -\alpha x^1$, $m\ddot{x}^2 = -\alpha x^2$. This is the simple model of a mass particle constrained to the $(x^1, x^2)$ plane whose motion is determined by elastic forces. One can, however, generalize this model by assuming both the particle mass and the elastic constant to be of tensor character instead of scalars. Such a generalization is reduced to the introduction of additional parameters: the motion (Newtonian) equation becomes $m_{ik}\ddot{x}^k = -\alpha_{ik}x^k$, $k = 1,2$, $m_{ik} = m_{ki}$, $\alpha_{ik} = \alpha_{ki}$. Of course, this system can be multidimensional, $k = 1, \ldots, n$; $n = 2$ is taken just for illustrative purposes. We can put for notational simplicity

$$m_{ik} = \begin{pmatrix} m_{11} & m_{12} \\ m_{21} & m_{22} \end{pmatrix} \equiv m \begin{pmatrix} a & b \\ b & c \end{pmatrix}, \qquad \alpha_{ik} = \begin{pmatrix} \alpha_{11} & \alpha_{12} \\ \alpha_{21} & \alpha_{22} \end{pmatrix} \equiv \alpha \begin{pmatrix} a & b \\ b & c \end{pmatrix},$$

where $a, b, c$ are constant, then the Lagrangian of the 2d harmonic oscillator takes a more general form:

$$L(\boldsymbol{\rho}) = L(x^1, x^2) = \frac{1}{2}\left[m\left(a(\dot{x}^1)^2 + 2b\dot{x}^1\dot{x}^2 + c(\dot{x}^2)^2\right) - \alpha\left(a(x^1)^2 + 2bx^1x^2 + c(x^2)^2\right)\right].$$

We shall assume that $b^2 - ac \neq 0$ so that the system (mass) matrix is non-degenerate. The Newton-Lagrange equations are

$$m_{11}\ddot{x}^1 + m_{12}\ddot{x}^2 + \alpha_{11}x^1 + \alpha_{12}x^2 = 0, \qquad m_{21}\ddot{x}^1 + m_{22}\ddot{x}^2 + \alpha_{21}x^1 + \alpha_{22}x^2 = 0$$

or, in the $a, b, c$ notation,

$$m(a\ddot{x}^1 + b\ddot{x}^2) + \alpha(ax^1 + bx^2) = 0$$

$$m(b\ddot{x}^1 + c\ddot{x}^2) + \alpha(bx^1 + cx^2) = 0$$

or in the matrix form

$$m \begin{pmatrix} a & b \\ b & c \end{pmatrix} \begin{pmatrix} \ddot{x}^1 \\ \ddot{x}^2 \end{pmatrix} + \alpha \begin{pmatrix} a & b \\ b & c \end{pmatrix} \begin{pmatrix} x^1 \\ x^2 \end{pmatrix} = 0.$$

Notice that if we multiply this matrix equation on the left by inverse matrix $m_{ik}^{-1} = \begin{pmatrix} a & b \\ b & c \end{pmatrix}^{-1}$, which exists because we assumed the determinant of $m_{ik} = \begin{pmatrix} a & b \\ b & c \end{pmatrix}$ to be non-zero, we get the usual form of the harmonic oscillator equations, $\begin{pmatrix} \ddot{x}^1 \\ \ddot{x}^2 \end{pmatrix} + \omega_0^2 \begin{pmatrix} x^1 \\ x^2 \end{pmatrix} = 0$, where $\omega_0^2 \equiv \alpha/m$ is the eigenfrequency of the oscillator. Notice also that the above Lagrangian is not the only one to produce the required motion equations, one can find a subset of equivalent Lagrangians, for instance, $\bar{L} = \frac{1}{2}\left[m\left((\dot{x}^1)^2 - (\dot{x}^2)^2\right) - \alpha\left((x^1)^2 - (x^2)^2\right)\right]$ corresponding to $a = 1, b = 0, c = -1$ or $\tilde{L} = m\dot{x}^1\dot{x}^2 - \alpha x^1 x^2$ corresponding to $b = 1, a = c = 0$ (probably the simplest one). One can see that here the equivalent Lagrangians have the same $|\det m_{ik}| = 1$.

We can also note in passing that even such a trivial example as the harmonic oscillator naturally leads to a rather sophisticated concept of mathematical physics such as that of equivalent Lagrangians. Remarkably, this concept is not uniquely defined: one can require either that equivalent Lagrangians should result in the same evolution (in particular, motion) equations or in the same solution set, the latter requirement being stronger. In general, as we have seen, for a classical mechanical system having a finite number of degrees of freedom $n$, the motion equations are produced by applying the Euler-Lagrange variational derivative operator to a Lagrangian function i.e. $\frac{\delta L}{\delta q^i} = 0, i = 1, \ldots, n$, where $\frac{\delta}{\delta q^i} = \frac{d}{dt}\frac{\partial}{\partial \dot{q}^i} - \frac{\partial}{\partial q^i}$. One can see that if two Lagrangian functions $L$ and $\tilde{L}$ differ by a total time derivative i.e., $\tilde{L} = \beta L + \frac{df}{dt}, \beta = \text{const}$, then Lagrangians $L$ and $\tilde{L}$ lead to the same equations of motion since condition $\delta S = 0$ coincides with $\delta\tilde{S} = 0$ ($S = \int_{t_1}^{t_2} L dt$ is the action for a mechanical system), see [60].

A possible Lagrangian for a 3d dissipative oscillator system

$$m\dot{\mathbf{v}} + \gamma\mathbf{v} + \nabla V(\mathbf{x}) = 0,$$

where $\mathbf{v} = \dot{\mathbf{x}}$ and damping coefficient $\gamma \geq 0$, may be represented as

$$L = e^{\gamma t/m}\left(\frac{1}{2}m\mathbf{v}^2 - V(\mathbf{x})\right).$$

Here $\nabla V(\mathbf{x}) \equiv \mathbf{F}(\mathbf{x})$ is the vector field of conservative forces driving the oscillator.

A nearly obvious but important generalization of the damped oscillator model is

$$m\ddot{q} + m\int_0^t D(t - \tau)\dot{q}(\tau)d\tau + \frac{\partial V(q)}{\partial q} = 0, \tag{9.2.4.}$$

where the damping kernel $D(t)$ accounts for time-delayed dissipative influence of the environment (frictional memory effects) and $V(q)$ is a potential (often having a "metastable" barrier $V_0$). If the oscillator is also driven by an external force $F(q, t)$, this force would enter the right-hand side of equation (9.2.4.) instead of 0. When the memory effects can be neglected i.e., if $D(t) = \gamma\delta(t)$, then the friction term is reduced to the usual Ohmic dissipation $m\gamma\dot{q}(t)$. A representation nonlocal in time (9.2.4.) is used, in particular, in macroscopic quantum mechanics, more specifically, in models describing the dissipative quantum tunneling. As shown in the series of classical papers by Caldeira and Leggett ([37]), linear equation (9.2.4.) can be obtained from the following Hamiltonian

$$H = \frac{p^2(q)}{2m} + V(q) + \sum_{i=1}^{N}\left[\frac{p_i^2}{2m_i} + \frac{m_i}{2}\left(\omega_i x_i + \frac{c_i}{m_i\omega_i}q\right)^2\right]$$

describing the coupling of the model thermal bath consisting of $N$ oscillators to the dissipative system in question, provided the limit

$$\lim_{N \to +\infty} \sum_{i=1}^{N} \frac{c_i^2}{m_i \omega_i^2} \cos \omega_i t$$

can be identified with $D(t)$ i.e., the above sum strives to a Fourier cosine series of $D(t)$. Here $(x_i, p_i)$ are the coordinate and momentum of the i-th bath oscillator, coefficients $c_i$ couple the bath oscillators with the system (of mass $m$) in question, $m_i$ are masses of the hypothetical bath oscillators. We may note here that the idea that dissipation can be modeled by coupling to a bath of oscillators had been expressed by R. P. Feynman and F. Vernon [60] twenty years before A. O. Caldeira and A. J. Leggett succeeded in revitalizing it [37].

One might be tempted to think that the classical model of harmonic oscillations is just a toy archetype having nothing to do with contemporary science or engineering. This is far from the truth. In engineering, harmonic oscillations are used to model buildings (specifically in seismic regions), tower cranes and other heavy lifting equipment, automotive design, etc. In quantum mechanics, which has become today a practically viable engineering discipline, the model of oscillations, specifically when treated in the form of a dynamical system, is convenient to describe the transfer of excitation between quantum objects. To understand the idea of using the oscillator model in quantum mechanics one can write the 1d Schrödinger equation in the form

$$\psi'' + k^2(x)\psi = 0, \qquad k^2(x) = \frac{2m}{\hbar^2}(E - V(x)).$$

Introducing vector $\Psi = \begin{pmatrix} \psi_1 \\ \psi_2 \end{pmatrix}, \Psi \in \mathbb{R}^2$, where $\psi_1 = \psi, \psi_2 = \psi_1' = \psi'$, we get the dynamical system

$$\Psi' = A(x)\Psi, \qquad A(x) = \begin{pmatrix} 0 & 1 \\ -k^2(x) & 0 \end{pmatrix}, \qquad A(x) \colon \mathbb{R}^2 \to \mathbb{R}^2$$

or in components $\psi_1' = \psi_2, \psi_2' = -k^2(x)\psi_1$. This form of Schrödinger equation proves to be convenient for a number of problems, especially in condensed matter and nanoscience.

One of the most important practical applications of modeling in terms of harmonic oscillators is the calculation of electromagnetic response of materials, in particular, semiconductors. For example, the dielectric function determining the electromagnetic response of semiconductor structures such as GaAs/Al$_x$Ga$_{1-x}$As can be quite accurately modeled by a series of harmonic oscillators (7-9 oscillators can be sufficient to obtain a fair fit in the near infrared domain, see, e.g., Terry, F. L. A modified harmonic oscillator approximation scheme for the dielectric constants of Al$_x$Ga$_{1-x}$As. J. Appl. Phys. v.70(1), 409-417 (1991) and the references therein). The physical reason for this modeling is that the local polarization of matter responds to an electron field as a harmonic oscillator. Recall that GaAs-AlGaAs and similar systems (e.g., InGaAsP) are used to produce numerous semiconductor devices such as lasers, light-emitting diodes (LED), solar cells and integrated circuits (IC). Modeling the electromagnetic response of semiconducting systems with the help of harmonic oscillators enables us to optimize technological parameters such as composition, layer thickness, etc. for the devices based on GaAs and analogous structures.

## 9.3. Normal modes

Motion of a linearized Lagrangian system near a stable equilibrium is described by Lagrangians $L(\mathbf{q}, \dot{\mathbf{q}}, t)$, where for a "natural" system $L = T - V$. In the case of small oscillations kinetic $T$ and potential $V$ energies are represented as quadratic forms $T = \frac{1}{2}(\dot{\mathbf{q}}, M\dot{\mathbf{q}}) = \frac{1}{2}M_{ij}\dot{q}^i\dot{q}^j$ and $V = \frac{1}{2}(\mathbf{q}, K\mathbf{q}) = \frac{1}{2}K_{ij}q^iq^j, i, j = 1, \ldots, n$ i.e. $n \times n$ matrices $M_{ij}$ and $K_{ij}$ are real and symmetric. Form $T$ is assumed positive-definite. Using the standard techniques of linear algebra one can simultaneously diagonalize this pair of quadratic forms (to bring them to the "principal axes" – the problem of the pair of forms). Geometrically, simultaneous diagonalization is achieved by a common linear change of coordinates $\mathbf{z} = S\mathbf{q}$, where $S$ is an orthogonal transformation. Coordinates $\mathbf{z}$ can be so chosen as to reduce form $T$ to a scalar quadrate $T = \frac{1}{2}(\mathbf{z}, \mathbf{z}) = \frac{1}{2}z_iz^i$. In these coordinates $V = \frac{1}{2}\sum_{i=1}^{n}\lambda_iz_i^2$. The advantage of simultaneous diagonalization of the matrices corresponding to kinetic and potential energies is enormous: we can solve $n$ independent differential equations separately rather than a system of intertwined ODEs with $n$ variables. In other words, a physical system in which small oscillations are excited (e.g., a molecule or a crystal) can be reduced by an appropriate choice of coordinates to a direct product of independent one-dimensional oscillators.

Recall that diagonalization of a pair of $n \times n$ matrices $M$ and $K$ consists in the following: one must find the solution to a generalized eigenvalue problem i.e., eigenpairs $\lambda_k, q_k$ which are the solutions to matrix equation $Kq = \lambda Mq$. If $M$ is nonsingular, then the solutions are given by the eigenpairs of $C \equiv M^{-1}K, Cq = \lambda q$. Eigenvalues $\lambda_k$ satisfy the characteristic equation $\det|K - \lambda M| = 0$. Since in the mechanical problem matrices $M > 0$ and $K$ are symmetric all $\lambda_k$ are real. Assume that there exists a matrix $S$ such that $S^TMS = \Lambda$ and $S^TKS = I$, where $\Lambda$ is a diagonal matrix, $I$ is the unit matrix, then one can say that $M$ and $K$ are simultaneously diagonalized, with diagonal matrix elements of $\Lambda$ being the generalized eigenvalues. One can prove (see, e.g., [65]) that the matrix $S$ exists if $K$ is symmetric ($K = K^T$) and $M$ is positive-definite. Eigenvalues $\lambda_k$ obtained from the generalized characteristic equation $\det|K - \lambda M| = 0$ define eigenfrequencies $\omega_k^2$, and the corresponding solutions (eigenstates) are often called in physics *normal modes*: such solutions describe the harmonic motion along each particular coordinate $z^i$ in a special coordinate system. In new coordinates $z^i$ the linearized dynamical (Lagrangian) system with $n$-dimensional coordinate space ($n$ degrees of freedom) splits into $n$ independent single-dimensional oscillations $\ddot{z}^i + \lambda z^i = 0$. Normal frequencies $\omega_k$ of a system are those at which the system such as an engineering structure prefers to vibrate. Coordinates $z^i, i = 1, \ldots, n$, form an orthogonal system in the metric defined by tensor $M_{ij}$. Thus, the total oscillation in a linear physical system can be represented as a sum of independent normal modes or, in other words, can be expanded over the normal modes. In other words, each normal mode is a superposition of elementary vibrations of individual oscillators, particles or atoms. This fact has been of fundamental importance in 20[th] century physics mostly based on the concept of vector (Hilbert) space and the superposition principle. In particular, the concepts of photon, phonon and, in general, of elementary excitation (the latter being probably the most essential in 20[th] century physics) emerged from an analogy with normal modes in a linearized Lagrangian system. Notice that the superposition of normal modes (harmonic oscillations) is not necessarily a periodic function. Notice also that the initial state under very general conditions can be represented as a superposition of normal modes (the Fourier series) (Figure 12).

In practical terms, one can find normal modes according to the following prescriptions: at first look for solutions of the form $\mathbf{q}(t) = \mathbf{A}e^{-i\omega t}$, where $\mathbf{A}$ are amplitudes[80], then inserting this ansatz into the Lagrange equations $d/dt(\partial L/\partial \dot{\mathbf{q}}) = \partial L/\partial \mathbf{q}$, $L = T - V = \frac{1}{2}(\dot{\mathbf{q}}, M\dot{\mathbf{q}}) - \frac{1}{2}(\mathbf{q}, K\mathbf{q})$ we get the motion equations $d/dt(M\mathbf{q}) = K\mathbf{q}$ or $(K - \omega^2 M)\mathbf{A} = 0$. This is a system of linear equations with respect to vector amplitudes $\mathbf{A}$. From the characteristic equation $\det|K - \omega^2 M| = 0$, we find $n$ eigenfrequencies $\lambda_k = \omega_k^2$ defining orthogonal eigenvectors $\mathbf{A}_k$. The general solution is given by a superposition over normal modes $\mathbf{q}(t) = \sum_{k=1}^{n} C_k \mathbf{A}_k e^{-i\omega_k t}$. To emphasize the requirement that $\mathbf{q}(t)$ should be real, one can take $\mathrm{Re}\,\mathbf{q}(t)$ or use trigonometric functions $\sin \omega_k t$ and $\cos \omega_k t$: $\mathbf{q}(t) = \sum_{k=1}^{n} C_k \left( \mathbf{A}_k \cos \omega_k t + \mathbf{B}_k \sin \omega_k t \right)$.

Explicit application of this procedure easily demonstrates the orthogonality property of vector amplitudes $\mathbf{A}_k$ corresponding to different eigenfrequencies $\omega_k$ in the metric defined by $M_{ij}$ or $K_{ij}$ i.e., $(\mathbf{A}_k, M\mathbf{A}_l) = (\mathbf{A}_k, K\mathbf{A}_l) = 0$. Indeed, we have the following equations for the amplitude components: $-\omega_k^2 M_{ij} A_k^j + K_{ij} A_k^j = 0$; $-\omega_l^2 M_{ij} A_l^j + K_{ij} A_l^j = 0$, where $\mathbf{A}_k = (A_k^1, ..., A_k^n)^T$, $\mathbf{A}_l = (A_l^1, ..., A_l^n)^T$ are vector amplitudes corresponding to eigenvalues $\lambda_k = \omega_k^2$ and $\lambda_l = \omega_l^2$, respectively, and there is no summation over eigenvalue indices $k$ and $l$. Multiplying the first equation by $A_l^i$, the second by $A_k^i$ and subtracting, we have, due to matrix symmetry $M_{ij} = M_{ji}$, $K_{ij} = K_{ji}$, $(\omega_l^2 - \omega_k^2) M_{ij} A_l^i A_k^j = 0$ i.e. for $\lambda_l \neq \lambda_k$ $M_{ij} A_l^i A_k^j = K_{ij} A_l^i A_k^j = 0$ or $(\mathbf{A}_k, M\mathbf{A}_l) = (\mathbf{A}_k, K\mathbf{A}_l) = 0$. Thus, vector amplitudes for nondegenerate eigenvalues ($\lambda_l \neq \lambda_k$) are mutually orthogonal in the $M_{ij}$ and $K_{ij}$ metric (i.e., when the inner product is defined through metric tensors $g_{ij} \equiv M_{ij}$ and $g_{ij} \equiv K_{ij}$. If eigenvalues (eigenfrequencies) $\lambda_k = \omega_k^2$ are degenerate ($\omega_k^2 = \omega_l^2$), one can still choose such amplitudes $\mathbf{B}_k$ which are linear combinations of amplitudes $\mathbf{A}_k$, $\mathbf{B}_k = \sum_{s=1}^{\nu} \alpha_{k,s} \mathbf{A}_k$, where $\nu$ is the multiplicity of $\lambda_k$ that quadratic forms $T$ and $V$ would be reduced to a diagonal representation and the solution would look as $\mathbf{q}(t) = \mathrm{Re}\sum_{k=1}^{n} C_k \mathbf{A}_k e^{-i\omega_k t}$. The difficulty in the case of degenerate frequencies is mainly notational: one needs to be careful with the forest of indices. An important thing is that in this case there are no secular (polynomial in time $t$) terms as in resonance, despite the fact that frequencies coincide ($\omega_k^2 = \omega_l^2$).

Notice that the situation with degenerate frequencies in mechanical oscillation theory is very close to that of degenerate energy levels in quantum perturbation theory: in both theories, the multiplicity of eigenvalues is caused by some additional symmetry to be fully understood by the means of group theory. Notice also that the problem of linear oscillations in the systems with $n$ degrees of freedom as much as computing corrections in the quantum perturbation theory is more the one of linear algebra rather than of physics.

Nevertheless, transition to normal modes[81] has a deep physical meaning. Introduction of normal coordinates, which is a purely algebraic transformation, in the oscillation theory has become a very important method in 20th century physics. This is an example of "the unreasonable effectiveness of mathematics in the natural sciences" [159]. The physical meaning of normal modes is collectively excited motions in the medium. Quasiparticles, elementary excitations, phonons, photons, excitons, polarons, plasmons, etc. are the main concepts spawning modern physical models. The concepts of quasiparticles, elementary excitations and collective modes pervade the whole of modern physics, and without these concepts the description of macroscopic properties of the matter would hardly be possible. For example, photons are elementary excitations of the ubiquitous electromagnetic field. Recall also that quantization of an ensemble of independent oscillators led to the development of modern quantum theory, both in its mechanical and statistical parts. In classical physics, normal modes are wavelike collective excitations whereas in quantum theories normal modes have particle-like features. Moreover, the study of collective motions and, specifically, of their instabilities has eventually resulted in the advanced theories of evolution, phase transitions, cooperative behavior, bifurcations, and self-organization. We have seen that applications of these theories and mathematical models are much wider than physics, present also in biology, economics (econophysics), financial modeling, structure formation, social dynamics, etc.

---

[80] More generally, $\mathbf{q}(t) = A_+ e^{i\omega t} + A_- e^{-i\omega t}$. Notice that the same approach is used to explore instabilities in oscillating physical systems, e.g., a very important case of linear plasma instabilities.

[81] Sometimes also known as "natural modes".

## 9.4. Classical chain of oscillators

Almost every traditional course of solid-state physics begins with a discussion of the simplest model of oscillations in the crystal lattice. It is customary to use in such a discussion the toy model (first considered probably by Max Born and Theodore von Kármán) of a one-dimensional chain with elastically interconnected masses $m_i$ (atoms) that can be displaced from their equilibrium positions $X_i$. For simplicity, we shall consider all the masses and elastic constants $k_{ij}$ determining the forces between atoms equal, $m_i = m$, $k_{ij} = k$. The one-dimensional chain of pointlike masses is a toy model of the crystal. The Hamiltonian function for such a chain can be written as

$$H = \frac{1}{2m} \sum_j p_j^2 + \frac{K}{2} \sum_j (u_j - u_{j+1})^2 \qquad (9.4.1.)$$

This form of the Hamiltonian assumes that each atom in the chain (1d lattice) interacts only with the nearest neighbors. The 1d model can be without much effort generalized to 3d; the main difficulties here are notational. The 3d model leads to the quantum theory of lattice vibrations (phonons)[82]. From (9.4.1.), we can obtain the classical motion equations (e.g., by writing the potential energy term as $\frac{K}{4}[\sum_j (u_{j+1} - u_j)^2 + \sum_j (u_j - u_{j-1})^2]$ ) as differential-difference equations

$$m\ddot{u}_j = -\frac{\partial H}{\partial u_j} = K(u_{j+1} - u_j) + K(u_{j-1} - u_j), \qquad j = 1,2,\dots. \qquad (9.4.2.)$$

One can look for solutions to (9.4.2.) of the form $u_j(t,k) = u_0 \exp(ik \cdot ja - i\omega t)$, where $a$ is the lattice period. Inserting this expression into (9.4.2.), we get instead of the system of differential-difference equations a single algebraic equation

$$-m\omega^2 = K[(e^{ika} - 1) + (e^{-ika} - 1)] = -4K \sin^2 \frac{ka}{2}.$$

We have seen that, in the wave motion, algebraic equations that interrelate frequency and propagation constant (wave number or wave vector) are known as the dispersion law $\omega(k)$ (one should not confuse $K$ which is the strain (elastic force) constant and $k$ which characterizes the lattice displacements (labels representations of the translation group). In the case of 1d monoatomic chain, we have $\omega(k) = 2\sqrt{K/m}\left|\sin\frac{ka}{2}\right|$ (we ignore negative frequencies). Frequencies of normal modes (collective excitations) change periodically from zero (at $k = 2\pi n/a, n = 0, \pm 1, ...$) to $2(K/m)^{1/2}$ (at $k = k_{max} = \pi n/a, n = \pm 1, ...$). Here $(K/m)^{1/2} = \omega_0$, the eigenfrequency of a single harmonic oscillator.

---

[82] The notion of a phonon was introduced in 1930s by I. Tamm, an outstanding Russian physicist.

Thus, the dependence $\omega(k)$ is periodic with period $G = 2\pi/a$, $G$ being the period of the dual (or reciprocal) lattice. The meaning of this periodicity is that the displacements of all the atoms in the chain remain intact when we replace wave number $k$ by $k' = k + nG, n = 0, \pm 1, \pm 2, ...$ Indeed, $u_j'(t,k) \equiv u_j(t, k + nG) = u_0 \exp[i(k + nG)ja - i\omega t] = u_0 \exp[ik \cdot ja + in2\pi j - i\omega t] = u_j(t,k)$ so that the physically observable quantities $u_j(t)$ do not change under the $k \to k' = k + nG$ transformation. Therefore, one can limit oneself to only considering wave numbers $k \in (-\pi/a, \pi/a]$ i.e., the set of points in $k$-space that completely characterize the solutions to (9.4.2.). In 2d and 3d models of crystalline solids, the interval $(-\pi/a, \pi/a]$ becomes the first Brillouin zone which is especially important when one considers the electronic subsystem (energy bands) of the solid state. Symmetry points corresponding to $k_{max} = -\pi/a, \pi/a$ (often denoted as $\Gamma$) define the zone boundaries, notice that the zone boundary representations are equivalent.

Recall that solid state physics, now regarded as a part of condensed matter physics, has brought a large number of engineering achievements, the most noticeable of them being transistor and microprocessor. Without such devices, the modern technological environment, which is today taken for granted, would not have existed. Moreover, solid state physics is indispensable for the development of energy conversion devices such as solar cells, based on the study of optical processes in condensed matter.

Notice that the wavelike modes $u_j(t,k) = u_0 \exp(ik \cdot ja - i\omega t)$ are just particular solutions. To obtain the general solution, one has to make a superposition of these elementary solutions and to use the boundary conditions. If the chain of $N$ atoms is finite, the boundary conditions are dictated by fixing the positions of the first and the last atom in the chain, e.g., to zero: $u_{j=0} = u_{j=N} = 0$. Such conditions lead to a superposition of the solutions propagating in the opposite directions i.e., corresponding to wave numbers $\pm k$. Boundary condition $u_{j=0} = 0$ produces standing wave $u_j(t,k) = u_0 \sin(kja)\exp(-i\omega(k)t, j = 0, \pm 1, ...$ In principle, conditions $u_{j=0} = 0$ and $u_{j=N} = 0$ are equivalent and assume symmetric solutions under $k \to -k$ since it does not matter from what end to start counting the atoms. Condition $u_{j=N} = 0$ gives $kNa = \pi n, n \in \mathbb{Z}$, therefore $k = \pi n/Na$. The number of normal modes within the first Brillouin zone in the chain with fixed ends is $k/\Delta k = \frac{\pi/a}{\pi/Na} = N$. However, one should throw out the normal modes corresponding to $k = 0$ and $k = \pi/a$ since they produce zero solutions ($u_j = 0$). Thus, in the chain with fixed ends, the number of normal modes is reduced to $N - 1$, and since the number of degrees of freedom (the number of independent displacements $u_j$) is also $N - 1$, we have one normal mode for a degree of freedom. This coincidence is an indication of the completeness of the system of normal modes, i.e., any solution can be represented as their superposition.

Other boundary conditions naturally select different solution types. For instance, very popular in solid state theory, periodic (cyclic) boundary conditions, $u_{j=0}(t) = u_{j=N}(t)$ or, more generally, $u_{j=0}(t) = u_{j=N \cdot l}(t)$, where $l \in \mathbb{Z}$ defines the order of proper rotation, produce traveling wave solutions, $u_j(t,k) = u_0 \exp(ik \cdot ja - i\omega(k)t)$. The cyclic boundary conditions imply $ikN \cdot la = 2\pi n$ i.e., wave vector $k = 2\pi n/(N \cdot la), n, l, N \in \mathbb{Z}$. Integers $l$ and $n$ label representations of the cyclic group; in physics one usually restricts oneself to one-fold rotations ($l = 1$).

Here, one should make the following important note. The observed fact that the number of normal modes, which are the wavelike elementary excitations in a multiatomic chain, coincides with the number of degrees of freedom is more an exception rather than a strict rule. The quasiparticle description of excited states emerging in a system that contains many interacting bodies does not necessarily possess the same number freedoms as in the exact description. Often, excited many-particle states can be characterized by fewer numbers of parameters as would be required by considering the respective dynamical system. It is exactly this reduction of variables necessary for description that makes the representation of a many-body system in terms of elementary excitations economic and convenient. However, the cost for this convenience is an approximate picture since an elementary excitation is, in general, not a stationary state (eigenstate) of the system, but only a superposition of eigenstates with narrow energy dispersion. Such a superposition, when modeling a particle through elementary excitation (quasiparticle), tends to spread in complete analogy to the spreading of a wave packet that models a particle in elementary quantum mechanics. A wave packet is a propagating localized disturbance that is constructed as a Fourier superposition of many (in principle, infinitely many) wavelike modes, each of them traveling with its own velocity. Spreading of a superposition of eigenmodes eventually results in the smoothing out of disturbance i.e., the dissipation of elementary excitations. Notice that just as in the case of a wave packet representing a particle or a pulse, the construction of a quasiparticle as a concentrated object from wavelike modes only makes sense when certain conditions are fulfilled. In particular, for a wave packet the frequency band width $\Delta\omega$ (energy band) or wave vector region $|\Delta\mathbf{k}|$ in the dual to configuration space ($\mathbf{k}$-space) must be small compared with, respectively, the carrier frequency $\omega$ (energy) or wave vector $\mathbf{k}$ (momentum).

Likewise, the description of many-particle states in terms of elementary excitations is reasonable if the energy width of the superposition of eigenstates $\Delta\mathcal{E}$ is small compared with the energy of excitation $\mathcal{E}(\mathbf{p})$: otherwise, the dissipation time (lifetime) of an elementary excitation becomes comparable with the period $\hbar/\mathcal{E}$ characterizing the main state with energy $\mathcal{E}$. In the language of oscillations, it means that the amplitude decay ($a^{-1}$, see formula (9.4.1.)) is of the order of the oscillation period $T = 2\pi/\omega$. Technically, the assumption that a many-body system can sustain elementary excitation leads to the system's description in terms of Green's functions which are similar to Green's functions for the damped linear oscillator (9.4.1.), with energy and dissipation of elementary excitations being determined respectively by the real and imaginary parts of the pole of Green's function.

It is instructive to consider certain limiting cases, for example, large wavelengths, $ka/2 \ll 1$ i.e., $\lambda \gg \pi a$. In this case, the dispersion law becomes linear in $k$, $\omega(k) = \sqrt{K/m}\, ka = \omega_0 ak$ which is typical of acoustic waves. Therefore, one can conclude that in the long wave case the collective oscillator mode corresponds to acoustic excitation, with $s \equiv (K/m)^{1/2}a = \omega_0 a$ being the classical sound wave velocity. However, when the wavelength of collective excitation becomes comparable with the lattice period, $\lambda \sim a$, we cannot replace $\sin(ka/2)$ by its argument and the dispersion law differs significantly from the acoustic one. While discussing wave motion, we have seen that for the linear dispersion law (in particular, for sound waves) phase and group velocities coincide. Here, when the excitation wavelength falls and eventually becomes of the order of lattice spacing $a$, group velocity $v_g = \omega_0 a \cos(ka/2)$ is lower than phase velocity $v_{ph} = \omega_0 a \sin(ka/2)/(ka/2)$ (since for $0 \le x < \pi/2$, $\sin x/x \ge \cos x$). At $k = k_{max} = \pi(2n+1)/a, n = 0, \pm 1, \ldots$ i.e., for the minimal within the first Brillouin zone wavelength $\lambda = 2a$, when the excited normal mode has nodes at each lattice center (atom), group velocity vanishes, $v_g = 0$. Physically, it means that excitation does not propagate through the chain of oscillators, and we have a standing wave instead of a traveling one. In solid state theory, this phenomenon corresponds to the so-called Bragg reflection, when the entire crystal works as a coherent scatterer of waves.

Let us now assume that that we have a chain containing two different atoms within each periodic link of size $a$. Such a chain is a toy model of a crystal having several atoms in an elementary cell. Let the displacements of atoms of sorts 1 and 2 from their equilibrium positions be respectively $u$ and $v$, their masses $m_1$ and $m_2$ and elastic constants $K_1$ and $K_2$. In general, when there are $s$ atoms in an elementary cell, position of the $j$-th atom, $j = 1, \ldots, s$, is defined by vector $\mathbf{r_{ns}} = \mathbf{r_n} + \boldsymbol{\rho_{nj}}$, where $\mathbf{r_n}$ labels an array of points defining the equilibrium structure of the whole crystal (the Bravais lattice), $\mathbf{r_n} = n_1\mathbf{a_1} + n_2\mathbf{a_2} + n_3\mathbf{a_3}, n_i \in \mathbb{Z}$ is the position of the $n$-th elementary cell whereas $\boldsymbol{\rho_{nj}}$ defines the basis inside this cell. The motion equations for a chain with two atoms in a 1d "elementary cell" (masses $m_1$ and $m_2$ in the link $a$) are

$$m_1\ddot{u}_j = K_1(v_{j-1} - u_j) + K_2(v_j - u_j), \qquad m_2\ddot{v}_j = K_1(u_{j+1} - v_j) + K_2(u_j - v_j). \qquad (9.4.3.)$$

The standard method of looking for solutions to this differential-difference equation again consists in inserting ansatz $(u_j, v_j)^T = (u_0, v_0)^T \exp(ik \cdot ja - i\omega(k)t)$ into (9.4.3.) which produces the system of homogeneous algebraic equations for amplitudes $(u_0, v_0)$:

$$[-m_1\omega^2(k) + (K_1 + K_2)]u_0 - (K_1 e^{-ika} + K_2)v_0 = 0,$$

$$-(K_1 e^{ika} + K_2)u_0 + [-m_2\omega^2(k) + (K_1 + K_2)]v_0 = 0.$$

The det $= 0$ requirement gives biquadratic equation

$$\omega^4(k) - (K_1 + K_2)\left(\frac{1}{m_1} + \frac{1}{m_2}\right)\omega^2(k) + \frac{2K_1 K_2}{m_1 m_2}(1 - \cos ka) = \omega^4 - \frac{K}{\mu}\omega^2 + \frac{4K_1 K_2}{m_1 m_2}\sin^2\frac{ka}{2} = 0,$$

where $K \equiv K_1 + K_2, \mu = \frac{m_1 m_2}{m_1 + m_2}$ (a reduced mass). The solution is the dispersion law

$$\omega_{\pm}^2(k) = \frac{K}{2\mu} \pm \sqrt{\frac{K^2}{4\mu^2} - \frac{4K_1 K_2}{m_1 m_2}\sin^2\frac{ka}{2}} \qquad (9.4.4.)$$

which has in this case two branches, lower – acoustic (-) – and upper – optical (+). The acoustic branch is characterized by $\omega_-(k) \to 0$ linearly with $k \to 0$ i.e., $\omega_-(k) = sk$ for small $k$ (long waves), where $s$ is the sound velocity. For $ka/2 \ll 1$, we may replace $\sin$ by its argument so that

$$\omega_{\pm}^2(k) \approx \frac{K}{2\mu}\left[1 \pm \left(1 - \frac{4K_1 K_2 m_1 m_2}{(K_1 + K_2)^2(m_1 + m_2)^2}(ka)^2\right)^{1/2}\right].$$

Since factor $B(m_1, m_2, K_1, K_2) \equiv \frac{4K_1 K_2 m_1 m_2}{(K_1 + K_2)^2(m_1 + m_2)^2} \delta 1$, we can write

$$\omega_{\pm}^2(k) \approx \frac{K}{2\mu}\left[1 \pm \left(1 - \frac{2K_1 K_2 m_1 m_2}{(K_1 + K_2)^2(m_1 + m_2)^2}(ka)^2\right)\right]$$

and the dispersion law for the acoustic mode $\omega_-^2(k) = \frac{K_1 K_2}{(K_1 + K_2)(m_1 + m_2)}(ka)^2$. The sound velocity is $s = \lim_{k \to 0}(\omega_-(k)/k) = a\left[\frac{K_1 K_2}{(K_1 + K_2)(m_1 + m_2)}\right]^{1/2}$. In a particular case of identical elastic constants for both atoms in a cell, $K_1 = K_2 \equiv K_0$, we have $s = a[K_0/2(m_1 + m_2)]^{1/2}$. Notice that the sound velocity in the chain with two sorts of atoms in a cell coincides with the one in the monoatomic chain if we put $m_1 = m_2$ and $a \to a/2$.

The upper branch $\omega_+(k)$ corresponding to the so-called optical mode does not go through zero: it starts at $k = 0$ from frequency $\omega_+(0) = (K/\mu)^{1/2}$ and falls down when wave number $k$ runs through the first Brillouin zone to its edges, $k = \pm\pi/a$. The term "optical" has been universally adopted due to the fact that oscillations corresponding to the upper mode effectively couple with optical (mainly infrared) radiation. Notice that the expression for optical frequency $\omega_+(0)$ is very similar to frequency $\omega_0$ for the monoatomic chain, with the difference that actual values for the mass and the elastic constant are replaced by some effective ones. This is not a mere coincidence since optical oscillations at $k = 0$ are just coherent motions of all atomic pairs within the cells, with accompanying strains, forces and accelerations. In real life, optical oscillations often correspond to periodic displacements of sublattices with respect to each other, e.g., in ionic crystals such as alkali halides (sodium chloride NaCl and other important compounds). Since each sublattice in such compounds contains ions carrying a definite charge such as $Na^+$ and $Cl^-$ in sodium chloride, macroscopic (long wave) oscillations of sublattices with respect to each other produce a large oscillating dipole moment, which leads to intensive coupling with electromagnetic radiation.

From the system (9.4.4.) of linearly dependent homogeneous algebraic equations for amplitudes, we can find their ratio in the long wavelength limit ($ka \to 0$):

$$\frac{u_0}{v_0} = \frac{K_1 e^{-ika} + K_2}{K_1 + K_2 - m_1 \omega^2(k)}. \tag{9.4.5.}$$

For the acoustic mode, the ratio of amplitudes tends to unity whereas for the optical mode when $\omega_+(k \to 0) \to (K/\mu)^{1/2}, u_0/v_0 \to -m_2/m_1$ which expresses momentum conservation within the elementary cell. If $m_2 \equiv m \ll m_1 \equiv M$, then $u_0/v_0 \approx -m/M \ll 1$: the heavier atom practically does not move acting as a steady wall between the cells.

At $k = \pm\pi/a$, we get from (9.4.5.) the minimum value of the optical mode frequencies

$$\min \omega_+^2(k) = \omega_+^2(k = \pm\pi/a) = \frac{K}{2\mu} + \sqrt{\frac{K^2}{4\mu^2} - \frac{4K_1 K_2}{m_1 m_2}}$$

and the maximum value of the acoustic mode

$$\max \omega_-^2(k) = \omega_-^2(k = \pm\pi/a) = \frac{K}{2\mu} - \sqrt{\frac{K^2}{4\mu^2} - \frac{4K_1 K_2}{m_1 m_2}}.$$

The difference $\Delta\omega = \omega_+(k = \pm\pi/a) - \omega_+(k = \pm\pi/a)$ gives the frequency gap between two normal modes, optical and acoustic. For $K_1 = K_2 \equiv K_0$ and $m_1 \equiv M, m_2 \equiv m \ll M$ i.e., when atoms of one sort are much heavier than atoms of the other sort, we have $\mu \approx m$ and $\min \omega_+^2(k) = 2K_0/m, \max \omega_-^2(k) = 2K_0/M$ so that $\Delta\omega = (2K_0)^{1/2}\left[(m)^{-1/2} - (M)^{-1/2}\right]$. Note that in this case $(m \ll M)$ dispersion of the optical mode is almost negligible i.e., the mode frequency $\omega_+(k)$ very weakly depends on wave number $k$. Indeed, for $K_1 = K_2 \equiv K_0$ and $m \ll M$ we have from (9.4.4.)

$$\omega_+^2(k) = \frac{K}{2\mu} + \sqrt{\frac{K^2}{4\mu^2} - \frac{4K_1 K_2}{m_1 m_2} \sin^2 \frac{ka}{2}} = \frac{K_0}{m}\left(1 + \sqrt{1 - \frac{4m}{M}\sin^2 \frac{ka}{2}}\right) \approx \frac{2K_0}{m}\left(1 - \frac{m}{M}\sin^2 \frac{ka}{2}\right)$$

so that $\omega_+(k) = \sqrt{\frac{2K_0}{m}}\left(1 - \frac{m}{2M}\sin^2 \frac{ka}{2} + \mathscr{O}\left(\frac{m}{M}\right)\right)$, an almost straight line parallel to the $k$-axis. Physically, this is quite understandable: light atoms can oscillate in their cells independent of one another, as in separate rooms, since oscillations of one atom cannot affect those of a neighboring atom through an almost steady wall (heavy atom).

## 9.5. Creation and annihilation

Transition to normal modes allowed the "founding fathers" of quantum mechanics, P. A. M. Dirac, W. Heisenberg, P. Jordan, H. A. Kramers, W. Pauli and other masters of modern physics, to construct quantum mechanics and especially quantum electrodynamics, a consistent theory of interaction of light with matter. The idea was to express the Hamiltonian function as the sum of non-interacting terms corresponding to independent harmonic oscillators. For example, in the classical field theory $H = \sum_{\mathbf{k},\lambda} H_{\mathbf{k},\lambda}$, $H_{\mathbf{k},\lambda} = \frac{1}{2}\left(p_{\mathbf{k},\lambda}^2 + \omega_{\mathbf{k},\lambda}^2 q_{\mathbf{k},\lambda}^2\right)$, where indices $\mathbf{k}, \lambda$ enumerate normal modes: wave vector $\mathbf{k}$ corresponds to spatial translations whereas parameter $\lambda$ enumerates transversal spatial modes, e.g., polarization. Since each term in this sum has the form of a one-dimensional oscillator, one usually says that the field is decomposed into oscillators. Recall that the classical motion equations of a 1d oscillator are

$$\dot{q} = \frac{\partial H}{\partial p} = p, \qquad \dot{p} = -\frac{\partial H}{\partial q} = -\omega^2 q.$$

If we now introduce complex amplitudes $a = \frac{(\omega q + ip)}{\sqrt{2\omega}}$, $a^* = \frac{(\omega q - ip)}{\sqrt{2\omega}}$ which are just linear combinations of $p$ and $q$, $p = \frac{i\omega}{\sqrt{2\omega}}(a^* + a)$, $q = \frac{1}{\sqrt{2\omega}}(a^* - a)$, $H = \omega a^* a$, then multiplying the first of the motion equations by $\pm\omega$, the second by $i$ and combining them we get $\frac{da^*}{dt} = -i\omega a^*$, $\frac{da}{dt} = i\omega a$ i.e., complex amplitudes $a^*, a$ are evolving independently. One can interpret these amplitudes as complex vectors having a fixed length but rotating in the opposite directions, $a^*(t) = a^*(0)e^{-i\omega t}$, $a(t) = a(0)e^{i\omega t}$. Thus, complex amplitudes $a^*(t)$ and $a(t)$ give an irreducible representation of an oscillator through normal modes. This representation of harmonic oscillators was one of the main techniques of 20th century physics and was extensively studied in modern mathematics (a particular case of the Heisenberg-Weil algebra). In quantum theory, normal modes are replaced by Hermitian-conjugate (adjoint) operators $a^+(t)$ and $a(t)$ which are called in physics creation and annihilation operators and in mathematics raising and lowering operators. These metaphors are easy to understand since, e.g., for oscillator the Hamiltonian

$$H = \frac{1}{2}\sum_{\mathbf{k},\lambda}\left(p_{\mathbf{k},\lambda}^2 + \omega_{\mathbf{k},\lambda}^2 q_{\mathbf{k},\lambda}^2\right) = \sum_{\mathbf{k}=1}^{n}\sum_{\lambda}\omega_{\mathbf{k},\lambda}\left(a_{\mathbf{k},\lambda}^+ a_{\mathbf{k},\lambda} + \frac{1}{2}\right)$$

consists of $3n$ independent quanta. If $a^+$ is applied to the oscillator, then the number of quanta increases by 1 (creation) whereas $a$ reduces their number (annihilation). Operators $a^+$ and $a$ play a crucial role in quantum field and particle theories, with $n(k) = a_k^+ a_k$, $(k^2 = k_0^2 - \mathbf{k}^2, k_0 = \omega/c)$ being operator of the number of particles.

## 9.6. Driven oscillator

The general solution for a driven system $\dot{\mathbf{x}} = \mathbf{y}, \dot{\mathbf{y}} = -\Omega^2 \mathbf{x} + \mathbf{F}(t), \mathbf{F} = (F^1, \dots, F^n)^T$ with the same initial conditions as for a free harmonic oscillator $\mathbf{x}_0 = \mathbf{x}|_{t=t_0}, \mathbf{y}_0 = \mathbf{y}|_{t=t_0}$ may be written as

$$\mathbf{x}(t) = \mathbf{x}_0 \cos \Omega(t - t_0) + \Omega^{-1} \mathbf{y}_0 \sin \Omega(t - t_0) + \Omega^{-1} \int_{t_0}^{t} \sin \Omega(t - t') \, \mathbf{F}(t') \mathrm{d}t',$$

where the combination $G(t, t') = \Omega^{-1} \sin \Omega(t' - t_0)$ is known as the matrix Green's function.

In conclusion to this section, we may bring the dimensionless form for a driven damped oscillator. The dimensionless form, as already mentioned (see Section 5.8. "Many ways of model simplification"), is especially useful for the numerical treatment of oscillator-based models. If we make a transformation $x = x(t), t = h(\tau), \tau = h^{-1}(t) \equiv g(t)$, then $\frac{dx}{d\tau} \frac{d\tau}{dt} = \frac{dx}{d\tau}(t) g'(t)$ and $\frac{d^2 x}{dt^2} = \frac{d}{dt}\left(\frac{dx}{d\tau}\frac{d\tau(t)}{dt}\right) = \frac{d^2 x}{d\tau^2}\left(\frac{d\tau(t)}{dt}\right)^2 + \frac{dx}{d\tau}\frac{d^2 \tau}{dt^2}$. If we now put $\tau(t) \equiv g(t) = \omega(t)t + \alpha(t)$, where $\omega(t)$ is the instantaneous frequency and $\alpha(t)$ is the instantaneous phase, then we have $\frac{d\tau(t)}{dt} = \omega(t) + t\frac{d\omega(t)}{dt} + \frac{d\alpha(t)}{dt}$, $\frac{d^2 \tau(t)}{dt^2} = t\frac{d^2 \omega(t)}{dt^2} + 2\frac{d\omega(t)}{dt} + \frac{d^2 \alpha(t)}{dt^2}$. In the dimensionless $\tau$-variable, we have $\frac{d^2 x}{d\tau^2}\left(\frac{d\tau(t)}{dt}\right)^2 + \frac{dx}{d\tau}\frac{d^2 \tau}{dt^2} + \omega_0^2 x = F(t)$, and in the linear time-independent model $\tau(t) = \omega t + \alpha, \omega = \text{const} \neq 0, \alpha = \text{const}$, so that $\frac{d^2 \tau}{dt^2} = 0$ and the second (dissipative) term disappears. Then we get

$$\frac{d^2 x(t(\tau))}{d\tau^2} + \left(\frac{\omega_0}{\omega}\right)^2 x(t(\tau)) = F(t(\tau)).$$

If frequency $\omega$ and phase $\alpha$ can change in time, then, after some simple algebra, we would have

$$\frac{d^2 z}{d\tau^2} + \beta(\tau)\frac{dz}{d\tau} + \Omega(\tau)z = G(\tau),$$

where $z(\tau)$ is a map $z(\tau) := x(t(\tau))$ and for some brevity it is defined

$$\beta(\tau) := \left.\frac{t\frac{d^2 \omega(t)}{dt^2} + 2\frac{d\omega(t)}{dt} + \frac{d^2 \alpha(t)}{dt^2}}{\left[\omega(t) + t\frac{d\omega(t)}{dt} + \frac{d\alpha(t)}{dt}\right]^2}\right|_{t=g^{-1}(\tau)}, \Omega(\tau) := \left.\frac{\omega_0^2}{\left[\omega(t) + t\frac{d\omega(t)}{dt} + \frac{d\alpha(t)}{dt}\right]^2}\right|_{t=g^{-1}(\tau)},$$

$$G(\tau) := \left.\frac{F(t)}{\left[\omega(t) + t\frac{d\omega(t)}{dt} + \frac{d\alpha(t)}{dt}\right]^2}\right|_{t=g^{-1}(\tau)}$$

In the linear model $\tau(t) \equiv g(t) = \omega(t)t + \alpha$, this expression reduces to

$$\frac{d^2x}{d\tau^2} + \frac{2\gamma}{\omega_0^2}\frac{dx}{d\tau} + x \approx \frac{1}{\omega_0^2}F\left(x, \frac{\tau}{\omega_0}\right), \tau = \omega(t)t, \frac{\omega_0^2}{4\gamma\tau} \ll 1, \omega(t) \approx \omega_0 + \gamma t.$$

The inverse transformation $t(\tau)$ is given by $\omega(t)t = \tau$ i.e., $t_\pm = -\frac{\omega_0}{2\gamma} \pm \left(\frac{\omega_0^2}{4\gamma^2} + \frac{\tau}{\gamma}\right)^{1/2}$. Notice that the dimensionality of $\gamma$ is $t^{-2}$.

## 9.7. Conclusion

In science, oscillation models appear repeatedly, in particular, at the frontiers of physics. For example, one of the most intriguing questions in elementary particle physics is the already mentioned problem of neutrino oscillations. The experimentally observed fact is that neutrinos of different kinds (flavor) turn into each other, and such transformations have an oscillatory character. If we take, for simplicity, just two neutrino flavors, electron and muon neutrinos, $\nu_e$ and $\nu_\mu$ (there exist frequently situations when only two neutrino flavors mix, leaving the tau-neutrino $\nu_\tau$ out) the problem of neutrino oscillations closely reminds us of the system of two coupled harmonic oscillators ([16], § 23).

The neutrino states $\nu_1, \nu_2$ at time $t$ (more exactly, the eigenstates of mass matrix with eigenvalues $m_1^2, m_2^2$) are $|\nu_1\rangle = |\nu_e\rangle\cos\theta + |\nu_\mu\rangle\sin\theta$; $|\nu_2\rangle = -|\nu_e\rangle\sin\theta + |\nu_\mu\rangle\cos\theta$, where $|\nu_e\rangle$ and $|\nu_\mu\rangle$ constitute the flavor basis of neutrinos. The inverse transformation expressing $|\nu_e\rangle, |\nu_\mu\rangle$ through $|\nu_1\rangle, |\nu_2\rangle$, $|\nu_e\rangle = |\nu_1\rangle\cos\theta - |\nu_2\rangle\sin\theta$; $|\nu_\mu\rangle = |\nu_1\rangle\sin\theta + |\nu_2\rangle\cos\theta$, where $\theta$ is mixing angle ($\alpha \equiv \sin\theta, \beta \equiv \cos\theta$ are mixing parameters) gives the following representation of neutrino oscillations, e.g., computed as the probability of the presence of the muon neutrino for each emitted electron neutrino at spacetime point $(x, t)$. Assume that at time point $t = 0$ an electron neutrino was born in a certain weak interaction event i.e., $|\nu_1(0)\rangle = \alpha|\nu_e(0)\rangle \equiv |\nu_e(0)\rangle\cos\theta$, $|\nu_2(0)\rangle = \beta|\nu_e(0)\rangle \equiv |\nu_e(0)\rangle\cos\theta, \alpha^2 + \beta^2 = 1$. These states propagate along the $x$ direction with speed $v \approx c$, where $c$ is the light velocity. Then the evolution of mass states $|\nu_1\rangle, |\nu_2\rangle$ may be represented as $|\nu_j(x, t)\rangle = g_j(x, t)|\nu_j(0)\rangle, j = 1,2$, where $g_j(x, t) = \exp(ik_j(\omega)x - i\omega t)$ is the linear propagation (Green's) function, $k_j(\omega) = \sqrt{(\omega/c)^2 - (m_jc/\hbar)^2}$ is the propagation constant (recall that $p^2c^2 + m^2c^4 = E^2$, combination $\hbar/m_jc$ is the Compton length of the j-th particle). This representation is very close to the one describing pulse propagation in dispersive media. Since states $|\nu_1\rangle, |\nu_2\rangle$ (normal modes) regarded as wave packets experience in general different dispersion, the ratio between their mixing parameters $\alpha \equiv \sin\theta, \beta \equiv \cos\theta$ varies so that the probability to observe the other sort of neutrino at spacetime point $(x, t)$ ($x$ is the detector position) is $w_{\nu_\mu}(x, t) = \left|\langle\nu_\mu|\nu_\mu\rangle\right|^2 = \left|\nu_\mu(x, t)\right|^2 = |g_1(x, t)\sin\theta + g_2(x, t)\cos\theta|^2$, where each state is a superposition of normal modes (see next section). One can write the final answer in various forms, one of the possible representations for ultrarelativistic particles ($E \gg m_jc^2$) i.e. $\hbar\omega \gg m_jc^2$ and $k_j(\omega) \approx \omega/c - (m_jc/\hbar)^2(c/2\omega)$ is $w_{\nu_\mu}(x, t) = \sin 2\theta\cos[(\omega x/4c)(m_2c^2/\hbar\omega)^2(1 - m_1^2/m_2^2)]$ or, in natural units $\hbar = c = 1$, $w_{\nu_\mu}(x, t) = \sin 2\theta\cos[(\omega x/4)(m_2/\omega)^2(1 - m_1^2/m_2^2)]$. The specific manner of representation of neutrino oscillations is not as important as the fact of oscillations themselves. In particular, the so-called solar neutrino puzzle (an inexplicable deficit of $\nu_e$) can be explained by neutrino oscillations.

So, the oscillator model is virtually omnipresent. In classical theory, it allows one to trace the main features of dynamical, in particular, Hamiltonian systems: the phase space and geometrical structures on it, the energy function (Hamiltonian), the flow, the time evolution, the role of initial data, integrals of motion, etc. It is easy to see on the oscillator example that the energy function is invariant under the group of time-independent symmetries such as, e.g., a group of one-parameter diffeomorphisms of classical mechanics, leaving invariant both the Hamiltonian and the symplectic structure (Hamiltonian symplectomorphisms). Extensive use of the oscillator model is one more illustration of the fact that many complex phenomena can be understood only in the context of a narrow model designed to specify them.

## Section 10. Classical mechanics as a modeling tool

The main task of classical mechanics is to study the evolution of a mechanical system for given initial conditions. Newton's mathematical laws of motion and gravity provided powerful tools to deal with this task in a regular way and to describe the motion of material bodies, in particular, the paths of planets and comets and the ballistic motion of projectiles. One can even assume that Newtonian mechanics survived against rival models (such as Aristotle's and Ptolemy's) because it best served two main activities of the time: sea trade and wars, as well as the combined activity – colonial wars.

In many areas, classical mechanics remains a standard basis for modeling and computation. For instance, in astrophysics and, in particular, in celestial dynamics, the methods of classical mechanics are still extremely efficient, although there exist newer and more advanced mathematical models, unless one neglects extremely fine astrophysical phenomena or small discrepancies between theory and observations such as, e.g., the perihelion advance of Mercury where general relativity was required to describe the effect. (In this sense, one can consider general relativity to be a more successful mathematical model than classical mechanics so that the latter should be used with care for astrophysical computations.)

Below we shall briefly discuss the use of Newtonian mechanics for modeling the flight and delivery of projectiles (the "Ballistics" subsection) and the elementary model of planetary motion in the Lagrangian context. Although Lagrangian dynamics is more general than Newtonian, in our modeling examples one can obtain the same results when using both frameworks. Now, to illustrate the use of the Lagrangian method for modeling, we may tackle the $CO_2$ (carbon dioxide) molecule whose role has recently been extensively discussed, even in political circles, in connection with the presumed anthropogenic component of global warming. The $CO_2$ molecule is a linear triatomic one, with two polar C=O ($d \approx 2.3D$) bonds being oriented in opposite directions so that their dipole moments cancel each other. Thus, the whole $CO_2$ molecule is non-polar i.e., it does not have a constant component of the dipole moment yet excitation of the vibrational degrees of freedom can break the symmetry of the molecule and enable it to absorb atmospheric radiation in the infrared region. Let us now build a simple spectroscopic model of the vibrational modes in the $CO_2$ molecule. Assume that the C and O atoms lie on the $x$ axis and have coordinates $x_O = x_1, x_3,\ x_C = x_2,\ x_1 < x_2 < x_3$. Now we can write the Lagrangian

$$L = T - V, \qquad T = \frac{1}{2}(m_O \dot{x}_1^2 + m_C \dot{x}_2^2 + m_O \dot{x}_3^2), \qquad V = \frac{k}{2}\left(\left(x_2 - x_1 - \frac{l}{2}\right)^2 + \left(x_3 - x_2 - \frac{l}{2}\right)^2\right),$$

where the carbon atom is in the center and the oxygen atoms are vibrating near their equilibrium positions. The length of the molecule is $l$; $k$ is the "spring" constant. Let us now introduce more convenient (generalized) coordinates, $q_1 = x_1 + l/2, q_2 = x_2, q_3 = x_3 - l/2$. For simplicity, we pay here no attention to the difference between contra- and covariant coordinates. Notice that the generalized coordinates are not necessarily curvilinear. We can write quadratic forms $T$ and $V$ in the new coordinates as

$$T = \frac{1}{2}(m_O \dot{q}_1^2 + m_C \dot{q}_2^2 + m_O \dot{q}_3^2), \qquad V = \frac{k}{2}((q_2 - q_1)^2 + (q_3 - q_2)^2).$$

We see that only conservative forces are acting in this model i.e., dissipative terms containing $\dot{q}_i, i = 1,2,3$, are not present. The fundamental matrix of the linear homogeneous system of equations has the form

$$A := \begin{pmatrix} m_O\omega^2 - k & k & 0 \\ k & m_C\omega^2 - 2k & k \\ 0 & k & m_O\omega^2 - k \end{pmatrix}.$$

Setting the determinant of this matrix to zero $\Delta(A) = [(m_O\omega^2 - k)^2(m_C\omega^2 - 2k) - k^2] - k^2(m_O\omega^2 - k) = 0$, we get the eigenfrequencies of molecular oscillations: $\omega_1 = (k/m_O)^{1/2}, \omega_2 = 0, \omega_3 = \left(\frac{k(2m_O + m_C)}{m_O m_C}\right)^{1/2}$ so that the solutions $q_i(t), i = 1,2,3$ can be written as

$$q_1(t) = q_0 + \dot{q}_0 t - k^2(a_1\cos(\omega_1 t - \varphi_1) - a_3\cos(\omega_3 t - \varphi_3)),$$
$$q_2(t) = q_0 + \dot{q}_0 t - 2k^2 a_3 \frac{m_O}{m_C}\cos(\omega_3 t - \varphi_3),$$
$$q_3(t) = q_0 + \dot{q}_0 t + k^2(a_1\cos(\omega_1 t - \varphi_1) + a_3\cos(\omega_3 t - \varphi_3)).$$

One may notice that one of the eigenfrequencies ($\omega_2$) is zero, therefore $q_2(t)$ (the carbon atom) oscillates only with the hybrid frequency $\omega_3$. Constants $a_1, a_3, q_0, \dot{q}_0$ and phases $\varphi_1, \varphi_3$ should be determined from initial conditions at $t = t_0 = 0$. Constant $\dot{q}_0$ is the center-of-mass velocity, $\dot{q}_0 \equiv u$.

The primary feature of models based on classical deterministic (dynamical) systems is causality, i.e., in such models the effect cannot precede the cause and the response cannot appear before the input signal is applied. One may note that causality does not follow from any deeply underlying equation or a theory, it is simply a postulate, a result of human experience.

Causality in general may be understood as a statement related to general properties of space and time. It manifests itself in many areas of knowledge, although it does not follow from any theorem warranting that causality should be omnipresent. Non-causal systems would allow us to get signals from the future or to influence the past - a marvelous possibility extensively exploited by phantasy writers. Figuratively speaking, in the noncausal world you could have killed your grandmother before she had been born. Models based on a dynamical systems approach are always causal, i.e., effect cannot precede the cause and the response cannot appear before the input signal has been applied. In the special theory of relativity, causality is interpreted as a consequence of the finite speed of signal or interaction, with acausal solutions being rejected - it is impossible to influence the past, it is presumed constant.

Causality is closely connected with time-reversal non-invariance (the arrow of time). The time-invariance requires that direct and time-reversed processes should be identical and have equal probabilities. Most mathematical models corresponding to real-life processes are time non-invariant (in distinction to mechanical models). There is a wide-spread belief that all real processes in nature, in the final analysis, should not violate time-reversal invariance, but this presumption seems to be wrong.

Physicists often treat classical mechanics with an element of disdain or at least as something alien to "real" physics, say condensed matter physics. However, classical mechanics is extremely useful for any part of physics, for many great ideas are rooted in classical mechanics. Examples of Gibbs and Dirac who persistently looked for opportunities to transfer methods of classical mechanics to other fields are very persuasive. One may notice that the ideas from dynamical systems theory have been present in mechanics from the time of Lyapunov and Poincaré but could not overcome the barrier between classical mechanics and physics until the 1970s when nonlinear dynamics all of a sudden became of fashion.

Why is classical mechanics so important? The matter is that classical mechanics has been the primary collection of mathematical models about the world, mostly about the motion of its objects. The principal aim of classical mechanics is to describe and explain this motion, especially under the influence of external forces. This verbal construct of Aristoteles can be translated into the mathematical language as the first-order differential equation, with the state of a moving body (or particle) being described by three coordinates $(x, y, z)$ that change under the influence of an external force

$$\frac{d\mathbf{r}}{dt} = \mathbf{f}(\mathbf{r}), \mathbf{r} = (x, y, z), \qquad \mathbf{f} = (f_x, f_y, f_z) \tag{10.1.}$$

This is really the simplest model of motion. One can immediately see that the crucial difference of Aristotle's model based on the first-order vector equation with the second-order system corresponding to Newton's model is, primarily, in the irreversible character of the motion.

Aristotle's considerations were rooted in everyday experience: the trolley should be towed to be in motion; if the muscle force stops acting, the trolley comes to rest. In such a model the state of a system would be given by positions alone – velocities could not be assigned freely. One might observe in this connection that even the most fundamental models reflecting everyday reality are not unique and often controversial.

The contrast of Aristotle's model to that of Newton is readily seen when one starts thinking about acceleration (as probably Einstein did). In Newton's model of single-particle dynamics, the general problem is to solve the equation $\mathbf{F} = m\mathbf{a}$, where

$$\mathbf{a} := \frac{d^2\mathbf{r}}{dt^2}, \qquad \mathbf{r} := (x, y, z), \qquad \mathbf{r} \in \mathbb{R}^3$$

when the force $\mathbf{F}$ is given.

One might, however, ask: was Aristotle always wrong? The trolley stops due to friction: $\mathbf{F}_R = -\alpha\mathbf{v}$. The Newton's equations for this case may be written in the form

$$\frac{d\mathbf{r}}{dt} = \mathbf{v}, \qquad m\frac{d\mathbf{v}}{dt} = \mathbf{F} - \alpha\mathbf{v}, \tag{10.2.}$$

where $\mathbf{F}$ is the towing force and $m$ is the mass of the trolley. When $\mathbf{F}$ is nearly constant, i.e., the towing force, as it is frequently the case, slowly varies with time, the mechanical system is close to equilibrium, so the inertial term $m\frac{d\mathbf{v}}{dt}$ is small compared to other terms. In this case, we get the equilibrium (stationary) solution $\mathbf{v} = \frac{d\mathbf{r}}{dt} = \frac{\mathbf{F}}{\alpha}$, which has Aristotle's form. This solution form is valid only when the motion is almost uniform, i.e., the acceleration is negligeable, and the friction is sufficiently large, $\left|\frac{d\mathbf{v}}{dt}\right| \ll |\alpha\mathbf{v}|$. One can, in principle, imagine the world constructed on Aristotle's principles: the Aristotle model would correspond to the Universe immersed in an infinite fluid with a low Reynolds number.

It is curious that Aristotle's model is in fact extensively used in contemporary physics, engineering, and even in everyday life. An example is Ohm's law, $\mathbf{j} = \sigma\mathbf{E}$, $\mathbf{v} = \mathbf{E}/ne\rho$, where $e$ is the electron charge, $\mathbf{E}$ is the electric field (acting force), $n$ is the charge density, and $\mathbf{v}$ is the average velocity of the charge carriers. Ohm's law is a typical macroscopic stationary model, when the driving force is compensated by resistance. Stationary models are typical of classical physics: in fact, classical physics dealt only with slow and smooth motions, e.g., planetary movement. Models describing rapid and irreversible changes, resulting in multiple new states, appeared only in the 20[th] century. Stationary and quasi-stationary models serve as a foundation of thermodynamics whose principal notion – temperature – may be correctly defined only for equilibrium.

## 10.1. Newtonian gravity

As already mentioned, classical mechanics was primarily used to calculate the orbits of planets circling around the Sun. One can model the planetary motion as the two-body (Kepler) problem, when a point mass $m \neq 0$ moves in the gravitational field[83] of mass $M$ (typically, it is assumed that $M \gg m$ so that one can use the coordinate system with the origin at the Sun). We shall discuss this problem on the simplest possible level, using Newton's theory of gravitation (Newtonian limit). The planet is treated in this model as a test particle that does not distort gravity created by mass $M$. The other planets of the solar system can, of course, affect the two-body motion, but in the zero order of the corresponding perturbation theory one completely neglects their presence, concentrating on the main factor, gravitational attraction to the Sun. In this approximation, the problem is spherically symmetric so that the angular momentum $\boldsymbol{\mathcal{L}}$ is a conserved quantity. Notice that $\boldsymbol{\mathcal{L}}$ is a vector, therefore it follows from its conservation, even without calculations, that mass $m$ moves in a plane that can be defined by angle $\theta = \pi/2$ to the polar axis coinciding with the direction of $\boldsymbol{\mathcal{L}}$. We can write the Lagrangian for the problem in spherical coordinates $(r, \theta, \varphi)$ as

$$L = \frac{1}{2}m(\dot{r}^2 + (r\dot{\varphi})^2) + \frac{GmM}{r},$$

where $G$ is the gravitation constant. We immediately see that $\varphi$ is the cyclic coordinate so that $\frac{d}{dt}\left(\frac{\partial L}{\partial \dot{\varphi}}\right) = 0$ and $\mathcal{L} = \mathcal{L}_z = mr^2\dot{\varphi} = \text{const}$. It is also evident that energy is preserved too, since the Lagrangian $L$ is invariant under time translations (autonomous system):

$$L = \frac{1}{2}m(\dot{r}^2 + (r\dot{\varphi})^2) - \frac{GmM}{r} = E = \text{const}.$$

This integral of motion means, as usual, that the phase trajectories are confined to the manifold of constant energy, and we can integrate the problem further without using the motion equations. Writing $\dot{r} = \frac{dr}{d\varphi}\dot{\varphi}$, we have

$$\frac{1}{2}m\left[\left(\frac{dr}{d\varphi}\right)^2\frac{\mathcal{L}^2}{(mr^2)^2} + \frac{\mathcal{L}^2}{(mr)^2}\right] - \frac{GmM}{r} = E,$$

---

[83] Mass $m = 0$, e.g., photon or any other particle moving along zero-geodesics cannot be handled as Kepler's problem and requires general relativity in the classical sector and quantum electrodynamics (QED) for the quantum treatment.

and making the obvious substitution $\frac{1}{r} = u$, $\frac{du}{d\varphi} = -r^{-2}\frac{dr}{d\varphi}$, $\frac{dr}{d\varphi} = -u^{-2}\frac{du}{d\varphi}$, we get

$$\frac{1}{2}\frac{\mathcal{L}^2}{m^2}\left[\left(\frac{du}{d\varphi}\right)^2 + u^2\right] - GmMu = E.$$

We can now differentiate this relationship over $\varphi$, obtaining the equation

$$\left(\frac{du}{d\varphi}\right)\left[\frac{d^2u}{d\varphi^2} + u - \frac{GMm^2}{\mathcal{L}^2}\right] = 0.$$

This equation has two fixed points in the $(u, \varphi)$ plane defined by equations $du/d\varphi = 0$ and $\frac{d^2u}{d\varphi^2} + u = B$, $B \equiv \frac{GMm^2}{\mathcal{L}^2} = $ const. The first equation describes (for finite $r$) a stationary state $r(\varphi) = $ const. This state corresponds to immovable planet and the corresponding equilibrium is unstable. The second equation describes the oscillator with unit frequency, driven by a constant force $B$ (note that oscillations and rotations are very close to each other). The solution to this equation can be represented in the form $u = B(1 + \varepsilon\cos(\varphi - \varphi_0))$, where $\varepsilon$ and $\varphi_0$ are constant (one can put $\varphi_0 = 0$). Constant $\varepsilon$ can, in principle, take arbitrary values defining the trajectory corresponding to a conical section. For $0 < \varepsilon < 1$, the planetary orbit is an ellipse ($\varepsilon$ is the eccentricity) with the perihelion at $\varphi = \varphi_0$. For $\varepsilon = 0$, the elliptic orbit degenerates into a circle, $u = B$ i.e., $r = \mathcal{L}^2/GMm^2$. For example, Earth's orbit is very close to circular, its eccentricity being 0.01671123<<1. One can find a more accurate exposition of Kepler's problem in the famous "Mechanics" textbook by L. D. Landau and E. M. Lifshitz [93].

We have already seen (in the section on Lagrangian mechanics) that the Newtonian law for gravity looks very similar to Coulomb's law for electrostatics, which fact resulted in numerous claims that gravity can be treated exactly as electromagnetism. This is, however, not true even in the nonrelativistic approximation since gravitational acceleration $a_g = GM/r^2$ does not involve the mass $m$ of a test particle in the gravitation field (notice that we implicitly use the principle of equivalence here). Contrariwise, the acceleration in an electric field $a_E = \left(\frac{e}{m}\right)q/r^2$, where $e$ is the test charge and $q$ is the charge that creates the electric field – more exactly the static spacetime of a single point charge – essentially depends on the $e/m$ ratio.

## 10.2. Ballistics

Looking back in the history of human civilization, one can observe that the military component of history (probably the most important history-shaping factor) is determined by the range of warfare agent delivery. Periods in military history are typically classified according to the warfare development (ancient, medieval, gunpowder, industrial, post-industrial or scientific). As we scrutinize the evolution of warfare from ancient times to the 21st century, we see that the delivery range was growing with increasing speed, from the human body scale to global reach i.e., from sticks and truncheons to intercontinental missiles carrying many nuclear warheads. The crucial point in warfare evolution was the exteriorization of warfare agents capable of inflicting serious damage to an adversary far beyond the personal contact, and to survive in the new environment of autonomously moving dangerous projectiles, one badly needed to understand the laws of their motion. Ballistics studies the motion of such projectiles, either accelerated to a high velocity within a very short time and afterwards moving inertially in some resistant medium or being self-propelled, when the projectile velocity can be changed or sustained due to additional momentum conveyed by a boost engine.

Consider, as a simple example of modeling real-life situations, the motion of a body near the Earth's surface, taking into account planetary rotation. This is one of the typical modeling problems in missile, satellite and artillery ballistics as the Earth's rotation affects the trajectories of missiles and shells in flight. Let us assume the body's size to be much smaller than any other length in the problem (e.g., Earth's radius, orbit parameters, etc.) so that the body in flight can be viewed as a material point with mass $m$. The motion equation of the body $m\ddot{\mathbf{r}} + 2m\boldsymbol{\omega} \times \dot{\mathbf{r}} = m\mathbf{g}$ takes the form $\dot{\mathbf{v}} + 2\boldsymbol{\omega} \times \mathbf{v} = \mathbf{g}$, where $\mathbf{v}$ is the body's velocity with respect to Earth's surface and $\boldsymbol{\omega}$ is the Earth's rotation angular velocity (a well-defined constant parameter). Notice that this equation, although written completely in Newtonian approximation, implicitly involves the equivalence principle of general relativity. One might also note that this motion equation is a very particular case of general equation (S2.5.). We can represent the non-inertial (Coriolis) term through a linear operator $\Omega$: $2\boldsymbol{\omega} \times \mathbf{v} = \Omega\mathbf{v}$ so that $\dot{\mathbf{v}} + \Omega\mathbf{v} = \mathbf{g}$. This vector equation has the standard form of a linear inhomogeneous dynamical system $\dot{\mathbf{y}} = A\mathbf{y} + \mathbf{f}(t)$, where $\mathbf{y} = (y^1, \ldots, y^n)$, $\mathbf{f} = (f^1, \ldots, f^n)$ and whose solution is

$$\mathbf{y} = Ce^{At} + \int_{t_0}^{t} e^{A(t-\tau)}\mathbf{f}(\tau)d\tau = \mathbf{y}_0 e^{A(t-t_0)} + \int_{t_0}^{t} e^{A(t-\tau)}\mathbf{f}(\tau)d\tau.$$

Applying this formula to our problem, we obtain

$$\mathbf{v} = \mathbf{v}_0 e^{-\Omega t} + \int_0^t \mathbf{g}e^{-\Omega(t-\tau)}\,d\tau \ , \mathbf{r}(t) = \mathbf{r}_0 + \int_0^t \mathbf{v}_0 e^{-\Omega\tau}\,d\tau + \int_0^t dt_1 \int_0^{t_1} dt_2 \mathbf{g}e^{-\Omega(t-t_2)} \quad (t_0 = 0),$$
$$\mathbf{v}_0 = \mathbf{v}(t = 0), \mathbf{r}_0 = \mathbf{r}(t = 0)$$

These expressions give the formal solution to the problem of projectile motion taking into account the deflection due to the Earth's rotation. However, one may wish to represent them in the form convenient for producing numerical results. Recall that operator exponent is defined as $e^{At} = I + A\frac{t}{1!} + A^2\frac{t^2}{2!} + \cdots$ and inserting $A = -\Omega\mathbf{v} = 2\boldsymbol{\omega} \times \mathbf{v}$, we get for trajectory $\mathbf{r}(t) = \mathbf{r}_0 + \mathbf{v}_0 t + \frac{1}{2}\mathbf{g}t^2 + \delta\mathbf{r}(\boldsymbol{\omega}, t)$, where $\delta\mathbf{r}(\boldsymbol{\omega}, t)$ is a correction to the trajectory introduced by the Earth's rotation, $\delta\mathbf{r} = -\boldsymbol{\omega} \times \left(\mathbf{v}_0 t^2 + \frac{1}{3}\mathbf{g}t^3\right) + \boldsymbol{\omega} \times \left[\boldsymbol{\omega} \times \left(\frac{2}{3}\mathbf{v}_0 t^3 + \frac{1}{6}\mathbf{g}t^4\right)\right] + \cdots$ (since $\Omega^2\mathbf{v} = 4\boldsymbol{\omega} \times (\boldsymbol{\omega} \times \mathbf{v})$, $\Omega^3\mathbf{v} = -2\boldsymbol{\omega} \times \Omega^2\mathbf{v} = 8\omega^2(\boldsymbol{\omega} \times \mathbf{v})$, etc.). In most cases, the angular velocity $\boldsymbol{\omega}$ of planetary rotation can be considered a small parameter, $|\mathbf{v}_0| = v \ll \omega R_\oplus \cos\vartheta$, where $\vartheta$ is the latitude, i.e., the velocity of the rotating coordinate system is assumed to be lower than that of the flying object, e.g., the missile (typically 6-7 Mach) so that the higher-order terms in $\omega$ can be disregarded. For the Earth, $\omega \approx 7 \cdot 10^{-5} s^{-1}$, and one can, in principle, get an increased boost "for free" using the Earth's rotation. For instance, France launches missiles from French Guiana, and in general one tends to establish the missile launch sites near the equator.

### 10.3. Modeling of weakly formalized processes

Despite more than three centuries of development of modern physics and mathematics, such disciplines as medicine, biology and behavioral or social sciences have not come to grips with the worldview and methodology of natural sciences. Note, however, that nearly all the achievements of modern biology and medicine are based on the techniques developed in experimental physics, e.g., magnetic resonance imaging, electronic paramagnetic resonance (EPR) analytics, X-ray and ultrasound diagnostics, various sensors, etc.

Likewise, in modern social dynamics a rapidly growing interest has recently been shown on the part of physicists to approach diverse interdisciplinary fields that had been traditionally very far from the physicists' domains of exploration. Thus, in the fields seemingly far from physics, for example in social dynamics, astonishingly physics-like phenomena occur, e.g., spontaneous self-organization effects such as the emergence of social, political and support groups, establishing of hierarchies, migration and group mobility, crowd collective behavior, accidental occurrence of a common culture or language or a consensus on a certain question. Order-disorder transitions as well as scaling and universality can be occasionally observed, although in social disciplines one deals not with particles as the fundamental elements, but with much more complex entities interacting with a relatively small number of their peers. Nonetheless, in modeling social phenomena these entities are usually regarded as comparatively simple. The modeled phenomena often remind us of those studied in modern statistical physics being known as macroscopic physical phenomena and complex systems.

In weakly formalized problems, the mathematics part gets messy and the considered processes fuzzy. We have already noted that weakly formalized disciplines are not based on solid mathematical theories, they only contain models whereas physics contains both theories and models. Application of the methods developed in physics and mathematics to poorly formalized processes, in particular to those occurring in social systems, is mostly associated with trying to establish the generic macroscopic rules of behavior (similar to physical "laws"). It is assumed that such rules can be guessed from observing the evolution of "typical" quantities that one can choose to characterize the system behavior. However, such systems as economics, social organizations and biological assemblies may have their own fundamental laws of evolution that cannot be necessarily derived from the microscopic physical laws of motion. One can hardly be considered a wise person if one persistently states that the solutions to all financial crises lie in appropriate mathematical models.

For instance, social dynamics is complex, and one may start from the assumption that the system to be described remains in a stationary state, and the "typical" quantities characterizing the system just take their average values over time and space. Then the description of social or other weakly formalized systems begins to remind us of the well-known statistical physics approach consisting in attributing the macroscopic statistical laws to microscopic properties of a system such as pair interactions, multiparticle correlations or emerging network links. One can find today increasingly more examples of the intersection of physics and sociology. Of course, the question always arises whether this approach universally adopted in statistical mechanics can adequately reflect the complex rules governing social systems. Even in the relatively simple problem of individual mobility one has to analyze great masses of data (e.g., GPS monitoring of vehicles) to find out statistically significant rules of collective behavior such as the frequencies of visiting specific destinations, preferred directions, traffic fluxes imposed on the road networks, path length and driving time distributions as well as other "typical" quantities related to individual mobility. While considering more intricate processes, where social interactions modified by individual attitudes are involved, even greater sophistication would be necessary, for example, the analysis of the data collected through monitored mobile devices, social networks and even interpersonal communications.

Applying the techniques of reduced (e.g., stochastic) description of complex non-equilibrium processes well developed in physics and mathematics is currently the main interdisciplinary trend in mathematical modeling of open systems (see section 6.4.). Just mentioned chaotic and self-organization phenomena manifest themselves not only in physics or chemistry, but also in bio-medical, behavioral and socio-economic disciplines. The crucial point is how to gain mathematical fluency in a range of seemingly disparate cultures by trying to put together different pieces and borrowing principles from a number of diverse fields.

The quantitative science of phenomena on the human scale, in particular those considered in medical, psychological and social disciplines, many of them being of acute interest to scholars and laymen alike, has been relegated into an insignificant, marginal position in the course of orthodox scientific development in the 20$^{th}$ century. For instance, in physics, the human-scale phenomena were claimed to be insufficiently fundamental and grossly neglected in most physical schools such as the famous Landau and Bogoliubov schools in Moscow, Russia.

An open question is: are the professionals in the so-called exact sciences capable of contributing to the development of working social models, without the help of the professional historians, sociologists, social psychologists and other experts in the humanities? Can the "exact" scientists provide a politically non-engaged social analysis and ideologically neutral prognosis? On the other hand, the trouble is that social sciences and the humanities are poorly protected against dilettantism: they erect neither cognitive firewalls nor language barriers as mathematics in physics, Latin and esoteric abbreviations in medicine or peculiar terminology in biological disciplines. Incompetence in the humanities and social disciplines is much harder to recognize than in science. It is far beyond the scope of this book to address these subjects in detail, yet a few matters deserve some attention.

## 10.4. Climate variability models

One of the weakly formalized and highly discussed processes is the climate. Prior to discussing climate variability, one has to define the term "climate". At least as the climate is defined through long-term temperature trends and, possibly, by the precipitation (e.g., rainfall) level, the climate so defined is bound to constantly change.

Mathematical models of climate dynamics are typically based on two main approaches: that of nonlinear deterministic dynamical systems and that of linear PDEs, generally of stochastic nature. There are currently attempts to unify these two approaches through the theory of stochastic dynamical systems, which combines the well-developed geometric methods used to analyze the behavior of dynamical systems with the concepts of measure and probability theory generalizing the treatment of individual phase trajectories. This unified approach leads to the notion of random attractors that universalize the typical attractors of deterministic dynamical systems, when noise and fluctuations are explicitly taken into account. Notice that when considering complex systems interacting with the environment, and climate is one of them, one cannot avoid its fluctuating influence as well as measurement errors, observational inaccuracies, ever-present limits of computational resources, random perturbations such as noise introduced by external agents (in the climate case, by the Sun or due to other uncontrollable astrophysical factors). Thus, the infinitely thin trajectories appearing in mathematical models based on deterministic dynamical systems should be smeared to allow for unavoidable noise or measurement uncertainties. Yet notice that even in the fully deterministic approximation, climate as a dynamical system cannot stay at rest (unless it resides in the vicinity of an equilibrium point), and if it does, some fluctuation or external influence would induce its changes.

Models designed to explain climate change are mostly centered around the dynamics of global mean air surface temperature (GMAST) whose allegedly rapid increase has produced a widespread popular debate. Climate research has challenged not only the academic community, but also the political circles and the population at large. We shall not discuss social, economic or political issues in climate modeling (although the debate persists), but solely the physical models; already the latter are contested enough. Mathematical modeling serves mainly for understanding, not for forecasting – a maxim that is often forgotten when modeling complex systems such as the Earth's climate. In section 12.1., we will note that complex systems tend to behave chaotically, at least in some domains of parameters, i.e., they exhibit irregular behavior. Chaotic attractors manifest extreme sensitivity to parameter variations, e.g., to slight changes in initial conditions; this mathematical fact is known as the "butterfly effect" which is a metaphor for strong structural instability. The trouble with dynamical systems having chaotic attractors is that corresponding mathematical models cannot provide trustworthy forecasts for the modeled system's behavior even on the qualitative level.

Computer modeling may only appear accurate, but genuinely accurate modeling of real processes is very difficult. Ignoring the very limited predictive capability of complex models brings about controversial end-of-the-world scenarios. Climate dynamics is a complex and fundamentally nonlinear physical problem, its complexity and nonlinearity contribute to the situation when science plays in climatic issues a subordinate role (sometimes verging on sheer servility). There are too many political and financial interests involved to restrain the issue of climate evolution to a purely scientific problem. Besides, social perceptions skewed by propaganda campaigns ("carbon fever") and economic considerations ("wartime speed of global investment to battle climate change"), one has to believe the results of computer modeling. Thus, the outcomes of numerical modeling become drivers of an obscure system of ideological beliefs. Unfortunately, irrational "pro" and "contra" biases can affect not only lay people and politicians, but also scientists. The trouble with climate studies is that there are too many "soft facts" in this field. Soft facts are not always wrong, they merely have not been properly documented or are not intended to be confirmed by corroborative evidence. A salient collection of soft facts is in oriental medicine, telepathy and other paranormal or "extrasensory" techniques. In this respect, soft facts, being on the "not even wrong" level, are different from the hard facts that are firmly documented and whose reproducible physical substance can be experimentally verified.

### 10.4.1. Fluid dynamics equations and climate modeling

One may note that the Navier-Stokes system describes the dynamics of a single-phase flow. To study the dynamics of multiphase fluids which are most often present in nature and technology, one should deal with more complicated systems of equations. For instance, a straightforward application of the above Navier-Stokes system to climate modeling is questionable since even such primary climatological and meteorological objects as clouds, raindrops, snow, water vapor, carbon dioxide and other greenhouse gases (GHG), etc. cannot be directly described by (S3.1).

Recall that the atmospheric "greenhouse effect" that is assumed to underlie the hypothesis of catastrophic global warming is actually an interplay of a number of effects related to the radiation transfer, so that to elucidate the problem of climate dynamics the Navier-Stokes system must be coupled with kinetic equations governing the propagation of electromagnetic radiation in different spectral regions. One can illustrate this coupling by the following parallel: when a greenhouse is built, its walls and roof are typically made of glass, the latter being transparent in the visible and near infrared ($\lambda \sim 3\mu m$), but opaque for longer wavelengths (that is why we can see through the glass). In other words, glass lets electromagnetic radiation in, but its longer wavelengths are unable to pass through the glass into the outer space, and thus the thermal infrared emission from the heated objects such as various devices, sun-warmed surfaces and humans is trapped inside the greenhouse. This radiation entrapment is governed by the radiation transfer (kinetic) equations that should be combined with the fluid dynamics equations describing the air motion, e.g., convection. Phenomenologically, one can model the combined heat transfer and convection in the atmosphere with the Navier-Stokes equation supplemented by the following modification of the usual heat transfer equation

$$\rho c_p \partial_t T + \partial_i \left( -k^{ij} \partial_j T + \rho c_p T u^i \right) = Q,$$

where $Q$ is the heat source density (describing heat generation within the atmosphere), $c_p$ is the heat capacity, $k^{ij}$ is the thermal conductivity tensor, $\rho$ is the fluid (air) density, $T = T(\mathrm{r}, t)$ is its local temperature and $u^i(\mathrm{r}, t)$ is its velocity. In general, heat transfer in the atmosphere is not necessarily isotropic, with directed streams, rotations and crosswind diffusion playing a significant role, so that the $k^{ij}$ tensor cannot be a priori reduced to $k\delta^{ij}$. The total heat flux $q^i := \left( -k^{ij} \partial_j T + \rho c_p T u^i \right)$ consists of two terms: conductive $-k^{ij} \partial_j T$ and convective $\rho c_p T u^i$ that enter the equations additively in the linear models[84].

Although one usually talks about consensus among scientists about global warming, actually little consensus has been achieved about the final formulation of climate models.

## 10.5. Modeling of natural catastrophes

In the study and forecast of natural catastrophic events, e.g., hurricanes, tornadoes or tsunamis, diversified physical and mathematical techniques are progressively employed. For example, hurricane and tornado models can be based on the global weather fluid dynamics approach. Such models belong to the class of dynamical models. Apart from these, there are also statistical models in which physical parameters are not point functions, but are "smeared" i.e., statistically distributed. An intermediate case is the class embracing the so-called ensemble models, which are basically dynamical ones, but with continuously differing initial conditions. Ensemble modeling is quite sensible since the initial state is never perfectly known. There are well-known limitations in mathematical and physical modeling of natural catastrophes. Thus, physics-based catastrophe modeling helps to understand the event generation and potential hazards, but usually does not predict individual damage parameters. For instance, dynamical modeling of earthquakes can give the amplitude of ground shaking, but not the consequences for the buildings: this is more structural mechanics and seismically safe design than mathematical modeling. In other words, given a hazard obtained by physically based modeling (e.g., expressed in terms of ground acceleration, seismic wave velocity, rupture propagation characteristics, wind velocity, etc.), the damage inflicted on buildings or other man-made objects must be determined by engineering methods.

One might note here that an earthquake is an extremely complex natural phenomenon that does not admit, at least so far, an adequate description – even at a phenomenological level, without speaking about a consistent theory.

Natural catastrophic events such as earthquakes, volcano eruptions, hurricanes, tornadoes, floods, tsunamis, landslides, forest fires, etc. have led to a tremendous number of fatalities throughout human history. Can it be that there is something basically wrong in nature if it admits such deadly events? One might recall the description of how the large island of Atlantis suddenly sank, which is speculated to be the effect of a devastating tsunami generated by an earthquake (see, e.g., http://en.wikipedia.org/wiki/Atlantis). The environmental effects of the broadly publicized Fukushima Daiichi nuclear plant accident are negligible compared with the consequences of the tsunami that preceded the Fukushima reactor destruction. Loss of life due to the tsunami was estimated to be over 20000 people, with over 10000 losing their homes (these numbers were incomparable with the casualties associated with the ensuing nuclear accident that had produced great political resonance and media hype). Yet, in spite of numerous attempts, forecasting and modeling of natural catastrophes remain a difficult and, in most cases, unrealistic task. Predominantly statistical methods, mainly driven by insurance companies, are employed whereas forward-projected dynamic modeling based on physical principles does not seem to be reliable so far. If, however, such modeling and nearly deterministic prediction techniques could be successfully applied, it would be a fair example of the practical value of mathematics.

---

[84] In an electric field $E_i = \partial_i \varphi$, where $\varphi$ is the electric potential, there is also heat transfer along the field lines, so that the total flux $q^i$ acquires the term proportional to the field $q^i := \left( -k^{ij} \partial_j T + \rho c_p T u^i + \sigma^{ij} E_i \right)$, where $\sigma^{ij}$ is the conductivity tensor proportional to electrical conductivity. More generally, in the inhomogeneous electrically charged environment, the total flux $q^i$ is proportional to the gradient of the chemical potential $\mu = \mu_0 + \varphi$ ($\mu_0$ is the local value of the chemical potential without the electric field), see, § 26.

## 10.5.1. Tsunami modeling

As an example of a severe catastrophic event, we may consider the tsunami wave. Tsunamis result from the sudden (in time $\tau$ short compared with the inverse wave frequency $\omega$ and damping $\gamma$) displacement of a large volume of water, usually following an underwater earthquake, volcano eruption, asteroid impact or strong blast, e.g., nuclear explosion. Tsunamis occur not only in the ocean, but they may also be produced in a sea or lake as well. From the physical viewpoint, a tsunami is a nonlinear wave or a wavetrain having the wavelength spectrum centered around very long values, typically of the order of $10^2$ km. For such long waves, the ocean is shallow waters, $kh \ll 1$, where $k = \omega/v_p$ is the wave number, $h$ is the depth and $v_p$ is the phase velocity of the wave i.e., the traveling speed of a particular point (phase) of the wave, e.g., its peak (crest) or trough. Notice that the usual wind-excited gravity waves on the water surface have typical wavelengths ~1 m so that for them the opposite limit holds, $kh \gg 1$ (deep waters). As a result, shallow and deep-water waves have completely different behavior even in the linear approximation (see e.g., [100]). For example, their propagation velocities are given by different expressions. Indeed, the general relationship between $\omega$ and $k$ (dispersion relation) for the waves propagating over the surface of a fluid having depth $h$ is $\omega^2 = gk \tanh kh$. For deep waters, $kh \gg 1$, one has $\tanh kh \to 1$ and thus one obtains $\omega^2 = gk$ so that $v_p = \omega/k = (g/k)^{1/2}$ whereas for shallow waters, $kh \ll 1$, $\tanh kh \approx kh - (1/3)(kh)^3 + \cdots$ and $\omega(k) = (gh)^{1/2}(k - (1/6)k^3h^2 + \cdots)$ so that $v_p \approx (gh)^{1/2} = v_g$, where $v_g = \partial\omega/\partial k$ is the group velocity. In deep waters, the group velocity i.e., roughly the speed at which energy of the wave is transmitted (not always the case) is expressed as $v_g = \frac{1}{2}(g/k)^{1/2} = \frac{1}{2}v_p = \frac{1}{2}(g\lambda/2\pi)^{1/2}$. Both phase and group velocities in deep waters grow with the wavelength i.e., longer waves propagate faster. Therefore, deep-water waves are dispersive in contrast with non-dispersive shallow-water waves whose velocity only depends on the depth. Using these classic expressions, we can roughly estimate the speed of a tsunami as it travels across the ocean (and tsunamis can travel over distances ~$10^3$ km from the epicenter of perturbation that produced them). For the typical depth of 1-10 km, the tsunami speed is $v_g = v_p \sim (1 - 3) \cdot 10^2$ m/s = 360-1000 km/h which is a considerable value in the human scale. As the tsunami wave approaches shallow waters, e.g., a bay, its velocity is decreased approximately as $h^{1/2}$, and its amplitude (height) rises roughly as $h^{-1/4}$ to ensure the energy flux conservation (this effect is known as shoaling). For example, if a tsunami wave has $\zeta_0 \sim 1$m height in the open ocean with depth $h_0 \sim 10$ km, its height near the coastline would be $\zeta \sim \zeta_0(h_0/h)^{1/4} \sim 6$ m. Large tsunamis were observed to reach 20-25 m height near impact areas.

So, when a tsunami is generated (e.g., by an underwater earthquake, volcano eruption or landslide), the tsunami wave propagation may be analyzed by classical fluid dynamics methods. Yet the prediction of tsunami heights in the real conditions of coastline irregularities (fractal geometry) leads to significant computational challenges for specific locations. The recent tragic events, e.g., 2004 Indian Ocean tsunami and 2011 Fukushima disaster drew attention to forecasting and modeling natural catastrophes as well as creating an early warning instrumental infrastructure.

As it is usual in linear wave propagation theory, the profiles of tsunamis and hence their destructive impact are given by oscillatory integrals whose asymptotes can be found analytically, e.g., by the stationary phase method.

So far, we discussed the linear model of tsunami, which is good for estimates but does not provide an adequate description of this phenomenon. A mathematical model of tsunami as a nonlinear wave can be constructed using the elementary soliton theory based on a single-dimensional convective equation, $\zeta_t + u(\zeta, x, t)\zeta_x = 0$, where $\zeta$ denotes the oceanic surface elevation as a function of spacetime variables $x, t$. Note that the wave velocity $u$ depends on elevation $\zeta$ which makes the problem nonlinear. The term "soliton" manifests a particle-like, concentrated behavior of this object: when, e.g., two solitons collide they remain intact, only their phases may, in general, change. In the particular case when $u = \text{const} \equiv c$, all waves propagate with the same velocity $c$ i.e., the wave equation admits traveling wave solutions, $\zeta(x, t) = \zeta(x - ct)$, which means that the initial wave or pulse profile reproduces itself in the process of propagation. In the general situation, $u = u(\zeta, x, t)$, the wave or pulse shape changes.

The peculiar and unique feature of a soliton is that the spread due wave dispersion exactly balances the wave or pulse compression caused by nonlinearity. It is this compensation that results in the localized traveling wave solutions known as solitary waves. However, in the case of a really propagating tsunami, dispersion and nonlinearity are not necessarily counterbalanced, one or the other may dominate. In qualitative terms, if nonlinearity dominates, the soliton gets the vertical asymptote at the forefront as in the shock wave. This is the most dangerous feature of a tsunami, mainly occurring in the shallow waters near the shoreline. The vertical wall of water traveling at considerable speed can have great energy density and the corresponding smashing effect. In the deep oceanic waters, dispersion usually dominates so that the propagating soliton tends to be transformed into a wave train (Zug) containing a number of "normal" oscillating waves. The competition of dispersion and nonlinearity is clearly seen from the Korteweg-de Vries (KdV) equation which is the main mathematical model for localized traveling waves. In the dimensionless form, the KdV equation may be written as

$$\zeta_t + \alpha\zeta\zeta_x + \beta\zeta_{xxx} = 0, \qquad (10.5.1.1.)$$

where the first two terms correspond to the nonlinear Hopf equation and the last term accounts for dispersion. Notice that an analogous equation may be written for the horizontal velocity component. Numeric coefficients $\alpha$ and $\beta$ characterize the relative roles played by the nonlinear and dispersive terms, respectively. Since it is a priori unclear which of the two processes would dominate in a specific situation, one can consider $\alpha$ and $\beta$ to be of the same order of magnitude i.e., $\alpha \sim \beta \sim 1$. One often puts, for convenience, $\alpha = 1, \beta = 1/3$, although by choosing appropriate units one can make $\alpha = \beta = 1$. In general, however, one must treat $\alpha$ and $\beta$ as parameters of the problem (like the Reynolds number for viscous fluids). For instance, nonlinearity in tsunami-related tasks is small, but always finite ($\alpha > 0$). Using scaling, we can put $\alpha = 1$ and in this case the KdV equation has a solution

$$\zeta = \frac{\zeta_0}{\cosh^2\left(\sqrt{\zeta_0/2\beta}\ \xi\right)},$$

where $\xi = x - \lambda t + \varphi$ and $\zeta_0 = 3\lambda, \varphi$ is a soliton's phase. This solution satisfies the asymptotic condition $\zeta(-\infty) = \zeta(+\infty) = 0$ for any velocity eigenvalue $\lambda > 0$. The theory of the KdV equation is thoroughly developed, and we shall not reproduce it here referring the reader to, e.g., the book by M. Lashmanan. Solitons, Tsunamis and Oceanographical Applications (Encyclopedia of Complexity and Systems Science, Part 19), Springer, Heidelberg, 2009.

Notice that in equation (10.5.1.1.), one more important process characterizing wave perturbations was not taken into account, namely dissipation. Although wave motion generally subsists in large spacetime regions outside the source domain, dissipation is unavoidable. A mathematical model unifying all three processes – nonlinearity, dispersion and dissipation – can be represented by the following equation

$$\partial_t u + \alpha u \partial_x u = \beta \partial_{xxx} u + \gamma \partial_{xx} u \qquad (10.5.1.2.)$$

and its multidimensional analog (e.g., 3d). When $\gamma = 0$, we have the KdV equation whereas for $\beta = 0$ we get the so-called Bürgers equation describing the propagation of nonlinear waves in a dissipative medium. Solutions to the Bürgers equation have the form $u = u_0 \tanh\left(\frac{\alpha}{4\gamma} u_0 (x - vt)\right)$, where $v$ is some average velocity. We can understand the meaning of parameter $\gamma$ by addressing the Navier-Stokes equation. Recall that the Navier-Stokes equation is the main mathematical model describing the motion of incompressible viscous fluid (see Supplement 3). Writing equation (10.5.1.2.) for elevation $\zeta$ and comparing with the Navier-Stokes equation

$$\partial_t \mathbf{u} + (\mathbf{u}\nabla)\mathbf{u} = -\frac{\nabla p}{\rho} + \frac{\mu}{\rho}\Delta\mathbf{u}$$

or, in dimensionless form,

$$\partial_t \mathbf{u} + (\mathbf{u}\nabla)\mathbf{u} = -\nabla p + \frac{1}{\text{Re}}\Delta\mathbf{u},$$

we may identify $\gamma$ with 1/Re (at least in the one-dimensional case). Here pressure $p$ is measured in $\rho(\bar{u})^2$, $\rho$ is the fluid density, $\bar{u}$ is its average velocity. The Reynolds number $\text{Re} = \rho\bar{u}L/\mu$, where $\mu$ is viscosity (internal friction) in the fluid, $L$ is a characteristic linear dimension of the fluid motion, determines the relative role of inertia compared to viscosity: at low Reynolds numbers viscous effects dominate, and fluid motion tends to be regular and smooth (laminar).

A more refined description of tsunami can be reached through using the Boussinesq-type equations

$$\partial_{tt}\zeta - \partial_{xx}\zeta = \sigma\partial_{xx}(\zeta^2) - \partial_{xxxx}\zeta,$$

where $\sigma$ is some numerical parameter (often $\sigma = 3/2$). This equation corresponds to an approximation taking into account the influence of the ocean bottom (shoaling). The physical picture underlying the Boussinesq approximation consists in the gradual variation of fluid particle trajectories as the wave approaches the coastline. Such trajectories are in general ellipses whose semi-axes depend on water depth, and in deep waters ($kh \gg 1$, where $k$ is the wave number, $h$ is the ocean depth) fluid elements move over the circles. In shallow waters, fluid particle trajectories are ellipses whose semi-axes depend on water depth; near the bottom, fluid particles move almost horizontally (locally parallel to the bottom). Incidentally, this almost horizontal motion of fluid elements explains the drawback effect of the water just before the tsunami hits the shore: the ocean recedes, which is the most salient symptom of tsunami coming. For the long waves over the bottom with appreciable depth variation, the wave amplitude (height) increases as the wave enters shallow waters to compensate for the diminishing group velocity $v_g$ i.e., energy flux $j_E = v_g w$ must remain constant in steady-state situation, $dj_E/dx = 0$.

There are a number of simplified variants of the Boussinesq-type equations, one of the most popular is:

$$\partial_{tt}\zeta + \partial_x(\zeta\partial_x\zeta) + \partial_{xxxx}\zeta = 0,$$

with the traveling-wave solution

$$\zeta(x,t) = -3p^2 \frac{1}{\cos\left[\frac{p}{2}(x \pm pt) + \phi\right]},$$

where $p$ and $\phi$ are arbitrary constants. There are other forms of solution to the Boussinesq equation (see, e.g. [44]). One can also note that the Boussinesq equation served as a good example for producing solutions of nonlinear equations by the inverse scattering method [2].

## 10.5.2. Earthquake modeling and prediction

One can model an earthquake as an oscillatory motion or an aperiodic slip along the fault surface. The latter is understood as a rough interplate boundary surrounded by an aseismic or smoothly sliding area. Prediction of an earthquake is a very difficult problem that has not been solved so far. The goal of earthquake prediction is to make reproducible estimates that a seismic event of magnitude $M$, $M_{min} \leq M \leq M_{max}$, will occur within the given spacetime domain $A \times T$, with the temporal accuracy $\Delta T$ of 10-100 hours and well-defined position of epicenter. In spite of significant efforts, funding and certain achievements (for well-studied geophysical conditions, e.g., faults), this goal remains elusive. One of the main problems with earthquake prediction is that the triggering mechanism of earthquakes remains obscure, which makes them completely unexpected.

Commonly, seismological research sees its goal in statistical assessment of the earthquake hazard in a given area within a certain time interval. At best, this risk assessment must provide quantitative estimates for the earthquake magnitude to be expected, with the defined accuracy corridor. One usually distinguishes between long-term (over a year), mid-term (months to years) and short-term (~10-100 hours) prediction. The long-term prediction is mainly based on statistical data and historical records i.e., can only give the mean figures, at least so far. The mid-term prediction is currently based on geophysical computer simulations, where the input data are taken from seismic monitoring. Today, the earthquake-forecasting computer codes and the underlying physical models are under development in a number of countries (China, Japan, Russia, USA and others). The short-term earthquake prediction is currently based on analyzing multiple precursory phenomena in real time, some of them being rather exotic or relying on irreproducible observations (such as changes in animal behavior). Some physically measured phenomena are also believed to constitute the set of the earthquake precursors, for example, increased radon and hydrogen emissions, change in velocity ratio $v_p/v_s$ of P (pressure) and S (shear) seismic waves, GPS-registered crust motions, variations of the electromagnetic wave propagation conditions in the ionosphere, stress-induced electric signal transmission, luminescence and so on. However, the usefulness of such physically based techniques for earthquake forecasting still remains controversial, just like biological observations.

The main problem with earthquake prediction is that the physics of seismic phenomena is still poorly understood. The tectonic plate hypothesis only accounts for large-scale motion underlying the earthquake, yet the physical mechanisms and macroscopic interactions in the plate boundary areas remain unclear. Moreover, the dynamics of tectonic plates seems to be currently discussed on a speculative level, relying more on consensus rather than reproducible verification. Empirical information about Earth's interior is mostly obtained from the analysis of seismic signals, which is a difficult inverse problem, in particular due to complex and often poorly known crustal structure. Thus, deep ruptures (at depths ~100 km) are hardly possible to observe directly so that their maps can only be produced with low accuracy.

The tectonic plates that make up the earth's crust are assumed to move as rigid bodies, experiencing stress and strain due to mutual interactions. However, both the nature and spacetime dependence of driving forces setting tectonic plates in motion are not well understood, which makes the deterministic modeling and prediction of earthquakes a difficult problem. For example, geophysical modeling of seismic events in Europe is quite complex because of the simultaneous interaction of several plates: Eurasian, African, Arabian as well as Anatolian and Aegean microplates. Although Europe, especially Western Europe, is commonly perceived as seismically calm region, there are indications at significant seismic hazard for densely populated European terrain (http://earthquake.usgs.gov/earthquakes/recenteqsww/Maps/region/Europe_eqs.php). One can recall in this connection the devastating Basel (1356), L'Aquila (1703), Lisbon (1755), Granada (1884) and Provence (1909) earthquakes as well as the 1915 Abruzzi event. Besides, historical information on the past earthquakes that may serve as an input for model construction is rather scarce and physically imprecise, e.g., it contains no mapping of seismically active faults. The inferences based on historical data can also be unreliable, for instance, in accordance with historical data, the fault corresponding to the 2008 Wenchuan earthquake was considered to present a low risk.

As to the attempts to introduce a deterministic component into seismic modeling and prediction, there is a phenomenological hypothesis that earthquakes possess a certain recurrence characteristic for each fault i.e., occur at quasiregular time interval. In other words, this hypothesis implies that one can describe seismicity as an almost-periodic process, with a finite number of characteristic frequencies being excited. The physical model underlying this hypothesis consists in the view that the energy and time needed to build up the stress or mechanical pressure to be released in the earthquake are essentially constant for each fault i.e., can be thought of as its physical characteristics. This speculation also presupposes that the magnitude of an earthquake is related to the recurrence period, although accurate paleoseismic records on a $10^3$ year scale are not available. The quasiperiodic recurrence assumption has obvious consequences for seismic event prediction. If the paleoseismic records had corroborated the presence of dominant frequencies for the given faults and amplitudes, then the occurrence rate of possible earthquakes could be extrapolated into the future. This would be of vital importance for seismic risk assessment. The dependence of the earthquake frequency on released energy[85] is typically assumed to be logarithmic or obeying the power-law proportionality in magnitude. This statement is known as the Gutenberg-Richter relationship which simply means that the number $N(M)$ of earthquakes with magnitude $M$ per year is proportional to $10^{a-bM}$ or $\log N = a - bM$. This power law distribution for earthquake occurrences implies that there is no characteristic energy in earthquake development and, respectively, no energy scale.

---

[85] The energy release during a typical earthquake is $E \sim 10^{18}$ J $\sim Gt$ (gigaton) TNT; 1 ton TNT of released energy approximately corresponds to $4.2 \cdot 10^9$ J.

A number of mechanical models simulating seismic processes have been proposed. One of the most popular among them was the Burridge-Knopoff slider block model [35]. This model replaces distributed seismicity with lumped sliding or oscillating bodies: in the original version, two sliding blocks of masses $m_1, m_2$ coupled by a spring $k$ can move in the horizontal plane (actually along the straight line) under the influence of a driving force exerted from a plate moving with constant velocity $v$ parallel to the line (Figure 13). The motion of the blocks is impeded by friction forces $F_1$ and $F_2$. This seemingly elementary mechanics problem may lead to a rather complex motion accompanied by chaos, self-organized criticality and even appearance of unstable traveling waves (when the number $N$ of considered masses $m_j, j = 1, ..., N$ is large). In this class of models, $N > 2$ point masses (blocks) can move along a straight line, each mass being subject to the force that is proportional to the distance $x_j(t)$ from this mass equilibrium position. The motion of each mass is also affected by the friction force $F(\dot{x}_j) \equiv F_j$. Such models are usually described by systems of differential-difference equations

$$m_j \ddot{x}_j = k(x_{j-1} - 2x_j + x_{j+1}) - k_0(x_j - vt) - F(\dot{x}_j), \qquad (10.5.2.1.)$$

reflecting the competition between rupture-inducing stresses and stopping friction forces that resist the motion of the unruptured plate.

Of course, the relevance of the Burridge-Knopoff class of models to real earthquakes should be questioned. One can, however, notice that the Burridge-Knopoff model, originally intended to describe fault dynamics during earthquakes i.e., stress in the Earth leading to rupture, has a broader meaning than just providing a caricature on frictional sliding of two (or more) tectonic blocks. This model, in fact, relates to excitable media with linear (elastic) coupling. Excitable media are encountered in a number of important disciplines, e.g., in biology, chemistry, medicine, physics, sociology and exhibit complex spacetime behavior. In the Burridge-Knopoff model, this behavior critically depends on friction forces $F_j$ as functions of relative velocity. The friction forces between the rough surface and sliding blocks ensure the dissipation of the driving energy. Notice that the fact that stick-slip motion exhibits the excitable medium features (like, e.g., ignition of explosives or nervous pulse transmission) is rather nontrivial.

In the continuum limit (the number of blocks, $N \gg 1$), we have the field equation

$$c^{-2}\partial_{tt}\psi - \partial_{xx}\psi = \alpha(\psi - vt) - \beta\varphi(\dot{\psi}), \qquad (10.5.2.2.)$$

where field $\psi$ denotes distributed deformation (having dimension of length) and $\varphi$ the density of friction forces. One may notice that the friction law $F(\dot{x}_j)$ and, respectively, $\varphi(\dot{\psi})$ are the only nonlinearities in the Burridge-Knopoff model. Introducing new variables $\chi = \dot{\psi}$ and $\eta = c^2\partial_{xx}\psi + \alpha(\psi - vt)$, we arrive at the following dynamical system:

$$\dot{\chi} = \eta - \beta\varphi(\chi), \qquad \dot{\eta} = c^2\chi_{xx} + \alpha(\chi - vt) \qquad (10.5.2.3)$$

One can, of course, rewrite (10.5.2.3) in the dimensionless form measuring deformation $\psi$ in terms of some characteristic length, e.g., spacing between the blocks. Then one can put parameter $\alpha$ determining the pulling effect to unity and obtain the following dynamical system

$$\dot{\chi} = \gamma(\eta - \varphi(\chi)), \qquad \dot{\eta} = \gamma^{-1}(c^2\chi'' + \chi - v), \gamma \equiv \frac{\beta}{\alpha} \qquad (10.5.2.4.)$$

There exist also other parameterizations of this model used by different authors. For example, to study the propagation of traveling or shock waves, one often writes the continuous Burridge-Knopoff model in the form

$$\psi_t = v, \qquad \psi_{tt} = \psi_{xx} - \psi - \gamma(\psi_t + \sigma v),$$

where friction is assumed to be subordinated to a linear law, $F(\dot{x}) = a\dot{x}$, typical of kinematic friction (Stokes' law). As we know from elementary mechanics, the friction force can be separated into at least two parts: static (rest) friction and dynamic (sliding or rolling) friction. The latter is the force exerted on the moving body from the medium and directed against the velocity relative to the medium. The absolute value of this force is usually modeled (e.g., in ballistics) as a polynomial on the absolute value of the relative velocity, $F(\dot{x}) = a_0 + a_1\dot{x} + a_2\dot{x}^2 + \cdots$. An alternative model of friction relies on expression ([84]):

$$F(\dot{x}) = \frac{F_0(1-\sigma)}{1 + 2\alpha\dot{x}/(1-\sigma)}, \qquad \dot{x} > 0.$$

Evolution described by the Burridge-Knopoff model is usually explored numerically, e.g., by using the Runge-Kutta method, and there has been a considerable amount of work invested into the numerical study of the model behavior under various assumptions, see, e.g., [117]. Viewing from more general positions, the Burridge-Knopoff model has become one of the favorite examples of driven dissipative systems in mathematical modeling.

## 10.6. Collective and individual motion

The above description of oscillations in a many-body system in terms of normal modes naturally leads to the concept of collective motion. This is a very important and universal concept which extends far beyond physics. Collective effects are manifest in many disciplines, not necessarily based on firm mathematical foundations, such as economics, sociology, social psychology, etc. In fact, collective motion describes the behavior of an individual in the medium when this behavior is strongly affected by a vast number of other individuals or elements of the medium. In the models of social systems, the role of interparticle forces is played by interpersonal communication whereas mass media and etatist coercion correspond to the fields or external forces acting on a physical system. In many situations, both in physics and social disciplines, the collective behavior can dominate over the individual dynamics, and then the system properties are determined only by certain macroscopic characteristics although, in the final analysis, such macroscopic characteristics are due to microscopic interactions between individuals or particles making up the system. For example, thermodynamic behavior of a physical system consisting of a large number of molecules can be satisfactorily described by just three parameters: temperature, pressure and the average number of particles (or the chemical potential). Notice, however, that this drastically reduced description makes thermodynamics a fully phenomenological discipline which can be built independently of the knowledge of underlying physical interactions. In dynamical situations when temporal evolution ought to be accounted for, collective motion is described by specific collective variables whose prototype is given by the normal modes.

Collective motion is ubiquitous in nature; it can be observed, for example, in large schools of fish, flocks of birds, swarms of insects, human groups, etc. In physics, collective motions often arise due to coherent interaction as in periodic lattices, antenna arrays or modulated electron beams, while in biologic or social systems the rules governing the cooperative behavior can be much more intricate than the phase coherence, even at the level of simple models of collective migration or velocity alignment. Such models usually start from a dynamic description of an unconstrained individual behavior such as that of a Brownian particle experiencing rare collisions. A nontrivial fact in this description is that correlations between the particles may result both in clustering and in spreading out of their spatial distributions. Indeed, take, for example, a random crowd of Brownian particles characterized by pair correlations with the nearest neighbor (one of the simplest models). Each of the individual particles reacts to the presence of another particle within some correlation radius, e.g., the field of vision of a biological object, with two main motions: fleeing and chasing. Accordingly, the particle motion with velocity $\mathbf{v}$ is naturally subdivided into two hemispheres, the forward ($\mathbf{rv} > 0$) and the rear ($\mathbf{rv} < 0$) ones. If the particle is approached from behind, it tries to flee by increasing its speed component in the $\mathbf{v}$-direction in order to evade the possible attack. On the other hand, if the particle notices another individual moving in the $\mathbf{v}$-direction, it tends to pursue this individual also by increasing its speed in this direction. Thus, a particle in this model has two principle states of motion, runaway and pursuit. The former motion tends to smooth out the particle distribution whereas the latter reaction can lead to bunching or clustering. One can observe the 1d runaway-pursuit competition on highways, where jams and rarefactions are formed: these are the collective effects in the medium made up of self-propelled particles such as vehicles. Anyway, the flee-pursuit dichotomy determines the macroscopic state of the medium, depending on the competition between the repulsive (escape) and the attractive (chase) forces [131]. In the medium comprised of biological objects, this competition can be directly attributed to the aggression level (predator-prey behavior).

From the standpoint of dynamical systems, the order observed in collective motion, e.g., in coherently moving schools of fish, flocks of birds, swarms of insects, human columns or political attitudes may correspond to some attractors, although the formation of cohesive aggregations does not necessarily lead to an advantageous state as compared to individual behavior, given the lack of available resources. The detailed calculations of dynamical evolution to aggregations do not appear to be known.

One could notice that before social turnovers such as, e.g., in France 1789, in Russia 1917 or in Eastern Europe in 1989, there was a distinctive "confusion and vacillation" accompanying the dawn of a turnover. In physical language, this phenomenon can be expressed as the development of internal fluctuations with rising amplitudes marking the pending phase transition.

## Section 11. Electromagnetic Fields and Waves

A wave is understood as a special case of a disturbance propagating with finite speed through a medium and progressively transferring energy from point to point. The propagating disturbance may have a variety of shapes, from an infinite harmonic wave to an ultrashort pulse. A characteristic feature of wave motion is that if one records a signal in some place $\mathbf{r}$ at moment $t$, then after some time $\tau$ one will observe a similar signal at another place $\mathbf{r} + \boldsymbol{\xi}$. Thus, wave motion can be described by periodic or quasiperiodic functions on spacetime $(\mathbf{r}, t)$.

When the primary and the repeated signals are completely identical, which corresponds to a strictly periodic case, then wave motion does not involve dissipation or dispersion (or these two phenomena accompanying wave processes cancel each other). The difference between the initial and the repeated signals can be noticed either by the amplitude change – diminishing in the case of dissipation and growing when the wave is focused on a smaller domain – or through the varying signal shape, e.g., by its spread over larger domain due to dispersion. In short, wave motion leads to many interesting and practically important questions that can be studied either on mathematical models or experimentally.

We may note that the concept of waves is essentially classical and pertaining more to statistical physics than to mechanics, although wave theory is traditionally adjacent to mechanics courses. Indeed, the wave is a pattern describing the behavior of many particles through their average characteristics which is the typical approach of physical statistics and thermodynamics. Since there is hardly any means to define more exactly what can be called in general a wave process, a very broad variety of problems is mentioned in this chapter. From the standpoint of dynamical systems theory, waves are stable (or almost stable) nonequilibrium modes appearing as solutions in a dynamical system. Such modes can also be observed in many PDE types and in systems of PDEs with an arbitrary number of variables (for instance plane waves, spherical waves, cylindrical waves, spiral waves, etc.) The medium sustaining the wave is not necessarily filled with matter, it can also be a vacuum as in the case of electromagnetic or gravitational waves as well as for the wave function in quantum mechanics (although this "probability wave" strictly speaking carries no energy). Waves can be classified into longitudinal and transverse: those in which the disturbance produced in the medium is perpendicular to the propagation direction are called transverse waves whereas those in which the disturbance of the medium is directed parallel (or antiparallel) to the wave travel course (e.g., specified by wave vector $\mathbf{k}$) are called longitudinal. Examples of transverse waves are electromagnetic waves, waves on a string and seismic S-waves (and also Love waves); examples of longitudinal waves are sound waves, pressure waves in elastic medium and seismic P-waves which are produced by earthquakes and explosions.

The waves described by the balance laws, the latter just being the equations relating the density and the flux of a conserved quantity, are actually of kinematic character i.e., do not explicitly depend on the forces producing the oscillating motion of particles in the media where the wave propagates. Examples of kinematic waves are given by spreading disturbances in traffic flows or by the local rise of water in rivers; in these situations, such extensive quantities as the number of vehicles and the mass of water are conserved. Waves readily become nonlinear: one can easily observe nonlinear waves in water. Nonlinearity of waves is extremely important in dealing with plasma, which manifests itself not only in laboratories, but also in everyday life. Thus, in the early days of broadcasting the so-called Luxemburg effect, when a receiver tuned to a certain station picked up a radio signal from a totally different frequency band, was one of the main causes of radio-signal disturbances. This cross-modulation could not be explained by the linear theory of wave propagation since according to it waves of different frequencies should travel through the same medium without interaction (the superposition principle). The Luxembourg effect is due to the electromagnetic wave propagation in the ionosphere and means that the superposition principle (i.e., linearity) does not, in general, hold for waves in plasma: in this specific case the powerful Luxembourg transmitter modified the wave propagation characteristics of the ionosphere.

A useful approximation for many media supporting the propagated waves is the system of equations of nonlinear geometric optics [25] and [29]. In the simplest form such equations can be represented as

$$\partial_t k^j + \left(v_g^i \partial_i\right) k^j = 0,$$
$$\partial_t E_{\mathbf{k}} + \partial_i \left(v_g^i E_{\mathbf{k}}\right) = 0, \qquad (11.1.)$$

where $k^j = k^j(\mathbf{x}, t), j = 1,2,3$ are the wave vector components, $v_g^i = \partial \omega / \partial k^i$ is the group velocity, $E_{\mathbf{k}}$ is the spectral energy density, $E_{\mathbf{k}} \sim (a_{\mathbf{k}})^2$, $a_{\mathbf{k}}$ is the amplitude of the wave packet, the latter being assumed "spectrally narrow" or quasi-monochromatic i.e., its amplitude, phase and frequency change negligibly during the period corresponding to the main (central) frequency. It is easy to see that both equations are a particular case of the $n \times n$ system of hyperbolic conservation laws, e.g., for 1d wave motion along the x-axis

$$\partial_t u + \partial_x f(u) = 0, \qquad u(x, 0) = u_0(x) \tag{11.2.}$$

or, more generally,

$$\partial_t u^j + \partial_i f^{ij}(\mathbf{u}) = 0, \qquad \mathbf{u} = \{u^i\}, i, j = 1, \dots, n \tag{11.3.}$$

see also section 6.1. ("Balance laws"), e.g., equation (6.1.6.). An example of such a system of hyperbolic conservation laws (balance laws) is the $2 \times 2$ system of equations for the flow of a compressible fluid, e.g., with the adiabatic state equation $p(\rho) = B\rho^\gamma, \gamma > 1, B > 0$, where $p$ is the pressure in fluid, $\rho$ is the fluid density:

$$\partial_t u + \partial_x \left(\frac{1}{2}u^2 + \frac{p(\rho)}{\rho}\right) = 0, \partial_t \rho + \partial_x(\rho u) = 0, \qquad \mathbf{u} = \{u, \rho\}^T.$$

One can see here a parallel (see section 3. "Expected properties of mathematical and computer models") between fluid dynamics and wave propagation materialized through the same type of equations used for mathematical modeling of the situation (hyperbolic balance laws). One can explore and solve such systems of equations, in particular, with the help of perturbation series, e.g., using for the nonlinear tensor flux $f^{ij}(\mathbf{u}), \mathbf{u} = \mathbf{u}_0 + \mathbf{v}$, the expansion around the constant state $\mathbf{u}_0$:

$$f^{ij}(\mathbf{u}_0 + \varepsilon\mathbf{v}) = f^{ij}(\mathbf{u}_0) + \varepsilon A^i(\mathbf{u}_0)v^j + \frac{\varepsilon^2}{2!}\partial_k\partial_l f^{kl}(\mathbf{u}_0)v^i v^j + \mathscr{o}(\varepsilon^2).$$

Here the constant state $\mathbf{u}_0(x)$ is a trivial solution of the balance law equations.

One can readily observe that wave motions described by linear and nonlinear equations are significantly different. To see it, compare two simple one-dimensional models: linear $\partial_t u + c\partial_x u = 0$, where $c > 0$ is constant, and quasilinear $\partial_t u + u\partial_x u = 0$ (the Hopf equation[86]). Suppose we have the same initial value $u(x, t = 0) = \varphi(x)$, where $\varphi(x)$ is some smooth function that will be assumed to have a compact support (i.e., it settles to zero outside a finite certain interval; some time ago such functions were known as finite, but this terminology seems to be obsolete). This assumption physically means that the initial disturbance is localized within some finite region. Any solution to our linear model has the form of a traveling wave $u(x, t) = \varphi(x - ct)$, where $c$ may be interpreted as the speed of disturbance propagation. The graph of $u(x, t)$ is produced from the one of $\varphi(x)$ by a linear time shift $ct$. This solution can be obtained either by introducing new variables $\xi = x - ct, \tau = t$ or using the method of characteristics. An interesting and somewhat surprising property of the Hopf model is that it does not have traveling wave solutions: indeed, assuming $u(x, t) = f(x - ct)$, we get the equation $ff' - cf' = 0$ i.e., either $f' = 0$ or $f = c$. Anyway, $f = \text{const}$.

If $c$ is not constant, but a function of the spacetime point, $c = c(x, t)$, which physically corresponds to waves in matter, then we also have the traveling wave solution (e.g., obtained by the same method of characteristics), but with velocity $c$ having a local character, $c = c(x_0, t_0)$ at point $(x_0, t_0)$. Such cases are often called in physics "parametric", when coefficients of the linear differential depend on its variables as parameters. For example, the concepts of parametric oscillations and parametric resonance (a child on a swing) is based on the dependence of some parameters of the oscillating system on time. From the standpoint of the theory of dynamical systems, this situation corresponds to a non-autonomous system. Note that such systems are encountered not only in classical mechanics, but also in optics and quantum mechanics: for example, the one-dimensional Schrödinger equation describing the motion of a particle in potential $V(x)$ may be interpreted as a parametric equation of oscillatory motion

$$\frac{d^2\psi}{dx^2} + k^2(x)\psi = 0, \qquad k^2(x) = \frac{2m}{\hbar^2}\big(E - V(x)\big) \equiv k_0^2 + q(x), \qquad (11.4.)$$

where $E$ is the particle energy. We shall deal with this form of the Schrödinger equation in the section devoted to quantum modeling. Here we can only notice how remarkably close quantum mechanics is to the classical wave and oscillation theory (but not necessarily to classical analytical mechanics[87]. Indeed, the mathematical model of a vibrating string of length $l$

$$u_{tt} = u_{xx} + q(x)u,$$

where $u(x, t)$ is the string displacement, e.g., supplemented with the condition of fixed ends, $u(0, t) = u(l, t) = 0$, after separation of variables results in solutions of the form $u(x, t) = \exp(ik_0 t)\,\psi(k_0, x)$. To find functions $\psi(k_0, x)$, we arrive at the boundary-value problem

$$\psi''(k_0, x) + \big(k_0^2 + q(x)\big)\psi(k_0, x) = 0, \qquad \psi(k_0, 0) = \psi(k_0, l) = 0,$$

which has the form of the Schrödinger equation (e.g., for one-dimensional electron). It is basically the same model, and its analysis in quantum mechanics is mathematically the same as in the linear vibration theory but embellished with peculiar quantum vocabulary. One might notice that from the rudimentary mathematical viewpoint the main content of quantum mechanics reduces to the theory of linear operators in Hilbert space i.e., to basic applications of linear algebra.

---

[86] Some authors call this PDE simply Euler's equation since it is indeed a one-dimensional version of Euler's equation of fluid dynamics.

Just as in quantum mechanics it is difficult to speak in the language of trajectories, wave motion does not naturally involve the notion of trajectory. Nonetheless, both in the theory of wave motion and in quantum mechanics there exist approaches – some of them being quite successful – based on the concept of trajectories. The most fundamental of such trajectory-based theories are the Feynman path integral theory and the bunch of asymptotic theories, collectively known as semiclassics or quasiclassics. On a somewhat less fundamental level, the Bohmian trajectory-based version of orthodox quantum mechanics has been developed, aiming to provide an intuitive connection with classical dynamics.

Of course, the analogy between wave motion and quantum mechanics is still a rough comparison; it can only illustrate the basic idea of quantum-mechanical modeling. Recall that at the turn of the 20th century there were several experimental facts that led to the concept of quanta and eventually led to the emergence of quantum mechanics in the first third of the century. Niels Bohr received in 1922 the Nobel Prize for a phenomenological model of the hydrogen atom (and other single-electron atoms), this model being based on a direct similarity with the Solar System. Bohr's intention was to explain the observed discrete character of atomic spectra, in particular, the emission lines of atomic hydrogen.

---

[87] Strictly speaking, classical mechanics is not the limit of quantum mechanics for $\hbar \to 0$. As already mentioned, quantum and classical mechanics are the theories of two different types: a linear wave theory vs. a nonlinear dynamical systems theory, and in general connecting them by a single limit is hardly possible. Notice in relation to the intricate issue of the $\hbar \to 0$ limit that transition to a limit in equations does not necessarily imply such a transition in solutions.

One of the main sets of empirical evidence underpinning quantum mechanics consisted of the observed discrete character of atomic spectra. A mathematical tool providing discrete spectra had already been known: it was the method of eigenfunction expansions used in the linear equations of mathematical physics, in particular, the wave equation and the Helmhotz equation derived from it – both needed to describe classical radiation and wave propagation processes. Besides, such important mathematical models leading to second order PDEs as the Laplace equation used in electrostatics and geophysics as well as heat transfer and diffusion equations making up the foundation of heat and mass transfer engineering models also produced discrete spectra, and the respective problems could be solved by eigenfunction expansions. Thus, applying such tools one could surely produce discrete spectra, e.g., enumeration of the allowed energy values. Other physical quantities could also be quantized, especially when the dimensionality of the problem was more than one. Nevertheless, linear eigenvalue problems are not the only method to obtain discrete quantities interpreted as the allowed values of observables. For example, one can require that such values should correspond to the roots of some algebraic equation: such a method could lead to a different model of quantization. Or one can postulate a certain functional defined on a set $\Sigma$ of measurable physical quantities and taking some combinations of them into integer numbers, $\Sigma \to \mathbb{Z}$ (the prototype of this method was Bohr's model of the atom) [122]. One might speculate that, had quantum mechanics appeared a century later, it could have used other mathematical tools to generate discrete spectra, e.g., taken from nonlinear mathematics instead of the theory of linear operators in Hilbert space.

One should not confuse parametric and nonlinear models (it happens sometimes), although both may lead to intricate solutions as compared to the linear case with constant coefficients: this latter case can rather be attributed to linear algebra than to the theory of differential equations. In the truly nonlinear models, the coefficients of the corresponding equations depend on the unknown function in an arbitrary way. For instance, the eikonal equation $(k \nabla S)^2 = 1$ describing light propagation in the shortwave (ray) approximation, where $k = k(x) > 0$ is inversely proportional to the wave velocity in the medium, is a truly nonlinear PDE. The eikonal equation is also very important in quantum mechanics, specifically to treat the scattering of fast particles. Recall that in quasilinear PDEs, the coefficients are allowed to depend on the unknown function, but its partial derivatives enter the equation only linearly. If we consider, for example, the Hopf equation $\partial_t u + u \partial_x u = 0$ which is one of the simplest quasilinear models, with the same initial condition $u(x, t = 0) = \varphi(x)$ as for the above linear model, then we also can obtain the solution by the method of characteristics, $dx/dt = u, du/dt = 0$ (or, formally, $du/0 = dx/u$). The first integrals are $\varphi_1 = u, \varphi_2 = x - ut$. Therefore, any solution is given by an implicit function, $\Phi(\varphi_1, \varphi_2) = \Phi(u, x - ut) = 0$, and in case one can resolve this equation with respect to $u$, we get $u = \varphi(x - ut)$. Notice that this solution is *not* the traveling wave: it cannot be produced by a Galilean transformation i.e., affine translation of the initial disturbance along the $x$-axis by $ct$, even when the wave phase velocity depends on a spacetime point, $c = c(x, t)$. Indeed, it is easy to see "physically" (i.e., quantitatively) that the nonlinear wave $u = \varphi(x - ut)$ obtained from the Hopf equation is necessarily deformed during its propagation. Assume for illustrative purposes that initial disturbance $\varphi(x)$ is a convex, non-negative, compact-support function (such as, e.g., $\cos \pi x, -1/2 \leq x \leq 1/2$; of course other pulses, not necessarily having a compact support, will also suit: for example, the Lorentzian profile $\sigma/\pi(\sigma^2 + x^2)$, the Gaussian profile $(1/\sigma\sqrt{\pi})\exp\left(-\frac{x^2}{\sigma^2}\right)$ or just the forward front model $(1/\pi)\cot^{-1}(x/\sigma)$). For our cosine-pulse at $t = 0$ the wave velocity has a maximum at $x = 0$, and with rising $t$, points of the wave profile lying to the left of maximum (the back front, initially $x \in [-1/2,0)$) are more and more lagging behind it, as they move with lower velocity, whereas the maximum begins overtaking the points belonging to the forward front, $x \in (0,1/2]$). Each point on the wave profile moves to the right ($\varphi > 0$) the faster, the higher the corresponding value of $\varphi$. At some critical time $t^*$, faster particles will overtake the slower ones, and eventually there appear points with a vertical tangent on the forward wavefront. For $t > t^*$, the function $u = u(x,t)$ ceases to be single-valued, which necessitates the study of discontinuous (generalized) solutions instead of considering the multivalued functions. In wave motion problems, discontinuous solutions typically represent shock waves and nonlinear pulse propagation. One sometimes defines shock waves as discontinuous solutions. The modern incarnation of such phenomena is catastrophe theory (more exactly, the theory of singularities of differentiable maps), providing general methods of studying discontinuous changes, jumps, folds and cusps. Note that the spontaneous formation of discontinuous solutions from continuous initial data is an essentially nonlinear effect which is never encountered in the linear world.

All the above is just a verbal description of point map $(x,u) \mapsto (x + ut, u)$ i.e., of phase flow which is a diffeomorphism of the phase plane. The physical meaning of the Hopf nonlinear model will be clear as soon as one considers the corresponding motion equation in Lagrangian representation. The Hopf equation describes a single-dimensional medium consisting of massive particles that are in inertial motion along the $x$-axis i.e., there is no external influence that might change their velocities. The particles also do not interact with each other. Let $u(x,t)$ denote the velocity at spacetime point $(x,t)$ of the $a$-th particle starting its path at $(x_a, t_a)$. Then, as the one-dimensional law of motion is $x(t,t_a)$ and since the motion equation for each particle is $\ddot{x} = 0$, we have $\dot{x} = u(x(t),t) = u(x(t,t_a),t,t_a)$ and, in the Eulerian picture, $\ddot{x} = \partial_t u + \dot{x}\partial_x u = \partial_t u + u\partial_x u = 0$. Naturally, if such non-interacting particles are moving in the field of forces $F(x,t)$, one will have $\partial_t u + u\partial_x u = F(x,t)$. For each individual non-interacting particle, however, the law of motion is an affine map $x - x_a = u(x_a)(t - t_a) = \varphi(x_a)(t - t_a)$, this is actually the Lagrangian picture of motion. The condition for the emergence of vertical regions at the forward front is $dx/dx_a = 0$ i.e., $1 + \varphi'(x_a)(t - t_a) = 0$. Combining the last two equations, one can try to exclude the starting point $x_a$ (this may be difficult analytically since one has to solve a transcendental equation) and obtain the critical values $(x^*, t^*)$, when the solution becomes multivalued (vertical tangent). The appearance of the vertical tangent means that the wave profile $u(x,t)$ becomes non-unique for $t > t^*$ which is physically meaningless. Notice that for any fixed $t$, factor $(t - t_a)$ is simply a proportionality coefficient. One can find the critical values from the following heuristic considerations: the values $x^*, t^*$ satisfy the relationship for the front $\partial x/\partial x_a = 1 + u'(x_a)(t - t_a) = 0$. For critical values, the front curvature must be zero (inflection point), i.e., $u''(x_a^*) = 0$ i.e., $t^* = t_a - 1/u'(x_a^*)$. Using equation $x = x_a + u(x_a)(t - t_a)$ (particles in inertial motion), we get $x^* = x_a^* - u(x_a^*)/u'(x_a^*)$. Notice that since inertial motion of particles leads to the linear relationship between position $x$ and time $t$, each characteristic for starting (Lagrangian) value $x_a$ is a straight line, with $x_a$ serving as a parameter of a family of such lines. Assuming certain analytic properties of $u(x_a)$, we can differentiate all the obtained relationships over parameter $x_a$ and get the envelope of the family of characteristics at critical points $x^*, t^*$.

## 11.1. Waves in dispersive media

When the speed of propagation of the wave perturbation does not depend on the frequency (or on the wavelength), we call such medium non-dispersive. When the speed of propagation does depend on the frequency, we call such medium dispersive. Function $\omega(\mathbf{k})$ expressing frequency $\omega$ as a function of wave vector $\mathbf{k}$ is called the dispersion relation. More generally, the dispersion relation for linear waves is given by some condition $\Delta(\omega, \mathbf{k}) = 0$; from this relation one gets complex functions $\omega(\mathbf{k})$ or $\mathbf{k}(\omega)$ depending on the setting of the problem (initial or boundary value problem). Since waves having different frequencies travel with different speeds, a question arises how to characterize the rate of propagation of a wave perturbation in the medium. Traditionally, wave propagation is characterized by two speeds: phase $\mathbf{v}_{ph}$ and group $\mathbf{v}_g$ velocities. By definition, $\mathbf{v}_{ph} = \omega(\mathbf{k})\mathbf{k}/k^2$ (i.e., $v_{ph}^i = (\omega/k)(k^i/k)$ and $\mathbf{v}_g = \partial\omega/\partial\mathbf{k}$ (i.e., the gradient of the $\omega(\mathbf{k})$ surface in the wave vector space). In scalar representation, one can produce the following relationship between phase and group velocities:

$$v_g = \frac{\partial\omega}{\partial k} \equiv \frac{\partial}{\partial k}\left(k\frac{\omega}{k}\right) = \left(1 + k\frac{\partial}{\partial k}\right)v_{ph}. \tag{11.1.1.}$$

One can consider $v_g$ to be a function of $\omega$. In electromagnetic theory and optics, one usually characterizes wave propagation by the wavelength in matter $\lambda = 2\pi/k = 2\pi v_{ph}/\omega$, refraction index $n$ and light speed in vacuum $c$ so that $k_0 = 2\pi/\lambda_0 = \omega/c$, $k = k_0 n$, $v_{ph} = c/n$, $k\partial/\partial k = -\lambda\partial/\partial\lambda$, $\lambda_0 = 2\pi c/\omega$ is the wavelength in vacuum, and (11.1.1.) is equivalent to a chain of relations:

$$v_g = \frac{1}{\frac{\partial}{\partial\omega}\left(\frac{\omega}{c}n\right)} = \frac{c}{n + \omega\frac{\partial n}{\partial\omega}} = \frac{v_{ph}}{1 + \frac{v_{ph}}{c}\omega\frac{\partial n}{\partial\omega}} = \frac{v_{ph}}{1 + \frac{\omega}{n}\frac{\partial n}{\partial\omega}} = \left(1 - \lambda\frac{\partial}{\partial\lambda}\right)v_{ph} = v_{ph} - \lambda\frac{\partial}{\partial\lambda}\left(\frac{c}{n}\right)$$

$$= v_{ph} + \frac{c}{n^2}\frac{\partial n}{\partial\lambda} = v_{ph}\left(1 + \frac{\lambda}{n}\frac{\partial n}{\partial\lambda}\right) = v_{ph}\left(1 - \frac{k}{n}\frac{\partial n}{\partial k}\right) = \frac{c}{n - \lambda_0\frac{\partial n}{\partial\lambda_0}}. \tag{11.1.2.}$$

One can see from (11.1.1.) that $v_g$ becomes different from $v_{ph}$ as the dispersion relation $\omega(k)$ from a linear function. In principle, $\omega(k)$ (more generally, $\omega(\mathbf{k})$) obtained from the dispersion relation $\Delta(\omega, \mathbf{k}) = 0$ can be an almost arbitrary function in the complex plane. If differential operator $1 + k\partial_k$ acting on $v_{ph} = \omega(k)/k$ produces negative values, then group and phase velocities are antiparallel. Physically it means that the group of waves moves e.g., backward whereas the phase of the wave i.e., a particular locus such as peak is displaced forward (or vice versa). If we address expression (11.1.2.) for group velocity, we may notice that in the medium with $\partial n/\partial\omega < 0$ i.e., when the high-frequency components in the wave group travel faster through the medium than the low-frequency components group velocity can exceed the speed of light. In the normal case, refraction index $n$ (or its mechanical, acoustic, etc. analogs) slightly grows with frequency which is the reason why red components of light are less deflected or scattered that the blue ones (to the last analysis, this explains the blue color of the sky).

The fact that the direction of group velocity $\mathbf{v}_g$ does not in general coincide with that of phase velocity $\mathbf{v}_{ph}$ can be expressed in the tensor form, $v_g^i = \alpha_{ij}v_{ph}^j$, where sign(det $\alpha_{ij}$) shows whether the group velocity is positive (i.e., its direction is parallel to phase velocity, $k^i/k$) or negative (antiparallel). In an isotropic medium tensor $\alpha_{ij}$ is diagonal so that $\mathbf{v}_g$ and $\mathbf{v}_{ph}$ are either parallel or antiparallel. The direction of wave vector $\mathbf{k}$ is usually taken as positive so that waves in which $\mathbf{v}_g = -|\alpha|\mathbf{v}_{ph}$, $\alpha \equiv$ Tr $\alpha_{ij}$ are known as having negative group velocity. Such waves have interesting properties, e.g., the Doppler effect for the waves with negative $\mathbf{v}_g$ i.e., ($\mathbf{v}_g \downarrow\uparrow \mathbf{v}_{ph}$) would be opposite with respect to the "normal" Doppler effect: the frequency of waves emitted by a moving source will be lower, not higher as compared with the source at rest; the Vavilov-Cherenkov effect will produce backward radiation, etc.

One usually identifies group velocity $\mathbf{v}_g$ with the speed of energy or information transfer [16] (which is in general not true). For instance, we saw that phase and group velocities can be antiparallel, but it does not mean that energy or information are propagating back in time. Specifically, when, e.g., $\partial \log n / \partial \log k > 1$, energy or information can hardly be carried by the wave back to the source. The notion of the group velocity does not work well when one cannot retain just one term in the exponent of the Fourier representation of Green's function describing pulse propagation in material media (see above).

Notice that the concept of phase velocity also cannot be universally applied. The notion of a phase itself is actually a modeling idealization applied to an "unphysical" periodic process. Indeed, to carry energy or information, the propagating perturbation cannot incessantly repeat itself, but rather be a "signal" whose profile is concentrated in space and time. For such a signal, the notion of a phase is badly defined in mathematical (operational) terms; even representation of a perturbation through analytical signal makes sense when one can center the signal around some main frequency $\omega_0$. Indeed, what is the period of the wave for an ultrashort pulse? The notions of amplitude, phase and frequency of the wave are not at all trivial and can be correctly discussed in terms of the Hilbert transform and analytic signal, especially for spread spectrum signals used in telecommunications, laser technology, modern electromagnetism and acoustics.

In some materials $\mathbf{v}_{ph}$ and $\mathbf{v}_g$ can be noncollinear and antiparallel by design as, e.g., in the so-called metamaterials. These are the media with negative refraction index $n$ as well as dielectric and magnetic functions ($\varepsilon < 0, \mu < 0, n < 0$). Of course, when the angle between the group and phase velocities is large, it is difficult to speak of the wave in its traditional sense as a periodic (time-harmonic) perturbation in space and time. Moreover, for a large frequency spread as, e.g., in the case of an ultrashort pulse propagating in a dispersive material, the pulse shape may rapidly change (at times comparable with the pulse duration), which makes the intuitive concept of a "wave group" somewhat unclear. Accordingly, group velocity $\mathbf{v}_g$, making sense for a small number of harmonic wave components, becomes an undetermined or even totally useless notion.

An interesting thing about wave motion is that water waves stand somewhat differently from electromagnetic, acoustic and seismic waves, as one could probably see above in the description of tsunami modeling.

## 11.2. Rays instead of waves

Rays appear in optics, acoustics, seismics or any other wave theory as an asymptotic limit of very large frequencies so that wave oscillations become indiscernible and must be averaged over. More accurately, the ray approximation is based on the assumption that the refraction index $n(\mathbf{r})$ or its analogs, characterizing the properties of the medium, as well as the wave field, vary slowly over the scale of wavelength $\lambda = \lambda_0/n$. In our daily life we mostly deal with rays, noticing wave effects (e.g., diffraction) as small corrections. Rays in nondispersive media can be described as the curves that are normal to the surfaces of constant phase; this is, however, not true for dispersive media, where rays are not necessarily orthogonal to the constant phase surfaces. One can cluster the rays producing ray tubes through which energy flows. The image of energy fluxes in the ray tubes is depicted in physical terms, this representation is, of course, utterly intuitive, but it helps to grasp the physical meaning of the concept "ray" which is rather nontrivial and can be approached from both mathematical[88] and physical standpoint [5].

The mathematical model of a ray is given by writing the wave field i.e., a solution to the Helmholtz equation $\Delta u + k^2 u = \Delta u + k_0^2 n(\mathbf{r})u = 0$ describing stationary wave propagation in the form $u(\mathbf{r}) = A(\mathbf{r})\exp\big(i\phi(\mathbf{r})\big)$, $\phi(\mathbf{r}) = k_0 n(\mathbf{r})\xi(\mathbf{r})(\mathbf{vl}) \approx k_0 n\xi(\mathbf{r})$. Here unit vectors $\mathbf{v}$ and $\mathbf{l}$ define respectively the direction of propagation ($\mathbf{v} = \mathbf{k}/k$) and trajectory ($\mathbf{l} = \boldsymbol{\xi}/\xi$). Quantity $\phi(\mathbf{r})$ is called eikonal and is often represented in the form $\phi = \phi_0 + k_0 \int_0^\mathbf{r} n(\mathbf{r})d\sigma$, where $d\sigma$ is a line element on the ray. One can also introduce the ray parameter $\tau$: $d\tau = d\sigma/n$ in order to represent rays of a wave field in the form of a dynamical system

$$\frac{d\mathbf{r}}{d\tau} = \mathbf{p}, \qquad \frac{d\mathbf{p}}{d\tau} = \frac{1}{2}\nabla n^2,$$

where $\mathbf{p} = \nabla\xi$ (notice that $\mathbf{p}$ is dimensionless here). In other words, if a ray propagates through the medium over distance $\tau$, its trajectory is given by equation $\mathbf{r}(\tau) = \mathbf{r}_0 + \mathbf{p}\tau$, the relation that is a base of the ray tracing techniques in computer graphics. The Helmholtz equation describes the stationary propagation of the waves of various nature, e.g., light, sound, electromagnetic signals, seismic waves, etc., if one puts $k_0 = \omega/c_0$, $n(\mathbf{r}) = c_0/c(\mathbf{r})$, where $c_0$ is some characteristic value of phase velocity the given medium and $c(\mathbf{r})$ is its local value. For the motion of a nonrelativistic quantum particle with mass $m$ and energy $E$ in potential $V(\mathbf{r})$, we have the stationary Schrödinger equation

$$(\Delta + k_0^2)\psi(\mathbf{r}) = \left(\frac{2mV(\mathbf{r})}{\hbar^2}\right)^{\frac{1}{2}}\psi(\mathbf{r}),$$

which has the form of the Helmholtz equation with $k_0 = \sqrt{2mE}/\hbar$, $n(r) = \sqrt{1 - V(\mathbf{r})/E}$ so that $k_0 n(\mathbf{r}) = \sqrt{2mE - V(\mathbf{r})}/\hbar$.

On the example of a ray, one can see the difference between physical and mathematical models: a mathematical ray $\mathbf{r} = \mathbf{r}(\sigma)$, where $d\sigma$ is the line element, is an infinitesimally thin spatial curve while a physical ray has a finite thickness $b$ determined by cross-sections of its Fresnel volume, $b \sim \sqrt{z/nk_0}$. Here $z$ is the coordinate along the direction of propagation (observation point). The ray theory in optics and the corresponding concepts of ray tubes and Fresnel volume are discussed in the classical book [28]. One can also find in this book a more physically motivated description of a ray, achieved in terms of a spatial (or spacetime) two-point correlation function defining the correlation radius. In this description, rays are treated as beams characterized by specific intensity which is expressed through the wave field variables. Anyway, rays are associated with the wave field and interpreted as its backbone system. Indeed, rays are defined by the normal vector to a small area on the wavefront (recall that a wavefront is an imaginary surface on which the wave field has a fixed phase with respect to some reference (source) point). To explore the wave field propagation in terms of rays is much easier than to deal directly with waves as spacetime excitations. Usually, information contained in the complete wave field description is immaterial and must be eventually integrated or averaged over. The ray representation disregards the superfluous information about fast changing instantaneous phase. Passing to the ray representation from the full wave field exposition is analogous to the transition to classical (semiclassical) description from quantum mechanics which is a typical wave theory.

---

[88] One should not, of course, confuse rays propagating in physical (Euclidean) space with rays in the functional-analytical sense as one-dimensional complex subspaces of the Hilbert space.

When passing to the ray representation, the wave field $u = Ae^{i\phi} = Ae^{ik_0 n\xi}, k_0 = \omega/c, \xi$ is real can be expanded into asymptotic series for $k_0 \to \infty$

$$A = A_0 + \frac{A_1}{ik_0} + \cdots = \sum_{m=0}^{\infty} \frac{A_m}{(ik_0)^m}. \qquad (11.2.1.)$$

(This expansion is sometimes called the "ray series"). As far as the phase $\phi$ goes, it is assumed that its gradient i.e., wave vector $\mathbf{k} = \nabla\phi = k_0\nabla(n\xi)$ in the medium is also a slow changing function of coordinates and can be expanded into asymptotic series like (11.2.1.). As a result, we get the representation of the wave field in the form of the product of slow varying amplitude by rapidly oscillating exponent:

$$u(r) = A(r)e^{i\phi(r)} = A(\mu r)e^{i\frac{\phi(\mu r)}{\mu}}, \qquad (11.2.2.)$$

$\mu = \frac{1}{k_0 L}$ is a dimensionless small parameter, $L$ is a characteristic scale on which amplitude $A(\mathbf{r})$, wave vector $\mathbf{k}(\mathbf{r}) = \nabla\phi(\mathbf{r})$ and refraction index $n(\mathbf{r})$ change. Representations (11.2.1.)-(11.2.2.) are convenient when the wave field i.e., amplitude $A$ and wave vector $\mathbf{k} = \nabla\phi$ change very slowly over the characteristic distance $L$ of the medium and on the local wavelength $\lambda(r) = 1/|k(r)|$ i.e. $\lambda|\nabla A| \ll |A|, \lambda|\nabla k_i| \ll |k_i|, \lambda|\nabla n| \ll n$. We may note that using equations (11.2.1.)-(11.2.2.) is not the only strategy of obtaining the ray representation: one can also use the Feynman integral techniques or the Lagrangian method [60], [59].

Because of the rapidly changing phase, the ray representation of the wave field is closely connected with the short wave (or high frequency) approximation that leads to asymptotic expansion of integrals of the kind

$$J(k_0, \boldsymbol{\alpha}) = \int_D A(x^1, \dots, x^n, \boldsymbol{\alpha})e^{ik_0\xi(x^1, \dots, x^n, \boldsymbol{\alpha})}dx^1 \dots dx^n$$

$$= \int_D A(x^1, \dots, x^n, \boldsymbol{\alpha})e^{\frac{iS}{\hbar}(x^1, \dots, x^n, \boldsymbol{\alpha})}dx^1 \dots dx^n, \qquad (11.2.3.)$$

as $k_0 \to \infty$ ($\hbar \to 0$), where functions $A, \xi$ and $S$ are assumed smooth, $\xi$ and $S$ are real, and integration is performed over some compact domain $D$ in $\mathbb{R}^n$. Integrals of the kind (11.2.3.) get their values near the stationary phase points $\mathbf{x} = \mathbf{x}_m, m = 1, 2, \dots$ defined by equations $\partial\psi/\partial\mathbf{x} = 0, \psi = \mathbf{k_0}\boldsymbol{\xi}$ or $\psi = S/\hbar$. One usually considers the Hessian $H_{ij}(\boldsymbol{\alpha}) = \partial^2\psi/\partial x^i\partial x^j$ to be nondegenerate ($\det H_{ij} \neq 0$). However, when parameter $\boldsymbol{\alpha} = (\alpha^1, \dots, \alpha^p)$ varies, so do isolated points $\mathbf{x}_m$, and for some values of $\boldsymbol{\alpha}$ these points may conflate forming a caustic defined by equation $\det H_{ij} = 0$. Caustics usually require more complicated treatment that is closer to the full wave field analysis.

The shortwave and high frequency asymptotic method results in a greatly simplified picture of wave propagation due to *localization*: solutions in a given point $\mathbf{r}$ of the medium depend only on its local properties expressed by $n(\mathbf{r})$, where $\mathbf{r}(\sigma)$ is located on the ray trajectory. This method is very close to the semiclass techniques in quantum mechanics. We shall discuss these techniques of solution of differential equations below (by necessity, on a rather primitive level).

A good illustration of asymptotic expansions in wave motion is given by the familiar Bessel functions $\mathcal{J}_m(x) = \int_{-\pi}^{\pi} e^{-i(m\varphi - x\sin\varphi)} \frac{d\varphi}{2\pi}$ for large order $m$ and large values of argument $x$. Recall that the Helmholtz equation has a solution in terms of Bessel functions $u(\mathbf{r}) = e^{\pm im\varphi}\mathcal{J}_m(kr)$, and their asymptotic representation ($kr \gg m$) corresponds to wave propagation far away from the source.

## 11.3. Classical Fields and Waves

One may easily notice that in the Newtonian picture of classical mechanics particles interact through instantaneous forces, $\mathbf{F} = m\ddot{\mathbf{r}}$. Forces $\mathbf{F}$ are always produced by other bodies or particles. But writing down the differential equation with forces is clearly not the whole story. So already in the first half of the 19th century, mostly due to experiments of M. Faraday, it became obvious that there was something else in the physical world besides bodies and particles. It is presumed that Faraday was the first to call this something a "field", although I failed to find this term in sources available to me on Faraday's works, see however http://en.wikipedia.org/wiki/Field (physics). Then J. C. Maxwell and O. Heaviside constructed the classical theory of the electromagnetic field.

The classical theory of fields is a remarkable subject because it unifies extremely abstract and mathematically advanced areas of modern theoretical physics and down-to-earth engineering. This intermediary, bridging position of classical electrodynamics is even more pronounced than in the case of classical mechanics. Although there exist a number of very good books on electromagnetic theory [81, 143], we still prefer the textbook by Landau and Lifshitz [96] and will often cite it. Almost all of the material contained in this chapter can be found in [96] except for some comments of mine, still I write out the main expressions – for the reader's convenience and because some of them may be needed further for more complex subjects. One may think that the classical field theory (CFT) in empty space is not a difficult subject – indeed, a great many of its results and occasional derivations appear boring and trivial, but this impression is deceptive. There exist a lot of refined implications of classical field theory and classical electrodynamics is full of understatements. But the most important thing is that CFT provides a solid foundation, a perfect model for all other field theories, and while touching upon them we shall recall CFT with gratitude.

## 11.4. The Maxwell Equations

The classical theory of fields [89] is based on Maxwell's equations which we have already discussed in other contexts. Since these equations are very fundamental, Let us write them once more. The inhomogeneous pair:

$$\nabla \mathbf{E} = 4\pi\rho, \qquad \nabla \times \mathbf{H} - \frac{1}{c}\frac{\partial \mathbf{E}}{\partial t} = \frac{4\pi}{c}\mathbf{j};$$

and the homogeneous pair:

---

[89] Here, we are limiting our discussion to an electromagnetic field.

$$\nabla \mathbf{H} = 0, \qquad \nabla \times \mathbf{E} + \frac{1}{c}\mathbf{H} = 0$$

When discussing dualities, we have mentioned the magnetic monopole. Modern physics does not exclude the possibility that the magnetic monopole might exist, and eventually it may be discovered. If the magnetic monopole really exists, then the right-hand side of the first equation from the homogeneous pair namely that expressing the solenoidal character of the magnetic field, $\nabla \mathbf{H} = 0$, should be not zero, but proportional to the density of magnetic monopoles. At present, however, this possibility is highly hypothetical, and we shall not take monopoles into account.

The electric and magnetic components of the electromagnetic field can be more conveniently written using the potentials[90]. By introducing the vector and scalar potentials, $A$ and $\varphi$, with

$$\mathbf{E} = -\nabla\varphi - \frac{1}{c}\frac{\partial \mathbf{A}}{\partial t}, \qquad \mathbf{H} = \nabla \times \mathbf{A} \tag{11.4.1.}$$

we get the wave-type equations for the potentials

$$\Delta \mathbf{A} - \nabla\left(\nabla \mathbf{A} + \frac{1}{c}\frac{\partial \varphi}{\partial t}\right) - \frac{1}{c^2}\frac{\partial^2 \mathbf{A}}{\partial t^2} = -\frac{4\pi}{c}\mathbf{j}$$
$$\nabla\varphi + \frac{1}{c}\nabla\frac{\partial \mathbf{A}}{\partial t} = -4\pi\rho \tag{11.4.2.}$$

Below, we shall often use the symbol $\square := \Delta - \frac{1}{c^2}\frac{\partial^2}{\partial t^2}$ (the D'Alembertian).

These are four coupled linear partial differential equations. Nevertheless, this system of equations with respect to four scalar functions is obviously simpler than the initial system of Maxwell equations for six field components. In fact, the matter is even more complicated, since one has to account for the motion of charged particles in the electromagnetic field. This is a self-consistency problem because these particles, on the one hand, are experiencing forces from the field and, on the other hand, themselves contribute to the field. The charge and current densities on the right-hand side are defined as

$$\rho(\mathbf{r}, t) = \sum_a e_a \delta\big(\mathbf{r} - \mathbf{r}_a(t)\big)$$

and

$$j(\mathbf{r}, t) = \sum_a e_a \dot{\mathbf{r}}_a(t)\delta\big(\mathbf{r} - \mathbf{r}_a(t)\big),$$

where $\mathbf{r}_a(t), \dot{\mathbf{r}}_a(t)$ are unknown quantities (here summation goes over all charged particles in the considered system). Thus, in order to close the self-consistent system of equations, we must provide equations for these quantities. In the classical case, for example, we may supplement the Maxwell equations with the Newton equations containing the Lorentz force:

$$m\ddot{\mathbf{r}}_a(t) = e_a \left[ \mathbf{E}(\mathbf{r}_a(t)) + \frac{1}{c}\left(\dot{\mathbf{r}}_a(t) \times \mathbf{H}(\mathbf{r}_a(t))\right) \right]$$

---

[90] It is worth noting that the four expressions that we know as the Maxwell equations are probably the creation of O. Heaviside who reduced the original system of J. C. Maxwell consisting of twenty equations to four vector equations, see, e.g., http://en.wikipedia.org/wiki/Maxwell's equations. We also highly recommend the beautiful book about O. Heaviside by a prominent Russian physicist and a very gifted writer, Prof. B. M. Bolotovskii. Unfortunately, this book is in Russian and I don't know whether its translation into other European languages exists.

(we neglect in these equations of motion close interparticle interactions, possibly of non-electromagnetic character). The solution of such self-consistent electromagnetic problems is usually a difficult task, it is usually treated in plasma physics and while considering the interaction of beams of charged particles with matter. Moreover, self-consistency leads in its limit to self-interaction of charges which has always been a source of grave difficulties in early attempts of electromagnetic field quantization. However, in classical field theory particles, in particular, electric charges, due to relativistic requirements, must be considered point-like (see [96], §15), which results in an infinite self-energy of the particle. Therefore, classical field theory becomes self-contradictory (at least at small distances) and should be replaced by a more advanced theory [96], §37, see also below.

The charge and current densities automatically satisfy the continuity equation

$$\frac{\partial \rho}{\partial t} + div\mathbf{j} = 0, \tag{11.4.3.}$$

which can be actually obtained from the Maxwell equations by taking the divergence of the $\nabla \mathbf{H}$ and using the Coulomb law equation $\nabla \mathbf{E} = 4\pi\rho$. The fact that the continuity equation, which expresses the electric charge conservation, is not independent of Maxwell's equations means that the latter are constructed in such a way as to be in automatic agreement with the charge conservation law. More than that, one can even say that charge conservation is a more fundamental concept than the Maxwell equations, since it results in their generalizations (see below).

Before we produce solutions to the equations for electromagnetic potentials, let us discuss some properties of the Maxwell equations. When speaking about differential equations, the first thing to pay attention to is their invariance (symmetry) properties. In many cases the requirement of an invariance of physically relevant equations, e.g., the motion equations with respect to some group of transformations allows one to single out the mathematical model from a wide class of available equations. For instance, one can prove [63] that among all systems of the first-order partial differential equations for two vector-functions $\mathbf{E}(\mathbf{r}, t)$, $\mathbf{H}(\mathbf{r}, t)$ there exists a unique system of equations invariant under the Poincaré group. This system is the Maxwell equations.

When discussing dualities in physics, we have already noticed that the Maxwell equations are invariant with respect to a dual change of functions performed by Heaviside [74]

$$\mathbf{E} \to \mathbf{H}, \qquad \mathbf{H} \to -\mathbf{E}. \tag{11.4.4.}$$

Later, this symmetry was generalized by to a single parameter family of plane rotations, $R(\theta)$:

$$\begin{pmatrix} \mathbf{E}' \\ \mathbf{H}' \end{pmatrix} = R(\theta) \begin{pmatrix} \mathbf{E} \\ \mathbf{H} \end{pmatrix},$$

where

$$R(\theta) = \begin{pmatrix} \cos\theta & \sin\theta \\ -\sin\theta & \cos\theta \end{pmatrix}$$

or

$$\mathbf{E}' = \mathbf{E}\cos\theta + \mathbf{H}\sin\theta$$
$$\mathbf{H}' = -\mathbf{E}\sin\theta + \mathbf{H}\cos\theta.$$  (11.4.5.)

The Lie-group analysis performed by a well-known Russian mathematician N. Ibrahimov shows that the ultimate local symmetry group for the system of Maxwell's equations in empty space is a 16-parameter group $C(1,3) \otimes SO(2,R)$ where $SO(2,R)$ is represented by $R(\theta)$ above.

## 11.5. Gauge Invariance in Classical Electrodynamics

Probably, the most important invariance property of the Maxwell equations is connected with the transformation of the potentials ([96], §18)

$$\mathbf{A} \to \mathbf{A}' = \mathbf{A} + \nabla\chi, \qquad \varphi \to \varphi' = \varphi - \frac{1}{c}\frac{\partial\chi}{\partial t},$$  (11.5.1.)

which is called the gauge transformation. It is clear that the fields $\mathbf{E}, \mathbf{H}$ do not change under this transformation. In other words, different potentials $(\varphi, \mathbf{A})$ and $(\varphi', \mathbf{A}')$ correspond to the same physical situation. This fact is known as gauge invariance. The term gauge refers to some specific choice of potentials, or what is synonymous, to a fixed attribution of the function $\chi$ which is usually called the gauge function. When one considers quantum-mechanical particles described by the wave function $\psi$, coupled to the electromagnetic field, one must extend the transformations of the electromagnetic potentials with the aid of gauge function $\chi$ by including the change of the phase of wave function

$$\psi \to \psi' = \psi\exp\left(\frac{ie\chi}{\hbar c}\right)$$  (11.5.2.)

In recent years, it has been established that the gauge invariance is not only a fundamental property of classical field theory, but it defines other fundamental interactions in nature, perhaps even all possible interactions. In particular, the principle of gauge invariance plays a decisive role in the so-called "Standard Model" that has been designed to describe electroweak and strong interactions between elementary particles. In the Standard Model of particle physics, all particles (other than the Higgs boson), transform either as vectors or as spinors. The vector particles are also called "gauge bosons", and they serve to carry the forces in the Standard Model. The spinor particles are also called fermions, and they correspond to the two basic constituent forms of matter: quarks and leptons. In this section, however, we shall only deal with electromagnetic forces – the low-energy classical limit of electroweak interactions. In the classical description of electromagnetic fields, gauge invariance brings about supplementary and seemingly unnecessary degrees of freedom which may be called the gauge degrees of freedom. In classical electromagnetic theory, we may interpret the electromagnetic potentials, $(\varphi, \mathbf{A})$, as a tool introduced to simplify the Maxwell equations which, however, does not have any direct experimental relevance. Immediately observable quantities of classical electromagnetic theory are the electric and magnetic fields, $\mathbf{E}, \mathbf{H}$, corresponding to gauge equivalence classes of the potentials $(\varphi', \mathbf{A}') \sim (\varphi, \mathbf{A})$ giving the same electric and magnetic fields. If one takes in the canonical formalism the electromagnetic potentials as phase space variables in order to obtain the Euler-Lagrange equations from the least action principle (see below "Equations of Motion for the Electromagnetic Field"), then one sees that the dimensionality of the classical phase space is reduced[91].

As is well known (see [96] , §46), in those situations when relativistic (Lorentz) invariance should be explicitly ensured one can use the Lorentz gauge, i.e., the condition

$$\nabla \mathbf{A} + \frac{1}{c}\frac{\partial \chi}{\partial t} = 0 \qquad (11.5.3.)$$

imposed on the potentials. In this case, the above equations for the potentials take the form of usual (linear) wave equations

$$L\mathbf{A} = -\frac{4\pi}{c}\mathbf{j}, \qquad L\varphi = -4\pi\rho, \qquad L := \Delta - \frac{1}{c^2}\frac{\partial^2}{\partial t^2}, \qquad (11.5.4.)$$

$L$ being the wave operator which coincides here with the D'Alembertian (this is not always the case since there may be many wave operators).

It is important to note that in many physical situations the manifest Lorentz invariance is irrelevant since one typically uses a privileged reference frame, the one being fixed with respect to the observer or the media. Thus, in considering the interaction of an electromagnetic field with atoms other gauges are more convenient. The most popular gauge in physical situations with a fixed frame of reference is the so-called Coulomb gauge defined by the transversality condition $\nabla \mathbf{A} = \mathbf{0}$. The terms "Coulomb gauge" and "transversality" can be easily explained. Suppose we managed to impose the condition $\nabla \mathbf{A} = \mathbf{0}$ – a little further we shall prove that it is always possible. Then we obtain the following equations for the potentials

$$l\Delta\mathbf{A} - \frac{1}{c^2}\frac{\partial^2 \mathbf{A}}{\partial t^2} - \frac{1}{c}\frac{\partial}{\partial t}\nabla\varphi - \frac{4\pi}{c}\mathbf{j} \qquad (11.5.5.)$$

$$\Delta\varphi = -4\pi\rho. \qquad (11.5.6.)$$

The last equation is exactly the same as the one from which the electrostatic potential can be determined. For the model of point particles, $\rho(\mathbf{r}, t) = \sum_a e_a \delta(\mathbf{r} - \mathbf{r}_a(t))$, the scalar potential becomes

$$\varphi(\mathbf{r}, t) = \sum_a \frac{e_a}{|\mathbf{r} - \mathbf{r}_a(t)|},$$

which again coincides with the form of elementary electrostatic equation. The solution for potential $\varphi$ may be written through the Coulomb Green's function

$$G_0(\mathbf{r}, \mathbf{r}') = -\frac{1}{4\pi}\frac{1}{|\mathbf{r} - \mathbf{r}'|}$$

---

[91] This phase space reduction may be accompanied by a change in geometry, the latter becoming more complicated, curved, and containing, e.g., holes. No canonically conjugated coordinates such as electromagnetic potentials and conjugated to them "momenta" may be globally possible, which makes a direct canonical quantization of the electromagnetic field quite difficult.

This purely electrostatic form of the scalar potential which is represented as the superposition of Coulomb potentials created by individual charges, explains the name of the Coulomb gauge. Let us now try to explain why this gauge is also called transversal. To do this we can, for example, expand the potentials into plane waves, as in [96] , §§51,52:

$$\mathbf{A}(\mathbf{r}, t) = \sum_{\mathbf{k}, j=1,2} \mathbf{A}_{\mathbf{k}, j}(t) \exp(i\mathbf{k}_j \mathbf{r}),$$

where index $j$ denotes two possible polarizations of the vector plane waves normal to the wave vector $\mathbf{k}$. We know from elementary functional analysis that plane waves form a full system of functions so that they may be taken as a basis and a series over them is legitimate. Below, when discussing the field quantization, we shall consider some details of expansions over the system of plane waves. Now, all we need is just a general form of such an expansion.

Inserting the expression for $\mathbf{A}(\mathbf{r}, t)$ into $\nabla \mathbf{A} = \mathbf{0}$, we see that

$$\nabla \mathbf{A}(\mathbf{r}, t) = i \sum_{\mathbf{k}, j=1,2} \left( \mathbf{k} \mathbf{A}_{\mathbf{k}, j}(t) \right) \exp(i\mathbf{k}_j \mathbf{r}) = 0$$

may hold for arbitrary $\mathbf{A}$ only if $\mathbf{k} \mathbf{A}_{\mathbf{k}, j} = 0$, which means that all the modes $\mathbf{A}_{\mathbf{k}, j}$ should be transversal to the wave vector. If we similarly expand the scalar potential

$$\varphi(\mathbf{r}, t) = \sum_{\mathbf{k}, j=1,2} \varphi_{\mathbf{k}, j}(t) \exp(i\mathbf{k}_j \mathbf{r}),$$

then we shall be able to represent the electric field as the sum of the longitudinal and the transversal components corresponding to the terms

$$\mathbf{E}_{\parallel} = -\nabla \varphi = -i \sum_{\mathbf{k}, j=1,2} \mathbf{k} \varphi_{\mathbf{k}, j}(t) \exp(i\mathbf{k}_j \mathbf{r}),$$

and

$$\mathbf{E}_{\perp} = -\frac{1}{c} \frac{\partial \mathbf{A}}{\partial t} = -\frac{1}{c} \partial_t \mathbf{A} = \sum_{\mathbf{k}, j=1,2} \dot{\mathbf{A}}_{\mathbf{k}, j}(t) \exp(i\mathbf{k}_j \mathbf{r})$$

The magnetic field comprised of components $\left( \mathbf{k}_j \times \mathbf{A}_{\mathbf{k}, j} \right)$ is obviously transversal ($\nabla \mathbf{H} = 0$). Thus, the Coulomb gauge allows one to separate the entire electromagnetic field into the transversal component corresponding to the vector potential $\mathbf{A}$ and the longitudinal component described by the scalar potential $\varphi$. This separation is usually quite convenient, at least for nonrelativistic problems – for instance, when treating the interaction of radiation with matter. In this class of problems, one may describe forces between the particles of the medium by the scalar potential $\varphi$ whereas the radiation field is described by the vector potential $\mathbf{A}$ alone.

This is all more or less banal, nevertheless, there are some nontrivial questions around such a decomposition of the electromagnetic field into transversal and longitudinal components. To begin with, we see that the electrostatic form of the scalar potential in the Coulomb gauge implies an instantaneous nature of the longitudinal field. How then can it be that the electromagnetic signal is causal and propagates with the speed $c$, as prescribed by relativity theory? This is obviously an apparent paradox, but its discussion leads to interesting representations of the gauge function.

## 11.6. Four-Dimensional Formulation of Electrodynamics

Many elementary results formulated above in the conventional language of vector analysis can be cast into a more elegant form using the invariance properties of the Maxwell equations, so this section in fact does not contain unfamiliar expressions. In principle, the Maxwell equations admit a number of various formulations (see below), but the most popular one – the four-dimensional formulation – is based on the use of four-component function $A_\mu = (\varphi, \mathbf{A})$ often called the four-dimensional potential [96], §16, or simply the 4-potential. This function is connected with the fields $\mathbf{E}, \mathbf{H}$ by the familiar relations which we write here, for future usage, through the momentum operator $\mathbf{p} = -i\nabla, p_\mu = -i\partial_\mu$:

$$\mathbf{E} = \frac{\partial \mathbf{A}}{\partial x_0} - i\mathbf{p}A_0, \qquad \mathbf{H} = i\mathbf{p} \times \mathbf{A}$$

Then the homogeneous Maxwell equations (see the next section) are satisfied identically, if one introduces the so-called electromagnetic field tensor as a four-dimensional rotation:

$$F_{\mu\nu} = -F_{\nu\mu} = \partial_\mu A_\nu - \partial_\nu A_\mu.$$

Indeed,

$$\partial_\lambda F_{\mu\nu} + \partial_\mu F_{\nu\lambda} + \partial_\nu F_{\lambda\mu} = 0$$

or, in terms of the dual tensor [96], §26, $F^{*\mu\nu} = \frac{1}{2}\epsilon^{\mu\nu\alpha\beta}F_{\alpha\beta}, \partial_\mu F^{*\mu\nu}$. The inhomogeneous Maxwell equations are

$$\partial_\mu F^{\mu\nu} = -\frac{4\pi}{c}j^\nu,$$

which follows from the variational principle with the Lagrangian density, see [96], §28,

$$\mathcal{L} = -\frac{1}{c}A_\mu j^\mu - \frac{1}{16\pi}F_{\mu\nu}F^{\mu\nu}.$$

We have seen that for a given electromagnetic field $A_\mu$ is not unique, since the gauge transformation $A_\mu \to A'_\mu = A_\mu + \partial_\mu \chi(x)$ leaves $F_{\mu\nu}$ unchanged:

$$F_{\mu\nu} \to F'_{\mu\nu} = \partial_\mu A'_\nu - \partial_\nu A'_\mu = \partial_\mu(A_\nu + \partial_\nu \chi) - \partial_\nu(A_\mu + \partial_\mu \chi) = F_{\mu\nu} + (\partial_\mu \partial_\nu - \partial_\nu \partial_\mu)\chi = F_{\mu\nu}.$$

If we raise index $\mu$ in $A_\mu$, we can write

$$\partial_\mu A'^\mu = \partial_\mu(A^\mu + \partial^\mu \chi) = \partial_\mu A^\mu + \partial_\mu \partial^\mu \chi$$

or, if we choose the gauge function $\chi$ to satisfy the equation

$$\partial_\mu \partial^\mu \chi = -\partial_\mu A^\mu$$

(we have seen above that we can always do it due to gauge invariance) or, in three-dimensional representation,

$$\Box \chi = -\frac{\partial A^0}{\partial x_0} - \nabla A = \frac{1}{c}\frac{\partial \varphi}{\partial t} - \nabla A,$$

then we get the familiar Lorentz gauge $\partial_\mu A^\mu = 0$. This supplementary condition reduces the number of independent components of $A_\mu$ to three, yet it does not ensure that $A_\mu$ is uniquely defined. From the above formulas for transition from $A'_\mu$ to $A_\mu$ it becomes clear that if $A_\mu$ satisfies the Lorentz gauge, so will $A'_\mu$ provided $\Box \chi \equiv \partial_\mu \partial^\mu = 0$. Thus, the ambiguity due to gauge invariance persists, and one needs a more restrictive constraint to eliminate it. We have seen that we can, for example, impose the condition

$$\partial_0 \chi = \partial x_0 = \frac{1}{c}\frac{\partial \chi}{\partial t} = -\varphi,$$

then $A'_0 = A_0 + \partial_0 \chi = A_0 - \varphi = \varphi - \varphi = 0$, i.e., we may put $\varphi = 0$ and $\nabla \mathbf{A} = 0$ – the Coulomb or radiation gauge discussed above. This gauge reduces the number of independent components of $A_\mu$ to only two, that is one more reason why working in the Coulomb gauge is usually *always* more convenient than in the Lorentz gauge unless it is indispensable to retain the explicit Coulomb invariance.

Let us now make a remark of a purely technical character. Some authors introduce the vector potential as a contravariant quantity, $A^\mu = (\varphi, \mathbf{A})$ which gives the corresponding skew-symmetric electromagnetic field tensor, $F^{\mu\nu} = -F^{\nu\mu} = \partial^\mu A^\nu - \partial^\nu A^\mu$. It is of course just a technicality and a matter of taste, since one can raise or lower indices with the help of the metric tensor which in the Minkowski (flat) background is simply $g_{\mu\nu} = \gamma_{\mu\nu} = diag(1, -1, -1, -1)$. So, both definitions differ by the sign of $\mathbf{A}$ in $A^\mu = (\varphi, \mathbf{A})$. However, such a definition is less convenient than $A_\mu$, specifically when introducing the covariant derivative $\partial_\mu \to \partial_\mu + (ie/c)A_\mu$. Moreover, the 1-form $A_\mu dx^\mu$, which is the connection form in the coordinate basis, may be used to write the particle-field coupling term in the action $S$ (see below)

$$S_{pf} = -\frac{1}{c}\sum_a \int e_a A_\mu(x_a)dx_a^\mu,$$

where the integral is taken over the world lines of particles, or in the "current" form

$$S_{pf} = -\frac{1}{c^2}\int j^\mu A_\mu d^4x.$$

Recall that the total action for the electromagnetic field, $S = S_p + S_f + S_{pf}$ is the additive combination of terms corresponding to particles (charges) without field, field with no particles, and particle-field coupling, respectively.

Using the four-component vector potential $A_\mu$ enables us to be well-prepared for field quantization in quantum field theory. Inserting $A_\mu$ and the corresponding expression for the fields into the Maxwell equations, we get the unified system of equations for the potentials $(\varphi, \mathbf{A})$ obtained previously in three-dimensional form:

$$p_\mu p^\mu A_\nu - p_\nu p_\mu A^\mu = \frac{4\pi}{c} j_\nu.$$

This is, as we have seen, the system of four linear equations for $A_\mu$ instead of eight Maxwell's equations. Here, this wave-like system is intentionally written through momenta, $p_\mu = -i\partial_\mu$. One can solve such a system – a linear system always can be solved for given source terms $j^\mu$ – and then obtain the experimentally observable fields $\mathbf{E}$ and $\mathbf{H}$. Thus, using the potentials results in a considerable simplification.

Now we may exploit gauge invariance, which permits, as we have seen, a certain arbitrariness in the choice of $A_\mu$. Indeed, we have already discussed that the Maxwell equations expressed through the potentials are invariant with respect to the substitution

$$A_\mu \to A'_\mu = A_\mu + \partial_\mu \chi = A_\mu + i p_\mu \chi,$$

where $\chi$ is some arbitrary function. In the four-dimensional formulation it is natural to use the Lorentz constraint, $p_\mu A^\mu = 0$, which allows one to uncouple the above system of equations for $A_\mu$.

Here, a remark about the use of potentials may be pertinent. Highschool and even some university students typically don't completely understand the meaning of potentials as well as the idea of introducing them. Only by trying to solve vector problems directly, with a lot of tedious computations, people begin to value a clever trick with a single scalar function, the electrostatic potential, from which the fields can be obtained by a simple differentiation. The fact that the electrostatic potential is defined up to a constant then becomes self-evident. Less evident, however, is that this fact is a manifestation of some hidden invariance which we now call the gauge invariance. Decoupling the system of equations for $A_\mu$ is basically the same thing, more generality is reflected in replacing the electrostatic relationship $\partial c / \partial \mathbf{r} = 0$, where $c = const$, by, e.g., $p_\mu A^\mu = 0$.

## 11.7. Classical Electromagnetic Field without Sources

This section is entirely dedicated to an elementary description of the classical electromagnetic field in the absence of electric charges and currents ($\rho = 0, \mathbf{j} = 0$). In this case, the Maxwell equations in vacuum take the homogeneous form

$$\nabla \mathbf{E} = 0, \qquad \nabla \times \mathbf{H} - \frac{1}{c} \frac{\partial \mathbf{E}}{\partial t} = 0, \qquad \nabla \mathbf{H} = 0, \qquad \nabla \times \mathbf{E} + \frac{1}{c} \mathbf{H} = 0$$

(As usual, we are writing this system as an array of two parts.) In terms of the field tensor (see [96] , §23), $F_{\mu\nu} = \partial_\mu A_\nu - \partial_\nu A_\mu$, take in four-dimensional notations a particularly compact form

$$\varepsilon^{\mu\nu\rho\sigma} \partial_\nu F_{\rho\sigma} = 0$$

or

$$\partial_\sigma F_{\mu\nu} + \partial_\nu F_{\sigma\mu} = 0$$

the first pair of Maxwell's equations and $\partial_\mu F^{\mu\nu}$ – the second pair. Recall that $\varepsilon$ denotes, as we have seen several times, the absolute antisymmetric tensor of the fourth rank whose components change sign with the interchange of any pair of indices and $\varepsilon^{0123} = 1$ by convention, see [96], §6 [92]. When discussing dualities in physics, we have already dealt with four-dimensional notations in electromagnetic theory, and we shall discuss this formulation later in connection with some geometric concepts of field theory. For many practical problems, however, such as the interaction of radiation with matter the four-dimensional formulation seems to be inconvenient, requiring superfluous operations. One may notice in this context that as long as we limit ourselves to the domain of special relativity (flat or Minkowski space), no distinction between contra- and covariant components is, strictly speaking, necessary and no metric tensor $g_{\mu\nu}$ can be introduced. We, however, shall be using the Minkowski space metric tensor $\gamma_{\mu\nu}$ from time to time, its only function will be to raise or to lower indices.

Introducing explicitly the vector and scalar potentials, $\nabla \times \mathbf{A} = \mathbf{H}, -\nabla\varphi - \frac{1}{c}\frac{\partial \mathbf{A}}{\partial t} = \mathbf{E}$, we get the wave equations for the potentials

$$\Delta \mathbf{A} - \nabla\left(\nabla \mathbf{A} + \frac{1}{c}\frac{\partial\varphi}{\partial t}\right) - \frac{1}{c^2}\frac{\partial^2 \mathbf{A}}{\partial t^2} = 0 \qquad (11.7.1.)$$

and

$$\nabla\varphi + \frac{1}{c}\nabla\frac{\partial \mathbf{A}}{\partial t} = 0 \qquad (11.7.2.)$$

or, in the four-dimensional notation

$$\frac{\partial^2 A_\mu}{\partial x_\nu \partial x^\mu} - \frac{\partial}{\partial x^\mu}\left(\frac{\partial A^\nu}{\partial x^\nu}\right) \equiv \partial_\nu\partial^\mu A_\mu - \partial_\mu\partial_\nu A^\nu = 0, \qquad (11.7.3.)$$

where, as usual, $A^\nu = \gamma^{\nu\mu}A_\mu$. Being a direct consequence of the Maxwell equations, these wave equations are of extreme importance in their own right, since they lead to the representation of the Maxwell field as a set of oscillator equations, which is the only equation that has an equidistant spectrum, and it is this equidistant spectrum, as we have already noticed, that allows one to interpret the field as consisting of independent particles – electromagnetic quanta or photons. Therefore, two accidentally occurring reasons Maxwell's equations leading to harmonic oscillator and the unique equidistant character of its spectrum, being combined render an essential part of the conceptual frame for entire contemporary physics. Physically speaking, since the charges interact with each other, and if they did not interact it would be impossible to observe them, the electromagnetic field becomes a necessity, and hence photons appear as an inevitable consequence of electromagnetic field quantization (based on the oscillator model). Thus, photons invoke the idea of interaction carriers. This idea has later evolved into the concept of gauge bosons.

## 11.8. Equations of Motion for the Electromagnetic Field

We have already discussed that the equations of motion for any field may – in fact must – be derived from an appropriate Lagrangian density using the least action principle. This fact is, in our terminology, the backbone of modern physics. In this approach, however, one has to correctly provide the Lagrangian density. It is not difficult to demonstrate (see [97]) that the Lagrangian density of the form

$$\mathcal{L}_0 = \frac{E^2 - H^2}{8\pi} = -\frac{1}{16\pi} F_{\mu\nu} F^{\mu\nu} \equiv -\frac{1}{16\pi} (\partial_\mu A_\nu - \partial_\nu A_\mu)(\partial^\mu A^\nu - \partial^\nu A^\mu) \qquad (11.8.1.)$$

can suit our purposes. Indeed, the Euler-Lagrange equations for a Lagrangian density $\mathcal{L}$

$$\frac{\partial}{\partial x^\mu} \frac{\partial \mathcal{L}}{\partial \left(\frac{\partial A^\nu}{\partial x^\nu}\right)} - \frac{\partial \mathcal{L}}{\partial A^\mu} \equiv \partial_\mu \frac{\partial \mathcal{L}}{\partial (\partial_\nu A^\nu)} - \frac{\partial \mathcal{L}}{\partial A^\mu} = 0$$

in the case of $\mathcal{L} = \mathcal{L}_0$ are reduced to the first term only, $\partial_\mu \frac{\partial \mathcal{L}}{\partial(\partial_\nu A^\nu)}$, that is

$$\frac{\partial \mathcal{L}_0}{\partial(\partial_\mu A_\nu)} - \frac{1}{16\pi} \left\{ \frac{\partial}{\partial(\partial_\mu A_\nu)} \left[ (\partial_\rho A_\sigma - \partial_\sigma A_\rho)(\partial_\rho A_\sigma - \partial_\sigma A_\rho) \right] \right\}$$

$$= -\frac{1}{8\pi} \left[ \frac{\partial}{\partial(\partial_\mu A_\nu)} \left( \partial_\rho A_\sigma \partial_\rho A_\sigma - \partial_\rho A_\sigma \partial_\sigma A_\rho \right) \right]$$

$$= \frac{1}{4\pi} (\partial_\mu A_\nu - \partial_\nu A_\mu) \qquad (11.8.2.)$$

---

[92] Not always, sometimes $\varepsilon_{0123} = 1$ is taken. This discrepancy of conventions, though unimportant, leads to confusion.

Inserting this expression into the Euler-Lagrange equation, we get the already familiar wave equation derived from the Maxwell equations

$$\partial_\nu \partial^\nu A_\mu - \partial_\mu \partial_\nu A^\nu = 0.$$

Thus, the wave equation may be regarded as the motion equation for an electromagnetic field.

One must note that $\mathcal{L}_0$ is not the unique Lagrangian density that produces the Maxwell equations. Indeed, if we add to $\mathcal{L}_0$, for example, the four-divergence of some vector $\chi^\mu$ (which may be a function of $A_\mu, x^\mu$), the equations of motion obtained from the Euler-Lagrange equations for the new Lagrangian density $\mathcal{L} = \mathcal{L}_0 + \partial_\mu \chi^\mu$ will coincide with those for $\mathcal{L} = \mathcal{L}_0$ (see [97]). More generally, any two Lagrangian densities that would differ by terms vanishing while integrated over spacetime, produce the same equations of motion.

This subject of field invariants is beautifully exposed in §25 of [93], and the only justification of repeating it here would be to provide some fresh views or interpretations.

## 11.9. Hamiltonian Formalism in Electromagnetic Theory

In this section, we shall deal with the Hamiltonian approach applied to classical electrodynamics. The intention is to represent electrodynamics in a form close to mechanics. Here, using the Hamiltonian formalism in electromagnetic theory is little more than a heuristic catch; this formalism emphasizing evolution in time does not seem to be perfectly tuned to relativistic problems. However, to gain more insight into field theory, we shall briefly discuss its Hamiltonian description.

When trying to apply the standard Hamiltonian formalism to a free electromagnetic field one can start, for example, from the gauge invariant Lagrangian density

$$\mathcal{L}_0 = -\frac{1}{16\pi c}F_{\mu\nu}F^{\mu\nu} = -\frac{1}{16\pi c}F_{\mu\nu}^2 = -\frac{1}{8\pi c}(\mathbf{H}^2 - \mathbf{E}^2).$$

Recall that the total action for the electromagnetic field is [96], §27

$$S = -\sum_a \int m_a c\, dl - \sum_a \int \frac{e_a}{c}A_\mu dx^\mu - \frac{1}{16\pi c}\int F_{\mu\nu}^2 d^4x.$$

Taking, as usual, the field potentials $A_\mu$ as the generalized coordinates, we obtain the canonically conjugate momenta

$$\Pi_\mu = \frac{1}{c}\frac{\partial \mathcal{L}}{\partial\left(\frac{\partial A^\mu}{\partial x^0}\right)}$$

so that the Hamiltonian density is

$$\mathcal{H}_0 = c\Pi_\mu \partial_0 A^\mu - \mathcal{L}_0.$$

One may immediately notice that by virtue of the relationship

$$\frac{\partial \mathcal{L}}{\partial\left(\partial_\mu A_\nu\right)} = -\frac{1}{4\pi c}F^{\mu\nu} = -\frac{1}{4\pi}\left(\partial^\mu A^\nu - \partial^\nu A^\mu\right)$$

the momentum $\Pi_0$ conjugate to the scalar potential $\varphi = A^0$ is identically zero. This is the old difficulty, well known to relativistic field theorists, see, e.g. [136]. There exist a number of ways to circumvent this difficulty, but the ultimate cause for it is the limited adequacy of the Hamiltonian formalism for relativistic field problems. Then we may sacrifice the Lorentz invariance anyway and choose the Coulomb gauge $div\mathbf{A} = 0, \varphi = 0$. In this gauge we can write

$$\mathcal{L}_0 = \frac{1}{8\pi}\left[\frac{1}{c^2}(\partial_t\mathbf{A})^2 - \left(\partial_i A_j - \partial_j A_i\right)^2\right],$$

where $i, j = 1, 2, 3$. From this Lagrangian density we obtain the field momentum components

$$\Pi_i = \frac{\partial \mathcal{L}_0}{\partial\left(\partial_t A^i\right)} = \frac{1}{4\pi c^2}\partial_t A_i = -\frac{1}{4\pi c}E_i$$

and the Hamiltonian density is obviously

$$\mathcal{H}_0 = \Pi_i \partial_t A^i - \mathcal{L}_0 = \frac{1}{8\pi}\left[\frac{1}{c^2}(\partial_t\mathbf{A})^2 + \left(\partial_i A_j - \partial_j A_i\right)^2\right] = \frac{1}{8\pi}(\mathbf{E}^2 + \mathbf{H}^2).$$

Using the Hamiltonian approach in electrodynamics mainly appeals to human intuition which tends to gauge everything by the archetypes inherited from classical mechanics. We have seen that the Hamiltonian formalism in classical mechanics was developed as a geometric theory on a finite-dimensional symplectic manifold having an even (2n) dimensionality. While trying to apply this formalism to electromagnetic theory we are less interested in geometrical aspects of symplectic manifolds, mostly focusing on the form of the Hamiltonian operator convenient for calculations[93]. This was, e.g., the strategy adopted in the classical textbook by W. Heitler [75] as well as many other sources related to the early development of field theory. Today, more sophisticated methods are commonly in use, and the Hamiltonian formalism is not considered well-adapted to treat the electromagnetic field regarded as a Hamiltonian system. Simply speaking, a Hamiltonian system is described by pairs $\left(p_i, q^j\right)$ of canonically conjugated local variables living in some vector space $Z \coloneqq (p, q)$. Then a Hamiltonian function $\mathcal{H}\left(p_i, q^j\right)$ is supposed to exist, with the property that the equations of motion are produced from it as from some potential in the form of a symplectic gradient

$$\dot{p}_i = -\frac{\partial \mathcal{H}}{\partial q^i}, \qquad \dot{q}^i = \frac{\partial \mathcal{H}}{\partial p_i}, \qquad i = 1, \dots, n$$

One can observe that this formalism was initially developed for finite-dimensional systems; however, it can be generalized to their infinite-dimensional analogs. One of the possibilities (probably not unique) for such a generalization is to replace coordinates $q^i$, at least some of them, by fields $\varphi^i(x^\alpha)$. Thus for an ensemble of particles and fields the Hamiltonian function is just a sum of kinetic energies of the particles and the energy contained in the electromagnetic field. Recall that for a conservative system the Hamiltonian is represented as the energy of the system expressed through canonical variables, e.g., $p, q$. The energy of a system "particles + EM field" can be written as

$$\mathcal{E} = \frac{1}{2} \sum_a m_a v_a^2 + \int d^3 r \frac{\mathbf{E}^2 + \mathbf{H}^2}{8\pi}, \qquad (11.9.1.)$$

where index $a$ enumerates particles. Here, an interesting and not quite trivial question arises: where is the interaction between particles and fields? The answer depends on how we define an electromagnetic field. If we understand by $\mathbf{E}$ and $\mathbf{H}$ the total electromagnetic field and not only its free part, then the energy of electromagnetic interaction is hidden in the total field energy. Recall that here we forget about all other fields acting on particles except the electromagnetic one.

Notice that although expression (11.9.1.) gives the energy of the system "particles + EM field", it is still not the Hamiltonian function because this energy is expressed through the Lagrangian (TM) variables $\left(q^i, \dot{q}^i\right)$ i.e., through pairs $(\mathbf{r}_a, \mathbf{v}_a)$.

This was a somewhat naive and straightforward construction of the Hamiltonian function for electromagnetism. We have seen above that the Hamiltonian formalism makes the passage to quantum theory intuitive and convenient.

## 11.10. Limitations of Classical Electromagnetic Theory

There are, of course, many limitations in any classical theory, which require a consistent quantum treatment. Traditionally, such limitations of classical electrodynamics are discussed in the initial chapters of textbooks on quantum field theory, and we can also observe some deficiencies of the classical electromagnetic theory when considering the quantum world of electromagnetic phenomena. Here we will say a few words about completeness of classical electrodynamics based on the Maxwell equations in the classical domain.

---

[93] Due to this reason and to attain greater simplicity of notations, we shall temporarily disregard the difference between contra- and covariant coordinates, although this may be considered a crime by geometrically-minded people.

The classical electromagnetic theory based on the Maxwell equations is a linear model – probably the simplest one possible in the Minkowski background – in which the scalar and vector potentials are to a large extent arbitrary and must be specified by fixing the gauge and the boundary conditions. The potentials in the electromagnetic theory are usually regarded as having mostly mathematical and not physical meaning. This is, however, not true. It would be a triviality to say that in quantum theory, especially in quantum field theory (QFT), the electromagnetic potentials have a clear physical significance. Yet even in the classical domain, the electromagnetic potentials, i.e., vector fields $A_\mu(x^\nu), \mu, \nu = 1, ..., 4$ are gauge fields that really act as local to global operators mapping the global space-time conditions to local electromagnetic fields[94]. These potentials produce physically observable effects, especially salient when one departs from the conventional interpretation of the electromagnetic theory as a linear model having a simple $U(1)$ gauge symmetry. (Above, in section 5.5., we discussed gauge symmetry in some detail.) This old interpretation of the Maxwell theory goes back to such prominent physicists and mathematicians as H. Bateman, O. Heaviside, H. Hertz, G. F. Fitzgerald, L. V. Lorenz, H. A. Lorentz, J. C. Maxwell, and W. Thomson. Mathematically, as I have just mentioned, this interpretation treats electrodynamics as a linear theory of $U(1)$ symmetry with Abelian commutation relations. Now we know that the electromagnetic theory can be extended to $SU(2)$ and even higher symmetry formulations. One might note in passing that elementary vector algebra corresponds to the $U(1)$ symmetry, whereas the $SU(2)$ group of transformations may be associated with more complicated algebraic constructions such as quaternions (see in this connection simple examples from sections on vector fields in Chapter 8).

When describing electromagnetic waves propagating through material media, such mathematical dynamic objects as solitons quite naturally appear. Solitons are in fact pseudoparticles, they can be classical or quantum mechanical, and, in principle, they can be both nonlinear and linear (at least piecewise). Other families of pseudoparticles include magnetic monopoles, magnetic charges (if defined separately from monopoles as e.g., dyons) and instantons. The standard electromagnetic $U(1)$ theory cannot describe solitons, so in order to be able to do that one must go beyond the linear electrodynamics based on the Maxwell equations, in other words, this theory must be extended (at least to $SU(2)$). Although we shall not deal methodically with nonlinear PDEs – this is a separate topic requiring serious treatment, we shall try to explain later what a symmetry extension means. Now we just want to notice that some physical effects unambiguously indicate that the conventional Maxwellian field theory cannot be considered complete.

For example, leaving apart the well-known difficulties with the classical electron radius, such phenomena as the Aharonov-Bohm (AB) effect or the Altshuler-Aronov-Spivak (AAS) effect [9] are based upon direct physical influence of the vector potential on a charged particle, free or conducting electrons respectively. It is interesting that both effects may be interpreted as breakdown of time-reversal symmetry in a closed-loop trajectory by a magnetic field (see about the time-reversal symmetry in Chapter 3.4). Recall that the vector potential $A_\mu$ has been commonly regarded as a supplementary mathematical entity introduced exclusively for computational convenience. It is mostly in view of this interpretation that the conventional electromagnetic theory may be considered incomplete – it fails to define $A_\mu$ as operators. Besides the AB and AAS effects, other important physical phenomena also depend on potentials $A_\mu$, for instance, the Josephson effect (both at the quantum and macroscopical level), the quantum Hall effect, de Haas-van Alphen effect and even the Sagnac effect (known since the beginning of the 20th century). Recently, the effects associated with the topological phase properties – and all the above phenomena belong to this class – have become of fashion, and we shall devote several pages to elucidate the essence of such phenomena.

---

[94] Here we are using notations as in [94].

One origin of the difficulties with classical electromagnetism lies in the obvious fact that the electromagnetic field (like any other massless field) possesses only two independent components, but is covariantly described by the *four*-vector $A_\mu$. However, in many models, for example, in radiation-matter interaction we usually break this explicit covariance. Likewise in nonrelativistic quantum theory, by choosing any two of the four components of $A_\mu$ (e.g., for quantization), we also lose the explicit covariance. Contrariwise, if one desires to preserve the Lorentz covariance, two redundant components must be retained. This difficulty is connected with the important fact that for a given electromagnetic field potentials $A_\mu$ are not unique: they are defined up to the gauge transformation $A_\mu \to A'_\mu = A_\mu + \partial_\mu \chi(x)$ (see above).

One may note that random processes (see Chapter 7) are a direct generalization of deterministic processes considered in classical mechanics. Historically, the first random processes to have been studied were probably the Markov processes: a random process is called a Markov process, if for any two $t_0$ and $t > 0$ the process probability distribution $w(t)$ (in general, under the condition that all $w(t)$ for $t \le t_0$ are known) depends only on $w(t_0)$. While for deterministic processes the state of the system at some initial moment uniquely determines the system's future evolution, in Markov processes the state (probability distribution) of the system at $t = t_0$ uniquely defines probability distributions at any $t > 0$, with no new information on the system's behavior prior to $t = t_0$ being capable of modifying these distributions. Here one may notice a hint at a preferred direction of time or time-reversal non-invariance, which is typical of real-life processes and does not exist in unprobabilistic mechanical theories (we do not consider the Aristotelian model here). Time-reversal invariance requires that direct and time-reversed processes should be identical and have equal probabilities. Thus, purely mechanical models based on Newton's (or Lagrangian) equations are time-reversible, whereas statistical or stochastic models, though based ultimately on classical (reversible) mechanics, are time-noninvariant. Likewise, mathematical models designed to describe real-life processes are mostly irreversible.

## 11.11. A Brief Remark on Integral Equations in Field Theory

For some reason, integral equations have become a subject partly alien to physicists. There are people – otherwise highly qualified – who say that differential equations are more than enough to cover all the principal areas of physics and to construct the models in other disciplines, so why should one attract new and rather complicated concepts? It is unnecessary and contradicts the principle of Occam's razor.

However, the statement that knowing only the differential equations is sufficient for physics is wrong: there are areas which cannot be studied (or at least become extremely difficult to study) without integral equations, and such areas are not rarely encountered in science and engineering. Take, for example, antenna theory and design. To obtain the required value of the electromagnetic field one has to compute or optimize the electric current, which is an unknown quantity being integrated over the antenna volume or surface. In scattering problems, both for waves and particles, integral equations are a naturally arising concept. In the problem of electromagnetic field scattering by a bounded inhomogeneity, the scattered field appears due to the induced currents which start flowing in the inhomogeneity under the influence of the incident field. More exactly, these currents may be viewed as a response to the total field present in the volume occupied by the inhomogeneity, this total field being the sum of the incident field and the one generated by induced currents. This is a typical self-consistence problem that leads to the volume integral equation where the unknown field stands under the integral. In general, self-consistence problems like scattering, quite often result in integral equations; now one more class of problems requires an extensive knowledge of integral equations, namely the inverse problems. We shall dwell on the subject of integral equations in association with the models which can be compartmentalized to each of these classes of problems. So integral equations may connect several seemingly unrelated topics, therefore every person interested in physmatics – a connected framework of physical and mathematical knowledge – should be familiar with integral equations. Here is a very brief overview.

In mathematics, integral equations are considered as a natural part of analysis linking together differential equations, complex analysis, harmonic analysis, operator theory, potential theory, iterative solutions, regularization methods and many other areas of mathematics. Someone who studies integral equations at the university, may be surprised to discover their multiple connections with other areas of mathematics such as various vector spaces, Hilbert spaces, Fourier analysis, Sturm-Liouville theory, Green's functions, special functions you name it.

## 11.12. Phenomenological Electrodynamics

The ideal of theoretical physics is a microscopic theory of everything. This is probably an unreachable star, yet some microscopic models built in physics, for example in non-relativistic quantum mechanics or quantum electrodynamics, have been quite successful. This success is, however, a precious rarity. More common are the phenomenological models. The collection of such models related to electromagnetic phenomena is aggregately called macroscopic electrodynamics. To better understand what it is let us consider a simple school-time example. Assume that we need to compute the electric field between a capacitor's plates. Then we would have to write the equations for electromagnetic fields produced by all the particles constituting the plates and the dielectric between them. We would also have to add to these field equations the ones describing the motion of the particles in the fields. The self-consistence problem thus obtained is, in principle, quantum mechanical, and any attempt to solve it would be a hopeless task.

Such an approach is usually called microscopic – because it considers phenomena on an atomic scale – it is too complicated and in most cases superfluous. Solving problems microscopically usually produces a great lot of immaterial data. It is much more reasonable to formulate general rules for the systems containing many particles i.e., for macroscopic bodies. Such rules have, by necessity, the average character, totally disregarding atomistic structure of the matter. An electromagnetic theory dealing with macroscopic bodies and fields between them is called phenomenological electrodynamics. Thematically, phenomenological electrodynamics belongs more to matter rather than to fields. However, it is more convenient to discuss this part of electromagnetic theory in connection with Maxwell's equations which should be properly averaged than in the context of, e.g., laser-matter interaction.

There may be many phenomenological theories related to the same subject, they are in fact comprehensive models placed between fundamental microscopic laws and ad hoc results describing a given phenomenon. When dealing with a phenomenological theory one cannot be sure that its equations are unique and correctly describe the entire class of phenomena considered unless a well-established procedure of obtaining these equations from microscopic theory is provided. In this respect, macroscopic electrodynamics and, e.g., thermodynamics are "lucky" phenomenological theories: they both can be derived – although with considerable difficulties – from underlying microscopic theories. On the other hand, empirical phenomenologies flourishing in medical, social, economic and even engineering research cannot – at least so far – be derived from first-principle theories and thus the limits of applicability for the respective phenomenological concepts are undetermined. In principle, the question of applicability of any phenomenological theory is very intricate, and later we shall try to illustrate this fact even on the examples of two privileged phenomenological theories: macroscopic electrodynamics and hydrodynamics.

Besides, most of what we call microscopic theories are in fact phenomenological. Take, for instance, Newton's laws of motion. They are considered exact, however Newton's laws can be with certainty applied only to macroscopic bodies i.e., to those composed of a very large number of atomic particles in slow, smooth motion. The Newtonian model will eventually lose validity if we continuously dissect such macroscopic bodies. It is, however, not always easy to distinguish between microscopic and phenomenological theories. Thus, the Schrödinger equation, which is also considered exact microscopic, is nothing more than a phenomenological model devised to describe the nonrelativistic motion of an atomic particle. The Schrödinger equation becomes invalid when one starts scrutinizing the interaction between particles through fields. The Newtonian theory of gravity is also phenomenological, this is the mathematical model stating that the attracting force between any two bodies does not depend on their composition, matter structure (crystalline, amorphous, liquid, plasma, etc.) and other constitutional details important from the physical point of view. This force depends only on some aggregate (and to some extent mysterious) coefficient called mass. Newtonian gravity is a very nontrivial model, gravitation could have been totally different, for example, inertial and gravitational masses could have been nonproportional to each other so that "light" objects would fall slower than "heavy" ones in the gravitation field. Phenomenological theory of gravitation, due to independence of physical details, allows astronomers to predict the motion of celestial bodies ignoring physical processes in them. In general, phenomenological theories usually neglect the effects of microscopic quantities or represent them by a set of numbers.

So philosophically speaking, most equations of physics are phenomenological. How can one try to establish validity of phenomenological theories? We have seen that the two basic concepts of microscopic – atomistic – models, in contrast to phenomenological theories (an extreme case is thermodynamics), are particles and fields. Phenomenological theories typically (but not always!) do not treat separate particles, they tend to regard objects as continuous without rapidly fluctuating local quantities such as true densities. This approach appears to be quite reasonable from the practical point of view: microscopic values vary in spacetime in a very complicated manner, and it would be merely meaningless to follow their instantaneous local values. In other words, any compelling theory should only operate with smoothed values, the fluctuations being averaged out. This way of thought naturally leads to the possibility of obtaining the phenomenological description by averaging the corresponding microscopic theories. This sounds simple – especially for linear theories – yet the question arises: what is actually "averaging"?

There does not seem to be a universal answer to this question, nor a universal recipe for the transition to the phenomenological version of a given microscopic theory. Each specific case is different, and "phenomenologization" can be quite intricate, not reducing to formal averaging. To illustrate the emerging difficulties let us get back to the relationship between microscopic and macroscopic electrodynamics. The system of Maxwell equations constituting the backbone of microscopic electrodynamics is linear, and thus it presumably can be averaged in a straightforward fashion i.e., one can simply substitute into the Maxwellian system average values for the field $\bar{\mathbf{E}}$ and $\bar{\mathbf{H}}$ in place of their genuine values $\mathbf{E}, \mathbf{H}$ containing fluctuations. However, here two problems arise. Firstly, the charge and current densities, $\rho$ and $\mathbf{j}$, representing inhomogeneities in the Maxwell equations should also be averaged, but nobody knows a priori what the relationship between the averaged fields and the averaged currents would be, and one badly needs this relationship to close the system of averaged equations of macroscopic electrodynamics. This is an important problem, and we shall discuss it later.

The second problem is the very meaning of the averaging operation and the corresponding mathematical procedure. Physicists traditionally relied in this issue on the concept of the so-called physically infinitesimal volume and declared averaging over this volume. The whole procedure of averaging over the infinitesimal volume is described in detail in the classical textbook [97]. Below I shall briefly reproduce this standard procedure accentuating the points that were for L. D. Landau and E. M. Lifshitz too obvious. Nevertheless, we may consider everything based on the "physically infinitesimal" misleading and poorly suitable for many practically important cases (such as X-ray optics, molecular optics, plasma physics, etc.). Besides, the very notion of the physically infinitesimal volume is badly defined – "on the hand-waving level" – so that it would be difficult to indicate the accuracy of the averaging procedure. One must be satisfied with the formally written average fields (overlined **E** and **H**), the question of their deviations from microscopic fields being irrelevant. One can only assume that fluctuations of the fields averaged over a "physically infinitesimal volume" are such that one consider these macroscopic fields as real statistical averages.

One might also remark that the difference between the fields averaged over a physically infinitesimal volume and statistically averaged is inessential. Right, it may be inessential for the description of quasistatic processes in simple model systems such as homogeneous dielectric. Even if we assume that averaging over physically infinitesimal volume is equivalent to averaging over all possible positions of scattering centers for the field (which is not at all obvious), in such averaging we neglect the motion of these centers. Moreover, it would be hardly possible to define a universal physically infinitesimal scale for all systems. For example, there exist many characteristic spatial and temporal parameters for plasmas, rarefied gases, turbulent fluids, superfluids, crystalline and amorphous solids, etc. The question of averaging over the physically infinitesimal volume is closely connected with the possibility of a universal definition of a continuous medium and leads to such nontrivial questions as dynamic irreversibility and time-noninvariance.

No matter what averaging method is applied to the Maxwell equations, the representation of the electromagnetic field as having exact (rapidly fluctuating) values at each spacetime point should be abolished. Bearing this in mind, one usually regards phenomenological electrodynamics as dealing only with the slow-varying fields, more specifically only with those having the wavelength $\lambda \gg n^{-1/3}$, where $n$ is the particle density in the medium. One may note that it is this condition that should exclude the X-ray range from macroscopic treatment (nonetheless, many problems of X-ray optics are actually handled by the tools of phenomenological electrodynamics). In general, along this way of thought it would be difficult to consider the short-wavelength phenomena in matter, e.g., those for large absolute values of dielectric function or refractive index. Besides, the matter response to an electromagnetic field may essentially depend on the motion of particles, the latter in reality "feeling" local and instantaneous fields in each spacetime point and not the mean fields averaged over the volume containing many particles. It is similar to the fact that the car driver reacts on the "here and now" traffic conditions more acutely than on smooth road curves, speed limiting signs, information tableaux and other factors of "macroscopic" character.

An alternative – and more correct – averaging method is not the one over the "physically infinitesimal volume", but a standard method of statistical physics: averaging over a statistical ensemble, e.g., in the case of equilibrium over the Gibbs distribution. Taking averages over the quantum state, with the wave function for the pure state or with the density matrix for the mixed state, will be discussed separately below. One may notice that in the statistical method there is no spatial averaging *per se*, and the fields can still remain quasilocal and quasi-instantaneous[95].

In the system of Maxwell equations

$$curl\mathbf{H} - \frac{1}{c}\frac{\partial \mathbf{E}}{\partial t} = \frac{4\pi}{c}(\mathbf{j} + \mathbf{j}_0) \qquad (11.12.1.)$$

$$div\mathbf{E} = 4\pi(\rho + \rho_0) \qquad (11.12.2.)$$

$$curl\mathbf{E} + \frac{1}{c}\frac{\partial \mathbf{H}}{\partial t} = 0 \qquad (11.12.3.)$$

$$div\mathbf{H} = 0, \qquad (11.12.4.)$$

where $\rho_0$ and $\mathbf{j}_0$ are external charge and current densities, the induced currents $\mathbf{j}$ and charges $\rho$ are the functions of fields in the matter, $\mathbf{j} = \mathbf{j}(\mathbf{E})$. In phenomenological electrodynamics, the quantities $\mathbf{E}, \mathbf{H}, \rho, \mathbf{j}$ are assumed to be averaged over either a "physically infinitesimal volume" or a statistical ensemble (see below). The relationship between the field and the current induced by it represents the matter response to an electromagnetic excitation and determines such essential quantities as the medium susceptibilities, both linear and nonlinear.

The phenomenological approach allows one to conveniently formulate mathematical problems for the "Maxwell operator". By introducing the tensor functions $\epsilon_{ij}(x)$ and $\mu_{ij}(x), x \in \Omega \subset \mathbb{R}^3$ i.e., dielectric permittivity and magnetic permeability of the medium (see below a detailed discussion of these quantities), we can write the homogeneous Maxwell equations in the operator form through the stationary operator $\mathcal{M}(\epsilon, \mu)$ acting on the pair $(\mathbf{E}, \mathbf{H})^T$ where $\mathbf{E} = \mathbf{E}(\omega, \mathbf{r}), \mathbf{H} = \mathbf{H}(\omega, \mathbf{r})$ are respectively electric and magnetic field vectors in the considered domain $\Omega$. More specifically, for a stationary electromagnetic field its eigenfrequencies correspond to the spectrum of $\mathcal{M}(\epsilon, \mu)$ acting on $(\mathbf{E}, \mathbf{H})^T$ according to the rule

$$\mathcal{M}(\epsilon, \mu)\begin{pmatrix}\mathbf{E}\\\mathbf{H}\end{pmatrix} = \begin{pmatrix}i\epsilon^{-1}\nabla \times \mathbf{H}\\-i\mu^{-1}\nabla \times \mathbf{E}\end{pmatrix}.$$

One usually assumes the solenoidal constraints $div(\epsilon\mathbf{E}) = 0, div(\mu\mathbf{H})$ and some boundary conditions, e.g.,

$$\mathbf{E}_\tau|_{\partial\Omega} = 0, (\mu\mathbf{H})_\nu|_{\partial\Omega} = 0.$$

Here index $\tau$ denotes the tangential and index $\nu$ the normal component of the respective vector on boundary $\partial\Omega$. In isotropic medium, permittivity and permeability may be reduced to positive scalar functions $\epsilon(\mathbf{r}), \mu(\mathbf{r})$. When these functions as well as boundary $\partial\Omega$ are sufficiently smooth, the spectral problem for the Maxwell operator can be solved. The Maxwell operator formulation is especially useful for calculating modes in electromagnetic resonators (see, e.g., [81, 143]).

---

[95] The microscopic electromagnetic fields are still bound from below by the nuclear scale ($\sim 10^{-13}$ cm). As for the macroscopic fields in the matter, it does not make sense to phenomenologically consider the distances smaller than the atomic ones ($\sim 10^{-8}$ cm)

In optical problems related to isotropic media, permittivity and permeability are also typically treated as smooth and positive scalar function determining the refraction index $n = (\epsilon\mu)^{1/2}$ and the velocity of light in the medium, $v_p h = c/n$. Then the "optical" metric

$$ds^2 = c^2 dt^2 - dx_i dx^i$$

turns, as already discussed, the considered domain $\Omega \subset \mathbb{R}^3$ into a Riemannian manifold.

## 11.12.1. The Traditional Averaging Procedure

The meaning of the averaging over a "physically infinitesimal volume" needs to be made clear for the sake of sound understanding of phenomenological electrodynamics. Let us now carry out explicitly the standard procedure of such an averaging. Apart from being a useful exercise, it is also a useful trick the details of which are not always given in the textbooks. In the standard averaging scheme of transition to macroscopic electrodynamics, one usually introduces the electric induction vector $\mathbf{D} = \mathbf{E} + 4\pi\mathbf{P}$ and the magnetic field strength $\mathbf{H} = \mathbf{B} - 4\pi\mathbf{M}$ where $\mathbf{P}$ is the polarization vector of the medium and $\mathbf{M}$ is the magnetization vector (by definition, it is the mean magnetic dipole moment for unit volume, $\mathbf{M}(r) = \sum_i N_i \overline{\mathbf{m}_i(\mathbf{r})}$, $\overline{\mathbf{m}_i(\mathbf{r})}$ is the average magnetic dipole of the elementary domain or cell in the vicinity of point $\mathbf{r}$, e.g., of a single molecule located at $\mathbf{r}$ and $N_i$ is the average number of such cells). One might complain that such phenomenological construction is far from being elegant. The average value of the magnetic field is usually denoted as $B$ and called the magnetic induction.

In the classical averaging scheme over the "physically infinitesimal volume", the polarization vector $\mathbf{P}$ is usually defined as the vector whose divergence equals (up to a sign) the average charge density in the medium, $\nabla\mathbf{P} = -\rho$ (see [97], §6). This definition is consistent with the general requirement of the matter electric neutrality, the latter being necessary for the stability of the matter. One can easily understand the physical meaning of the polarization vector $\mathbf{P}$ as the average dipole moment of the unit volume. Indeed, if for the zero-moment of the averaged charge density $\rho$ we have, due to neutrality,

$$\bar{\rho} = \int_\Omega \bar{\rho}(\mathbf{r}, t) d^3 r = 0$$

for all $t$, for the first moment we may write

$$\int_\Omega \mathbf{r}\bar{\rho}(\mathbf{r}, t) d^3 r = \int_\Omega \mathbf{P}(\mathbf{r}, t) d^3 r$$

where integration runs over the whole matter (e.g., a macroscopic dielectric body). One may notice that these relations are supposed to be valid not only in the static case. To prove the second relation, we can use the vector integral identity

$$\oint_\Sigma \mathbf{r}(\mathbf{P}d\sigma) = \int_\Omega (\mathbf{P}\nabla)\mathbf{r} d^3 r + \int_\Omega \mathbf{r}\nabla\mathbf{P} d^3 r,$$

where $\Sigma$ is any surface enclosing the matter volume $\Omega$. If this surface passes outside this volume (a material body), the integral over it vanishes, and since $(\mathbf{P}\nabla)\mathbf{r} = \mathbf{P}$ we get the required relationship for the first moment of $\bar{\rho}$. The above vector identity can be proved by a number of ways, e.g., by using the tensor (index) notations. The simplest, but possibly not very elegant, proof would consist in using the identity

$$\nabla(\varphi(\mathbf{r})\mathbf{P}) = \varphi\nabla\mathbf{P} + (\mathbf{P}\nabla)\varphi$$

which can be applied to any component of $\mathbf{r} = (x, y, z)$ and then integrating it over $\Omega$, with the left-hand side giving the surface integral.

Thus, we have the four vectors – two vector field pairs – of macroscopic electrodynamics, $\mathbf{E}, \mathbf{D}, \mathbf{H}, \mathbf{B}$, which must be supplemented by some phenomenological relations between these vectors. Besides, if the above four quantities must satisfy the equations of the type of Maxwell's equation for the microscopic vacuum values, the question of sources for the phenomenological quantities arises. The textbook relationships between the vectors within each pair, $\mathbf{D} = \epsilon\mathbf{E}$ and $\mathbf{B} = \mu\mathbf{H}$, complete the traditional construction of macroscopic electrodynamics.

This simple construction can be justified for static or slowly varying (quasistationary) fields, but it usually becomes totally inadequate for rapidly changing vector functions such as for short wavelengths or ultrashort pulses (USP), the latter being an increasingly popular tool in contemporary laser physics. One may note that such commonly used terms as "short wavelengths" or "hard electromagnetic radiation" are relative: for certain media the radiation typically perceived as "soft", e.g., infrared, may exhibit short-wavelength features whereas in some other type of matter the same radiation may be treated as quasistationary. Thus, in plasma, where the average distance between the particles, $n^{-1/3}$, $n$ is the particle density, may easily exceed the characteristic wavelength $\lambda$ of the electromagnetic field, using the phenomenological field equations obtained by averaging over the physically infinitesimal volume can easily become meaningless.

We have already mentioned that the conventional approach to macroscopic electrodynamics, corresponding to the averaging of microscopic fields over "physically infinitesimal volume", consists in the additive decomposition of the total induced current and charge densities, $\mathbf{j}$ and $\rho$ (also averaged over the physically infinitesimal volume), into "physically different" parts, e.g.,

$$\mathbf{j} = \mathbf{j}_c + \mathbf{j}_p + \mathbf{j}_m,$$

where $\mathbf{j}_c$ represents the current of conductivity electrons, $\mathbf{j}_p$ is the polarization current, $\mathbf{j}_p = \partial\mathbf{P}/\partial t$, where $\mathbf{j}_m$ is the magnetization current, $\mathbf{j}_m = c\, curl\,\mathbf{M}$. (I would recommend reading carefully the respective material in the textbook by L. D. Landau and E. M. Lifshitz [97].) One does not need to perform the same decomposition procedure for the charge density because of the constraint given by the continuity equation 11.4.3. which remains valid after any kind of averaging (due to its linearity).

## 11.12.2. Ensemble Averaging of Fields and Currents

Paying attention to the difference between averaging over a "physically infinitesimal volume" and ensemble averaging may be essential to understanding of the fundamentals involved. The Maxwell equations in the medium obtained by the traditional averaging procedure (i.e., over the "physically infinitesimal volume") and having the form ([97], §75)

$$curl\mathbf{H} - \frac{1}{c}\frac{\partial \mathbf{D}}{\partial t} = \frac{4\pi}{c}\mathbf{j}_0 \qquad (11.12.2.1.)$$

$$div\mathbf{D} = 4\pi\rho_0 \qquad (11.12.2.2.)$$

$$curl\mathbf{E} + \frac{1}{c}\frac{\partial \mathbf{B}}{\partial t} = 0 \qquad (11.12.2.3.)$$

$$div\mathbf{B} = 0, \qquad (11.12.2.4.)$$

being supplemented by the "material equations" relating the quantities $\mathbf{D}, \mathbf{B}$ and $\mathbf{E}, \mathbf{H}, \mathbf{D} = \epsilon\mathbf{E}$ and $\mathbf{B} = \mu\mathbf{H}$ can be used without reservations only for relatively slow-varying fields (static and quasistationary). For fast changing electromagnetic fields and pulses as well as for the media where spatial field variations can be shorter than the average distance between the constituent particles (e.g., in plasmas), these equations become inconvenient or merely inadequate. Indeed, it seems to be nearly obvious that, even leaving aside the dubious and in general mathematically incorrect procedure of averaging over the physically infinitesimal volume, breaking down the total current into presumably non-overlapping components cannot be unambiguous for high frequencies. It may be easily seen that the currents excited in the medium due to free and to bound electrons cannot be separated already for optical frequencies or for rapid variations of an external field (ultrashort pulses). One may illustrate this fact by a simple example of an atomic electron in an external field $\mathbf{E}$. For simplicity, we may consider here classical (non-quantum and non-relativistic) motion, $m\ddot{\mathbf{r}} = e\mathbf{E}(t)$, $e, m$ are the electron charge and mass respectively, then the characteristic displacement or oscillation amplitude of an electron in the field $\mathbf{E}(t)$ of an electromagnetic pulse or wave is $r_0 \sim eE\tau^2/m$ or $r_0 \sim eE/m\omega^2$ where $\tau$ is the ultrashort pulse duration. Such a displacement can be readily scaled down to atomic distances (Bohr's radius, $r_B \sim 10^{-8}$cm) even for rather strong fields, say only an order of magnitude lower than atomic fields, $E_{at} \sim e/a_B^2 \sim 10^9$V/cm$^2$.

This simple example (and many others, too) demonstrates that the difference between the conductivity, polarization and magnetization currents, the latter being due to the charge motion along closed trajectories, rapidly becomes very vague with the decreased wavelength of the radiation field. Starting from ultraviolet light, unambiguous decomposition of current into these three components becomes virtually impossible. At least, such a decomposition at optical frequencies and for ultrashort electromagnetic pulses is pretty arbitrary. Thus, the optimal possibility we have in the case of fast varying fields is to consider the total current $\mathbf{j}(\mathbf{r}, t)$ incorporating all kinds of charge motions caused by an electromagnetic field. This current, in the situation of thermal equilibrium, should be averaged over a statistical ensemble i.e., with the Gibbs distribution (or equilibrium density matrix in the quantum case). Technically, it is often convenient to introduce the total polarization $\mathcal{P}(\mathbf{r}, t)$ (see [137]) instead of the total current:

$$\mathcal{P}(\mathbf{r}, t) = \int_{-\infty}^{t} \mathbf{j}(\mathbf{r}, t')dt', \qquad \mathbf{j}(\mathbf{r}, t) = \partial_t \mathcal{P}(\mathbf{r}, t)$$

It is important to remember that total polarization $\mathcal{P}(\mathbf{r}, t)$ includes all currents, due both to free and to bound charges, and not only the displacement current, as in the intuitive scheme of averaging over the physically infinitesimal volume. One can also see that the total polarization accumulates current contributions starting from the remote past ($t \to -\infty$), but not from the domain $t' > t$. This is a manifestation of the causality principle which is one of the crucial assumptions in physics (see also Chapters 3,10). In the distant past, $t \to -\infty$, polarization is assumed to be absent.

Introducing the total polarization enables us to write down the total induction, $\mathbf{D}(\mathbf{r}, t) = \mathbf{E}(\mathbf{r}, t) + 4\pi \mathbf{P}(\mathbf{r}, t)$, and not only induction only owing to displaced bound charges, as in traditional theory of dielectrics. Using the total induction, we may write the averaged Maxwell equations in the medium as

$$curl\mathbf{E} + \frac{1}{c}\frac{\partial \mathbf{H}}{\partial t} = 0 \qquad (11.12.2.5.)$$

$$div\mathbf{H} = 0 \qquad (11.12.2.6.)$$

$$curl\mathbf{H} - \frac{1}{c}\frac{\partial \mathbf{D}}{\partial t} = \frac{4\pi}{c}\mathbf{j}_0 \qquad (11.12.2.7.)$$

$$div\mathbf{D} = 4\pi\rho_0, \qquad (11.12.2.8.)$$

where the last equation is in fact a consequence of the continuity equation 11.4.3. Indeed, from

$$\frac{\partial \rho}{\partial t} + \nabla \mathbf{j} = \frac{\partial \rho}{\partial t} + \nabla \frac{\partial \mathcal{P}}{\partial t} = 0$$

we have

$$\rho = -\nabla \mathbf{P} + \rho_{-\infty} = \frac{1}{4\pi}\nabla(\mathbf{D} - \mathbf{E}),$$

where $\rho_{-\infty}$ is the integration constant that can be put to zero (we assume that there were no polarization in the distant past). Then we have $4\pi\rho = -\nabla(\mathbf{D} - \mathbf{E})$, but $\nabla\mathbf{E} = 4\pi(\rho + \rho_0)$ where $\rho_0$ is, as before, the density of external charges introduced into the medium. Thus, we get $\nabla\mathbf{D} = 4\pi\rho_0$.

One may notice that in this "optical" approach one does not need to introduce, in addition to a magnetic field, such a quantity as magnetic induction and, consequently, magnetic susceptibility [137]. So, there is no distinction between $\mathbf{B}$ and $\mathbf{H}$. Indeed, the total polarization already contains a contribution from the magnetization (circular) currents, e.g., arising from nearly closed trajectories of electrons moving in the electromagnetic field of elliptically polarized light.

One often asks: why are currents and polarization in the medium considered the functions of the electric field $\mathbf{E}$ alone, with magnetic field $\mathbf{H}$ being disregarded both in the dielectric relationship $\mathcal{P} = \chi\mathbf{E}$ and in the conductivity relationship $\mathbf{j} = \sigma\mathbf{E}$? Or, to put it slightly differently, why is it always implied that the induction $\mathbf{D}$ is only proportional to $\mathbf{E}$ and not to $\mathbf{H}$? One might encounter several versions of answering this question. One of the versions is that $\mathbf{H}$ should be excluded because (1) it is an axial vector ($\mathbf{H} = \nabla \times \mathbf{A}$), whereas $\mathbf{E}$ is a "true" vector so that they both cannot be simply superposed, and (2) $\mathbf{H}$ is not a time-invariant quantity. The second argument implies, however, an a priori requirement imposed by our intuitive perception of the world: nobody can be sure that all electromagnetic phenomena must be strictly invariant under time reversal (even in the static case). As to the first argument, it may be "neutralized" by introducing a pseudoscalar proportionality coefficient between $\mathbf{P}$ and $\mathbf{H}$ (or $\mathbf{j}$ and $\mathbf{H}$ and $\mathbf{D}$ and $\mathbf{H}$). In reality, we may disregard the magnetic field in the "material relations" because it can be expressed through the electric field using Maxwell's equations. Besides, even if we wished to explicitly involve the magnetic field, it would make little sense unless we considered ultrarelativistic motion of charges, which is an exotic case in the medium. The point is that factor $v/c$ always accompanying a direct effect of a magnetic field on moving charges would make its influence hardly noticeable for the particles in matter whose characteristic velocities are of the atomic scale ($v_{at} \sim e^2/\hbar$) i.e., at least two orders of magnitude lower than the speed of light.

Now the main problem is: how to link the current induced in the medium to an external field exciting this current? It is clear that such a problem is in general extremely difficult and can hardly be solved for an arbitrary matter containing a macroscopic number ($N \sim 10^{23}$) of particles placed in a fortuitous field. Nonetheless, there are several approaches to this universal problem. One of such approaches has been developed within the framework of nonlinear optics (NLO). This is a comparatively new discipline which emerged in the 1960s, following the advent of lasers. Before that time optics was essentially linear, and probably the only attempt to consider nonlinear electromagnetic effects were made in quantum electrodynamics while treating the scattering of light by light [7] and [152], which is in fact vacuum breakdown process[96]. Linearity of electrodynamics required that the polarization of the medium and induced current should be written respectively as $\mathcal{P}_i - \chi_{ij} E_j$, where $\chi_{ij}$ is the susceptibility tensor, and $j_i \sigma_{ij} E_j$, where $\sigma_{ij}$ is the conductivity tensor[97]. A more general, nonlocal, linear expression linking the induced current to an electric field may be written as

$$j_i(\mathbf{r}, t) \equiv j_i^{(1)}(\mathbf{r}, t) = \int \sigma_{ij}(\mathbf{r}, \mathbf{r_1}; t, t_1) E_j(\mathbf{r_1}, t_1) d^3 r_1 dt_1 \qquad (11.12.2.9)$$

Here we did not indicate integration limits implying that they are ranging from $-\infty$ to $+\infty$; however, it is necessary to comment on this point. If one assumes that the causality principle holds for all types of medium, then one must consider polarization and currents, observed at time $t$, depending only on the field values related to the preceding moments of time, $t_1 < t$. Therefore, integration over $t_1$ goes only from $-\infty$ to $t$. Furthermore, relativistic causality requires that spatial integration may spread only over spacelike domains $|\mathbf{r_1} - \mathbf{r}| < c|t_1 - t|$ because an electromagnetic field existing at points outside this domain cannot produce an excitation of the medium at a spacetime point $(\mathbf{r}, t)$. In fact, however, integral expressions for currents and polarization determining the response of the medium essentially depend on the properties of the response functions $\sigma_{ij}(\mathbf{r}, \mathbf{r_1}; t, t_1)$, $\chi_{ij}(\mathbf{r}, \mathbf{r_1}; t, t_1)$ serving as kernels in the integral transformations realizing nonlocal maps. Experience as well as physical considerations show that the response functions are essentially different from zero at much shorter distances than those required by relativistic causality. Thus, one may consider integration limits really infinite, which is true not only in the linear case, but also for nonlinear response functions.

It may be rewarding to clarify the physical reason for spatial and temporal nonlocality in (11.12.2.9). For integration over time, the origin of nonlocality is pretty obvious: it is the retardation of response. For example, a particle at point $\mathbf{r}$ of its trajectory $\mathbf{r}(t)$ feels the influence of the force $\mathbf{F}(\mathbf{r}, t)$ not only taken at time-point $t$, but also related to preceding moments, $t' < t$. Indeed, directly from Newton's equation we have

$$\mathbf{r}(t) = \int_{t_0}^{t} dt_1 \int_{t_0}^{t} dt_2 \frac{1}{m} \mathbf{F}(\mathbf{r}(t_2), t_2) + \mathbf{v}_0(t - t_0) + \mathbf{r}_0$$

where $\mathbf{r}_0 := \mathbf{r}(t_0), \mathbf{v}_0 := \dot{\mathbf{r}}(t_0) \equiv \dot{\mathbf{r}}_0$ are integration constants having the meaning of the starting point and initial velocity of the trajectory; $t_0$ is the initial time point for the motion. In many physical situations one is not much interested in parameter $t_0$ so that, e.g., for many particles one can average over all possible $t_0$ (as well as over initial conditions, which amounts to a simple statistical approach) or, in mechanical problems, one may take the limit $t_0 \to -\infty$ assuming $\mathbf{r}_0 = \mathbf{r}_{-\infty} = 0$ and $\mathbf{v}_0 = \mathbf{v}_{-\infty} = 0$ so that

$$\mathbf{r}(t) = \int_{t_0}^{t} dt_1 \int_{t_0}^{t} dt_2 \frac{1}{m} \mathbf{F}(\mathbf{r}(t_2), t_2).$$

---

[96] The usual electrodynamics was perceived as essentially linear so that even in the classical textbook by L. D. Landau and E. M. Lifshitz [96], §77, small intensities of fast changing fields are always assumed, see the text before formula (77.3). The development of lasers has radically changed this perception.

[97] Here we disregard the difference between contra- and covariant components; such a difference is inessential for our - rather superficial - intuitive discussion.

We shall also discuss the role of the initial moment of time $t_0$ in many-particle systems in connection with the Liouville equation and the choice of supplementary (e.g., boundary) conditions to it (Chapter 7). In many cases it is appropriate to shift the non-physical parameter $t_0$ to $-\infty$.

If the particle considered is, e.g., a free electron, then the force $\mathbf{F}$ may be exerted by an electromagnetic wave or a pulse, $\mathbf{F}(\mathbf{r}, t) \approx e\mathbf{E}(\mathbf{r}, t)$, whereas for a bound electron the force acting on it is a superposition of the incident and atomic fields

$$\mathbf{F}(\mathbf{r}, t) \approx e\mathbf{E}(\mathbf{r}, t) - \nabla V(\{\mathbf{R}\}, \mathbf{r}, t).$$

Here $V(\{\mathbf{R}\}, \mathbf{r}, t)$ is the atomic potential depending, besides spacetime variables $\mathbf{r}, t$, on a set of atomic parameters denoted by symbol $\{\mathbf{R}\}$. When external fields are much lower than the atomic Coulomb fields, $E_{at} \sim e/r_B^2$, the force acting on electron is mainly determined by the atomic potential and the time interval of force averaging is essentially determined by atomic (or interatomic) parameters i.e., is of the order of atomic time $\tau_{at} \sim \hbar/I \sim 10^{-16}$cm where $I$ is the typical atomic energy of the order of ionization potential, $I \sim e^2/r_B$. Physically, this is the time of the atomic velocity change, $v_{at} \sim r_B/\tau_{at} \sim r_B I/\hbar$. For a free electron, the integration time interval is of the order of the velocity relaxation time.

There may be situations when one has to take into account initial correlations between the particles entering the medium or the domain occupied by a field at times $t_{0a} \to -\infty$ ($a$ enumerates particles). Then simple averaging or taking the limit $t_{0a} \to -\infty$ does not hold and should be replaced by an integration (convolution) with the correlation functions.

In strong electromagnetic fields such as produced by lasers, polarization and currents in the medium may become nonlinear functions of the field. This nonlinearity can be easily understood by using one of our favorite models: that of an oscillator, but a nonlinear one (see below). For example, in the case of quadratic nonlinearity we shall have

$$j_i^{(2)}(\mathbf{r}, t) = \int \sigma_{ijk}(\mathbf{r}, \mathbf{r}_1, \mathbf{r}_2; t, t_1, t_2) E_j(\mathbf{r}_1, t_1) E_k(\mathbf{r}_2, t_2) d^3 r_1 d^3 r_2 dt_1 dt_2$$

and, in general,

$$j_i^{(n)}(\mathbf{r}, t) = \int \sigma_{j_1 \ldots j_n}(\mathbf{r}, \mathbf{r}_1, \ldots, \mathbf{r}_n; t, t_1, \ldots, t_n) E_{j_1}(\mathbf{r}_1, t_1) \ldots E_{j_n}(\mathbf{r}_n, t_n) d^3 r_1 \ldots d^3 r_n dt_1 \ldots dt_n$$

Continuing this approximation process, we obtain an expansion of the current[98] (or polarization) in powers of electric field:

$$\mathbf{j}(\mathbf{r}, t) = \sum_{n=1}^{\infty} j^{(n)}(\mathbf{r}, t).$$

Now the question naturally arises: what is the actual (dimensionless) parameter of such an expansion? Plausible physical discourse gives a cause for a belief that in wide classes of condensed matter such as dielectric solids and non-conducting liquids the expansion parameter should be $\eta_E = E/E_{at}$, where $E_{at} \sim e/r_B^2 \sim 10^9 V/\text{cm}$ is a characteristic atomic field. Indeed, for $\eta_E \ll 1$ an external electromagnetic field cannot strongly perturb the matter, and an expansion on $\eta_E$, with retaining only a few initial terms, must be fully adequate. However, one has to bear in mind that although parameter $\eta_E$ can be used for expansion over external field in many types of condensed media and even in gases, there exist other types of media, in particular containing free charges such as plasmas, metals, and even semiconductors, where parameters of nonlinearity can be different. At least, one needs to carry out a special analysis for each of these media. This is an important point, and we shall return to it later.

## 11.13. A Brief Note on Plasma

One of the very interesting collections of mathematical models describing real-life processes is comprised in plasma theory. The study of plasmas covers a wide spectrum of physical phenomena: from mechanics and classical electrodynamics through special relativity to quantum theory.

Physical and mathematical models of plasma are especially important when one considers its interaction with external electromagnetic fields. As a rule, to model such interaction processes it is sufficient to use classical theory, more specifically, classical mechanics coupled with classical electrodynamics. Here, we shall only touch upon the ultimate Coulomb systems i.e., the fully ionized plasma. In fact, this is typical of high-temperature plasmas whereas the definition of plasma as a state of matter involves any degree of ionization. Plasma is usually regarded as a fluid, although fully ionized plasma is noticeably different from such fluids as gas or liquid as concerns the role played by particle collisions.

One can start the discussion of mathematical models of plasma physics with one the most important subjects: interaction of plasma with electromagnetic (EM) fields. Even the most primitive models such as those describing the behavior of a single classical charged particle (for definiteness, we shall speak about an electron) in an external EM field bears many essential features of the plasma-field interaction. The classical motion equation for this case is

$$\frac{d\mathbf{p}}{dt} = e\left[\mathbf{E}(\mathbf{r}, t) + \frac{1}{c}[\mathbf{v}\mathbf{H}(\mathbf{r}, t)]\right],$$

where $\mathbf{p} = \gamma m\mathbf{v}$ is the particle momentum, $e$ and $\mathbf{v}$ are its charge and velocity, $\mathbf{E}(\mathbf{r}, t)$ and $\mathbf{H}(\mathbf{r}, t)$ are the electric and magnetic fields in which the charge moves, $\gamma = (1 - v^2/c^2)^{-1/2}$ is the relativistic factor.

---

[98] Recall that here we are talking about microscopic currents, also called microcurrent, induced by an external macroscopic field in matter.

## Section 12. Nonlinear world

Nonlinear phenomena can be encountered under many guises such as chaos, self-organization, synergetics, systems far from equilibrium, spatio-temporal pattern formation, emergent behavior and, most recently, complex systems. The field of nonlinear phenomena is so vast that it would not be possible to cover them all even on a superficial level. The list of important subjects is too long for this manuscript.

In the not-so-distant past, the nonlinear world was often ignored: it was considered too difficult and the respective equations insoluble. Understanding the importance of nonlinear dynamical systems for modeling real-world processes was a paradigm shift, especially in physics. This paradigm shift has brought a new set of multipurpose tools that can be applied both in science and society. Chaos, bifurcations, period doubling, coherent structures, pattern formation, self-organization, complexity, integrable systems, solitons, fractals and so on have become indispensable notions used in the models that attempt to describe the dynamical processes in real life. Nonlinear dynamics may be considered the core of modern science producing an interdisciplinary impact. It has already been mentioned that all many-body problems must be, in principle, nonlinear. An indirect manifestation of this fact is the existence of complex physical objects exhibiting highly nonlinear properties, for example, continuum media phenomena such as gas and plasma flows, nonlinear acoustics, nonlinear waves in material media, gravitation on cosmological scales, etc. Such phenomena are generally described by nonlinear equations, e.g., the Navier-Stokes, Bürgers, Korteweg-de Vries, sine-Gordon, nonlinear Schrödinger equations as well as Einstein's gravitation field equations. Nonlinear PDEs are the equations that govern most real-life processes such as weather and climate systems, turbulent motion, natural catastrophes, etc. The subclass of linear systems discussed above is a great idealization of the real world which is basically nonlinear. We have seen that for linear systems the principle of superposition holds i.e., a linear combination of the solutions to a system is also a solution to the system. This principle is not valid for nonlinear systems and respectively no simple unifying concepts analogous to vector spaces and linear operators exist in nonlinear theories. Mainly because of this, there are no regular methods of dealing with nonlinear models. In particular, nonlinear field equations are the most complicated to solve (for example, Einstein's equations in general relativity).

Although linear systems can always be exhaustively explored, they can be regarded as an exception rather than a norm. In most cases nonlinearities appear naturally when one tries to model real-world processes. For example, in the socio-economic disciplines and in general in many-body systems, interaction between the agents constituting a system naturally produce nonlinearities.

Nevertheless, despite significant difficulties in treating nonlinear problems, some new techniques emerged, specific for nonlinear science and modeling. For instance, a powerful method of approximate averaging (the Krylov-Bogoliubov-Mitropolsky or KBM method) which allows one to reduce a nonlinear problem to a process of successive refinements. For example, the first – "rough" – approximation such as simple harmonic motion $x = a\cos(\omega t + \varphi)$ in the problem of nonlinear oscillations can be refined to the desirable accuracy by consecutively taking into account the higher harmonics. In general, current nonlinear trends in science and engineering are accompanied by the development of powerful novel mathematical structures, mostly of geometrical and topological nature. One can name in this connection the KAM (Kolmogorov-Arnold-Moser) theory, which is focused on conservation of nonlinear invariants, the concept of dynamic entropy (Kolmogorov-Sinai entropy), nonlinear waves and solitons, chaotic dynamics, coarse graining, etc.

It is interesting to notice that most fields of science and engineering tend to develop their own brands in the description of nonlinear phenomena. Thus, approaches to nonlinear modeling usually differ across such disciplines as nonlinear optics, acoustics, electrical, electronic and radio-engineering. The most versatile nonlinear medium seems to be plasma, where nonlinear regimes are quite easily attained and linear approaches (very popular in the traditional plasma theory) quickly become inadequate. The study of plasma is an interesting interplay of mathematical and computer models: it combines particle and fluid dynamics, electrodynamics, theory of linear and nonlinear waves, stochastics, kinetics, stability theory and models of turbulence. In short, we are immersed in the natural world of nonlinear events. For example, our emotional reactions, our likes and dislikes, seem to be highly nonlinear with respect to the input stimuli. From the modeling standpoint, human behavioral responses to heat, cold, sound and light intensity, colors, local pressure and other stimuli are mostly subordinated to the Weber-Fechner law: the magnitude $R$ of psycho-physiological response is proportional to the logarithm of magnitude $J$ of physical stimulus i.e., $R = A \ln(J/J_0)$, $J > J_0$, $R = 0, J \le J_0$ ($J_0$ is threshold).

One often says that the 20$^{th}$ century was the one of physics whereas the 21$^{st}$ century will be the one of biology. The latter discipline mostly deals with nonlinear phenomena; linear effects are exceptionally rare in the biological world so that the respective mathematical models should, by necessity, be nonlinear. On the macroscopic scale, it can be illustrated by a simple variant of the Lotka-Volterra competition model (two species, usually denoted as predator and prey, are struggling for existence). This biological process is modeled by a two-dimensional dynamical system giving the time rate of the species' evolution.

The main feature of the nonlinear world is that it is typically not possible to make long-term predictions about the behavior of a nonlinear system, even if it is perturbed only slightly.

## 12.1. Complex systems

The study of complex systems has lately become one of the major parts of interdisciplinary research which is now significantly changing the whole approach to science: many scientific disciplines can no longer be confined to isolated, compartmentalized subunits, although interdisciplinary communities still tend to be amorphously shaped with their tasks unclearly stated. Today, even such important concepts as control and feedback that are the precursors of complex systems are of interdisciplinary nature.

We may remark here that the concept of interdisciplinarity is viewed with some contempt by many specialists that consider themselves genuine professionals. They argue that the so-called interdisciplinarity is a trench for dilettantes or, at least, a tool to get financing. This is, however, a very limited view which is far from reality. One does not mean by interdisciplinary approach the situation when a poor physicist talks to a poor biologist and the supervising bureaucrat hopes that something good necessarily comes out. Boundaries between disciplines were demarcated by people and usually have little to do with scientific criteria or quantitative parameters, and the sooner such boundaries and the accompanying thresholds (the latter being more of social than of scientific nature) are removed or massively penetrated, the more unexpected technological directions and nontrivial scientific results shall we see. One can consider numerous examples of interdisciplinary cross-fertilization, e.g., two major developments, one in computer science, the other in genomics, being partly merged, resulted in the appearance of bioinformatics which is now radically changing the face of medicine, in particular, of its preventive chapter. The use of physical models in financial analysis, customarily treated with skepticism by the old-school economists, is gradually becoming an established branch of quantitative economics (e.g., asymptotic techniques in financial analysis). For better or for worse, the process of interdisciplinary development is unstoppable, and the study of complex systems is one of its examples.

Complex systems are especially important for biomedical, environmental and social sciences, where the crossroads with mathematics, physics, computer science and engineering rapidly proliferate. People active in the first group of disciplines, traditionally considered "soft" as compared with the second group of "hard sciences", seem to be more and more dissatisfied with the fact that one can compute the missile trajectory with a great accuracy or predict the eclipses of the Sun a thousand years ahead, but cannot reliably forecast the weather for a weekend, the next day's stock exchange outcome, or their best friend's behavior, without mentioning floods, earthquakes, tsunamis, etc. All such poorly predictable and sometimes tragic events testify that the future of complex systems, i.e., in a highly nonlinear world, is far from being given by the system's current state. The task of forecasting catastrophes such as floods, earthquakes and tsunamis is currently more and more reduced to mathematical modeling of complex (i.e., highly nonlinear) systems. Sometimes, highly nonlinear complex systems exhibit such an intricate behavior that some people are inclined to endow it with elements of mystics (the latter can be defined as those aspects of human experience that contradict physics).

Most real-world processes can be modeled by nonlinear evolution equations. In particular, propagation of traveling waves constitutes an important class of nonlinear physical phenomena. Probably, the greatest importance in mathematical modeling of wave propagation has the well-known Korteweg-de Vries (KdV) equation

$$u_t + uu_x + \beta u_{xxx} = 0,$$

where $\beta \in \mathbb{R}$ (or $\beta \in \mathbb{C}$ ) is a non-zero constant (see also section 10.5.1.). Each term in the KdV has its own meaning. Thus, the term $u_t$ in the KdV equation accounts for the temporal evolution of wave, the second (quasilinear) term $uu_x$ describes the steepening of propagated waves tending to localize them whereas the dispersive term $\beta u_{xxx}$ corresponds to the spreading of waves. It is the competition of these two terms that determines either the runaway growth (e.g., up to the monster wave) or quiet fading of wavelike perturbations. An exact balance between quasilinear steepening and dispersive spreading determines a soliton, one of the few exactly defined notions in the rather vague nonlinear discipline known as the study of complex systems.

We know that the solutions to nonlinear problems in general cannot be represented as the sum of partial solutions and can therefore be said to have unexpected properties. Thus, it is very hard to foresee, in contrast with linear or mechanical systems, the state of the human body and the future development of social, economic and political systems, which sometimes makes life unnecessarily hard. Existence of unexpected irregularities and surprising events, which existed in classical physics only in the form of sudden (random) collisions, always brings in the issue of predictability and feasibility of long-term forecasts i.e., beyond a certain temporal horizon usually determined by the current state of technology. For instance, predictions in meteorology are time-limited in accordance with the available computing power. Moreover, it is not the future parameters as time-dependent curves that can be predicted, but the range of future parameters. In the study and forecast of natural catastrophic events, exact time points of an event are never obtained, at best only the time and magnitude intervals. For instance, when a tsunami is generated by an underwater earthquake, the tsunami wave propagation is analyzed by classical fluid dynamics methods whose accuracy is finite. The prediction of tsunami heights in the real conditions of coastline irregularities (fractal geometry) leads to significant computational challenges for specific locations. Besides, predictions in complex systems are based on observational data, and one cannot create a device that measures physical wave quantities with zero error so that the instability in nonlinear systems can produce a significant spread of the parameters to be predicted. One should never give the people who are using mathematical models of complex systems (such as, e.g., atmosphere, climate or weather) false comfort about the accuracy of such models.

However, there is a certain observation that can help in mathematical modeling of large classes of complex systems, consisting in the concept that the real-life processes tend to evolve in such a way as to choose the optimal route among a number of alternatives. A sufficiently complex system usually behaves as if it could explore – figuratively speaking, sniff – the integral trajectory of the corresponding dynamical system serving as a mathematical model. In more mathematical terms, it means that the global character of solutions to a dynamical system may sometimes be guessed or even known in advance due to some guiding principles (such as variational principles, optimal control, etc.). Furthermore, complex systems can in certain cases adapt to unexpected pitfalls and barriers (e.g., modeled by attractors and repellers in a corresponding dynamical system) like in a computer game. Complex systems can also make a series of trials, for instance, in the evolution to complexity (which is illustrated already on the onset of a Rayleigh-Bénard system of cells).

Such catastrophic events as instability, chaos, bifurcation, extreme sensitivity to initial conditions, etc. arising as the response of a system to the change of a control parameter did not enter the repertoire of classical science, where the dependence on parameters was nearly always assumed smooth. Complex systems are necessarily nonlinear and therefore develop in the presence of feedback. Due to feedbacks, a complex system beyond some critical value of the control parameter totally forgets its initial state and may respond as an amplifier drastically incrementing the external influence or internal fluctuations, sometimes to catastrophic amounts.

## 12.2. Nonlinear oscillator

When discussing the harmonic oscillator, we restricted ourselves to a very special choice of the phase velocity vector field $\boldsymbol{w} = (\dot{x}^1, \dot{x}^2)^T = (F, G)^T$ as a linear function of phase variables. This is rarely the case when modeling the processes in real-life systems, e.g., in chemistry, biology, ecology, engineering, etc. The world is unfortunately more complex than prescribed by the sterile model of the harmonic oscillator. By introducing nonlinearity in the archetypal oscillator model, one can observe primary manifestations of nonlinear phenomena such as the dependence of amplitude on frequency and the emergence of self-excitation.

One of the simplest models of nonlinear oscillations is given by the equation containing a polynomial nonlinearity $P_3(x) = \alpha x + \beta x^3$ i.e., oscillations in the field of force $f(x) = -P_3(x) + F(x, t)$:

$$\ddot{x} + \gamma \dot{x} + \alpha x + \beta x^3 = F(x, t), \tag{12.2.1.}$$

where $F(x, t)$ denotes an external forcing, e.g., $F(x, t) = A \cos \omega t$ (harmonic driving force) or $F(x, t) \equiv 0$ (a free Duffing oscillator) and so forth.

Introducing, as usual, the plane space variables $x = x^1$ and $\dot{x} = x^2$ we can rewrite the free equation $+\frac{1}{2}\left(\frac{D^2 u}{Du}\right)^2 + \frac{1}{2}\left(\frac{D^2 s}{Ds}\right)^2$ notice (12.2.1.) (known as the Duffing equation) in the form of a dynamical system

$$\begin{pmatrix} \dot{x}^1 \\ \dot{x}^2 \end{pmatrix} = \begin{pmatrix} x^2 \\ -\gamma x^2 - \alpha x^1 - \beta (x^1)^3 \end{pmatrix}.$$

To elucidate the role of nonlinearity we shall simplify the equation, primarily concentrating on the cubic term and putting dissipation $\gamma = 0$. The oscillator is nonlinear when $\beta \neq 0$, and if $\beta \ll 1$ the system can be solved by perturbation techniques. However, when $\beta \sim 1$ the system becomes highly nonlinear, and new behavior such as chaos or bifurcation may manifest itself. In such cases, perturbation techniques generally fail, and one must resort to numerical methods. Although equation (12.2.1.) is one of the simplest specimens of a nonlinear system, it can be used as a tool to study all typical complications related with nonlinearity. We can write the dissipationless equation (12.2.1.) with a simple harmonic forcing in the form

$$\ddot{x} + \alpha(1 + \varepsilon x^2)x = A \cos \omega t,$$

where $\varepsilon \equiv \beta / \alpha$, to observe that the cubic-nonlinear (Duffing) oscillator represents a harmonic oscillator with a variable eigenfrequency, $\omega_0^2(x) = \alpha(1 + \varepsilon x^2)$. Since in the mechanical model of oscillator eigenfrequency is expressed as $\omega_0^2 = k/m$, where $k$ is a spring constant, $m$ is the oscillator mass, or as $\omega_0^2 = mgl/I$ (a pendulum), where $I = ml^2$ is the moment of inertia, $l$ is the pendulum's length, we can assume that local changes of eigenfrequency are either due to a variable spring constant $k = k(x)$ or due to a variable pendulum length $l = l(x)$. One more possibility would be a variable mass, $m = m(x)$, but such a model would require special properties of the medium supporting the oscillations (e.g., a periodic one) so that we shall not consider this case. Physically, oscillations with a variable elastic constant or a pendulum length mean that the oscillation period can even be modified within a single oscillation cycle. In general, we can observe that the frequency variations within a single period are an attribute of nonlinear oscillations. Notice that the so-called generation of harmonics, which is typical of nonlinear wave phenomena, can be physically interpreted through the waveform distortion resulting from instantaneous frequency changes within a single oscillation period. It is these instantaneous frequency changes leading to mode coupling and waveform distortions that make the powerful Fourier transform largely inadequate in the nonlinear domain, and one has to invent other techniques to deal with nonlinear oscillations and waves. As to integral transforms, perhaps the simplest one to deal with instantaneous quantities such as amplitudes, phases and frequencies is the Hilbert transform widely used in signal analysis [72] and [87].

### 12.2.1. Pendulum: an example of a dynamical system

Let us illustrate the motion of a nonlinear oscillator by the model of a pendulum as a simple dynamical system. A fully nonlinear 1d oscillator is usually known as a planar pendulum. This is a system with 2d phase space $\mathbb{R}^2$ and dynamics $\dot{x}^1 = x^2$, $\dot{x}^2 = -\sin x^1$, where $x^1 = \varphi$ is the pendulum deflection angle and $x^2$ is its angular velocity $\dot{\varphi} \equiv \omega$ or $x^2 = \mathcal{L}_\varphi = I\omega$, $I = ml^2$ is the moment of inertia, $l$ is pendulum's length, $\mathcal{L}_\varphi$ is its angular momentum. In these dynamic equations, units are so chosen as to make $x^1, x^2$ dimensionless ($\omega^2 = mgl/I = 1$). The physical motion equation of a free pendulum is $I\ddot{\varphi} = -mgl \sin \varphi$. In fact, dynamical variable $x^1$ does not run throughout the entire $\mathbb{R}$, this set is the covering space of the configuration manifold for the physical variable, angle $\varphi$. Since values of $x^1$ and $x^1 + 2\pi m, m = 0,1,2, ...$ correspond to the same state of the pendulum, one can take as a phase space the cylinder $P = (x^1 \mod 2\pi, x^2)$ instead of the $\mathbb{R}^2$ plane (Figure 14). The phase trajectories are then closed curves on the cylindrical surface (with the exception of separatrices between phase curves manifesting different properties i.e., between oscillating and rotational regimes as well as of lower and upper equilibrium positions). It is clear that the energy $E = \frac{(\dot{x}^1)^2}{2} - \cos x^2 \equiv H(x^1, x^2)$ is preserved by the flow which remains on the level set of $H(x^1, x^2)$. This flow is symplectic which is tantamount in this case (for 2d phase space) to area preservation.

One can use this occasion to expand a bit more on the configuration space and generalized coordinates. The configuration space $Q$ of a physical system (with $n$ degrees of freedom) may be interpreted as a set of all its possible states, with generalized coordinates just being functions that take numerical values and are defined on $Q$. Thus, the configuration space is understood as a manifold $Q = Q^n$ on which coordinates $q^i$ have been chosen. One usually assumes that a collection of $n$ more or less arbitrary functions enables us to define uniquely all the points in the configuration space of an arbitrary physical system, which is in fact a bold assumption. For example, some difficulties with the uniqueness may arise, in particular, when one chooses angles or phases as the generalized coordinates. Nevertheless, if we assume that $n$ coordinate functions uniquely depict all the points in a configuration space $Q$, we may consider $Q$ to be a subset of the real space of numbers i.e., $Q \subseteq \mathbb{R}^n$.

After this general picture, we may briefly focus on some details. The vector field of the pendulum $\dot{x}^1 = x^2$, $\dot{x}^2 = -\omega^2 \sin x^1$, $\mathbf{v} = (v^1, v^2)^T = (x^2, -\omega^2 \sin x^1)^T$ has singular points $x^1 + \pi m, m = 0, 1, \ldots$, $x^2 = 0$ i.e., located equidistantly on the $x^1$ axis. The harmonic oscillator appears when we restrict the pendulum motion to small amplitude, $|x^1| \ll 1$, and in this case we easily obtain the exact solution expressing the dynamical system evolution in terms of elementary functions, e.g., $x^1(t) = a\cos \omega t$, $x^2(t) = -a\omega \sin \omega t$. The pendulum model can also be solved exactly, however not through elementary but special functions. Indeed, using "physical" variable $\varphi$ instead of $x^1$, we have the equation $\ddot{\varphi} + \omega^2 \sin \varphi = 0$, $\omega^2 = mgl/I$. Integration gives $\frac{1}{2}\dot{\varphi}^2 - \omega^2 \cos \varphi = \omega^2 \cos \varphi_m$, where $\varphi_m$ is the maximal angle i.e., $\dot{\varphi}|_{\varphi = \varphi_m} = 0$. One usually denotes for convenience parameters $\alpha \equiv \sin(\varphi_m/2)$ and $\vartheta$, $\alpha \sin \vartheta \equiv \sin(\varphi/2)$, then we shall have $\cos \varphi_m = 1 - 2\alpha^2$, $\cos^2(\varphi_m/2) = 1 - \alpha^2$, $\sin \varphi = 2\alpha \sin \vartheta \sqrt{1 - \alpha^2 \sin^2 \vartheta}$, $\cos \varphi = \cos \varphi_m + 2\alpha^2 \cos^2 \vartheta = 1 - 2\alpha^2 \sin^2 \vartheta = 1 - 2\sin^2(\varphi/2)$ and $\dot{\varphi} = \sqrt{2}\omega(\cos \varphi - \cos \varphi_m)^{1/2} = 2\omega\alpha \cos \vartheta$. From the relationship $\sin \varphi = 2\alpha \sin \vartheta \sqrt{1 - \alpha^2 \sin^2 \vartheta}$ we obtain the connection between $d\varphi$ and $d\vartheta$, $\cos \varphi d\varphi = 2\alpha \cos \vartheta \frac{1 - 2\alpha^2 \sin^2 \vartheta}{\sqrt{1 - \alpha^2 \sin^2 \vartheta}} d\vartheta$, $d\varphi = 2\alpha \frac{\cos \vartheta}{\sqrt{1 - \alpha^2 \sin^2 \vartheta}} d\vartheta$ so that $dt = \frac{d\varphi}{2\omega\alpha \cos \vartheta} = \frac{1}{\omega} \frac{d\vartheta}{\sqrt{1 - \alpha^2 \sin^2 \vartheta}}$ and the law of motion is given by the parametric function $t = \frac{1}{\omega} \int_0^{\vartheta} \frac{d\vartheta}{\sqrt{1 - \alpha^2 \sin^2 \vartheta}}$. The oscillation period can be expressed via elliptic functions as $T = \frac{4}{\omega} \int_0^{\pi/2} \frac{d\vartheta}{\sqrt{1 - \alpha^2 \sin^2 \vartheta}} = \frac{4}{\omega} E(\alpha, \frac{\pi}{2})$ or, equivalently, by series $T = T_0 \left(1 + \left(\frac{\alpha}{2}\right)^2 + \left(\frac{1 \cdot 3}{2}\right)^2 \left(\frac{\alpha}{2}\right)^4 + \left(\frac{1 \cdot 3 \cdot 5}{2 \cdot 3}\right)^2 \left(\frac{\alpha}{2}\right)^6 + \cdots \right) = T_0(1 + \frac{1}{16}\varphi_m^2 + \cdots)$, where $T_0 = 2\pi/\omega$ denotes the period of harmonic oscillations. There exist also other parameterizations of the pendulum period, e.g., directly in terms of angle $\varphi$ (using the hyperbolic relationship $\dot{\varphi} = 2\omega(\alpha^2 - \sin^2(\varphi/2))^{1/2}$, but integration in such cases is less convenient, see [93].

Let us now briefly discuss the symmetry of the pendulum model. The corresponding dynamical system has an obvious symmetry under rotation by $k\pi, k = 0, 1, 2, \ldots$ which is in fact a discrete symmetry $(x^1, x^2) \to (-x^1, -x^2)$ mapping the flow into itself. Another discrete symmetry, $(x^1, x^2) \to (x^1, -x^2)$ reverses the motion by flipping the vector field and is equivalent to time reversal operation $t \to -t$ under which the phase portrait preserves its form (this reflects the time-reversal symmetry in autonomous Hamiltonian systems). It would be probably pertinent to mention that popular numerical techniques such as Euler's, Runge-Kutta, multistep and similar methods (all of them actually based on Taylor expansions) do not care much about preserving the phase portrait, symmetry or other geometric properties of the flow. This inattention to qualitative features of motion can result in additional errors of nonphysical nature. For example, if a system (e.g., a pendulum) stays on a cylinder, it should remain on this manifold for all discretized time in process of numerical integration, otherwise some spurious degrees of freedom may appear.

The pendulum model, although more general and closer to reality than that of the harmonic oscillator, is still based on a number of "mathematical" assumptions. Thus, the rod of the pendulum is assumed absolutely rigid (its length $l = const$) and massless; there is no dissipation, no deviation from planar motion, the gravitation field is absolutely homogeneous and strictly one-dimensional ($\mathbf{g} = g_z\mathbf{e}_z, g_z = g = const$), etc.

### 12.2.2. Transition between pendulum and harmonic oscillator

The pendulum is a nonlinear model, and its linearization leads to the harmonic oscillator. Physically, it means that the motion, as in most linear models, occurs near the equilibrium position. If, the other way round, we have a harmonic oscillator, one may ask: what will happen when its oscillations are no longer small? To explore the oscillations that gain a considerable swing, we can try to slightly correct the model of harmonic motion in the following way

$$\dot{x}^1 = x^2 + \varepsilon h^1(x^1, x^2), \qquad \dot{x}^2 = -x^1 + \varepsilon h^2(x^1, x^2), \qquad \varepsilon \ll 1 \qquad (12.2.2.1.)$$

or, more generally, $\mathbf{x} = \mathbf{v}(\mathbf{x}) + \varepsilon\mathbf{h}(\mathbf{x})$. Here, for simplicity, time is dimensionless i.e., oscillation frequency $\omega = 1$. It is usually assumed that vector $\mathbf{x}$ is restricted to some compact area $D$ of radius $d, |\mathbf{x}| \leq d$; in our two-dimensional example $r^2 = (x^1)^2 + (x^2)^2 \leq d$. Obviously, for $\varepsilon = 0$ we get the harmonic motion $x^1 = a\cos(t - t_0)$, $x^2 = -a\sin(t - t_0)$ that is represented on the phase plane $(x^1, x^2)$ by a family of circular trajectories with variable radius $a$. When $\varepsilon \neq 0$, the phase point can deviate from these trajectories yet for $\varepsilon \ll 1$ stays nearby. Using the language of stability theory (see above) we may conclude that a non-dissipative pendulum as well as undamped harmonic oscillator is Lyapunov stable, but not asymptotically stable, and both models acquire the status of asymptotic stability when even an infinitesimal dissipation is accounted for. The perturbed trajectories, however, are not necessarily closed since the phase point does not in general stay on the energy surface $H(x^1, x^2) = \frac{(x^1)^2 + (x^2)^2}{2} = E$. In this sense, the linear oscillator loses its conservative properties, although remaining, for time-independent correction vector $\mathbf{h}(\mathbf{x}, t) = \mathbf{h}(\mathbf{x})$, an autonomous system. The deviation of the phase trajectories from the energy manifold physically signifies that the oscillator can gain or lose energy. Accordingly, one can quantify these gains or losses by taking the time derivative of energy along the phase trajectories i.e., in the vector field direction. This time derivative is

$$\dot{E} = \frac{dH(x^1, x^2)}{dt} = \varepsilon\big(x^1 h^1(x^1, x^2) + x^2 h^2(x^1, x^2)\big). \qquad (12.2.2.2.)$$

However, this derivative – like any derivative – only gives a local rise or fall, and to find the gain or loss in energy for the whole cycle we have to integrate (12.2.2.2.) along the deviated phase path that is actually unknown. Nevertheless, we can perform this integration approximately over the orbit corresponding to the harmonic oscillation period $T = 2\pi/\omega$ (in our system of units $T = 2\pi$), retaining only the first-order terms in small parameter $\varepsilon$. Then we have

$$\Delta E = \int\limits_{t_0}^{t_0+2\pi} \dot{E}\, dt$$

$$= \int\limits_{t_0}^{t_0+2\pi} \frac{dH(x^1, x^2)}{dt}\, dt = \int\limits_{t_0}^{t_0+2\pi} \frac{dH(a\cos(t-t_0), -a\sin(t-t_0))}{dt}\, dt + \sigma(\varepsilon)$$

$$= \int\limits_{0}^{2\pi} \frac{dH(a\cos t, -a\sin t)}{dt}\, dt + \sigma(\varepsilon)$$

$$= \varepsilon \int\limits_{0}^{2\pi} \left(x^1 h^1(x^1, x^2) + x^2 h^2(x^1, x^2)\right) dt + \sigma(\varepsilon) = \varepsilon G(a) + \sigma(\varepsilon),$$

where $G(a) \equiv \oint_C (h^1 dx^2 - h^2 dx^1)$ is the symplectic integral over circular contour $C = \left((a\cos t, -a\sin t)\right)$.

Now, we can have a qualitative judgement: if $G(a) > 0$, then, for small $\varepsilon > 0$, energy grows and phase trajectories expand i.e., $\dot{r} > 0$ and $dr/d\varphi > 0$, $r = (x^1)^2 + (x^2)^2$ whereas for $G(a) < 0$, $\dot{r} < 0$ and phase curves spiral towards the origin $(x^1, x^2) = (0,0)$. It means that oscillations are damped, although there is no explicitly dissipative term in the dynamical equations. One can also imagine the third possibility, when function $G(a)$ changes its sign i.e., passes through zero. In this case, we see that roots $a_i, i = 1,2, \ldots$ of equation $G(a) = 0$ correspond to closed orbits in the phase space so that the motion looks as if it were confined to the manifold of constant energy, $\dot{E} = dH(x^1, x^2)/dt = 0$ i.e., is similar to simple harmonic motion. There is, however, one crucial difference: closed orbits corresponding to the roots of $G(a)$ are isolated whereas the family of closed trajectories corresponding to harmonic oscillations continuously depends on initial data that determine the solution coefficients. Such isolated closed loops with radii $a_i$ (amplitudes) are known as limit cycles of a dynamical system. A limit cycle, in general, reflects some rhythmic activity, and for a limit cycle, totally different initial conditions give rise to the same amplitude of periodic oscillations. So, the perturbed oscillator ($\varepsilon \neq 0$) is no longer conservative, which leads to qualitatively new phenomena, the emergence of limit cycles being one of them. In a 3d continuous dissipative system, a limit cycle is represented by the following signs of Lyapunov exponents: $(\Lambda_1 = 0, \text{sign}\, \Lambda_2 = -1, \text{sign}\, \Lambda_3 = -1) \equiv (0, -1, -1)$.

A limit cycle can be stable, ($G'(a_i) < 0$), when all neighboring trajectories approach it with $t \to \infty$, or unstable ($G'(a_i) > 0$), when such trajectories spiral towards it for $t \to -\infty$ i.e., they go away from the limit cycle as the physical time passes[99]. However, phase trajectories can never join or even touch the limit cycles since it would signify the violation of the existence-uniqueness theorem. A limit cycle physically corresponds to some periodic behavior; it means that the process repeats itself. For example, in biological modeling, such repetitive processes are heartbeat or breathing, some intrinsic cycles in economic modeling, periodic chemical reactions. There is an important and still a controversial problem: how many limit cycles can a polynomial (in particular, quadratic) system have? Thus, the above-discussed Lotka-Volterra model is a quadratic system and exhibits a behavior typical for such systems; how many limit cycles can this model in principle possess?

---

[99] Some people prefer calling unstable limit cycles just closed trajectories.

One can mention the Bendixson criterion and the Poincaré-Bendixson theorem in association with limit cycles. The Bendixson criterion is a negative statement telling us when there are no limit cycles (closed periodic orbits) for an autonomous dynamical system in the plane (2d). This criterion is an almost obvious qualitative statement that can be formulated as follows: if in a connected two-dimensional domain $D$, the flow $\mathbf{v}$ has a divergent property, div $\mathbf{v} \neq 0$ (here $\mathbf{v}$ is the vector field), then there are no limit cycles in $D$. One of the popular proofs of the Bendixson criterion is based on Green's theorem: there can be no closed trajectories in $D$ since it would lead to a contradiction with this theorem. The Poincaré-Bendixson theorem states that any bounded solution $\mathbf{x}(t)$ of an autonomous dynamical system $\dot{\mathbf{x}} = \mathbf{v}(\mathbf{x})$ in the plane (2d) converges for $t \to \infty$ either to an equilibrium point ($\mathbf{v}(\mathbf{x}) = 0$) or to a limit cycle. Thus, the Poincaré-Bendixson theorem effectively excludes chaos in 2d continuous dynamical systems (flows); it has already been noted that one needs at least three dimensions for chaos. For discontinuous systems (maps), chaos can appear even in 1d, as we have seen in logistic map.

The study of limit cycles is a very important part of mathematical modeling based on the dynamical systems approach, leading to intriguing mathematical problems. We, however, will not engage in discussing this subject in depth in order to keep the present book within reasonable limits. The only example we shall briefly touch upon in connection with limit cycles is the famous van der Pol oscillator since this model is ubiquitous in physics, biology, electrical engineering and in a number of other disciplines. The van der Pol oscillator is described by equation $\ddot{x} - \varepsilon\dot{x}(1 - x^2) + x = 0$ which for $\varepsilon = 0$ is reduced to the harmonic oscillator. Writing the van der Pol equation in the form for small $\varepsilon$, we shall have $h^1(x^1, x^2) = 0, h^2(x^1, x^2) = x^2(1 - (x^1)^2)$. Accordingly, $G(a) = -\oint_c h^2(x^1, x^2)dx^1 = \int_0^{2\pi} a^2 \sin^2 t(1 - a^2 \cos^2 t)dt = \pi a^2(1 - a^2/4)$. At $a^2 = 4$ i.e., $a \equiv a_0 = 2, G = 0$ (the value $a = -2$ has no physical meaning). For $a = a_0 + \delta$ ($\delta > 0$), $G < 0$, for $a = a_0 - \delta, G > 0$ which means that $G'(a_0) < 0$ and $G(a_0) = 0, a_0 = 2$ corresponds to a stable limit cycle i.e., closed circular orbit $(x^1)^2 + (x^2)^2 = 4$ in the phase plane that portrays the "auto-oscillations" (whose amplitude does not depend on the initial conditions).

### 12.2.3. Many linked pendulums

It is instructive to compare the model of many linked pendulums with the linear chain model (9.4.1.), where normal modes i.e., elementary excitations appear, in that case similar to phonons in solid state physics. In the linked pendulums model, other types of excitations – nonlinear ones – appear. If we consider $N$ identical mathematical pendulums elastically coupled to one another whose pivots are placed equidistantly along the horizontal $x$-axis, then we can model such a structure by the following motion equations:

$$ml^2\frac{d^2\varphi_i}{dt^2} = -mgl\sin\varphi_i + k(\varphi_{i+1} - \varphi_i) + k(\varphi_{i-1} - \varphi_i), \qquad (12.2.3.1.)$$

where $\varphi_i$ is the angle with the vertical axis $y, m$ is the pendulum mass, $l$ is its length, $k$ is the elastic coupling (spring) constant. For simplicity, all such parameters are considered equal for all the pendulums. If we introduce an elementary length $h := L/N$ where $L$ is the length of the whole chain of pendulums and let this parameter tend to zero, we shall arrive at the continual limit ($\varphi_{i+1} - \varphi_i) - (\varphi_i - \varphi_{i-1}) \to h^2(\partial^2\varphi/\partial x^2)$ so that we get a nonlinear PDE for angle $\varphi$:

$$\frac{\partial^2\varphi}{\partial t^2} = -\omega_o^2\sin\varphi + c^2\frac{\partial^2\varphi}{\partial x^2}, \qquad (12.2.3.2.)$$

where $\omega_0^2 \equiv \frac{g}{l}$, $c \equiv \frac{h}{l}\sqrt{\frac{k}{m}} = \frac{L}{Nl}\sqrt{\frac{k}{m}}$ (note that elastic constant $k$ has the dimensionality of energy in this model). Equation (12.2.3.2.) is the famous sine-Gordon PDE, where parameter $c$ has the meaning of the speed of sound wave. The sine-Gordon equation is one of the most important nonlinear models, producing solitary wave solutions such as solitons, kinks, antikinks, breathers and the like. A number of important problems in physics, chemistry, biology and various engineering branches (specifically telecommunications and geophysical risk assessment) lead to the models based on the sine-Gordon equation. This nonlinear hyperbolic PDE has solutions describing the usual sound waves, but the most interesting are the soliton solutions. The latter may be constructed using the mapping $\varphi = 4\tan^{-1} u$. Differentiating, we get

$$\frac{\partial \varphi}{\partial x} = \frac{4}{1+u^2}\frac{\partial u}{\partial x}; \; \frac{\partial^2 \varphi}{\partial x^2} = \frac{4}{1+u^2}\frac{\partial^2 u}{\partial x^2} - \frac{8u}{(1+u^2)^2}\left(\frac{\partial u}{\partial x}\right)^2;$$

and separating the variables, $u(x,t) = f(x)/g(t)$ we can seek solutions of the form $f(x) = \exp(x/a)$, $g(t) = \exp(Vt/a)$, where $a$ and $V$ are some constant parameters. Finally, we obtain the soliton solution

$$\varphi(x,t) = 4\tan^{-1}\left(\exp\left(\frac{x-Vt}{a}\right)\right).$$

Notice that in the case of small amplitudes, $\varphi \ll 1$ i.e. $\sin\varphi \approx \varphi$, equation (12.2.3.2.) is reduced to the linear Klein-Gordon equation, $\varphi_{tt} - c^2\varphi_{xx} + \omega_0^2\varphi = 0$, having solutions of the form $\varphi(x,t) = \varphi_0 \cos(kx - \omega t)$ with $\omega^2 = \omega_0^2 + c^2 k^2$.

## 12.3. A Brief Remark about the Lorenz model

Chaos in nonlinear systems is often illustrated on the famous Lorenz model [109]. The behavior of a system described by a simple-looking system of three ODEs

$$\begin{cases} \dfrac{dx}{dt} = -\sigma(x-y) \\ \dfrac{dy}{dt} = -xz + rx - y \\ \dfrac{dz}{dt} = xy - bz \end{cases} \qquad (12.3.1.)$$

with positive parameters $\sigma, b, r$. The behavior of nonlinear system (12.3.1.) i.e., its trajectories looks extremely irregular (e.g., for $\sigma = 10$, $b = 8/3$ and $r = 28$ – an example used by E. Lorenz himself), although dynamical system (12.3.1.) is completely deterministic and even quasilinear. More specifically, this system has an attractor (the limit set for almost all nearby trajectories) which is neither periodic nor asymptotically periodic[100]. Moreover, since the trajectories exhibit extreme sensitivity to initial data, the Lorenz attractor is chaotic.

The Lorenz system is a 3d dynamical system obtained by a reduction of the model for Rayleigh-Bénard instability in fluid dynamics.

## 12.4. Instabilities and chaos

The most outstanding fact in the theory of dynamical systems is that fully deterministic systems depending on only a few variables can exhibit chaotic behavior which is similar to that formerly encountered only in many-body systems. Many nonlinear deterministic systems, although looking quite simple are observed to behave in an unpredictable, seemingly chaotic way. The term "chaotic" is commonly attributed to evolutions exhibiting an extreme sensitivity to initial data, but this is not the whole story. Chaos is also understood as an aperiodic – close to random – behavior emerging from a totally deterministic environment i.e., described by a dynamical system involving no random parameters or noise. Such a behavior appears only in nonlinear systems and manifests an extreme sensitivity to initial conditions (in general, also to external parameters). Therefore, nonlinear systems can be viewed as one of the most interesting subjects in mathematics since even in the fully deterministic case they envisage a certain unpredictability.

Thus, classical chaotic systems, although they may have only a few degrees of freedom and be mathematically represented through deterministic equations, have to be described by probabilistic methods. Nonetheless, a distinction from the traditional probability theory is one of the main issues of chaos theory. Chaos is also sometimes called "deterministic noise".

The notion "aperiodic" mathematically means that the flow paths do not converge for $t \to +\infty$ to fixed points or to periodic (quasiperiodic) orbits such as limit cycles. We have just mentioned that the extremely sensitive dependence of the system trajectories on initial conditions is the distinguishing mark of deterministic chaos. But this is just a verbal expression that has to be quantified. One of the possible ways to estimate this sensitivity is to find a variational derivative $\delta x(t, x_0)/\delta x_0$, where $x(t) = x(t, x_0) = g_{t,t_0} x_0$, $x_0 \equiv x(t = t_0)$ is the motion generated by flow (the dynamical system) $g_t$. In simple cases, we can replace the variational derivative by the usual one to have

$$\frac{dx(t, x_0)}{dx_0} = A \exp\left(\frac{t}{\tau_0}\right),$$

where $\tau_0 \approx \Lambda_0^{-1}$ is the predictability horizon, $\Lambda_0 > 0$ is the greatest positive Lyapunov exponent which expresses instability in the form of "stretching" (and, perhaps, also "folding" leading to chaos). Thus, the necessary (not sufficient!) condition for deterministic chaos is that dynamics should be unstable. In particular, instability means that small perturbations in the initial conditions bring large uncertainties in dynamics after the predictability horizon.

---

[100] It is curious that the now famous paper by Edward Lorenz "Deterministic nonperiodic flow" had been published nine years before the mathematical and physical communities became fully aware of it. This is a lesson to be learned by those who doubt the usefulness of interdisciplinary studies.

By remarking that extreme sensitivity to initial data is not the whole story, we wanted to hint that such sensitive dependence can be found in very simple, e.g., linear systems. Take, for example, map $x_{n+1} = 2x_n, x_n \in \mathbb{R}, n \in \mathbb{Z}$, which has an unfortunate property to explode (see also below, "The logistic model: the bugs are coming"). However, the explosive divergence of nearby trajectories in this case alongside the blow-up of an initial data discrepancy to infinity has nothing to do with deterministic chaos. The latter, combining the sensitivity to initial data with unpredictable behavior, appears only if the trajectories are bounded which is possible only in nonlinear dynamical systems: in linear ones there can be either bounded trajectories or sensitive dependence on initial conditions but not both, so that nonlinearities are necessary to have both effects.

The word "chaos" is intuitively associated with a disorganized state, completely without order. This is deceptive: chaos in dynamical systems is not a total disorder but corresponds to irregular variations of the system's variables controlled by rather simple rules. Probably, no mathematical definition of chaos has been universally accepted so far, but the following descriptive explanation of what chaos is can be used to work with this phenomenon. Chaos is a durable irregular (i.e., aperiodic) behavior emerging in deterministic systems which become extremely sensitive to slight variations of parameters (in particular, initial conditions). Notice that this verbal description of chaos contains three components:

1. Durable irregular (aperiodic) behavior implies that the phase paths do not asymptotically ($t \to \infty$) stabilize either to a point or to periodic orbits.
2. The term "deterministic" implies that the system is not described by stochastic differential (or other) equation i.e., there are no random parameters or noise present.
3. Sensitivity to slight variations of parameters (in particular, initial conditions) implies that the integral trajectories, at first very close to one another, diverge exponentially fast with time, the divergence rate being governed by the Lyapunov exponents (with at least one of them positive).

Thus, if one abstracts oneself from certain fine points such as irregular trajectories, uncorrelated behavior in close time points and so on, chaos can be basically viewed as an absence of Lyapunov stability. Although "chaos" has become the code word for nonlinear science, there is nothing particularly exotic or intricate about chaos. The fact that chaotic phenomena practically had not been studied until the 1970s, when chaos suddenly came into fashion together with its interdisciplinary applications, can only be attributed to an absolute dominance, both in science and technology, of successful linear theories such as classical electrodynamics, quantum mechanics, linear oscillations, plasma instabilities, etc. Ubiquitous linear input-output models in electrical engineering that have resulted in many practically important technologies also left little room for studying somewhat exotic nonlinear models. The main feature of chaos in finite-dimensional dynamical systems (known as deterministic chaos to distinguish it from molecular chaos in many-particle systems) is the exponential divergence of trajectories. The quantitative measure of this divergence is the so-called K-entropy (the Kolmogorov-Krylov-Sinai entropy [139] and [61]). The value of K-entropy is positive for chaotic states, which corresponds to mixing and exponential decay of correlations.

Many people are still inclined to think that one should not treat chaos in deterministic systems as a special subject, and the word itself is just a poetic metaphor for the long familiar instability. This is wrong: chaos is not completely synonymous with instability; it incorporates also other concepts (such as irregularity of behavior and decorrelation in time series). In distinction to instabilities, chaos points at unpredictability of behavior in the systems described by completely deterministic and even primitive looking equations. Instability in dynamical systems can be viewed as a symptom of the possible transition to chaotic motion. One can observe on some examples, in particular on deterministic models of growth that may exhibit instability (i.e., have diverging integral trajectories), but cannot be viewed as chaotic; the simplest example of such a system is the model of exponential growth $\dot{x} = ax$. Conversely, one should not think that the chaotic state is always visibly unstable: for example, turbulence is a stable chaotic regime. Thus, the often-repeated view that chaos is just an extreme case of instability i.e., in chaotic regimes paths in the phase space that start arbitrarily close to each other diverge exponentially in time whereas in regular (nonchaotic) regimes two nearby trajectories diverge not faster than polynomial, typically linear in time, is not quite accurate. If we compute the distance $d(t)$ between two phase paths whose initial separation is $d_0 \equiv d(0)$, then exponential divergence of trajectories, $d(t) = d_0 e^{\Lambda t}$, where $\Lambda$ is the Lyapunov characteristic exponent, is a necessary but not sufficient condition for chaos.

Notice that the concept of dynamical systems as evolving quasi-closed parts of the world does not exclude their chaotic behavior, when a system's evolution (i.e., dependence of its observable quantities $x^i$ in time) looks like a random process; at least it is totally unpredictable beyond a certain "horizon of predictability" (recall weather forecasts). One should not, however, think that unpredictability in chaotic systems is identical to the true randomness, but the difference is rather academic since chaos looks quite like a random process and can also be described by a probability measure (distribution functions). One can crudely say that there are at least two types of randomness: one is due to an unobservable quantity of interacting subsystems (as, e.g., in gas), the other – usually known as deterministic chaos – is due to our limited ability to formulate the rules of behavior under the conditions of drastic instability. Such poorly known rules must govern the processes that basically arise from irreversibility in dynamical systems.

The main feature of chaos in dynamical systems is an extreme sensitivity to small perturbations, e.g., to tiny inaccuracies in input data such as initial conditions. It is this property that makes it impossible to forecast the state of a dynamical system for the time exceeding some characteristic predictability horizon amenable to the state-of-the-art numerical computation. The time scale for exponential divergence of nearby trajectories, $\Lambda^{-1}$, may serve as an estimate for this predictability horizon. It is important that this time scale usually does not depend on the exact value of initial conditions. Anyway, large Lyapunov exponents are symptomatic of the onset of chaos. Recall that the Lyapunov characteristic exponent manifests the expansion rate of linearized dynamical system along its trajectory.

Nevertheless, one can distinguish between dynamical chaos in deterministic systems and physical chaos in many-particle models. For example, evolution to thermodynamic (statistical) equilibrium is directed to the most chaotic and disordered state characterized by maximal entropy. The difference between "dynamical" and "physical" chaos has been reflected in long-standing debates about the origin of stochasticity in physical systems. There were historically two distinct trends of thought: (1) stochasticity arising due to dynamic instability of motion in nonlinear deterministic systems and (2) necessity of statistical description due to the enormous number of degrees of freedom i.e., huge dimensionality of the phase space in realistic (many-particle) physical systems, resulting in practical irreproducibility of solutions (integral trajectories). These two trends were poorly compatible because they required different approaches to physical statistics. In the section devoted to statistical mechanics and thermodynamics, we shall comment on the possibility to reconcile the two manners of description.

One can produce many examples illustrating the importance of chaotic behavior in real life. Thus, when an asteroid or a comet approaches a planet (e.g., Earth), the planet's gravity perturbs the comet's trajectory so that small changes of the latter can be amplified into large and poorly predictable deflections. Since a comet or an asteroid trajectory is affected by numerous close encounters with planets, small variations in the initial parameters of the trajectory may result in practically complete unpredictability of the body's eventual motion (for large enough time – outside the predictability horizon). One can call this few body process a trivial chaotization of the comet or asteroid path. Because of such sensitivity to small variations of parameters, trajectories of small celestial bodies can diverge and cross the planetary orbits, possibly with devastating consequences. The information about the fate of a comet or an asteroid is practically lost after several Lyapunov time scales, and the resulting unpredictability may be the source of meteorite hazard for the Earth.

## 12.4.1. Chaos in dissipative systems

Chaotic behavior is quantified by the presence and the value of Lyapunov exponents that must be positive in order for the chaotic regime – and the complexity associated with it – to emerge. Recall that the notion "chaotic" is related to the bounded motions displaying an extreme sensitivity to initial data. If there is no dissipation i.e., the dynamical system is conservative, the sum of all Lyapunov exponents $\Lambda_i$ must be zero – to ensure that a volume element of the phase space remains intact along the phase trajectory (see 8.4.1. "Phase space and phase volume"). If the system is dissipative, the sum of all Lyapunov exponents should be negative, and if $\sum_i \Lambda_i < 0$, at least one Lyapunov exponent must exist in a dissipative system[101].

We observed from expression (1) that a free pendulum eventually stops: it is damped down to a stable state corresponding to the minimum of potential energy in the gravitational field. Such a stable state is referred to today as an attractor. Although there seems to be no universally accepted definition of attractor, we may intuitively understand it as a limit set for which any nearby orbit regardless of its initial conditions ends up (has its limit points) in the set. The simple example of a pendulum has, however, a deep meaning. Firstly, it demonstrates that a dynamical system can radically change its behavior when energy is withdrawn from the system. Secondly, for large times ($t \to \infty$), trajectories of a dissipative system tend to a low-dimensional subset of the phase space which is an attractor. Thirdly, the evolution of a dissipative system, after a certain time, is restricted to this attractor forever.

There may be more than one attractor in a dynamical system, and the latter can wander between its attractors as the value of some control parameter varies (we shall observe such transitions shortly on a simple example of the logistic map). Many physical systems exhibit such a behavior. For example, addressing again the meteorite hazard, we may note that small celestial bodies such as asteroids or comets (sometimes called "planetesimals") move around the Sun on almost Keplerian orbits experiencing a slight drag due to the presence of interplanetary gas and solar radiation. This small resistance forces the small body trajectories to be damped down to the Sun, and the ensuing radial component of the motion results in crossing the planetary orbits. Then small bodies can be trapped by the gravity field of a major planet such as Earth or Jupiter, usually in some resonance motion – a far analogy to Bohr's model of the atom. The drag force plays the role of control parameter: with the increased drag, equilibrium paths of asteroids or comets can bifurcate (as in the logistic map, see below). For large enough drag, the system may become chaotic so that an asteroid or a comet no longer stays on a fixed trajectory relative to the planet, but wanders between seemingly random distances, down to a number of close approaches to the planet within a short time interval on an astronomical scale. So, the dissipative chaotic motion may bring celestial bodies dangerously close to the Earth's surface. However, exact mathematical modeling of the dissipative motion of small celestial bodies is rather involved and requires numerical integration on high-performance computers rather than analytical calculations: the cascade of chaotic returns to close encounters produces a complex pattern that can hardly be explored analytically. The uncertainty cloud of the initial conditions of the asteroid or comet, taken from observational data, must be propagated throughout decades to detect possible intersections with the Earth's orbit (see, e.g., [129]).

---

[101] Contrariwise, a positive Lyapunov exponent reflects stretching along the corresponding axis.

In conclusion to this section, one can make the following nearly obvious remark about instabilities and chaos. We have seen that the theory of continuous-time differentiable dynamical systems is largely based on the linearization of the respective differential equations near equilibrium points. Therefore, this theory may fail while attempting to explain the global behavior of highly nonlinear systems, in particular, the complexity of chaotic motions. Notice that, in the phase portrait, chaotic motion is typically imaged as "clouds" of the phase points.

## 12.5. Nonlinear Science

We have seen that the theory of dynamical systems is just another name for nonlinear dynamics, at least these two fields are inseparable. Nonlinear dynamics is, in fact, as old as classical mechanics. In general, the dynamics of planetary motion turns out to be nonlinear, but, fortunately, two-body systems are integrable and can be treated exactly. Three-body problems are of course also nonlinear, but in general non-integrable and very complicated. For example, the question originally posed in the XIX century: "is the Solar System stable?" naturally leads to a nonlinear description of the paths of three bodies in mutual gravitational attraction. It is this ancient problem which was simple to formulate but extremely difficult to solve that resulted in many modern ideas of nonlinear science, e.g., chaos. It is curious that with the advent of quantum mechanics and displacement of accent on atomic and nuclear physics nonlinear dynamics almost disappeared from physical books up to the 1970s. There was very little or no discussion of nonlinear science in the popular courses on methods of mathematical physics during the prewar and postwar (WW-2) period, when quantum mechanics and linear electrodynamics dominated science and engineering. It is also curious that in the 1980s nonlinear science gained a compensatory extreme popularity, to the extent that many people in the physics community began considering linear models as too primitive and unrealistic.

The great difference that exists between linear and non-linear problems is one of the most important and, perhaps, subtle features of mathematics. One always tries to linearize whenever possible, because linear problems are enormously easier to solve. Unfortunately, the world is not linear, at least to a large extent, so we have to learn how to deal with non-linear problems.

Nonlinearity in general can be well understood as the opposite of linearity – actually its negation. The main features of linearity are additivity and homogeneity, which result in linear superpositions. In linear theories such as quantum mechanics or classical (pre-laser) electrodynamics superposition is the key feature: an infinity of solutions may be constructed provided a finite set of solutions is known. It is true that pure linearity rarely occurs in the mathematical description of real-life phenomena, including physical systems. The example of quantum mechanics, which is an essential linear theory, although extremely important, may be considered an exception – yet to a certain extent, since the superposition principle for the wave function implies the infinite dimensionality of the respective vector space (see Chapter 5). We shall see that many finite-dimensional nonlinear systems can be mapped, with the help of some transformations, to linear systems with infinite dimensionality. In other words, nonlinear systems, i.e., those which should be modeled by nonlinear differential (integro-differential) equations or nonlinear discrete maps are ubiquitous, while linear ones really seem to be exceptions or approximations.

One obvious and strong motivation to study nonlinear dynamical systems is the rich variety of applications of such studies covering such apparently different areas in mathematics, physics, chemistry, biology, medical sciences, engineering, economics, political and military planning, financial management, etc. Yet, the ubiquity of nonlinear systems is counterbalanced by their complexity, so another motivation to study nonlinear dynamics is their intricate behavior. Examples are bifurcations, solitons, strange attractors, and fractal structures. There are no counterparts of these manifestations in the linear world; one can say that the extreme complexity of nonlinear structures marks an essential difference between linear and nonlinear phenomena.

The main source of difficulties in the nonlinear world is the fact that it is in general not feasible to describe a nonlinear system by dividing it into parts which are treated independently or blockwise – a favorite trick in linear system theory. As near as I know, no general techniques have been invented so far to foresee even the qualitative properties of a nonlinear system. For instance, it is rarely possible to predict a priori whether a dynamical system would exhibit regular or chaotic behavior. We shall illustrate this difficulty even on the simplest example of the logistic equation.

There exist, of course, a multitude of textbooks, journal articles and other sources where nonlinear dynamics in general and chaotic behavior in particular are beautifully described. We shall try to discuss only some cases of nonlinear dynamics which are quite important. There are, of course, other cases, e.g., of nonlinear PDEs (elliptic, parabolic and hyperbolic), which are very important in modern mathematical physics and, besides, present an intrinsic theoretical interest. However, the theory of these equations as well as related functional spaces and operator theory for nonlinear analysis are not included in this book. We shall discuss Euler and Navier-Stokes equations for incompressible fluids, but this topic seems to be inexhaustible, so the discussion may be considered limited.

It is widely believed that the $20^{\text{th}}$ century was the century of physics and the $21^{\text{st}}$ is the century of biology. The latter deals mostly with nonlinear phenomena, and the respective models should by necessity be nonlinear. On a macroscopic scale, it has been illustrated above by Lotka-Volterra model of the struggle for existence between two competing species. This dynamic situation (a predator-prey model) is described by nonlinear differential equations giving the time rate of evolution. This is a very simple model, of course, as compared with the level of complexity typically encountered in biology, but it provides a good foundation for more sophisticated mathematical modeling in this field.

We are immersed in the natural world of nonlinear events. For instance, our emotional reactions and our likes and dislikes are probably highly nonlinear. Our behavioral reactions to heat and cold, to colors and sounds, to local pressure and other stimuli are mostly subordinated to the Weber-Fechner law: the magnitude $R$ of psychological response is proportional to the logarithm of magnitude $J$ of physical stimulus, $R = A\ln(J/J_0), J = J_0, R = 0$ for $J < J_0$ (threshold).

At the end of this section, we supply a few references to some interesting and useful books in the field of nonlinear science. Some of these sources deal predominantly with mathematical concepts of nonlinear dynamics, such as [1, 103], whereas others accentuate the physical aspects [26, 140, 129, 119]. Considering the abundant literature on dynamical systems and nonlinear dynamics, the curious reader should start from simple, physically motivated examples which can be found in any textbook or in the Internet (see, e.g., http://www.faqs.org/faqs/sci/nonlinear-faq/).

# Section 13. Some outstanding models and case studies

Now that we have looked at many well-examined models, we will look at some that are very much in flux. These models are also quite important as test tools: for instance, many problems of classical and quantum theories have been critically tested on the oscillator model (which is also the crucial model in quantum field theory and, to some extent, in modern string theories). The main ideas are usually exemplified in simple models before being formulated as general theories.

## 13.1. Exponential growth

Let us start from one of the simplest, but a very important model: that of exponential growth and decay. In section 5.4., the exponential growth model $dN/dt = bN$, $b = 1/\tau > 0$, where $\tau$ is a characteristic time constant, was briefly mentioned. This model assuming the population growth rate proportional to the actual number of individuals (i.e., a linear model) has an explicit solution $N = N_0 \exp[b(t - t_0)]$, where $N_0 = N(t_0)$ is the initial quantity (e.g., initial population). This solution means that the population grows exponentially with $t \to \infty$ and vanishes exponentially with $t \to -\infty$. The direction field of this model lies in the upper right quadrant, $N > 0, t > 0$, and there are no vertical asymptotes which means that infinite values of the quantity $N$ cannot be reached in finite time. The value $N = 0$ is also unreachable in this model. The model of exponential growth can describe, for example, the rising number of scientific publications in the 20th-21st centuries. According to this model, quantity $N$ (for definiteness the population) is doubled independent of its actual state $N$ at constant periods $T_2 = \tau \log 2$. If one applies this model to the entire human population, then one sees that the Earth's population doubles during the period $T_2 = b^{-1} \log 2 \approx 60$ years, corresponding to the world average current growth rate of 1.14 percent i.e., $b = 1.14 \cdot 10^{-2}$/year. Of course, different countries have very unequal growth rates: thus, Afghanistan currently has growth rate $b \approx 4.8 \cdot 10^{-2}$/year which corresponds to approximately 14.5 years doubling time. If this growth rate remains the same for a rather long period, the country's projected population will reach about 1 billion by 2070. This seems to be an unrealistic expectation, which demonstrates certain deficiencies of the simple exponential growth model. Namely, this model does not take into account possible interaction between individuals, e.g., in the form of competition for resources.

Contrariwise, the Czech Republic currently has a negative growth rate ($b \approx -0.1$/year) so that one cannot calculate the doubling time (as the population is shrinking). However, for $b < 0$ one can determine the "half-time" $T_{1/2} = |b|^{-1} \log 2$ corresponding not to doubling, but to halving of the initial quantity $N_0$. In the theory of radioactive decay, time period $T_{1/2}$ is known as half-life. So, the mathematical model for a quantity that diminishes with a constant rate is given by changing the sign of the growth factor $b = 1/\tau$, $dN/dt = -bN, N > 0$, $b > 0$. This equation, in particular describing radioactive decay, has become a valuable tool to verify geophysical, astrophysical and biological data. The same equation results in the so-called barometric formula $\rho/\rho_0 = p/p_0 = \exp(mgh/T)$, where $\rho_0$ and $p_0$ are respectively the air density and the pressure at the sea level, $m$ is the molecular mass and temperature $T$ is written in energy units (the Boltzmann constant $k_B = 1$). In engineering, the barometric formula is typically represented as $\rho/\rho_0 = p/p_0 = \exp(Mgh/RT)$, where $M = N_A m$ is the mass of a mole and $R = N_A k_B$ is the gas constant, $N_A$ is the Avogadro constant. The combination $RT/Mg = k_B T/mg$ has the dimensionality of length and is called in meteorology the scale height (numerically about 8 km). Using the barometric formula, one can determine, for example, the air density or pressure at a certain height or, conversely, estimate at what height $h$ the air density is, e.g., half of the sea level density, $\rho(h) = 0.5\rho_0$: $h = (T/mg) \log 2 \sim (\overline{v^2}/2g) \log 2 \sim (s^2/2g) \log 2$, where $\overline{v^2}$ is the mean squared thermal velocity of the air molecules, $s$ is the sound velocity, temperature $T$ is assumed constant. Such estimates (giving about 5 km) are rather crude, but one cannot require high accuracy in using the barometric formula which is obtained under rather unrealistic assumptions (uniform temperature throughout the atmosphere, perfect gas in equilibrium, linearity, etc.). The barometric formula can also be considered a very particular case of the Boltzmann-Gibbs distribution of statistical physics.

One can notice that the model of exponential growth or decay demonstrates a very important idea of linearity: each process is linear for small increments i.e., locally. It is the same idea that underlies classical calculus. Exponential functions appear everywhere when the increment of a quantity is directly proportional to its actual value, thus exponential functions are always important when looking at growth or decay. Examples from economics are the value of an investment that increases by a constant percentage in a fixed period (such as rent from a property or interest on a bond), sales of a company that increase at a constant percentage during each period, models of economic growth, etc. In epidemiology, the spread of an epidemic, especially at its initial stage, is described by the exponential growth model. A very interesting object is the growing bacterial population. From the physical viewpoint, such populations are non-equilibrium systems interacting with each other and their environment.

Up to approximately the 1930s, science was exponentially expanding with the annual growth rate of about 6 percent (some researchers of science even assert that its exponential growth was observed until the 1960s, however with a lower rate). If this expansion of science were extrapolated into the $21^{st}$ century, then by the end of it nearly the entire world population would be engaged in scientific research, most computer networks would be dedicated to data and article exchange between scientists, and the planetary forests would not be able to restore after having been consumed by paper manufacturers supplying countless scientific publishers. Luckily, the model of simple exponential growth appears to be no longer valid; the logistic model seems more adequate, and now we are observing a slowdown of scientific growth as compared to exponential one.

## 13.2. The logistic model: the bugs are coming

Imagine a bunch of insects reproducing generation after generation so that initially there were $B$ bugs, and after $i$ generations there will be $N_i$ of them (to bug us: the total mass of insects on the Earth grows faster than that of humans). When the population in a given region is sufficiently large, it can be represented by a real continuous variable $N(t) > 0$. If we assume that there is a maximum sustainable population in the region (an equilibrium state) and that that the population dynamics can be described by a single autonomous equation, $\dot{N}(t) = \varphi(N)$, then we can produce a simple model assuming that the iteration function has a quadratic polynomial form, $\varphi(N, a) = aN(1 - kN)$. This is a real-valued function of two variables, $a > 0$ and $0 \leq N \leq 1/k$. Notice that the logistic model $\varphi(N, a)$ is not invertible with respect to $N = N(\varphi)$ since all the states in $(0, N)$ have two sources in $\varphi$-domain (preimages) so that each value of $\varphi$ corresponds to a pair of different values of population. Therefore, information is lost in the course of inversion. In the Bourbaki nomenclature, the logistic model is not injective; topologically oriented people prefer calling such maps non-continuous because they do not take an open set into an open set.

Equation $\dot{N}(t) = aN(1 - kN)$ is called the logistic model. Here parameter $a > 0$ is called a control or growth parameter (in theoretical ecology, this parameter, interpreted as the growth rate per individual, is usually denoted as $r$)[102] and parameter $k > 0$, determining the competition for some critical resources, is often called in ecology and population biology an inverse "carrying capacity" of the environment: it defines the equilibrium population when the competition decreases the growth rate to such an extent that the population ceases to rise. In electrical engineering, the same equation describes the energy of $E$ of a nonlinear oscillator in the self-excitation regime (in the first Bogoliubov-Mitropolsky approximation). A discrete alternative to the continuous (differential) logistic model is the so-called logistic map: $N_{i+1} = aN_i(1 - kN_i)$, $0 \leq N_i \leq 1/k$, $i = 0,1,2,...$, where the discrete variable $i$ plays the role of time in the continuous model. This variable can be interpreted as years or other characteristic temporal cycles, e.g., naturally related to the reproductive behavior of respective species. Recall that the term a "map" or "mapping" $\varphi$ usually refers to the deterministic evolution rule with discrete time and continuous state space $X, \varphi: X \to X$. Then evolution is synonymous with iteration: $x_{n+1} = \varphi(x_n)$. The logistic map, in contrast with the logistic equation, is a model with discrete time i.e., corresponds to snapshots taken at time points $t = n\tau, n = 0,1,...$ (the elementary time step $\tau$ may be put to unity). It is interesting that discrete-time dynamical systems can be produced from flows described by continuous-time differential equations, an example is a stroboscopic model provided by the Poincaré sections.

---

[102] In a more general context of dynamical systems (8.4.), we denoted the control parameter as $\mu$.

A discrete version of the logistic model can also have a direct physical or biological meaning, for example, in the cases when generations are separated (non-overlapped) in time. Thus, some insects just lay their eggs and die; they do not interact with the next generations. One more famous (appeared around 1200) discrete-time dynamical system, which was initially also a mathematical model of biological reproductive behavior, is the Fibonacci sequence, $a_{k+1} = a_{k-1} + a_k, k = 1,2, ..., a_0 = 0, a_1 = 1$. To describe this process in terms of evolution, we can introduce matrices

$$x_k = \begin{pmatrix} a_{k-1} \\ a_k \end{pmatrix}, A = \begin{pmatrix} 0 & 1 \\ 1 & 1 \end{pmatrix}$$

so that the Fibonacci sequence will be represented by a discrete-time cascade $x_{k+1} = Ax_k = A^k x_1$. Here the phase space is $X = \mathbb{R}^2$. The Fibonacci map, in distinction to the logistic map, is a linear transformation, and we can directly apply the standard prescriptions of linear algebra. The characteristic equation of the Fibonacci map is $\det(A - \lambda I) = \det \begin{pmatrix} -\lambda & 1 \\ 1 & 1-\lambda \end{pmatrix} = \lambda^2 - \lambda - 1$ so that eigenvalues $\lambda_{1,2} = \frac{1 \pm \sqrt{5}}{2}$ give eigenvectors $v_{1,2} = (1, \lambda_{1,2})^T$. Then $x_1 = C_1 v_1 + C_2 v_2$ and $x_{k+1} = \lambda_1^k C_1 v_1 + \lambda_2^k C_2 v_2$. Using initial conditions $x_1 = \begin{pmatrix} 0 \\ 1 \end{pmatrix} = C_1 \begin{pmatrix} 1 \\ \lambda_1 \end{pmatrix} + C_2 \begin{pmatrix} 1 \\ \lambda_2 \end{pmatrix}$ we get $C_1 = -C_2, C_1 \lambda_1 + C_2 \lambda_2 = 1$ so that $C_1 = \frac{1}{\lambda_1 - \lambda_2} = \frac{1}{\sqrt{5}} = -C_2$ and $x_{k+1} = \begin{pmatrix} a_k \\ a_{k+1} \end{pmatrix} = \frac{1}{\sqrt{5}} \left[ \begin{pmatrix} 1 \\ \lambda_1 \end{pmatrix} \lambda_1^k - \begin{pmatrix} 1 \\ \lambda_2 \end{pmatrix} \lambda_2^k \right]$ which gives for Fibonacci numbers $a_k = \frac{1}{\sqrt{5}} \left( \lambda_1^k - \lambda_2^k \right)$.

One can notice that the logistic map is a simple case of polynomial maps. Of course, the logistic map can be represented in dimensionless form $x_{i+1} = ax_i(1 - x_i)$, where all $x_i$ are numbers interpreted as the population density (or expectation value), and it is required that $0 \leq x_i \leq 1$, although certain values of intrinsic growth parameter $a$ and initial data $x_0$ may lead to negative population densities, which fact indicates that the logistic map should not be taken literally as a demographic model. In other words, the logistic map takes the interval [0,1] into itself. Despite its apparent simplicity, the logistic map is very general since any function having a nondegenerate extremum behaves in the latter's neighborhood like this map near $x = 1/2$. An obvious extension of the logistic map depending on parameter $a$ is the relationship $x_{i+1} = f(x_i, a)$ usually called the Poincaré map; here generations corresponding to $i = 0,1,2, ...$ can be interpreted as periods of motion. Notice that in the case of maps with discrete time, phase trajectories are given by a discontinuous sequence of points.

The fixed point of the map ($x = f(x)$) does not change with generations, $x_{i+1} = x_i$ so that it would be natural to define $\mu_i := dx_{i+1}/dx_i$ as a multiplier. It is clear that the maximum value in the sequence $x_i$ is reached when $\mu_i = a(1 - 2x_i) = 0$ and is $x_i = 1/2$. Therefore, the maximum iterated value $x_{i+1}(a) = a/4$, which means that to hold the logistic map in the required domain $0 \le x_i \le 1$ one must restrict the growth parameter to segment $0 \le a \le 4$.

In general, for one-dimensional Poincaré recurrences, the iteration function $f$ and control parameter $a$ can be selected and normalized in such a way as when the initial data $x_0$ (known as the seed) are taken from a finite interval $P = (\alpha, \beta)$, the iterates $x_1, \ldots, x_n$ also belong to $P$ (in the logistic map $P = (0,1)$ for small values of the control parameter, see below). It means that function $f$ maps interval $P$ into itself i.e., this map is an endomorphism (the term "into" means that the iterations may not fill the whole interval). Single-dimensional endomorphisms are often not invertible, which physically means that the past cannot be uniquely reconstructed from the current data, so that one might say that in such cases we are dealing with the systems having an unpredictable past (isn't it typical of human history?). When, however, an endomorphism has a smooth inverse, then the map is a diffeomorphism.

The discrete logistic model may exhibit a somewhat unusual behavior of the population. Thus, for small values of the growth (multiplication) parameter $a$, the initial value of the population produces a rather little effect on the population dynamics, the latter being mostly controlled by parameter $a$. Nevertheless, when this parameter is increased, the population (e.g., of bugs, mosquitoes or other insects and, apart from the latter, of rats, mice, reptiles, leeches, etc.) start to change chaotically, and in this chaotic regime one can observe an extreme sensitivity to the concrete number of initially present members $N(t_0) = B$ (or to the "seed" $x_0$). For large enough values of the growth parameter, e.g., for $a > 3$ in the simple logistic map $x_{i+1} = ax_i(1 - x_i)$, the population bifurcates into two, so that the colony of insects acts as if it were "attracted" by two different stable populations (such asymptotic solutions are called attractors). This is a primary example of the so-called period doubling. As this process has become very popular in nonlinear dynamics (starting from approximately 1976) and has generated a great lot of papers most of which are easily available (see the list of literature) and because of the lack of space, we shall not reproduce here the respective computations, restricting the current exposition to a catalogue of the basic facts. We only mention here that period doubling occurs in many models and in a variety of disciplines: in the Navier-Stokes equation (turbulence), in meteorology (the Lorenz system), in chemical reactions, in nerve pulse propagation, etc. In all such cases, we see that for different values of the control parameter $a$ the system's behavior alters drastically: from settling down to a point (the population dies out or tends to a non-zero fixed value), through quasi-regular oscillations, then irregular oscillations, and finally to chaos i.e., totally decorrelated process. Recall that the dynamical systems, described by ODEs, in general may have four types of solutions: equilibrium states, regular (periodic) motion, quasi-periodic motion and chaos. These solution types are associated with four attractor varieties: stable equilibrium, limit cycle, d-dimensional torus and chaotic attractor.

One usually studies the properties of the logistic map with the help of a computer, giving the seed $x_0$ and the growth parameter $a$ as inputs. The task is to find the asymptotic value of $x_i$ for $i \to +\infty$; one should of course bear in mind that computer modeling gives no genuine asymptotics, but only the values corresponding to large finite $i$-numbers. Yet one can obtain the graph $(x, a)$ which is known as a bifurcation diagram for the logistic map. This diagram plots a sequence of generations (population densities $x_i$) vs. growth parameter $a$, actually representing the limit solutions i.e., the same and cyclically repeated (after some number $m$ of steps) values of $x$. Such cycles appear following the attempts to find the already mentioned fixed (also known as stationary) points of the map $x_{i+1} = \varphi(x_i, a)$ i.e., limit solutions of equation $x = \varphi(x, a)$ mapping point $x$ onto itself. Fixed points can be both stable and unstable, in particular, depending on the values of the control parameter $a$. Specifically, for the logistic map, when $0 < a < 1$ the only limit value is $x = x^{(1)} = 0$, and at $a = 1$ the first bifurcation occurs: now there are two solutions i.e., besides $x^{(1)} = 0$ a new solution $x^{(2)} = 1 - 1/a$ appears. Indeed, the search for fixed points gives $x = ax(1 - x)$ which produces $x^{(2)}$ for $x \neq 0$. In the range $1 < a < 3$, the computed limit value corresponds to this solution i.e., fixed points coincide with $x^{(2)} = 1 - 1/a$ since it is stable whereas the first solution $x^{(1)} = 0$ is unstable. We can test the stability of both solutions using the standard linearization procedure. Putting $x = x^* + \delta x$, where $x^*$ is either $x^{(1)}$ or $x^{(2)}$, we have after linearization $\delta x_{i+1} = a(1 - 2x^*)\delta x_i$, and we see that when $x^* = 0$, this solution is stable for $a < 1$ and unstable for $a > 1$. If $x = x^{(2)}$, then $\delta x_{i+1}/\delta x_i = 2 - a$ i.e., $\delta x_i = (2 - a)^i \delta x_0$ and $\delta x_i$ converges to zero for $|2 - a| < 1$ and diverges for $|2 - a| > 1$. Thus, solution $x^{(1)}$ becomes unstable for $a > 1$ and solution $x^{(2)}$ for $a > 3$; within interval $1 < a < 3$, $x^{(2)}$ remains stable. At point $a = 3$, the second bifurcation occurs: apart from solutions $x^{(1)} = 0$ and $x^{(2)} = 1 - 1/a$, both of which become unstable, a stable two-cycle ($m = 2$) fixed point emerges. One can predict the population for this two-cycle attractor e.g., by requiring that generation $(i + 2)$ has the same number (density) of individuals as generation $i$: $x_i = x_{i+2} = ax_{i+1}(1 - x_i)$. Combining this relationship with the logistic map, we get $x = a^2 x(1 - x)(1 - ax + ax^2)$. One may notice that solutions $x^{(1)} = 0$ and $x^{(2)} = 1 - 1/a$ satisfy this algebraic equation. To find two other solutions, one can divide the equation by $x - x^{(1)} = x \neq 0$ and $x - x^{(2)} = x - (1 - 1/a)$ to obtain the quadratic equation $x^2 - (1 + 1/a)x + (1/a)(1 + 1/a) = 0$ whose solutions are

$$x = \frac{1 + a \pm \sqrt{a^2 - 2a - 3}}{2a}.$$

Putting here the value $a = 3$ that delimits the new bifurcation interval in the diagram, we get the exact solution $x = 2/3$ (which can also be verified directly from the logistic map). It is remarkable that a simple computer iteration procedure demonstrates rather slow convergence to this exact solution. It is also interesting that one can already observe the onset of oscillations: when $2 < a < 3$, the population density $x_i$ begins to fluctuate near the solution $x^{(2)} = 1 - 1/a$. With the growth parameter exceeding $a = 3$, oscillations become more and more pronounced, at first between two values (depending on $a$) and then, with a further increase of the growth parameter, between $4, 8, 16, \ldots, 2^m$ values. The size ratio of subsequent bifurcation intervals on the diagram converges to the so-called Feigenbaum constant $\delta = 4.6692 \ldots$ that has been computed up to more then $10^3$ decimal places. So, with the growth parameter being increased, period doubling bifurcations occur more and more often, and after a certain value $a = a^*$ (sometimes called an accumulation point) has been reached, period doublings transit to chaos. One can also notice that different parts of the bifurcation diagram contain similar plots i.e., they are self-similar.

The physically limiting case $a = 4$, i.e., $x_{i+1} = 4x_i(1 - x_i)$ is especially often encountered in the literature on logistic maps because one can find an exact solution in this case. One usually makes the substitution $x_i = (1/2)(1 - \cos 2\pi\theta_i) \equiv \sin^2 \pi\theta_i$ to obtain the following form of the logistic map: $\cos 2\pi\theta_{i+1} = \cos 4\pi\theta_i$. One can then define the principal value of $\theta_i$ as belonging to interval $[0, 1/2]$ so that $\theta_{i+1} = 2\theta_i$ and $\theta_i = 2^i\theta_0$. Thus, depending on the initial value $\theta_0$, the map admits a countable set of cyclic (periodic) solutions and a continuum of aperiodic solutions. Periodic solutions can be found by putting successively $\theta_0 = 1$ ($x = 0$); $1/2$ ($x = 0$); $1/3$ ($\theta_i = 1/3, 2/3, 4/3, 8/3, \ldots$ i.e., stationary solution $x = 3/4$); $1/5$ (double cycle $\theta_i = 1/5, 2/5, 4/5, 8/5 \ldots \rightarrow 1/5, 2/5$ i.e., $x_1 \approx 0.345, x_2 \approx 0.904$); $1/7$ (triple cycle $1/7, 2/7, 4/7$ i.e. $x_1 \approx 0.188, x_2 \approx 0.611, x_3 \approx 0.950$), etc. Using the standard trick of linearization, one can see that all such solutions are exponentially unstable, $\delta\theta_i = 2^i\delta\theta_0$ so that the actual solutions fluctuate between such unstable ones: the situation known as chaos (or at least quasi-chaos, when some solutions still remain stable).

This example is quite instructive since one can observe on it the emergence of simple statistical concepts. Indeed, although chaos is a dynamical notion arising in deterministic systems (there are no random variables in the starting equations), the continuum chaotic states can only be adequately described by introducing a probability measure i.e., probability to find the population between $x$ and $x + dx$. It is this indispensable emergence of probabilistic description that is usually known as stochasticity. To find the respective distribution function, let us, in accordance with the archetypal model of statistical mechanics, consider the simplified logistic map $\theta_i = 2^i \theta_0$ as an ensemble of solutions labeled by different initial conditions. A dynamic map $x = g_t x_0$, where $g_t$ is a one-parameter group (or semigroup) of translations along vector field trajectories, takes these initial conditions into the actual state of the system. For example, in the case of continuous time and smooth vector field $v(x)$ on an $n$-dimensional manifold $M$, $g_t$ is a measure-preserving diffeomorphism (see the above section on dynamical systems) i.e., if $\mu$ is a measure that in any local coordinate system is represented through the probability density, $d\mu = \rho(x)dx, x = \{x^1, \dots, x^n\}$, then measure $\mu$ is invariant under the action of $g_t$ if density $\rho(x)$ satisfies the Liouville equation, $\partial_i(\rho v^i) \equiv \text{div}(\rho \mathbf{v}) = 0$ (the Liouville theorem, see section 7.9). Such a measure is an integral invariant of flow $g_t$ (i.e., of dynamical system). A similar probability distribution for the logistic map with growth parameter $a = 4$ for fixed $i$ is $\rho(\theta_i) = d\theta_i/d\theta_0 = 2^i = \text{const}$. But from substitution $x_i = (1/2)(1 - \cos 2\pi\theta_i)$ we have

$$\frac{d\theta_i}{dx_i} = \frac{1}{\pi \sin 2\pi\theta_i} = \frac{1}{\pi\sqrt{1 - \cos^2 2\pi\theta_i)}} = \frac{1}{\pi\sqrt{1 - (1 - 2x_i)^2}} = \frac{1}{2\pi\sqrt{x_i(1 - x_i)}}$$

and, in particular, $\frac{d\theta_0}{dx} = \frac{1}{2\pi\sqrt{x(1-x)}}$. Defining $d\mu = \rho(x)dx := \rho(\theta)d\theta$, we get $\rho(x) = \frac{\text{const}}{\sqrt{x(1-x)}}$, where the constant can be determined from the normalization condition, $\int_0^1 \rho(x)dx = 1$ which gives const $= 1/\pi$. Finally, we have the distribution function for the probability to find the logistic system (e.g., the population) between $x$ and $x + dx$

$$\rho(x) = \frac{1}{\pi\sqrt{x(1-x)}} \tag{13.2.1.}$$

Notice that this distribution function does not depend on the starting value $0 \le x_0 \le 1$ i.e., is universal. This is a specific feature of the physically limiting case of the logistic map with $a = 4$.

One can show that the logistic map for $a = 4$, which is in this case chaotic for almost all initial conditions, may be related to the Lorenz attractor appearing in the three-dimensional meteorological model constructed in 1963 by E. Lorenz. This mathematical model is represented by a dynamical system with 3d phase space and is fully deterministic since it is represented by three ODEs $\mathbf{x} = \mathbf{f}(\mathbf{x}, a), \mathbf{x} = (x^1, x^2, x^3)$ (with quasilinear vector field $\mathbf{f}$). Nonetheless, the model demonstrates chaotic behavior i.e., abrupt and apparently random changes of state for some set of control parameters $a$.

We have already noted that there are many interesting things about chaos and one of them is that it had not been discovered much earlier. The logistic map, for instance, could well have been explored by the brilliant Enlightenment mathematicians, but probably, after the creation of Galileo-Newton's mechanics in the late 17$^{\text{th}}$ century, scientists were preoccupied with continuous-time mathematical analysis and differential equations describing the objects that change smoothly with time. Even today, in the epoch of digital technologies, many "old-guard" scientists are much better familiar with differential equations than with their discrete counterparts.

Another chrestomathic example of a discrete-time system $x_{n+1} = f(x_n), n \in \mathbb{Z}$ besides the logistic map is given by piecewise linear (!) maps known as the Bernoulli shift $B(x)$

$$f: [0,1), f(x) \equiv B(x) = \begin{cases} 2x, 0 \le x < 1/2 \\ 2x - 1, 1/2 \le x < 1 \end{cases} = 2x \bmod 1.$$

This simple map produces rather complex dynamics which is characterized by an extreme sensitivity to initial conditions which can be demonstrated, as usual, by computing the Lyapunov exponents. Indeed, for two paths beginning in two nearby points $x_0$ and $x_0' = x_0 + \varepsilon_0$ with displacement $\varepsilon_0 \ll 1$ we shall have $\Delta x_n := |x_n' - x_n| = 2\varepsilon_{n-1} = 2^2\varepsilon_{n-2} = \cdots = 2^n\varepsilon_0 \equiv e^{n \ln 2}\varepsilon_0$. We see that two closely located points diverge with the rate $\lambda = \ln 2 > 0$, which is the Lyapunov exponent for map $B(x)$. Since $\lambda > 0$ the map exhibits an exponential instability and can be viewed as chaotic.

Discrete time maps $x_{n+1} = f(x_n), n \in \mathbb{Z}$ are also known as fixed point iterations. In general, single-dimensional discrete time maps $f(x): x \in X \subseteq \mathbb{R}, X \to X$ that may be represented as $x_{n+1} = f(x_n)$ (an obvious generalization of the logistic map), even very primitive ones, can also exhibit an exponential dynamical instability and thus may be called chaotic (in the Lyapunov sense). Expression $x_{n+1} = f(x_n)$ supplied with initial condition $x(0) = x_0$ (as the initial population in the logistic map) can be viewed as an equation of motion for our 1d deterministic dynamical system with discrete time.

Let us now return to the continuous-time case (such systems are the most interesting for physics and technology). In accordance with the idea of linearity, the model of exponential growth, discussed in the preceding section, can be applied when the number $N$ of individuals is relatively small so that the correlation effects between them can be disregarded. Correlations between individuals can be observed on both the pairwise and collective level[103]. Pair correlations give rise, e.g., to the hyperbolic or explosion model leading to the sharpening regime with vertical asymptotes meaning that the population will tend to infinity in finite time (this model was briefly discussed in 5.4.), whereas collective correlations manifest themselves, for example, in the competition for resources (such as food, water, living space, energy, information, position in the hierarchy, political power, etc.) within the population. With the rising number of individuals, competing for resources tends to impede the growth rate i.e., growth factor $b = b(N), db/dN < 0$. The simplest model of an impeded growth would be putting rate $b$ to fall linearly with growing population, $b = a - kN, a > 0, k > 0$. This mathematical model is quite natural since one can approximate any smooth function by a linear one for sufficiently small values of its argument (this is, by the way, the main idea of calculus and differentiation), in the present case, for small enough $N$. Then we have the equation expressing, in the continuous case, the logistic model:

$$\frac{dN}{dt} = (a - kN)N = aN\left(1 - \frac{N}{N_0}\right), \tag{13.2.2.}$$

where $N_0 \equiv a/k$ is the equilibrium population (in ecological literature this parameter is usually denoted as "carrying capacity" $K$). One-dimensional equation (13.2.2.) is known as the logistic equation; this is an apparently primitive, but very rich and important mathematical model which may be interpreted as a combination of the exponential and hyperbolic models. The meaning of the logistic model is that the reproductive rate is assumed to be proportional to the number of individuals, whereas the mortality rate is proportional to the frequency (probability) of pair encounters (collisions). Of course, by properly scaling time $t$ and number of individuals $N$, e.g., $t \to \tau := at$, $N \to kN/a = N/N_0 := x$, we can reduce (13.2.2.) to a dimensionless form $\dot{x} = x(1 - x)$ which is convenient to study the main dynamical properties of the model in the $(x, \tau)$ plane, but we shall rather stick to the form (13.2.3.) in order to better understand the role of parameters $a$ and $N_0$.

---

[103] The patterns of pairwise vs. collective correlations is quite often encountered in many-particle physical models, e.g., in statistical and plasma physics.

The logistic equation, though being a nonlinear ODE, is simple in the sense that it can be easily integrated (since it is autonomous and variables in it can be separated). There are many ways to integrate the logistic equation; for demonstration purposes, we can choose probably the simplest one, representing the logistic equation as

$$\frac{dN}{dt} = (a - kN)N = aN - kN^2, \qquad N(t_0) = B, a > 0, k > 0, \qquad (13.2.3.)$$

$B, a, k$ are constants. For $N \neq a/k$, $t - t_0 = \int_B^N dN(aN - kN^2)^{-1}$. This integral exists only when both $N$ and $B = N(t_0)$ i.e., the current and the initial values of the population lie in intervals $(0, a/k)$ or $(a/k, +\infty)$. In other words, there exist no solutions that cross the straight line $a - kN = 0$. Integration gives

$$t - t_0 = \frac{1}{a} \int_B^N dN \left( \frac{1}{N} + \frac{k}{a - kN} \right) = \frac{1}{a} \log \frac{N(a - kB)}{B(a - kN)}$$

Solving this equation with respect to $N$, we have

$$N(t) = \frac{aB\exp[a(t - t_0)]}{a - kB + kB\exp[a(t - t_0)]} = \frac{aB}{kB + (a - kB)\exp[-a(t - t_0)]}$$

One can see that for an initial population lower than the equilibrium value, $N(t_0) < N_0$ i.e., $0 < B < a/k$, $N(t)$ is defined for all $t$, $0 < t < +\infty$, while for $N(t_0) > N_0$ i.e., $B > a/k$, $N(t)$ is only defined for

$$t > t_0 - \frac{1}{a} \log \frac{kB}{kB - a}$$

In the case $N(t_0) \equiv B = a/k$, the solution is a constant, $N(t) = a/k$, since in this case $dN/dt = 0$.

One can also see that the solution converges to the constant value $a/k$. For $B < a/k$, $N(t) < N_0 \equiv a/k$ for all $t$ so that $a - kN(t) > 0$ and $dN/dt > 0$ which means that if the initial population is lower than the equilibrium one, the number of individuals monotonously increases. In the opposite case, when the starting population exceeds the equilibrium one, $B > a/k$, we have $N(t) > N_0$ for all $t$ and $dN/dt < 0$ which means that the population is monotonously shrinking.

The linear regime of the logistic model corresponds to the case when one can neglect the second term (proportional to $N^2$) in the logistic equation. This is correct for small probabilities of pair correlations and for short observation times. More accurately, $(kB/a)(\exp[a(t - t_0)] - 1) \ll 1$ which gives for the short period of observation, $a(t - t_0) \ll 1$, the following restriction on the death rate in the population, $kB(t - t_0) \ll 1$. An obvious drawback of the logistic model is that it does not contain spatial variables, which means that it cannot be applied to spatially inhomogeneous situations.

The most frequently used form (13.2.2.) of the logistic equation introduces explicitly the asymptotic value $N_0 = k/a$. Then the solution with the initial condition $N(t = t_0) \equiv B$ can be written as

$$N(t) = \frac{BN_0 e^{a(t-t_0)}}{N_0 + B(e^{a(t-t_0)} - 1)} = N_0 \frac{N(t_0)}{N(t_0) + (N_0 - N(t_0)e^{-a(t-t_0)})} \qquad (13.2.4.)$$

So, by directly integrating the logistic equation (which is a rare occasion in the world of nonlinear equations), we get the explicit expression for integral curves that are usually called the logistic curves whose family depends (in the deterministic case) on three parameters, $a, N_0$ and $B = N(t_0)$. The graph of this solution produces an S-shaped curve which can be represented in dimensionless units by the function $N(t) = 1/(1 + \exp(-at))$, in which one can easily recognize the Fermi distribution of fermions over single-particle energy states in quantum statistical physics. The process described by the logistic model (the logistic process) has two equilibrium points, $N = 0$ and $N = N_0$, with the first being unstable, since small population $\delta N$ grows near $N = 0$ ($\dot{N} > 0$), and the second stable ($\dot{N} < 0$ for $N > N_0$ and $\dot{N} > 0$ for $N < N_0$ near $N = N_0$ (which can be seen already from equation (13.2.2.) i.e., without even integrating it). In other words, the population dynamics evolves to the equilibrium value $N = N_0$. More exactly, the process tends asymptotically for $t \to +\infty$ to the stable equilibrium $N = N_0$ at any starting value $N(0) = B > 0$. For $t \to -\infty$, the process asymptotically converges to the state $N = 0$. Thus, the logistic model describes the transition from the unstable state $N = 0$ to the stable state $N = N_0$, occurring in infinite time. We shall see the examples of similar transitions when addressing quantum-mechanical models below. The integral curves have vertical asymptotes $t = const$ for any $t > 0$. The logistic process is very close to the exponential (Malthusian) growth for small $N$, ($N \ll N_0$), but begins to fall behind approximately at $N \approx N_0/2$. This saturation effect is a manifestation of correlations within the population.

One usually considers a single control parameter in the logistic model, the growth parameter $a$. In fact, however, there are at least two control parameters which can be both important, especially if the logistic model is not necessarily applied to describe the population growth, when all the parameters and variables are positive by default. Renaming the constants, e.g., $a = kN_0^2, b = a/kN_0$ (see (13.2.3.)-(13.2.4.)), we arrive at equation $\frac{dx}{dt} = f(x, a, b) = ax(b - x)$ whose solutions depend on two parameters $a$ and $b$, not necessarily strictly positive. In other words, the dynamical evolution occurs in 3d space $(x, a, b)$ without restricting the motion to domain $x > 0, a > 0, b > 0$ as in population models. For $b > 0$, point $x = 0$ is unstable whereas for $b < 0$ it is stable; at $b = 0$ stability is changed to instability both at the stable point $x = b$ and the unstable point $x = 0$ of the traditional logistic model – a simple example of the bifurcation.

One can emphasize that there is a significant contrast between the solutions to differential equations and the ones to the respective difference equations (the logistic map). If, e.g., we take differential equation $\dot{x} = ax(b - x)$, its solution (an S-curve) can be easily obtained:

$$x(t) = \frac{x_0 e^{at}}{1 - x_0(1 - e^{at})}, \qquad x_0 = x(0)$$

(see above, here $t_0 = 0$), whereas the solution to the logistic map $x_{i+1} = ax_i(1 - x_i)$ obtained by iterating this difference equation has a totally different character reflecting much more complicated behavior. This behavior has been studied by many researchers and is still hardly totally understood.

From a more general viewpoint, the logistic model is a very particular case of the evolution of an autonomous system described by vector equation $\dot{\mathbf{x}} = \mathbf{f}(\mathbf{x}, a)$, where vector field $\mathbf{f}$ depends on parameter $a$ (it can also be a vector). As we have seen, in certain cases, variations of the control parameter $a$ can radically change the system's motion, for instance, result in chaotic behavior. Note that the logistic model is not necessarily identical with the population model. When the logistic equation is not interpreted as describing the population growth, one can explore the behavior of solutions as parameter $k$ (see (13.2.2.)-(13.2.3.)) is varied.

## 13.2.1. Extensions of the logistic model

Although the logistic model was initially devised to describe mathematically the population dynamics, in particular, the survival conditions, this model has a more general meaning. We have already mentioned the similarity between the continuous-time logistic model and the self-sustained oscillations in laser generation. Indeed, one often writes the logistic equation in the extended form $dN/dt = aN(1 - kN) = (\alpha - \gamma - \beta N)N$, where parameters $\alpha$ and $\gamma$ define the birth and mortality rates, respectively, whereas the nonlinear term $\beta N$ corresponds to population shrinking due to intraspecific competition for resources. This form of the logistic equation is analogous to one of the forms of the Van der Pol equation written for the oscillation energy $E = \left(\frac{m}{2}\right)(v^2 + \omega_0^2 x^2)$ (in fact the Lyapunov function)

$$\frac{dE}{dt} = (\alpha - \beta E)E, \qquad \alpha = \alpha_0 - \gamma,$$

where $\alpha_0$ is the feedback parameter, $\gamma$ and $\beta$ are friction coefficients (linear and nonlinear). Thus, the birth rate in the logistic equation characterizes the positive feedback, while the mortality rate accounts for friction.

This analogy enables one to link the logistic model with the nonlinear theory of Brownian motion. This latter theory is quite universal and has a number of important physical, chemical and biological applications, in particular, in electrical and chemical engineering (e.g., catalysis), radio physics, laser technology, biological self-organization, etc. Mathematically, the analogy between the logistic model and the nonlinear Brownian motion may be expressed on the level of the Fokker-Planck equation (see Section 7.11.) with the nonlinear diffusion coefficient, $D(N) = \gamma + \beta N$:

$$\frac{\partial f(N, t)}{\partial t} = \frac{\partial}{\partial N}\left(D(N)N\frac{\partial f(N, t)}{\partial N}\right) + \frac{\partial}{\partial N}[(-\alpha + \gamma + \beta N)Nf(N, t)],$$

where $f(N, t)$ is the distribution function characterizing the probability of the population to be in a state with $N$ individuals at time $t$. Notice that the description of population dynamics both in terms of the continuous-time logistic equation and the Fokker-Planck equation is valid for a large number of individuals, $N \gg 1$.

Both the logistic map and the logistic model can be extended and improved. Thus, the growth (control) parameter $a$ in the logistic map may depend on discrete time i.e., iteration step $i$ so that we shall have $x_{i+1} = a_i x_i(1 - x_i)$. This is a discrete-time version of the optimal control problem, and the corresponding extremal properties similar to Pontryagin's maximum principle in continuous-time dynamical systems can be established. In a more general situation than the one-dimensional logistic map, the discrete-time iterated process may have a vector character, $\mathbf{x}_{i+1} = \boldsymbol{\varphi}(\mathbf{x}_i, \mathbf{a}_i)$, where $\mathbf{x}_i = \{x_i^1, \ldots, x_i^p\}$, $\mathbf{a}_i = \{a_i^1, \ldots, a_i^q\}$ ($\mathbf{a}$ is known as a decision vector), and index $i$ enumerates iteration steps.

The simplest logistic model with scalar control is of course linear, e.g., $a_i = a_0(1 + \sigma i), i = 1,2,...$ One can rewrite the logistic map for the variable growth parameter as $x_{i+1} - x_i = a_i x_i (X_i - x_i)$, where $X_i \equiv 1 - 1/a_i$ (recall that this quantity coincides, for $a_i = a = $ const, with one of the main fixed points of the logistic map, $X = x^{(2)}$). If we assume that $\Delta a_i = a_{i+1} - a_i \ll a_i$, then we can approximate the logistic map by the continuous-time equation, $x_{i+1} - x_i \equiv \frac{x_{i+1} - x_i}{1} \approx \frac{dx}{dt}$ and $a_i \approx a(t)$ so that $\frac{dx}{dt} = a(t)x(X(t) - x)$ or, in the form (13.2.2.) $\frac{dx}{dt} = a^*(t)x\left(1 - \frac{x}{X(t)}\right)$, where $a^*(t) \coloneqq a(t)X(t)$. We see that the logistic map under the assumption of small variations of the growth parameter between two successive generations is reduced to the continuous-time logistic model with variable growth parameter and drifting stable equilibrium point. The drift on a slow time scale of equilibrium population $X$ is quite natural since the ecosystem carrying capacity may change with time due to human technological impact, biological evolution or geophysical variations (such as climate change).

One can consider two subcases here: one corresponding to adiabatic variations of parameters $a^*(t)$ and $X(t)$: i.e., they both change little during the characteristic time $t \sim 1/a^*$ of the process i.e., $|da^*(t)/dt| \ll \left(a^*(t)\right)^2$, $|dX(t)/dt| \ll a^*(t)X(t)$ on some set of temporal values $t$; the other reflects the opposite situation of abrupt changes. In the linear model of growth parameter variation, $a^*(t) = a_0^*(1 + \sigma t)$ (one can omit the asterisks for simplicity), the adiabatic regime can be described by a series over small parameter $\sigma$ (the characteristic time of slow control parameter variation is $1/\sigma$). Writing the logistic equation in the form $x = X - \frac{\dot{x}}{ax}$ and noticing that $\dot{x} \sim (a_0 \sigma t)x$, we may obtain the expansion $x = X + \delta x^{(1)} + \delta x^{(2)} + \cdots$, where $\delta x^{(1)} = -\frac{\dot{X}}{a_0 X} \lesssim \sigma t$, $\delta x^{(2)} = -\frac{1}{a_0 X}\left(\dot{X}\frac{\delta x^{(1)}}{X} + \frac{d \delta x^{(1)}}{dt}\right) = \frac{\ddot{X}}{(a_0 X)^2}$, etc. The meaning of the adiabatic regime is that the stable fixed point $X$ (e.g., interpreted as the eventually reached population) is slowly drifting and the solution $x(t)$ (current population) adjusts to this smooth evolution of parameters. In this adaptation process a band of values is formed instead of pointlike solution $x$. The positions of bifurcation points also change a little so that the initial bifurcation diagram is slightly distorted.

In the other limiting subcase of abrupt changes i.e., occurring at times $\tau \ll 1/a$, one can consider all the model parameters $(a, X, ...)$ to remain constant during the change so that the solution will jump between two levels (or two bands). This situation is close to the ones treated within the framework of perturbation theory in quantum mechanics (recall that the quantum-mechanical perturbation theory had its origins in the methods of finding solutions to ordinary differential equations).

One can of course consider the system of two or more interacting populations, each being described by the logistic model. For two populations we might have the following system of coupled logistic equations

$$\frac{dx_1}{dt} = a_1 x_1 \left(1 - \frac{x_1 + x_2}{N}\right), \qquad \frac{dx_2}{dt} = a_2 x_{21} \left(1 - \frac{x_1 + x_2}{N}\right). \qquad (13.2.1.1.)$$

We can explore the stability of solutions to this model (with a two-dimensional phase space in the same way as for a single population) and, e.g., draw the phase portrait. The model of two coupled populations belongs to the class of competition models which we shall discuss shortly.

We have already noted that the logistic model does not account for spatially inhomogeneous processes. In general, mathematical and computer models that are constructed by averaging out the spatial[104] information such as territorial population distribution and migration and keeping only the time variable as the vector field parameter hide many essential characteristics. To overcome this obvious drawback of the logistic model, one can use the simple trick of introducing one more parameter. Namely, we can write the initial condition $N(t_0) \equiv B$ (see equation (13.2.4.)) in the form $N(x, t_0) \equiv B(x) = 1/(1 + \exp(-px))$, where $x$ is interpreted as a spatial variable and treated as a supplementary parameter. One can regard variable $x$ as completely hidden due to some averaging, with only the time variable remaining directly accessible. The spatially dependent initial condition is assumed to restore the hidden variable and to evolve with time into a space-dependent solution

$$N(x, t) = \frac{1}{1 + e^{-p\xi}}, \qquad \xi = x + wt$$

(here, we put for simplicity $t_0 = 0$). In this extension of the logistic model, one can interpret $N(x, t_0) = N(x, 0)$ as a spatially inhomogeneous distribution that spreads with time as a kinematic wave: its propagation is due to nonlinear terms and is not caused by dynamical interactions. If we put $\xi = 0$, we get $x = wt$ i.e., a steady profile. In other words, condition $\xi = 0$ corresponds to choosing the coordinate frame moving with transfer velocity $w = x/t$.

The mentioned shortcoming of the model – absence of spatial information – can also be surmounted by using the logistic model combined with partial differential equations, $Lu = \varphi(u)$, where $L$ is some operator (not necessarily linear) acting on the space on which function $u$ is defined, $\varphi(u) = au(1 - u)$. Here for brevity the dimensionless form is used. Operator $L$ is assumed to contain a spatial part, for instance, $L = \partial_t - \partial_{xx}$ (diffusion) or $L = \partial_{tt} - \partial_{xx}$ (linear waves), $L = \partial_t - (\eta(u))_{xx}$ (nonlinear diffusion), etc. In such models, each zero of $\varphi(u)$ corresponds to a stationary solution of the associated PDE, $Lu = 0$. When $L$ is the diffusion operator, one can interpret the problem simply as supplementing the logistic model with a spatially dependent spreading part (which reminds one of the Navier-Stokes equation)

$$u_t = a\Lambda^2 u_{xx} + \varphi(u) \qquad (13.2.1.2.)$$

so that solutions now depend on both space and time variables, $u = u(x, t)$. Here quantity $\Lambda$ is the so-called diffusion length, by an appropriate choice of scaling one can set $\Lambda^2$ to unity. Models similar to (13.2.1.2.) are rather popular in biological sciences and biomathematics, for example, Fisher's model of biological adaptation[105]. One can interpret this diffusion model as a strategy to compute varying distributions of gene frequencies within the population – an approach known today as population genetics. Equation (13.2.1.2.) is also said to describe nonlinear diffusion and is known as the reaction-diffusion equation. One often ascribes the probabilistic meaning to the right-hand side $\varphi(u)$ (it is interpreted as being proportional to the product of probabilities $p$ that an event occurred and $q = 1 - p$ that it did not), but mathematically this is still the function forming the logistic equation. The corresponding PDE has two stationary solutions $u = 0$ and $u = 1$ and a traveling wave solution, $u(x, t) = F(x - wt)$. Note that equation (13.2.1.2.) is invariant under reflection $x \to -x$ so that velocity $w$ can be both positive and negative. It is also clear that (13.2.1.2.) is invariant under affine translations of $x$ and $t$, therefore one can add any arbitrary constant to $z := x - wt$. Inserting the ansatz $u = F(z)$ into PDE (13.2.1.2.), we get the second order ODE

$$F'' + \frac{w}{a\Lambda^2}F' + \frac{1}{a\Lambda^2}\varphi(F) = 0, \qquad (13.2.1.3.)$$

where the prime denotes differentiation over $z$. This ODE is of the nonlinear damped oscillator type (sometimes equations of this type are called the Liénard equations), and after solving it, we can find, if possible, asymptotic solutions for $x \to \pm\infty$ and the transitions between two stationary points $u_1 = u(-\infty) = F(-\infty)$ and $u_2 = u(+\infty) = F(+\infty)$ (in the case of the underlying logistic model, $u_1 = 0, u_2 = 1$). Such asymptotic requirements play the role of boundary conditions for (13.2.1.3.); now the problem is to determine the values of $w$ (one may consider them eigenvalues) for which solution $F \geq 0$ satisfying these asymptotic conditions exists, and if it does, then for what initial states $0 \leq u(x,0) \leq 1$. Will the solution $F$ take the form of a traveling wave for any initial distribution $u(x,0)$? This question proved to be rather nontrivial and stimulated extensive research. One of the natural ways to treat this problem is by representing (13.2.1.3.) in the form of a dynamical system

$$F' = y, y' = -\frac{w}{a\Lambda^2}y - \frac{1}{a\Lambda^2}\varphi(F),$$

then we can investigate the solution in the phase plane $(y, y')$ or, more conveniently, $(u, y) \equiv (y_1, y_2)$. The phase trajectories are obtained, as usual, by excluding the vector field parameter (here $z$), and we have[106]

$$\frac{dy}{du} = -\frac{1}{a\Lambda^2 y}\big(wy - \varphi(u)\big) \equiv -\frac{1}{a\Lambda^2 y_2}\big(wy_2 - \varphi(y_1)\big). \qquad (13.2.1.4.)$$

---

[104] Also, information provided from other underlying spaces, not necessarily of geometrical nature, such as biological, ecological, epidemiological, immunological, physiological, social, etc. character.

[105] R. Fisher was a prominent British biologist and statistician. Fisher's model was explored in detail by A. N. Kolmogorov, I. G. Petrovsky and N. S. Piskunov and is sometimes called the Fisher-KPP model.

This equation has two equilibrium (also known as fixed, singular or critical) points in the $(u, y) \equiv (y_1, y_2)$-plane: $(0,0)$ and $(1,0)$. In the vicinity of point $(0,0)$, one can linearize equation (13.2.1.4.), which corresponds to the transition from logistic to exponential growth model, obtaining

$$\frac{dy}{du} \approx -\frac{1}{a\Lambda^2 y}(wy - au) = \frac{u}{\Lambda^2 y} - \frac{w}{a\Lambda^2} \qquad (13.2.1.5.)$$

or, identically,

$$\frac{dy_2}{dy_1} = \frac{y_1}{\Lambda^2 y_2} - \frac{w}{a\Lambda^2}$$

When exploring the behavior of solutions, one can observe here the competition between two terms in the right-hand side, therefore the phase portrait changes its character for velocity $w$ greater or smaller some critical value ($w = w_c$). By solving (13.2.1.5.), one can show that $w_c \sim a\Lambda$. To determine the type of fixed points and thus to characterize the phase flow, we may proceed according to the standard prescriptions of two-dimensional dynamical systems. Writing $y_1' = y_2, y_2' = -\frac{1}{\Lambda^2}y_1 - \frac{w}{a\Lambda^2}y_2$ and looking for a solution of the form $\exp(\mu z)$, we obtain the system matrix

$$A = \begin{pmatrix} 0 & 1 \\ -\dfrac{1}{\Lambda^2} & -\dfrac{w}{a\Lambda^2} \end{pmatrix}$$

---

[106] Here, to avoid confusion we are placing vector indices below. The difference between vectors and covectors is usually immaterial in the context of two-dimensional dynamical systems.

(notice that quantity $w/a\Lambda^2$ is formally analogous to the dissipation coefficient). The trace of matrix $A$ is $\mathrm{Tr}\,A = -w/a\Lambda^2$ and $\det A = 1/\Lambda^2$ so that the characteristic equation is $\mu^2 - \mu\mathrm{Tr}\,A + \det A = \mu^2 + \mu w/a\Lambda^2 + 1/\Lambda^2 = 0$, and the roots are real and different when discriminant $D = (w^2 - 4a^2\Lambda^2)/4a^2\Lambda^4 > 0$ i.e. $|w| > 2a\Lambda$. This last value can be identified with $w_c$. Assume at first that the roots of the characteristic equation have the same sign, then the real solutions are $y_1 = y_{10}e^{\mu_1 z}, y_2 = y_{20}e^{\mu_2 z}$, where $y_{10}, y_{20}$ are arbitrary constants. One can, as before, get rid of parameter $z$ and obtain either the family of parabola-like[107] orbits $|y_1| = c|y_2|^{\mu_1/\mu_2}$, where $c$ is some constant (independent of $z$), or $y_1 = 0$. Recall that critical point of this kind is called a node; if $\mu_1, \mu_2 < 0$ i.e. $w > 2a\Lambda$, then point $(0,0)$ is a positive attractor i.e. there exists a neighborhood of $(0,0)$ such that all the paths starting at this neighborhood at some $z = z_0$ finish at $(0,0)$ with $z \to +\infty$ (this property is almost obvious due to the exponential character of solutions). Likewise, for $\mu_1/\mu_2 > 0$, the critical point $(0,0)$ is a negative attractor. From the physical viewpoint, condition $w > 2a\Lambda$ signifies that the disturbance for $t > 0$ moves in the direction $x > 0$ (recall that $u = F(x - wt)$). If $0 < w < 2a\Lambda$, the singular point $(0,0)$ becomes an unstable focus, and the degenerate case $w = 0$ (i.e., a stationary disturbance; from the positions of dynamical systems, this fixed point is a center) leads to "unphysical" solutions $u < 0$ (recall that we need to have solutions in the band $0 \le u \le 1$). One may note that in general equilibrium near the coordinate origin, when one can disregard nonlinearity, is an unstable focus.

In the same way one can explore the other singular point $(1,0)$. In the vicinity of this point $u \approx 1$ so that the linearized equation, instead of (13.2.1.5.), is

$$\frac{dy}{du} \approx -\frac{1}{a\Lambda^2 y}\big(wy - a(u-1)\big) = \frac{u-1}{\Lambda^2 y} - \frac{w}{a\Lambda^2}.$$

Proceeding just as before, we see that point $(1,0)$ is a saddle for all $w \ge 0$ as long as we consider growth factor $a > 0$ and constant and diffusion length $\Lambda$ real.

A more complicated mathematical setting of the Fisher-KPP model is the Cauchy problem for a nonlinear parabolic equation (parabolic equations express the most popular models arising from biological studies)

$$u_t(x,t) = (\eta(u))_{xx} + \varphi(u)$$

---

[107] The ratio $\mu_1/\mu_2$ not necessarily equals 2 or ½.

in domain $D := \{x \in \mathbb{R}, 0 \leq t < +\infty\}$ with initial condition $\lim_{t \to +0} u(x,t) = u_0(x)$ and, in some versions, asymptotic boundary values $\lim_{x \to -\infty} u(x,t) = 0$, $\lim_{x \to +\infty} u(x,t) = 1$, with a non-negative and continuously differentiable on $[0,1]$ source function $\varphi(u)$ such as $\varphi(0) = \varphi(1) = 0, \varphi'(0) > 0$. The initial function $0 \leq u_0(x) \leq 1$ can be piecewise continuous in $\mathbb{R}$. When $\eta(u) = u$, we have a linear diffusive process with a nonlinear source.

## 13.2.2. Applications of the logistic model

Both the logistic model and the logistic map have many applications in science, society and engineering. The general idea leading to the logistic model – simple growth limited by self-interaction – may be applied to many real-life processes. For instance, epidemics and the spread of rumors can be modeled by the logistic equation. We can take a typical microeconomic situation as another example: what will be the output of certain goods produced by a company (e.g., car manufacturer) over several years? The simplest model describing the annual growth of the output, with the imposed constraints of limited resources and market saturation would be a logistic model. What would be the total revenue (and respectively the profit) of a company over several years, if the annual revenue growth is $a$? The revenue for the $(i + 1)$-th year will be $x_{i+1} = ax_i$. However, for rather high revenues the latter are restrained, e.g., by the market share, and we arrive again at the logistic model. These examples suggest a natural generalization: any human activity subordinated to the imposed constraints of external factors and/or limited resources can be described by a logistic model. The nonlinear term that limits growth is sometimes metaphorically interpreted as "influence of the future".

If we make an affine transformation $x_i = pz_i + q$, where $p, q$ are as yet undetermined parameters, we get from the logistic map a quadratic recurrence equation

$$z_{i+1} = -a \left[ pz_i^2 - (1 - 2q)y_i - \frac{q}{p}(1 - q) \right],$$

and putting $p = -1/a, q = 1/2$, we obtain the quadratic map $z_{i+1} = z_i^2 + c, c \equiv a/2 - a^2/4$ which produces Julia sets. Quadratic Julia sets are probably the best-known examples of fractals that are generated by this quadratic map for almost any value of $c$ (although $c = 0$ and $c = -2$ are exceptional: the produced sets are not fractals). It is interesting that the above quadratic map was commercially used in the graphical industry to obtain rather beautiful ornaments which are fractals: this is an example of direct market value of mathematics.

It is remarkable that the logistic map can serve as a rough model of the transition to turbulence, when the regular (laminar) or almost periodic character of fluid motion is destroyed after some critical value of the flow parameter (the Reynolds number $Re$) has been reached. Like the eventual population in most logistic models, turbulence practically does not depend on the initial state of the fluid. The Reynolds number is a control parameter in the models of turbulent flow and plays the role of intrinsic growth parameter $a$. Multiplier $\mu$ passes in the transition to turbulence the value $+1$.

The logistic map manifests such common features of discrete-time algorithms as stability and chaotic behavior. From this viewpoint, it is interesting for numerical techniques and generally for computational science and engineering. As far as engineering applications of the continuous-time logistic model go, we have already mentioned that the equation describing the energy evolution of a nonlinear oscillator in the self-excitation mode has the form $\dot{E} = aE(1 - E)$ (in dimensionless units). Here the growth parameter $a$ is close to 1.

One can consider an example of estimating the population growth with the help of the logistic model. We may assume the total current (2010) human population to be $6.9*10^9$, the growth factor to be $a = 0.029$, the annual population growth to be 0.011 year$^{-1}$ (more or less standard demographic data). Then we have

$$\frac{dN(2010)/dt}{N(2010)} = \frac{d}{dt}\log(N(2010)/N(t_0 = 2010) = 0.011 = a - kN(2010)$$
$$= 0.029 - k \cdot 6.9 \cdot 10^9$$

which can be considered an equation to find the attenuation factor $k \approx 2.32 \cdot 10^{-12}$. Then the projection for the equilibrium population will be $N_0 = a/k \approx 11.1 \cdot 10^9$ i.e., the world population tends to converge to approximately 11 billion people. This result is not very sensitive to the slight changes of the constant $a$ and the annual population growth rate.

We can comment here on the general concept of the control parameter whose variation results in altering the evolution regime of a dynamical system. Choosing the control parameter may present a special problem when modeling a complex (multiparametric) process, in particular, when the transition to chaos in an open system should be explored. For instance, in medico-biological studies the concentration of the prescribed medicine can play the role of control parameter. In surgery, the control parameter may be identified with the pre-planned invasion path. In fluid motion problems, one can define the control parameter as the pressure gradient between the flow boundaries, e.g., pipe ends. In laser technology, the control parameter can correspond to the pumping level i.e., energy input producing the population inversion. Using the engineering language, one can define an analogous control parameter as the feedback level in classical generators.

One often uses the logistic model in practical ecology to control the population. For example, a slight modification of this model (harvesting) is applied in fisheries. Harvesting means mathematically a transition from the population freely evolving in accordance with internal processes (balance of the birth, death, and migration) to introducing an external pressure $q(N,t)$ i.e., the model equation will be $\dot{N}(t) = \varphi(N) - q(N,t)$, where term $q(N,t)$ signifies the removal of $q$ individuals per unit time (say, each year). In fisheries, for example, this external influence corresponds to fishing quotas that can be constant, $q(N,t) \equiv q$ or differential, $q(N,t) = q(N)$. The latter case may be interpreted as the simplest manifestation of a feedback, which is a milder model than the one corresponding to $q = $ const. Indeed, $q = q(N)$ depends on the actual state of the system. It is usually important to select quotas $q$ in such a way as to ensure sustainable fishing. Sometimes, as e.g., during genocides, external influence $q(N,t)$ cannot even be exactly known and should be treated as a perturbation, not necessarily small. One can in such cases figuratively call function $q$ the murder rate.

More complicated population models arise when pair bonding is considered. The effectiveness of survival mechanisms tends to decrease as the population density falls since it becomes increasingly difficult for sexually reproducing species to find appropriate mates. The respective mathematical models are in general spatially dependent, at least at the level of any given individual, since extended mobility and higher mate detection capability (greater identification distance) can to a certain degree compensate low population density. However, there are stringent physiological limitations for excessive mobility since it requires higher metabolism rates that are only possible under the conditions of ample resource availability (just as increased consumption in the rich population groups of human society enhances mobility and selectiveness). Considering pair stability issues further complicates the model.

To focus on the role of harvesting we can use, for simplicity, the dimensionless form of the logistic model that can be, in particular, achieved by scaling time $t$ as $\tau = aN_0t$ (see equation (13.2.3.) and below). Formally, we can put the growth parameter $a$ and the "equilibrium population" $N_0$ equal to unity (it is trivial to notice that for $a \neq 1, q \to q/a$). Then we have $\dot{x} = x(1-x) - q \equiv f(x,q)$, and when quota $q$ is constant, its critical value is $q = 1/4$. In fisheries, this critical point is usually called maximum sustainable yield (MSY), and its evaluation is vital for sustainable fishing. For $0 < q < 1/4$, there are two equilibrium points ($\dot{x} = 0$) corresponding to two roots, $x_1, x_2$ of the resulting quadratic equation. The lower equilibrium point $x_1$ is unstable which means that if, for some reason, the population falls under $x_1$, then the entire population dies out in finite time. When $q > 1/4$, both equilibriums disappear, but what is more important, $f(x,q) < 0$ for all values of population $x$ i.e., *the population necessarily becomes extinct.* On the contrary, for $q < 1/4$ the population never dies out, and at the bifurcation point $q = 1/4$, the vector field $f(x,q) = 1/2$ i.e., equilibriums $x_1$ and $x_2$ merge, which, for sufficiently big initial population, can ensure that it will asymptotically end up near the $x = 1/2$ value. However, just a small drop of the population below this value would result in its extinction in a finite time. In this sense, bifurcation point $q = 1/4$ is unstable. When the growth parameter $a \neq 1$, bifurcation occurs at critical value $q = a/4$.

The population evolution accompanied by forceful elimination of some part of the individuals represents simple feedback mechanisms. For example, one can exercise the ecological control over the quantities of certain species (e.g., mosquitoes), in such cases $q = q(x) > 0$ and $dq/dx > 0$ for all $x$. One can assume $q(x)$ to be a polynomial with positive coefficients. Let us consider the simplest case of a feedback scenario, $q(x) = bx$. In this case there are two stationary values, $x_1 = 0$ (unstable) and $x_2 = 1 - b$, $0 < b < 1$ (stable). We see that equilibrium point $x_2$ corresponds to elimination, e.g., harvesting or hunting, quota $bx_2 = b(1-b)$ with sustainable maximum (optimum) $b = x_2 = 1/2$. When $b \to 1, x_2 \to 0$ i.e., both equilibrium points merge, and the population becomes extinct in finite time ($\dot{x} = -ax^2 \to x(t) = 1/a(t - t_0)$). Physically, this signifies excessive harvesting, hunting or killing and geometrically to the disappearing crossing point of parabola $y = x(1-x)$ and straight line $y = bx$. For $b > 1$ equilibrium point $x_2$ becomes negative. It is easy to see that when the growth parameter $a \neq 1$, we must scale $b \to b/a \equiv \tilde{b}$ so that a stationary state corresponding to an optimum is reached with $b = a/2$.

The case of quadratic elimination quota is analyzed similarly to the case of the linear one i.e., $\dot{x} = ax(1-x) - bx - cx^2$ so that the stationary points are $x_1 = 0$ and $x_2 = \left(1 - \frac{b}{a}\right) / \left(1 - \frac{c}{a}\right)$, $0 < b < a$, $0 < c < a$. The optimum point is $b = a/2$.

The same model describes bankruptcy of companies and, with slight modifications, the downfall of political entities such as groups, parties, unions, states, etc. In certain models, e.g., taxation models in some economies, where $x$ denotes a tax and $\varphi(x)$ the taxation base, quotas $q(x)$ can be negative and piecewise continuous. The meaning of the model is that one should not harvest more than a certain threshold, otherwise the system will devour itself in finite time, regardless of what it is: fisheries, small businesses, emigration of scientists and specialists. One can also interpret this model as the manufacturer-consumer equilibrium, where the logistic part $ax(1-x)$ corresponds to production and the harvesting part $-q(x) = bx + cx^2 + \cdots$ to the consumption of goods.

The lesson learned from studying the logistic model and logistic map is that apparently simple and completely deterministic dynamical systems can exhibit very complex motion that can be perceived as chaotic and stochastic. It may be instructive to make the computer implementation of the logistic model, in particular, trying to produce some computer codes corresponding to it. One can also readily apply Euler's method to the logistic equation (see section on scientific computing below).

## 13.3. Competition models

Ecological models based on simple dynamical systems such as single-species population dynamics may lead to unreliable predictions, even on the qualitative level. Thus, ignoring interaction between populations in the ecosystem can provoke ecologically detrimental decisions, e.g., overestimation of safe catch in fisheries (this actually occurred with cod in 1992). However, when new populations are included in the model, its phase space dimensionality is increased along with the model complexity. New bifurcations, chaotic attractors and catastrophes can appear, making credible forecasts nearly impossible.

There exist many species in nature, and none of them is isolated: species interact with each other and with the environment. In the logistic model, interaction is accounted for phenomenologically by the nonlinear (collision) term $\sim N^2$ corresponding to the interaction within the given species or by the harvesting or hunting term $q$ describing the elimination of individuals of the given species by other species. This term manifests ecological pressure on the given species whose dynamics evolves according to the logistic model. Notice that one can introduce competition into a single-component exponential (Malthusian) model not necessarily through the logistic equation; one might as well consider the following dynamical system

$$\frac{dx}{dt} = axe^{-bx} \equiv a(x)x, \qquad a > 0, b > 0, \tag{13.3.1.}$$

where exponent $e^{-bx}$ reduces growth factor $a$ due to the competition between individuals when the population increases. This model, however, is less popular, probably because it cannot be as conveniently explored as the logistic one. Indeed, solution of (13.3.1.)

$$\int_{x_0}^{x} \frac{e^{bu}}{u} du = a(t - t_0) \tag{13.3.2.}$$

can be expressed through the exponential integral function $\text{Ei}(x)$ which is a special function whose values are tabulated and must be studied numerically, which requires more time and effort than for the logistic model that has a straightforward analytic solution. We can, however, produce a Taylor expansion in (13.3.2.):

$$\log\frac{x}{x_0} + \frac{bx}{1 \cdot 1!} + \frac{(bx)^2}{2 \cdot 2!} + \frac{(bx)^3}{3 \cdot 3!} + \cdots \Big|_{x_0}^{x} = a(t - t_0)$$

and obtain, e.g., up to the second-order terms for $bx < 1$ the following transcendental equation defining population $x$ as a function of time $t$ along each 1d "path", counting from initial position $x_0$: $a(t - t_0) \approx \log\frac{x}{x_0} + b(x - x_0) + b^2(x^2 - x_0^2)$ which can be solved, e.g., by iterations.

Models that are more sophisticated explicitly describe the interaction between a number of species; the respective dynamical systems are usually called competition models. The phase space in such models is no longer one-dimensional, its dimensionality is equal to the number of the considered interacting species. Usually, the number of species i.e., the dimensionality of the phase space $n$ is considered constant. One can interpret this requirement as the assumption that the whole assembly of populations makes up a stable system, at least on the time scale much greater than the maximal characteristic period of population changes. In ecology, one tacitly presupposes that the greater the number of interacting species, the more stable is the assembly of populations. Intuitively, one tends to believe that a vast collection of species can more successfully react to the environmental challenges than just one or two species. It follows from this assumption that one can use the quantities characterizing the biodiversity such as entropy to measure the ecosystem stability. For instance, the standard definition of entropy $S = -\sum_{i=1}^{n} p_i \log p_i$, where $p_i = N_i/N, N = \sum_i^n N_i$, $N_i$ is the number of individuals of the $i$-th species, seems to be applicable to specify both the diversity and the stability of an assembly of populations. Yet this intuitive application of the notion of entropy can lead to discrepancies. Indeed, the peak stability then must be attained at equilibrium when entropy reaches a maximum, but this state corresponds to constant frequencies $p_i = N_i/N = 1/n$ i.e., all the species are symmetrically distributed and equirepresent in the ecosystem. There exists neither hierarchy nor dominance among the species. However, this equipartition and egalitarianism are not observed in nature; on the contrary, dominance and hierarchical structure seem to be very pronounced in stable ecosystems (as well as human societies). It means that standard entropy can manifest the stability of ecosystems only very crudely.

The simplest model of the competition type is limited to a pair of species (2d phase space) and is known as the predator-prey or Lotka-Volterra (LV) model. One can construct this model from very intuitive considerations. In ecosystems, individuals interact both with other representatives of the same species and with individuals of other species. Biomass is transferred between the species through the uptake of one kind of organism by the other. Imagine a closed ecosystem (biotope) supporting two species, with their numbers (or biomass) being respectively $x$ and $y$. We assume that species $y$ are preying on species $x$ whereas the latter feed on natural resources of the biotope. If there were no predators $y$, then the number of species $x$ would grow exponentially, $\dot{x} = \alpha x$. However, when $y \neq 0$, according to the harvesting model we must have $\dot{x} = \alpha x - bxy$. The number of predators diminishes exponentially when there is no food so that the predators die out without prey as $\dot{y} = -\beta y$. However, when there is enough food i.e., prey species, then predators begin to proliferate with the rate proportional to consumed prey i.e., $\dot{y} = -\beta y + cxy$. Quantities $x, y$ are non-negative by their meaning so that the phase space of the model is defined by the domain $x \geq 0, y \geq 0$.

Thus, the compound "predator-prey" model is often formulated as a system of first order ODEs

$$\frac{dx}{dt} = \alpha x - p(x)y; \quad \frac{dy}{dt} = -\beta y + q(x)y \qquad (13.3.3.)$$

or one can assume the proportionality $q(x) = kp(x)$, where $x(t), y(t)$ are the populations of "preys" and "predators", respectively. Spatial variations are disregarded in this model, and only time-dependence of the populations is considered. Function $p(x)$, subordinated to the condition $p(0) = 0$, is the number (or biomass) of prey individuals consumed by a single predator per unit time, this function is known as the trophic function. For modeling purposes, one typically approximates the trophic function by a polynomial, $p(x) = a_1 x + a_2 x^2 + \cdots + a_n x^n$. The main assumption of the model is that when there are no predators the population of the prey species grows exponentially (with rate $\alpha$) whereas predators die out in the absence of prey with rate $\beta$; $k$ is some constant that may need to be tuned for the model in different situations (ecological, biological, economic, social, military, etc.). One can observe that the standard predator-prey model does not necessarily imply nonlinear mapping (in contrast with the logistic model), when the populations of prey and predator species are decoupled. We see that the Lotka-Volterra model is expressed by an autonomous system on a plane (2d phase space) so that it can be easily explored, especially in its simplest polynomial form

$$\frac{dx}{dt} = \alpha x - pxy; \quad \frac{dy}{dt} = -\beta y + qxy. \tag{13.3.4.}$$

The popularity of this model is partly due to its intuitive nature and ease of treatment (although this ease is deceptive). One can interpret this dynamical system as a weakly-nonlinear (second-order nonlinearity) model, when the rate of prey destruction by the predators is proportional to the frequency of predator-prey encounters with some constant coefficient $p$. The growth rate of the predator population is also proportional to this frequency, with coefficient $q$ (usually, $p \geq q$). One can notice that the bilinear character of nonlinearity in the Lotka-Volterra model is the same as in chemical reactions, epidemiology, mathematical economics or, e.g., dynamical meteorology. Biological population dynamics may be regarded as a kind of "chemistry", when the living organisms and not the molecules interact and even form new compounds. Environmental resources are consumed to be converted into the new biomass, individuals replicate and proliferate, and predators annihilate prey. One can also notice that the dynamics of interacting populations significantly differs from that of isolated species. The phase portrait of the Lotka-Volterra model reminds us that of harmonic oscillator ($dx/dt = ay, dy/dt = -bx$), see the next section, but in distinction to harmonic motion the phase trajectories are not elliptic (nor circular for $ab \equiv \omega^2 = 1$, in particular, when $a = b = 1$). Nevertheless, the phase trajectories of the LV model are closed, which reflects the oscillations of the predator and prey numbers. Such oscillations, however, are not harmonic in the sense that, due to nonlinearity, the period begins to depend on amplitude, with this non-isochronicity being more and more pronounced as the trajectories run away from fixed point $(x_0, y_0) \equiv (\beta/q, \alpha/p)$. Indeed, the family of phase trajectories in the LV model is determined by equation

$$\frac{dy}{dx} = \frac{q - \frac{\beta}{x}}{-p + \frac{\alpha}{y}} = -\frac{qx}{py} S(x, y)$$

which differs from the harmonic oscillator loops by formfactor $S(x, y) \equiv \frac{y}{x} \frac{1 - \frac{\beta}{qx}}{1 - \frac{\alpha}{py}}$ . Obviously, dynamical system (13.3.3.) has two fixed points: $(0,0)$ and $(\beta/q, \alpha/p)$, the first of them trivial and devoid of practical interest whereas the second corresponds to stationary populations of predators and prey being in dynamic equilibrium with one another. The population of predators remains constant due to the natural balance of births and deaths, while the prey population is preserved because of predators' activity. Phase trajectories of the Lotka-Volterra model are represented by a family of loops encircling the equilibrium point $(x_0, y_0) \equiv (\beta/q, \alpha/p)$ of the center type. This point determines the equilibrium population vector $(x_0, y_0)^T$. The direction field $P(x, y) = \alpha x - pxy, Q(x, y) = -\beta y + qxy$ is, by its ecological interpretation, only defined in the positive quadrant $x \geq 0, y \geq 0$. The family of phase trajectories (loops) satisfies the implicit equation

$$\Phi(x, y, \alpha, \beta, p, q) \equiv qx - \beta \log x + py - \alpha \log y = \text{const.} \tag{13.3.5.}$$

Near fixed point $(\beta/q, \alpha/p)$, solutions to (13.3.3.) are very close to harmonic oscillations with frequency $\omega = (\alpha\beta)^{1/2}$ i.e., oscillations of both populations near equilibrium are isochronous, their period $T = 2\pi/(\alpha\beta)^{1/2}$ does not depend on amplitude. The implicit equation (13.3.5.) hints at a natural change of variables, $u = \log qx$, $v = -\log py$, then the model becomes Hamiltonian:

$$H(u, v) = \beta u - e^u - \alpha v - e^{-v}, \qquad \frac{du}{dt} = -\frac{\partial H}{\partial v}, \qquad \frac{dv}{dt} = -\frac{\partial H}{\partial u}, \tag{13.3.6.}$$

with the conserved quantity $H$ in variables $(u(x,y), v(x,y))$ i.e., also in variables $(x,y)$, and it can be easily seen that the trajectories are closed. It means that the populations of both prey and predators vary periodically, provided the initial populations do not coincide with the equilibrium value $(x_0, y_0) \equiv (\beta/q, \alpha/p)$. In reality, the trajectories of the LV model are not necessarily closed due to "dissipative" effects, when the model loses its Hamiltonian character, e.g., when resource limiting factors or intraspecies competition (self-interaction terms proportional to $x^2, y^2$) come into play. The graph of (13.3.6.) gives the Hamiltonian surface of the Lotka-Volterra model.

To better understand the Lotka-Volterra model, we can ultimately simplify it by putting all the parameters of the problem equal to 1: $\alpha = \beta = p = q = 1$. Then we have the equilibrium point $(x_0, y_0) = (1,1)$. The dynamical system for the LV model in this case has the form

$$\frac{dx}{dt} = x - xy; \quad \frac{dy}{dt} = -y + xy \tag{13.3.7.}$$

whose Jacobi matrix is

$$A = \frac{\partial f^i}{\partial x^j} = \begin{pmatrix} 1 - y & -x \\ y & -1 + x \end{pmatrix}$$

which gives at critical point $(1,1)$ a symplectic matrix $\begin{pmatrix} 0 & -1 \\ 1 & 0 \end{pmatrix}$ corresponding to the linear oscillator $\dot{x} = -y, \dot{y} = x$ (one can of course interchange variables $x$ and $y$ corresponding to prey and predator populations to get the more common form of the oscillator matrix, see next section). The phase portrait near the equilibrium point consists of concentric circles around the center at $(1,1)$ – prey and predator populations oscillate out of phase.

The system of equations (13.3.7.) can be solved directly to obtain the phase curves. Eliminating $t$ we have $\frac{dy}{dx} = \frac{y(1-x)}{x(-1+y)}$ so that variables separate $\frac{y-1}{y} dy = \frac{1-x}{x} dx$ and we get after integration $y - \log y = \log x - x + C_1$, where $C_1 = $ const. Exponentiating this expression we have $\frac{e^y}{y} = C_2 x e^{-x}$ or $xe^{-x} y e^{-y} = C$ i.e., phase curves are the contour (level) curves of function $\chi(x,y) = xy e^{-(x+y)}$ which is an exponentiation of (13.3.5.). Here constant $C$ is the integral of motion. One can easily represent such curves graphically, they are a 2d extension of the graph of function $\varphi(u) = ue^u$ which has a maximum at $u = 1$, see Figure 15. One can easily see that the contour curves are closed.

In a general case corresponding to (13.3.5.) the Jacobian matrix $A$ of the model is

$$A \equiv A(x, y; \alpha, \beta, p, q) = \begin{pmatrix} \alpha - py & -px \\ qy & qx - \beta \end{pmatrix},$$

and the maximum value of constant $C$ is reached at the equilibrium point $(x_0, y_0) \equiv (\beta/q, \alpha/p)$. Matrix $A_0 \equiv A(x_0, y_0)$ i.e., the Jacobian matrix at the equilibrium point is the symplectic one

$$A_0 = \begin{pmatrix} 0 & -\dfrac{\beta p}{q} \\ \dfrac{\alpha q}{p} & 0 \end{pmatrix}$$

with the purely imaginary eigenvalues $\pm i\sqrt{\alpha\beta}$. Note that since these eigenvalues have zero real parts the equilibrium point $(x_0, y_0) \equiv (\beta/q, \alpha/p)$ is not hyperbolic and hence one cannot apply the important Hartman-Grobman theorem of the dynamical systems theory so that one cannot in general characterize the dynamics of the initial nonlinear system near $(x_0, y_0)$. This equilibrium point is a center with a bunch of concentric orbits going around it. Recall that a center is in general a critical point for which there exists an infinite sequence of periodic orbits around it. If the linearized system has a center, it does not necessarily mean that the equilibrium point of the initial nonlinear system is a center: it may be one, but it can also be an asymptotically stable or unstable spiral point.

The (0,0) equilibrium point is an unstable saddle so that it cannot be an attractor resulting in eventual extinction of both prey and predator populations. Indeed, the Jacobian matrix of the Lotka-Volterra model at the origin (0,0) is $A(0,0) = \begin{pmatrix} \alpha & 0 \\ 0 & -\beta \end{pmatrix}$ with the obvious eigenvalues $\lambda_1 = \alpha$, $\lambda_2 = -\beta$. Since one of the assumptions of the model is that $\alpha > 0, \beta > 0$, the eigenvalues always have different signs which corresponds to a saddle point. Extinction of species in the Lotka-Volterra model can only follow an artificial setting of either of the populations to zero.

One can transform the Lotka-Volterra dynamical system to polar coordinates, $(x, y) \rightarrow (r, \varphi)$, which is hardly useful since the resulting equations are rather clumsy, e.g., the family of loops is given by

$$\frac{dr}{d\varphi} = \frac{r(p \sin\varphi - q \cos\varphi)}{\frac{\alpha}{r}\cot\varphi + \frac{\beta}{r}\tan\varphi - p\cos\varphi + q\sin\varphi} + \frac{\alpha + \beta}{\frac{\alpha}{r}\cot\varphi + \frac{\beta}{r}\tan\varphi - p\cos\varphi + q\sin\varphi}.$$

In reality, however, the behavior of interacting species is more complex than described by the Lotka-Volterra model. For example, prey can partly poison the predators with its metabolites so that the oscillations become unstable: centers are displaced from the line $\text{Tr}\, A = 0$ and the evolution is described by spiral sinks. Moreover, the predictions of the Lotka-Volterra model can hardly be regarded as ecologically realistic even within the sterile class of dynamical systems on a plane (2d phase space). Indeed, assume that there are no predators in the considered ecosystem ($y = 0$), then the prey population grows exponentially, $x = x_0 e^{\alpha t}$. Likewise in the absence of prey species ($x = 0$), the population of predators dies out exponentially, $y = y_0 e^{-\beta t}$. Such purely exponential regimes have not been observed in nature.

Despite its apparently restricted character, predator-prey oscillations have given rise to a number of new concepts such as synchronization in complex systems, spatio-temporal pattern formation, chaotic regimes in ecology and so on. Numerous refinements of the original Lotka-Volterra model produce rather complicated dynamics and are currently studied in a variety of disciplines (ecology, economics, demography, etc.)

### 13.3.1. Extensions of the Lotka-Volterra model

One can try to improve the simple Lotka-Volterra model by introducing perturbations, e.g.,

$$\frac{dx}{dt} = x\big(\alpha - py + \varepsilon U(x, y)\big), \qquad \frac{dy}{dt} = y\big(\beta - qy + \varepsilon V(x, y)\big). \qquad (13.3.1.1.)$$

We can expect that the initial predator-prey system does not substantially change for $\varepsilon \ll 1$ (functions $U$ and $V$ are assumed smooth). One can prove (see [17], §2) that the position of equilibrium points $(x_k, y_k)$ of a 2d dynamical system $\dot{x} = F(x, y, \varepsilon), \dot{y} = G(x, y, \varepsilon)$ smoothly depends on parameter $\varepsilon$ provided the Jacobian $J = \frac{D(F,G)}{D(x,y)}\Big|_{(x_0,y_0,\varepsilon=0)} \neq 0$. System (13.3.1.1.) depends on one more parameter in addition to the unperturbed predator-prey model so that both populations may become unstable for some values of parameter $\varepsilon$. One can find an interesting analysis of this situation in the cited book by V. I. Arnold [17].

### 13.3.2. Kolmogorov's model

System (13.3.1.1.) is a special case of the Kolmogorov model $\dot{x} = xf(x,y), \dot{y} = yg(x,y)$ generalizing the Lotka-Volterra model. Functions $f(x,y)$ and $g(x,y)$ determine per capita growth of the prey and predator populations. When system (13.3.1.1.) is interpreted in ecological terms, variable $x$ may denote the number (or density) of the prey species whereas variable $y$ that of predators. As usual, both variables are defined in the ecological problem in the quadrant $x > 0, y > 0$. One may notice that the main feature of both Kolmogorov and Lotka-Volterra models is that the growth rate of each species is proportional to the actual size of the population, which reflects the fact that when $x = y = 0$, no reproduction of the species is possible. There is no spontaneous generation of either of the species. The Kolmogorov model is usually accompanied by certain ecologically natural assumptions, for example, $\partial f(x,y)/\partial y < 0$ (growth of the predator population inhibits prey); $\partial g(x,y)/\partial x > 0$ (growth of the prey population favors predators); $\partial g(x,y)/\partial y \leq 0$ (there is an intraspecific competition within the predator population); there exists $K > 0$ (prey carrying capacity such that $f(K,0) = 0$ and $f(x > K, y) < 0$ (the prey population reaches equilibrium if there are no predators); there exists $J > 0$ such that $g(J,0) = 0$ (predators cannot survive if the prey population is too small); there exists $Y > 0$ such that $f(0, Y) = 0$ (prey cannot survive if there are too many predators). There may also be other modeling assumptions, for example, $\partial f(x,y)/\partial x > 0$ for small $x$ and $\partial f(x,y)/\partial x < 0$ for large $x$ i.e., the prey population reproduction rate has a maximum for some $x = x_*$. Analogously for predators: $\partial g(x,y)/\partial y > 0$ for small $y$ and $\partial g(x,y)/\partial y < 0$ for large $y$ i.e., maximum at $y = y_*$. In the language of the logistic model for a single species, it means that the growth parameter $a$ has a maximum for some optimal population density.

In the famous work of Volterra (Volterra, V. Leçons sur la théorie mathématique de la lutte pour la vie. Gauthier-Villars, Paris, 1931, 1990), the two-dimensional predator-prey model was generalized to $n$ interacting species

$$\frac{dN_i}{dt} = \alpha_i N_i + \frac{1}{\gamma_i} \sum_{j=1}^{n} p_{ij} N_i N_j, \qquad N_i \geq 0,$$

where $N_i$ is the number of individuals of the $i$-th species. Coefficients $\alpha_i$ determine the evolution of the $i$-th population when other species are absent or do not affect this evolution (the wolves are miraculously indifferent to the sheep). In principle, $\alpha_i$ can be both positive and negative, but they cannot all have the same sign in real ecosystems since in this case all populations would boundlessly grow with time (if $\alpha_i > 0, i = 1, \ldots, n$) or disappear ($\alpha_i < 0, i = 1, \ldots, n$). The bilinear terms express the interaction of the $i$-th species with all other ones. In the Volterra reasoning, matrix $p_{ij}$ is antisymmetric, $p_{ij} = -p_{ji}$ which corresponds to a perfect balance: the removed biomass of species $i$ (interpreted as the prey) is exactly compensated for by the growing biomass of species $j$ (predators). In reality, however, an exact compensation of the consumed biomass never occurs, and this dictates the introduction of the phenomenological correction coefficients $\gamma_i$. For $n$ interacting species, the Kolmogorov model may be written as $\dot{x}_i = x_i f_i(\mathbf{x})$, $i = 1, \ldots, n, x = \{x_1, \ldots, x_n\}$ (no summation over repeated indices). For example, in the case of three interacting species one often considers the bilinear Kolmogorov model

$$\dot{x}_1 = x_1(a + a_{11}x_1^2 + a_{12}x_1x_2 + a_{13}x_1x_3 + a_{22}x_2^2 + a_{23}x_2x_3 + a_{33}x_3^2$$

$$\dot{x}_2 = x_2(b + b_{11}x_1^2 + b_{12}x_1x_2 + b_{13}x_1x_3 + b_{22}x_2^2 + b_{23}x_2x_3 + b_{33}x_3^2$$

$$\dot{x}_3 = x_3(c + c_{11}x_1^2 + c_{12}x_1x_2 + c_{13}x_1x_3 + c_{22}x_2^2 + c_{23}x_2x_3 + c_{33}x_3^2,$$

$a, b, c > 0$. One can apply this model even to such exotic situations as the survival of small businesses approached by organized criminal groups.

An intermediate case is represented by the generalized Lotka-Volterra system $\dot{x}_i = x_i f_i(\mathbf{x})$, where vector $\mathbf{f}$ is given by a linear relationship, $\mathbf{f} = \mathbf{a} + A\mathbf{x}$. Here components $a_i$ of vector $\mathbf{a}$ reflect the birth and death rate balance in the i-th population whereas matrix $A$ manifests the relationships between the species. For example, if matrix element $a_{ij} > 0$ but the transposed element $a_{ji} < 0$ then species $i$ is preying or parasitizing on species $j$ since the growth of the i-th population negatively affects the j-th one. Fixed points are determined from the linear system $\mathbf{x} = -A^{-1}\mathbf{a}$ which may produce both positive and negative values for components $x_i$. In the latter case, the problem requires a somewhat special study as there may be no ecologically meaningful stable attractor.

A model closely related to the Lotka-Volterra model describing the competition phenomena among many participants is the Wilson-Cowan model for excited and inhibited neural clusters [160]. One can represent both models through a dynamical system

$$\dot{x}_j = a_{ij}x_j + \omega(x_k)(b_{ij}x_j + c_j), \qquad i, j, k = 1, \dots, n,$$

where $b_{ij}$ is the connectivity (or competition) matrix and coefficient $\omega(x_k)$ represents the intensity of competition. For an overview of numerous attempts to improve the Lotka-Volterra model, the reader may refer to a comprehensive overview of the dynamics of interacting ecosystems in [18], [45].

### 13.3.3. Effect of harvesting

We have already discussed the effect of harvesting in the logistic model. In the context of the Lotka-Volterra model, the term "harvesting" denotes removing a certain fraction of the prey or predator or both populations at a constant rate $k$. One can imagine, e.g., fishing with nets that do not discriminate between predators (e.g., pike, tuna or shark) and their prey such as crucian carp, (Carassius vulgaris) for the pike. This example seems to be chrestomathic for compact ecosystems such as ponds. The two-component predator-prey system is described by equations

$$\frac{dx}{dt} = \alpha x - pxy - kx; \quad \frac{dy}{dt} = -\beta y + qxy - ky, \qquad k > 0 \qquad (13.3.3.1.)$$

or $\dot{x} = (\alpha - k)x - pxy; \;\; \dot{y} = -(\beta + k)y + qxy$. We see that the Lotka-Volterra model remains the same, but its coefficients are renormalized due to harvesting, $\alpha \to \tilde{\alpha} = \alpha - k, \beta \to \tilde{\beta} = \beta + k$. Accordingly, the position of the second critical point $(x_0, y_0) \equiv (\beta/q, \alpha/p)$ is changed to $(\tilde{x}_0, \tilde{y}_0) \equiv \left(\frac{\beta + k}{q}, \frac{\alpha - k}{p}\right)$, and since the harvesting rate $k$ is positive the equilibrium population of predators is reduced whereas that of the prey rises. The ecological balance is shifted, due to harvesting, towards the disadvantage of the predator population, which result is not quite as intuitive as one could assume that indiscriminate harvesting would affect both populations equally.

One can infer from the predator-prey model with uniform harvesting that ecological strategies should not be as straightforward as they seem. For instance, there is a ubiquitous opinion that since DDT (an acutely toxic substance used as an insecticide) may more efficiently kill mosquitoes than any other mosquito repellent and thus prevent the spread of malaria, isn't it worthwhile to use DDT on mosquito-plagued ecosystems, e.g., in sub-Saharan Africa? Proponents of this viewpoint argue that malaria in hard-hit African regions is much worse than DDT. Yet the mathematical model says that such a strategy could be counterproductive. Indeed, DDT is known to have a major negative effect on birds and fish who feed on mosquitoes, and if one indiscriminately sprays this compound, the anti-malaria program may fail since the equilibrium in the environment can be shifted to the advantage of the mosquito population.

### 13.3.4. Spatially inhomogeneous populations

An obvious generalization of the Lotka-Volterra dynamical system, taking into account migration of the species from high population density to low-density areas, is the PDE-based reaction-diffusion model

$$\begin{cases} \dfrac{\partial u}{\partial t} = k_1 \Delta u + u(a_1 - b_1 u + c_1 v) \\ \dfrac{\partial v}{\partial t} = k_2 \Delta v + v(a_2 - b_2 u + c_2 v). \end{cases} \qquad (13.3.4.1.)$$

Here $u = u(\mathbf{x}, t)$ and $v = v(\mathbf{x}, t)$ denote spatially-dependent population densities of two competing species, $k_{1,2}$ are the diffusion coefficients reflecting the migration rate of the species, $a_{1,2}$ denote the birth rates of species $u, v$, and the competition is accounted for by cross-factors $c_1$ and $b_2$. In this model, logistic coefficients $b_1, c_2$ manifesting the self-limited growth are retained, in contrast with the Lotka-Volterra system. Usually, all $k_i, a_i, b_i, c_i, \ i = 1,2$ are assumed to be positive constants. If model (13.3.4.1.) is interpreted in the ecological or biological sense, then quantities $u$ and $v$ can only take non-negative values. System (13.3.4.1.) can be defined on some smooth bounded domain $\Omega \subset \mathbb{R}^n$ and should be, as usual, supplemented by some boundary conditions, e.g., of Dirichlet ($u|_{\partial\Omega} = v|_{\partial\Omega} = 0$), Neumann $\left.\dfrac{\partial u}{\partial \mathbf{n}}\right|_{\partial\Omega} = \left.\dfrac{\partial v}{\partial \mathbf{n}}\right|_{\partial\Omega} = 0$) or of more general mixed $\left.\left(\dfrac{\partial u}{\partial n^i} + \kappa_i\right)\right|_{\partial\Omega} = \left(\dfrac{\partial u}{\partial n^i} + \kappa_i\right)\Big|_{\partial\Omega} = 0$ type.

Detailed analysis of this problem leads to rather involved calculations and would require much space in the book; therefore, we may refer the reader to authoritative sources [14] and [34].

When treating the models similar to (13.3.4.1.), one is often interested in the so-called coexistence states that are stationary (time-independent) solutions with $u > 0, v > 0$ i.e., the competing, e.g., predator and prey populations do not simultaneously disappear. This simplification reduces the model to the following elliptic system

$$\begin{cases} \Delta u + (a_1 - b_1 u + c_1 v)u = 0 \\ \Delta v + (a_2 - b_2 u + c_2 v)v = 0, \end{cases}$$

where coefficients $a_i, b_i, c_i, \ i = 1,2$ are properly normalized and $\mathbf{x} \in \Omega \subset \mathbb{R}^n$.

In conclusion of this section, we may note that population size and population density are those rare social variables that can be directly measured. Besides, the population dynamics is basically described by the balance conservation laws $\frac{\partial x^i}{\partial t} = F^i(x^j, t) - G^i(x^j, t) + \nabla \mathbf{C}^i, i, j = 1, \ldots, n$ widely used in physics and related disciplines (see section 6.1.). Here $x^i$ denotes the population density (or the number of the $i$-th species), $F^i$ is the number of births and $G^i$ is the number of deaths among the same sort of species per unit time, and vector $\mathbf{C}^i$ designates the migration flux of the i-th species (in a spatially inhomogeneous system). Therefore, the mathematical description of the population dynamics has become one of the favorite modeling subjects in mathematically-oriented social disciplines (or protosciences).

## 14. Conclusions and suggested readings

The subject of the present book was an attempt to analytically describe physical models for the processes accompanying the development of various complex phenomena including natural processes. The physical mechanisms of such processes are presented in such a way as to emphasize the mathematical content of the research and not the empirical facts lying at its foundations. The author's goal was to achieve a certain degree of methodical understanding, without limiting himself to the description (however meticulous and accurate) of experimental facts. Nevertheless, it should be noted that all the results presented in the monograph are based on scientific investigations carried out by multiple research groups who had carefully studied empirical foundations allowing one to build more or less simplified mathematical models. Since the main effort was directed towards the basic physical principles and crude mathematical modelling of the reviewed intricate processes, many necessary complexities and parameters have been omitted from the consideration.

Attempts to work with axiomatic concepts applied to real-world phenomena usually fail before model representations are attempted. The main thing about modeling is that one builds mathematical models by neglecting details that are declared "insignificant". It would be extremely difficult to invent a theory that would explain all possible phenomena in the world, even if we restrict ourselves only to physical processes.

Most of the models discussed in this book are just simple derivations from first physical principles. Ancient philosophers believed that air, earth, fire, and water were the basic elements of the world, and one only needs to combine them properly in order to explain any phenomenon in the universe. That concept was the prototypic theory of everything. Nowadays the question: "Is there a single theory that can describe everything?" is also answered affirmatively by many scientists and philosophers. Many bold people – not necessarily completely crazy – tried to devise such a theory and failed. In the meanwhile, a tribe of researchers breaks the image of the world into pieces and invents a partial theory for each piece. Such partial theories are often called models: they are indeed human-contrived models of the world, and in case they are expressed in mathematical terms they are called mathematical models. These latter may be more practical than the general theory of everything, at least when one needs to quickly find an engineering solution.

A large class of mathematical models, namely quantum and statistical ones, have not been touched upon in this book; such models require a separate treatment. In particular, statistical models combine mathematical or computer simulation with data analysis. Such combinations have become increasingly popular lately for the description of processes in living nature, e.g., biochemical phenomena. Some people assert that if life can be treated with the help of mathematical modeling at all, the respective models must be statistical i.e., including the data analysis. A new area of application of statistical modeling is "physical economy" or "econophysics".

Deterministic dynamical models are only capable of giving a crude framework of specific economic or financial processes. Note that unlike classical equations of dynamics or, e.g., the Dirac equation for the electron, economic and financial models are farther from experimental facts than it is considered standard in natural sciences. It does not mean, however, that complex systems such as in economics or social disciplines can only be modeled through statistical approaches. Many of the conventional economic and social stereotypes can acquire a precise meaning in deterministic modeling. In general, complex processes can be described both by deterministic and statistical methods, in particular, by stochastic equations. As far as quantum modeling is concerned, apart from traditional problems (description of atomic and molecular systems, solid state and its engineering applications, laser design, nuclear fusion, etc.), new areas have recently been opened such as nanotechnology and quantum computer. New computing schemes inherited from quantum mechanics are expected to evolve into an engineering discipline providing a new hardware basis for fast computations.

In many complex situations, mathematical modeling is by necessity a synthetic endeavor, it interpolates between mathematics, physics, computer science, engineering, biology, economics and a number of social disciplines. Thus, efficient mathematical modeling, despite its compartmentalized character, is based on the transfer of ideas between different domains of knowledge. This process is quite complex, very subjective and depends on the cultural background and individual attitudes of each modeler. Therefore, an ill-defined quality that we usually call intuition is an indispensable component of mathematical modeling, specifically when it involves interdisciplinary links. There exist numerous examples of when models can be significantly enriched due to multidisciplinary work, although it is one more concept regarded with suspicion by narrow-focused professionals. An important example of a theory that cuts across disciplines is the description of self-organization phenomena. Despite the fact that such phenomena are ubiquitous in nature and society, most of their mathematical models have been developed only recently and are far from being exhaustive. Accordingly, most of the models in this book were not treated profoundly enough: we only intended to demonstrate how mathematical models could be processed analytically or numerically and implemented on computers. Many instructive models had to be omitted, in some cases because of the lengthy analytical calculations or mathematical complexity, in other cases due to the sheer size of numerical procedures, all this being inadmissible in a rather short book. For example, climate models, if treated correctly, would require a considerable volume; the same applies to missile dynamics, flight of an aircraft, passive tracer tracking in ecology, military actions accompanied by terrorist activities, various potential threats to humanity, etc.

The surrounding world seems to be united and can hardly be understood through separate mathematical models, however thoroughly defined. In this book, we wanted to pay attention to synergetic interaction between different fields of physics, mathematics, and other disciplines where mathematical models, in particular those constructed according to a physical pattern, are extensively used. Scientific disciplines, in particular physics, are in perpetual development, conquering new areas and incorporating new results – new nodes, in networking terminology. The networking approach, increasingly popular these days, allows one to migrate between the physical and mathematical subjects with comparative ease.

Networking reasoning is a good instrument to unify diverse environments. Modern physics began with Einstein's attempt to reconcile electrodynamics, mechanics and thermodynamics in 1905 and his later endeavor to unify the specific relativity and Newtonian theory of gravitation. In a more general social context, the synthesizing approach of Einstein means a convergent scientific change – a retreat from austerity concepts of Max Weber who insisted on "Zweckrational" action rejecting all unnecessary circumstances. Einstein was rather thinking in the Renaissance spirit or more lenient

"communicative rationality" program, encouraging critical reassessment of "everything knows that" concepts and the establishment of mutual understanding between diverse scientific communities. As a consequence, Einstein's program proved to be very successful as compared to rival endeavors, not because it could explain more "facts" or was more powerful mathematically. It was better than rival programs probably because it provided a wide basis of interpenetration and communication between several main paradigms of 19th century physics. And in the long run, of course, the Einstein theory culminated in multiple empirical successes.

New endeavors in science will be inevitably dealing with unexpected links between apparently disjoined, isolated physics-based mathematical models rather than with the stand-alone and ad hoc constructs. In this sense the compartmentalization of mathematical models may diffuse giving way to the network of interlinked concepts. Such an approach most probably will call for an integrated type of scientific skills and education. But this is, of course, a vision.

## Supplement 1. Mathematical glossary

In this subsection, a few well-known mathematical facts and basic notions are rendered. The main purpose is to ensure a more or less firm ground to painlessly read the sections where such facts and notions are mentioned or used (e.g., about Hamiltonian mechanics, dynamical systems, quantum models, etc.). Readers proficient in contemporary mathematics can omit this section or at least use it as needed. Of course, only a cursory coverage of relevant concepts is provided here, and only basic definitions and notations used throughout the paper are presented.

### S1.1. Sets

A set has traditionally been considered the most fundamental concept in mathematics, although some modern mathematicians of different schools argue that there exist still more fundamental notions (such as, e.g., category, relations, equivalence, logical operations, etc.) The notion of a set is known to survive without a formal definition, at least it would be difficult to encounter any. A set can be loosely understood as a collection of objects usually called elements. Other terms for a set may be a variety, a class, a family, an assemblage or an assortment. Synonyms for the things called elements may be, for example, members, constituents or even points. Some sets have standard notations, e.g., $\emptyset$ is an empty set (containing no elements), the set of all real numbers is denoted $\mathbb{R}$, the set of all complex numbers is $\mathbb{C}$, the set of all integers is $\mathbb{Z}$, the set of all natural numbers is $\mathbb{N}$. A set of elements $x_1, x_1, ...$ will be denoted here by curly brackets $\{x_1, x_1, ...\}$. In case each element $x_i$ belonging to a set $A$, $x_i \in A$, is also an element of a set $B$, then $A \subseteq B$. Obviously, when both relationships $A \subseteq B$ and $B \subseteq A$ are valid, then $A = B$.

There are simple rules to operate with sets (see any textbook on the set theory). This theory seems at first very abstract, yet it may help solve practical problems, specifically in computer science. One can easily construct an example of a real-life situation when the set theory reasoning can help. Suppose you have to analyze the test results for a group of students being examined on three subjects: mathematics, physics, and German. There were $N$ persons to undergo the tests, with $n_m$ successful in math, $n_p$ in physics and $n_G$ in German. We know that $a_{mp}$ failed both in mathematics and physics, $a_{mG}$ in mathematics and German, $a_{pG}$ could pass neither physics nor German; $a_{mpG}$ failed in all three subjects. The question is, how many students, $n_{mpG}$, did successfully pass all three tests? In this example, just drawing the usual pictures (set diagrams) would help to solve this problem.

*Simple examples*. The first objects we encounter in school mathematics are numbers: we learn how to add and multiply them. At the University, when studying higher algebra, we learn that the set of numbers of a certain kind is known in mathematics as a field (one should not confuse this mathematical notion with the field in physics).

## S1.2. Maps

Now, to make the discussion of sets more concrete, one can review basic mathematical (algebraic and geometric) structures and operations. A vague notion of a "mathematical object" does not necessarily denote a set endowed with a certain algebraic or geometric structure, but if it does, then a mathematical object can be understood as some structure, in particular, group (a set with a single operation), ring (a set with two operations), or a vector space. Here we shall briefly discuss mappings (or maps) between the sets. We can just mention that in most applications such as mathematical modeling the sets endowed with operations are typically studied, which makes the following question relevant: what happens with the set structure, i.e., with the set elements and operations on them, when one set is mapped to another?

A rudimentary map is a 1D function: $X \to Y, x \in X, x \mapsto y, X \subset \mathbb{R}, Y \subset \mathbb{R}$ that "eats" a variable from $X$ and produces a number from $Y$. Of course, one can extend $X$ and $Y$ to $n$-dimensional manifolds. Thus, in the study of sets, the most important concept seems to be mapping. Assume there are two sets, $X$ and $Y$, then a rule $f$ assigning $y \in Y$ to each $x \in X$, $X \to Y$ i.e., $x \mapsto y$ usually written as $y = f(x)$ is called a mapping or a map. In human language, this sequence of symbols means that a mapping $f$ from set $X$ into set $Y$ associates to every element $x \in X$ an element $y = f(x) \in Y$ which is conventionally considered uniquely defined (not always the case).

In the general algebraic context one can define such mappings as isomorphisms which are typically understood as bijective homomorphisms. For the above examples of the most popular algebraic structures such as groups, rings or vector spaces, homomorphisms are defined as some specific, context-dependent mappings, e.g., a linear operator for the vector space, group and ring homomorphisms respectively for groups and rings. In any case, a homomorphism is one of the basic morphisms, it is a mapping that preserves the algebraic structure of two sets or algebraic objects to be mapped from one to another, in particular, from the domain $X$ into the target image of a function.

## S1.3. Manifolds

The concept of a manifold is quite fundamental. A manifold looks locally as a vector space, but it is essentially different from the latter when observed globally. To get an idea of a manifold one can think about spheres (also multidimensional), tori and higher-order surfaces.

One can introduce a coordinate system in the vicinity of point x in manifold $M$ as a diffeomorphic mapping $f$ of the neighborhood of the origin (zero-point) in Euclidean space $\mathbb{R}^n$ on the neighborhood of x $\in M$.

Manifolds play a significant role in physics: thus, classical mechanics, special and general relativity are formulated in the language of manifolds. In particular, spacetime in relativity theory is understood as a manifold $M$ endowed with Lorentzian metric $g_{ik}$ which is a symmetric, smooth, non-degenerate ($\det g_{ik} \neq 0$) (0,2)-tensor field on $M$ that contains all relevant information about spacetime. Two spacetimes $M$ with metric $g_{ik}$ and $N$ with metric $h_{ik}$ are considered identical from the physical standpoint, if there exists a diffeomorphism between these manifolds that maps these two metrics.

One can illustrate the importance of the concept of a manifold by the example of a velocity manifold $TQ$ which is indispensable for the Lagrangian formulation of mechanics. If we consider two points moving over the surface of a sphere $\mathbb{S}^2$, the difference of their velocities $\mathbf{v_{12}} \equiv \mathbf{v_1} - \mathbf{v_2}$ does not lie in any tangent plane to the sphere, in contrast with velocities $\mathbf{v_1}, \mathbf{v_2}$ of individual particles, but represents a rather complicated surface embedded in $\mathbb{R}^3$. In less primitive cases, the presence of such an ambient space is not at all ensured. For example, if one considers a simple mechanical system such as a double pendulum which already has a more complicated configuration manifold $Q = \mathbb{S}^2 \times \mathbb{S}^2$, than a particle moving on spherical surface, it would not be trivial to find an ambient space both for $Q$ and $TQ$. So, it would be highly desirable to learn how to treat the problems in terms of manifolds and their intrinsic properties, rather than in terms of the ambient vector spaces.

## S1.4. Metric spaces

A metric space may be considered as the simplest setting to study the relationship between objects, in particular between points. This property is achieved when a set of points in some space is endowed with a notion of a distance between them, thus the set becomes a metric space. Once the distance (the metric function) has been defined, such basic concepts of analysis as limit, convergence, continuity, derivative, compactness, etc. can be introduced.

## S1.5. Groups

A group G is basically a set $g_1, \dots, g_m$ endowed with the product $g_i g_k, i, k = 1, \dots, m$ and the inverse $g_i^{-1}, i = 1, \dots, m$ for each element, together with an identity element $I$ (or $E$) which is unique. This is how these group operations work: $gI = Ig = g, \ gg^{-1} = g^{-1}g = $I$, g \in $ G$, g_1(g_2 g_3) = (g_1 g_2)g_3$, $g_i \in $ G$, i = 1,2,3$.

The concept of a group is probably the simplest and the most basic among all mathematical structures used in physics or in mathematical modeling. At least it is an example of a very efficient mathematical structure. A group is an algebraic structure consisting of a set of objects together with a binary operation on this set. In other words, an operation is defined which produces a third object of the same kind from any two. Historically, the notion of a group has originated from studying the symmetry transformations in geometry on a plane (what we used to call planimetry in high school). Algebraic generalizations of such transformations have led to classifying symmetry properties through the operations regarded as elements of certain sets. After the invention of matrices[108] and especially following their systematic study by A. Cayley in the 1850s, matrices began playing the key role in algebra, specifically in linear algebra. Accordingly, matrix representations of groups emerged, a tool especially widely used in quantum mechanics. The discovery of Lie groups (named after Sophus Lie) has fomented the idea of applying continuous groups to the solution of differential equations.

More formally, a set G of elements of arbitrary origin is known as a group if a group operation (composition) $\otimes$ is defined and the following rules are fulfilled.

Closure: for any two elements $a, b \in $ G, there exists an element $c = a \otimes b, c \in $ G.

---

[108] It seems to be difficult to trace who really has introduced matrices in mathematical calculations.

This operation is associative: for any three elements $a, b, c \in \mathrm{G}$, $(a \otimes b) \otimes c = a \otimes (b \otimes c)$.

There exists a neutral element $e$ such that for any $a \otimes e = e \otimes a = a$.

For each element $a \in \mathrm{G}$ there exists an inverse (symmetric) element $a^{-1}$ also belonging to G such that $a \otimes a^{-1} = a^{-1} \otimes a = e$.

Note: if the group operation $\otimes$ is called multiplication (usually denoted as $\times$ or $\cdot$), the element $c$ is called a product of $a$ and $b$, the neutral element $e$ is called unity, and the group itself is known as multiplicative. The group operation (composition) symbol in multiplicative groups is often omitted i.e., $a \otimes b \equiv ab$. If the group operation $\otimes$ is understood as addition, the element $c$ is usually written as $c = a + b$, the neutral element is called zero, and the inverse element to $a$ is also called the opposite one, being written as $-a$. The group itself is known as additive. If, for any two elements $a, b \in \mathrm{G}$, $a \otimes b = b \otimes a$, the group is called Abelian or commutative.

## S1.6. Group representations

A representation of an abstract group is often understood as a realization according to the group axiomatic of linear automorphisms of a vector space.

One can easily prove that G is a group defined on a set $F$, $f \in F$ of all invertible transformations $y = f(x)$, where $x, y \in U$ ($U$ is some set, e.g., a differentiable manifold). Indeed, all group axioms are valid, e.g., $e(x) = x \in U$ so that $e(x) \in \mathrm{G}$. Analogously one can prove that if we have a map $y = f(x)$ then an inverse $x = f^{-1}(y)$, $f^{-1}f = ff^{-1} = e$ belongs to G for any $f \in \mathrm{G}$. A product $h := g \otimes f \equiv gf = g(f(x))$ of any $f, g \in \mathrm{G}$ also belongs to group G, $h \in \mathrm{G}$, so that the closure axiom is fulfilled.

## S1.7. Lie groups

The Lie groups are simultaneously algebraic structures and differential (smooth) manifolds as geometric objects (obeying the group properties). Both types of structures are consistent with one another in the sense that the group operations are differentiable. There exists an infinite number of Lie groups and even of so-called simple Lie groups, with most of them being some generalizations of rotation groups. Important examples are matrix groups such as the general linear group $\mathrm{GL}(n, \mathbb{R})$ (and $\mathrm{GL}(n, \mathbb{C})$), special linear group $\mathrm{SL}(n, \mathbb{R})$ ($\mathrm{SL}(n, \mathbb{C})$), orthogonal groups $\mathrm{O}(n)$, $\mathrm{SO}(n)$, etc. Thus, the substantial advantage of Lie groups is that one can study them by using ordinary calculus of infinitesimals. Incidentally, Lie himself was known to call such groups infinitesimal ones.

## S1.8. Example of a transformation group: the rotation group

The abstract notion of a transformation group is related to a set $G$ of transformations $y = f(x)$, where $x, y \in U$ are some abstract objects or variables and mapping $y = f(x)$ is assumed invertible. The set $G$ has the following properties:

The identity transformation $Id := I(x) = x$ belongs to $G$, $I \in G$.

An inverse $x = f^{-1}(y)$, $f^{-1}f = ff^{-1} = I$ belongs to $G$ for any $f \in G$.

A product $h := g \cdot f$, $h(x) := g(f(x))$ of any $f, g \in G$ also belongs to $G$ i.e., $h \in G$.

One can thus prove that $G$ is a group defined on a set $F \in f$, of all invertible transformations $y = f(x)$, where $x, y \in U$.

These elementary properties of transformation groups are illustrated by a group of rotations. Studying the rotation group is indispensable for many quantum-mechanical applications such as the theory of angular momentum, atomic spectra and nuclear physics. There is one more strong motivation to become familiar with the rotation group: it is a wonderful example of general Lie groups, and the latter serve as a natural tool for treating symmetry in physics (for instance, the Lie group SU(3)×SU(2) ×U(1) plays the essential part in the Standard Model of high energy physics). Note that group SU(2), the special unitary group of the second order (or degree), is homomorphic to the rotation group SO(3).

The simplest transformation group is probably the one of rotations of a circle. Although one can probably find even simpler examples, rotation of a circle is interesting both to physicists and to mathematical modelers since it is a trivial analog of the SO($n$) group which corresponds to rotations of $n$-dimensional Euclidean space $\mathbb{E}^n$. Let $g_\alpha$ denote the rotation by angle $\alpha$ so that the representation $R(g_\alpha)$ i.e., operator acting in a space of functions $F(\varphi)$ of the form $F(\varphi) := f(\cos\varphi, \sin\varphi)$ is defined by expression $R(g_\alpha)F(\varphi) = F(\varphi + \alpha)$. Such an operation symbolizes a rotation by angle $\alpha$. One can slightly generalize this expression, writing $R(g)f(\mathrm{x}) = f(g^{-1}\mathrm{x})$.

All proper rotations of 3d space (det = +1) constitute a group which is denoted SO(3), the special orthogonal group of order (or degree) 3. This group is very important for physics because its operations are isometries i.e., they preserve lengths and angles between directions (vectors). Indeed, since the scalar (inner) product of two vectors can be expressed through their lengths, $\mathrm{a} \cdot \mathrm{b} = |\mathrm{a}||\mathrm{b}| \cos\varphi = \frac{1}{2}(|\mathrm{a} + \mathrm{b}|^2 - |\mathrm{a}|^2 - |\mathrm{b}|^2)$, any length-preserving operation automatically preserves angle $\varphi$ between vectors $\mathrm{a}, \mathrm{b}$. One can arrive at the SO(3) group by the following simple considerations. Take a plane figure $S$ that can be rotated about an arbitrary axis whose position in space is fixed. Then the rotated state can be uniquely determined by three vectors: $\mathrm{e}_1$ (determining the rotation axis), $\mathrm{e}_2$ (lying in the plane of $S$ and orthogonal to $\mathrm{e}_1$, i.e., $\mathrm{e}_1 \cdot \mathrm{e}_2 = 0$), and $\mathrm{e}_3 = \mathrm{e}_1 \times \mathrm{e}_2$. In a generic local 3d basis, all matrices $g_{ij} \equiv \mathrm{e}_i\mathrm{e}_j$ (Gram matrices) form a 9d space $\mathbb{R}^9$, however, the orthonormality conditions, $\mathrm{e}_i\mathrm{e}_j = \delta_{ij}, i, j = 1,2,3$ give 6 constraints defining two 3d manifolds (with det = $\pm1$). Group SO(3) corresponds to proper rotations i.e., matrices with det = +1. Thus, each element of SO(3) is represented by a $3 \times 3$ orthogonal matrix. By a transformation $R \in$ SO(3), point x (e.g. of a rigid body) is mapped to $\mathrm{x}' = R\mathrm{x}, \det R = +1$. One can say that SO(3) is a subset of $\mathbb{R}^9$ (embedded in $\mathbb{R}^9$). In general, all possible positions (i.e., the coordinate space) of a mechanical or engineering system represented as a rigid body are given by the SO(3) group, and the phase space of such a system is the tangent bundle of SO(3). The so-called Euler angles are usually chosen as local coordinates in SO(3), and the motion of the resulting dynamical system is described by Euler's equation for a rigid body.

## S1.9. Green's functions

Green's functions have already been mentioned a few times in this manuscript. The method of Green's functions is in fact a new way of obtaining solutions to differential equations.

$$AG(t, t') = \delta(t - t')$$

As an example, let us consider Green's function of a linear oscillator. We have seen that the Hamiltonian and the Lagrangian functions for the one-dimensional linear oscillator are, respectively,

$$H(p, q) = \frac{p^2}{2m} + \frac{m\omega^2 q^2}{2}, \qquad L(q, \dot{q}) = [p\dot{q} - H(p, q)]_{\dot{q}(p,q) = \partial H/\partial p} = \frac{m\dot{q}^2}{2} - \frac{m\omega^2 q^2}{2}.$$

These Hamiltonian and Lagrangian result in the classical motion equation, $m\ddot{q} = -m\omega^2 q$. Let us now consider equation

$$\left(\frac{d^2}{dt^2} + \omega^2\right) G(t, t') = \delta(t - t').$$

Function $G(t, t')$ satisfying this equation is known as Green's function for differential operator $A = A(\omega) = d^2/dt^2 + \omega^2$.

One can notice that Green's function in autonomous case i.e., when the coefficients in the differential operator $A$ do not depend on time, is invariant under time shift, $G(t, t') = G(\tau)$, where $\tau = t - t'$. Green's function $G(\tau)$ of operator $A(\omega) = d^2/dt^2 + \omega^2$ defined by $A(\omega)G(\tau) = \delta(\tau)$ can be written as

$$G(\tau) = -\frac{i}{2\omega} \left[ \theta(\tau) e^{-i\omega\tau} + \theta(-\tau) e^{i\omega\tau} \right]$$

and can thus be interpreted as the superposition of a retarded and an advanced signal. Such an interpretation produces associations with the Feynman-Dirac representation of particles as superpositions of particles and corresponding antiparticles, the first having positive energy and propagating forward in time whereas the second having negative energy and propagating backward in time.

We can also notice that $G(\tau)$ can be analytically continued onto imaginary values of $\omega$ by rotating the integration contour (Wick's rotation, see Section 5.4.) without meeting any singularity. The absence of singularities in the course of such extrapolation is rather important for calculations since it allows us to introduce the "Euclidean" representation $(\tau \to i\tau)$ when the integrands no longer oscillate. This formal trick is frequently used to compute Feynman integrals in QED.

One can mention a useful formula here that is not very often encountered in physical literature:

$$\frac{1}{2\pi i} \lim_{\varepsilon \to 0} \left(\frac{1}{(x-a) - i\varepsilon}\right) - \left(\frac{1}{(x-a) + i\varepsilon}\right) = \delta(x - a)$$

This formula is related to the so-called incomplete distributions, $\delta_+(x)$ and $\delta_-(x)$ functions defined as

$$\delta_+(x) = \int\limits_0^\infty e^{ipx} \frac{dp}{2\pi}, \qquad \delta_-(x) = \int\limits_{-\infty}^0 e^{ipx} \frac{dp}{2\pi} = \int\limits_0^\infty e^{-ipx} \frac{dp}{2\pi}$$

Indeed, denoting $F(x, \varepsilon) := \frac{1}{2\pi i}\left(\frac{1}{x - i\varepsilon} - \frac{1}{x + i\varepsilon}\right)$, we have

$$F(x, \varepsilon) = \frac{\varepsilon}{\pi(x^2 + \varepsilon^2)} = \int\limits_{-\infty}^\infty e^{-|p|\varepsilon} \cos px \frac{dp}{2\pi}$$

(the Lorentz representation) and

$$\lim_{\varepsilon \to 0} F(x, \varepsilon) = \int\limits_{-\infty}^{\infty} e^{ipx} \frac{dp}{2\pi} = \delta(x) = \delta_+(x) + \delta_-(x).$$

The incomplete delta-functions are widely used both in physics (e.g., in the radiation and scattering theory) and in signal theory. They usually correspond to the real and imaginary parts of the spectral amplitude which are interrelated by the Hilbert transform.

## S1.10. Symplectic geometry and complex structures

Symplectic geometry (more exactly the geometry of symplectic manifolds) is based on the concept of a symplectic structure in a vector space $V$. The basic idea of a symplectic space seems to be very natural: one studies the geometric structure characterized by a skew-symmetric bilinear form $\omega(x, y)$ instead of a symmetric one as in Euclidean space. In other words, a symplectic structure is similar to an Euclidean structure with the following difference: the symplectic structure is a skew-symmetric function of a pair of vectors from $V$, linear in each of them. Simply speaking, one replaces the symmetric scalar (inner) product condition for the Euclidean structure, $(x, y) = (y, x)$ by $(x, y) = -(y, x)$. One can say that each vector in a space $V$ endowed with the symplectic structure is orthogonal to itself. It is thus clear that vector space $V$ must be even-dimensional: symplectic geometry can be applied to essentially even-dimensional spaces, while odd-dimensional spaces would be incompatible with symplectic structures.

It is easy to see that symplectic geometry is conceptually close to the geometry of complex numbers. Indeed, a symplectic form $\omega(u, v)$ on vector space $V = \mathbb{R}^{2n}$, $u, v \in V$ generalizing the antisymmetric condition $(x, y) = -(y, x)$ is represented by the Poisson matrix

$$\omega := \begin{pmatrix} 0 & I_n \\ -I_n & 0 \end{pmatrix},$$

where $I_n$ denotes an $n \times n$ unit matrix. On the other hand, a complex structure on a vector space $V$ is a linear map $J: V \to V$ with $J^2 = -I$, where symbol $I$ stands for the identity transformation (often denoted as Id). Here map $J$ symbolizing the complex structure is identified with multiplication by $i$, $i^2 = -1$, simply speaking $J(z) = iz$. Thus a complex structure $J$ is equivalent to a vector space over $\mathbb{C}$ i.e., $\mathbb{C}^n$ with coordinates $z^k = x^k + iy^k$, $i^2 = -1$, and multiplying by $i$ we get, for example, $i\partial x^k = \partial y^k$, $i\partial y^k = -\partial x^k$. All such expressions can be represented as a constant map acting on $V = \mathbb{R}^{2n}$ (more exactly, for the latter expressions on the tangent space of $\mathbb{R}^{2n}$):

$$J(u) := \begin{pmatrix} 0 & J^2 \\ -J^2 & 0 \end{pmatrix} u \equiv Ku,$$

where the Gram matrix[109] $K$ has the same structure as the symplectic matrix $\omega$. One can also notice that $\omega(u, v) = v^T K u$, $u, v \in \mathbb{R}^{2n}$ (this expression gives the imaginary part of a complex number)

---

[109] Recall that the Gram matrix defines an inner product: for a set of $n$ vectors $\mathbf{u}_i$ the Gram matrix is defined as $G_{ij} = (\mathbf{u}_i, \mathbf{u}_j)$. One might notice that the entries of the Gram matrix can be identified with the components of metric tensor $g_{ij}$. An important property of the Gram matrix is that it gives a criterion of linear independence of a family of vectors $\mathbf{u}_i$, $i = 1, \ldots, n$, in a vector space $V_n$: these vectors are linearly independent if and only if their Gram matrix is nonsingular. In

which reflects the fact that the symplectic matrix can be identified with the imaginary number $i$: one can easily verify that $K^2 = -1$. In general, a complex structure is provided by introducing a linear operator $J$ such that $J^2 = -1$. In the simplest case of plane vectors $(x, y)$ (or, in the complex plane representation, $z = x + iy$), one can easily see that operator $J$ represented by matrices $K$ or $\omega$ transform any plane vector $(x, y)^T$ into $(-y, x)^T$ i.e., rotates $(x, y)^T$ counterclockwise, just like the complex operator $i = e^{i\pi/2}$ extensively used in electrical and radio engineering modeling.

One can express the same fact in a different way by stating that matrix $e^{K\varphi}$, where $K = \begin{pmatrix} 0 & 1 \\ -1 & 0 \end{pmatrix}$, is the rotation matrix by angle $\varphi$ in a plane $\begin{pmatrix} \cos\varphi & \sin\varphi \\ -\sin\varphi & \cos\varphi \end{pmatrix}$ that corresponds to complex function $e^{i\varphi}$. It is also easy to verify that the matrix of multiplication by a complex number $z = x + iy$ with respect to basis $e_1 = 1, e_2 = i$ is symplectic of the form $Z = \begin{pmatrix} x & -y \\ y & x \end{pmatrix}$. If we in general consider the SU(2) group of symplectic matrices of this form subjected to condition $|x|^2 + |y|^2 = 1$, then setting $x = a + ib$ and $y = c + id$ we see that the above normalization condition defines a three-dimensional sphere in a 4d space with coordinates $a, b, c, d$. Recall in passing that sphere $S^n$ is by definition a submanifold of Euclidean space $\mathbb{R}^n$, $S^n \subset \mathbb{R}^n$ i.e., $S^n := x^j x_j \equiv \sum_{j=1}^{n} (x^j)^2 = 1$. Thus, $S^0$ consists of just two points, $\pm 1$ belonging to the real axis $\mathbb{R}^1$, $S^1$ is a unit circle on the $\mathbb{R}^2$ plane, the habitual $S^2$ sphere is a smooth manifold in Euclidean space $\mathbb{R}^3$ in which we used to live.

One can naturally identify $\mathbb{C}^n$ with $\mathbb{R}^{2n}$ i.e., introduce a complex structure $J$ with $J^2 = -I$, $I \equiv \mathrm{Id}$ in $\mathbb{R}^{2n} \to \mathbb{R}^{2n}$, $z = (z^1, \dots, z^n) = (x^1 + iy^1, \dots x^n + iy^n) = \big((x^1, y^1), \dots (x^n, y^n)\big)$. The symplectic geometry of Hamiltonian mechanics is naturally related to the complex geometry if we identify $V = \mathbb{R}^{2n}$ with $\mathbb{C}^n$. In particular, for $n = 1$ one would study the area (defined by a complex curve) instead of the length of a real line. However, extending symplectic geometry to higher dimensions may be not as trivial as, say, for Euclidean or Riemannian geometries: for example, it is already complicated to consistently define an oriented area for our everyday three-dimensional space (and in general for odd-dimensional spaces $\mathbb{R}^{2k+1}, k = 0, 1, \dots$).

We can finally note that there exists a version of geometry lying at the intersection of complex, symplectic and Riemann geometries, and this version is known as the Kähler geometry.

## Supplement 2. The necessity of geometric formulation

Mathematical study of mechanical motion began with Newton's exposition of calculus that enabled him to represent the universal laws of motion in the form of infinitesimal changes (differential equations). Integrating the latter allowed Newton and his successors to explain the crucial facts of physics and astronomy such as, e.g., Kepler's laws, mechanical energy and momentum conservation, etc. However, some basic concepts of Newtonian mechanics such as inertial frames, absolute time, trajectories and surfaces of motion, invariance under some classes of transformations remained unexplained or not properly explained – one rather had to take such concepts for granted – and haunted the study of mechanical motion for over two centuries. The most embarrassing and intriguing

---

other words, the Gram determinant $G = \det G_{ij}$ is zero iff the set $\mathbf{u}_i$ of vectors is linearly dependent whereas for a set of linearly independent vectors $G > 0$.

had then been the geometric objects of mechanics (we have already met some of them such as the phase space).

Geometric objects that are of interest in classical mechanics have scalar, vector or tensor nature. In general, all such objects are represented as tensors. A tensor $T$ defined at point $x$ is related only to this point. More generally, one deals with tensor fields (scalar and vector fields are just specific cases so that we do not treat them separately), when a tensor, in particular corresponding to some physical quantity, is defined at each point $x$ of manifold $M$. Examples of tensor fields are well-known in physics: continuum density $\rho(x)$ in fluid dynamics (see Section 6.11.) is usually represented as a 3d scalar field; spacetime of special relativity is described as a 4d = 3d + 1 pseudo-Euclidean vector space (manifold $M^{3+1}$), with vector and tensor fields such as four-velocities $u^\mu$, electromagnetic field $F_{\mu\nu}(x)$, etc. being defined on $M^{3+1}$.

When beginning to study mechanics one sees at once that mechanical motion gives rise to geometric concepts. A free particle is known to move along a geodesic (for a given connection) i.e., along the curve $\gamma := x^i(t) \in M$ on which the covariant derivative (see section 6.1.) of the tangent vector field $\dot{x}^i(t)$ vanishes:

$$\nabla_\gamma \dot{x}^j(t) = \dot{x}^i(t)\nabla_i \dot{x}^j(t) = \dot{x}^i(t)\big(\partial_i \dot{x}^j + \Gamma_{li}^j \dot{x}^l\big) = \frac{d^2 x^j}{dt^2} + \Gamma_{li}^j \frac{dx^l}{dt}\frac{dx^i}{dt} = 0, \qquad (S2.1.)$$

where functions $\Gamma_{li}^j$ are the Christoffel symbols

$$\Gamma_{jk}^i = \frac{1}{2} g^{il}\big(\partial_k g_{jl} + \partial_j g_{kl} - \partial_i g_{jk}\big) \qquad (S2.2.)$$

We can see that the Christoffel symbols essentially depend on metric $g_{ik}(x^j)$ i.e on the underlying geometry. One calls spaces, where Christoffel symbols $\Gamma_{jk}^i \neq 0$, to be endowed with an affine connection. Equations of the (S2.1.) type describe the motion in arbitrary spacetimes and not only in Euclidean (or Cartesian) frame when $\Gamma_{jk}^i = 0$, as it is often assumed in traditional formulations of Newtonian mechanics. The Christoffel symbols can be interpreted as the connection coefficients in local coordinates, and as such they can be used for practical computations in differential geometry and general relativity. For example, in the latter theory the Christoffel symbols assume the role of the gravity force whereas in curvilinear coordinates i.e., in a curved space or manifold the Christoffel symbols manifest "inertial forces". Equation (S2.1.) may be interpreted as the disappearance of the covariant 4d acceleration $\frac{D^2 x^j}{Dt^2} = 0$.

Geodesic curves (or colloquially geodesics) are the trajectories of some special – inertial – motion of a body when there are no forces acting on it. This inertial motion is some standard one of a purely geometrical nature. In the special case of Euclidean space (or spacetime) these trajectories are reduced to straight lines along which the body moves with a constant speed. In a more general (and more realistic) case of a manifold, e.g., in a Riemann space, geodesics can be of a complicated form that should be found by solving the geodesic equation (S2.1.) containing the Christoffel symbols (S2.2.). Note that (S2.1.) in an $n$-dimensional manifold is actually a system of $n$ ODEs for $n$ coordinates so that if the conditions for the applicability of the Cauchy-Peano (or Picard-Lindelöf) theorem are fulfilled (in particular, the Lipschitz continuity) this system of equations can always be solved, provided the initial conditions are stipulated.

Geodesics have the property that they minimize the length between two sufficiently close points on a manifold i.e., locally (we shall clarify the meaning of geodesics in connection with the Lagrangian version of mechanics). In the Euclidean space $\Gamma_{li}^j = 0$, and geodesics, as already mentioned, are straight lines so that a free particle moves along them. The Christoffel symbols are substantially different from zero when the metric tensor $g_{ik}$ varies from point to point which evidently never occurs in the Euclidean space. Notice that the metric tensor is associated with gravity, while non-zero Christoffel symbols (connections) $\Gamma_{jk}^i \equiv \begin{pmatrix} i \\ jk \end{pmatrix}$ give rise to inertia. Thus, inertia is rooted in geometry.

It might be interesting to recall here that the great mathematician David Hilbert who in 1900 formulated 23 challenging problems in mathematics suggested in his $4^{th}$ problem to describe all metric geometries where geodesics would be straight lines ("Problem of the straight line as the shortest distance between two points"), see, for example, http://www.seas.harvard.edu/courses/cs121/handouts/Hilbert.pdf. Some mathematicians, however, argue that the $4^{th}$ Hilbert problem is formulated too vaguely (e.g., in the form "construct all metrics where geodesics are straight lines") to state whether it has been solved or unsolved.

One might say that Euclidean geometry is just kinematics of a particle (or of a rigid body) when the spacetime curvature vanishes i.e., gravitation is absent, and the motion is considered in the non-accelerating frame of reference. If the geodesic motion is affected by the vector field of forces $F: Q \to \mathbb{R}^n$ (here $Q$ is the configuration space, $n$ is finite: for a single particle $n = 3$), the particle motion is described by equation:

$$\ddot{q}^j + \Gamma_{ik}^j \dot{q}^i \dot{q}^k = \frac{1}{m} F^j \qquad \text{(S2.3.)}$$

or

$$m_{ik}\left(\ddot{q}^k + \Gamma_{jl}^k \dot{q}^j \dot{q}^l\right) = F_i, \qquad m_{ik} := mg_{ik}, \qquad F_i = g_{ik}F^k. \qquad \text{(S2.4.)}$$

Expression (S2.6.) generalizes Newton's second law to the motion in a curved space, in particular, in an accelerated frame or gravitation field. In engineering the terms with $\Gamma_{ik}^j \neq 0$ play a significant part in robotics (see, e.g., [64]). One can rewrite (S2.3) in the form, more appealing to "physical intuition"

$$m_{ik}\ddot{q}^k = F_i - mg_{ik}\Gamma_{jl}^k \dot{q}^j \dot{q}^l. \qquad \text{(S2.5.)}$$

When there is no external forcing ($F_i = 0$) the motion is induced only by inertia and/or gravitation, as, in fair approximation, the motion of the Earth around the Sun or, in general, of a body in a free fall. In a very special case of "inertial frame" (i.e., when $\Gamma_{ik}^j = 0$ and $F_i = 0$) the body moves along a straight line with a constant velocity.

Formula (S2.5.) might still seem intimidating, but in fact it is a simple generalization of well-known expressions for "fictitious" forces arising in non-inertial mechanical motion, e.g., centrifugal and Coriolis forces that appear, for example, when Newton's laws are transformed to a rotating frame of reference. The physical meaning of expressions (S2.3.) is that the body with a nonzero mass strives to retain its instantaneous position as much as possible. From the standpoint of an observer in an accelerating – in this example rotating – frame, the body begins accelerating, and the observer has to apply the force equal to $-mg_{ik}\Gamma_{jl}^k \dot{q}^j \dot{q}^l$ (the second term in (S2.5.)) to prevent the body from accelerating. These terms are usually called "fictitious", although this is incorrect: they are absolutely real as any person who drives a car can testify. However, such terms have a manifestly geometric origin and cannot be simply attributed to the action of nearby bodies or fields. "Fictitious" forces also differ from the "normal" ones by the fact that the former do not in general obey the vector transformation properties. A fascinating cluster of mathematical models related to fictitious forces was stimulated by Mach's principle that was a suggestion that inertia and inertial forces emerge due to the totality of masses spread in the universe.

Recall that, for instance, Newtonian mechanics is based on a tacit assumption of working only in inertial coordinate (reference) frames; this assumption inevitably leading to the principle of covariance and associated geometric concepts such as the ideas of covariant derivative and connection. (The author, however, does not intend to enter the extensive debate about the impact of modern geometry on further development of physics.)

An external influence F shifts the body from one geodesic trajectory onto another. If field F is conservative i.e., there exists a scalar potential function $V(q): Q \to \mathbb{R}$ such that $\text{F} = -\nabla V$, then one can substitute in (S2.5.) $F_i = -\partial_i V$ ($\partial_i \equiv \partial/\partial q^i$). In most cases, one can additively separate conservative ($-\partial_i V$) and non-conservative ($F^i$) fields of forces, $\Phi^i = -g^{ik}\frac{\partial V}{\partial q^k} + F^i$, where $\Phi^i$ is the total force acting on a body.

Expression (S2.3.) for Newton's law is natural from geometrical positions since the commonly used formula $dp_i/dt = F_i$ or $dp_i/dt = -\partial V/\partial q^i$ equalizes a poorly defined derivative of a covector $p_i$ along a curve to a vector of forces or a covector (gradient). To provide the geometric (intrinsic) meaning to Newton's law, one can replace the ordinary derivative $d/dt$ by the covariant derivative $\nabla/dt$ so that instead $dp_i/dt = d(g_{ik}\dot{q}^k)/dt = -\partial V/\partial q^i$ or $d\dot{q}^k/dt = -g^{ki}\partial V/\partial q^i$ we get $\nabla\dot{q}^k/dt = -g^{ki}\partial V/\partial q^i$ or, in local coordinates,

$$\frac{d\dot{q}^k}{dt} + \Gamma_{ij}^k \dot{q}^i \dot{q}^j = -g^{ki}\frac{\partial V}{\partial q^i} + F^k. \qquad (S2.6.)$$

The very appearance of the metric tensor $g_{ik}$ and related quantities such as the Christoffel symbols in the motion equations is quite natural and can be illustrated on the example of rotational motion. Notice that rotational motion is very important for engineering applications, especially in biotechnology and nuclear engineering (see more on that below). If we consider the motion in the coordinate system rotating with constant angular velocity $\omega$ about z-axis:

$$\begin{pmatrix} x \\ y \end{pmatrix} = \begin{pmatrix} \cos\omega t & -\sin\omega t \\ \sin\omega t & \cos\omega t \end{pmatrix} \begin{pmatrix} x' \\ y' \end{pmatrix}, \qquad (S2.7.)$$

then, after some simple manipulations, we get for the squared distance element[110]

---

[110] In geometric terms of dynamical systems theory, (17) gives a flow (a diffeomorphism from $\mathbb{R}^2$ to $\mathbb{R}^2$) generated by a rotational vector field $X(x, y) = -y\partial_x + x\partial_y$ (up to a constant z-component of angular momentum) on manifold $M \coloneqq \mathbb{R}^2$. This flow is also a commutative (Abelian) one-parameter group of transformations isomorphic to SO(2) (the group of $2 \times 2$ real orthogonal matrices) and to U(1) (the multiplicative group of complex numbers with the unit absolute value, $e^{i\varphi}$).

$$ds^2 = dx^2 + dy^2 + dz^2 + (r\omega)^2 dt^2 - 2y\omega dxdt + 2x\omega dydt$$
$$= \left(1 - \frac{y^2}{r^2}\right)dx^2 + \left(1 - \frac{x^2}{r^2}\right)dy^2 + dz^2 + 2xydxd \qquad \text{(S2.8.)}$$

where $r^2 = x^2 + y^2$, and for the length along the smooth curve γ: $x^i(t)$

$$l_{1-2} = \int\limits_{t_1}^{t_2} |\mathsf{v}(t)| dt, \qquad \mathsf{v}(t) = \left(\dot{x}^1(t), \dot{x}^2(t), \dot{x}^3(t)\right)^T \equiv (\dot{x}(t), \dot{y}(t), \dot{z}(t))^T$$

we have

$$l_{1-2} = \int\limits_{t_1}^{t_2} (g_{ik}\dot{x}^i\dot{x}^k)^{1/2} dt.$$

That said, we see that the curve length is no longer represented by the sum of squared differentials of coordinates (the Pythagorean theorem) as in inertial systems. Notice that in different non-inertial or curvilinear systems one has different off-diagonal coefficients $g_{ik}$ before $dx^i dx^k, (i \neq k)$. Correspondingly, both the geodesics and the motion equations are changed. Notice also that coefficients $g_{ik}$ are functions of point r, $g_{ik} = g_{ik}(x,y,z)$ (in an $n$-dimensional space, $g_{ik}(x), x = \{x^j\}, i,j,k = 1, \dots, n$), which makes the Christoffel symbols non-zero. Recall that $dl^2 = g_{ik}(x)dx^i dx^k$ is often called metric (especially in the physical literature) and functions $g_{ik}(x)$ are called the components of a metric.

There exist standard and quite powerful methods of representing differential operators in arbitrary coordinate systems (see, e.g., [93]), yet it would be instructive to produce such a representation in a straightforward fashion. If we introduce basis vectors $\mathsf{a}_k$ corresponding to arbitrary (e.g., curvilinear) coordinates, we get new coordinate values defining each point $\mathsf{r} = (x^1, \dots, x^n) = x^k \mathsf{e}_k, k = 1, \dots, n$ (for simplicity we shall regard here $n = 3$). Here basis vectors $\mathsf{e}_k$ are related to Euclidean coordinates and are assumed constant. Notice that basis vectors $\mathsf{a}_k$ of curvilinear coordinate systems in general depend on point $\mathsf{a}_k = \mathsf{a}_k(\mathsf{r}), \mathsf{r} = (x^1, x^2, x^3)$ which is represented in the new coordinate system as $\mathsf{r} = (\xi^1, \xi^2, \xi^3) = \xi^k \mathsf{a}_k = x^j \mathsf{e}_j$, $\frac{\partial \mathsf{r}}{\partial \xi^k} = \frac{\partial (x^j \mathsf{e}_j)}{\partial \xi^k} = \mathsf{e}_j \frac{\partial x^j}{\partial \xi^k} = \frac{\partial (\xi^i \mathsf{a}_i)}{\partial \xi^k} = \mathsf{a}_k + \xi^i \frac{\partial \mathsf{a}_i}{\partial \xi^k}$. In other words, transition to a new coordinate system is accompanied with the change of variables $x^i \mapsto \xi^k$, where $x^i = \phi^i(\xi^k)$ and, conversely, $\xi^k = \psi^k(x^i)$, $\psi = \phi^{-1}$. We assume here as usual that the Jacobian matrix $J_{ij} := \frac{\partial^2 \psi}{\partial x^i \partial x^j}$ is non-singular.

Depending on the symmetry of the considered problem or model, radius-vector r and, accordingly, fields $\phi^i, \psi^k$ can be written in the most suitable coordinate system that can differ from the Cartesian coordinates. In particular, in the spherical system $(r, \varphi, \vartheta)$, $\mathsf{r} = \xi^r \mathsf{a}_r + \xi^\varphi \mathsf{a}_\varphi + \xi^\vartheta \mathsf{a}_\vartheta$. If we deal with the linear problem (see section 5.10. Linear models), differential operator $D$ usually includes a standard set of elementary classical differential operators such as gradient $\partial_i A$, divergence $\partial_i A^i$, curl $\varepsilon_{ijk}\partial^j A^k$, Laplacian $\partial_i \partial^i$, where $\partial^i = g^{ik}\partial_k$, $g^{ik}$ are the components of the inverse metric tensor, $g^{ik}g_{kj} = \delta^i_j$, $g^{ik} = g^{il}g_{lm}g^{mk}$, and the d'Alembert operator $\partial_\mu \partial^\mu, \mu = 0,1,2,3$, which is a generalization of the Laplace operator to the Minkowski space (e.g., (0,-1,-1,-1)). In curvilinear coordinates $\xi^k = \psi^k(x^i)$, in 3d $(\xi^1, \xi^2, \xi^3)$, the Laplacian can be represented as

$$\Delta = \partial_i \xi^\alpha \partial^i \xi^\beta \frac{\partial^2}{\partial \xi^\alpha \partial \xi^\beta} + \partial_i \partial^i \left( \xi^\alpha \frac{\partial}{\partial \xi^\alpha} \right).$$

A natural generalization of the Laplace operator to the manifolds more general than the Minkowski space leads to one more linear elliptic operator of the second order called the Laplace-Beltrami operator $\Delta_{L-B} = \frac{1}{\sqrt{|g|}} \partial_i \left( \sqrt{|g|} g^{ik} \partial_k \right), g \equiv \det g_{ik}, g^{ik} = g_{ik}^{-1}$.

If we now write the metric in both coordinate systems corresponding respectively to $\phi$ (Cartesian) and $\psi$ (generic, in particular curvilinear), we get for the squared distance element

$$ds^2 = \eta_{ij} dx^i dx^j = \eta_{ij} \frac{\partial \varphi^i}{\partial \xi^k} \frac{\partial \varphi^i}{\partial \xi^l} d\xi^k d\xi^l = g_{kl} d\xi^k d\xi^l,$$

where $\eta_{ij} = \mathrm{diag}(1, -1, -1, -1)$ is the Galilean (or Euclidean) metric tensor whereas $g_{kl} = \frac{\partial \phi^i}{\partial \xi^k} \frac{\partial \phi^i}{\partial \xi^l} = \mathrm{a}_k \mathrm{a}_l$ (since $\mathrm{a}_k = \mathrm{e}_j \frac{\partial \phi^j}{\partial \xi^k}$, as $d\mathbf{r} = \mathrm{a}_k d\xi^k = \mathrm{e}_j dx^j$).

Simply speaking, Cartesian coordinates are defined by the condition that $g_{ij} = (e_i, e_j) = \delta_{ij}$ everywhere in $\mathbb{R}^n$ (here $e_i, e_j, i, j = 1, \dots, n$ denote the orthonormal basis vectors); a more accurate statement would be that the system of coordinates is Cartesian if and only if the metric tensor

$$g_{ik}(\mathrm{x}) = \sum_{j=1}^n \frac{\partial x^j}{\partial q^i} \frac{\partial x^j}{\partial q^k} = \delta_{jl} \frac{\partial x^j}{\partial q^i} \frac{\partial x^l}{\partial q^k} = \delta_{ik}$$

i.e., the Jacobian matrix $J \equiv A_i^j := \frac{\partial x^j}{\partial q^i}$ is orthogonal and constant everywhere in $Q$. In other words, Cartesian (Euclidean, Galilean) coordinates have a global character and are linearly transformed everywhere to other coordinates of the same class with the help of orthogonal matrices.

Moreover, the Cartesian coordinates are unique in the analytical description of motion in the sense that velocities, accelerations and, when needed, higher order derivatives of the actual position of a considered object are simply determined by the time derivatives of the components $x^i$ of position vector $\mathrm{x} = (x^1, \dots x^n)$ i.e., of coordinates corresponding to each individual basis vector $e_i$, $\mathrm{x} = x^i e_i$, $i = 1, \dots, n$. In curvilinear coordinate systems, basis vectors are, in general, functions of the spacetime point, therefore to obtain spatial or temporal derivatives one has, in contrast with the Cartesian coordinates, to differentiate also the basis vectors. Let us see it as a simple example of time derivatives. If we denote, as previously, by $\mathrm{a}_i, i = 1, \dots, n$ the basis vectors in arbitrary, in particular curvilinear, coordinates and by $\mathrm{e}_j$ in the Cartesian ones, we shall have, due to linearity, $\mathrm{a}_i = A_i^j \mathrm{e}_j, i, j = 1, \dots, n$. The inverse transformation is $\mathrm{e}_j = \left( A_i^j \right)^{-1} \mathrm{a}_i = A_j^i \mathrm{a}_i$ (since the transformation matrix $A$ is an orthogonal one). We can rewrite these transformations in the matrix form as $\mathrm{a} = A\mathrm{e}$ and $\mathrm{e} = A^T \mathrm{a}$ where $\mathrm{a} = (a_1, \dots, a_n)^T, \mathrm{e} = (e_1, \dots, e_n)^T$ and matrix $A = (A_i^j)$ is orthogonal, $A^{-1} = A^T$. The time derivative is $\frac{d\mathrm{a}}{dt} = \frac{dA}{dt} \mathrm{e} + A \frac{d\mathrm{e}}{dt} = \left( \frac{dA}{dt} A^T \right) \mathrm{a} \equiv D\mathrm{a}$, where matrix $D := \frac{dA}{dt} A^T$ corresponds to the differentiation operator acting on the basis vector of a curvilinear coordinate system. Notice that matrix $D$ in general is time-dependent so that $\frac{dD}{dt} = \frac{d^2 A}{dt^2} A^T + \frac{dA}{dt} \frac{dA^T}{dt}$.

In order not to overload the text with cumbersome generic expressions (that are almost useless anyway), let us consider a particular case of some popular curvilinear coordinate system, e.g., spherical coordinates in 3d, $r = ra_r + \vartheta a_\vartheta + \varphi a_\varphi \coloneqq (r, \vartheta, \varphi)$. In this rather specific case, the basis transformation matrix

$$A = \begin{pmatrix} \sin\vartheta\cos\varphi & \sin\vartheta\sin\varphi & \cos\vartheta \\ \cos\vartheta\cos\varphi & \cos\vartheta\sin\varphi & -\sin\vartheta \\ -\sin\vartheta & \cos\vartheta & 0 \end{pmatrix}$$

gives rise to the basis differentiation matrix

$$D = \begin{pmatrix} 0 & \dot\vartheta & \dot\varphi\sin\vartheta \\ -\dot\vartheta & 0 & \dot\varphi\cos\vartheta \\ -\dot\varphi\sin\vartheta & -\dot\varphi\cos\vartheta & 0 \end{pmatrix}. \tag{S2.9.}$$

Let us now calculate the velocity $v = \dot{r}$ of a point characterized by coordinate $r = ra_r$ i.e., $v = \frac{d}{dt}(ra_r)$. We have

$$v = \frac{d}{dt}(r,0,0)\begin{pmatrix} a_r \\ a_\vartheta \\ a_\varphi \end{pmatrix} = [(\dot r,0,0) + (r,0,0)D]\begin{pmatrix} a_r \\ a_\vartheta \\ a_\varphi \end{pmatrix} = (\dot r, r\dot\vartheta, r\dot\varphi\sin\vartheta)\begin{pmatrix} a_r \\ a_\vartheta \\ a_\varphi \end{pmatrix}$$
$$= \dot r a_r + r\dot\vartheta a_\vartheta + r\dot\varphi\sin\vartheta a_\varphi.$$

Applying the same procedure of basis vector differentiation we can obtain higher order time derivatives of point position $ra_r$ as polynomials in operator $D$, $\frac{d^m}{dt^m}(ra_r) = \sum_{k=0}^{m}\binom{m}{k}\frac{d^{m-k}r(t)}{dt^{m-k}}\frac{d^k}{dt^k}a_r$, where $\frac{d}{dt}a_r = Da_r$ and higher derivatives are obtained by induction. For example, in the same particular 3d case $r = ra_r$, we have for acceleration

$$\ddot r = \frac{dv}{dt} = \frac{d}{dt}\left[(\dot r, r\dot\vartheta, r\dot\varphi\sin\vartheta)\begin{pmatrix} a_r \\ a_\vartheta \\ a_\varphi \end{pmatrix}\right]$$

$$= \left[(\ddot r, \dot r\dot\vartheta + r\ddot\vartheta, \dot r\dot\varphi\sin\vartheta + r\ddot\varphi\sin\vartheta + r\dot\varphi\dot\vartheta\sin\vartheta) + (\dot r, r\dot\vartheta, r\dot\varphi\sin\vartheta)D\right]\begin{pmatrix} a_r \\ a_\vartheta \\ a_\varphi \end{pmatrix}$$

$$= \ddot r - r\dot\varphi^2\sin^2\vartheta - r\dot\vartheta^2)a_r + (r\ddot\vartheta + 2\dot r\dot\vartheta - r\dot\varphi^2\sin\vartheta\cos\vartheta)a_\vartheta$$
$$+ (r\ddot\varphi\sin\vartheta + 2\dot r\dot\varphi\sin\vartheta + 2r\dot\varphi\dot\vartheta\cos\vartheta)a_\varphi,$$

where the matrix representation of operator $D$ is given by (S2.9). Similar expressions can be presented for other popular curvilinear coordinates, e.g., cylindrical, parabolic, spheroidal, toroidal, etc., each coordinate system being characterized by its own time-differentiation operator $D$. The time derivatives in curvilinear coordinates that are higher than those of the second order are written through more complex expressions, but luckily, they are seldom needed in practical problems, e.g., in mechanics. The same simplification applies to spatial derivatives of vector fields.

One can comment a little more on the role played by the additional term in formulas (S2.3.), as compared to school-time Newtonian equation $F = ma$, where $a$ is the body's acceleration. In general, it is by no means trivial that most of the currently known dynamical laws of motion have a rather simple mathematical form. Notice that in Newtonian mechanics formulated in Euclidean space, forces producing the acceleration of a test body are always caused by the presence of other bodies. Not so in curved spaces. Figuratively speaking, due to the additional (nonlinear!) term a body moving through a curved space can drift or glide across the vacuum without any need to be pushed or pulled by other bodies. This fact is closely related to the principle of equivalence between inertia and gravitation that was regarded by Einstein as fundamental and served as the base of general relativity. Recall that the effect of being in an accelerated frame of reference with the effective (renormalized) free-fall acceleration $g \rightarrow g - a$ lies at the foundation of general relativity.

The principle of equivalence leads to a generalization of the Euclidean (or pseudo-Euclidean) geometry of spacetime so that it becomes curved: the worldlines of inertially moving particles are in general relativity curved geodesics rather than straight lines as in the flat Euclidean space. In broad brush strokes, if special relativity states that there is no ether i.e., no preferred inertial frame, general relativity states that there are no inertial frames at all. Thus, the principle of equivalence and, in particular, general relativity exempted spacetime from the restriction of flatness. General relativity can be viewed as a dynamic geometry of the world described throughout the metric tensor, and persistent attempts to formulate a theory of gravitation as the model with external field had to be abandoned or at least modified in favor of Einstein's geometric formulation. In the latter scheme components of the metric tensor simultaneously serve as the gravitational field potentials which is uncommon for field theories when field variables are independent of the metric. In this respect, geometric formulation of general relativity stemming from the equivalence principle and based on the concepts of general covariance, geodesic motion and other notions of modern geometry radically differs from the conventional field theories.

One may note in passing that the metric is actually everywhere, even without general relativity. For example, in electromagnetic theory the main object – the tensor of the electromagnetic field $F_{\mu\nu} = \partial_\mu A_\nu - \partial_\nu A_\mu$ produces an invariant (the so-called Lorentz invariant which is, in fact, up to a constant the Lagrangian density needed to obtain the action-based formulation of classical electromagnetic theory) $F_{\mu\nu} F^{\mu\nu}$ that is transformed in the orthodox special relativity with the help of the Galilean metric tensor $\gamma_{\mu\nu}$ i.e., $F_{\mu\nu} F^{\mu\nu} = \gamma_{\mu\alpha} \gamma_{\nu\beta} F^{\alpha\beta} F^{\mu\nu}$. In a more general context of a curved spacetime, the metric may no longer be constant (Galilean or Minkowski), and the metric tensor $\gamma_{\mu\nu}$ should be replaced by the general one $g_{\mu\nu}(x)$.

In crude terms, special relativity is a universal theory: although there still exist a number of paradoxes[111] complicating its full understanding, it has been thoroughly tested experimentally in many fields of physics, and every literate physicist has to comply with it. In a first approach, general relativity is just a theory of gravitation, at least so far. The main equations of general relativity – Einstein's equations for the gravitation field – can be interpreted as a generalization of the Poisson equation in electrostatics, where the charge density is superseded by the mass density and $\sqrt{g_{00}(x)}$ ($g_{00}$ is the zeroth component of the spacetime metric) enters under the Laplacian instead of scalar potential $\varphi \leq 0$. In the Newtonian limit i.e., for slow ($v \ll c$) particles in the weak gravitation field, the metric coefficient $g_{00}$ is related to potential $\varphi$ as $g_{00}(x) = 1 + 2\varphi(x)/c^2 + \sigma(1/c^2)$ which results in the reduced time interval between two events ($d\tau \approx (1 + \varphi/c^2)dt$). One can prove (see, e.g., [96], Ch.XII, §99 in Latin) that Einstein's equations can be reduced in the Newtonian limit to the Poisson equation for potential $\varphi$.

---

[111] For example, the famous twin paradox, one traveling into space whereas the other remaining on the Earth. After returning to the Earth, the space traveler found the other (Earth-based) twin to have aged more than himself. The twin paradox is resolved in special relativity by noticing that the world lines (geodesics) of the Minkowski spacetime maximize the proper time between two events i.e. $t_{elapsed} = \int d\tau = \int \left( \gamma_{\mu\nu} \frac{dx^\mu}{d\lambda} \frac{dx^\nu}{d\lambda} \right)^{1/2} d\lambda = \int (\gamma_{\mu\nu} dx^\mu dx^\nu)^{1/2} d\lambda$.

The energy-momentum tensor standing on the right-hand side of Einstein's equations is directly related to the generalized Laplacian[112]. This is, of course, a very loose interpretation of the basic field equations of general relativity; there exist other commentaries. In principle, one can consider general relativity as the main physical illustration of geometric methods being applied to the field theory. In crude terms, Einstein's equations $R_{\mu\nu} - \frac{1}{2}R g_{\mu\nu} = \kappa T_{\mu\nu}, \kappa \equiv 8\pi G c^{-4}$ directly relate differential geometry (represented through the curvature) to physics (represented through energy-momentum i.e., stress-energy pseudotensor $T_{\mu\nu}$).

One can ask: what is the outcome of Einstein's equations? The answer is that Einstein's field equations have various spacetime metrics as their solutions. Since Einstein's equations are nonlinear, it is difficult to solve them exactly for an arbitrary configuration of masses in a spacetime (see the famous book [51]). Nonlinearity of Einstein's equations is manifested by the fact that the motion and the distribution of masses in the spacetime (expressed through tensor $T_{\mu\nu}$) influences the spacetime i.e., the structure of geodesics, which in its turn affects the motion and the distribution of masses. Metric $g_{\mu\nu}$ that should be produced as a solution is intended to describe the spacetime together with the gravitational (inertial) motion of masses in it.

It is interesting that one cannot automatically include gravity into mechanics, even when its relativistic version is considered, in complete analogy with electromagnetic forces. If one formally replaces electric charges by masses (interpreted as gravitational charges) and takes the "gravity vector potential" $G_i(x) \coloneqq (\varphi, 0,0,0)$ analogously to electromagnetic vector potential $A_i(x)$, where $\varphi = \varphi(x)$ is the usual potential of Newtonian gravity, this model can be mathematically consistent but leads to conflicts with observations (which was indicated already by mathematicians Hilbert and Poincaré who analyzed the motion of Mercury, see, e.g., [113], [32]). The correct results for the precession of Mercury's orbit obtained from apparently abstract theory based on the equivalence principle and spacetime metric were one of the most convincing confirmations of Einstein's general relativity. Notice that in Newtonian gravity, there is an action at a distance which is simultaneous and incompatible with special relativity, that assumes locality – the principle questioning action at a distance, with nothing in-between interacting objects that mediates the interaction.

We have already mentioned that Newtonian gravity is a classical model that is incompatible in two ways with special relativity: action at a distance and absolute simultaneity. We also noted that any sensible physical theory should include gravity into mechanics, even when its relativistic version is considered, in complete analogy with electromagnetic forces. One arrives at Newtonian gravity and gravitational corrections to Newtonian mechanics in the limit of weak gravitation field and slow massive particles moving in this field. It is remarkable that gravity in the $17^{\text{th}}$ century was universally perceived as a field i.e., people already had the notion of the field, though perhaps not mathematically detailed.

---

[112] The Laplacian $\Delta$ in curvilinear coordinates is usually known as the Beltrami-Laplace operator, for a scalar function $\varphi$ we have $\Delta\varphi = \frac{1}{\sqrt{g}}\partial_i(\sqrt{g}g^{ik}\partial_k\varphi)$. The Beltrami-Laplace operator has been extensively used lately even in engineering models, e.g., for calculating fields in bent waveguides and in nanotechnology.

It is important to emphasize once again that gravity is not regarded as an external force in the right-hand side of generalized Newton's law (S2.5.)-(S2.6.), but is merely the metric in 4d spacetime i.e., defines via the Christoffel symbols the coefficients of the system of motion equations. A particle (e.g., point mass $m \neq 0$) in the gravitation field is physically interpreted as a free particle moving over geodesic $\gamma(\tau) : x^i(\tau) \in M^4$ that lies in spacetime $M^4$ characterized by metric $g_{ik}, i, k = 0, \dots, 3$. As already commented, the geodesic is defined by Lagrangian $L = m g_{ik} \frac{dx^i}{d\tau} \frac{dx^k}{d\tau}, m \neq 0$, which ensures the transition to the usual explicit second-order Newtonian model of motion. Notice that if $m = 0$ as in the case of photon, graviton and possibly neutrino, the particle moves over the light geodesic.

All the above is related to point masses. It still remains to be seen how to correctly apply general relativity to the mechanical concept of rigid body. We, however, shall not deal with this problem, leaving those who are interested in dedicated courses of mechanics in general relativity.

Mathematically, the second-order derivative of the Newtonian model automatically leads to the appearance of additional terms when the motion is viewed in general non-Cartesian, albeit orthogonal, coordinate system $\{\xi^i\}$

$$d\mathrm{x} = \mathrm{e}_i d\xi^i, ds^2 = d\mathrm{x} d\mathrm{x} = g_{ik} dx^i dx^k, g_{ik} = g_{ik}(\mathrm{x}).$$

For instance, in the simple case of planar motion, $x(s) = (x^1(s), x^2(s))^T$, where $s$ is the natural parameter (arc length), $dx(s) = \left(dx^1(s), dx^2(s)\right)^T, (dx^1/ds)^2 + (dx^2/ds)^2 = 1$, we have for acceleration

$$\mathrm{a}(s) = \frac{d^2 s}{dt^2} \mathrm{v}_\tau(s) - \frac{1}{\rho} \left(\frac{ds}{dt}\right)^2 \mathrm{v}_n(s),$$

where $\mathrm{v}_\tau(s), \mathrm{v}_n(s)$ are the tangential and the normal unit vectors, respectively, $\rho$ is the radius of curvature $k$:

$$k = 1/\rho = d\varphi/ds = d^2 x^2/ds^2 (dx^1/ds) - d^2 x^1/ds^2 (dx^2/ds) = d^2 x^2/ds^2 (dx^1/ds)^{-1}$$
$$= -d^2 x^1/ds^2 (dx^2/ds)^{-1}, ds = \rho d\varphi, \cos \varphi = dx^1/ds, \sin \varphi = dx^2/ds$$

since the differentiation of identity $(dx^1/ds)^2 + (dx^2/ds)^2 = 1$ gives $(dx^1/ds) d^2 x^1/ds^2 + (dx^2/ds) d^2 x^2/ds^2 = 0$. In particular, in polar coordinates $\mathrm{x} = (r, \varphi)$, the acceleration is

$$\mathrm{a} = \ddot{\mathrm{x}} = (\ddot{r} - r\dot{\varphi}^2) \mathrm{v}_r + (r\ddot{\varphi} + 2\dot{r}\dot{\varphi}) \mathrm{v}_\varphi.$$

We see that acceleration a contains extra terms – not only the second time derivatives $\ddot{r}$ and $\ddot{\varphi}$, as in the Cartesian coordinates. The Christoffel symbols in polar (and cylindrical) coordinates are $\Gamma^1_{22} = -r, \Gamma^2_{12} = \Gamma^2_{21} = 1/r$, the rest are zero. The appearance of additional (inertial) terms giving rise to "fictitious forces" is a salient feature of the Newtonian model of motion distinguishing it from, e.g., the Aristotelian model. If we lived in a world where the changes of motion would be determined by the higher-order time derivatives, e.g., the third (in which case we would have to know not only the initial positions and velocities, but also the initial accelerations to find the law of motion), we would have more inertial forces. Even the geometric properties, which are in general consistent with the mechanical motion (curvature, geodesics, gravity, affine connection, covariant derivatives, etc.), in such a world would probably be completely different. An example of a classical world where higher-order derivatives would play a crucial part is the world with radiation reaction being explicitly taken into account and changing significantly the motion of charged particles.

One can observe from (S2.5.)-(S2.6.) that the motion equations written in the explicit (Newtonian) form include three types of terms, in accordance with the order of derivatives:

$$m_{ik}\ddot{q}^k + m_{ik}\Gamma^k_{jl}\dot{q}^j\dot{q}^l + \partial_i V = F_i. \qquad (S2.10.)$$

Here $F_i = g_{ik}F^k$, $m_{ik}$ is the mass matrix. The first term involving the second derivatives of generalized local coordinates with respect to time represents the essence of Newtonian theory of motion expressing the time-reversible change of state. The second term containing affine connection (the Christoffel symbols) is bilinear in $\mathrm{q} = \{\dot{q}^i\}$, with the coefficients of the bilinear form in general depending on position q. The diagonal terms i.e., those corresponding to $\Gamma^k_{jj}$ may be called generalized centrifugal forces whereas the off-diagonal ($j \neq l$) terms may be called generalized Coriolis forces. Terms of the third kind in the left-hand side (LHS) of (S2.10) are the functions of $\mathrm{q} = \{\dot{q}^i\}$ and possibly of $t$ and correspond to potential fields such as gravitational or electrostatic. The right-hand side (source) terms correspond to external non-conservative influence. Equations (q + 7) can be interpreted as the conventional Newtonian equations of motion for any choice of coordinates. The geodesic form of these equations corresponds to $V = \text{const}$ and $F_i = 0$.

We have mentioned that Newtonian mechanics perceived as a mathematical model (i.e., without absolutization) admits various extensions, some of them may be taken as modeling curiosities. One of such extensions can be to reverse the 3rd law of Newton i.e., to produce a model when one of the particles in a two-body system is attracted by the other, but the latter is repelled by the first. This unconventional situation can be encountered when modeling the behavior of some biological, ecological or social objects (primitively speaking, a predator is attracted by a prey whereas the latter is running away from the attacking raptor; see the Lotka-Volterra model below). In electrodynamics, the 3rd law of Newton (at least its instantaneous formulation) does not hold when one considers the retardation effects. This law may also fail if negative or tensor masses (accelerating against the applied force) enter the picture. The simplest one-dimensional two-body anti-Newtonian model with polynomial interaction can be written as (see [141]) $m_1\ddot{x}_1 = m_2\ddot{x}_2 = \alpha x|x|^p$, where $x = x_2 - x_1$, $m_1 > 0, m_2 > 0, \alpha > 0$ and coordinates $x_1, x_2$ denote the instantaneous positions of the particles. Of course, momentum is not conserved in such a system, as it is not invariant under spatial translations. The second law (dynamics of motion) is, nevertheless, left intact, and the model differs from the conventional one-dimensional motion only in the sign of $\ddot{x}_1$ (or $\ddot{x}_2$). One can formally replace $m_1$ by $-m_1$ to produce the usual Newtonian model on a straight line. Integrating the above system of equations by introducing parameter $v = v(x)$ (relative velocity as a function of interparticle separation $x$), $\ddot{x} = v\frac{dv}{dx}$, $\ddot{x} = \frac{\alpha}{\mu_-}x|x|^{p-1}$, where $\mu_- \equiv \frac{m_1 m_2}{m_2 - m_1}$, $m_1 \neq m_2$, we obtain $v^2(x) = v_0^2 + \frac{2\alpha}{\mu_-(p+1)}\left(x^{p+1} - x_0^{p+1}\right), p \neq -1$. In the special case $p = -1$ i.e., $\ddot{x} = \frac{\alpha}{\mu_-|x|}$, we have $v^2(x) = v_0^2 + \frac{2\alpha}{\mu_-}\ln\frac{x}{x_0}$. Now, one can consider different cases, for example $\mu_- > 0$ ($m_1 < m_2$) i.e., the prey is heavier than the predator, or $\mu_- < 0$ ($m_1 > m_2$) i.e., the attracting particle is lighter than the repelling (attacking) one. One can also consider the dissipative system $m_i\ddot{x}_i + \gamma_i\dot{x}_i = F(x), i = 1,2$; $x = x_2 - x_1$ that leads to a multiparametric variety of solutions even in a single dimension. One can further make the model more complex and realistic by considering $d$ spatial dimensions (e.g., $d = 2,3$) and a number $N$ of interacting particles, which makes the phase space of the resulting dynamical system multidimensional ($n = 2Nd$).

## S2.1. Lagrangian mechanics

The Lagrangian formalism of classical mechanics is based on a variational paradigm that determines the motion equations. Possibly this idea stemmed from Fermat's principle of optics. Soon it became clear that the Lagrangian formulation of mechanics has certain advantages over the Newtonian one due to much more convenient expounding and, as a consequence, extensive use of geometric concepts such as symmetries, group properties of the change of variables, then tangent bundles and, lately, path integral quantization.

The Lagrangian formulation of mechanics based on the principle of stationary action is valuable as a multidimensional variational problem. Such problems constitute the core of contemporary physics and are also of great importance for other disciplines where mathematical techniques of concrete calculations must be mastered, e.g., economics, optimal planning, uncertainty and sensitivity analysis, decision support, environmental and climate studies, treatment planning in medicine, etc. It is remarkable that all basic laws of physics admit a variational formulation and can be obtained from a single construction called action. The latter is a functional of the path $\gamma(x) \coloneqq x(t), t_1 \le t \le t_2$ and a function of the initial and final moments of motion, $t_1, t_2$. Here, it is implied that curve $\gamma(x)$ is represented through Cartesian coordinates $x = (x^1, \ldots, x^n) \in \mathbb{R}^n$ i.e., metric tensor $g_{ik}(x) = \delta_{ik}$.

Such a description is adequate, e.g., for a system of point masses $m_a, a = 1, \ldots, N$ (in mechanics, $N = 3n$) moving in the free space, but as soon as we have to consider constraints[113] or the motion restricted to surfaces, we must treat Lagrangian dynamics as evolving on manifolds that look as $\mathbb{R}^n$ only locally. Notice that the vector space $\mathbb{R}^n$ may be viewed as a manifold consisting of a single chart, with tangent space to $\mathbb{R}^n$, e.g., in point 0 (and then in any point) being naturally identified with $\mathbb{R}^n$. The word "locally" signifies in this context that for any arbitrary point $q$ of manifold $Q$, there exists an open neighborhood $U_q$ homeomorphic to an open set $X \subseteq \mathbb{R}^n$ so that one can introduce a local coordinate system identifying $U_q$ with $X$. Accordingly, instead of Cartesian coordinates $x^i$ on $\mathbb{R}^n$, we may introduce the "generalized coordinates" $q^j = q^j(x^i), i, j = 1, \ldots, n$; functions $q^j$ are assumed differentiable (smooth) yet in many cases it is sufficient to assume $q^j(x^i) \in C^2$ (or even $q^j(x^i, t) \in C^2$).

The statement that the laws of nature can be represented mathematically as variational principles is a generalization of experimental facts, and a rational explanation to this concept does not seem to be known. For example, rays of light somehow take the shortest way (which is compatible with the equally empirical Huygens principle in the wave theory) or a particle (more generally, a mechanical system) moves in such a manner as to make the action $S = \int L dt$ stationary ($\delta S = 0$). For a mechanical system comprised of particles, the Lagrangian function $L(q, \dot{q}, t)$ depends on the state $q \equiv \{q^1(t), \ldots, q^n(t)\}$ and its rate of change $\dot{q} \equiv \{\dot{q}^1(t), \ldots, \dot{q}^n(t)\}$, where state $q = q(t)$ belongs to a configuration space $Q$ which is often identified with some Riemannian manifold $M^n$ i.e., $q \in M^n$. In other words, the Lagrangian treated as a mathematical function (from calculus) is defined on the tangent bundle of the configuration space. Notice that commonly used notation $q$ for a state in configuration space $Q$ is an abbreviation.

---

[113] One can give a physical interpretation of the constrained motion as the one under the action of a very strong field of forces pinning the system to a prescribed curve or surface.

The derivative of the Lagrangian over generalized velocities $\dot{q} \equiv \{\dot{q}^1(t), \ldots, \dot{q}^n(t)\}$ i.e., $\partial L/\partial \dot{q} \equiv p$ is known as momentum. When the Lagrangian also depends on acceleration, $L = L(q, \dot{q}, \ddot{q})$, we shall have for the momentum $p = \partial L/\partial \dot{q} - d/dt(\partial L/\partial \ddot{q})$. Momentum can also be written as $p = \partial L/\partial \dot{v} - d/dt(\partial L/\partial a)$, where $v$ is velocity and $a$ is acceleration. In general, invariance under translations and rotations dictates that the Lagrangian $L$ becomes a function of $\dot{q}, \ddot{q}$ and possibly rotation parameter $\dot{\alpha} \equiv \omega$.

The variational principle for a mechanical system essentially maintains that the system's motion is compatible with physical reality iff $\delta S[q(t), \dot{q}(t), t] = 0$ i.e., the state $q(t)$ and its evolution rate $\dot{q}(t)$ must deliver a critical point of action $S$ (here square brackets denote a functional). Symbol $\delta$ denotes the variational operator that, after acting on action $S$ defines an associated Euler-Lagrange differential operator, leading, e.g., in classical mechanics to the Euler-Lagrange equations

$$\frac{\delta S}{\delta q(t)} = \frac{d}{dt}\frac{\partial L}{\partial \dot{q}^i} - \frac{\partial L}{\partial q^i} = 0.$$

There may be situations when variations in $q(t)$ leave the value of $S(q)$ (more generally, $S(\varphi)$, see below) intact which means that the action possesses some symmetries and, hence, the conserved quantities arise.

The Euler-Lagrange equations describe the critical (fixed) points of action $S$ with respect to variations in trajectory $q(t)$. This trajectory, often denoted as $\gamma$ is, as we know, a map $t \in [t_1, t_2] \to q = q(t)$, and the tangent vector $\dot{q}(t)$ has the squared length $v^2$ (more exactly, $\alpha v^2$, where $\alpha$ is some scaling coefficient) with respect to the Riemannian metric on $M^n$: $\alpha v^2 = ds^2/dt^2 = g_{ik}\dot{q}^i\dot{q}^k$ which is identified with kinetic energy $T$ of a mechanical system. The scaling coefficient $\alpha$ is denoted in mechanics as $m/2$, where $m$ is mass. One can also define on manifold $M^n$ some function $V(q): M^n \to \mathbb{R}$ that characterizes forces acting on a mechanical system.

The physical meaning of the Lagrangian function or Lagrangian is that it characterizes the slowing down of particles (or waves) in the process of propagation. This fact, which can be called the time-of-flight delay, becomes nearly obvious in the context of Hamilton's principle of mechanics applied to geometric optics if one writes the least action principle in the form

$$\delta S = \delta \int_a^b n\,(q^1, q^2, q^3)ds = \delta \int_{t_1}^{t_2} n\frac{ds}{dt}dt = \delta \int_{t_1}^{t_2} L dt = 0, \qquad (S2.1.1.)$$

where $L = n(\dot{q}_i\dot{q}^i)^{1/2}$, $n = n(q^1, q^2, q^3)$ is the refraction index, $ds = (dq_i dq^i)^{1/2}$, $i = 1,2,3$ is the arc length parameter (here, for simplicity, we consider the case of Euclidean metric; in the case of Riemannian metric $L = n(g_{ik}\dot{q}^i\dot{q}^k)^{1/2}$). In relativistic mechanics, where Lagrangian $L = -mc^2(1 - \beta^2)^{1/2}$, $\beta = v/c$ is proportional to the proper time interval $d\tau = (1 - \beta^2)^{1/2}dt$ which expresses the slowing down of time in the moving system, the time-of-flight delay comes out almost at the level of definition.

The action integral for a free particle on a flat (Euclidean) space

$$S = \frac{1}{2} \int\limits_{t_1}^{t_2} dt \, m \dot{x}_i \dot{x}^j = \frac{1}{2} \int\limits_{t_1}^{t_2} dt \, m (\dot{x}^i)^2$$

is in general written on a manifold as a curved space version

$$S = \frac{1}{2} \int\limits_{t_1}^{t_2} dt \, m g_{ik}(q) \dot{q}_i \dot{q}^k,$$

where coefficient $m$ can be regarded, as in Newtonian mechanics on manifolds, as a scale factor for the mass tensor $m_{ik} \equiv m g_{ik}$. One can also put it slightly differently: the metric term corresponds to the mass tensor of a body moving along a geodesic (which is just another formulation of the Maupertuis variational principle).

Lagrangian physics, i.e., a set of physical equations derived from the invariance requirement of the action functional, lies at the foundation of most modern theories, e.g., numerous field theories. The well-known empirical fact that the dynamical laws of physics can be described via variational principles (such as, e.g., Fermat's principle of light propagation or Hamilton's principle of mechanical motion) still has no rational explanations and is sometimes referred to as a "medical fact". Hamilton's principle, which is curiously more related to the Lagrangian than to the Hamiltonian formulation of mechanics has the form (S2.1.1.)

$$\delta S = \int_{t_1}^{t_2} L dt = 0$$

familiar to any person who studied analytical mechanics. Here $L : TQ \times \mathbb{R} \to \mathbb{R}$ is the Lagrangian function $L = L(q^i, \dot{q}_i, t)$ or simply Lagrangian defined on the tangent bundle $TQ$ of configuration space $Q$ ($TQ$ is sometimes also referred to as the velocity phase space). In field theories, the Lagrangian $L$ is also called the Lagrangian density.

One can write the main expression of the Lagrangian physics (i.e., action) in a general form as an integral

$$S(\varphi) = \int_M L(J^n \varphi) dV_M$$

over some manifold $M$, with $dV_M$ being its elementary volume. Here $\varphi$ is a function on $M$ (that can be, in principle, vector-valued). Symbol $J^n \varphi$ denotes elements of the space of so-called $n$-jets i.e., of the Taylor polynomials of the $n$-th degree. For instance, for a set of scalar fields $\varphi_a(x), a = 1, \dots, N_a$ defined on $x = x^0, x^1, x^2, x^3 \in M^4$ the Lagrangian density is $L(x) = L(\varphi_a(x), \partial_\mu \varphi_a(x), \dots)$, with action $S(\varphi)$ on a four-dimensional manifold being

$$S = S_4(\varphi) = \int_{M^4} d^4 z L(z) = \int_{M^4} d^4 z L(\varphi_a(z), \partial_\mu \varphi_a(z), \dots).$$

In the simplest case when the Lagrangian density contains a single scalar field i.e., $N_a = 1$ it is usually written in the form [93], §3.2

$$L = \frac{1}{2} \partial_\mu \varphi(x) \partial^\mu \varphi(x) - V(\varphi(x)),$$

where $V(\cdot)$ is a scalar function. One often plays with $V(\varphi(x)) = \frac{1}{2}(m\varphi)^2$, where $m$ is interpreted as mass of a free particle or with the so-called $\varphi^4$-Lagrangian ("quartic interaction")

$$L = \frac{1}{2} \partial_\mu \varphi(x) \partial^\mu \varphi(x) - V(\varphi(x)) - \frac{\lambda}{4!} \varphi^4, \lambda > 0,$$

which leads to nonlinear problems (such as theories with self-interaction). In the 1980s, $\varphi^4$-Lagrangians became popular in cosmological models, e.g., inflation and the like. Notice that the free-particle Lagrangian has a simple discrete symmetry $\varphi(x) \longleftrightarrow -\varphi(x)$ ($Z_2$-symmetry), whereas the $\varphi^4$-Lagrangians result in much more intricate symmetry concepts such as $SO(N_a)$ symmetries, spontaneous symmetry breaking and the famous Goldstone model. Thus, seemingly primitive considerations of Lagrangian mechanics lead to the most modern concepts of fundamental physics.

It is proved in most textbooks on classical mechanics that the equations for geodesics exactly coincide with the Euler-Lagrange equations following from the calculus of variations, which at first sight has nothing in common with differential geometry, provided the Lagrangian is of the form $L = g_{ik} \dot{q}_i \dot{q}^k$. Energy in this case is $E = T = L$. Recall that energy in mechanics (and not only in mechanics) may be defined through the Lagrangian $L$ as $E = \dot{q}^i \partial L / \partial \dot{q}^i - L = (\dot{q}^i p_k - \delta_k^i) \delta_i^k L$. Accordingly, we can write the action functional in the form

$$S = \int\limits_{t_1}^{t_2} dt \left[ p_i \dot{q}^i - H(p, q) \right],$$

where $H(p, q) \equiv H(p_i, q^k), i, k = 1, \ldots, n$ is the Hamiltonian. The first term in the above integral $S_0 = S_0[q(t)] = \int_{q_1}^{q_2} p_i dq^i$ is called the abbreviated action functional that lies at the foundation of the Maupertuis variational principle, $\delta S_0 = 0$. The latter principle differs from the Hamiltonian one $\delta S = 0$ (although the two are sometimes identified) since it requires energy conservation, $H(p, q) =$ const or $dH/dt = dE/dt = 0$, whereas the Hamiltonian variational principle leading to the Lagrange (Euler-Lagrange) equations does not. Accordingly, the abbreviated action and the Maupertuis principle allow one to determine the form of the trajectory (in coordinates $q^k$) but not the law of motion $q^k(t)$ that is obtained by solving the Lagrange equations.

One can see that $dH/dt = dE/dt = 0$ along the extremal paths (or simply extremals) i.e., along geodesics, provided the function $L$ does not depend explicitly on parameter $t$ (this is the Lagrangian statement of energy conservation). We can also verify momentum conservation if we choose variables $q^i$ as "cyclic" coordinates, e.g., $\partial L / \partial q^1 \equiv \partial_1 L = 0$. Then we have $\partial_t (\partial L / \partial \dot{q}^1) = \dot{p}_1 = 0$ along extremals (see a little below). Furthermore, since $dL/dt = 0$ together with $dL/dt = 0$, the speed of the system running along the extremals (coinciding with geodesics) is constant, which is usually a property only of the natural parameterization. One can emphasize that the Euler-Lagrange equations are very general and simple expressions whose origin is more geometrical rather than mechanical.

In classical mechanics, one usually considers the so-called *natural* systems i.e., whose Lagrangian function can be represented as the difference of kinetic and potential energies, $L := L(q^i, \dot{q}^i, t) = T - V$, where kinetic energy $T = m_{ik} \dot{q}^i \dot{q}^k$ is a quadratic form on a tangent space (manifold) $TQ$ and potential energy $V = V(q^i, t) \colon Q \times \mathbb{R} \to \mathbb{R}$ is a differentiable function. Thus, the Lagrangian is a function defined on the tangent bundle $TQ$ of configuration manifold $Q$. The concept of potential energy is associated with conservative forces encountered, for example, in electrostatic and Newtonian gravity. Recall that a vector field v (in particular, a field of forces F) has a potential function in 3d if and only if it is irrotational, curl v = 0. For the field of forces this criterion means that they have a potential (one often says "are potential") iff their work over the closed path vanishes. For example, all central fields, i.e., invariant under all the motions leaving the center intact, are potential fields (which is, strictly speaking, only true for Euclidean spaces of any dimensionality $n$). The just mentioned Newtonian gravitational and Coulomb electrical fields are central fields and therefore potential ones.

For a system of $N$ material points, the potential energy is assumed to be a function of $N$ vector variables $r_1, \ldots, r_N$ and some parameters characterizing the behavior of potential energy $V(\mathrm{r})$ with distance. If we consider the idealized model of a closed system[114] which physically means that all material points comprising the system are so far away from the rest of the world that they only interact with one another, then due to spatial homogeneity required by Galilean invariance we shall have $V(\mathrm{r}_1, \ldots, \mathrm{r}_N) = V(\mathrm{r}_i - \mathrm{r}_j), i \neq j, i, j = 1, \ldots, N$. In other words, potential energy in a closed system is a function of $N - 1$ independent variables. In many physically interesting cases, interactions between the elements of a system are independent of the presence or absence of other elements, which reflects the "do not interfere" principle. Moreover, such interactions often depend only on the distance between particles, which gives $V(\mathrm{r}_1, \ldots, \mathrm{r}_N) = V(\mathrm{r}_i - \mathrm{r}_j) = \sum_{i \neq j}^{N} V(|\mathrm{r}_i - \mathrm{r}_j|)$. One can easily prove (see, e.g., [16], §10) that if the forces between the particles depend only on interparticle distances $|\mathrm{r}_i - \mathrm{r}_j|$, then such forces have a potential. Moreover, if the system of particles evolves without external influence i.e., admits only mutual interactions, then the dynamics of such a closed system must be invariant with respect to translations of its center of mass $\mathrm{r}_c = M^{-1} \sum_{j=1}^{N} m_j \mathrm{r}_j$, where $M = \sum_{j=1}^{N} m_j$. In the Lagrangian language, the Lagrangian of such a system does not depend on $\mathrm{r}_c$. The total momentum $\mathrm{P} = M\dot{\mathrm{r}}_c = \sum_{j=1}^{N} m_j \dot{\mathrm{r}}_j$ of the system of particles is conserved and $\mathrm{V}_c \equiv \dot{\mathrm{r}}_c = const$ so that for a closed system one can always find such a frame of references, where both $\mathrm{r}_c = 0$ and $\mathrm{V}_c = 0$. Notice that Newton's third law is a direct consequence of translation invariance, provided the second law (Newtonian model of mechanical motion) is valid. Indeed, since $\dot{\mathrm{V}}_c = \ddot{\mathrm{r}}_c = M^{-1} \sum_{j=1}^{N} \dot{\mathrm{p}}_j = M^{-1} \sum_{i,j=1}^{N} \mathrm{F}_{ij} = 0$, the only possibility for a general solution would be $\mathrm{F}_{ij} = -\mathrm{F}_{ji}$ for any $\mathrm{F}_{ij}, i \neq j$.

If the motion of $N$ particles is considered in 3d Euclidean space, then the configuration space is $\mathbb{E}^{3N}$ and can be represented as a direct product of one-particle configuration spaces $\mathbb{E}^{3N} = \mathbb{E}^3 \times \ldots \times \mathbb{E}^3$ i.e., the product of $N$ copies of our 3d space (or $n$ copies of a real line). This total configuration space is also Euclidean, $\mathbb{E}^n = \mathbb{E}^{3N}$, with the appropriately defined inner product in $n$-dimensional space (making it a special case of a Hilbert space) providing the lengths of $n$-dimensional vectors, which is known as a normed space, and the distance between them (a metric space). A complete (i.e., where all Cauchy sequences converge) and normed linear space is known as a Banach space. Banach spaces are very general and play a fundamental role in both classical calculus and classical mechanics

---

[114] A closed system is an idealization already because the backward action related to the observation is completely disregarded.

whereas Hilbert spaces are indispensable for functional analysis and quantum theory. In calculus, one usually considers real Banach spaces such as, e.g., the vector space $\mathbb{R}$ (or $\mathbb{R}^n$).

Time, as usual in such cases, is considered as a parameter rather than a variable. For the sake of brevity, we shall not consider in this section Lagrangians depending explicitly on time (e.g., for the systems in external fields), relegating the treatment of such cases to the study of non-autonomous dynamical systems and certain quantum problems (when some quantum-mechanical models are briefly mentioned). Here, we shall only note that if the potential energy $V$ is time-dependent, then the motion equations are also explicitly time-dependent. For example, if a time-dependent external force $F(t) = \{F_i(t)\}$ is acting on a system (a very frequent case in mechanical, electrical and optical engineering), then we have $L(q^i, \dot{q}^i, t) = L(q^i, \dot{q}^i) + q^i F_i(t)$ so that the motion equations will be $m_{ik}\ddot{q}^k = -\partial_i V + F_i(t)$. One usually requires in classical mechanics that potential energy $V(q) \equiv V(q^1, \ldots, q^n)$ should be a differentiable function of $q^i, i = 1, \ldots, n$, but it is not necessarily required that the derivatives of $V(q)$ over $q^i$ should be continuous. Physically it means that forces may have jumps between different spatial domains. Note that in quantum mechanics the potential energy itself is often allowed to have jumps – a rare case in classical theory. This difference between classical and quantum modeling is important since it reflects the difference between classical and quantum ideologies: in classical mechanics smooth and slow motions are mostly considered whereas quantum theory is predominantly associated with quantum jumps and, in general, discontinuous processes. It has already been mentioned that the kinetic energy $T(q, \dot{q})$ is not necessarily a scalar quantity, in certain physical models it can be a tensor, $T^{ij}(q, \dot{q}) = \sum_{a=1}^{N} m_a q^i q^j$.

As already mentioned, it is generally assumed that each basic phenomenon in physics may be derived from a corresponding action principle i.e., that one can always construct a Lagrangian functional and find its stationary curves (or manifolds) for every physical process. In other words, there exists a default belief that all fundamental laws necessarily admit a global variational formulation. In particular, the famous "Course of Theoretical Physics" by L.D. Landau and E.M. Lifshitz is almost totally built on this belief, and in the conventional quantum field theory variational formulation is almost always the starting point. The Lagrangian formulation of classical mechanics [93] is probably the most demonstrative example of the usefulness of variational calculus. In the variational formulation of classical mechanics, the system's (e.g., particle) trajectories $\gamma := q^i(t)$ are extremizers (minimizers), or at least critical points, of the action integral with fixed endpoints $t_1, t_2$, $S = \int_{t_1}^{t_2} L(q, \dot{q}, t) \, dt$, where $L(q, \dot{q}, t) : \mathbb{R}^{2n+1} \to \mathbb{R}$ is the Lagrangian. If we consider Lagrangian dynamics in a vector space, the integral is taken over a real curve $\gamma$ in $\mathbb{R}^n \times \mathbb{R}$ defined by $x^i = x^i(t)$ (here $x$ denotes a point in $n$-dimensional vector space $\mathbb{R}^n$). Notice that quantity $t \in [t_1, t_2]$ again can be regarded here as any real parameter, not necessarily time, defined in such a way that values $t_1, t_2$ correspond to the fixed endpoints of the curve $\gamma$: $x^i(t), t \in [t_1, t_2]$ (for instance, $x^i(t_1) = a^i$ and $x^i(t_1) = b^i$ with vectors $\mathrm{a} = (a^1, \ldots a^n)^T$, $\mathrm{b} = (b^1, \ldots b^n)^T$ explicitly denoting the endpoints.

Recall that any system of differential equations, even of linear ones, admits a great number of solutions that are, in general, physically non-equivalent. To select which of these solutions would correspond to reality, one has to add to the equations some supplementary conditions traditionally known as initial and boundary. The procedure of imposing such supplementary conditions introduces an arbitrary element into a seemingly pure mathematical problem, but it is this arbitrariness that is a measure of reality specifying a mathematical model within the generic theory. Thus, although being usually called a functional, action $S$ along the concrete particle trajectory $\gamma$ is in fact a function depending on two spacetime points, $S = S(x_1, x_2) \equiv S(q_1, t_1; q_2, t_2) = \int_{t_1}^{t_2} L(q, \dot{q}, t) \, dt$. According to the least action principle, this integral between two fixed spacetime points assumes a stationary value so that its variational (functional) derivative $\delta S / \delta q(t) = 0$. This condition, also known as Hamilton's principle, expresses the requirement that the first-order variation must vanish for any possible perturbation $\eta(t)$ near path $\gamma(t)$ i.e., the true path followed by the system should be a maximum, a minimum or a saddle point. One can notice that when action $S$ is interpreted as a two-point function $S(x_1, x_2)$, there is no guarantee that such a function would exist for any pair $(x_1, x_2)$: for some pairs of spacetime points it might not exist or be multivalued.

One typically assumes the Lagrangian to be smooth and strictly convex in $\dot{q} \equiv v$ i.e., $\partial_{vv}L > 0$, however, this last condition may be wrong for the systems with negative mass (although mechanical models with negative mass are mostly of speculative character). The minimizing trajectories are then the solutions to the Euler-Lagrange equations

$$\frac{\delta S}{\delta q(t)} = \frac{d}{dt}\frac{\partial L}{\partial \dot{q}^i} - \frac{\partial L}{\partial q^i} = 0$$

or

$$\frac{\partial^2 L}{\partial \dot{q}^i \partial \dot{q}^j}\ddot{q}^j + \frac{\partial^2 L}{\partial \dot{q}^i \partial q^j}\dot{q}^j + \frac{\partial^2 L}{\partial t \partial \dot{q}^i} - \frac{\partial L}{\partial q^i} = 0,$$

where we have used the symbol of functional derivative[115]. Here we may also note that the Euler-Lagrange equations may be applied not only to "natural systems" i.e., to those characterized by a standard (in mechanics) form of Lagrangian function, $L = T - V$. Even in classical mechanics, there may be Lagrangians which do not have this form. For instance, we can use Lagrangian $L = (T - V)\exp(\beta t)$ to study the one-dimensional model of a damped linear oscillator

$$\dot{q}^1 = -\gamma_1 q^1 + \frac{1}{m}q^2 + F(q^1, q^2, t), \dot{q}^2 = -kq^1 - \gamma_2 q^2 + G(q^1, q^2, t), k \equiv m\omega_0^2$$

or $\begin{pmatrix} \dot{q}^1 \\ \dot{q}^2 \end{pmatrix} = A\begin{pmatrix} q^1 \\ q^2 \end{pmatrix} + \begin{pmatrix} F \\ G \end{pmatrix}$, where $A = \begin{pmatrix} -\gamma_1 & 1/m \\ -k & -\gamma_2 \end{pmatrix}$. Note that this Lagrangian is explicitly time-dependent, which leads to difficulties in the transition to quantum theory.

In most cases, the Lagrangian function $L(q, \dot{q}, t) \equiv L(q^i, \dot{q}^k, t)$ is subordinated to condition $\det J_{ik} \equiv \det\left|\partial^2 L/\partial \dot{q}^i \partial \dot{q}^k\right| \neq 0$ which allows one to bring the system of Euler-Lagrange equations to the normal form: one can resolve this system with respect to higher order derivatives so that they may be placed on the left-hand side whereas the functions of generalized coordinates $q^i$ remain on the right-hand side. If $\det J_{ik} = 0$, one cannot in general invert equations $p_i(\dot{q}^j, t) = \partial L(q^j, \dot{q}^i, t)/\partial \dot{q}^i$ as $\dot{q}^j = \dot{q}^j(p_i, t)$. In classical mechanics, this situation may occur when, e.g., the constraints are imposed on a system, $\Phi_a(q) = 0, a = 1, \dots, s$, then the rank of matrix $J_{ik}$ can be $n - s < n$, since new "coordinates" $\eta = (\eta^1, \dots, \eta^s)$ appear: $L \to \tilde{L} = L + \sum_{a=1}^{s}\Phi_a(q)\eta^a$ and the transformed Lagrangian $\tilde{L} = \tilde{L}(q, \dot{q}, \eta)$ does not depend on $\dot{\eta} = (\dot{\eta}^1, \dots, \dot{\eta}^s)$.

It is, by the way, a delusion to think that the Lagrangian systems, e.g., in physics, are necessarily non-dissipative. One can, for example, obtain the motion equations for a particle influenced by dissipative forces from Lagrangian $L = e^{\beta t/m}(T - V)$, where $\beta$ is the friction coefficient that is assumed constant, $T = \frac{1}{2}m\dot{q}^2$, $V = V(q, s)$, where $s$ denote a set of parameters, $m$ is the particle mass. Indeed, we have from this Lagrangian

$$\frac{d}{dt}\left(\frac{\partial L}{\partial \dot{q}}\right) = \frac{d}{dt}\left(e^{\beta t/m}m\dot{q}\right) = e^{\beta t/m}(m\ddot{q} + \beta\dot{q}), \frac{\partial L}{\partial q} = e^{\beta t/m}\left(-\frac{\partial V}{\partial q}\right).$$

---

[115] Notice that the term "functional derivative" seems to be more common in physics than in mathematics, where one usually talks about variational derivatives. In physics, one usually does not distinguish between these two types of derivatives; however strictly speaking, they may represent different concepts.

Thus, the motion equation for our 1d problem is $m\ddot{q} + \beta\dot{q} + \frac{\partial V}{\partial q} = 0$ i.e., the equation describing the one-dimensional particle motion under the influence of dissipative forces proportional to the particle's velocity. In the multidimensional situation, one has $T = m_{ij}q^i q^j$, $V = V(q^i, s^\mu)$, $i, j = 1, \dots, n, \mu = 1, \dots, P$ and the motion equation is $m_{ij}\ddot{q}^j + \beta_{ij}\dot{q}^j + \frac{\partial V}{\partial q^i} = 0$, where $\beta_{ij}$ is the tensor of dissipation. The $e^{\beta t/m}$ exponent should be written in this case as $\exp\{\beta_{ij}M^{ji}t\}$, where matrix $M$ is an inverse for the mass matrix $m_{ij}$. In Cartesian coordinates $m\ddot{\mathbf{r}} + \beta\dot{\mathbf{r}} + \nabla V = 0$.

A special case of the dissipative motion equation $m\ddot{\mathbf{r}} + \beta\dot{\mathbf{r}} + \nabla V = 0$ that would correspond to a free particle in a single-dimensional dissipative medium, $m\ddot{x} = -\beta\dot{x}^2$, i.e., when in equations (9.2.2.) (see section 9.2. "Damped oscillator") and (1) $b = 0, f(x, \dot{x}, t) = -\partial V/\partial x \equiv 0$ and the friction coefficient is $\beta = am$. We can write the Lagrangian for this case as

$$L = \frac{1}{2}m\dot{x}^2 e^{-2\beta x/m}.$$

Then $p = \partial L/\partial \dot{x} = m\dot{x}e^{-\frac{2\beta x}{m}}, H = \dot{x}p - L = p^2 e^{\frac{2\beta x}{m}}/2m$ and the Euler-Lagrange equations give $\frac{d}{dt}\left(m\dot{x}e^{-\frac{2\beta x}{m}}\right) - \frac{1}{2}m\dot{x}^2 e^{-2\beta x/m}\left(-\frac{2\beta}{m}\right) = 0$ or $m\ddot{x}e^{-\frac{2\beta x}{m}} + \beta\dot{x}^2 e^{-\frac{2\beta x}{m}} - 2\beta\dot{x}^2 e^{-\frac{2\beta x}{m}} = 0$ i.e., the motion equation is $m\ddot{x} = -\beta\dot{x}^2$. One can also write the equations of motion for this dissipative case in the form of a dynamical system

$$\frac{dx}{dt} = v, \qquad \frac{dv}{dt} = -\frac{\beta v^2}{m}$$

or in the Hamiltonian form (in variables $x, p$)

$$\frac{dx}{dt} = \frac{p}{m}e^{\frac{2\beta x}{m}}, \qquad \frac{dp}{dt} = -\beta\frac{p^2}{m^3}e^{\frac{2\beta x}{m}}.$$

In general, an ODE system not necessarily related to classical mechanics still can be interpreted in terms of the Euler-Lagrange equations, provided some Lagrangian function for the given ODE system can be found.

One might ask, why should one express a given system of differential equations, in particular a dynamical system, in the Euler-Lagrange form, specifically when being derived from some artificially constructed Lagrangian? The answer is hidden in comparing advantages and drawbacks: there are definitely certain advantages inherited from powerful mathematical techniques of classical mechanics such as the possibility of finding the conserved quantities in dynamical evolution that are the natural consequence of equations i.e., in the parlance of mechanics, the integrals of motion or first integrals. These constants of evolution are linked with the symmetry of equations so that while identifying the first integrals one has to study both symmetries and integrability of the physical, biological, engineering or economic system modeled by differential equations. Moreover, one builds a Hamiltonian function for this system (in particular, by using the Legendre transform, if the Lagrangian is non-degenerate i.e., its Hessian matrix $\partial^2 L/\partial \dot{q}^i \partial \dot{q}^k$ is non-singular), which allows one to construct a quantum version of the studied dynamical model.

One may note that the velocity vectors transform like position vectors, therefore the Lagrangian dynamics is comparatively simple to treat in a vector space $\mathbb{R}^n$. Not so, however, with the manifolds: the Lagrangian $L := L(q^i, \dot{q}^i, t), TQ \times \mathbb{R} \to \mathbb{R}$ depends (for fixed parameter $t$) on $2n$ independent curvilinear coordinates running over the tangent manifold $TQ$ ($Q$ is a configuration manifold). In general, the manifold $TQ$ is not a vector space or at least a vector space only locally. Velocities $\dot{q}^i$ lie along the tangential directions to all possible trajectories $\gamma(t) \in Q$ at each point $q \in Q, \gamma := q^i(t), q^i(t_0) = q_0^i$. In other words, Lagrangian variables $q^i$ belong to the coordinate manifold $Q$, and a whole vector space is attached to each point $q \in Q$, which is tangent space $T_q Q$ containing all possible velocities at this point. In case we consider Lagrangian dynamics on manifolds, $q, t$ viewed as local coordinates on a chart of a real, $n$-dimensional, $C^\infty$ differentiable (smooth) manifold $Q$, the associated coordinate functions for this chart are denoted by $q^i, i = 1, \ldots, n$ with fixed endings $q^i(t_1) = a^i, q^i(t_2) = b^i$.

The action integral of mechanics belongs to a more general class of Euler's functionals $F[\varphi] = \int_\Omega \varphi(x, y^{(\alpha)}(x))dx$, where $x = (x^1, \ldots, x^n), y(x) = (y^1(x), \ldots, y^m(x))$, symbol $\alpha$ denotes the derivatives of $y$ over $x$, e.g., $\partial^j y^k/\partial x^1 \ldots \partial x^j$. Due to its general character, the Eulerian-Lagrangian framework has many advantages and spreads well beyond classical mechanics and even beyond physics.

We have devoted some time and space to Lagrangian dynamics since it forms a base for a large domain of mathematical and computer modeling known as optimal control. The theory of optimal control was developed by the Russian mathematician L. S. Pontryagin and his collaborators in the late 1950s and has established itself as one of the major areas in engineering, biology, ecology, medicine, economics and finance. In medicine, for example, one can use optimal control to find the most efficient therapy (e.g., insulin treatment of diabetes or HIV healing) and to plan surgeries; in economics one seeks the optimal strategy (e.g., maximizing total consumption); in environment one finds optimal harvesting, etc.

It may be pertinent to make the following remark in connection with the variational approaches in physics. One might notice that the standard derivation of the laws of motion from the least action principle is based on certain assumptions. For example, why are the physical systems limited to the so-called natural ones i.e., described by the Lagrangians of the $L = T - V$ form? Moreover, why do the composite (1 + 2) systems always have Lagrangians of the $L = L_1 + L_2 - V_{12}$ type, where $L_1$ and $L_2$ are the Lagrangians of isolated individual subsystems 1 and 2 whereas $V_{12}$ corresponds to their interaction? Generally, it does not seem absolutely clear why we obtain the laws of motion by minimizing the action functional defined as an integral of Lagrangian $L$ (the action functional); there is an element of surrealistic prescriptiveness in the whole variational methodology.

### S2.1.1. Lagrangian mechanics and geometry

This section describes certain geometric aspects of Lagrangian mechanics and can be omitted at first reading. When one has sufficient grounds for worrying that the model in question (or the description of a part of the world) begins to be dependent on arbitrary choices, e.g., of a particular coordinate frame, geometrical considerations become vital. The very term "geometry" can be crudely interpreted as a multitude of points admitting certain classes of transformations[116].

In this section, natural links between Lagrangian mechanics as a multidimensional variational problem and elementary geometric properties of the space (manifold), where one observes mechanical motion, are roughly presented. Differential equation (S2.5) for the free motion $\ddot{q}^i + \Gamma^i_{jk}\dot{q}^j\dot{q}^k = 0$ or in coordinate-free representation $\nabla_v(v) = 0$, where $v = \{v^i\} = \{dq^i/dt\}$ is the velocity vector over curve $\gamma := \{q^i(t)\}$ defines the geodesic lines. In standard physical situations, e.g., those involving gravity or torsion-free accelerations, connection coefficients (the Christoffel symbols) $\Gamma^j_{ik}$ are symmetric with respect to lower indices and can be expressed through the metric tensor (formula (S2.9)). Hence the geodesic lines are uniquely determined by the metric, and in case the metric is related to mechanical motion (e.g., as we have seen above, the metric tensor is proportional to mass tensor, $g_{ik} = m_{ik}/m$), geometry is uniquely linked with mechanics. The properties of geodesics and of affine connections are thoroughly described in any contemporary course of geometry, therefore we shall not dwell on them here. It is, however, important to recall, in connection with variational problems, that geodesics are locally the shortest trajectories between two points on a manifold: the length of a geodesic does not exceed the length of any other curve connecting these points. The main geometric feature of a geodesic is that the parallel transport of a velocity (tangent) vector along this curve also produces a velocity vector.

A complete geometrization of the Euler-Lagrange techniques, e.g., using manifolds instead of vector fields can be mathematically nontrivial and cumbersome, therefore we shall evade refinements and generalities. Now, let $L(q,\dot{q}) \equiv L(q,v)$ be some function $TQ \to \mathbb{R}$ of variables $q = \{q^i\}$ and $v = \{v^i\}$. Take a pair of points $A = (q_1^i)$ and $B = (q_2^i)$ and all smooth curves $\gamma(t)$: $q^i(t), a \leq t \leq b, q^i(a) = q_1^i, \ q^i(b) = q_2^i$. Take the quantity $S[\gamma] := \int_a^b L(q(t),\dot{q}(t))dt$. On what curve $\gamma$ is the quantity $S[\gamma] = \min$? It can be easily proved that the variational derivative $\delta S[\gamma]/\delta q^i = \partial L/\partial q^i - d/dt(\partial L/\partial \dot{q}^i)$ along curve $\gamma$: $q^i = q^i(t)$ on which $S[\gamma]$ between a pair of points reaches minimal or, in general, stationary values (i.e., $\frac{d}{d\varepsilon}S[\gamma + \varepsilon\eta]_{\varepsilon \to 0} = 0$ for any smooth vector-function $\eta(t)$, $\eta(a) = \eta(b)$) should vanish, $\delta S[\gamma]/\delta q^i = 0$. Notice that one can define the variational derivative as $\frac{d}{d\varepsilon}S[\gamma + \varepsilon\eta]_{\varepsilon \to 0} = \int_a^b dt\eta^i \ \delta S/\delta q^i$. For so-called natural systems $L(q,\dot{q}) \equiv L(q,v) = \frac{m}{2}\sum_i(v^i)^2 - V(q)$, the geometric Euler-Lagrange equations for extremals $\delta S[\gamma]/\delta q^i = \partial L/\partial q^i - d/dt(\partial L/\partial \dot{q}^i) = 0$ give Newtonian equations for particle motion $m\ddot{a}^i = -\partial V/\partial q^i$ or $dp_i/dt = F_i$.

In the already considered example of the free motion, $L(q,v) = g_{ik}v^iv^k$, $S[\gamma] = \int_a^b g_{ik}\dot{q}^i\dot{q}^k dt$, and if the metric is Euclidean, then $g_{ik} = \frac{1}{2}m\delta_{ik}$ (it is more convenient to write $L(q,v) = \frac{1}{2}g_{ik}(q)v^iv^k$; in general $g_{ik} = g_{ik}(q)$) and we have $p_i = \partial L(q,v)/\partial v^i = g_{ik}v^k, F_i = \partial L(q,v)/\partial q^i = \frac{1}{2}\frac{\partial g_{jk}}{\partial q^i}v^jv^k$. The equations for extremals are $\frac{dp_i}{dt} = \frac{1}{2}\frac{\partial g_{ik}}{\partial q^k}v^iv^k$ or $g_{ik}\dot{v}^k + \frac{\partial g_{ik}}{\partial q^j}v^jv^k = \frac{1}{2}\frac{\partial g_{jk}}{\partial q^i}v^jv^k$ or

$$g_{ik}\ddot{q}^k + \frac{\partial g_{ik}}{\partial q^j}\dot{q}^j\dot{q}^k = \frac{1}{2}\frac{\partial g_{jk}}{\partial q^i}\dot{q}^j\dot{q}^k.$$

If we multiply this equation from the left by $g^{il} = g^{li}$, we get

$$\delta_k^l\ddot{q}^k + g^{li}\frac{\partial g_{ik}}{\partial q^j}\dot{q}^j\dot{q}^k = \frac{1}{2}g^{li}\frac{\partial g_{jk}}{\partial q^i}\dot{q}^j\dot{q}^k$$

or

$$\ddot{q}^l + g^{li}\left(\frac{\partial g_{ik}}{\partial q^j} - \frac{1}{2}\frac{\partial g_{jk}}{\partial q^i}\right)\dot{q}^j\dot{q}^k = 0.$$

Now, if we change the indices $j \rightleftarrows k$ in the term $g^{il}\dot{q}^j\dot{q}^k\frac{\partial g_{ik}}{\partial q^j} = \frac{1}{2}g^{il}\dot{q}^j\dot{q}^k\frac{\partial g_{ij}}{\partial q^k} + \frac{1}{2}g^{il}\dot{q}^j\dot{q}^k\frac{\partial g_{ik}}{\partial q^j}$, we obtain

$$\ddot{q}^l + \frac{1}{2}g^{jl}\dot{q}^i\dot{q}^k\left(\frac{\partial g_{ij}}{\partial q^k} + \frac{\partial g_{jk}}{\partial q^i} - \frac{\partial g_{ik}}{\partial q^j}\right) = 0 \qquad (S2.1.1.1.)$$

(we changed dummy index $j$ to $i$) or, denoting

$$\Gamma_{ik}^l \equiv \left(\frac{\partial g_{ij}}{\partial q^k} + \frac{\partial g_{jk}}{\partial q^i} - \frac{\partial g_{ik}}{\partial q^j}\right),$$

we get the equation for geodesic lines

$$\ddot{q}^l + \frac{1}{2}\Gamma_{ik}^l\dot{q}^i\dot{q}^k = 0.$$

This is the geodesic equation. Recall that geodesics are smooth curves $\gamma\colon [a, b] \to Q$ of minimal length, $|\gamma| \coloneqq \int_a^b \left(g_{ik}\dot{q}^i\dot{q}^k\right)^{1/2}dt$. From (S2.1.1.1.) we see that a particle in an arbitrary coordinate (reference) system – in general non-inertial – moves along a geodesic. We have also established this result within the framework of the Newtonian paradigm (formula (S2.1.1.1.)). We shall see in the next section that the same result follows from the Hamiltonian paradigm, which geometrically testifies in favor of equivalence of the description of mechanical motion in three principal versions of classical mechanics: Newtonian, Lagrangian and Hamiltonian.

One might note that curve $|\gamma_{ab}|_g = \int_a^b |v(t)|_g dt = \int_a^b \left(g_{ik}\dot{q}^i\dot{q}^k\right)^{1/2}dt$ can be reparametrized without changing its length – such a phenomenon is known in physics as degeneracy. The latter, however, can be fixed if one replaces the length functional with the "energy" on the curve $E(\gamma) \coloneqq \frac{1}{2}\int_a^b (v_k v^k)_{g(q)}dt = \frac{1}{2}\int_a^b g_{ik}\dot{q}^i\dot{q}^k dt$.

Thus, the Euler-Lagrange equations for a free particle with Lagrangian $L(q, v) = \frac{1}{2}g_{ik}v^iv^k = \frac{1}{2}m_{ik}\dot{q}^i\dot{q}^k$ tell us that the particle moves along geodesics. In general, any Lagrangian can be viewed as a geometric object. The free-particle energy $E = L = \frac{1}{2}m_{ik}\dot{q}^i\dot{q}^k = \frac{1}{2}m|\dot{q}|^2$ is preserved along geodesics. A particle in potential $V(q)$ i.e., with Lagrangian $L(q, \dot{q}) = \frac{1}{2} m_{ik} \dot{q}^i \dot{q}^k - V(q)$ moves under the influence of an external force $F_i(q) = -\partial V(q)/\partial q^i$ that tends to displace the particle from a given geodesic to another one.

We know in general that the vector field $\mathbf{v} = \{v^i\}$ determines a local one-parameter group of transformations $g_\tau(x) \colon v_\tau(x) = {}^d\!/_{d\tau}\, g_\tau(x)|_{\tau=0}$ i.e., $\mathbf{v} = v^i \partial/\partial x^i$ with $\{v^i\} = {}^d\!/_{d\tau}\, g_\tau x|_{\tau=0}$; this vector field is tangent to the trajectories of $g_\tau(x)$. As already noted, each transformation $g_t(x), g_{t_1+t_2} = g_{t_1} \cdot g_{t_2}, g_{-t} = g_t^{-1}$ is a diffeomorphism of manifold $M$ onto itself and the set of transformations $g_t, t \in \mathbb{R}$ is an Abelian group of diffeomorphisms of $M$ onto $M$. According to the Cauchy-Peano (Picard-Lindelöf) existence and uniqueness theorem[117] for any point $x \in M = M^n$, where $\mathbf{v}(x) \neq 0$ (non-singular point), one can find a domain $D$ and $I \subset \mathbb{R}$, where evolution $g_t(x), x \in D, t \in I, g_{t_1+t_2} = g_{t_1} \cdot g_{t_2}, g_{-t} = g_t^{-1}$ exists and is smooth for $|t| \leq I$. In other words, evolution transformations can be correctly defined everywhere in $D \times I$. The proof of the existence and uniqueness theorem is rather lengthy (although not especially laborious). One can find this proof in practically any textbook on ordinary differential equations, and it need not be reproduced here.

One can introduce new coordinates $\{z^i\}$ on $D$ so that $g_\tau x = \{x^1 + v^1(x)\tau, x^2, \dots, x^n\} = \{z^1 + \tau, z^2, \dots, z^n\}$ (which is merely a scaled form of $x = g_\tau x_0 = \{x_0^1 + v_0^1\tau, x^2, \dots x^n\}$), then the time derivative $dL/d\tau$ of Lagrangian $L(x, v) = L(x, \dot{x})$ characterizing its change can be written in coordinates $\{z^i\}$. If evolution $g_t$ preserves Lagrangian $L(x, v)$ i.e., $dL(g_t(x), g_{t*}(x))/d\tau = 0$ i.e., the Lagrangian is invariant under the action of the one-parameter group $g_t(x)$, then we get a number of conservation laws. Here, symbol $g_{t*}(x)$ designates a mapping between tangent spaces. In the geometric language, one says that the Lie derivative along vector field $\mathbf{v} = \{v^i\}$ vanishes:

$$\left. \frac{dL(x,y)}{d\tau} \right|_{\tau=0} = v^i \frac{\partial L}{\partial x^i} + \frac{\partial L}{\partial y^i} y^j \frac{\partial v^i}{\partial x^j} = 0. \qquad (S2.1.1.2.)$$

This condition is represented in the form of the vanishing infinitesimal generator of function $L(x, y)$ along curve $\gamma(t)$ on manifold $M$ such that $d\gamma(t)/dt = \mathbf{v}(\gamma(t))$. One can interpret equation (aa) as the fact that shifts along vector field $\mathbf{v}(x)$ preserve the Lagrangian $L(x, y)$, where temporary variable $y$ was introduced for tangent vector $\mathbf{v}$ for the sake of convenience – to avoid confusion and abuse of notation. We must emphasize that one has to calculate the Lie derivative of a function $L(x, y)$ of $2n$ variables. In physically habitual terms, we can write for the Lie derivative along vector field $\mathbf{v}(x)$ in the form $v^i \partial_i L + v^j \partial_j v^i \partial L/\partial v^i = 0$ ($\partial_i \equiv \partial/\partial x^i$). Notice that if the Lagrangian is invariant under the one-parameter evolution group $g_t(x)$, then the components of generalized momenta along vector field $\mathbf{v}(x)$ (associated with group $g_t(x)$) are conserved. This statement generalizes the usual intuitive representation of relating the system's invariance under certain operations to respective conservation laws [93], §9.

As an illustration, we can produce the conservation of the momentum component along the $x^1$ axis. The infinitesimal generator $dL/d\tau|_{\tau=0} = 0$ can be written through the above introduced coordinates $\{z^i\}$ on domain $D$ near non-singular $x \in M$ ($\mathbf{v}(x) \neq 0$) as

$$\left.\frac{dL}{d\tau}\right|_{\tau=0} = \frac{\partial L(x,\dot{x})}{\partial x^1}v_0^1 = 0$$

or using the Lagrange equations

$$v_0^1 \frac{d}{dt}\frac{\partial L(x,\dot{x})}{\partial \dot{x}^1} = 0$$

or $\frac{d}{dt}\left(v_0^1 \frac{\partial L(x,\dot{x})}{\partial \dot{x}^1}\right) = 0$. In general, $v\frac{dp}{dt} = 0$ i.e., $v^i \dot{p}_i = 0$ or $\frac{d}{dt}(v^i p_i) = 0$, where $v^i$ is the vector field of a dynamic system, $\{v^i\} = \frac{d}{d\tau}g_\tau x|_{\tau=0}$, $v = v^i \partial/\partial x^i$ which should not be confused here with the particle velocity.

---

[117] Sometimes also known as the Cauchy-Lipshitz theorem.

We have already mentioned that in distinction to the Newtonian model which limits mechanical evolution to the knowledge of initial positions and velocities allowing one to find the law of motion, one can imagine a world where also accelerations or even higher-order time derivatives would be needed to produce the integral curve and trajectory. In other words, a more detailed knowledge of initial variations is required for prediction, and one should use higher order differential equations, $\Phi(q,\dot{q},\ddot{q},\dddot{q},\dots q^{(m)},t) = 0$ to compute mechanical motion. Being translated into the variational Lagrangian framework, one has to consider higher-order derivatives in the Lagrangian and action $S[\gamma] = \int_\gamma L(q,\dot{q},\ddot{q},\dddot{q},\dots q^{(m)},t)dt$. The Euler-Lagrange equation for extremals is

$$\frac{\delta S[\gamma]}{\delta q} = \frac{\partial L}{\partial t} + \sum_{k=1}^{m}(-1)^k \frac{d^k}{dt^k}\frac{\partial L}{\partial x^{(k)}} = 0$$

so that if, e.g., the Lagrangian depends on acceleration, the momentum will be not simply $p = \partial L/\partial \dot{q} \equiv \partial L/\partial v$ but $p = \partial L/\partial \dot{q} - d/dt(\partial L/\partial \ddot{q}) \equiv \partial L/\partial v - d/dt(\partial L/\partial a)$ and accordingly the Euler-Lagrange equation for extremals acquires an additional term. The angular momentum $M := r \times p = r \times \partial L/\partial v - r \times d/dt(\partial L/\partial a) = r \times \partial L/\partial v - d/dt(r \times \partial L/\partial a) + dr/dt \times \partial L/\partial a = r \times p + v \times K + S$, where $K := \partial L/\partial a = mr$ is proportional, in the case of a collection of particles, to the center of mass position and $S := -d/dt(r \times \partial L/\partial a) = -d/dt(r \times mr)$.

## S2.2. Hamiltonian mechanics

Hamiltonian mechanics is a vast area, specifically when viewed from the perspective of modern geometry. However, in this section, only the Hamiltonian basics is given. Just as Lagrangian mechanics, Hamiltonian mechanics is based on variational principles, $\delta S = 0$, where quantity $S = \int_{t_1}^{t_2}[p_i\dot{q}^i - H(p,q,t)]$ is the action corresponding to Hamiltonian $H(p,q,t) \equiv H(p_i,q^j,t), i,j = 1,\dots,n = 3N$ ($N$ is the number of particles, $n$ is the number of degrees of freedom); abbreviations $p,q$ denote the collection of all $p_i, q^j$). Hamiltonian $H$ of a mechanical system is usually defined as a real function on phase space $T^*Q$ which is understood as a cotangent bundle of configuration space $Q$ ($T^*Q \to \mathbb{R}$) (in particular, the event space $\mathbb{R} \times Q$ of a mechanical system).

One can sometimes use complex Hamiltonians which are not self-adjoint so that the time evolution operator $U = \exp\left(-\frac{i}{\hbar}Ht\right)$ would no longer be unitary (the quantum-mechanical probability is no longer preserved). Such non-unitary models can be applied, in particular, to decaying states. In old-fashioned nuclear physics the model using complex Hamiltonians has been known as the optical model.

The cotangent bundle $T^*Q$ i.e., the phase space of a mechanical system is endowed in Hamiltonian mechanics with a symplectic form $dp_i \wedge dq^i$. The trajectories of a Hamiltonian system are integral curves of the Hamiltonian vector field (symbolically written as $u = u_i \partial^i + u^i \partial_i$) on $T^*Q$ that satisfy Hamiltonian equations $u^i = \partial^i H, u_i = -\partial_i H$ (see also below). One can notice that there is no notion of force in Hamiltonian mechanics.

Dynamic systems explored through the Hamiltonian approach are collectively known as Hamiltonian systems. Notice that variables $p_i$ and $q^j$ in the Hamiltonian formulation are assumed independent and are thus independently varied, $p_i(t) \rightarrow p_i(t) + \delta p_i(t), q^j(t) \rightarrow q^j(t) + \delta q^j(t)$ with $\delta \dot{q}^j = \frac{d}{dt} \delta q^j$. Under these assumptions, the Euler-Lagrange variational procedure for $\delta S = 0$ gives the Hamiltonian equations of motion. One can find these basic facts, almost banalities, in any contemporary textbook on mechanics. A less trivial fact is that in Hamiltonian mechanics any quantity $f$ that can be potentially interpreted as a dynamical variable is treated as a function of canonical variables.

There have been many discussions on what version of mechanics is more advanced and more convenient for applications: Lagrangian or Hamiltonian. Newtonian mechanics is tacitly regarded as an outdated and too elementary formulation. There exists a widespread view that the Lagrangian approach is more general than the Hamiltonian one. This belief is usually supported by relativistic arguments claiming that the Hamiltonian formalism is less convenient when Lorentz invariance should be explicitly ensured. Some people argue that Lagrangian mechanics is the most fundamental since it involves variational principles, and the latter are important everywhere, even in general relativity, which is a theory of a quite distinct type. In the classical book by V. I. Arnold [16], the opposite view is expressed: Lagrangian mechanics is included in Hamiltonian mechanics as a special case (beginning of Part 3). Others say that Hamiltonian mechanics is the most basic theory since it leads directly to quantum mechanics and uses the Hamiltonian function (energy) as the crucial quantity. Although it is difficult to rigorously establish the equivalence of Lagrangian and Hamiltonian versions since they express different concepts (variational calculus vs. symplectic geometry) and, accordingly, rely on different mathematical objects, one may observe that there also exists a symplectic formulation of Lagrangian mechanics making it more feasible to "compare" which version of mechanics is more fundamental. Each formalism possesses its own key features: in Lagrangian mechanics it is the action functional whereas for Hamiltonian version it is phase space and canonical transformations. Moreover, these two formulations of mechanics have different sets of observables: coordinates and velocities in Lagrangian mechanics and coordinates and momenta in Hamiltonian mechanics.

We have already mentioned that the Hamiltonian $H(p, q, t)$ is the Legendre-transformed Lagrangian, $L(q, \dot{q}, t) = L(q^i, \dot{q}^i, t), i = 1, \ldots, n = 3N$ i.e., $H = p_i \dot{q}^i - L(q^i, \dot{q}^i, t)$, where $\dot{q}^i = \dot{q}^i(p_j, t)$. In other words, generalized velocities must be expressed through generalized momenta. This resolution of velocities in terms of momenta $p_i = \partial L / \partial \dot{q}^i$ can be performed if $I := \det m_{ik} \neq 0$, where $m_{ik} = \partial L / \partial \dot{q}^i \partial \dot{q}^k$ (one usually requires $I > 0$). Equivalently, one can say that the Legendre transform enabling one to go back and forth between Lagrangian and Hamiltonian formulations of mechanics does not always exist: one must require the mass matrix $m_{ik} = \partial L / \partial \dot{q}^i \partial \dot{q}^k$ to be nonsingular (an invertible matrix) to transit to the Hamiltonian version from the Lagrangian one and the dispersion matrix $w^{ik} = \partial H / \partial p_i \partial p_k$ to be nonsingular to obtain the Lagrangian version from the Hamiltonian. If both $m_{ik}$ and $w^{ik}$ are nonsingular, the Hamiltonian and the Lagrangian approaches are completely equivalent.

It is not difficult to show that the Legendre transform, at least of functions on $\mathbb{R}^n$, reduces to identity when being repeated: this property of transformations is known as involutive (one often says "the transformations are in involution")[118]. We can note that the transition between the two approaches – Hamiltonian and Lagrangian – (given by the Legendre transform) is probably the most prominent example of natural dualities quite often encountered in physics. More specifically, the Legendre transform is an example of the so-called musical isomorphism (# and ♭) between the tangent bundle $TQ$, where $Q$ is the configuration manifold, and cotangent bundle $T^*Q$ i.e., in simple terms, switching between velocities (variables of Lagrangian mechanics) and momenta (variables of Hamiltonian mechanics). We shall not discuss this interesting subject in any detail referring the reader to a comprehensive textbook [16] and [47].

Because of the keen attention to the Hamiltonian systems in classical and quantum mechanics, one tends to think that there are predominantly Hamiltonian systems around us. This is not true. There are, roughly speaking, many more non-Hamiltonian systems in the world than Hamiltonian. The latter represents a very special case when the motion equations can be produced by differentiation (taking a symplectic gradient) of a scalar function. In simple words, it means that the motion equations have a first integral, the scalar function determining the system's energy. The existence of such a function is a rather strong requirement: the absolute majority of dynamical systems do not possess the Hamiltonian function i.e., their evolution, cannot be conveniently derived by simple differentiation of a scalar quantity. Accordingly, there is in general no energy manifold for an arbitrary dynamical system.

The most adequate language of Hamiltonian mechanics is currently that of smooth manifolds, vector fields, differential forms and symplectic geometry. The basic expressions in this language may be interpreted as grammar rules for mathematical models using the principles of the Hamiltonian approach, and such principles are applied far beyond mechanics or even physics. We have already mentioned Hamiltonian systems in the context of dynamical systems and the corresponding vector fields. In the Hamiltonian version, dynamics in classical mechanics is generated by the Hamiltonian flow i.e., a one-parameter group of diffeomorphisms $g_t \colon P \to P, g_{t+s} = g_t \circ g_s, g_{t=0} = 1$ preserving the symplectic structure of the phase space $P$. It would probably be pertinent to remember here once again that a configuration space of a mechanical system is manifold $Q$ whereas the phase space of the same system is the cotangent bundle $T^*Q$ of $Q$, where a symplectic form is provided. Any Hamiltonian system is of this type, at least locally.

Let $H = H(p_i, q^k, t) \colon T^*Q \times \mathbb{R} \to \mathbb{R}$ be a continuous differentiable function over all its variables (one often requires $H \in C^m, m \geq 2$, i.e., the Hamiltonian to be a smooth scalar function of phase variables (p, q) and parameter $t$). The system of $2n$ equations $\dot{p}_i = -\partial H/\partial q^i, \dot{q}^i = \partial H/\partial p_i$ is known as the system of Hamiltonian equations (also called canonical equations), and the function $H$ is called the Hamiltonian function or simply the Hamiltonian. The Hamiltonian system of equations serves as a mathematical model in many fields of physics and also outside physics (e.g., in economics).

One can write the Hamiltonian flow $g_t(H) \colon P \to P$ generated by the Hamiltonian vector field $X_H$ on the cotangent bundle of configuration space $Q$ i.e., $X_H \subset T^*Q$ (the latter is usually identified with phase space $P$) through the Poisson brackets:

---

[118] Elementary examples of involution i.e. of maps such as $f(f(x)) = x$ or $A(Ax) = x$ are reflections in a mirror and complex conjugation.

$$\{H,\cdot\} \equiv \frac{\partial H}{\partial p_i}\frac{\partial}{\partial q^i} - \frac{\partial H}{\partial q^i}\frac{\partial}{\partial p_i}.$$

This expression, which is actually the Lie bracket of Hamiltonian vector fields, means that the Hamiltonian flow belongs to the tangent bundle of the phase space (more exactly is a section of it), i.e., to the tangent bundle of the cotangent bundle of the configuration manifold $Q$, $g_t(H) \subset T(T^*Q)$.

In classical mechanics, for any smooth function $f$ of $q^i, p_i, i = 1, \dots, n \in P$ i.e., defined on symplectic manifold $P$ and, for non-isolated systems, also of time $t$ the time derivative of $f$ is

$$\dot{f} \equiv \frac{df}{dt} = \frac{\partial f}{\partial t} + \frac{\partial f}{\partial q^i}\dot{q}^i + \frac{\partial f}{\partial p_j}\dot{p}_j = \frac{\partial f}{\partial t} + \{f, H\},$$

where the Hamiltonian equations were used and symbol $\{.,.\}$ denotes the classical Poisson bracket $\{A, B\} = \frac{\partial A}{\partial p_i}\frac{\partial B}{\partial q^i} - \frac{\partial A}{\partial q^i}\frac{\partial B}{\partial p_i}, A = A(p, q, t), B = B(p, q, t)$. This is the Hamiltonian flow (the flow of a Hamiltonian vector field) $X_H f = \{f, H\}$. In canonical coordinates, $(q^i, p_j)$ such that $\{p_i, p_j\} = \{q^i, q^j\} = 0, \{q^i, p_j\} = \delta^i_j$ the Hamilton equations are a special case of the Hamiltonian flow. Notice that the Hamiltonian flow preserves $H$, $X_H f = \{H, H\} = 0$, which is the mathematical formulation of the conservation of mechanical energy. Analogous conservation laws exist for other quantities preserved under the Hamiltonian flow, $X_H f = \{f, H\} = 0$.

It is obvious that the standard symplectic structure of Hamiltonian mechanics can be extended by admitting non-vanishing Poisson brackets between all relevant quantities. A rather straightforward generalization of the Poisson brackets could consist in adding nonzero commutators between coordinates which would generalize the symplectic structure of Hamiltonian mechanics but can result in changes of the Lagrangian formulation. The primitively generalized Poisson brackets in Hamiltonian mechanics can look as

$$\{q^i, q^j\} = \alpha^{ij}, \qquad \{q^i, p_j\} = \delta^i_j, \qquad \{p_i, p_j\} = \beta_{ij},$$

where quantities $\alpha^{ij}$, $\beta_{ij}$, $\delta^i_j$ can, in principle, depend on $p_i$ and $q^j$. Recall that $q^j$ and $p_i$ are usually understood as holonomic coordinates on the tangent bundle $T^*Q$. We can denote the coordinates in the phase manifold[119] as $z^k \coloneqq (q^i, p_j), i, j = 1, \dots, n, k = 1, \dots, 2n$. Here all indices of local symplectic coordinates are put up since the notational distinction between the elements of vector ($TQ$) and covector or linear functional ($T^*Q$) spaces seems to be unimportant in the given context. Besides, there is no risk of confusion. Then we get $\{z^i, z^j\} = \eta^{ij}$, where $\eta^{ij}$ are the elements of a $2n$ symplectic matrix

$$\eta = \begin{pmatrix} \alpha_n & I_n \\ -I_n & \beta_n \end{pmatrix},$$

---

[119] At the quantum level, one can replace the Poisson brackets by commutators: $[q^i, q^j] = i\theta^{ij}$, $\{q^i, p_j\} = -i\delta^i_j$, $\{p_i, p_j\} = i\eta_{ij}, i^2 = -1$.

where $I_n$ is a unit matrix, $\alpha_n$ and $\beta_n$ are skew-symmetric matrices $\alpha_n = \begin{pmatrix} 0 & \cdots & \alpha^{1n} \\ \vdots & \ddots & \vdots \\ -\alpha^{1n} & \cdots & 0 \end{pmatrix}, \beta_n = \begin{pmatrix} 0 & \cdots & \beta_{1n} \\ \vdots & \ddots & \vdots \\ -\beta_{1n} & \cdots & 0 \end{pmatrix}$, where due to symplecticity $\alpha^{ij} = -\alpha^{ji}, \beta_{ij} = -\beta_{ji}, \alpha^{jj} = \beta_{jj} = 0, i, j = 1, \ldots, n$. For instance, in a rather special model of 2+1 dimensions (3), $\eta = \begin{pmatrix} 0 & \alpha & & \\ -\alpha & 0 & I_2 & \\ & -I_2 & 0 & \beta \\ & & -\beta & 0 \end{pmatrix}$.

If we, for instance, allow the Poisson bracket to be non-zero in the position space, then this non-commutativity, e.g., for single-particle variables, would result in nonlocalities and one can expect the Hamiltonian and the Lagrangian to depend on the particle acceleration and, possibly, on higher order time-derivatives. In other words, the emergence of higher-order time derivatives in the Hamiltonian or the Lagrangian (and hence in Newton's law) implies some non-locality in time as a penalty for admitting noncommutativity of coordinates. More exactly, Hamiltonians and Lagrangians that depend on a finite number of time derivatives taken at a current time point $t$ i.e., $q(t), \dot{q}(t), \ldots, q^{(n)}(t)$ would turn into genuinely non-local Hamiltonians or Lagrangians for $n \to \infty$. In this case a Hamiltonian or a Lagrangian would depend on all points of trajectory $q(t - \tau)$ i.e., for all values of $\tau \in \mathbb{R}$. A generalization of Hamiltonian theories with higher-order time derivatives leads to higher-derivative models in field and gravitation theories.

One more generalization of the Hamiltonian theory consists in the following modification of phase space variables: $q^i \to \varphi^{ij}(x), \ p_i \to \pi_{ij}(x)$, and the Poisson brackets can be given as $\{\varphi^{ij}(x), \pi_{kl}(y)\} = \frac{1}{2}\delta(x - y)\big(\delta_k^i \delta_l^j + \delta_l^i \delta_k^j\big)$, where $x, y$ are the spatial (or spacetime) points. In many cases, quantities $\varphi^{ij}(x)$ can be interpreted as the spatial (or spacetime) metric, $\varphi^{ij}(x) = g^{ij}(x) = g^{im} g^{jn} g_{mn}(x)$ (recall that indices are raised and lowered with the help of metric tensor $g_{ik}$ and its inverse $g^{ik}$).

One can naturally produce Hamiltonian equations, given the Lagrangian formulation of mechanics. To put it simply, one can write $n$ second-order Euler-Lagrange differential equations on the configuration manifold $\mathrm{q} = (q^1, \ldots q^n) \in Q^n \equiv Q$ as a dynamical system with $2n$-dimensional phase space $P$ on the cotangent bundle to $Q$: $T^*Q$, $\mathrm{x} = (p_1, \ldots p_n, q^1, \ldots q^n) \in P$. We have mentioned that one usually identifies the phase space with the cotangent bundle $T^*Q$, but strictly speaking they need not coincide. However, if manifold $Q$ may be interpreted as representing all possible positions of a dynamical (in particular, Hamiltonian) system, then the cotangent bundle $T^*Q$ can be regarded as the set of all possible positions and momenta i.e., the phase space. In the procedure of establishing two mutually inverse vector bundle isomorphisms, $TQ \leftrightarrow T^*Q$, one gets the above system of $2n$ first-order evolution equations, where the Hamiltonian $H = H(\mathrm{p}, \mathrm{q}, t): T^*Q \times \mathbb{R} \to \mathbb{R}$ is obtained from the Lagrangian $L(\mathrm{q}, \dot{\mathrm{q}}, t)$ defined on the tangent bundle $L(\mathrm{q}, \dot{\mathrm{q}}, t): TQ \times \mathbb{R} \to \mathbb{R}$ (or, in the autonomous case, $L(\mathrm{q}, \dot{\mathrm{q}}): TQ \to \mathbb{R}$) through the Legendre transform $H(\mathrm{p}, \mathrm{q}, t) = \mathrm{p}\dot{\mathrm{q}}(\mathrm{q}, \mathrm{p}, t) - L(\mathrm{q}, \dot{\mathrm{q}}(\mathrm{q}, \mathrm{p}, t), t)$. Here $p_j(\mathrm{q}, \dot{\mathrm{q}}, t) = \partial L(\mathrm{q}, \dot{\mathrm{q}}, t)/\partial \dot{q}^j$ is *by definition* the j-th component of canonical momentum. Notice that $L$ and $\mathrm{p}$ are defined as functions of $\dot{\mathrm{q}}$ so that to transit to the Hamiltonian picture in the phase space we ought to invert $\dot{\mathrm{q}}(\mathrm{q}, \mathrm{p}, t)$ with respect to $\mathrm{p}$ after the Legendre transform which is possible if (and only if) the map $p_j = \partial L/\partial \dot{q}^j: TQ \to T^*Q$ is invertible. Mathematically, kinetic energy in Hamiltonian mechanics i.e., written via momenta $p_j$ on cotangent bundle $T^*Q$ is an inverse inner product $(\cdot, \cdot)_{g^{-1}(q)}: T_q^*Q \times T_q^*Q \to \mathbb{R}$.

When considering autonomous systems, one can suppress (at least temporarily) the dependence on parameter $t$. One can easily prove that function $H = H(\mathrm{p}, \mathrm{q})$ is the first integral of the system of the Hamiltonian. Indeed, the vector field $\mathrm{v} = (\dot{p}_i, \dot{q}^k)$ corresponding to a $2n$-dimensional dynamical system $\dot{\mathrm{x}} = \mathrm{v}(\mathrm{x})$ is described by the Hamiltonian equations, and the Lie derivative along this vector field is

$$L_{\mathrm{v}} H = \frac{\partial H}{\partial p_i} \dot{p}_i + \frac{\partial H}{\partial q^i} \dot{q}^i = 0. \qquad (S2.2.1.)$$

The Lie derivative in the direction of vector field $\mathrm{v}(x)$, the latter being assumed to generate the flow, can be treated as the corresponding evolution semigroup, with the underlying flow $g_t$ (or, rather, the translation semiflow $g_t^+ x(t) = x(t + s)$ on halfline $\mathbb{R}^+$) to be given by the parameters of a dynamical system driving the evolution. Equation (S2.2.1.) expresses the conservation law (for Hamiltonian flow) of the kind described in section 6: each Hamiltonian system is characterized by a Hamiltonian function of symplectic (see below) variables $\mathrm{p}, \mathrm{q}$ which is constant along the system's solutions. This equation is usually formulated in more general terms as a particular instance of the property of symplectic forms to preserve certain geometric structures. The physical meaning of (.) is that if the Hamiltonian does not depend explicitly on time, then its specific value – the mechanical energy – is constant along the system's phase path: $H(\mathrm{p}(t), \mathrm{q}(t)) = H(\mathrm{p}(0), \mathrm{q}(0)) = E = \mathrm{const}$ – energy is preserved under the flow. In other words, the Hamiltonian function provides the energy value along a solution $\mathrm{q}(t)$ i.e., all the trajectories (integral curves) $\gamma: [t_1, t_2] \to \mathbb{R}^n$ are restricted to the $(2n - 1)$-dimensional convex surfaces of constant energy $H = E$. In contrast, the Lagrangian primarily defines the action.

We can rewrite the Hamiltonian motion equations $\dot{q}^i = \partial H / \partial p_i = g^{ik} p_k$, $\dot{p}_i = -\partial H / \partial q^i = -\frac{1}{2} \big( \partial_i g^{jk}(q) \big) p_j p_k - \partial_i V$, $\partial_i \equiv \partial / \partial q^i$ for Hamiltonian $H(p, q) = \frac{1}{2} g^{ik} p_i p_k + V(q)$ through velocities $v^i := \dot{q}^i = g^{ik} p_k$ to obtain $\ddot{q}^i + \Gamma^i_{jk} \dot{q}^j \dot{q}^k = -\partial_i V$, where $\Gamma^i_{jk} = \frac{1}{2} g^{im} \big( \partial_j g_{mk} + \partial_k g_{mj} - \partial_m g_{jk} \big)$ are the Christoffel symbols. When $\partial_i V = 0$ (a free particle), we get the geodesic equation $\ddot{q}^i + \Gamma^i_{jk} \dot{q}^j \dot{q}^k = 0$ (see formulas (S2.1.1.2.) and (S2.2.1.)).

In the non-autonomous case, when the Hamiltonian depends explicitly on time[120], this constant in general does not exist so that there is no conserved energy (we have already mentioned the non-autonomous case when discussing the evolution operator and the Liouville phase fluid). Thus, one should not always identify Hamiltonian and conservative systems: although in the autonomous case Hamiltonian systems always conserve energy, one can find non-Hamiltonian dynamical systems that would preserve a similar quantity (first integrals of dynamical systems). In other words, all autonomous Hamiltonian systems are conservative, but not all conservative systems are necessarily Hamiltonian. Moreover, as already mentioned, non-autonomous Hamiltonian systems do not in general preserve energy. In modeling long-term processes, Hamiltonian systems are seldom adequate because of fluctuations and dissipative effects, but there is no universal criterion of adequateness for a concrete model: its usefulness is context dependent.

In case the Hamiltonian (or the Lagrangian) depends explicitly on time, it means that the physical system is either placed into an external time-varying field or not closed (i.e., can exchange energy and matter with the environment). The very term "non-autonomous" implies that the system is controlled by external forces. When external influence is not explicitly considered, one can figuratively say that a non-autonomous system "breathes" i.e., its vector field $X$ depends on time $X = X(t)$. Recall that the Hamiltonian $H = H(\mathrm{p}, \mathrm{q})$ of an autonomous system is in most cases defined as a real function on phase space $T^*Q$ which is the cotangent bundle of configuration space $Q$. One of the popular methods to consider the non-autonomous case is to extend the Hamiltonian vector field $X$ on $T^*Q$ to vector field $\tilde{X} = (X, 1)$ on $T^*Q \times \mathbb{R}$ (see, e.g., [16]). We can see that time $t$ plays a dual role in the non-autonomous case. On the one hand, $t$ runs over the base manifold for the affine event space (more exactly, a fibre bundle over $\mathbb{R}$) and, thus, is the variable of integration in the least action principle. On the other hand, $t$ is a rightful argument in the Hamiltonian (or the Lagrangian) function which is on equal footing with cotangent (respectively tangent) bundle coordinates. By fixing $t$ we get instantaneous values of these functions which, however, do vary over the phase space. Accordingly, while the evolution of an autonomous system treated as a group or semigroup mapping only depends on elapsed time $\tau = t - t_0$, in the non-autonomous case both the starting instant $t_0$ and the current moment $t$ are important parameters of the problem.

---

[120] Non-autonomous Hamiltonian systems are sometimes referred to as having $n + 1/2$ degrees of freedom.

The minus sign in the Hamiltonian (canonical) equations is very important. It signifies that Hamiltonian systems are natural dynamical systems in the so-called symplectic geometry. Thus, to study the Hamiltonian version of mechanics, one usually needs some elementary notions from this geometry. The term "symplectic" stems from the Greek equivalent of the Latin word "complex", and this linguistic parallelism is not accidental (see below). Symplectic geometry characterizes the phase space of classical mechanics, where the canonically conjugated variables are met pairwise. Recall that the phase space of classical mechanics is a particular case of a state space, when the number of variables is even (typically $6N$, where $N$ is the number of particles) and can be broken into pairs in which one variable is proportional to the time derivative of the other, e.g., $x^i, p_i = m\dot{x}^i$ as in Newtonian mechanics. For example, the phase space of a free particle moving in $\mathbb{R}^3$ is $\mathbb{R}^6$ whose points are $(x^1, x^2, x^3, p_1, p_2, p_3)$. One can, however, notice that outside mechanics the state space variables are not necessarily pairwise related; this is one of the reasons that we mostly use the term "state space" instead of "phase space" throughout this book.

Now, we can sum up the main geometric differences between the Lagrangian and the Hamiltonian approaches. In Lagrangian mechanics, we operate in the configuration manifold $q \in Q$, e.g., $Q = (\mathbb{R}^n)^N$, $n$ is the dimensionality of the problem, typically $n = 1,2,3$, $TQ$ is its tangent bundle. The Lagrangian function $L: TQ \times \mathbb{R} \to \mathbb{R}$ generates the action functional $S: TQ \times [a, b] \to \mathbb{R}$, and the variational (Hamilton's) principle, $\delta S = 0$, $\delta \mathrm{q}(a) = \delta \mathrm{q}(b) = 0$ results in motion equations. In the Hamiltonian picture, one works in the phase space $P = T^*Q$, which is the cotangent bundle of $Q$, employing the Hamiltonian function,

$$H: T^*Q \times \mathbb{R}, \qquad H(q^i, p_k, t) = p_j v^j(q^i, p_k, t) - L(q^i, v^k(q^i, p_k, t), t), \qquad (S2.2.2.)$$

where function $v^j(q^i, p_k, t)$ is produced by solving the equation for conjugate momenta $p_k = \partial L(q^i, v^j, t)/\partial v^k$ corresponding to generalized velocities $\dot{q}^k$. Relationship (..) shows that the Hamiltonian is the Legendre transform of the Lagrangian; accordingly, we can write the action functional (for autonomous systems) in the form [40]

$$S = \int_{t_1}^{t_2} dt \left[ p_i(t)\dot{q}^i(t) - H(q^i(t), p_k(t)) \right],$$

and the principle of the least action in Hamiltonian form can be represented as

$$\delta S = \int_{t_1}^{t_2} dt \left[ p_i(t) \dot{q}^i(t) - H(q^i(t), p_k(t)) \right] = 0.$$

This form is convenient to use, e.g., in the path integral formulation of classical mechanics [40].

Transform (..) is feasible when the Hessian matrix of the Lagrangian $H_{ij} \equiv \partial^2 L(q^i, v^j, t)/\partial v^j \partial v^k$ is nonsingular i.e., $\det H_{ij} \neq 0$, which is usually guaranteed in classical mechanics since the mass matrix $M_{ij} = H_{ij}$ is non-degenerate; in such cases the Lagrangian is called regular. Physically speaking, classical mechanics in distinction to electrodynamics and quantum field theory, only considers particles with non-zero mass. We have already remarked that the transition between Lagrangian and Hamiltonian dynamics (given by the Legendre transform) is one of the most prominent examples of natural dualities, the latter being often provided by physics.

Hamiltonian dynamics is evolved on symplectic manifolds $(P, \Omega)$ i.e., differentiable manifolds equipped with a canonical symplectic two-form $\Omega = dq^i \wedge dp_i$. Notice that the Hamiltonian picture implies the concept of a propagator, $x(t) = g_t(x_0) \equiv g_t x_0, x(t) \equiv (q(t), p(t))$. Let us illustrate mechanical evolution looking for a formal solution of Hamiltonian equations for an $N$-particle system. The values of position and momentum of each ($\alpha$-th) particle at time $t, x_\alpha(t) \equiv (q_\alpha(t), p_\alpha(t)), \alpha = 1, \dots, N$ are defined in a deterministic dynamical system by the initial values of coordinates and momenta of all $N$ particles i.e., by the initial position of the corresponding phase point in the $n = 6N$ dimensional $\Gamma$-space, $x(0) = (x_1(0), \dots, x_N(0))$ i.e., by the flow $x_\alpha(t) = \varphi_\alpha(t, x(0)) = g_t^{(\alpha)} x(0)$, where functions $\varphi_\alpha(t, x(0)) = (q_\alpha(t, q(0), p(0)), p_\alpha(t, q(0), p(0)))$ obey the system of Hamiltonian equations

$$\dot{p}_\alpha(t, x(0)) = -\frac{\partial H}{\partial q^\alpha}, \dot{q}_\alpha(t, x(0)) = \frac{\partial H}{\partial p_\alpha}. \qquad (S2.2.3.)$$

In simplified notations, Hamiltonian equations describe the evolution of vector field $X, \dot{z} = X(z)$. In the Hamiltonian picture, dynamical evolution is completely determined by Hamiltonian $H: P \to \mathbb{R}, H = (p, q, t)$. Accordingly, a vector field $v := X_H$ that we denoted as $X$ is defined on a phase space through Hamiltonian equations, $X_H = \left( \frac{\partial H}{\partial p_\alpha}, -\frac{\partial H}{\partial q^\beta} \right), \alpha, \beta = 1, \dots, n$.

We emphasize that indices $\alpha, \beta$ enumerate the particles i.e., is not in general a vector component index. In local coordinates $q^\alpha, p_\beta, \alpha, \beta = 1, \dots, n$ the Hamiltonian equations can be represented as a dynamical system

$$\begin{pmatrix} \dot{q}^1 \\ \vdots \\ \dot{q}^n \\ \dot{p}_1 \\ \vdots \\ \dot{p}_n \end{pmatrix} = \begin{pmatrix} \partial H/\partial p_1 \\ \vdots \\ \partial H/\partial p_n \\ \partial H/\partial q^1 \\ \vdots \\ \partial H/\partial q^n \end{pmatrix}.$$

Using the Poisson bracket

$$\{F(x), G(x)\} \equiv \sum_{\alpha=1}^{N} \left( \frac{\partial F}{\partial q^\alpha} \frac{\partial G}{\partial p_\alpha} - \frac{\partial F}{\partial p_\alpha} \frac{\partial G}{\partial q^\alpha} \right),$$

we can write Hamiltonian equations in the form $\dot{x}_\alpha(t) = \{x_\alpha, H(x)\}$. The main advantage of using the algebra of Poisson brackets is that they are invariant under canonical transformations i.e., $\{x_\alpha, H(x)\} = \{x_\alpha(t, x(0)), H(x(0))\}$ so that one can write Hamiltonian equations as $\dot{x}_\alpha(t, x(0)) = \{x_\alpha(t, x(0)), H(x(0))\}$. This is the Hamiltonian flow defining the evolution of some arbitrary initial state $x(0)$. Since all differential operations in Hamiltonian equations (...) represented through the Poisson bracket are performed over initial variables $x(0)$ in which the initial time point can be arbitrarily chosen so that we may write $x(0) = x$, the formal solution for the Hamiltonian flow is $x_\alpha(t, x) = g_t^N x_\alpha$, where the evolution operator $U(t, t_0 = 0)$ for an $N$-particle system

$$U(t, t_0 = 0) = g_t^N = \exp(it A_N), A_N = -i \sum_{\alpha=1}^{N} \left( \frac{\partial H}{\partial p_\alpha} \frac{\partial}{\partial q_\alpha} - \frac{\partial H}{\partial q_\alpha} \frac{\partial}{\partial p_\alpha} \right)$$

is a self-adjoint operator on the phase space known as the Poisson bracket operator. If $t_0$ is kept fixed (we have put $t_0 = 0$) and $t$ is arbitrarily varied, then all the phase points $x_0 \equiv x(0)$ go over to $x(t)$ under the action of evolution operator $U(t, t_0)$. In deterministic dynamical, in particular, Hamiltonian systems, points $(x, x_0)$ uniquely determine each other (see the example of harmonic oscillator) so that the entire phase space $\Gamma$ uniquely (and smoothly) transforms into itself. Furthermore, since this transformation for an isolated system does not depend on the choice of initial point $x_0$ in the phase space, but only on the time difference $t - t_0$ i.e., $U(t, t_0) = U(t - t_0)$ the Hamiltonian flow is stationary – velocities of the phase points do not change with time, and the phase fluid is incompressible, see also sections 5.5.("Symmetry") and 7.9. ("The Liouville phase fluid").

Recall that the flow representation $x(t, x(0)) g_t x(0)$ holds for any dynamical system and not only for Hamiltonian ones (see section 8.4.2. "Flows and vector fields"). Moreover, one can write the Hamiltonian flow for any function $\Phi$ of phase variables $x_i$, $\Phi\left(x_1(t, x_1(0)), ..., x_N(t, x_N)\right) = g_t^N \Phi\left(x_1(0), ..., x_N(0)\right)$. From the physical viewpoint, the most important function on the phase space of a mechanical system is its total energy $H = H(x_1, ... x_N) = H(q_1, p_1, ..., q_N, p_N)$. One can see that Hamiltonian $H$ is the generator of time translation so that the canonical time translation operator is $e^{\tau H}$, where $\tau$ is the translation parameter (strictly speaking, infinitesimal). For the model of closed (isolated) system, this function has a constant value i.e., is an integral of the system of motion equations for $N$ particles. Therefore, for any constant $E$ the domain of phase space defined by constraint $H(x_1, ..., x_N) = E$ is an invariant manifold in the phase space usually known in physics as a surface of constant energy (see also section 8.4.1. "Phase space and phase volume"). More generally, we see that variable $f$ is preserved by the flow if it does not depend explicitly on time and its Poisson bracket with Hamiltonian $H$ is zero. There is one more possibility, namely quantity $f$ might be preseved even if it depends explicitly on time when there is an exact compensation of the Poisson brackets in $A_N$ and the rate of change is $-\partial f / \partial t$.

The Hamiltonian theory of dynamical evolution lies at the foundation of statistical mechanics, where a probabilistic treatment of dynamical processes is adopted. The canonical time translation operation performed on functions $f(p, q)$ defined on the phase space is

$$f(\tau) \equiv f\left(p(\tau), q(\tau)\right) = e^{\tau H} f\left(p(0), q(0)\right) \equiv e^{\tau H} f(0),$$

where the operator exponent is defined, as usual, through power series so that

$$f\big(p(\tau), q(\tau)\big) = f(0) + \tau[H, f(0)] + \frac{\tau^2}{2!}\big[H, [H, f(0)]\big] + \cdots$$

To finalize this section, we can emphasize the link between the general theory of dynamical systems and classical mechanics describing the motion of material bodies. Namely, one can show (see, e.g., [16, 93, 47]) that any dynamical system $dx^i/dt = v^i(x^j)$ admits, at least not in the vicinity of its critical points, a symplectic structure (i.e., Hamiltonian vector field) $\omega_{ij}(x), x = (x^1, \dots, x^n)$ so that one can write the dynamical system as $v^i\omega_{ij}(x) = \partial_j H(x)$, where $H(x)$ is the respective Hamiltonian function.

## Supplement 3. The Navier-Stokes system

The Navier-Stokes equations are in fact a version of Newtonian laws of motion supplemented by the continuity equation that expresses the conservation of mass, a mathematical tool to describe the motion of real (viscous) fluids. Amazingly, except for rather simple situations (see, e.g., [92]), no general solution to these equations is known so far so that the Navier-Stokes equations have to be solved by using numerical discretization schemes and often with the help of supercomputers [71] and [127]. The emerging set of coupled algebraic equations related to the nodes of numerical mesh covering the flow domain can be solved by linear algebra techniques to produce discrete values of, e.g., pressure and velocity at nodes [127]. Other relevant variables such as density, temperature, concentration of species, inclusion of other phases (solid particles, liquid droplets, bubbles, submerged jets and other multiphase admixtures) can be computed by using the linked transport equations. Yet the evolution of a given state of the fluid (at time instant $t = t_0$) obtained by some approximate methods from the Navier-Stokes equations does not produce a reliable forecast of the future behavior of the fluid, in contrast with the relative predictability of the motion of rigid bodies or point particles. We can observe the detrimental effects of the fluid motion unpredictability through such phenomena as freak ocean waves, tsunamis, tornadoes, etc. The Clay Mathematics Institute even formulated the problem of existence and analytical properties (such as smoothness) of solutions to the Navier-Stokes equation as one of the Millennium Prize Problems.

The $n$-dimensional Navier-Stokes equation for incompressible fluid can be written (in local coordinates) in the form

$$\partial_t u^i + u^k \partial_k u^i = \nu \Delta u^i - \partial^i p + f^i(\mathrm{x}, t), i = 1, \dots, n, \mathrm{x} = (x^1, \dots, x^n)^T, \qquad \text{(S3.1)}$$

containing $n + 1$ unknown functions $\mathrm{u}(\mathrm{x}, t) = (u^1(\mathrm{x}, t), \dots, u^n(\mathrm{x}, t)), p = p(\mathrm{x}, t)$ and parameter $\nu$ (the viscosity coefficient). Vector function $\mathrm{f}(\mathrm{x}, t) = (f^1(\mathrm{x}, t), \dots, f^n(\mathrm{x}, t))$ denotes local external forces. Notice that if $\mathrm{f} = 0$, then the system becomes mainly of a kinematic character. Vector variable f belongs to some manifold $M$ which may, e.g., coincide with $\mathbb{R}^n$ or be a compact domain $D$ with a boundary $\partial D$, not necessarily smooth. Here the units are so chosen as to assume density $\rho = 1$. This vector equation is in many cases supplemented by the incompressibility condition div $\mathrm{u} = \partial_i u^i = 0$ to obtain the Navier-Stokes system. However, the latter assumption becomes invalid for supersonic regimes. In many mathematically oriented papers, the viscosity coefficient is also put to unity, $\nu = 1$; we shall, however, retain this coefficient as a control parameter which is usually known as the inverse Reynolds number 1/Re. An interesting case is when $\nu \to 0$ (Re $\to \infty$). The Navier-Stokes system can be supplied by boundary conditions on $\partial D$, e.g., $\mathrm{u}(\mathrm{x}, t)|_{\partial D} = 0$ that are known as non-slip boundary conditions.

There exist many ways to derive the Navier-Stokes system of equations. We shall use one of the simplest of those based on physical considerations and used in [92] (§15) and [154]. This method, apart from its simplicity, allows one to evade considering strain rates and time derivatives and, correspondingly, the question what time derivative to use: partial $\frac{\partial \mathrm{u}(\mathrm{x},t)}{\partial t}$ or convective

$$\frac{D\mathrm{u}(\mathrm{x},t)}{Dt} = \frac{\partial \mathrm{u}(\mathrm{x},t)}{\partial t} + (\mathrm{v}\nabla)\mathrm{u}(\mathrm{x},t),$$

where $\nabla\mathrm{u}$ denotes the covariant derivative of vector field u, $\nabla_j u^i = \partial_j u^i + \Gamma^i_{jk} u^k$. We know that the class of practical applications when the effects of viscosity are negligible (perfect fluid), the equation of fluid motion reduces to Euler's equation. One can assume that turning on the viscosity leads to the inclusion of new terms in the motion (Euler's) equation that would come out only additively i.e., with no renormalization of already existing parameters. The Euler equation for perfect fluid motion is

$$\partial_t u^i + u^k \partial_k u^i = -\frac{1}{\rho}\nabla^i p + f^i(\mathrm{x},t), i = 1, \dots, n, \mathrm{x} = (x^1, \dots, x^n)^T, \nabla^i = g^{ij}\nabla_j.$$

In simple cases (inertial systems, rectangular and orthogonal coordinates – in general, when the connection coefficients $\Gamma^i_{jk} = 0$), $\nabla_j = \partial_j$ so that the motion equation for a perfect fluid is

$$\partial_t u^i + u^k \partial_k u^i = -\frac{1}{\rho} g^{ij} \partial_j p + f^i(\mathrm{x},t).$$

Here, vector $f^i$ denotes the density of external volume forces acting on a unit mass. Using the continuity equation

$$\partial_t \rho + \partial_k (\rho u^k) = 0,$$

we can write Euler's equation in the form

$$\partial_t (\rho u^i) = -\partial_k (T^{ik}),$$

where $T^{ik} = p\delta^{ik} + \rho u^i u^k$ is the momentum flux density. Tensor $T^{ik}$ describes momentum transfer in an inviscid ("dry") fluid, which is due to purely mechanical i.e., reversible motion of fluid elements under internal pressure forces. However, this dissipationless momentum transfer, in more realistic cases of a viscous ("wet") fluid, should be supplemented by the processes of dissipative exchange of momentum and energy. As soon as macroscopic fluid elements (i.e., containing many molecules) start moving with respect to one another, irreversibly exchanging energy and momentum, additional terms must appear in the motion equation, accounting for entropy increase or production in the fluid. Microscopically, stress tensor $T^{ik}$ can be represented through the one-particle distribution function $f = f(\mathrm{r}, \mathrm{u}, t)$ as $T^{ik} = \int d^3 p \, u^i u^k f = \int d^3 p \, u^i u^k (f_0 + f_1) \equiv \int d^3 p \, u^i u^k f_0 (1 + \varphi) = \rho u^i u^k - \sigma^{ik}$, where $f_0$ is some equilibrium distribution function (e.g., Maxwell's distribution for gas, $f_0(\mathrm{p}) = N/(2\pi m T)^{\frac{3}{2}} \exp\left(-\frac{p^2}{2mT}\right)$), $f_1 \equiv f_0 \varphi$ is a correction to the distribution function, either the equilibrium or the non-equilibrium one, and $\sigma^{ik} = -p\delta^{ik} + \tilde{\sigma}^{ik}$. Tensor $\tilde{\sigma}^{ik}$ is known as a viscous tensor. It is the presence of tensor $\tilde{\sigma}^{ik}$ that is due to irreversible corrections $\varphi$ to the equilibrium distribution function $f_0$, thus leading to the dissipative Navier-Stokes equation of fluid motion.

The majority of derivations of the Navier-Stokes equations are centered around choosing an appropriate form of viscous tensor $\tilde{\sigma}^{ik}$. Certain physical arguments (see, e.g., [92], §15) lead to the following generic form of the viscous tensor:

$$\tilde{\sigma}^{ik} = \eta\left(\partial^i u^k + \partial^k u^i - \frac{2}{3}\delta^{ik}\nabla u\right) + \zeta\delta^{ik}\nabla u = \eta\left(g^{ij}\partial_j u^k + g^{jk}\partial_j u^i - \frac{2}{3}\delta^{ik}\nabla u\right) + \zeta\delta^{ik}\nabla u,$$

where parameters $\eta$ and $\zeta$ are real and positive.

The motion equation of a viscous fluid is a generalization of Euler's equation to the dissipative one by additively including viscous stresses $\partial_k\tilde{\sigma}^{ik}$ i.e.,

$$\rho(\partial_t u^i + (u^k\partial_k)u^i) = -g^{ij}\partial_j p + \partial_k\tilde{\sigma}^{ik} + f^i(x,t)$$
$$= -g^{ij}\partial_j p + \partial_k\left[\eta\left(g^{ij}\partial_j u^k + g^{jk}\partial_j u^i - \frac{2}{3}\delta^{ik}\nabla u\right)\right] + \delta^{ik}\partial_k(\zeta\nabla u) + f^i(x,t)$$

If coefficients $\eta$ and $\zeta$ can be assumed nearly constant within the fluid so that $\partial_i\eta = \partial_i\zeta \approx 0$, then we have for the first term in the square brackets: $\eta g^{ij}\partial_j(\partial_k u^k)$, and for an incompressible fluid this term reduces to zero. Then we arrive at the Navier-Stokes system (S3.1).

## Supplement 4. Scientific computing and numerical modeling

One might ask: "Why should we invent and study theoretical models when we can have enormous computing power through HPC (high performance computing) with the current processing speed over 30 petaFLOPS (PFLOPS) and rapidly increasing? Can't we get all we need from direct numerical simulations? Isn't theoretical modeling a hopelessly old-fashioned approach?" The answers to such questions of the growing number of enthusiastic advocates of purely numerical simulations can be reduced to a simple fact that one cannot get anything out of computations without simple models based on subject matter. Furthermore, it is the theory and especially simple models that give meaning to numerical results. Computer modeling provides practical implementation of theoretical structures developed in mathematical modeling *per se*. Computer implementation of mathematical models usually follows the archetypal scheme: world → theory → model → algorithm → program, these phases reflecting the progressive abstraction more and more ignoring details. Using simple numerical algorithms and mathematical manipulators such as Mathematica or Maple, one can visualize the equations used for modeling, making the results more intuitive.

The unhappy outcome of numerical modeling can be the accumulation of a great amount of data accompanied by progressively worsening understanding, even despite beautiful pictures that modern visualization software can produce. Moreover, even despite increasing computational power, the gap between numerical modeling and understanding may widen.

Newton's scheme to solve equation $f(x) = 0$: $x = x^* = \lim_{n\to\infty} x_n$, $f(x_n) + Df(x_n)(x_{n+1} - x_n) = 0$, where $Df(x_n)$ denotes a derivative of function $f$ at point $x_n$ i.e., $D \equiv d/dx$. This scheme is based on a trivial geometric identity, yet it produces an extremely fast convergence: $\mathcal{O}(\epsilon^{2^n})$, where $\epsilon = |x - x^*|$ is the error.

### S.4.1. Object-oriented modeling

A fair illustration of focusing only on relevant aspects is the object-oriented approach to implementing models in software, when the world is fully described through objects that are defined by their states and behavior. In this – quite simplistic – approach, everything is viewed as an object: a number, a document, an individual, an electronic device, a car, a piece of machinery, a bank – you name it. Software developers treat such program components as menus, windows, icons, text boxes, buttons, scroll bars, database records, etc. as objects i.e., from a programmer's standpoint, objects are self-contained miniprograms embedded into the main code. An object is defined through its state and behavior: an object's state is specified by *fields* where the data components (they say what an object is) are contained whereas an object's behavior is specified by *methods* (they say what an object does). Due to these two state characteristics, an object is distinguishable from other objects i.e., it has an identity. Moreover, an object is in a certain sense stable: its state can only be changed due to interaction with other objects; one usually says that objects are "encapsulated".

One can naturally generalize the notion of object: the latter is an instance of a class which in practice means that different objects of the same class have similar fields and methods, the values of the fields i.e., specific data in general differ (for example, the mythological gods are described by different features). All humans have hair or skin color, but the hair or skin color of a given individual can be different from that of another.

Object-oriented design (OOD) is a powerful tool to convert real-world models into computer codes and produced much enthusiasm when it first appeared. Indeed, there are many substantial benefits in object-oriented modeling such as a fair mapping of the real world or a natural structure to ensure a modular design. OOD has recently gained wide acceptance in such important areas as engineering design and CAD/CAM. However, despite all the hype around object-orientation and, in particular, object-oriented programming (OOP), this approach is not free from essential limitations. For instance, arithmetic expressions are usually not processed optimally by the compilers of OOP languages so that their use to solve numerical problems may be inferior to using procedural languages such as FORTRAN and C. Furthermore, memory management may become an issue in object-oriented computer modeling since due to their very nature (separate and encapsulated) objects are typically placed at different main memory locations. Thus, the performance may severely suffer when countless small objects scattered over the entire memory ought to be addressed and accessed by the program written in the OOP language during its execution. Claims of universality of object-oriented modeling are mostly of ideological, not of technical character.

In connection with object-orientation, we can mention the so-called Unified Modeling Language (UML) that helps to specify, visualize and document the models of software systems. The UML is mostly based on a diagram technique (class diagrams, use case diagrams, sequence diagrams, package diagrams and so on – overall 14 types). Some software engineering proponents view the value of UML-based modeling in the capability to design software before coding (but not instead of coding). The UML enables one to standardize the process of visual modeling in the fields of class or object modeling, business process modeling and use cases, component modeling, interface engineering, reengineering, greenfield (from scratch), distribution and deployment of engineering systems, etc. See the details in http://www.uml.org/ and http://www.omg.org/.

### S.4.2. General algorithmic approach

In order to properly learn numerical techniques, one has to establish their connection with analytical mathematics. This is usually achieved by understanding the construction of algorithms, the latter serving as a bridge between mathematics and computing. For instance, one must know how to solve operator or matrix equations, e.g., $Au = f$, where $u$ is an unknown function and $f$ is an external (source) function into which operator $A$ takes function $u$. One of the possible ways to cope with this equation is based on introducing some parameter $\mu \neq 0$ and a new operator $B_\mu := I - \mu^{-1}A$. Then $Au = \mu Iu - \mu B_\mu u = f$ so that $u - B_\mu u = \mu^{-1}f$. This equation can be solved by iterations (see the next subsection).

**Internet resources**

# Finite symmetry transformations: decorations

- We can decorate a square:
- This figure still retains the full symmetry of the square

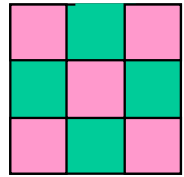

The figure below does not have the symmetry of a square:

The square is symmetric with respect to rotation about the center, reflection in diagonals and bisectors. The square is also symmetric with respect to an inversion through its center

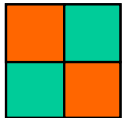

- this figure does not transform into itself after a rotation by $\pi/4$ about ist center

- We can extend the symmetry by introducing a more complicated transform: first rotation, then flip – turning orange into green and vice versa. Practical example: textile production (H.J.Woods, 1930 – "black and white groups" and "braids"); generalization – polychromatic groups; crystallography – Shubnikov groups; a physical example – magnetism (spin flip), L.D.Landau

Mathematical and Computer
Modeling in Science and Engineering

TUΠ

# Catastrophic behavior

- Catastrophe theory proved useful in applied science, e.g. biology, chemistry, macroscopic and laser physics, optics, even sociology – in those fields where bifurcations are observed. In such cases, catastrophe theory is an adequate language to describe nonlinearity

- Example (Whitney cusp): compare a smooth map of a plane ( $x_1, x_2$)
$$y_1 = x_1^3 + x_1 x_2, y_2 = x_2 \quad \text{with} \quad y_1 = x_1^2 - x_2^2, y_2 = 2x_1 x_2$$

The first one is structurally stable, the second structurally unstable (*structural stability* means that any perturbed map has the same singular points, at least locally – here near (0,0)):

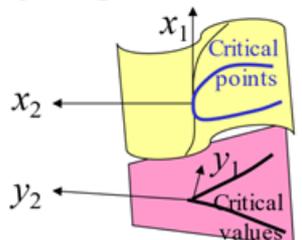

$$y_1 = x_1^2 - x_2^2 + \alpha x_1, y_2 = 2x_1 x_2 - \alpha x_2$$

Equation for critical points:
In this case crit. points
form a circle $x_1^2 + x_2^2 = \dfrac{\alpha^2}{4}$
$$\begin{vmatrix} \dfrac{\partial y_1}{\partial x_1} & \dfrac{\partial y_1}{\partial x_2} \\ \dfrac{\partial y_2}{\partial x_1} & \dfrac{\partial y_2}{\partial x_2} \end{vmatrix} = 0$$

Mathematical and Computer
Modeling in Science and Engineering

TUM    41

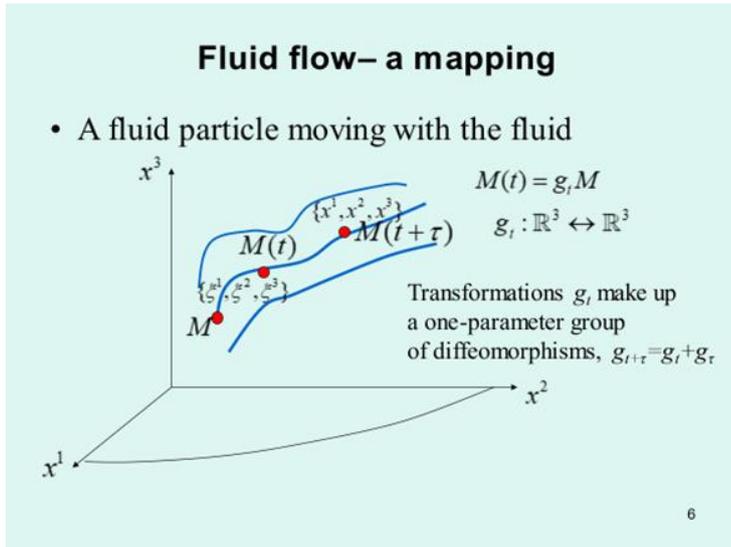

# Physical interpretation of the Euler and Lagrange descriptions

• Different physical interpretations

  <u>In the Euler's picture</u>, an observer is always located at a given position $\{x^i\}$ at a time t → observes fluid particles passing by

  In the Lagrange picture, an observer is moving with the fluid particle, which was initially at the position $\{\xi^i\}$ → views changes in the flow co-moving with the observer's particle

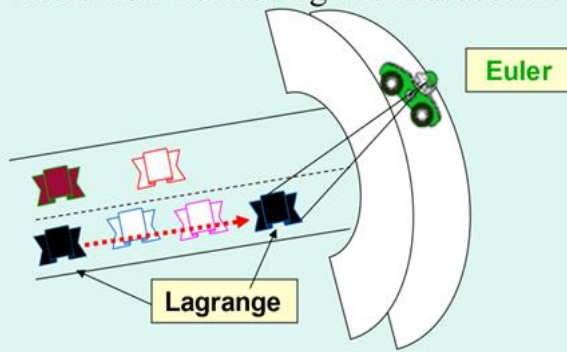

Euler

The Euler description – important for CFD;
The Lagrange description – important for ecological modeling (passive tracers, contamination spread, etc.)

Lagrange

Figure 5: The velocity field

# The velocity field

- Usually it is the fluid velocity $u^i(x^j, t)$ that is a measurable quantity. The velocity identifies a fluid particle (in terms of its initial position $\{\xi^i\}$)

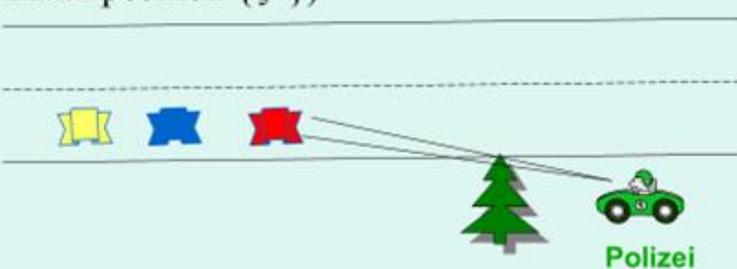

An observer is sitting in a position $\{x^i\}$ to measure the speed of moving particles (initially were at $\{\xi^i\}$)

Polizei

- The fluid velocity is a vector field (transformation properties!). Throughout all time, a set of velocity components $\{u^i(x^j, t)\}$ remains attached to a fixed position – the field variables

- <u>In numerics or in experiment:</u> a fluid velocity $u^i$ ($1d$, $2d$, $3d$, $4d$) for the time interval $\Delta t \rightarrow$ to determine $x^i \approx x^i(\xi^j, t)$, $\Delta x^i \approx u^i \Delta t$

Figure 6: Flow past an obstacle. Chaotization of fluid flow with increasing Reynolds number

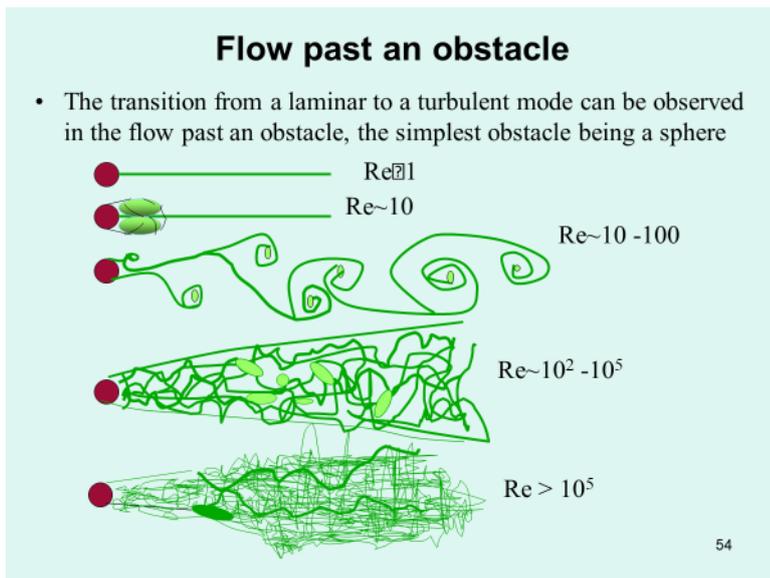

# Galilean structure

- The Galilei's space-time structure serves as fundamental symmetry of the classical world. It defines the class of so called inertial systems and includes the following components:

1) Our world – a 4D affine space $\mathbb{R}^4$ whose elements are events (called world points). Parallel transport of the world $\mathbb{R}^4$ gives 4D linear space $\mathbb{R}^4$

2) The classical time is a linear mapping of parallel transport of the world on real time axis: t: $\mathbb{R}^4 \to \mathbb{R}$

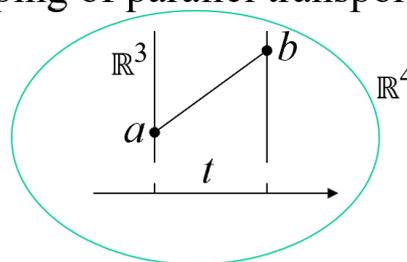

$a,b$ are events, $t(a,b)=t(b-a)$ is a time Interval between events. If $t(a,b)=0$, these events are simultaneous.

A set of simultaneous events constitutes 3D affine space $\mathbb{R}^3$

3) The distance between two simultaneous events: $\rho(a,b)=\|a\text{-}b\|$

A typical phase portrait showing the intermittency of fixed points in a single-dimensional dynamical system. Stable and unstable fixed points by necessity successively change each other. Here stable points are depicted as red and unstable ones as green bullets.

# Phase portraits of a classical system

- The equilibrium solution is represented by a point $x$=0,$v$=0

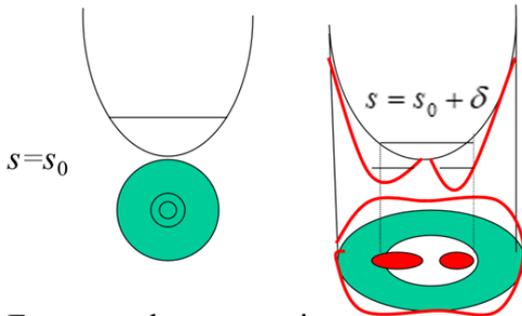

Equilibrium positions of the system are points ( $a$,$s_0$) with $\partial U(x,s)/\partial x\,|_{x=a}=0$
The gradient of the first integral $E(x,s)$ is $(\partial U/\partial x,0)$; it is zero only at equilibrium points.

For $s$=$s_0$ phase portraits are represented by ellipses, their size is growing with increased energy.

With the change of the parameter $s$, the form of the potential well is modified and new equilibrium positions appear. The equilibrium state ($x$=0,$v$=0, $s$=$s_0$) becomes unstable.

Mathematical and Computer
Modeling in Science and Engineering

TUM    200

## Figure 9: Hyperbolic equilibria and saddle points

<<Note to the publisher: This figure was taken from the Internet, so it needs to be redrawn before publication. https://en.wikipedia.org/wiki/Hyperbolic_equilibrium_point>>

Orbits near a two-dimensional saddle point, an example of a hyperbolic equilibrium.

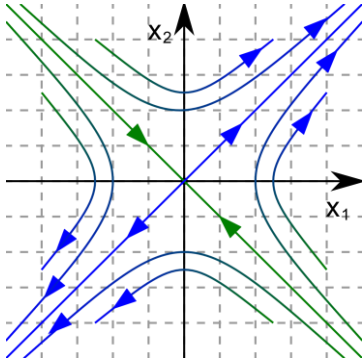

# The Lyapunov stability

- A solution $x=\varphi_0(t)$, $y=\psi_0(t)$ is called to have Lyapunov stability if for any $\varepsilon>0$ one can find $\delta=\delta(\varepsilon)>0$, so that for all solutions $x=\varphi(t)$, $y=\psi(t)$ satisfying $|\varphi_0(t_0)\text{-}\varphi(t_0)|<\delta$, $|\psi_0(t_0)\text{-}\psi_0(t_0)|<\delta$, for all $t>t_0$ holds: $|\varphi_0(t)\text{-}\varphi(t)|<\varepsilon$, $|\psi_0(t)\text{-}\psi(t)|<\varepsilon$

- If the solution possess the Lyapunov stability and if for small $\delta>0$ the following conditions are satisfied

$$\lim_{t\to\infty}\left|\varphi_0(t)-\varphi(t)\right|=0, \lim_{t\to\infty}\left|\psi_0(t)-\psi(t)\right|=0,$$

then the solution $\varphi_0(t)$, $\psi_0(t)$ is said to be asymptotically stable

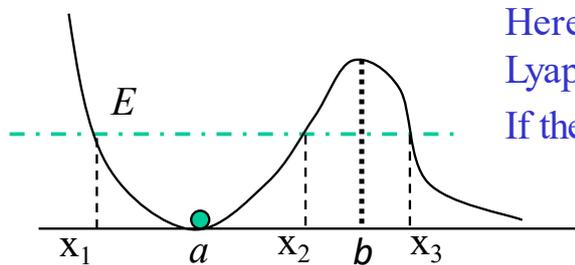

Here the equilibrium position $a$ is Lyapunov -stable if there is no friction.

If there is friction, then the ball motion will contract with time, i.e. the equilibrium $a$ is asymptotically stable. Equilibrium $b$ is unstable

Mathematical and Computer
Modeling in Science and Engineering

- Oscillations are finite motions that occur in the vicinity of equilibrium points
- If $a$ is a local minimum of the potential $U(x,s)$, then $x=a$ brings the Lyapunov stability, i.e. for initial conditions $\{p(0),x(0)\}$ sufficiently close to $\{0,a\}$ the whole phase trajectory $\{p(t),x(t)\}$ is close to $\{0,a\}$

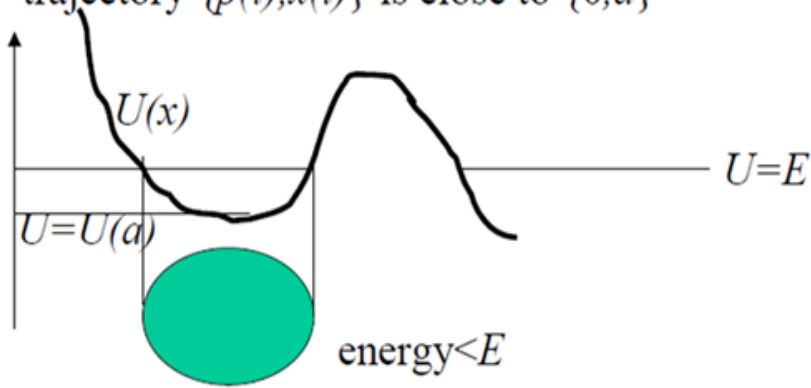

## Figure 12: One-dimensional harmonic oscillators

- There may be three cases for each one-dimensional system:

1. $\lambda = \omega^2 > 0$, the solution $Q = C_1 \cos \omega t + C_2 \sin \omega t$ - oscillations

2. $\lambda = 0$, the solution $Q = C_1 + C_2 t$ - "null space", $v$=const

3. $\lambda = -\beta^2 < 0$, the solution $Q = C_1 ch\beta t + C_2 sh\beta t$ - instability

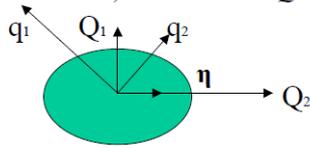

The system has $n$ natural modes alongside orthogonal direction $Q_k$. Any small oscillation may be represented as a sum of natural (normal) modes:

$$\mathbf{q}(t) = \mathrm{Re} \sum_{k=1}^{n} C_k e^{i\omega_k t} \boldsymbol{\eta}_k$$

## Figure 13: Burridge-Knopoff model

A model of frictional behavior built up by R. Burridge and L. Knopoff is intended to describe the response of sheer strain $x_j(t)$ to sheer stress. The latter is usually defined as the magnitude of forces $F_j$ divided by the area of the fault plane. The model consists of an array of slider blocks connected by strings. This simple model (the original Burridge-Knopoff model contained just two blocks) can help to understand the stop-and-go (stick-slip) motion typical of earthquakes. The dependence of friction forces $F_j$ on relative velocity of plates $v$ is of critical importance for the displacement behavior $x_j(t)$.

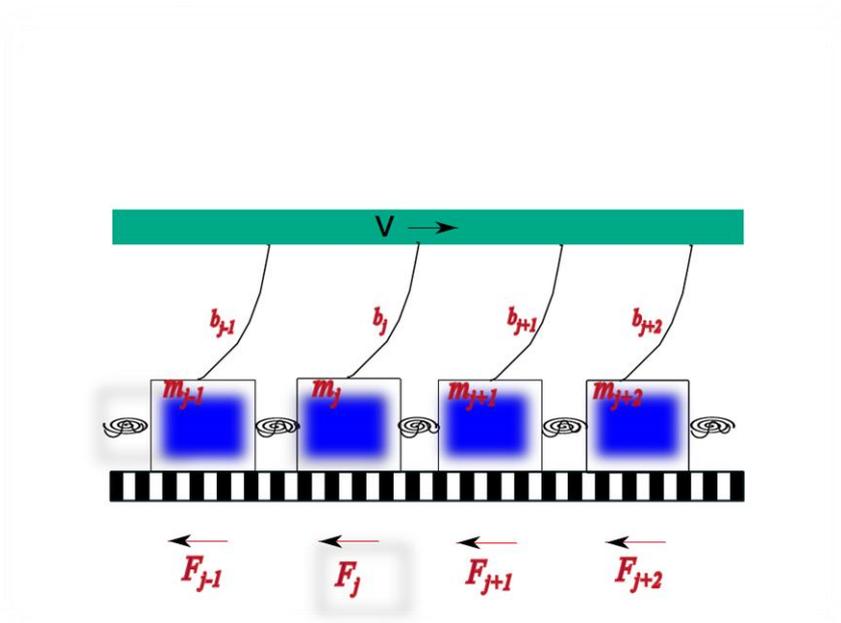

# Pendulum motion as a dynamical problem II

What happens when the pendulum total energy equals $mgl$?

• In this case $E=F$. Does the pendulum oscillate or rotate?

• The period of oscillations becomes infinite – motion over a separatrix (a curve that separates finite and infinite motion)

• The phase portrait contains two equilibrium points ($p=0$): one at the coordinate origin ($\theta = 0$), which is stable (elliptical) another at $\theta=\pm\pi$, which is unstable (hyperbolic)

• Phase trajectories look like ellipses in the neighborhood of an elliptic point and like hyperbolas near a hyperbolic one.

Pendulum as a dynamical system contains an infinite number of singular points: elliptic $p=0$, $\theta =2\pi n$ and hyperbolic: $p=0$, $\theta =(2n+1)\pi$

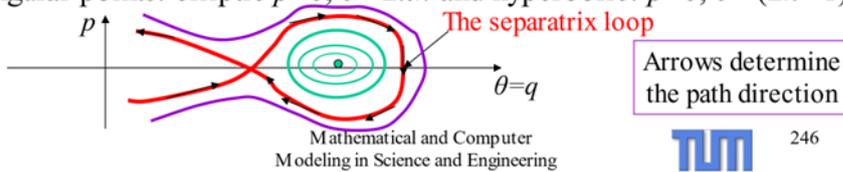

The separatrix loop

Arrows determine the path direction

Mathematical and Computer
Modeling in Science and Engineering

# Figure 15: Phase trajectories of the Lotka-Volterra model

Phase trajectories of the Lotka-Volterra model are concentric closed curves around the equilibrium point $(x_0, y_0) \equiv (\beta/q, \alpha/p)$. In the vicinity of this point, solutions of the Lotka-Volterra system are harmonic oscillations with frequency $\omega = (\alpha\beta)^{1/2}$. When the phase point moves away from the equilibrium position $(x_0, y_0)$, the nonlinearity becomes noticeable so that the oscillations are substantially different from harmonic: their frequency depends on amplitude.

<<Note to the publisher: This figure was taken from the Internet, so it needs to be redrawn before publication. https://en.wikipedia.org/wiki/Lotka%E2%80%93Volterra_equations>>

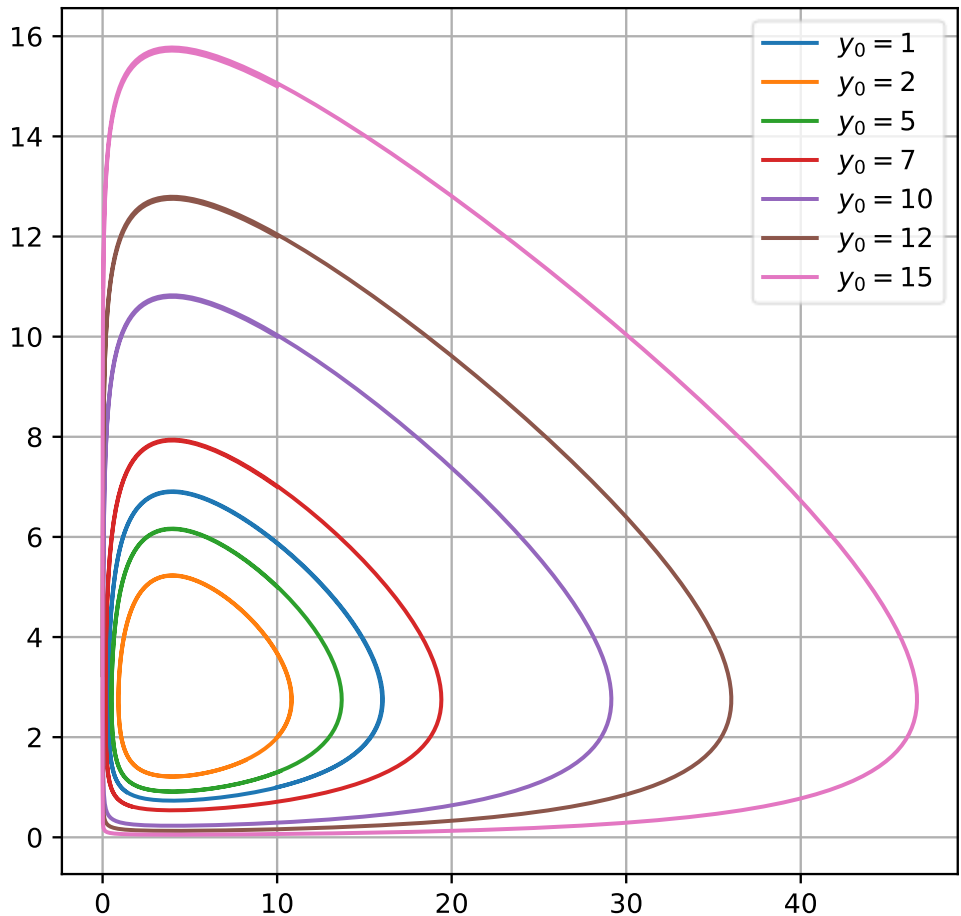